\pdfoutput=1
\documentclass[11pt,twoside,a4paper,cmspaper,final,collab]{cms-tdr}
\def\svnVersion{fde9e30}\def\svnDate{2025/05/21}\def\cmsCernNoTag{CERN-EP-2025-014}\def\cmsCernDate{\today}\def\cmsMessage{Published in Physics Letters B as \href{http://dx.doi.org/10.1016/j.physletb.2025.139556}{\doi{10.1016/j.physletb.2025.139556}.}}
\begin{document}\cmsNoteHeader{HIN-23-004}

\newcommand{\pp}{\ensuremath{\Pp\Pp}\xspace}
\newcommand{\PbPb}{\ensuremath{\text{PbPb}}\xspace}
\newcommand{\pPb}{\ensuremath{\Pp\text{Pb}}\xspace}
\newcommand{\ptjet}{\ensuremath{p_{\text{T,jet}}}\xspace}
\newcommand{\ptSuper}[1]{\ensuremath{\pt^{\text{#1}}}\xspace}
\newcommand{\ptCh}{\ptSuper{ch}}
\newcommand{\JEWEL}{\textsc{Jewel}\xspace}

\ifthenelse{\boolean{cms@external}}{\providecommand{\cmsLeftUpperCase}{Upper\xspace}}{\providecommand{\cmsLeftUpperCase}{Left\xspace}}
\ifthenelse{\boolean{cms@external}}{\providecommand{\cmsRightUpperCase}{Lower\xspace}}{\providecommand{\cmsRightUpperCase}{Right\xspace}}
\ifthenelse{\boolean{cms@external}}{\providecommand{\cmsLeft}{upper\xspace}}{\providecommand{\cmsLeft}{left\xspace}}
\ifthenelse{\boolean{cms@external}}{\providecommand{\cmsRight}{lower\xspace}}{\providecommand{\cmsRight}{right\xspace}}

\ifthenelse{\boolean{cms@external}}
{\providecommand{\suppMaterial}{the supplemental material, model comparisons~\cite{supplementaryMaterial}}}
{\providecommand{\suppMaterial}{Appendix~\ref{app:suppMat}}}

\cmsNoteHeader{HIN-23-004}
\title{Observation of nuclear modification of energy-energy correlators inside jets in heavy ion collisions}

\date{\today}

\abstract{
   Energy-energy correlators are constructed by averaging the number of charged particle pairs within jets, weighted by the product of their transverse momenta, as a function of the angular separation of the particles within a pair. They are sensitive to a multitude of perturbative and nonperturbative quantum chromodynamics phenomena in high-energy particle collisions. Using lead-lead data recorded with the CMS detector, energy-energy correlators inside high transverse momentum jets are measured in heavy ion collisions for the first time. The data are obtained at a nucleon-nucleon center-of-mass energy of 5.02\TeV and correspond to an integrated luminosity of 1.70\nbinv. A similar analysis is done for proton-proton collisions at the same center-of-mass energy to establish a reference. The ratio of lead-lead to proton-proton energy-energy correlators reveals significant jet substructure modifications in the quark-gluon plasma. The results are compared to different models that incorporate either color coherence or medium response effects, where the two effects predict similar substructure modifications.
}

\hypersetup{
pdfauthor={CMS Collaboration},
pdftitle={Energy-energy correlators from PbPb and pp collisions at 5.02 TeV},
pdfsubject={CMS},
pdfkeywords={CMS, jets, energy, resolution, correlator}}

\maketitle 

\section{Introduction}

\begin{figure}[h!]
  \centering{
    \includegraphics[width=0.41\textwidth]{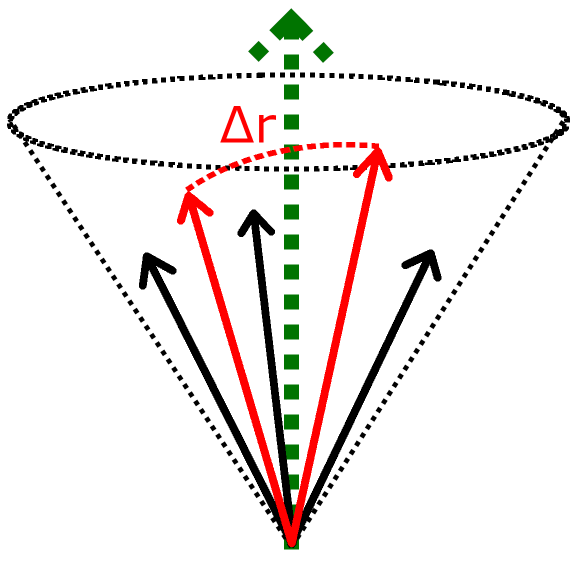}
    \includegraphics[width=0.45\textwidth]{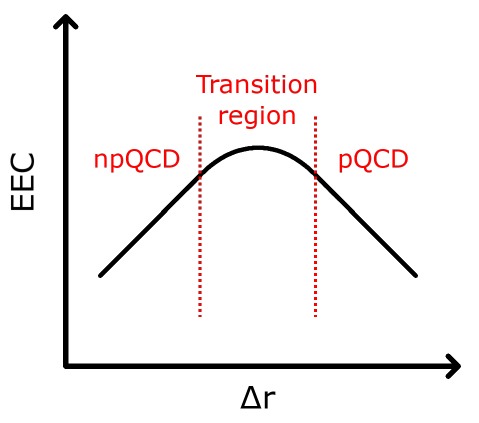} 
    \caption{\cmsLeftUpperCase: Illustration on how energy-energy correlators are constructed. The green dashed arrow represents the jet axis and the solid arrows represent charged particles. \cmsRightUpperCase: Schematic structure of an energy-energy correlator distribution in \pp collisions.}
  \label{fig:eecIllustration}
  }
\end{figure}

Energy-energy correlators are important observables for studying quantum chromodynamics (QCD)~\cite{Basham:1978bw, Basham:1978zq}. They can be calculated with great accuracy in asymptotically free perturbation theory and can be experimentally measured with excellent precision. The two-point energy correlator is defined as a weighted distribution of angular separation of all possible particle pairs within a jet cone, where the weights are the product of the transverse momenta (\pt) of the two particles in a pair, as illustrated in the \cmsLeft panel of Fig.~\ref{fig:eecIllustration}. Recent theoretical work has explored energy-energy correlators at LHC energies~\cite{Chen:2019bpb, Chen:2020vvp, Li:2021zcf, Craft:2022kdo, Chen:2022muj}, and theory comparisons have been made to the CMS open data from 7\TeV proton-proton (\pp) collisions~\cite{Komiske:2022enw, Lee:2022ige}. One of the most useful properties of the energy-energy correlator in \mbox{\Pep\Pem} and \pp collisions is that different time scales in jet evolution are imprinted in different angular scales of the correlator. It follows from QCD color coherence that daughter parton emissions to angles larger than the opening angle of the parent partons are strongly suppressed~\cite{angularOrdering1, angularOrdering2, angularOrdering3, angularOrdering4}, which leads to angular ordering of parton showers. In jet evolution, early times correspond to large angles, while late times correspond to small angles. A schematic illustration of how this translates to a structure of an energy-energy correlator distribution in \pp collisions is given in the \cmsRight panel of Fig.~\ref{fig:eecIllustration}, following the discussion in Ref.~\cite{Komiske:2022enw}. The energy-energy correlators at small angles are approximately proportional to the angle squared, which is expected from the propagation of uniformly distributed noninteracting hadrons. At large angles, the relevant dynamics to explain the scaling behavior are the perturbative interactions of asymptotically free quarks and gluons~\cite{Andres:2024xvk}. In between, there is a transition region that reflects the impact of hadronization.

While previously studied in \mbox{\Pep\Pem} collisions at LEP~\cite{ALEPH:1990eec}, many experimental collaborations are now undertaking measurements of these correlators in hadron collisions. These measurements include correlators within jets from \pp collisions at 200\GeV by the STAR Collaboration~\cite{STAR:ppEEC}, at 5.02\TeV by the ALICE Collaboration~\cite{ALICE:ppEEC}, and at 13\TeV by the CMS Collaboration~\cite{CMS:ppEEC}. The transition region between the perturbative and nonperturbative QCD regimes is found to scale inversely with the mean jet energy~\cite{ALICE:ppEEC}. The scaling factor is proportional to $\Lambda_{\mathrm{QCD}}$, the fundamental energy scale of QCD below which nonperturbative effects become dominant. With increasing jet momenta, the transition region moves to smaller angles. At larger angular distances, where perturbative QCD is probed, the correlators are sensitive to the strong coupling constant \alpS and to the various color factors associated with parton-initiated showers~\cite{CMS:ppEEC}. 

In heavy ion collisions, the interaction of scattered hard partons with quark-gluon plasma leads to jet quenching~\cite{CMS:firstJetQuenching, hotQCD2023, Cao:2020wlm, Apolinario:2022vzg}. This phenomenon has been studied for nearly two decades through measurements ranging from the suppression of high-\pt hadrons~\cite{PHENIX:2001hpc, STAR:2002ggv, CMS:2012aa} and jets~\cite{ATLAS:jetRaa, CMS:2021vui} to more differential observables, such as jet shapes~\cite{CMS:2013lhm, CMS:inclusiveJetShape, CMS:dijetJetShape} and other substructure observables~\cite{CMS:2012nro, CMS:2017qlm, CMS:2018fof, CMS:2020plq, ALICE:2021njq, ALICE:2023jetAxis, ATLAS:2022jetrg}. The CMS heavy ion program is heavily involved in these studies, with a summary description of previous results presented in Ref.~\cite{CMS:run2legacy}. Recent theoretical papers propose energy-energy correlators as complementary observables that may further enhance our understanding of the quark-gluon plasma~\cite{Andres:2022ndd, eecTheoryHeavyIon, Andres:2023xwr, Barata:2023qds, Barata:2023uoi, Barata:2023bhh, Yang:2023dwc}. In particular, energy-energy correlators in heavy ion collisions are predicted to be sensitive to color coherence effects~\cite{eecTheoryHeavyIon}, where only the gluons emitted above some critical angle can independently emit more gluons. They are also predicted to be sensitive to the ``jet wake'' effect~\cite{Yang:2023dwc}, where a high-\pt parton traversing the plasma drags medium particles with it. While the relative contribution of these two mechanisms of energy loss in the quark-gluon plasma has yet to be experimentally established, energy-energy correlators promise to be important new observables for their study~\cite{hybridEEEC}. 

In this Letter, two-point energy correlators within high-\pt jets are measured using charged particles in lead-lead (\PbPb) collisions at a nucleon-nucleon center-of-mass energy of $\sqrtsNN = 5.02\TeV$, recorded in 2018 by the CMS experiment. This is the first time these correlators have been measured in heavy ion collisions. The \PbPb results are compared to those from \pp collisions recorded in 2017 at the same center-of-mass energy to uncover possible modifications in the correlator shapes resulting from the presence of the quark-gluon plasma. The total integrated luminosities are 1.70\nbinv~\cite{CMS:PbPbLuminosity} for the \PbPb data set and 302\pbinv for the \pp data set~\cite{CMS:ppLuminosity}. The \PbPb energy-energy correlators are extracted in four ranges of nuclear overlap and jet \pt, two charged particle \pt thresholds, and with two different choices of \pt weights that offer different sensitivities to hard- and soft-parton interactions. Tabulated results are provided in the HEPData record for this analysis~\cite{hepdata}.

\section{The CMS experiment}
\label{sec:cms}

The central feature of the CMS apparatus is a superconducting solenoid of 6\unit{m} internal diameter, providing a magnetic field of 3.8\unit{T}. Within the solenoid volume are a silicon pixel and strip tracker, a lead tungstate crystal electromagnetic calorimeter, and a brass and scintillator hadron calorimeter, each composed of a barrel and two endcap sections. The silicon tracker measures charged particles within the pseudorapidity range $\abs{\eta} < 3.0$. During the LHC running period relevant to this Letter, the silicon tracker consisted of 1856 silicon pixel and $15\,148$ silicon strip detector modules. Two hadron forward steel and quartz-fiber calorimeters complement the barrel and endcap detectors, extending the calorimeter from the range $\abs{\eta} < 3.0$ provided by the barrel and endcap out to $\abs{\eta} < 5.2$. The hadron forward calorimeters are azimuthally subdivided into $20^\circ$ modular wedges and further segmented to form $0.175 \times 0.175$ ($\Delta\eta \times \Delta\varphi$) towers. The distribution of the sum of the transverse energies detected in the hadron forward detectors $(3.0 < \abs{\eta} < 5.2)$ is used to define the event centrality~\cite{centralityDetermination} in term of percentiles of the total distribution. Here, 0\% centrality corresponds to the largest overlap of the colliding nuclei. Muons are measured in the range $\abs{\eta} < 2.4$, with detection planes made using three technologies: drift tubes, cathode strip chambers, and resistive plate chambers. A detailed description of the CMS detector, together with a definition of the coordinate system used and the relevant kinematic variables, can be found in Refs.~\cite{CMSexperiment,CMS:detectorUpgrade}.

Events of interest are selected using a two-tiered trigger system. The first tier, composed of custom hardware processors, uses information from the calorimeters and muon detectors to select events at a rate of around 100\unit{kHz}~\cite{CMS:L1trigger2020}. The second tier, known as the high-level trigger, consists of a farm of processors running a version of the full event reconstruction software optimized for fast processing. It reduces the event rate to around 1\unit{kHz} before data storage~\cite{CMS:trigger2017}. 

\section{Event selection and simulated event samples}
\label{sec:eventSelection}

Events are selected using high-level triggers that require at least one calorimeter-based jet with high \pt. These jets are reconstructed using the anti-\kt jet clustering algorithm with a distance parameter of $R = 0.4$~\cite{antiKt}. For \PbPb collisions, we combine triggers with jet thresholds $\pt > 60\GeV$, $\pt > 80\GeV$, and $\pt > 100\GeV$. This combination achieves full trigger efficiency in the entire jet \pt range covered in this analysis, while minimizing the statistical uncertainty of the results. Heavy ion collisions produce many soft particles in addition to those resulting from the hard scattering processes associated with jets. This underlying event contribution is subtracted from the jets using an iterative method~\cite{puSubtraction} before comparing the jet \pt to one of the threshold values. For \pp collisions, a combination of triggers with thresholds $\pt > 60\GeV$ and $\pt > 80\GeV$ is used. No underlying event subtraction is applied for \pp collisions.

For \PbPb collisions, a minimum bias triggered sample is also used, where the aim is to select any hadronic interactions. The minimum bias trigger requires a signal on each side of the interaction point in the hadron forward calorimeters above the readout threshold of a fixed analog-to-digital converter count number that translates to 6--12\GeV, depending on the tower.

To reduce contamination from beam-gas collisions, vertex and noise filters are applied to both the \pp and \PbPb data following the example of previous analyses~\cite{CMS:2016xef}. At least two hadron forward calorimeter towers on each side of the detector are required to have an energy deposit of at least 4\GeV per tower. The primary vertex is required to have at least two tracks. In addition, it needs to be reconstructed within 15\unit{cm} of the nominal interaction point in the beam direction ($z$) and within 2\unit{cm} in the transverse direction. Additionally, in \PbPb collisions the shapes of the clusters in the pixel detector are required to be compatible with those expected at the measured vertex location.

The average number of collisions per beam bunch crossing (pileup) is about 2 in \pp and smaller than 0.01 in \PbPb collisions, respectively. As the pileup is low in both of the data samples, no pileup corrections are necessary.

Simulated event samples are used in the analysis to unfold the measurements to particle level, thus accounting for detector resolution effects. The hard jet events are generated with the \PYTHIA 8.230 event generator~\cite{pythia82} with tune CP5~\cite{pythiaCP5tune} and parton distribution function set NNPDF3.1 at next-to-next-to-leading order~\cite{nnpdf31}. As an alternative event generator for systematic uncertainty studies, the \HERWIG 7.2.2 generator~\cite{herwigManual, herwig7} with tune CH3~\cite{herwigCH3tune} and NNPDF3.1 is also used. The CMS detector response is simulated using the \GEANTfour toolkit~\cite{geant4}. The soft underlying event for the \PbPb collisions is simulated with the \HYDJET 1.9 event generator~\cite{hydjet}, which is tuned to reproduce global event properties, such as the charged-hadron \pt spectrum. The energy density in the \HYDJET simulation is tuned to match the data by shifting the centrality binning in the standard \HYDJET simulation by 4 percentage points upwards. This tuning is based on a random-cone study, where cones with a radius of 0.4 are placed in random directions in events, and the energy densities in the cones are determined by summing the particle transverse momenta within the cone. The above tuning provides the best agreement in random-cone energy density distribution shapes between data and \HYDJET. To simulate jet events in \PbPb collisions, \PYTHIA generated hard events are embedded into \HYDJET events. This sample is denoted as {\PYTHIA}+\HYDJET.

In \PbPb collisions, jets are produced more frequently in central events than in noncentral events because of a larger number of binary nucleon-nucleon collisions. A centrality-based weighting is applied to the {\PYTHIA}+\HYDJET sample to match the centrality distribution of the jet-triggered \PbPb data after event selection. An additional weighting procedure is performed to match the simulated vertex distributions with data for both the \pp and \PbPb samples.

\section{Jet and track reconstruction}
\label{sec:jetTrackReconstruction}

The track reconstruction used for \pp and \PbPb collisions is described in Ref.~\cite{trackResolution}. The charged particle tracks in \PbPb collisions are required to have at least eleven hits in the tracker layers and satisfy a stringent fit quality requirement $\chi^{2}/\mathrm{dof}/N_{\text{layers}} < 0.18$, where the $\chi^{2}$ and the number of degrees of freedom are related to the fit, and $N_{\text{layers}}$ is the number of tracker layers hit. To reduce the likelihood of counting nonprompt charged particles originating from secondary decay products in both \pp and \PbPb collisions, it is required that the distance of closest approach of a charged particle track to the primary vertex, divided by its uncertainty, is less than three. Furthermore, the relative \pt uncertainty $\delta\pt/\pt$ for the tracks must be less than 10\%. Finally, for \PbPb collisions, in order to reduce the contribution of misreconstructed tracks with very high \pt values, it is required that tracks with $\pt > 20\GeV$ have an associated energy deposit in the calorimeters corresponding to at least half of the track momentum. Corrections for tracking efficiency, detector acceptance, and misreconstructed tracks are obtained and applied following the procedures discussed in Ref.~\cite{CMS:2016xef}.

A particle-flow algorithm using an optimized combination of information from various elements of the CMS detector is used to reconstruct leptons, photons, and charged and neutral hadrons~\cite{particleFlow}. Jets in both \pp and \PbPb collisions are reconstructed using the anti-\kt algorithm with a distance parameter $R = 0.4$, as implemented in the \textsc{FastJet} framework~\cite{fastJet}, with particle-flow candidates as an input. The jets are first clustered using E-scheme clustering~\cite{fastJet} and then, using the same constituents, the jet axis is recalculated using the winner-take-all algorithm~\cite{wta1, wta2}. In the E-scheme clustering, particle pairs are iteratively combined to form pseudo-jets (an object that is a combination of particles or other pseudo-jets), with the direction of the new pseudo-jet given by the sum of the four-momenta of the particles. In the winner-take-all scheme, the direction of the new pseudo-jet in each iteration is aligned with the direction of the particle or pseudo-jet with the highest \pt value. It follows that the E-scheme axis for the final jet is given by the sum of pseudo-jet four-momenta, while the winner-take-all axis is determined by the direction of the hardest pseudo-jet. The axis recalculation is done to avoid an artificial dip in the particle density distributions close to the jet radius, which is a feature for an E-scheme jet axis.

In order to subtract the soft underlying event contribution from the jet energy in \PbPb collisions, a constituent subtraction method~\cite{constituentSubtraction} is employed. This involves a particle-by-particle approach that corrects the jet constituents based on the local average underlying event density. The energy density is estimated using an event-by-event, iterative algorithm~\cite{puSubtraction} that finds the mean value and dispersion of the energies from the particle-flow candidates in $\eta$-strips~\cite{CMS:firstJetQuenching,centralityDetermination}. To take into account the global hydrodynamic flow~\cite{ALICE:2011ab}, the azimuthal angle $(\varphi)$ distribution of the particle-flow candidates is fitted with second- and third-order Fourier components on an event-by-event basis. The background energy density is modulated by the results of this fit. In \pp collisions, where the underlying event level is negligible, jets are reconstructed without any underlying event subtraction.

\section{Energy-energy correlators}
\label{sec:analysisMethods}

Energy-energy correlators are constructed for charged particles in the vicinity of high-\pt jets. The analysis is done for jet \pt (\ptjet) ranges $120 < \ptjet < 140\GeV$, $140 < \ptjet < 160\GeV$, $160 < \ptjet < 180\GeV$, and $180 < \ptjet < 200\GeV$. Two different charged particle \pt (\ptCh) thresholds are used, with $\ptCh > 1\GeV$ and $\ptCh > 2\GeV$, respectively. While limiting the selection to higher-\pt particles reduces background, this also reduces part of the expected signal related to medium-induced radiation and soft-particle emissions within the jet. In \PbPb collisions, the measurement is done in four centrality ranges of  0--10\%, 10--30\%, 30--50\%, and 50--90\%. The jets are selected with $\abs{\eta} < 1.6$ to ensure stable detector performance within the jet cone. During the \PbPb run, there was a region in the tracker with a low track reconstruction efficiency. Since the analysis is not statistics limited, jets with $-0.1 < \varphi < 1.2$ in \PbPb collisions are not included in the analysis to avoid potential bias from the poor detector performance.

The energy-energy correlators are defined as
\begin{equation}
  \text{EEC}(\Delta r) = \frac{1}{W_{\text{pairs}}} \frac{1}{\delta r} \, \sum_{\text{jets} \in [p_{\mathrm{T},1},p_{\mathrm{T},2}]} \, \sum_{\text{pairs} \in [\Delta r_{a}, \Delta r_{b}]} \left( p_{\mathrm{T},i} \, p_{\mathrm{T},\,j} \right)^{n} ,
  \label{eq:eec}
\end{equation}
where ($i$,$j$) refer to paired particles, $\Delta r \equiv \Delta r_{ij} = \sqrt{\smash[b]{(\eta_{i} - \eta_{j})^{2} + (\varphi_{i} - \varphi_{j})^{2}}}$, $\Delta r_{a}$ and $\Delta r_{b}$ are $\Delta r$ bin boundaries, $\delta r = \Delta r_{b} - \Delta r_{a}$ is the bin width, and $p_{\mathrm{T},1}$ and $p_{\mathrm{T},2}$ are \ptjet bin borders. The particle pairs are constructed using all possible pairs of charged particles corresponding to the selected \ptCh threshold that are within $\Delta r_{\text{ch}}^{\text{jet}} = 0.4$ from a selected jet axis. One-dimensional histograms of $\text{EEC}(\Delta r)$ are produced by weighting each entry with the product of the particle transverse momenta $(p_{\mathrm{T},i}p_{\mathrm{T},j})^{n}$, with either $n=1$ or $2$. Similar to higher charged particle \pt thresholds, employing higher powers for the momentum weight exponent helps to reduce the underlying event background in heavy ion collisions. We note that in this analysis the $(p_{\mathrm{T},i}p_{\mathrm{T},j})^{n}$ weight for the particle pairs does not include the normalization factor $1/p_{\text{T,jet}}^{2}$ that is present in many theoretical publications because the single-particle \pt resolution is much better than that of \ptjet. Instead, the hard scale enters in Eq.~\eqref{eq:eec} through the choice of narrow \ptjet ranges. We have checked in simulation that the resolution effects for the weights are negligible if the \ptjet term is not included. Finally, the distribution is normalized by the weighted number of pairs $W_{\text{pairs}}$ such that the integral of the distribution in the analysis region is unity.

There are two kinds of particles within the jet cone. The fragmentation products of the hard-scattered parton, together with particles related to jet-medium interactions, are considered signal particles. All other particles, which mainly come from the underlying event and are not correlated with the signal particles, are considered background particles. When constructing the energy-energy correlators, we pair all selected charged particles with all other charged particles, resulting in three different types of pairings in our signal event:
\begin{itemize}
  \item Signal+Signal ($\mathrm{S}+\mathrm{S}$);
  \item Signal+Background ($\mathrm{S}+\mathrm{B}$); and
  \item Background+Background ($\mathrm{B}+\mathrm{B}$).
\end{itemize}

In \pp collisions, the \ptCh thresholds are sufficiently high to keep background components at negligible levels. For \PbPb collisions, we estimate the background distribution using the mixed cone method. For each signal event, we find two minimum bias events with similar underlying event distributions as the signal event. This is achieved by requiring that the primary vertex $z$ position is within 0.5\cm and the centrality within the same 0.5\% wide bin in the minimum bias events compared to those in the signal event. We then place a jet cone at the same $(\eta,\varphi)$ position as in the signal event to ensure the same detector performance. We call this cone in the minimum bias event a mixed cone. After this, three different pairings are constructed:
\begin{itemize}
  \item Signal Event + Minimum Bias Event 1 ($\mathrm{SE}+\mathrm{M1}$);
  \item Minimum Bias Event 1 + Minimum Bias Event 1 ($\mathrm{M1}+\mathrm{M1}$); and
  \item Minimum Bias Event 1 + Minimum Bias Event 2 ($\mathrm{M1}+\mathrm{M2}$).
\end{itemize}
For each of these pairings, each particle within the jet cone in the first event is paired with each particle within the jet cone in the second event. From these components, the background estimation is constructed as follows. The $\mathrm{SE}+\mathrm{M1}$ pairing provides a good description of the $\mathrm{S}+\mathrm{B}$ component, since the underlying event is uncorrelated from the signal. However, it does not describe the $\mathrm{B}+\mathrm{B}$ component properly. This occurs for two reasons. First, correlations in the underlying event are lost when pairing underlying event particles between two cones. Second, the $\mathrm{B}+\mathrm{B}$ component is overestimated when pairing all underlying event particles from the signal cone with those from the mixed cone. This overestimation is illustrated in a simple example. Assume you have ten underlying event particles in the signal cone. The number of unique particle pairs in the true $\mathrm{B}+\mathrm{B}$ background is $\binom{10}{2}=45$. However, if also the mixed cone has ten underlying event particles, pairing all underlying event particles from the signal cone with all underlying event particles in the mixed cone gives $10\times10=100$ unique pairings. Thus, we need to subtract this mismodeled $\mathrm{B}+\mathrm{B}$ component from the background estimator, and add a properly modeled $\mathrm{B}+\mathrm{B}$ component into it. This is done by using pairings from only minimum bias events. The probability that a random cone in a minimum bias event contains jet signal is negligible; hence, $\mathrm{M1}+\mathrm{M1}$ pairings provide an accurate description of the $\mathrm{B}+\mathrm{B}$ component as they retain correlations within the underlying event and avoid overcounting pairs. Conversely, $\mathrm{M1}+\mathrm{M2}$ pairings give the $\mathrm{B}+\mathrm{B}$ component that suffers from issues of losing correlations in the underlying events and overestimating background pairs. Using these pairings, we can construct the mixed cone background estimation with:
\begin{equation}
\mathrm{BG}_{\text{mixed}} = (\mathrm{SE}+\mathrm{M1}) + (\mathrm{M1}+\mathrm{M1}) - (\mathrm{M1}+\mathrm{M2}).
\label{eq:mixedConeBackground}
\end{equation}
This method accurately reproduces the generator-level background in the {\PYTHIA}+\HYDJET simulation for the entire kinematic region used in the analysis.

To correct for detector and background fluctuation effects that can affect the \pt range to which a given jet is associated, we apply an iterative D'Agostini unfolding method~\cite{DAgostini:BayesUnfold} with early stopping, as implemented in the \textsc{RooUnfold} software package~\cite{rooUnfold}. The response matrices are constructed using \PYTHIA simulation for \pp and {\PYTHIA}+\HYDJET simulation for \PbPb collisions. To take into account possible bin-by-bin statistical correlations in $\Delta r$, a covariance matrix determined from data is provided to the unfolding procedure to properly treat the statistical uncertainties. The number of iterations is chosen by calculating the $\chi^{2}$ value between consecutive unfolded distributions and determining the iteration number where the corresponding $p$-value~\cite{pValue} describing if the two distributions match becomes larger than 0.05. This method gives four as the best number of iterations for this analysis.

In addition to single-track efficiency corrections mentioned in Section~\ref{sec:jetTrackReconstruction}, track pair efficiency corrections were derived from \PYTHIA simulation for \pp and {\PYTHIA}+\HYDJET simulation for \PbPb collisions. These corrections account for remaining reconstruction effects for charged particles. The efficiency to reconstruct a pair of tracks drops sharply for opening angles $\Delta r \lesssim 0.015$. After applying the pair efficiency correction, the simulation shows good closure. Bin-to-bin resolution effects in $\Delta r$ are not significant above $\Delta r \gtrsim 0.008$, negating the need to unfold $\Delta r$ when restricting the analysis to this region. This choice does not impact the results, as the substructure modifications are expected to manifest themselves for $\Delta r \gtrsim 0.08$.

The analysis flow is as follows: first, construct the energy-energy correlators with tracking efficiency corrections; next, unfold the detector and background fluctuation effects that can lead to the migration of a \ptjet value across a range boundary; and finally, subtract the background from the unfolded distribution. The unfolding is done before background subtraction to correct for the higher average underlying event level within a jet cone compared to a random cone. In the background subtraction step, an additional scaling factor for $\mathrm{BG}_{\text{mixed}}$ is needed. Because of detector resolution effects, a true \ptjet value is likely to be slightly smaller than the corresponding reconstructed \ptjet value. Thus, the unfolding procedure applies a slight upward shift to the ranges of reconstructed \ptjet values considered, such that the defined \ptjet range boundaries correspond to true \ptjet values after unfolding. Having higher mean \ptjet corresponds to having either more signal particles or signal particles having a higher mean \pt. Because the underlying event level does not strongly depend on \ptjet, this leads to a better signal-to-background ratio for energy-energy correlators within higher \pt jets. Thus, we scale the $\mathrm{BG}_{\text{mixed}}$ distribution using a data-driven method to avoid over subtracting the unfolded distributions. This is done by noting that the peak position in the energy-energy correlator distribution depends on \ptjet, as previously measured in \pp collisions~\cite{CMS:ppEEC, ALICE:ppEEC}. Before unfolding, the mean \ptjet value in each \ptjet range is determined from the measured \ptjet distributions, and the $(\mathrm{S}+\mathrm{B})/\mathrm{B}$ ratio is estimated taking the ratio of integrals of the unsubtracted energy-energy correlator distribution and $\mathrm{BG}_{\text{mixed}}$. We then relate the $\Delta r$ positions of the energy-energy correlator peaks to the mean \ptjet values. Locating the peak positions after unfolding gives us the mean \ptjet values of the unfolded distributions. By interpolating between the $(\mathrm{S}+\mathrm{B})/\mathrm{B}$ and mean \ptjet values determined from the data before unfolding, the expected $(\mathrm{S}+\mathrm{B})/\mathrm{B}$ ratio at the unfolded mean \ptjet is found. Unfolding improves the $(\mathrm{S}+\mathrm{B})/\mathrm{B}$ ratio by 2--8\%, depending on the kinematic selection. We then scale the $\mathrm{BG}_{\text{mixed}}$ such that this ratio is met. Denoting the scaled background as $\mathrm{BG}_{\text{mixed}}^{\text{scaled}}$, the unfolded and fully corrected signal is extracted as
\begin{equation}
\mathrm{EEC}_{\text{signal}} = \mathrm{EEC}_{\text{unfolded}} - \mathrm{BG}_{\text{mixed}}^{\text{scaled}}.
\label{eq:eecSignal}
\end{equation}
We have verified that the final closure of this method in {\PYTHIA}+\HYDJET simulation is consistent with unity within statistical uncertainties.

\section{Systematic uncertainties}
\label{sec:systematics}

The systematic uncertainties of the energy-energy correlator distributions are categorized into two types: point-by-point uncertainties and shape uncertainties. To illustrate the sizes of uncertainties, we provide representative relative uncertainty values for low, intermediate, and high $\Delta r$ ranges from the 0--10\% central energy-energy correlator distributions with $n=1$ and $\ptCh > 1\GeV$ selection. The quoted relative uncertainty ranges reflect the span for different selections of \ptjet. The point-by-point uncertainty sources are:

\begin{itemize}
  \item \emph{Tracking efficiency}. Following the method in Ref.~\cite{CMS:2016xef}, it is estimated that tracking efficiency differences between data and simulation are as much as 5\% for \PbPb collisions and 2.2\% for \pp collisions. To estimate uncertainties arising from this, the tracking efficiency corrections are varied by these amounts up and down. The resulting differences to nominal values are found to be negligible, since they do not induce a shape change and the distributions are normalized to unity in the analysis region.
  \item \emph{Track selection}. For this uncertainty, we vary the track selections to be looser or tighter than the nominal ones. The looser selections are $\delta\pt/\pt < 0.15$ for \PbPb only, and the distance of closest approach to a vertex divided by its uncertainty smaller than five for both \pp and \PbPb. The tighter selections are $\chi^{2}/\mathrm{dof}/N_{\text{layers}} < 0.15$ for \PbPb only, $\delta\pt/\pt < 0.05$, and the distance of closest approach to a vertex divided by its uncertainty smaller than two for both \pp and \PbPb. The bigger difference to the nominal result is taken as a systematic uncertainty. Representative relative uncertainties are ${<}5\%$ in the low $\Delta r$ region, ${<}1\%$ in the intermediate $\Delta r$ region, and ${<}3\%$ in the high $\Delta r$ region.
  \item \emph{Track pair efficiency}. For the track pair efficiency, we vary the pair efficiency corrections to twice the amount of the single-track corrections. This is because the $\Delta r$ dependent efficiency might be different between data and simulation compared to the general tracking efficiency. Doubling the variation covers possible extra differences. The track pair correction is thus varied up and down by 10\% for \PbPb collisions and 4.4\% for \pp collisions. The larger variation in the resulting distributions with respect to the nominal one is taken to be the systematic uncertainty. Representative relative uncertainties are ${<}1\%$ in the low and intermediate $\Delta r$ regions, and 1--8\% in the high $\Delta r$ region.
  \item \emph{Response matrix statistics}. To estimate the uncertainty coming from limited statistics in the simulations used to construct the unfolding response, we create 50 randomized response matrices. Each point in the original response matrix is randomized with a Gaussian distribution where the mean is the original value and $\sigma$ is the uncertainty of this value. We then calculate the standard deviation of the unfolded results for each $\Delta r$ value and assign this as the systematic uncertainty. Representative relative uncertainties are 1--2\% in the low $\Delta r$ region, and ${<}1\%$ in the intermediate and high $\Delta r$ regions.
\end{itemize}

The shape uncertainty sources are:
\begin{itemize}
    \item \emph{Background subtraction}. The background subtraction uncertainty has two components. 
    
  First, there is an uncertainty for the background scaling factor that we use to obtain agreement with the expected signal-to-background ratio after unfolding. The nominal value is determined using an average of the scaling factors for the different jet \pt ranges, with the uncertainty taken as the largest difference of the individual values from the average. Representative relative uncertainties are ${<}1\%$ in the low and intermediate $\Delta r$ regions, and 15--21\% in the high $\Delta r$ region.
    
  Second, to estimate the uncertainty of the mixed cone background subtraction, we compare the results from this method to those from a reflected-cone method. In the reflected-cone method, we place a new jet cone in the same signal event, but reflect the $\eta$ value of the jet axis (or shift it by $2R = 0.8$ if we are close to midrapidity to avoid the cones overlapping). We then pair each particle from the signal cone with each particle from the reflected cone to estimate the background. To overcome the background pair overcounting issues discussed earlier, we apply a scaling factor determined from the ratio of generator-level background to the reflected-cone background in the Monte Carlo simulation. This gives a reliable background estimate in the region where the $\mathrm{B}+\mathrm{B}$ component of the background is small. For $n=1$ and $\ptCh > 1\GeV$, where the $\mathrm{B}+\mathrm{B}$ component is significant, we double the uncertainty found for the $n=1$, $\ptCh > 2\GeV$ selection, as the reflected cone fails to provide a reasonable background estimate in this lower-\pt threshold case. Representative relative uncertainties are ${<}1\%$ in the low and intermediate $\Delta r$ regions, and 11--17\% in the high $\Delta r$ region.

  For \pp collisions, where no background subtraction is done for the nominal results, we compare the nominal results to reflected-cone subtracted results. Here the scaling factors for the reflected cone are derived from the peripheral 85--95\% centrality bin using the {\PYTHIA}+\HYDJET simulation, where differentiating signal and background is easier compared to pure \PYTHIA. The reflected-cone background scaled using these factors is then subtracted from the unfolded distribution, and these results are compared to the nonsubtracted results to evaluate the uncertainty. Representative relative uncertainties are ${<}1\%$ in the low and intermediate $\Delta r$ regions, and 1--2\% in the high $\Delta r$ region.
  \item \emph{Jet energy resolution}. The uncertainties in the jet energy resolution are divided into centrality independent and centrality dependent parts. The centrality-independent part is applied for both \pp and \PbPb collisions. In the nominal result, the jet \pt response matrices for unfolding are constructed by increasing the jet \pt resolution by the estimated data-to-simulation difference. Different amounts of broadening are applied to account for uncertainties in data-to-simulation differences, and new response matrices are constructed for the different resolution values. The difference to nominal is taken as the uncertainty. Representative relative uncertainties are 1--3\% in the low $\Delta r$ region, ${<}2\%$ in the intermediate $\Delta r$ region, and 4--7\% in the high $\Delta r$ region.
  
  For the centrality-dependent uncertainty in \PbPb collisions, we construct response matrices using 2\% and 6\% centrality shifts and compare the results to the nominal 4\% shifted result. This degree of shift covers the possible difference in background energy density between data and simulation. Representative relative uncertainties are 1--4\% in the low $\Delta r$ region, ${<}3\%$ in the intermediate $\Delta r$ region, and 6--10\% in the high $\Delta r$ region.
  \item \emph{Jet energy scale}. To evaluate the uncertainties from the jet energy scale, we shift the jet \pt in the \PYTHIA or {\PYTHIA}+\HYDJET simulations up and down by the uncertainty of the jet energy corrections. The procedure to estimate the uncertainty of the corrections is explained in detail in Ref.~\cite{CMS:2016lmd}. The jet energy correction uncertainties vary between 3--7\% of the jet \pt, and are 3.6\% on average. After shifting the jet \pt, we construct new response matrices using the shifted \pt values, and compare the final unfolded results to the nominal ones. Representative relative uncertainties are 6--10\% in the low $\Delta r$ region, 3--5\% in the intermediate $\Delta r$ region, and 30--45\% in the high $\Delta r$ region.
  \item \emph{Unfolding prior}. The nominal response matrices for the analysis are constructed using the nominal \PYTHIA and {\PYTHIA}+\HYDJET simulations. To estimate the uncertainty related to the shape of the prior, we find a function that scales the reconstructed jet \pt spectrum in \PYTHIA to match the one in \pp data. Then both generator- and reconstructed-level jet \pt spectra in \PYTHIA and {\PYTHIA}+\HYDJET simulations are weighted with this function. The difference between the data unfolded with the nominal response matrices to the data unfolded with the weighted response matrices is taken as the systematic uncertainty of the prior distribution. Representative relative uncertainties are ${<}4\%$ in the low $\Delta r$ region, ${<}2\%$ in the intermediate $\Delta r$ region, and 3--6\% in the high $\Delta r$ region.
  \item \emph{Monte Carlo model}. To evaluate the dependence of the unfolding performance on the used Monte Carlo model, the full analysis procedure with corrections and unfolding matrices derived from \PYTHIA simulation was performed for \HERWIG simulation and a second, statistically independent \PYTHIA simulation. Both cases close within the relevant systematics and are compatible with each other, so no additional uncertainty from the Monte Carlo model is needed.
  \item \emph{Unfolding iterations}. The optimal number of iterations for the D'Agostini method is found to be four, as explained in Section~\ref{sec:analysisMethods}. As an uncertainty for this number, we use three or five iterations and compare the resulting distributions to those unfolded with the nominal number of iterations. Representative relative uncertainties are ${<}2\%$ in the low and intermediate $\Delta r$ regions, and 3--8\% in the high $\Delta r$ region.
\end{itemize}

To obtain the total point-by-point and shape systematic uncertainties, the uncertainties within each of these categories are added together in quadrature. These are then reported as two distinct sources of uncertainty. For \PbPb to \pp ratio plots, both categories of uncertainties are propagated as uncorrelated. Since the data-taking periods occurred in different years, uncertainties in the ratios do not cancel out. For double ratios of the \PbPb to \pp ratios with different \ptCh thresholds, the jet related uncertainties fully cancel, and the track related uncertainties cancel for the overlapping part. The uncertainties are the highest in central collisions for low-\pt jets. They strongly depend on $\Delta r$, with the largest relative uncertainties occurring at high $\Delta r$, with somewhat smaller uncertainties at low $\Delta r$, and significantly smaller uncertainties at intermediate $\Delta r$.  This happens as multiple sources of uncertainty ``tilt'' the distribution around a fixed central point, leading to an increase on one side and a decrease on the other. The relative uncertainties at high $\Delta r$ are much higher because the energy-energy correlator distributions are steeply falling towards larger $\Delta r$ values. The dominant source of systematic uncertainty is the jet energy scale, which is much larger than any other source. This results from the relatively narrow jet \pt ranges that are used in this analysis and from the dependence of the energy-energy correlator distribution shape on jet \pt.

\section{Results}
\label{sec:results}

\begin{figure*}[h]
  \centering{
    \includegraphics[width=0.98\linewidth]{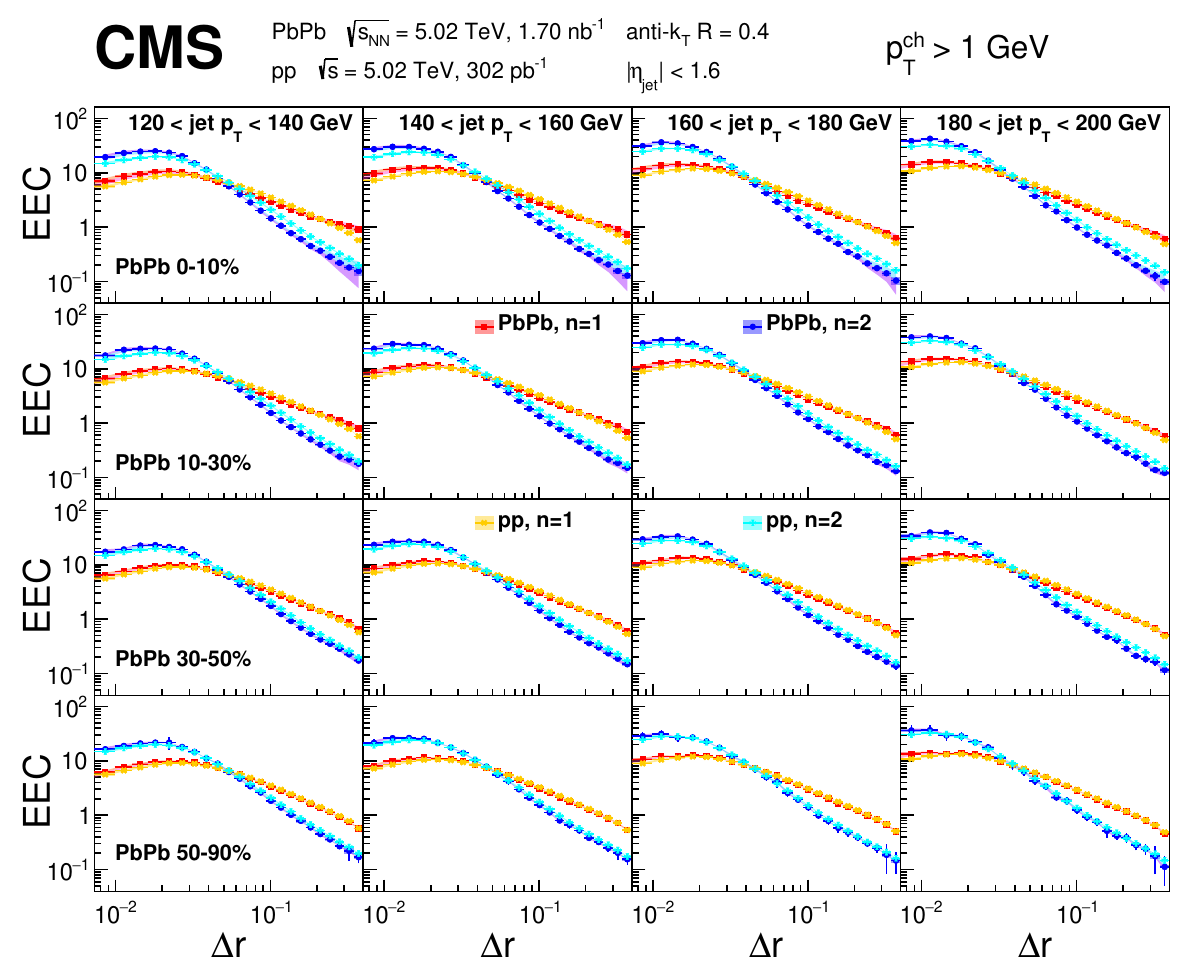} 
    \caption{Centrality and \ptjet dependent energy-energy correlators for $\ptCh > 1\GeV$. The red squares show the $n=1$ and the blue circles the $n=2$ distributions for \PbPb collisions. The \pp results for different centralities are identical, with orange crosses showing the $n=1$ and cyan triangle crosses the $n=2$ distributions. The error bars show statistical uncertainties, the point-by-point systematic uncertainties are shown in boxes, while the error bands show systematic uncertainties related to the shape of the distribution. The two colors illustrate that the shape uncertainties tend to tilt the distribution one way or another. All correlators have been normalized to unity in the plotted range.}
  \label{fig:bigEecCanvasPt1}
  }
\end{figure*}

\begin{figure*}[h]
  \centering{
    \includegraphics[width=0.98\linewidth]{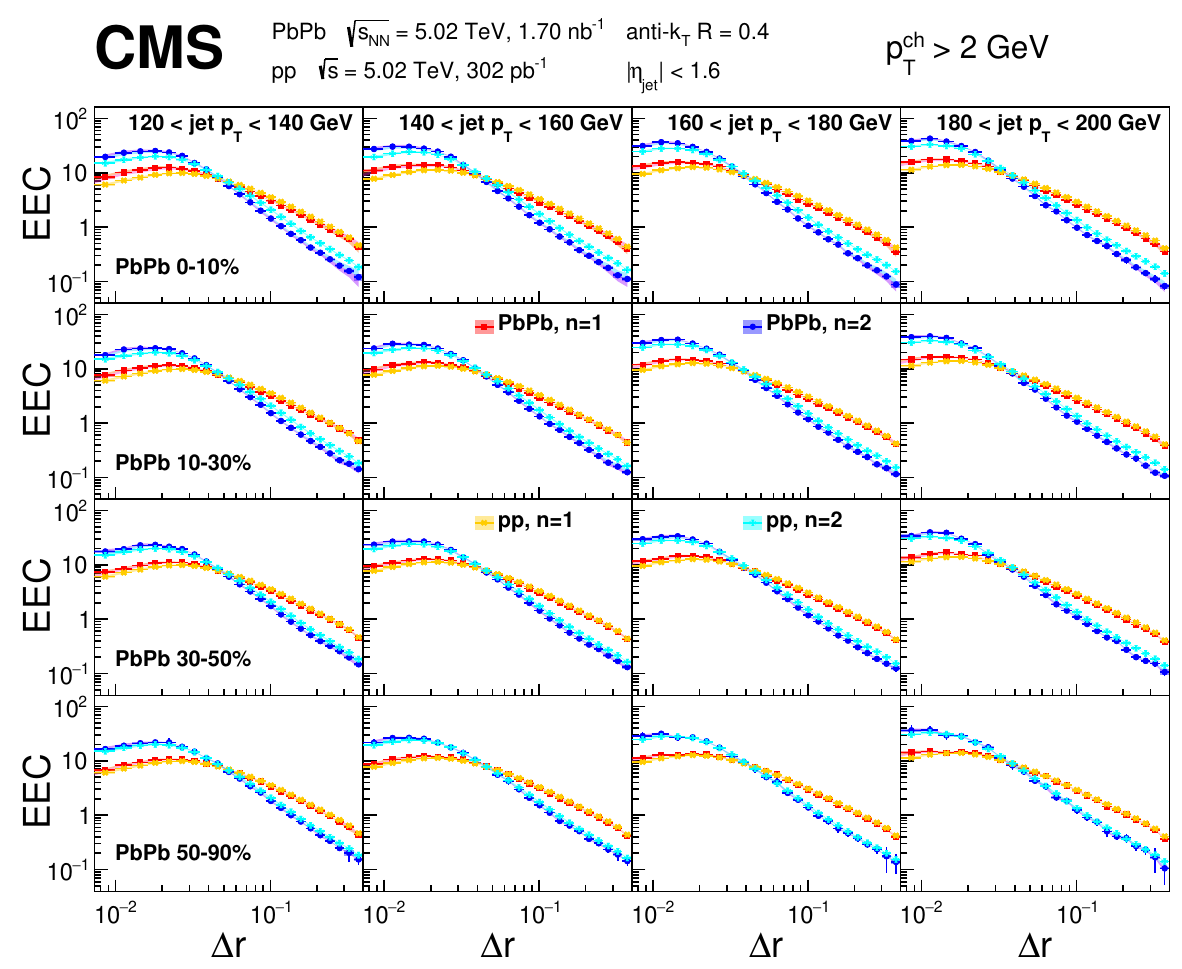} 
    \caption{Centrality and \ptjet dependent energy-energy correlators for $\ptCh > 2\GeV$. The red squares show the $n=1$ and the blue circles the $n=2$ distributions for \PbPb collisions. The \pp results for different centralities are identical, with orange crosses showing the $n=1$ and cyan triangle crosses the $n=2$ distributions. The error bars show statistical uncertainties, the point-by-point systematic uncertainties are shown in boxes, while the error bands show systematic uncertainties related to the shape of the distribution. The two colors illustrate that the shape uncertainties tend to tilt the distribution one way or another. All correlators have been normalized to unity in the plotted range.}
  \label{fig:bigEecCanvasPt2}
  }
\end{figure*}

The energy-energy correlators for $\ptCh > 1\GeV$ are presented in Fig.~\ref{fig:bigEecCanvasPt1} and for $\ptCh > 2\GeV$ in Fig.~\ref{fig:bigEecCanvasPt2}. First, comparing the shapes of the distributions in \PbPb collisions to those from \pp collisions, we observe the same general structure. An ascending slope marking the free-hadron region at small $\Delta r$, a peak related to a transition region at moderate $\Delta r$, and a descending slope interpreted as a free quark/gluon region at large $\Delta r$, as illustrated in Ref.~\cite{Komiske:2022enw} for \pp collisions, are all also clearly visible in the \PbPb data. Medium-induced emissions can break the angular ordering of emissions in vacuum QCD parton showers~\cite{angularOrderingInMedium}. However, the similarity in the energy-energy correlator shapes between the collision systems suggests that at small $\Delta r$, the shower in \PbPb collisions effectively follows angular ordering. The peak in the transition region consistently appears at lower $\Delta r$ values in \PbPb than in \pp data, with a larger difference in more central collisions. This experimental result could be explained by virtuality evolution~\cite{Domdey:2008gp}. If the jets in \PbPb collisions start with higher virtuality, they will have a longer shower duration to reach the scale of $\Lambda_{\mathrm{QCD}}$ and begin the hadronization process. Higher initial virtuality corresponds to higher initial jet \pt. Because jets in heavy ion collisions have, on average, experienced some amount of energy loss~\cite{CMS:2021vui}, the measured final-state jet \pt corresponds to higher initial jet \pt in \PbPb collisions compared to \pp collisions. Thus, the shift of the peak can be interpreted as an independent sign of jet quenching in \PbPb collisions. It also provides an additional handle of the initial virtuality of the jet without requiring a photon or a {\PZ} boson tag. The shifting of the peak position towards lower $\Delta r$ is also observed as a function of jet \pt. This follows naturally from the fact that higher \pt jets have higher initial virtuality.

\begin{figure*}[h]
  \centering{
    \includegraphics[width=0.98\linewidth]{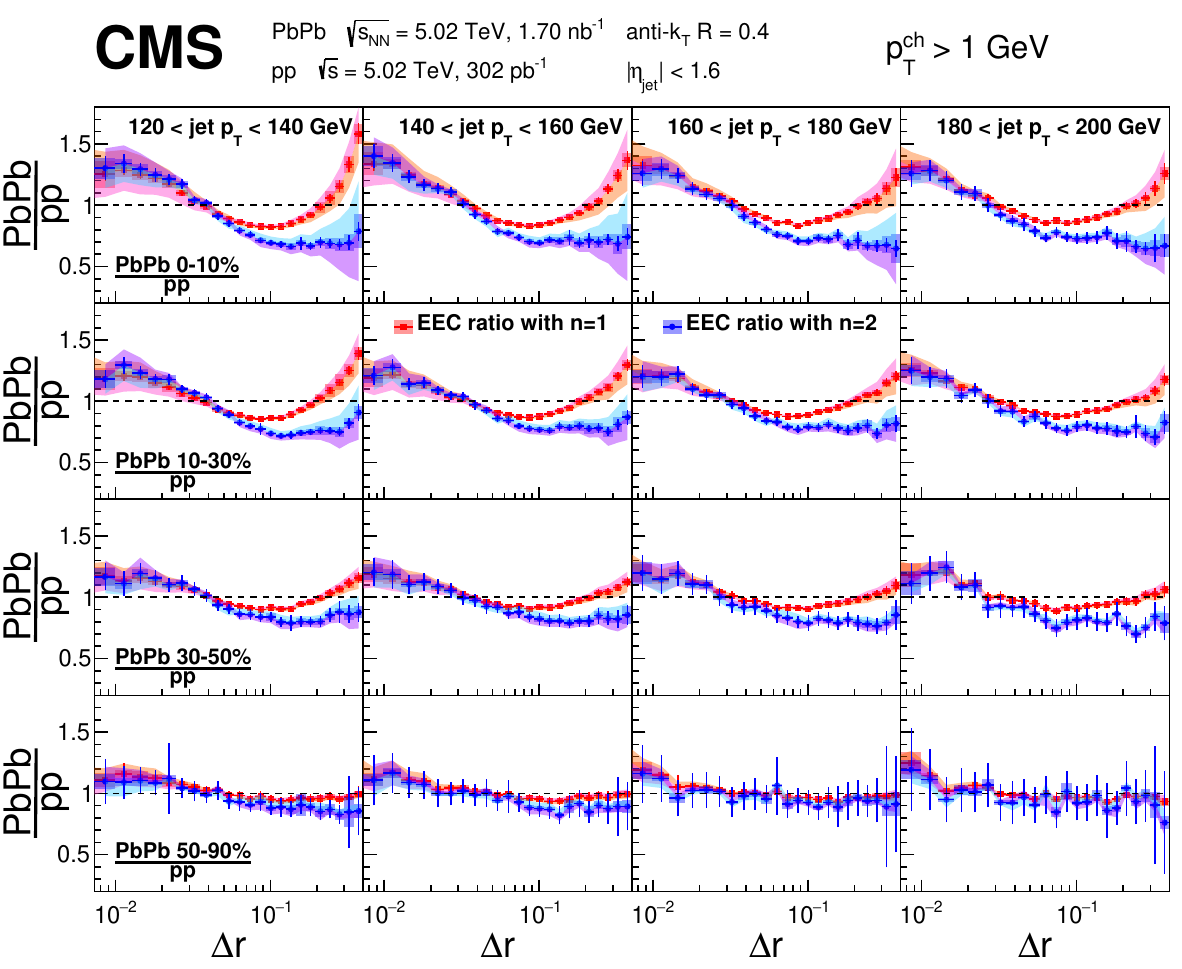} 
    \caption{Centrality- and \ptjet-dependent ratios of \PbPb to \pp energy-energy correlators with $\ptCh > 1\GeV$ and for both $n=1$ (red squares) and $n=2$ (blue circles). The error bars show statistical uncertainties, the point-by-point systematic uncertainties are shown in boxes, while the error bands show systematic uncertainties related to the shape of the ratio. The two colors illustrate that the shape uncertainties tend to tilt the ratio one way or another.}
  \label{fig:bigEecRatio1}
  }
\end{figure*}

\begin{figure*}[h]
  \centering{
    \includegraphics[width=0.98\linewidth]{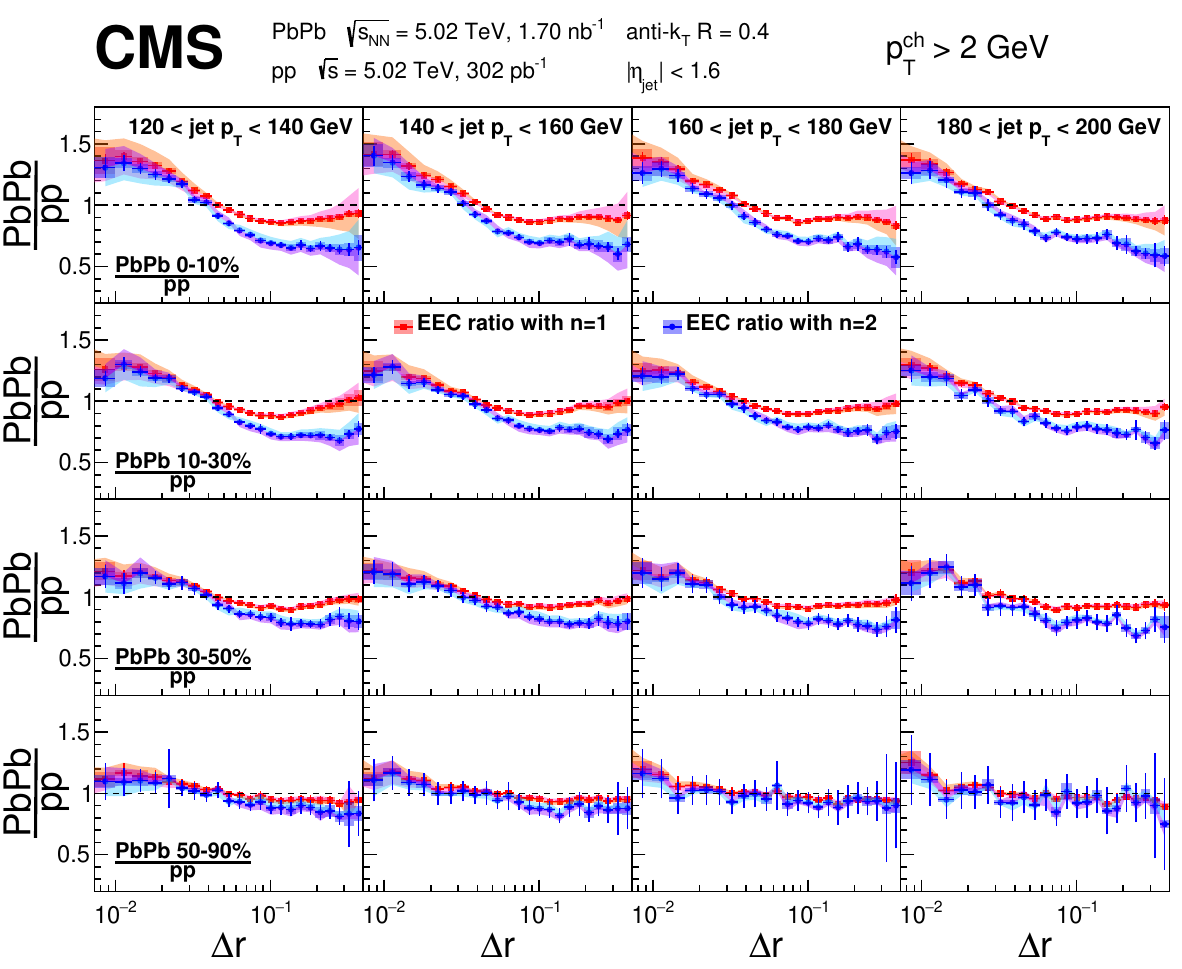} 
    \caption{Centrality- and \ptjet-dependent ratios of \PbPb to \pp energy-energy correlators with $\ptCh > 2\GeV$ and for both $n=1$ (red squares) and $n=2$ (blue circles). The error bars show statistical uncertainties, the point-by-point systematic uncertainties are shown in boxes, while the error bands show systematic uncertainties related to the shape of the ratio. The two colors illustrate that the shape uncertainties tend to tilt the ratio one way or another.}
  \label{fig:bigEecRatio2}
  }
\end{figure*}

To investigate predicted modifications to the jet substructure from interactions with quark-gluon plasma, we examine the ratios of the \PbPb distributions to \pp distributions from Figs.~\ref{fig:bigEecCanvasPt1} and \ref{fig:bigEecCanvasPt2} in Figs.~\ref{fig:bigEecRatio1} and~\ref{fig:bigEecRatio2}, respectively. The shift in the peak position observed in the distributions appears as an enhancement in the ratios at small $\Delta r$. The effects of selecting different initial populations of jets between collision systems have been previously observed as inclusive jet substructure distributions getting narrower by ALICE~\cite{ALICE:2023jetAxis, ALICE:2021groomedJetRadius} and ATLAS~\cite{ATLAS:2022jetrg}. A recent CMS $\gamma$-jet measurement shows that this narrowing occurs when selecting jets with minimal energy loss and vanishes when more quenched jets are included~\cite{CMS:gammaJetSubstructure}. All of these results highlight the importance of understanding the biases in jet selection when comparing jets between different systems.

A key observation from the ratios is the enhancement of the \PbPb to \pp ratio for central collisions at large $\Delta r$ with $n=1$ and $\ptCh > 1\GeV$. This is the kinematic region of the highest sensitivity to medium modifications. Interestingly, if either the charged particle \pt threshold is increased to $\ptCh > 2\GeV$ or the momentum weight is increased to $n=2$, this enhancement largely disappears. Both of these selections reduce the sensitivity to medium modifications by reducing the impact of soft particles at the periphery of the jet cone~\cite{CMS:inclusiveJetShape, CMS:dijetJetShape}. For $n=1$ with $\ptCh > 1\GeV$, where the enhancement is visible, the change of shape from descending slope at low $\Delta r$ to ascending slope at large $\Delta r$ happens around $\Delta r \approx 0.07\text{--}0.09$. Similar values have been recently suggested to be the relevant angular scales for the coherence length~\cite{Mehtar-Tani:2024jtd}. The size of the enhancement at large $\Delta r$ gradually decreases when going from central to more peripheral collisions, and is no longer visible in the 50--90\% centrality class. The enhancement is also stronger for lower jet \pt values compared to higher jet \pt selections.

We also focus on the low-$\Delta r$ region in the ratios. In particular, in the lower jet \pt ranges, the first few $\Delta r$ bins in the ratios correspond to the free-hadron regime. It can be observed that regardless of the \ptCh threshold or the value of the momentum weight $n$, the results are similar. Additionally, the \PbPb to \pp ratio in the low-$\Delta r$ region is relatively flat, before transitioning to a descending slope around the transition and free quark/gluon regimes. The flat ratio suggests a universal scaling behavior for free hadrons, even for jets that have propagated through quark-gluon plasma.

\begin{figure}[h]
  \centering{
    \includegraphics[width=0.48\textwidth]{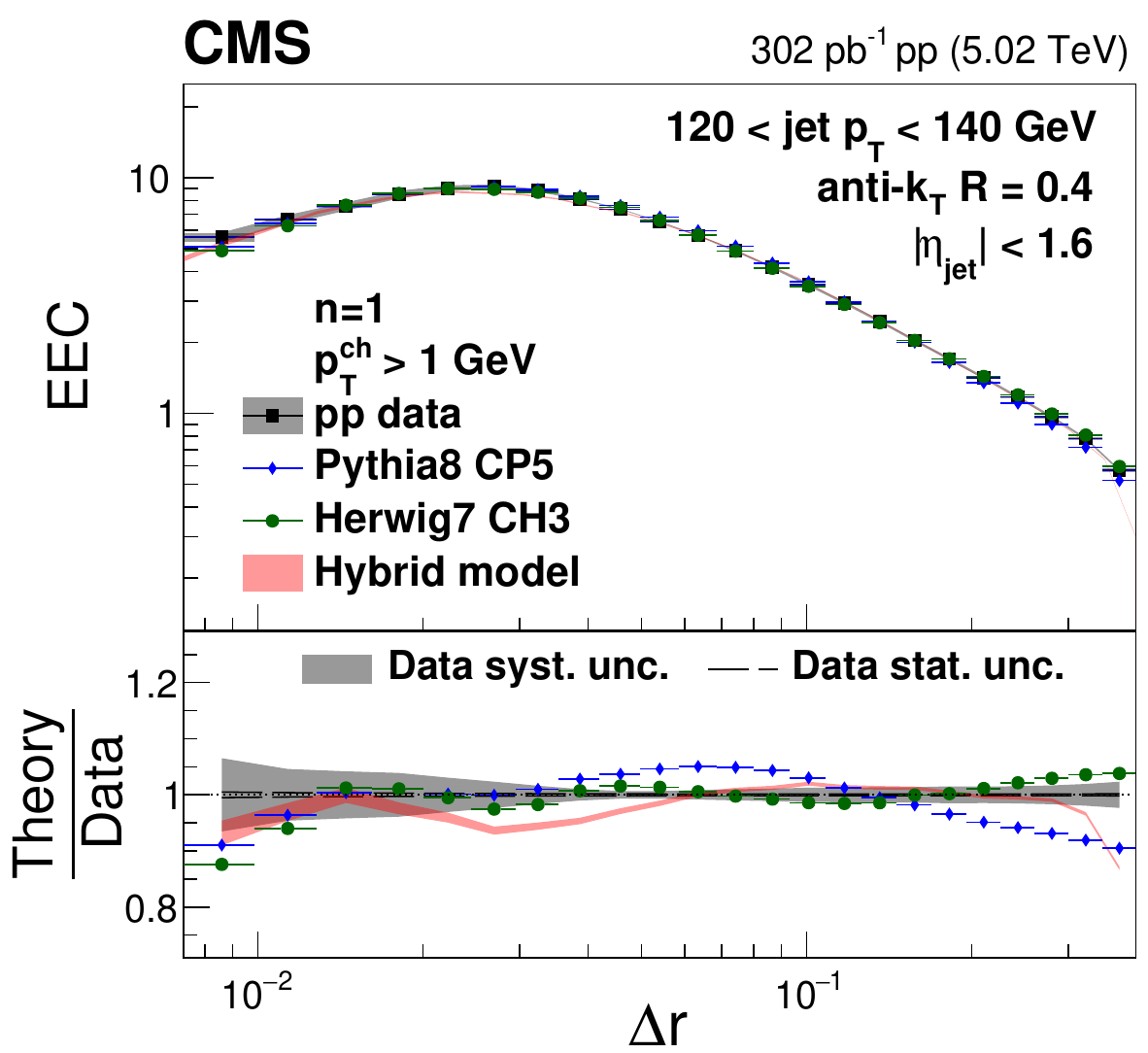}
    \includegraphics[width=0.48\textwidth]{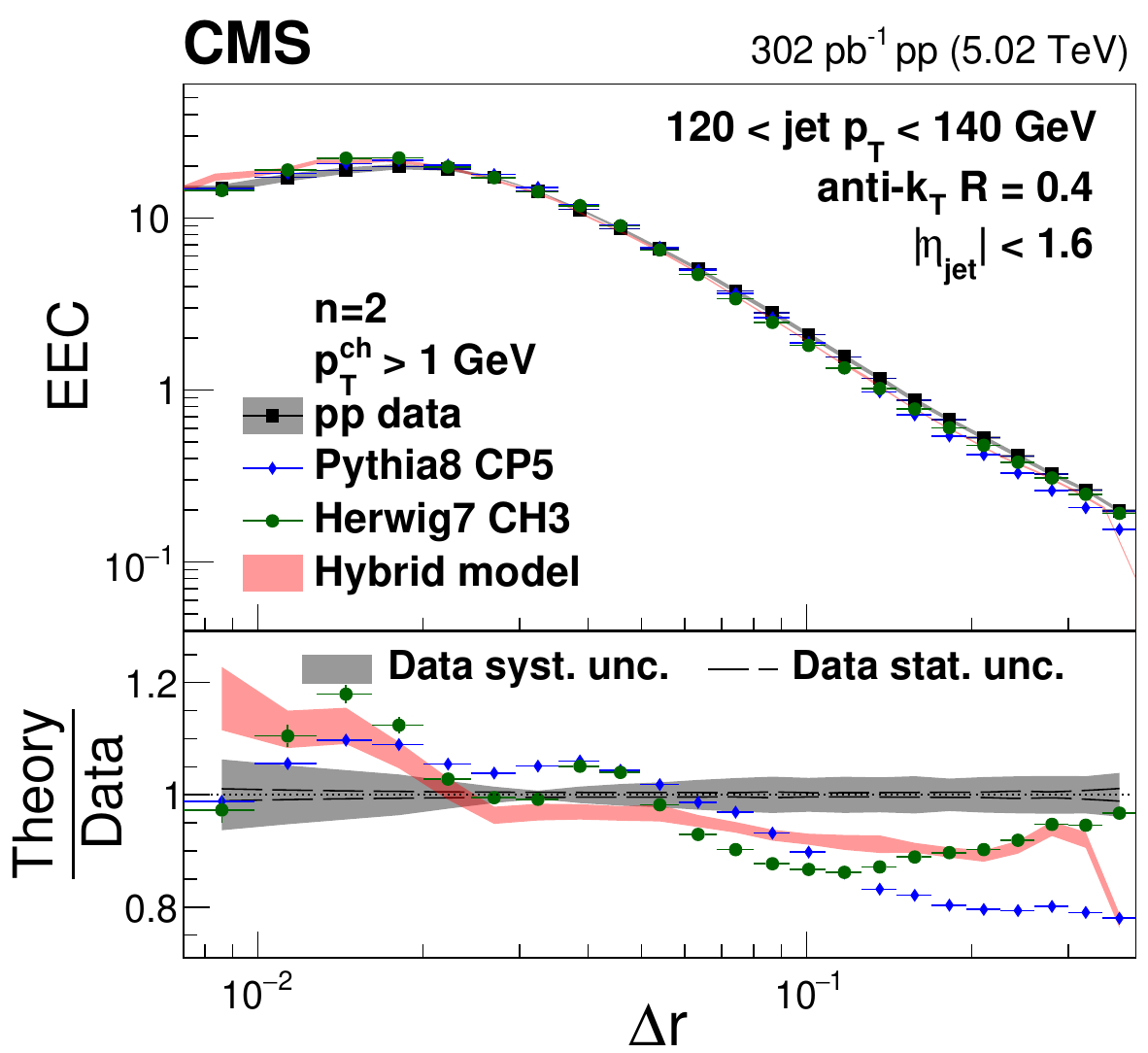} 
    \caption{Comparison of \PYTHIA~\cite{pythia82}, \HERWIG~\cite{herwigManual, herwig7}, and the hybrid model~\cite{hybridModel1, hybridModel2, hybridModel3} calculations to the observed energy-energy correlators for $\ptCh > 1\GeV$, $120 < \ptjet < 140\GeV$ and $n=1$ (\cmsLeft) and $n=2$ (\cmsRight) in \pp collisions. In the lower panels, the experimental uncertainties are indicated by the bands around unity.}
  \label{fig:ppMCcomparison}
  }
\end{figure}

Predictions from {\PYTHIA}8~\cite{pythia82} tune CP5~\cite{pythiaCP5tune}, {\HERWIG}7~\cite{herwigManual, herwig7} tune CH3~\cite{herwigCH3tune}, and a hybrid strong/weak coupling model~\cite{hybridModel1, hybridModel2, hybridModel3} are compared to the \pp results in Fig.~\ref{fig:ppMCcomparison}. The vacuum in the hybrid strong/weak coupling model is described by a specific tune of {\PYTHIA}8. The comparison is shown for $\ptCh > 1\GeV$, $120 < \ptjet < 140\GeV$, and for both momentum weights $n=1$ and $n=2$. Concentrating first on the comparison with $n=1$, all models are found within 5--10\% of the measurement, with \HERWIG and the hybrid model slightly outperforming \PYTHIA. Interestingly, the agreement with data gets worse for all of the models when the momentum weight is increased with $n=2$. In this case, the models predict an energy-energy correlator distribution that is too narrow.

\begin{figure}[h]
  \centering{
    \includegraphics[width=0.48\textwidth]{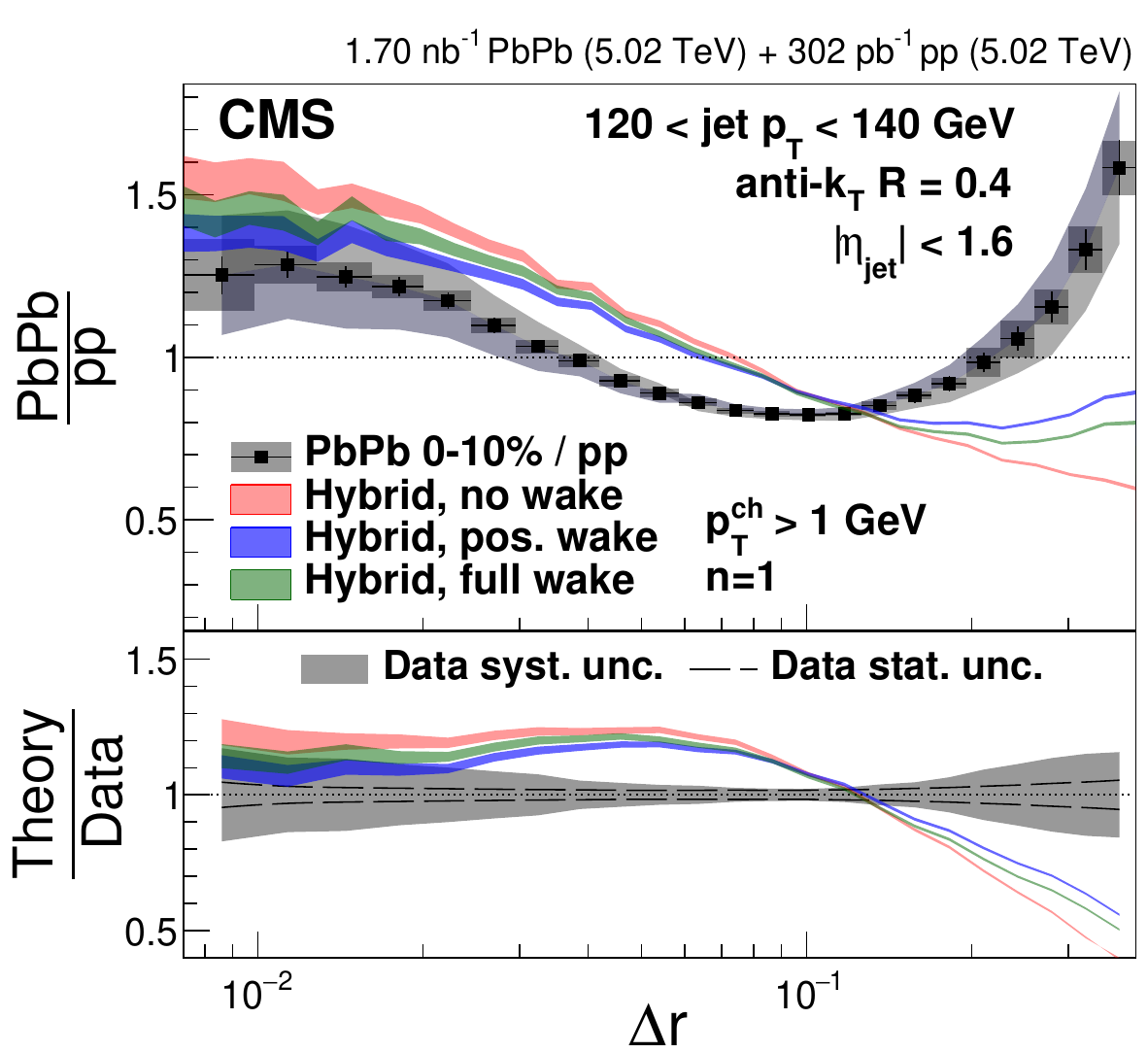}
    \includegraphics[width=0.48\textwidth]{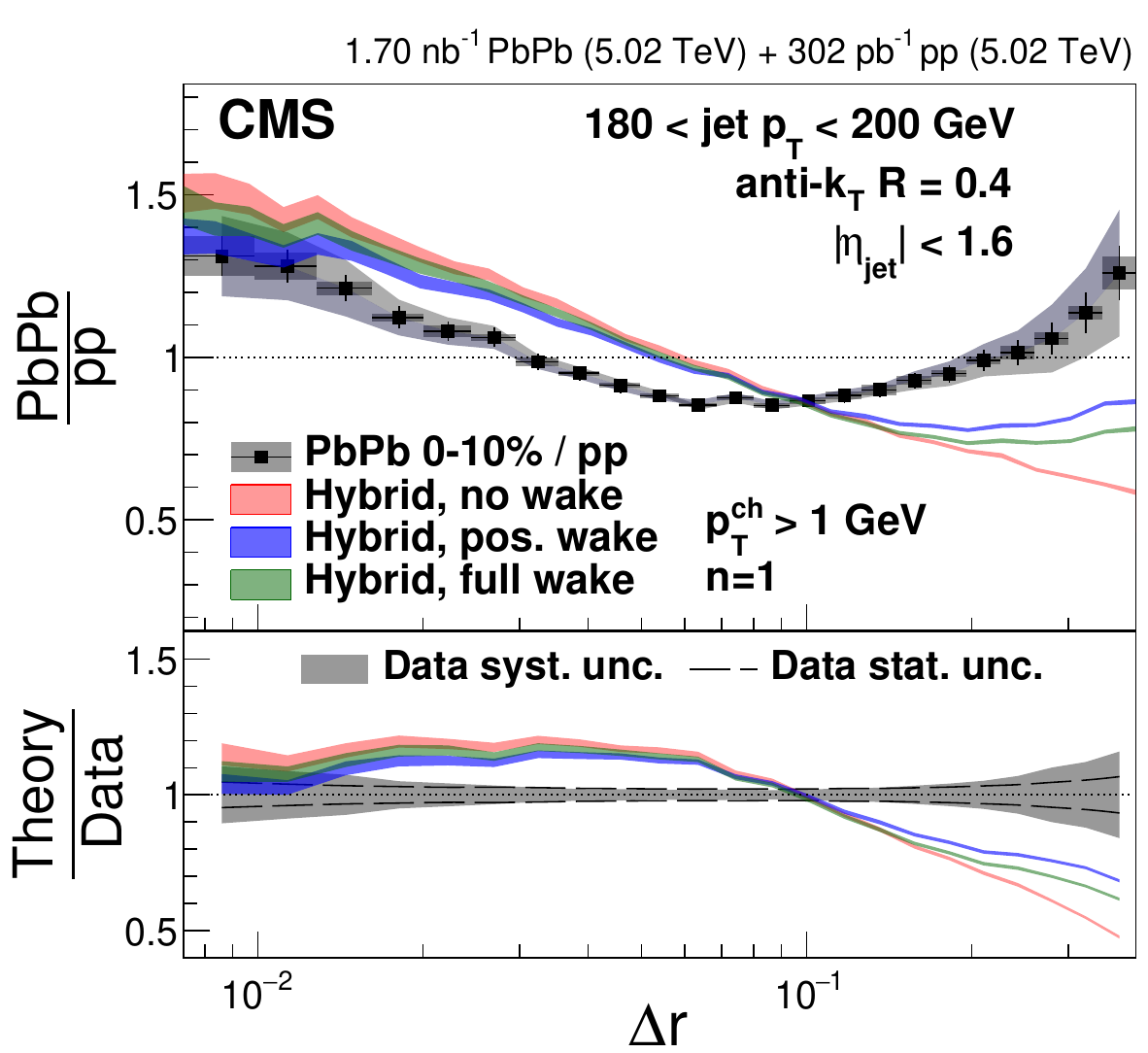}
    \caption{The \PbPb to \pp ratios of energy-energy correlators with $\ptCh > 1\GeV$ and $n=1$ are shown for $120 < \ptjet < 140\GeV$ (\cmsLeft) and $180 < \ptjet < 200\GeV$ (\cmsRight) together with the hybrid model~\cite{hybridModel1, hybridModel2, hybridModel3} predictions with three different wake settings. In the lower panels, the experimental uncertainties are indicated by the bands around unity.}
  \label{fig:hybridModelRatio}
  }
\end{figure}

To better understand the observed medium modifications to energy-energy correlators, we compare the results to different model predictions, focusing on the high-$\Delta r$ region. First, specifically to study the jet wake effects, we compare the \PbPb to \pp ratios with the hybrid model predictions in Fig.~\ref{fig:hybridModelRatio}. The hybrid model predictions are shown with three different wake settings: no wake, only positive wake, and full wake. The ``only positive'' configuration disregards the contribution from energy-momentum depletion the wake leaves to the quark-gluon plasma, while this is included in the ``full wake'' contribution. It can be seen that in both of the shown $\ptjet$ selections, the change of slope from descending to ascending in the ratio only happens if the wake contributions are included. However, the location of the turning point and the magnitude of enhancement are different in the hybrid model prediction compared to data. While the wake likely influences the modifications at large angles, it cannot fully explain the observed effects.

\begin{figure}[h]
  \centering{
    \includegraphics[width=0.48\textwidth]{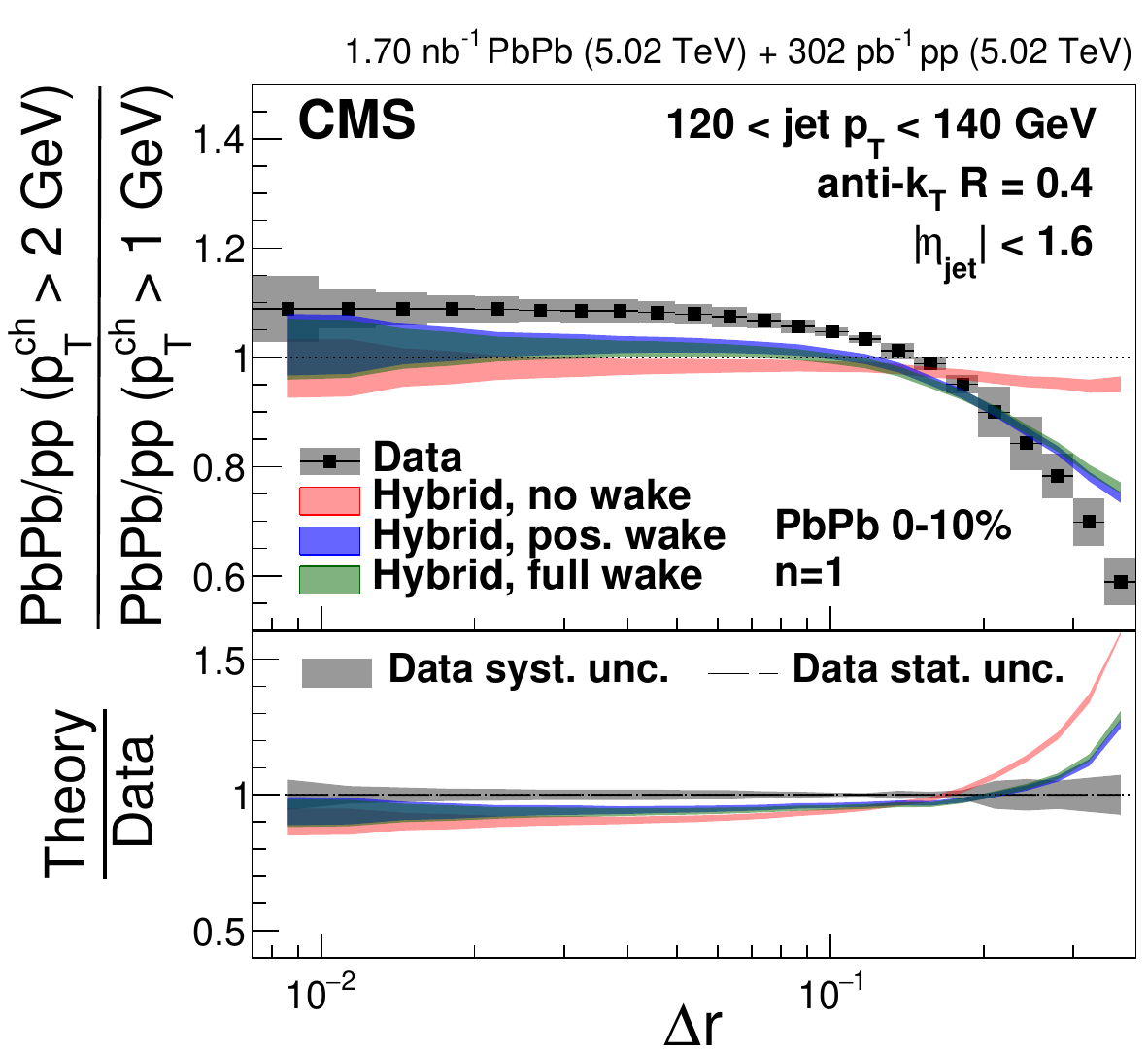}
    \includegraphics[width=0.48\textwidth]{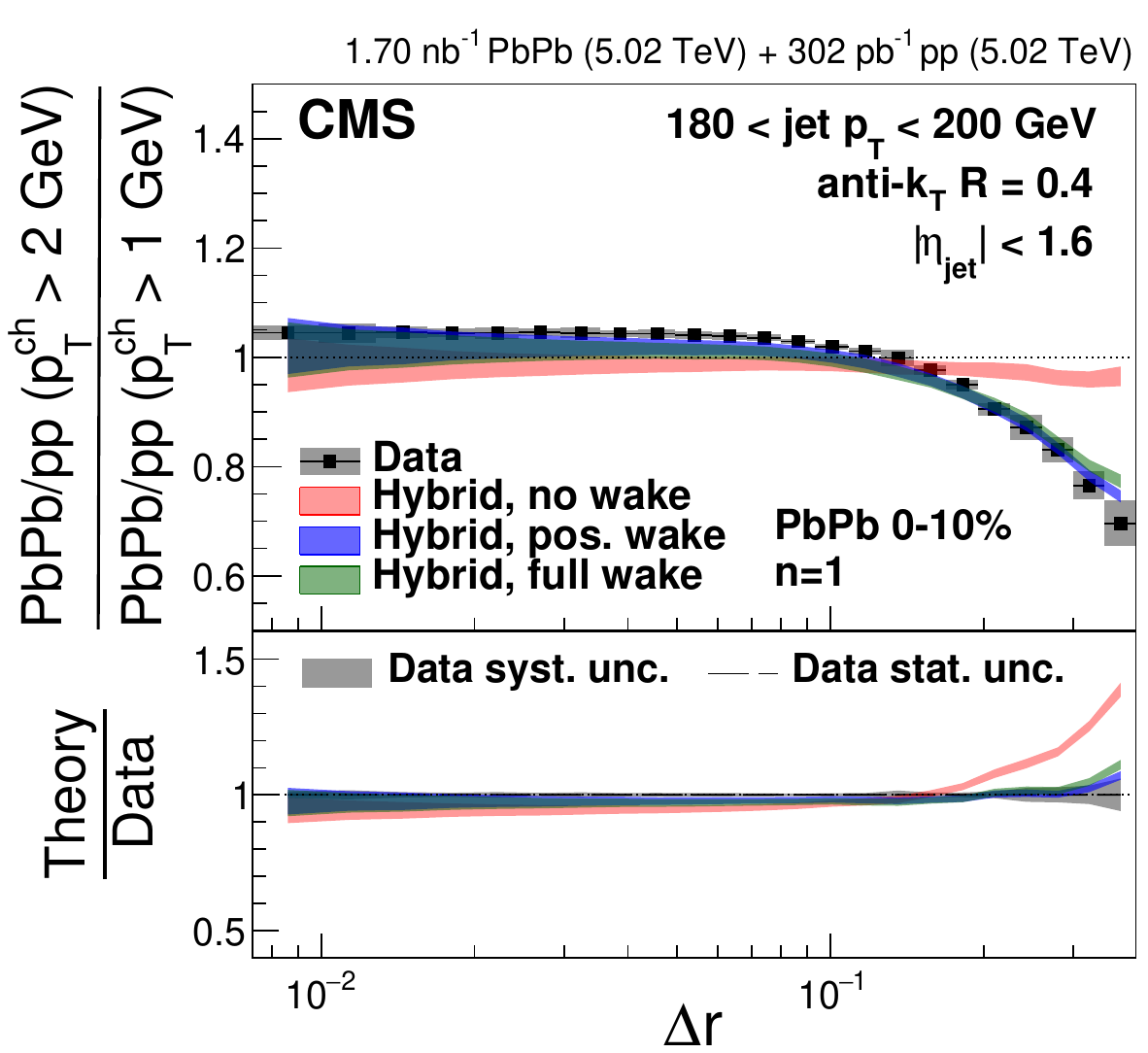}
    \caption{The double ratios of $n=1$ \PbPb to \pp single ratios with $\ptCh > 2\GeV$ to $\ptCh > 1\GeV$ are shown for $120 < \ptjet < 140\GeV$ (\cmsLeft) and $180 < \ptjet < 200\GeV$ (\cmsRight) together with the hybrid model~\cite{hybridModel1, hybridModel2, hybridModel3} predictions with three different wake settings. In the lower panels, the experimental uncertainties are indicated by the bands around unity.}
  \label{fig:hybridModelDoubleRatio}
  }
\end{figure}

We isolate the effects of the jet wake by taking a double ratio of \PbPb to \pp ratios between $\ptCh > 2\GeV$ and $\ptCh > 1\GeV$ thresholds, as shown in Fig.~\ref{fig:hybridModelDoubleRatio}. The jet wake is created by jet particles, which generally have high-\pt values, interacting with mostly low-\pt underlying event particles. Contributions from these low-high \pt pairs are highlighted in the double ratio. By isolating the wake behavior, the hybrid model predictions that include a wake show significantly better agreement with the experimental results.

\begin{figure}[h]
  \centering{
    \includegraphics[width=0.48\textwidth]{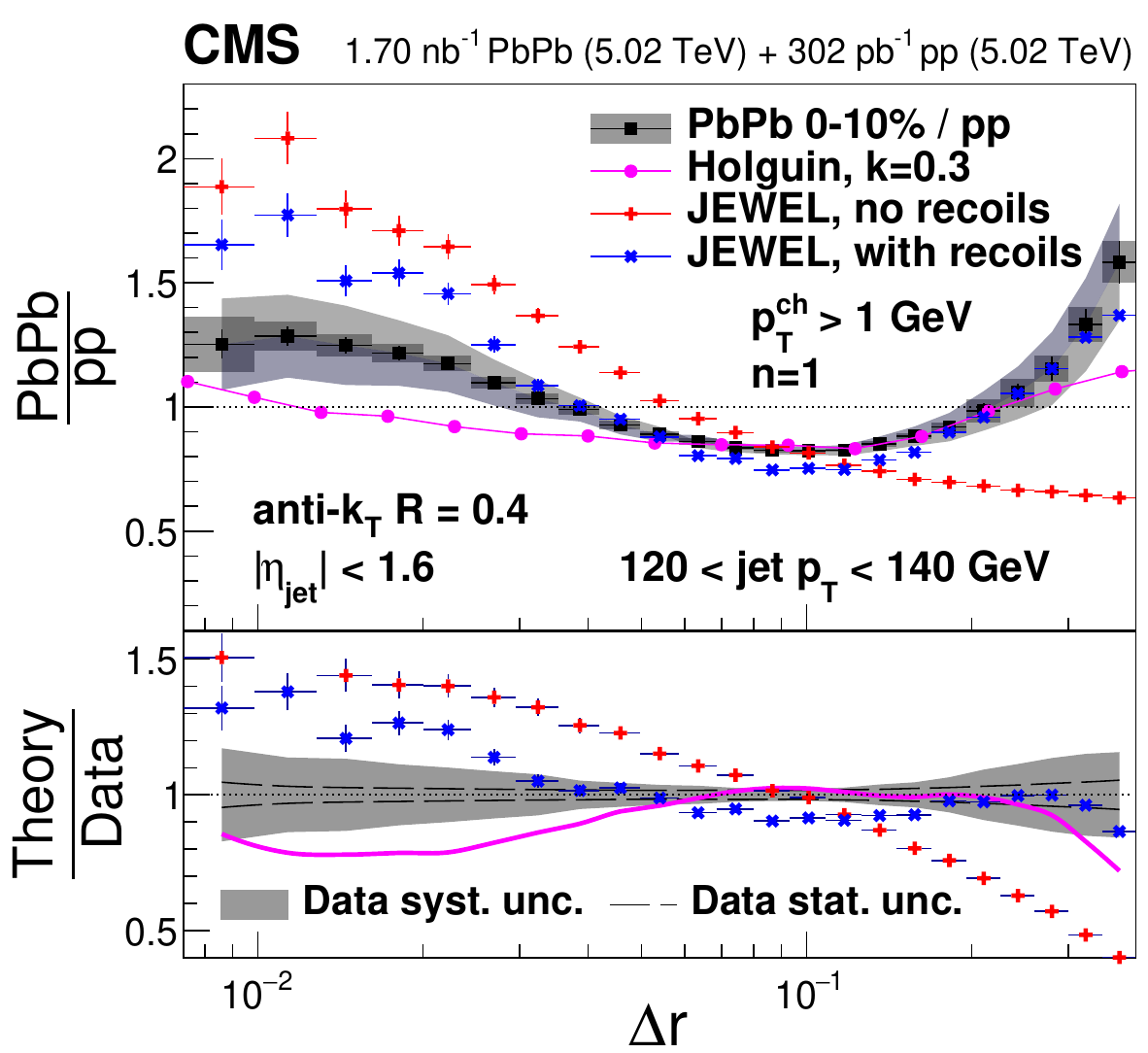}
    \includegraphics[width=0.48\textwidth]{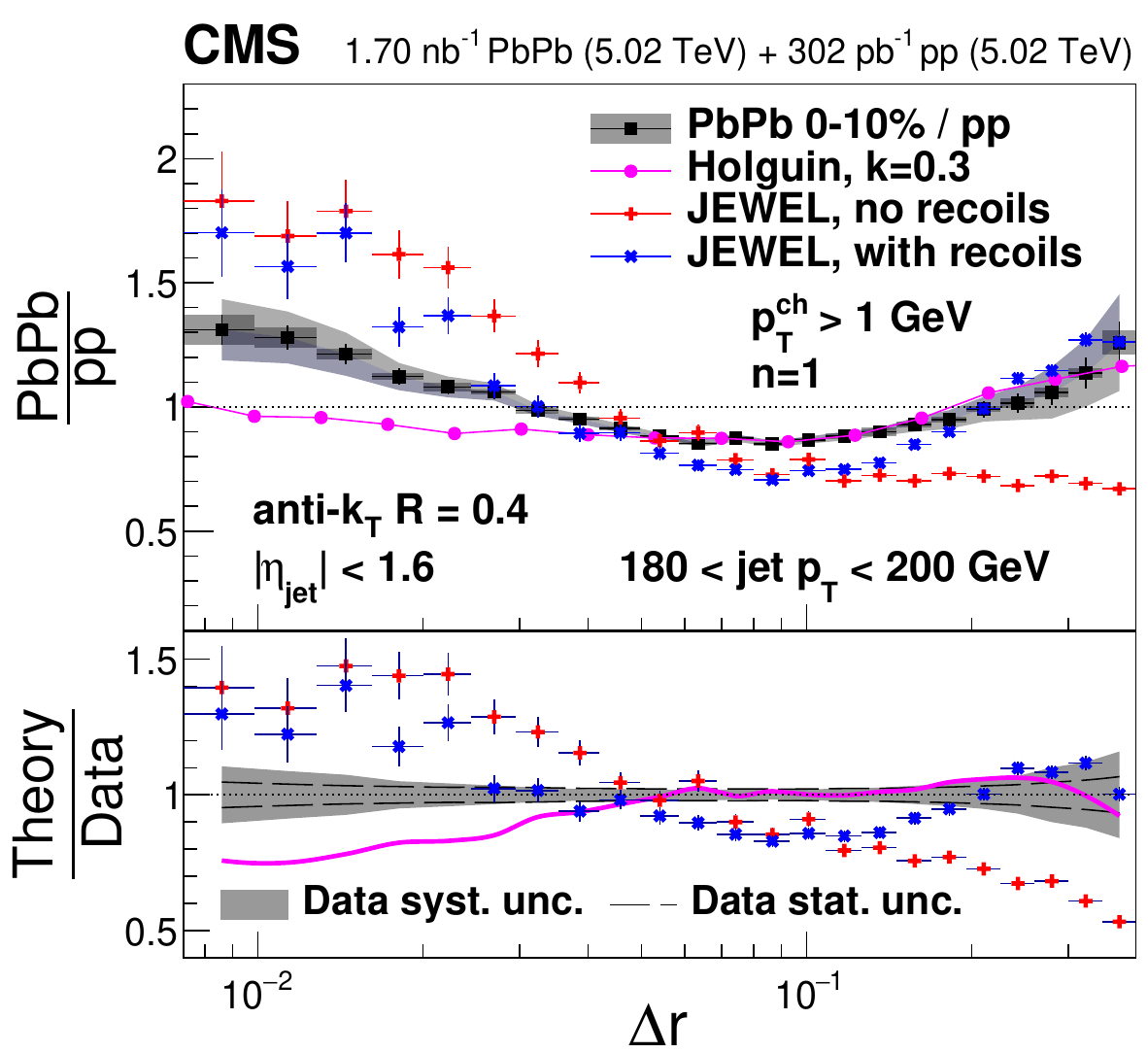}
    \caption{The \PbPb to \pp ratios of energy-energy correlators with $\ptCh > 1\GeV$ and $n=1$ are shown for $120 < \ptjet < 140\GeV$ (\cmsLeft) and $180 < \ptjet < 200\GeV$ (\cmsRight) together with predictions from perturbative calculation by Holguin and collaborators~\cite{holguinPrediction} and from \JEWEL simulation~\cite{jewel1, jewel2}. In the lower panels, the experimental uncertainties are indicated by the bands around unity.}
  \label{fig:holguinJewelRatio}
  }
\end{figure}

The measured \PbPb to \pp ratios are compared with a perturbative calculation by Holguin and collaborators~\cite{holguinPrediction} and with the \JEWEL 2.4.0 simulation~\cite{jewel1, jewel2} in Fig.~\ref{fig:holguinJewelRatio}. The perturbative calculation includes physics from color coherence effects. It has two free parameters that have been fitted to the measured data. First, there is an overall normalization that is determined by matching the integral to the data in the region $0.042 < \Delta r < 0.126$. The second parameter is $k$, the constant of proportionality between the linear density of the medium and the temperature. In practice, $k$ controls the amount of enhancement seen at large $\Delta r$. The small-$\Delta r$ region is described by a simple nonperturbative model, so the predicted values below $\Delta r < 0.042$ are not expected to be reliable. The most robust prediction of this calculation is the turn-on angle, when the descending slope in the ratio turns into an ascending one, since neither of the free parameters affects this value. It can be seen that this calculation predicts the turn-on angle well for both $\ptjet$ ranges shown in Fig.~\ref{fig:holguinJewelRatio}. This is a strong hint that this turn-on angle is indeed related to color coherence effects.

The medium response in \JEWEL is implemented by thermal partons that undergo an elastic scattering with the jet parton. The thermal parton will receive a four-momentum kick from the jet parton, which makes it recoil from its original trajectory. After this recoil, these partons do not further interact before hadronization. Comparison with data is shown with and without these recoils. It can be seen that the \JEWEL simulation with recoils and the associated thermal background subtraction has a descending slope with $\Delta r$, turning into an ascending slope around the same turn-on angle as in the data, although with poor agreement with the magnitude of the ratio. Without recoils, we see a falling trend in the ratio without a turn-on angle.

\begin{figure}[h]
  \centering{
    \includegraphics[width=0.48\textwidth]{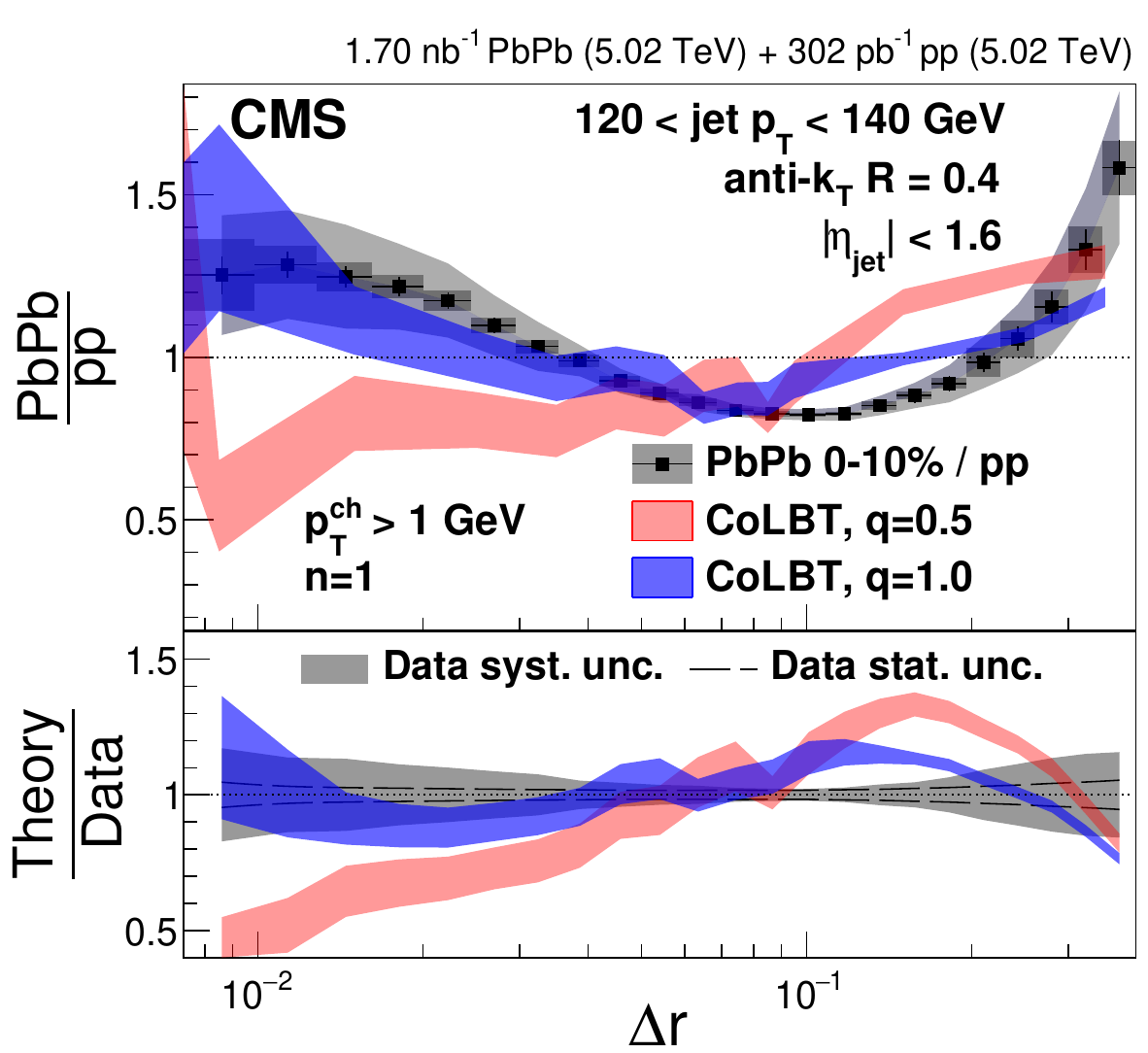}
    \caption{The \PbPb to \pp ratios of energy-energy correlators with $\ptCh > 1\GeV$ and $n=1$ are shown for $120 < \ptjet < 140\GeV$ together with predictions from CoLBT model~\cite{colbt1, colbt2, colbt3}. The q-values are shown for \PbPb collisions, the value is fixed to 0.5 for \pp. In the lower panel, the experimental uncertainties are indicated by the bands around unity.}
  \label{fig:colbtRatio}
  }
\end{figure}

The \PbPb to \pp ratio is compared to CoLBT model~\cite{colbt1, colbt2, colbt3} predictions in Fig.~\ref{fig:colbtRatio} for two different $q$ values. The $q$ parameter in the CoLBT model describes the minimum virtuality scale for the vacuum splittings, with $q = 0.5$ in the case of the \pp reference data. The CoLBT model does not describe the data qualitatively when $q$ is set to 0.5 for \PbPb collisions. In data, we see a clear shift in initial virtuality for \PbPb collisions compared to \pp collisions. This shift in virtuality in the CoLBT model is reproduced by increasing the value of the $q$ parameter, and we do indeed see much better qualitative agreement with $q=1$. However, the turn-on angle is smaller than what is seen in the data, and not enough enhancement is predicted at large angles. The enhancement at large angles in the CoLBT model is coming from the medium response and medium-induced gluon radiation contributions, with the medium response being more important. This response includes recoil partons from interactions with the jet and the underlying event. The energy-momentum depletion in the quark-gluon plasma due to recoils is modeled as negative back reaction partons.

Model comparisons for more kinematic selections are available in \suppMaterial.

\clearpage

\section{Summary}
\label{sec:conclusions}

For the first time, energy-energy correlators inside high transverse momentum (\pt) jets are measured in heavy ion collisions. The correlators from lead-lead (\PbPb) data are compared to those from proton-proton (\pp) collisions at the same nucleon-nucleon center-of-mass energy of 5.02\TeV obtained using the CMS detector. A highly differential measurement of the two-point energy correlators is presented as a function of separation in azimuth and pseudorapidity of two charged particles within a jet cone, $\Delta r$.

The time evolution of the jet fragmentation is imprinted in the shapes of the energy-energy correlator distributions. For both \PbPb and \pp collisions, we see similar large scale trends. There is a negative slope at large $\Delta r$ in the data, which is interpreted to correspond to the dynamics of the parton shower phase. The positive slope at low $\Delta r$ is expected to reflect the propagation of noninteracting hadrons. The transition region between these two limits is conjectured to be sensitive to information regarding the hadronization process. Within the same kinematic selections, the transition region in \PbPb collisions is shifted to lower $\Delta r$ values compared to \pp collisions, consistent with parton energy loss. A pronounced enhancement at large $\Delta r$ is observed in the ratio of the \PbPb to \pp distributions when using a low charged particle \pt threshold of $\ptCh > 1\GeV$ and taking the product of the \ptCh of the two particles as the momentum weight. This enhancement disappears if the sensitivity to low-\pt particles is decreased by increasing the \ptCh threshold to 2\GeV or by taking the square of the product as the momentum weight. A comparison of our data to several models suggests that medium response is partially accountable for the observed enhancement for large $\Delta r$ values. A perturbative calculation with color coherence effects also appears to favor enhancement in this ratio at large $\Delta r$ values. This measurement establishes a powerful new tool for studying the influence of medium response and color coherence on jet substructure and helps to elucidate how jets are affected by interactions with the quark-gluon plasma.

\begin{acknowledgments}
  We congratulate our colleagues in the CERN accelerator departments for the excellent performance of the LHC and thank the technical and administrative staffs at CERN and at other CMS institutes for their contributions to the success of the CMS effort. In addition, we gratefully acknowledge the computing centers and personnel of the Worldwide LHC Computing Grid and other centers for delivering so effectively the computing infrastructure essential to our analyses. Finally, we acknowledge the enduring support for the construction and operation of the LHC, the CMS detector, and the supporting computing infrastructure provided by the following funding agencies: SC (Armenia), BMBWF and FWF (Austria); FNRS and FWO (Belgium); CNPq, CAPES, FAPERJ, FAPERGS, and FAPESP (Brazil); MES and BNSF (Bulgaria); CERN; CAS, MoST, and NSFC (China); MINCIENCIAS (Colombia); MSES and CSF (Croatia); RIF (Cyprus); SENESCYT (Ecuador); ERC PRG, RVTT3 and MoER TK202 (Estonia); Academy of Finland, MEC, and HIP (Finland); CEA and CNRS/IN2P3 (France); SRNSF (Georgia); BMBF, DFG, and HGF (Germany); GSRI (Greece); NKFIH (Hungary); DAE and DST (India); IPM (Iran); SFI (Ireland); INFN (Italy); MSIP and NRF (Republic of Korea); MES (Latvia); LMTLT (Lithuania); MOE and UM (Malaysia); BUAP, CINVESTAV, CONACYT, LNS, SEP, and UASLP-FAI (Mexico); MOS (Montenegro); MBIE (New Zealand); PAEC (Pakistan); MES and NSC (Poland); FCT (Portugal);  MESTD (Serbia); MICIU/AEI and PCTI (Spain); MOSTR (Sri Lanka); Swiss Funding Agencies (Switzerland); MST (Taipei); MHESI and NSTDA (Thailand); TUBITAK and TENMAK (Turkey); NASU (Ukraine); STFC (United Kingdom); DOE and NSF (USA).

\hyphenation{Rachada-pisek} Individuals have received support from the Marie-Curie program and the European Research Council and Horizon 2020 Grant, contract Nos.\ 675440, 724704, 752730, 758316, 765710, 824093, 101115353, 101002207, and COST Action CA16108 (European Union); the Leventis Foundation; the Alfred P.\ Sloan Foundation; the Alexander von Humboldt Foundation; the Science Committee, project no. 22rl-037 (Armenia); the Fonds pour la Formation \`a la Recherche dans l'Industrie et dans l'Agriculture (FRIA-Belgium); the Beijing Municipal Science \& Technology Commission, No. Z191100007219010 and Fundamental Research Funds for the Central Universities (China); the Ministry of Education, Youth and Sports (MEYS) of the Czech Republic; the Shota Rustaveli National Science Foundation, grant FR-22-985 (Georgia); the Deutsche Forschungsgemeinschaft (DFG), among others, under Germany's Excellence Strategy -- EXC 2121 ``Quantum Universe" -- 390833306, and under project number 400140256 - GRK2497; the Hellenic Foundation for Research and Innovation (HFRI), Project Number 2288 (Greece); the Hungarian Academy of Sciences, the New National Excellence Program - \'UNKP, the NKFIH research grants K 131991, K 133046, K 138136, K 143460, K 143477, K 146913, K 146914, K 147048, 2020-2.2.1-ED-2021-00181, TKP2021-NKTA-64, and 2021-4.1.2-NEMZ\_KI-2024-00036 (Hungary); the Council of Science and Industrial Research, India; ICSC -- National Research Center for High Performance Computing, Big Data and Quantum Computing and FAIR -- Future Artificial Intelligence Research, funded by the NextGenerationEU program (Italy); the Latvian Council of Science; the Ministry of Education and Science, project no. 2022/WK/14, and the National Science Center, contracts Opus 2021/41/B/ST2/01369 and 2021/43/B/ST2/01552 (Poland); the Funda\c{c}\~ao para a Ci\^encia e a Tecnologia, grant CEECIND/01334/2018 (Portugal); the National Priorities Research Program by Qatar National Research Fund;  MICIU/AEI/10.13039/501100011033, ERDF/EU, "European Union NextGenerationEU/PRTR", and Programa Severo Ochoa del Principado de Asturias (Spain); the Chulalongkorn Academic into Its 2nd Century Project Advancement Project, and the National Science, Research and Innovation Fund via the Program Management Unit for Human Resources \& Institutional Development, Research and Innovation, grant B39G670016 (Thailand); the Kavli Foundation; the Nvidia Corporation; the SuperMicro Corporation; the Welch Foundation, contract C-1845; and the Weston Havens Foundation (USA).
\end{acknowledgments}

\bibliography{auto_generated}

\ifthenelse{\boolean{cms@external}}{}{
  \clearpage
  \appendix
  \numberwithin{figure}{section}
  \numberwithin{table}{section}
  \section{Supplemental material: model comparisons\label{app:suppMat}}
  Comparison between theoretical predictions and measured data in all kinematic ranges where predictions have been provided are presented in this supplemental material. The comparison between different models and \pp distributions are presented in Figs.~\ref{fig:ppDistributionsNominal}--\ref{fig:ppDistributionsJewelSquared}. The comparisons between different models and \PbPb distributions are in Figs.~\ref{fig:PbPbDistributionsNominal50-90}--\ref{fig:PbPbDistributionsJewelSquared}, while the \PbPb over \pp ratios are presented in Figs.~\ref{fig:PbPbRatiosNominal50-90}--\ref{fig:PbPbRatiosSquaredJewel}. The comparison between measured double ratios and hybrid calculations are in Figs.~\ref{fig:doubleRatioHybridJet50-90}--\ref{fig:doubleRatioHybridJet0-10}, while the same comparison with the JEWEL simulation is in Fig.~\ref{fig:doubleRatioJewel}. 

\begin{figure*}[h]
  \centering{
    \includegraphics[width=0.4\textwidth]{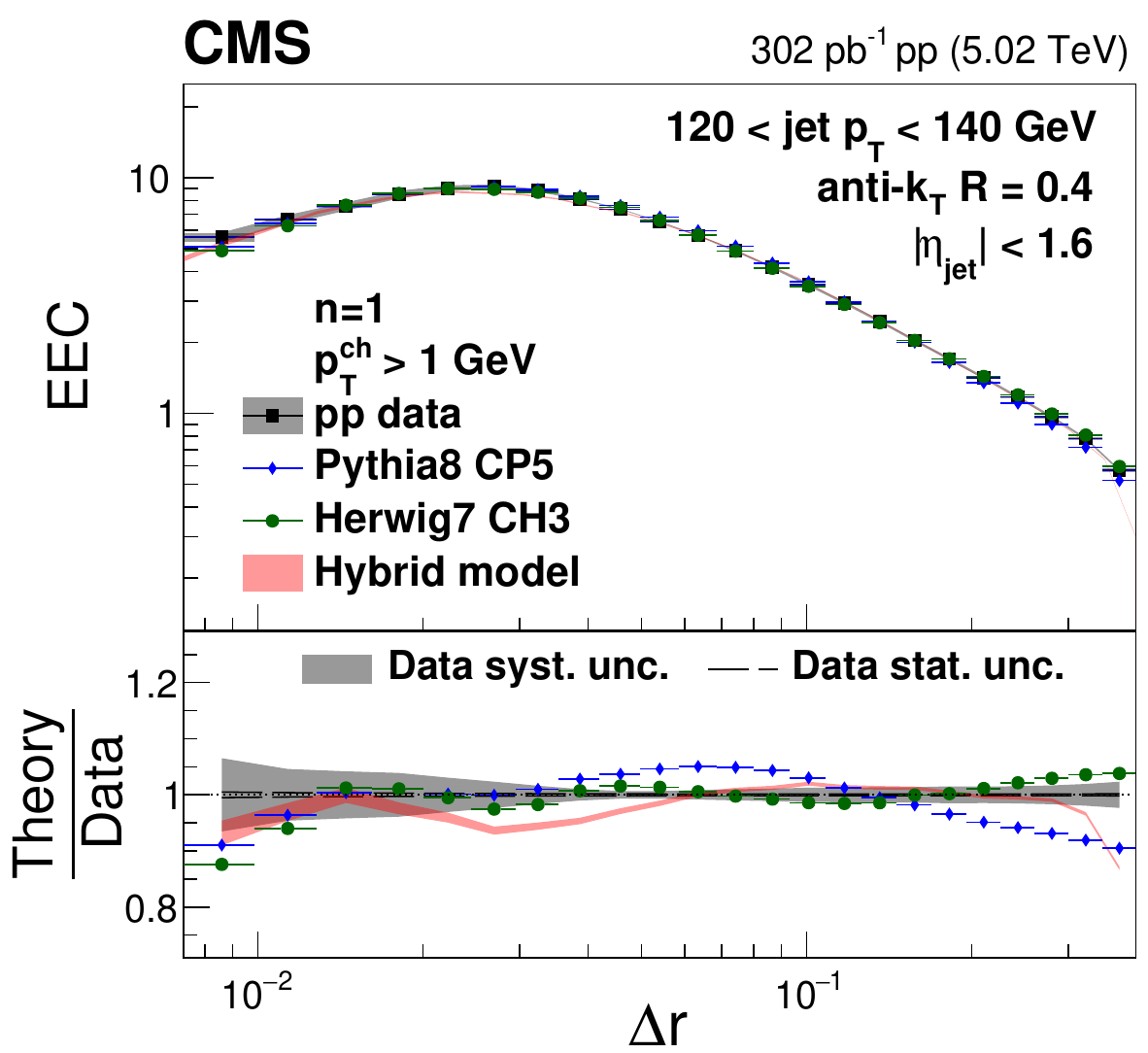}
    \includegraphics[width=0.4\textwidth]{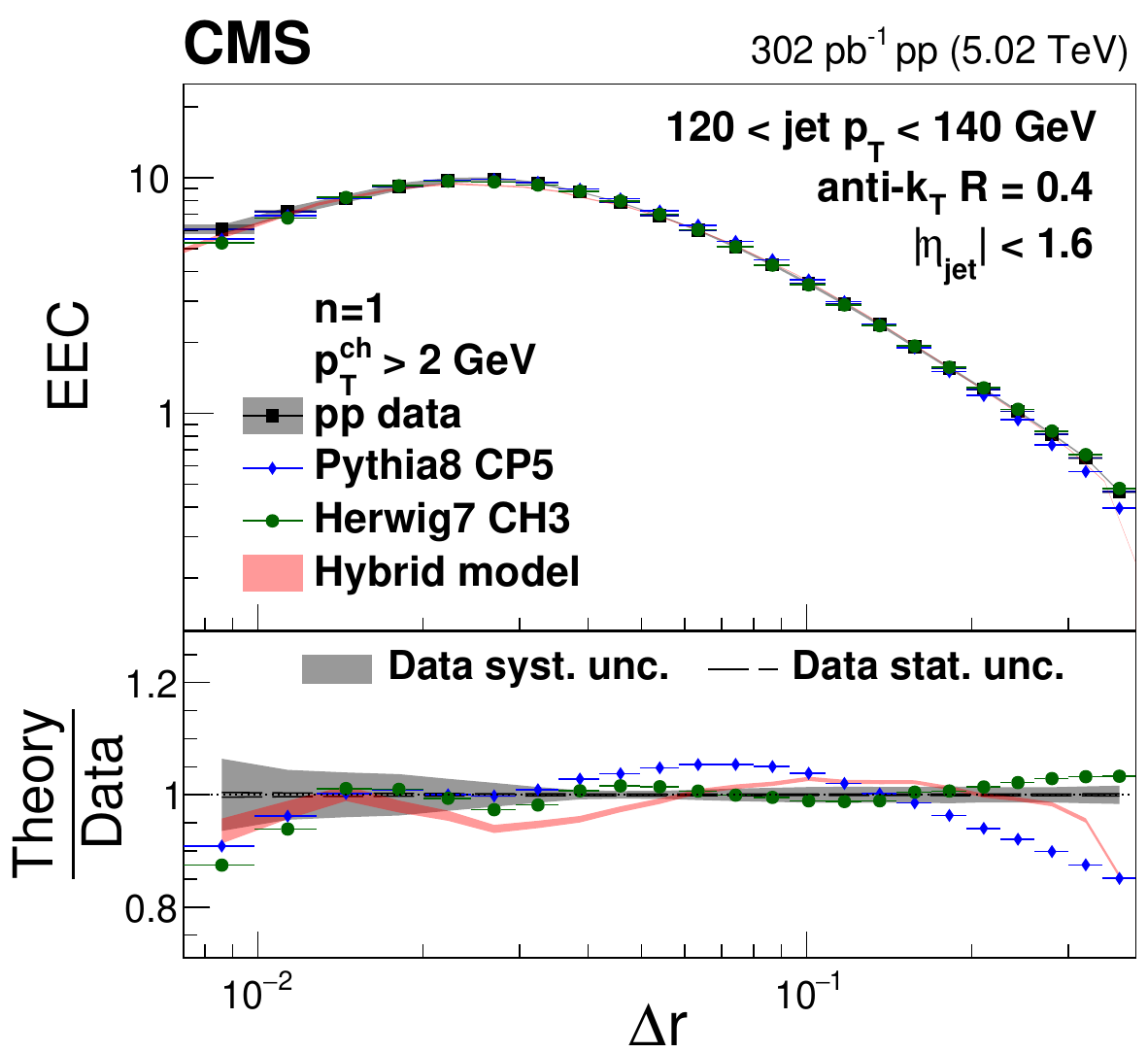}  
    \includegraphics[width=0.4\textwidth]{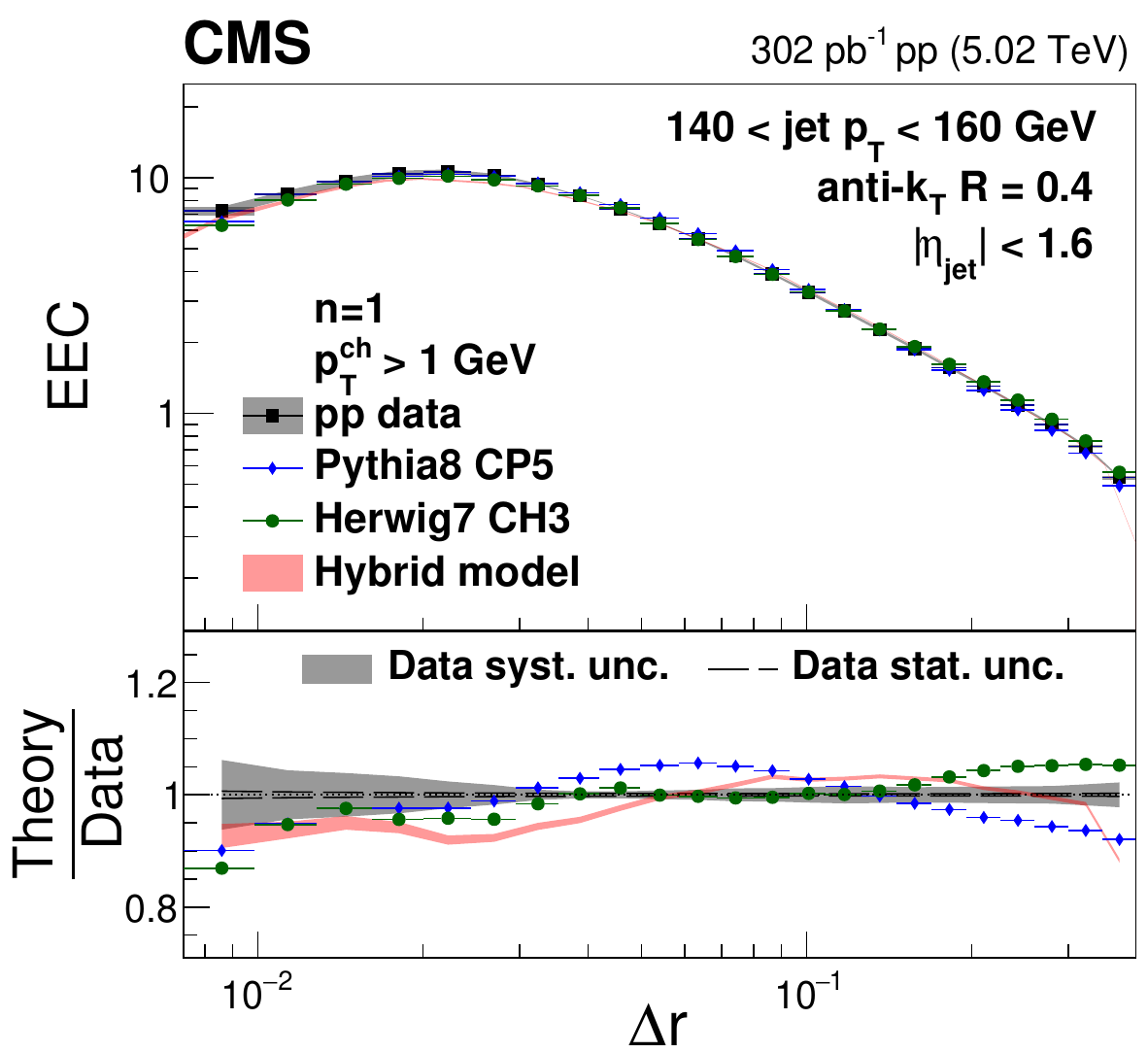}
    \includegraphics[width=0.4\textwidth]{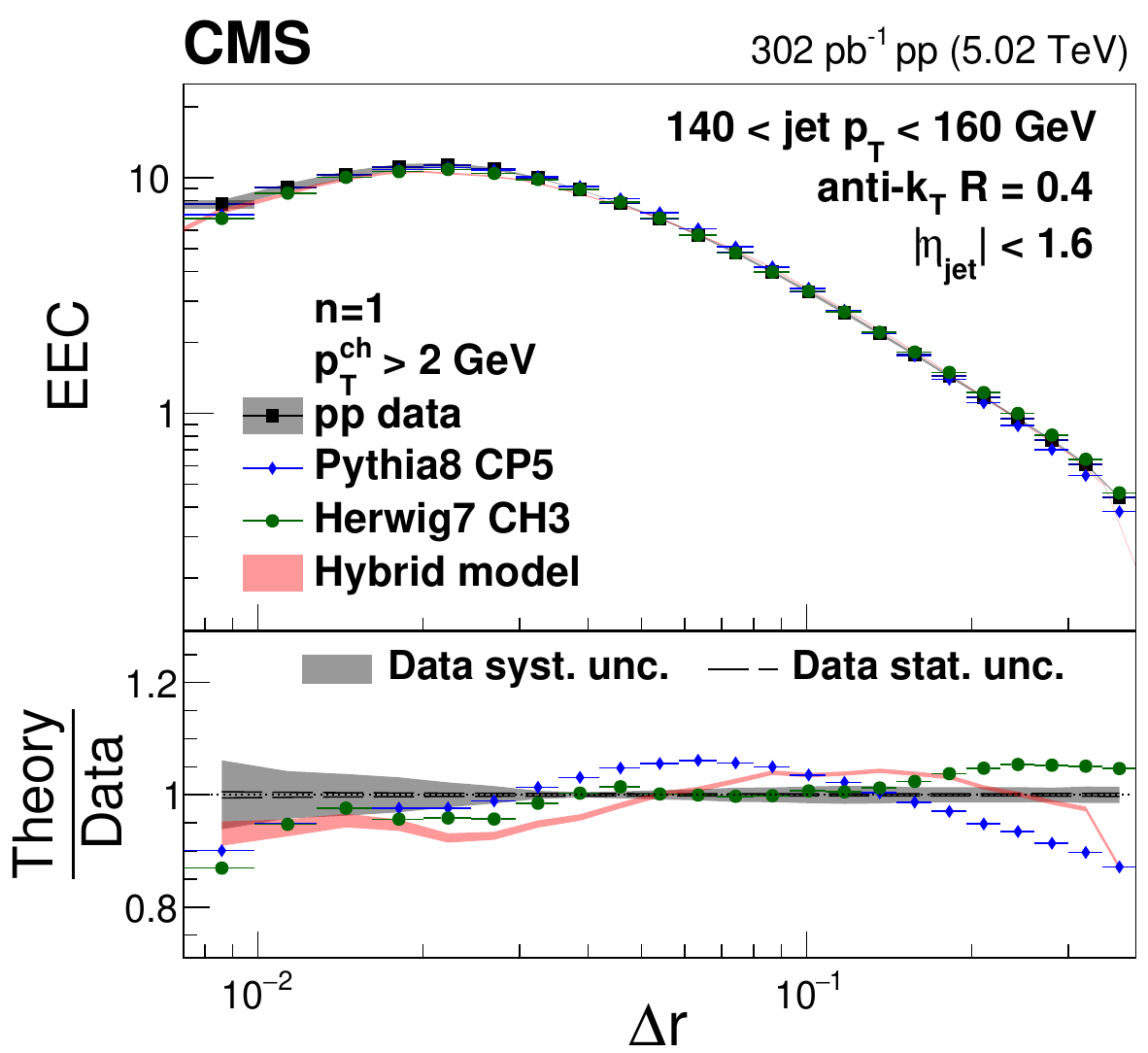}
    \includegraphics[width=0.4\textwidth]{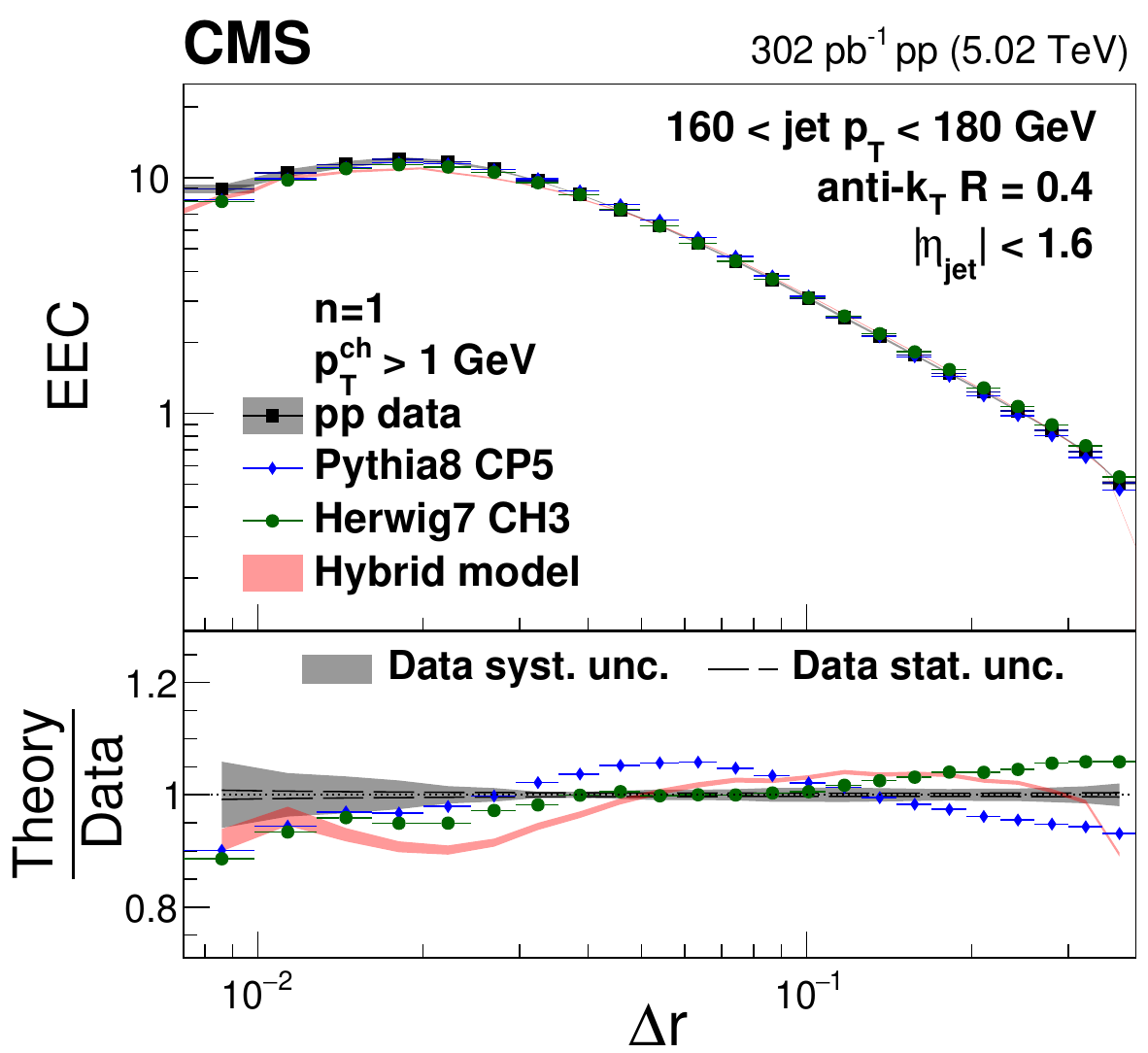}
    \includegraphics[width=0.4\textwidth]{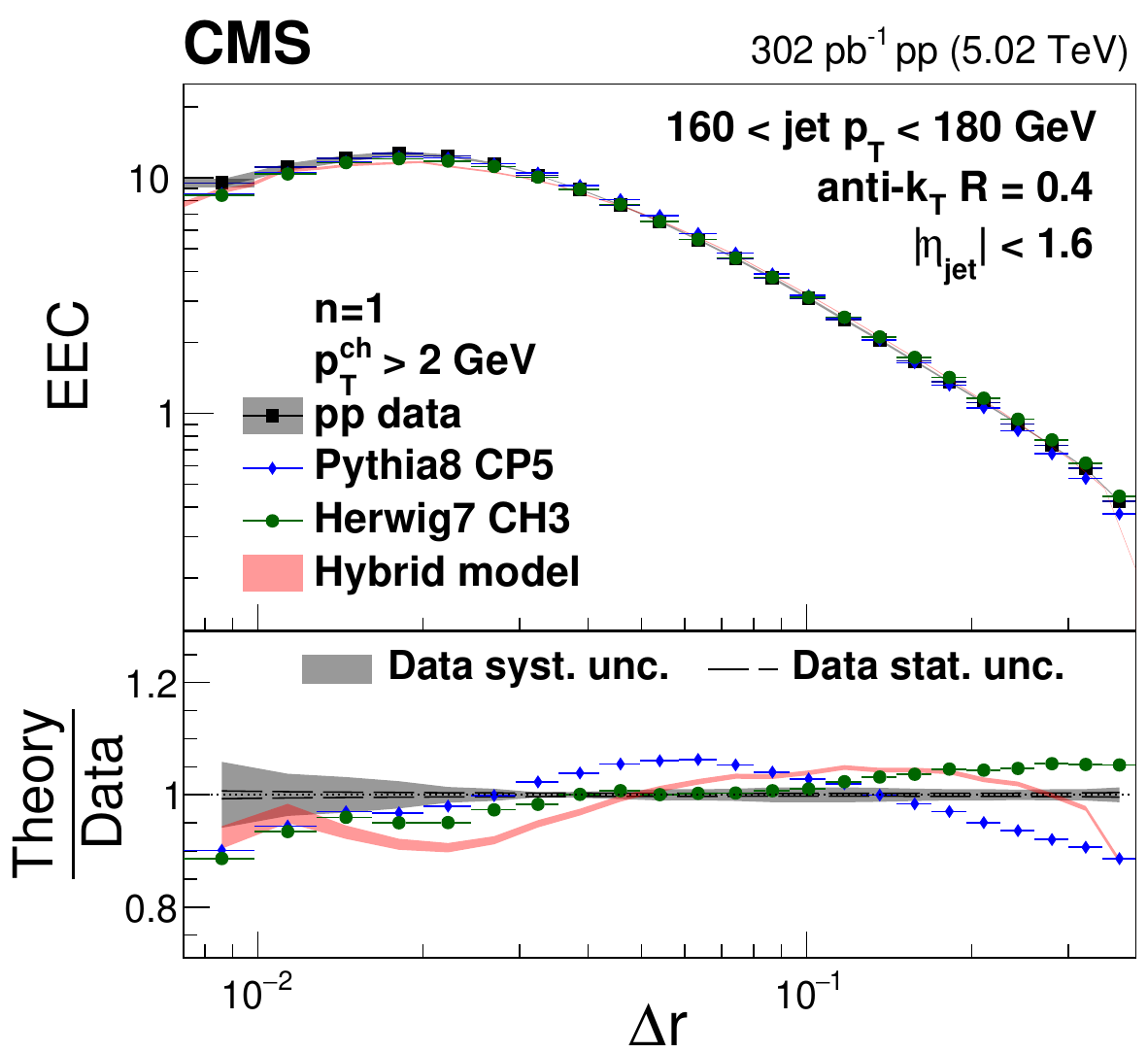}
    \includegraphics[width=0.4\textwidth]{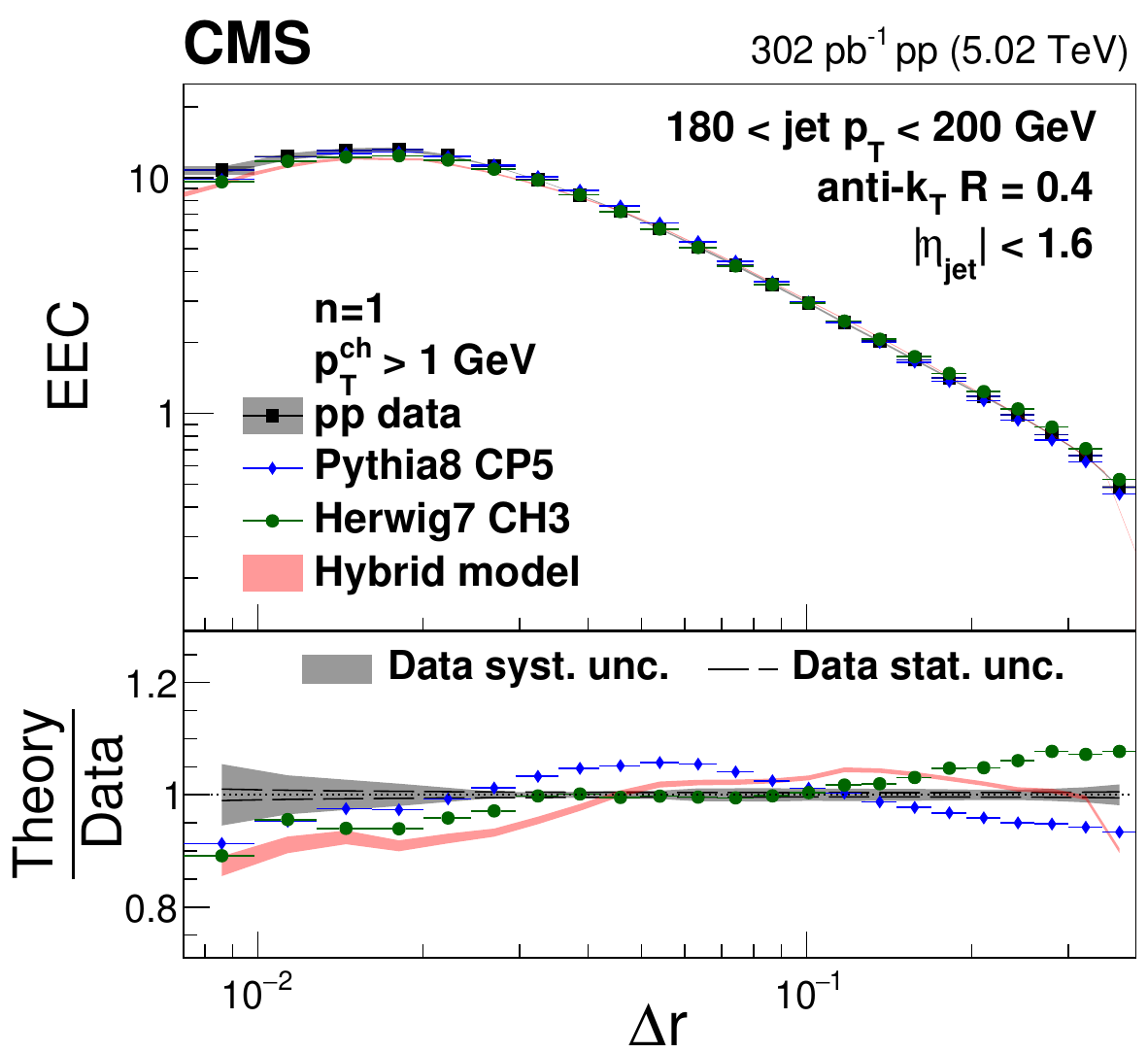}
    \includegraphics[width=0.4\textwidth]{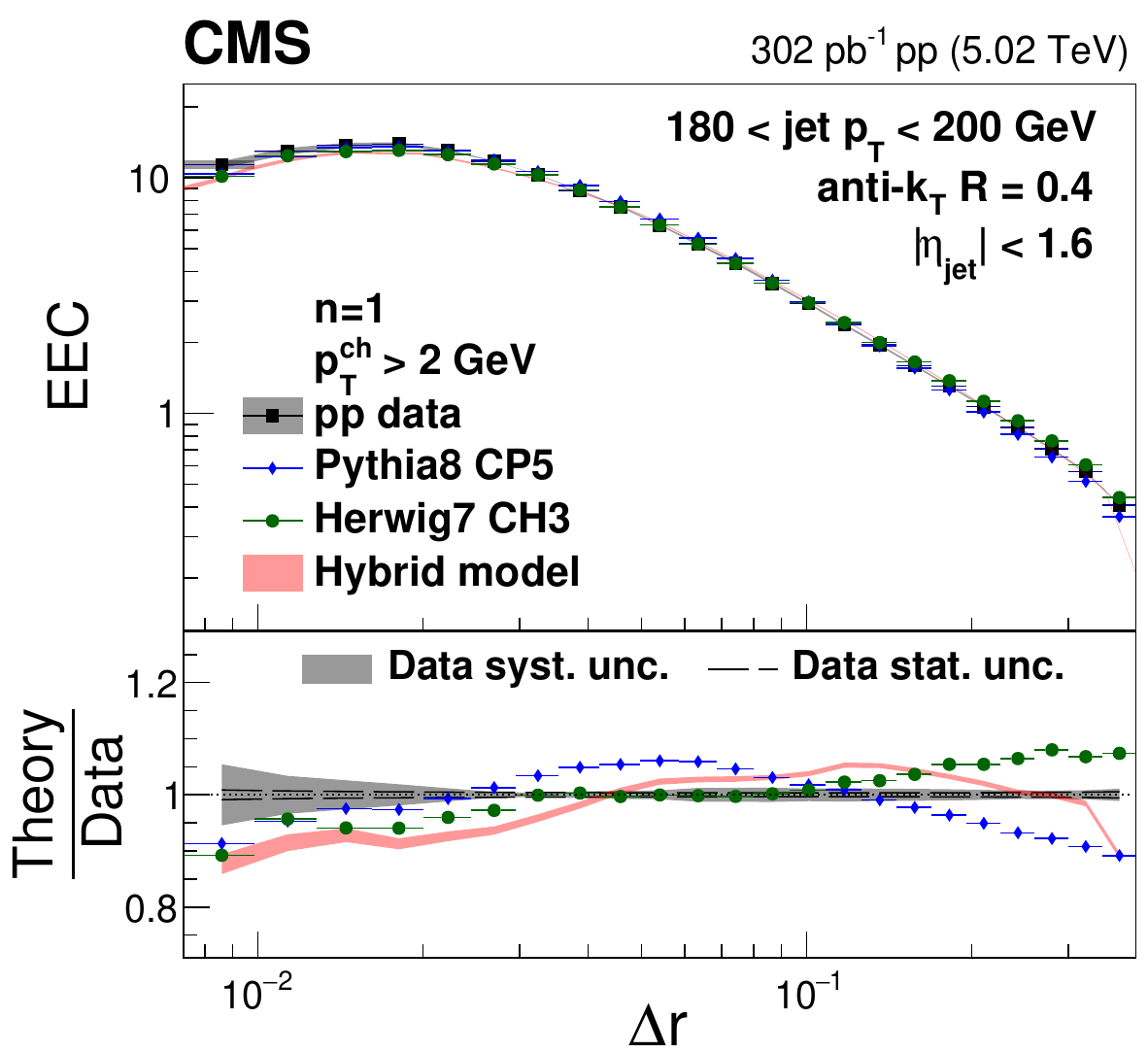}
    \caption{Energy-energy correlator distributions from pp collisions with $n=1$ in different \ptCh and \ptjet bins compared to predictions from \PYTHIA, \HERWIG, and the hybrid model.} 
    \label{fig:ppDistributionsNominal}
  } 
\end{figure*}

\begin{figure*}[h]
  \centering{
    \includegraphics[width=0.4\textwidth]{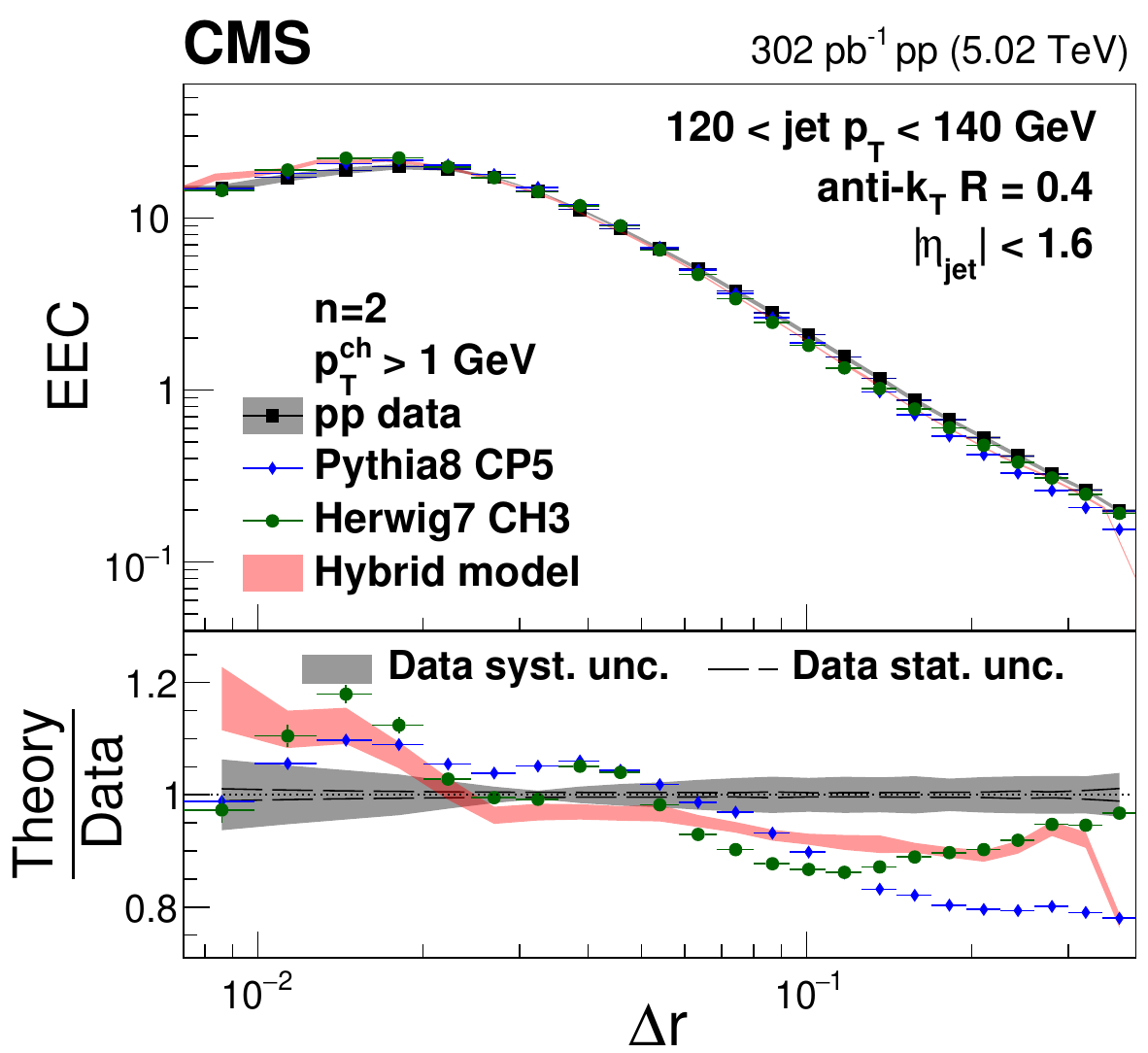}
    \includegraphics[width=0.4\textwidth]{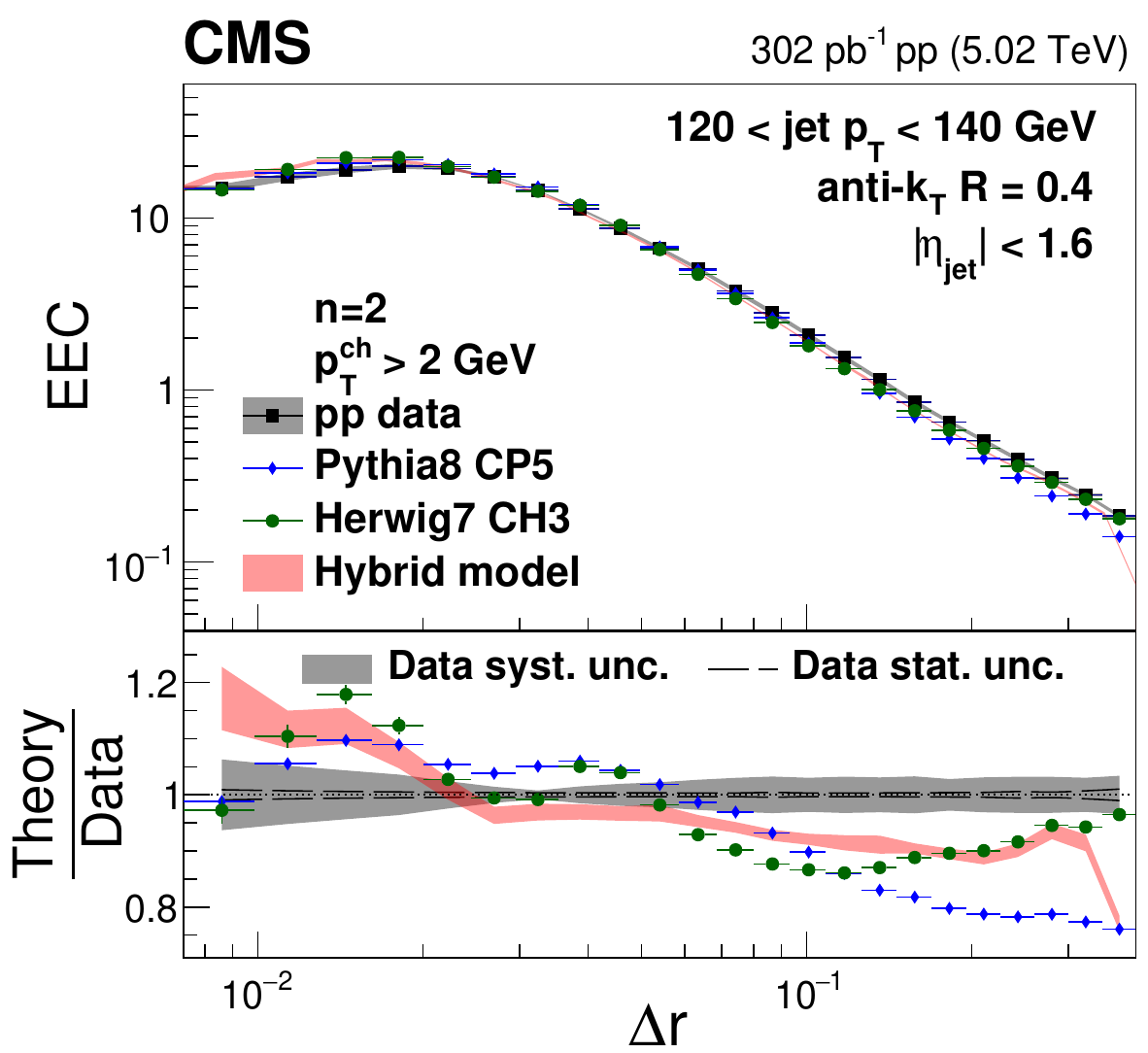}  
    \includegraphics[width=0.4\textwidth]{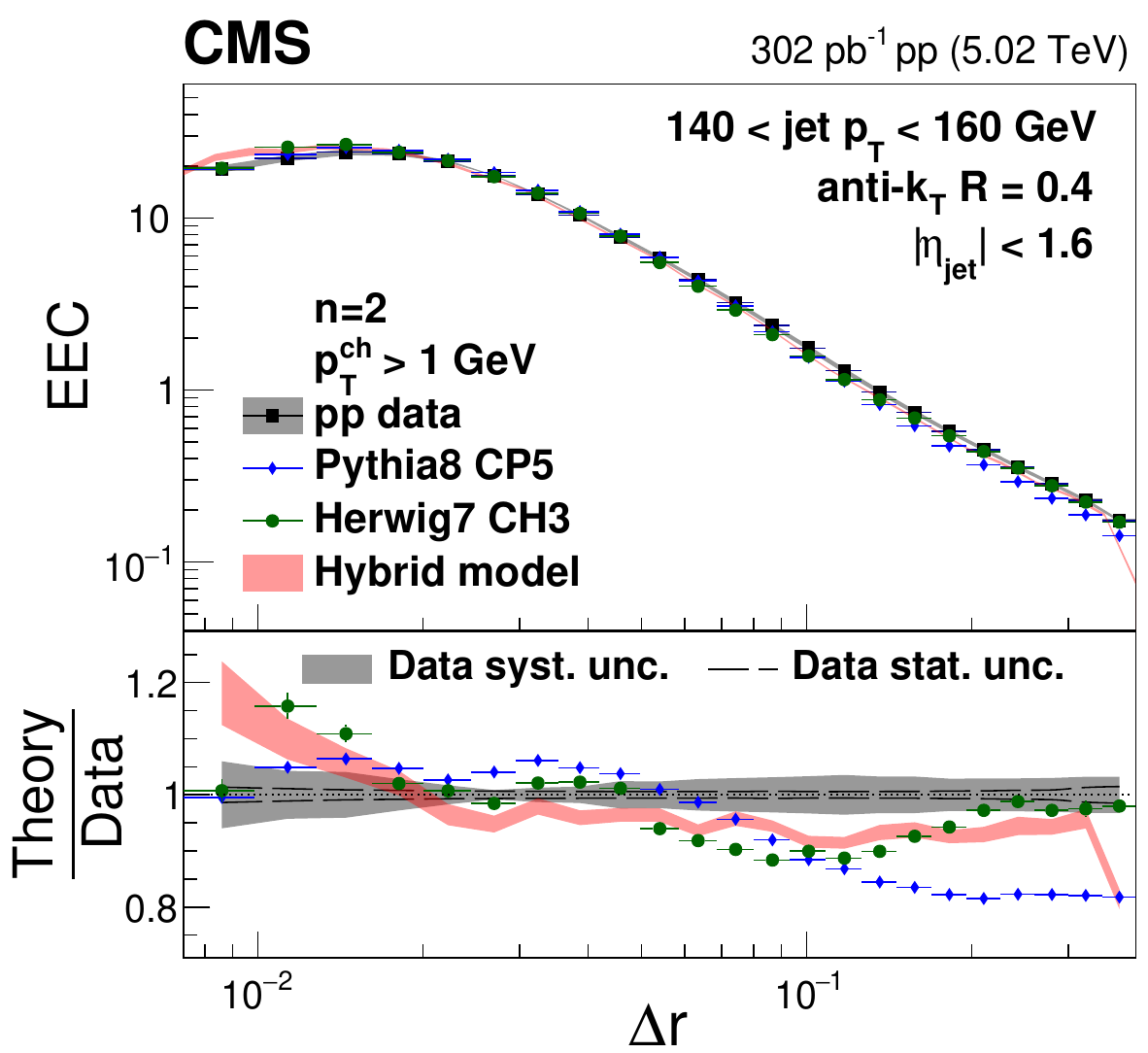}
    \includegraphics[width=0.4\textwidth]{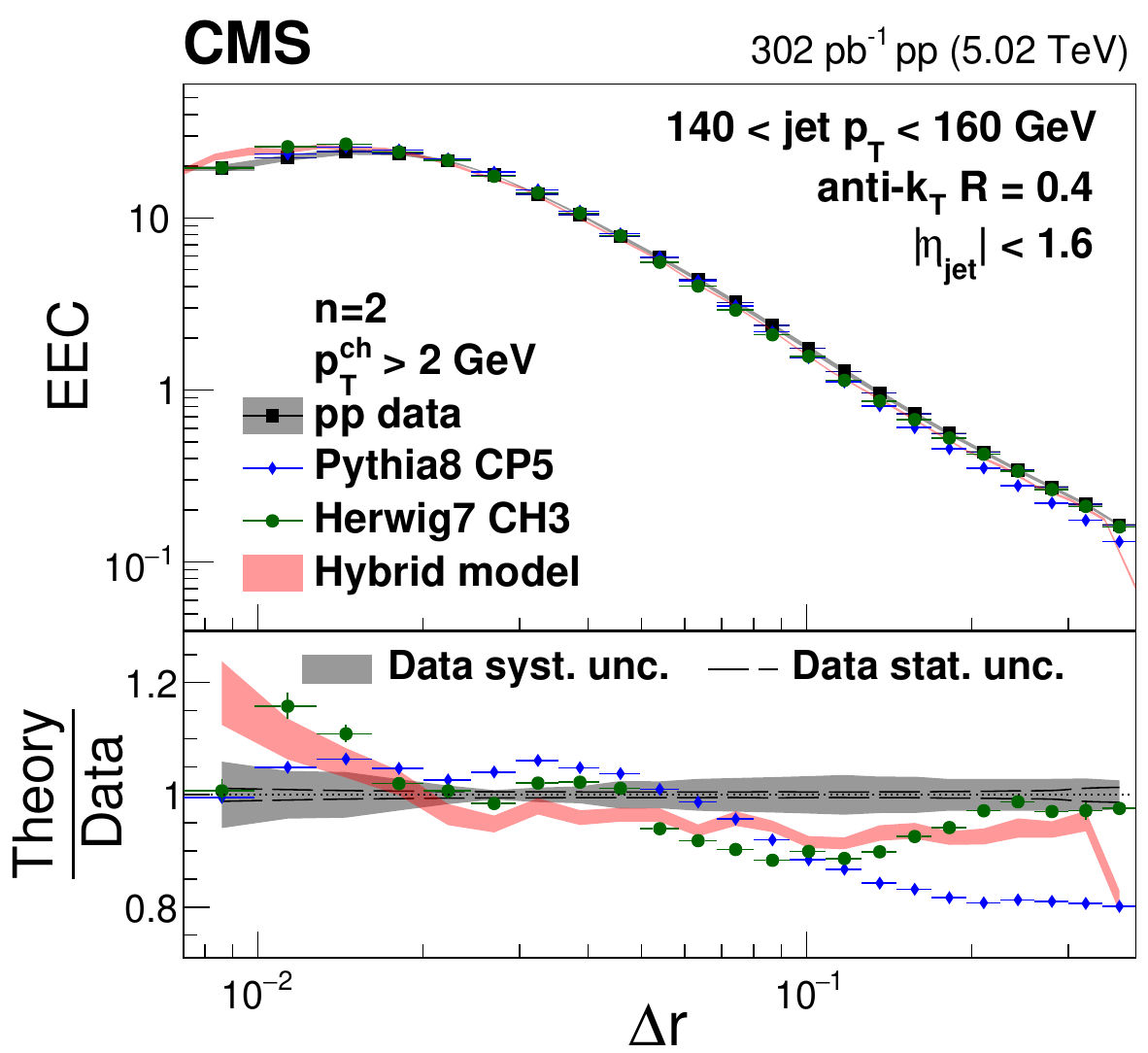}
    \includegraphics[width=0.4\textwidth]{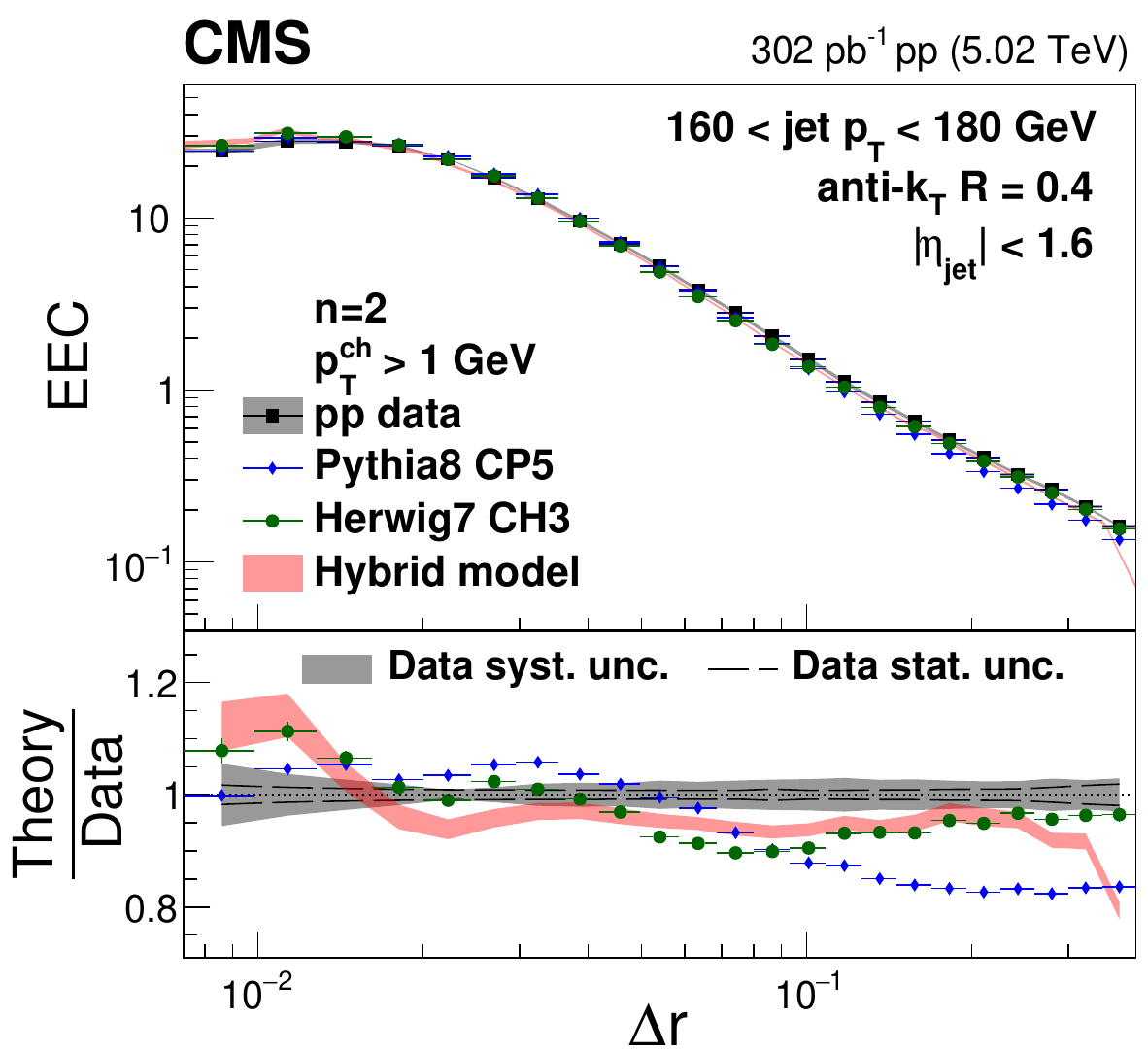}
    \includegraphics[width=0.4\textwidth]{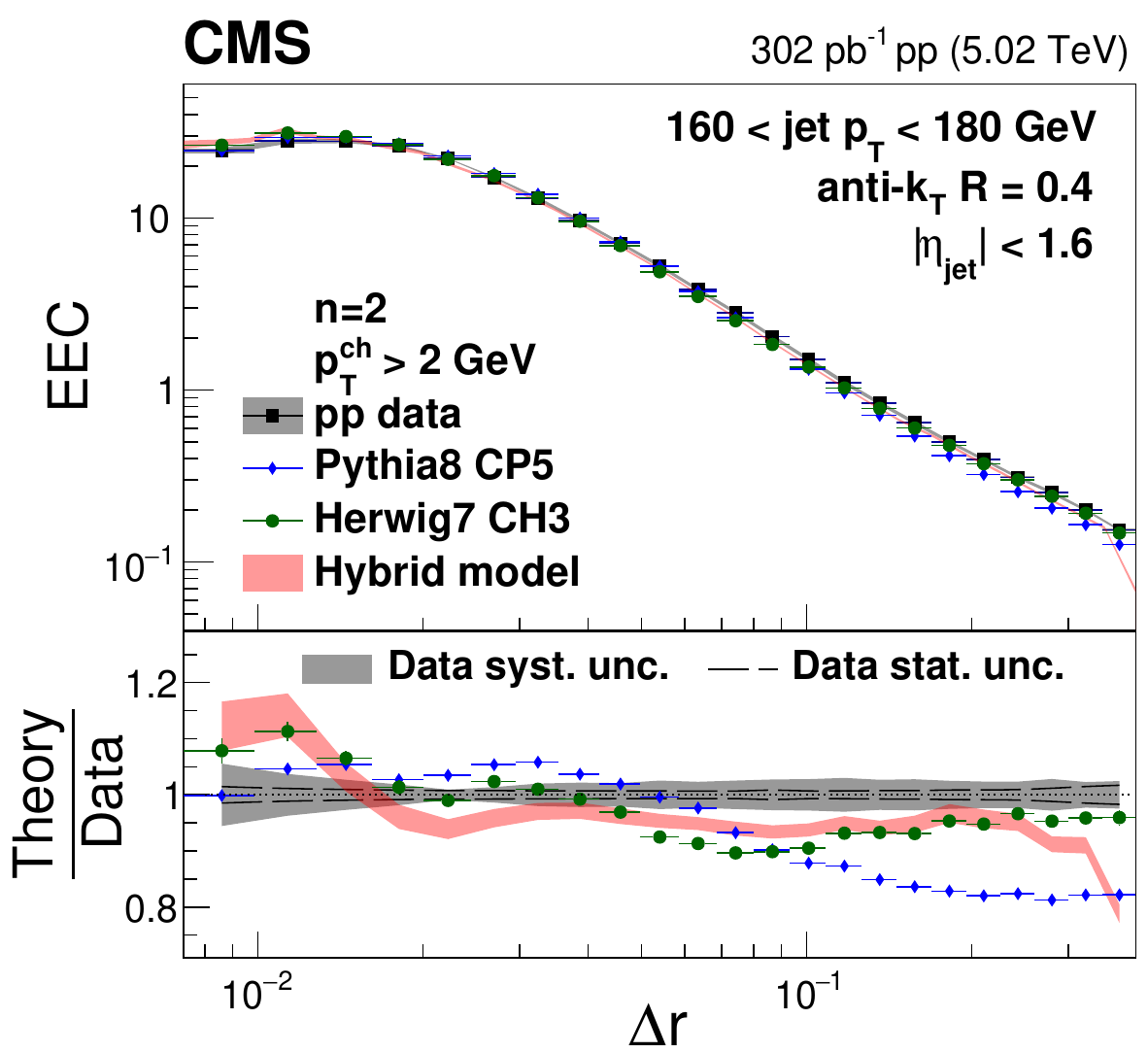}    
    \includegraphics[width=0.4\textwidth]{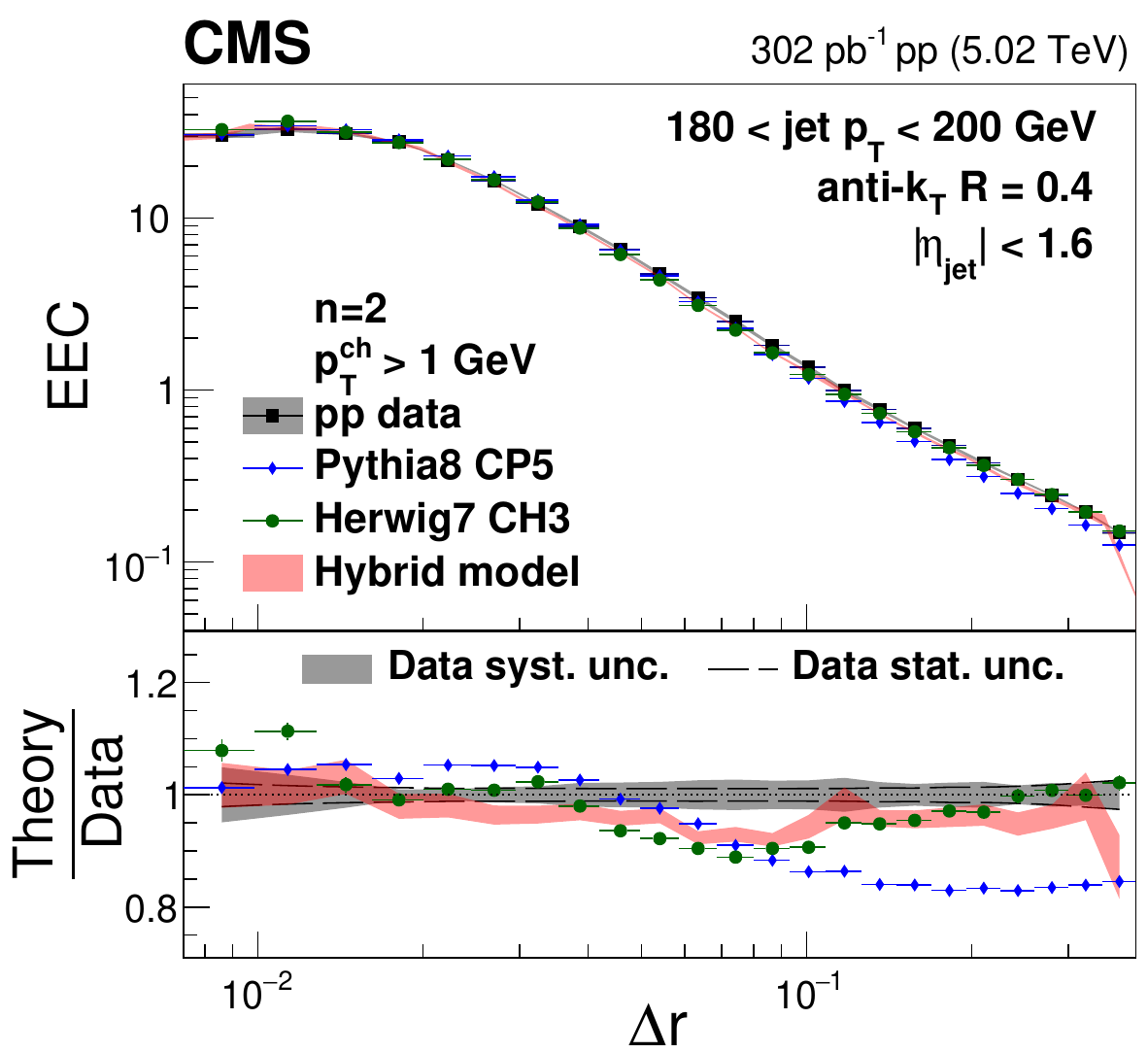}
    \includegraphics[width=0.4\textwidth]{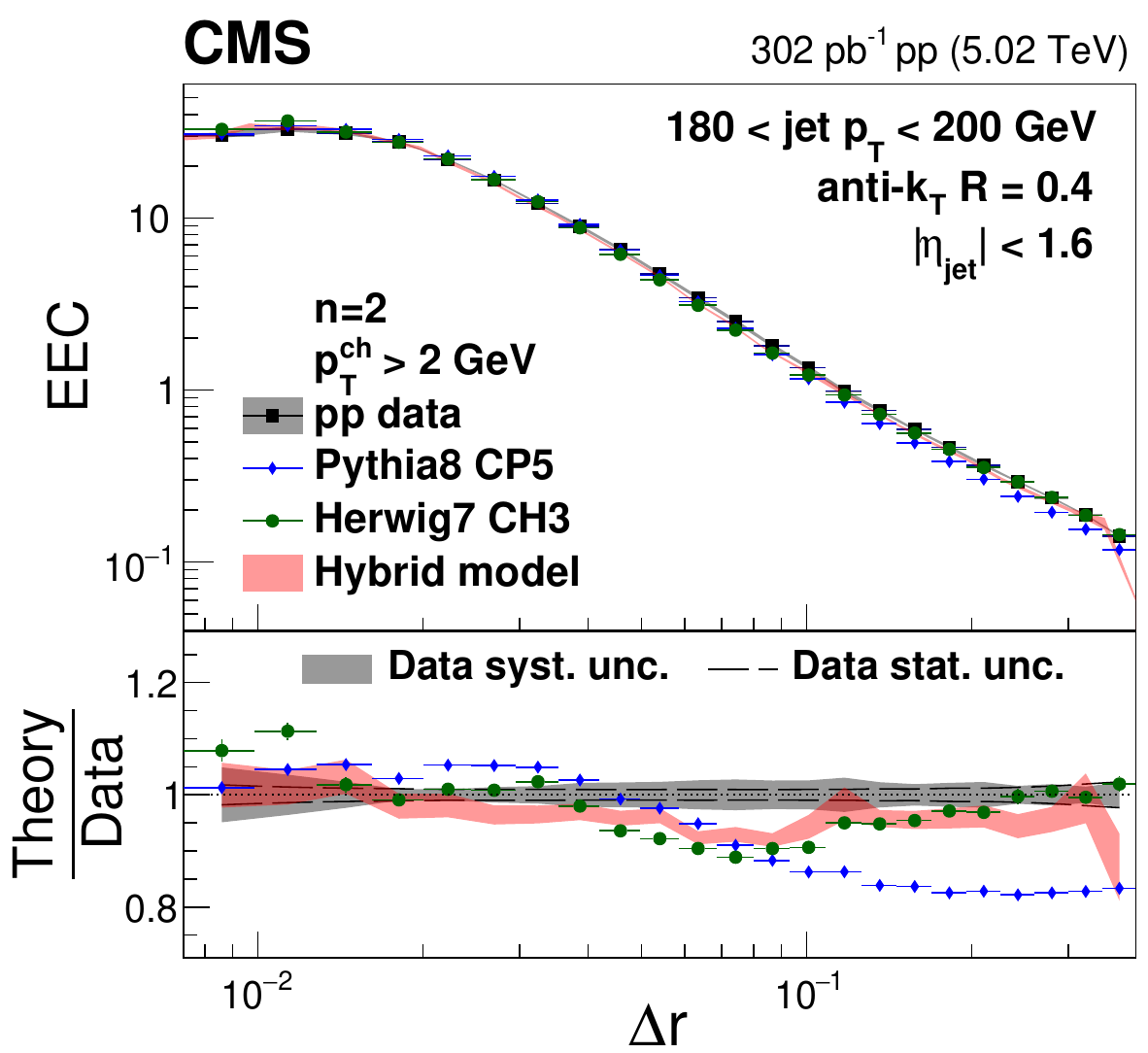}
    \caption{Energy-energy correlator distributions from \pp collisions with $n=2$ in different \ptCh and \ptjet bins compared to predictions from \PYTHIA, \HERWIG, and the hybrid model.} 
    \label{fig:ppDistributionsSquared}
  } 
\end{figure*}

\begin{figure*}[h]
  \centering{
    \includegraphics[width=0.4\textwidth]{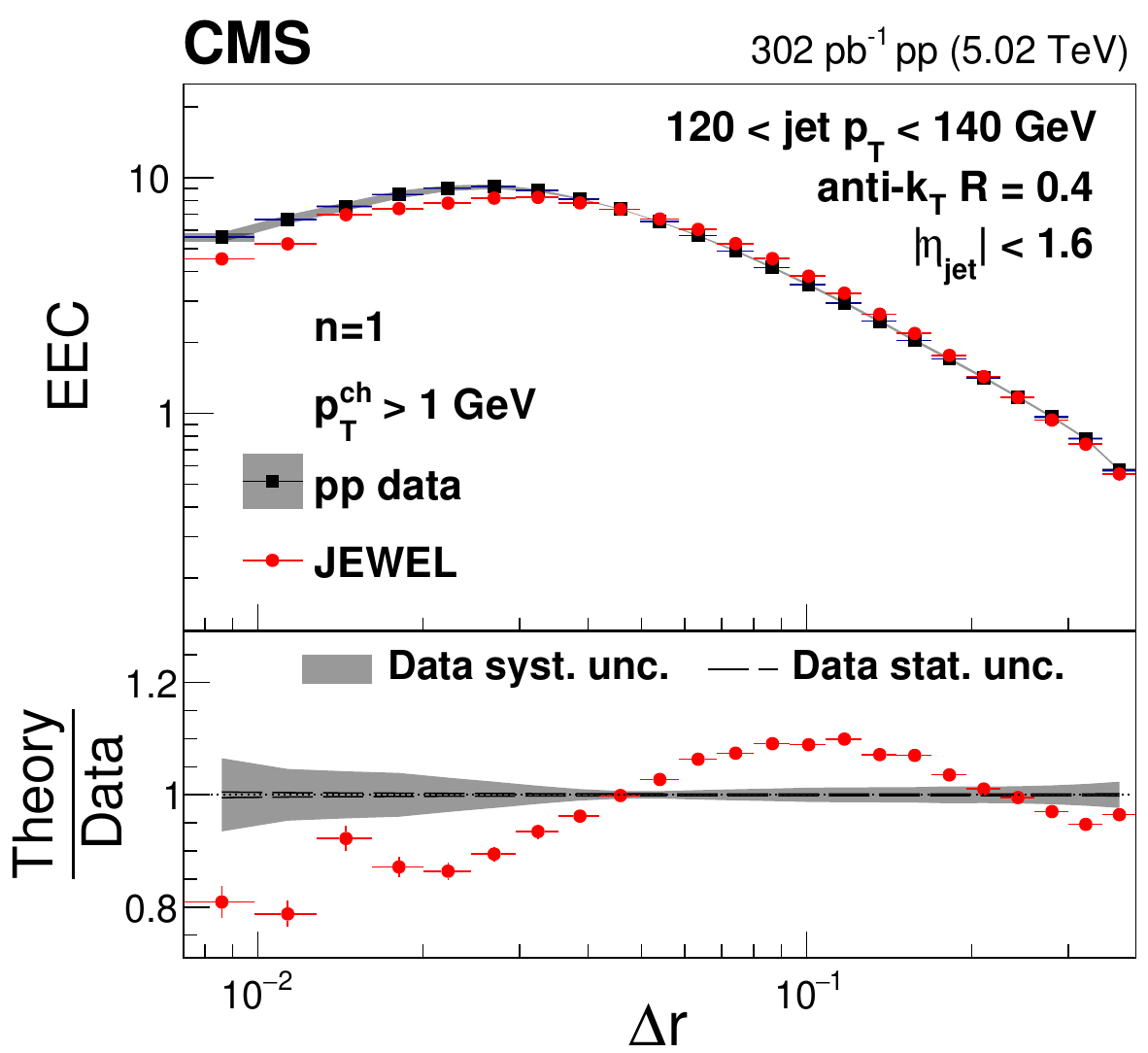}
    \includegraphics[width=0.4\textwidth]{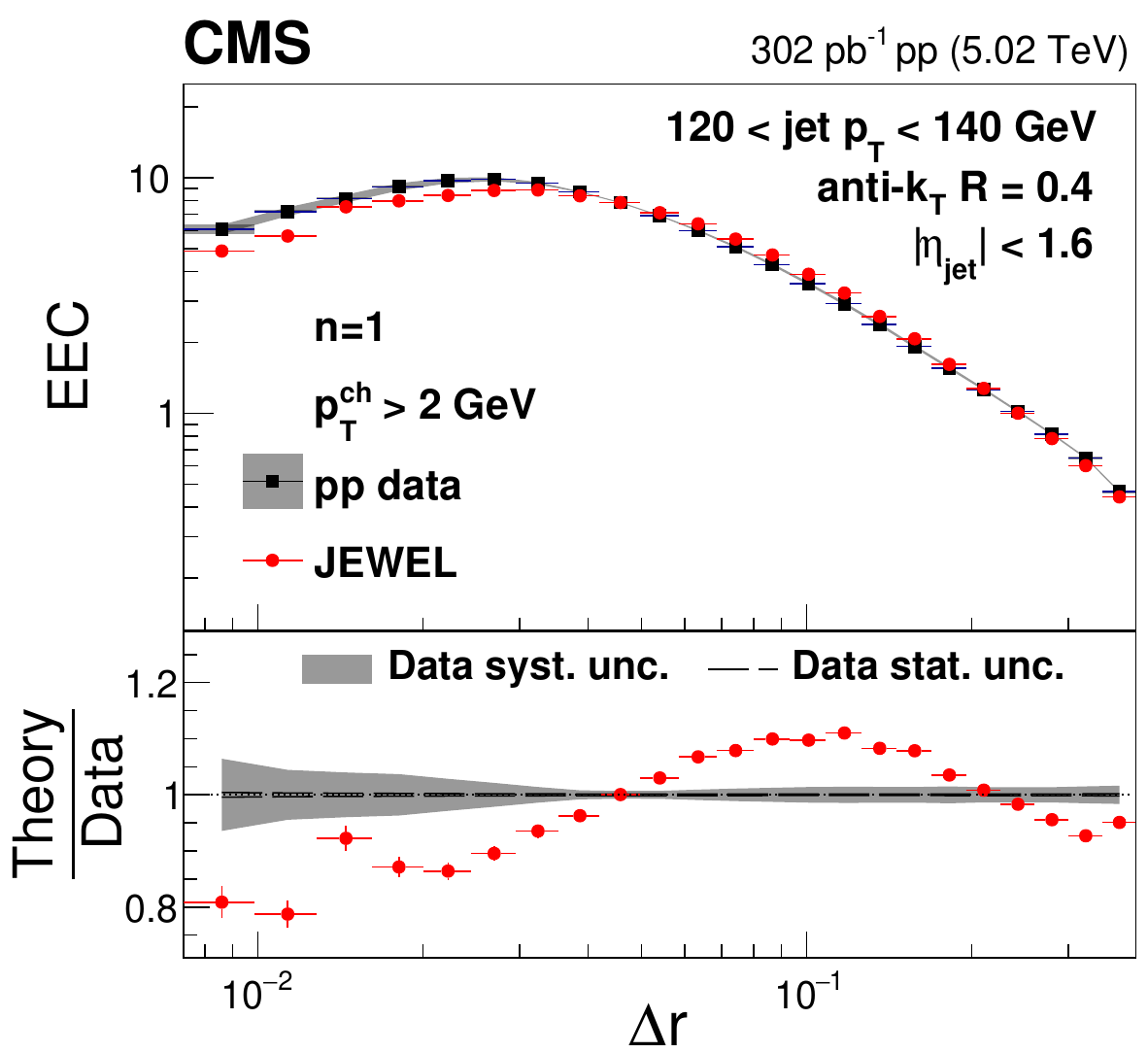}  
    \includegraphics[width=0.4\textwidth]{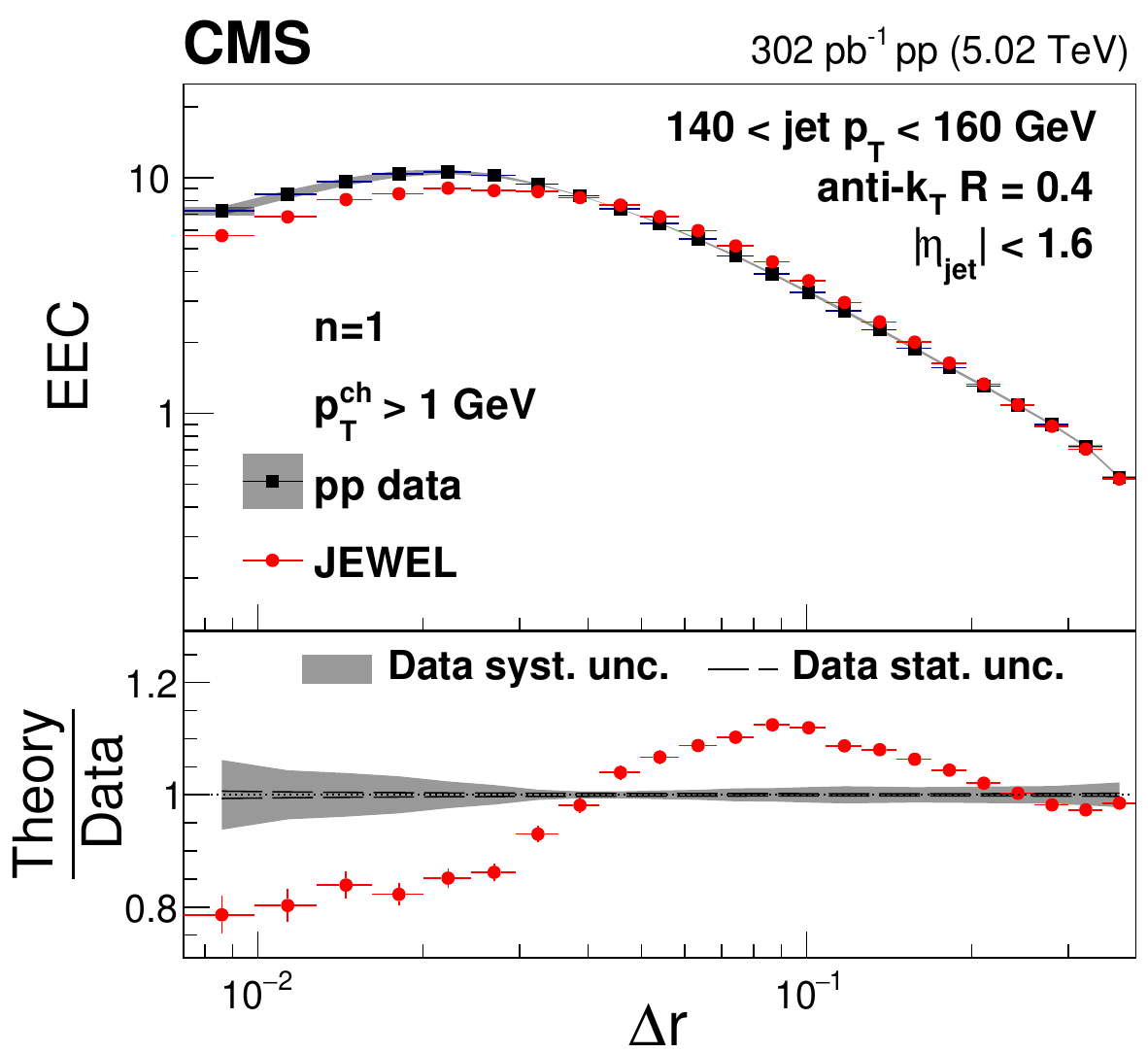}
    \includegraphics[width=0.4\textwidth]{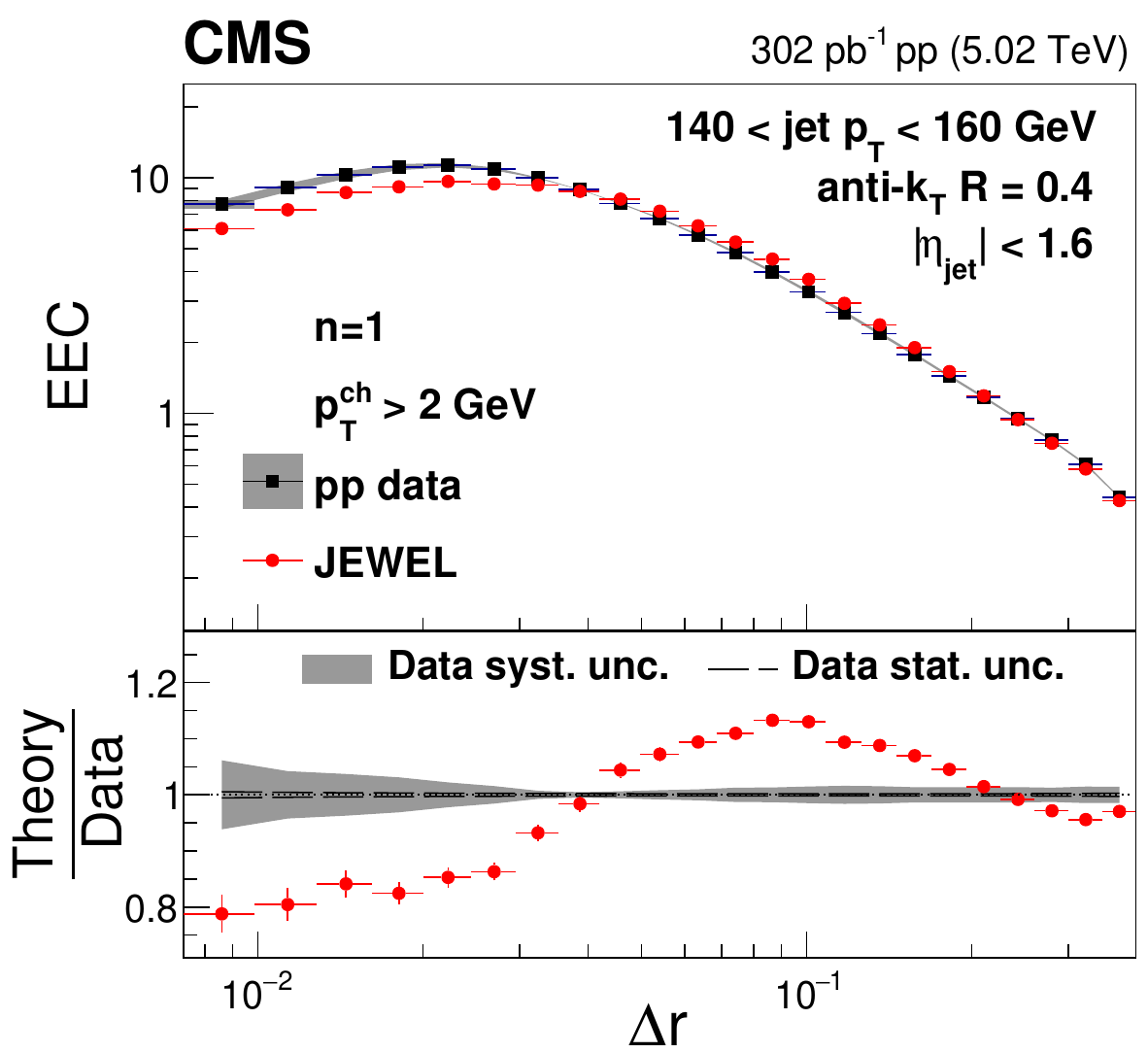}
    \includegraphics[width=0.4\textwidth]{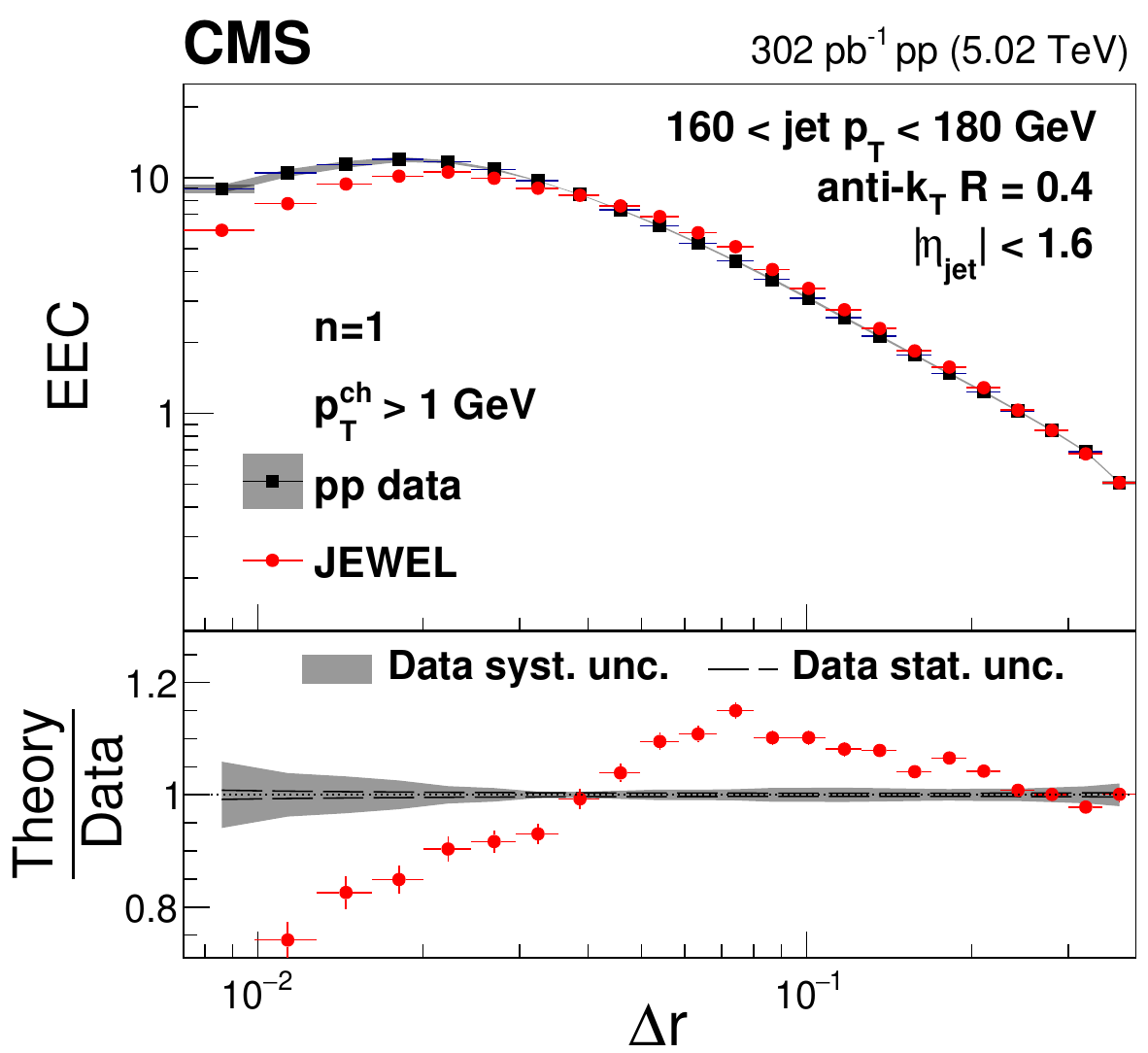}
    \includegraphics[width=0.4\textwidth]{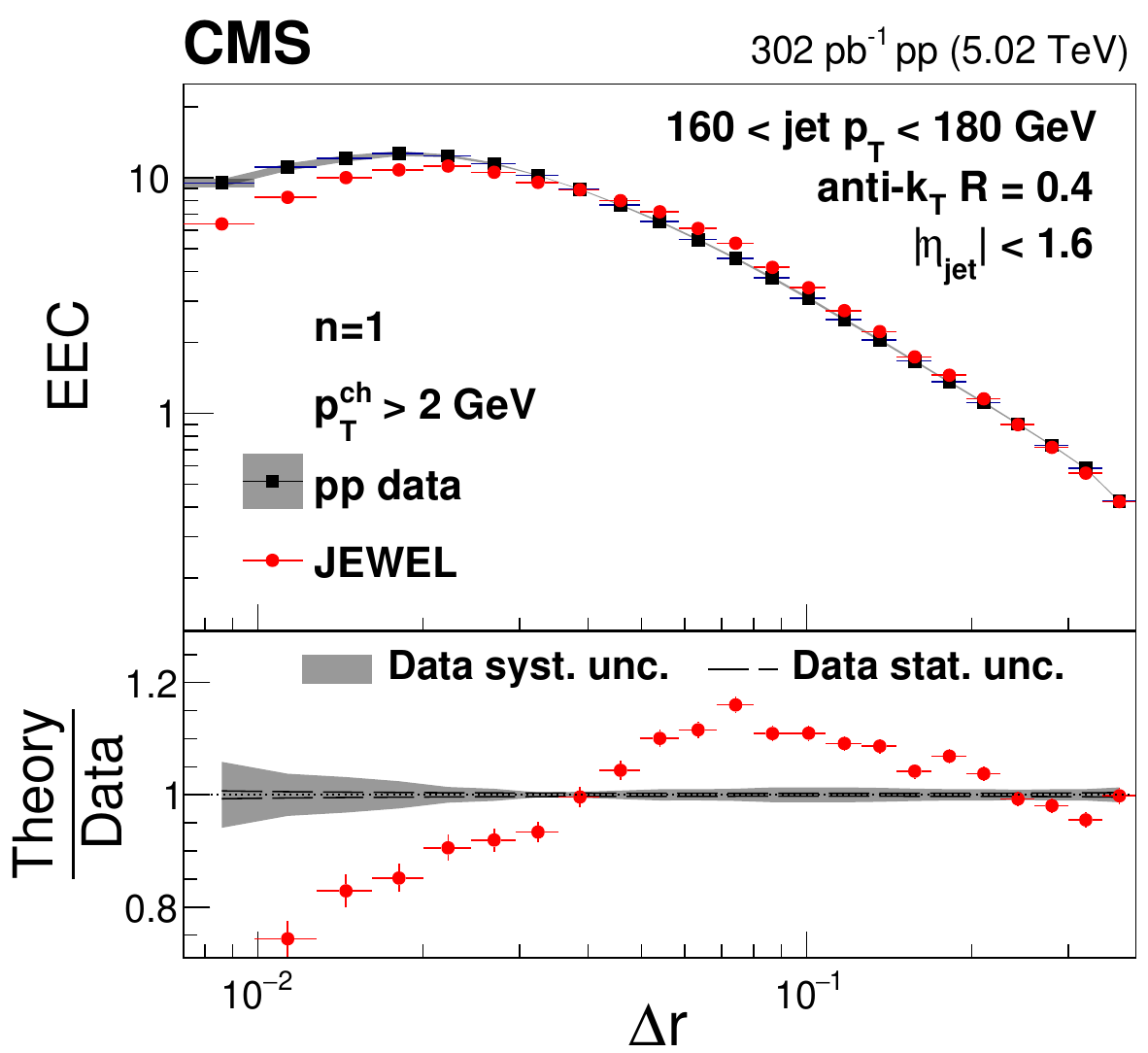}
    \includegraphics[width=0.4\textwidth]{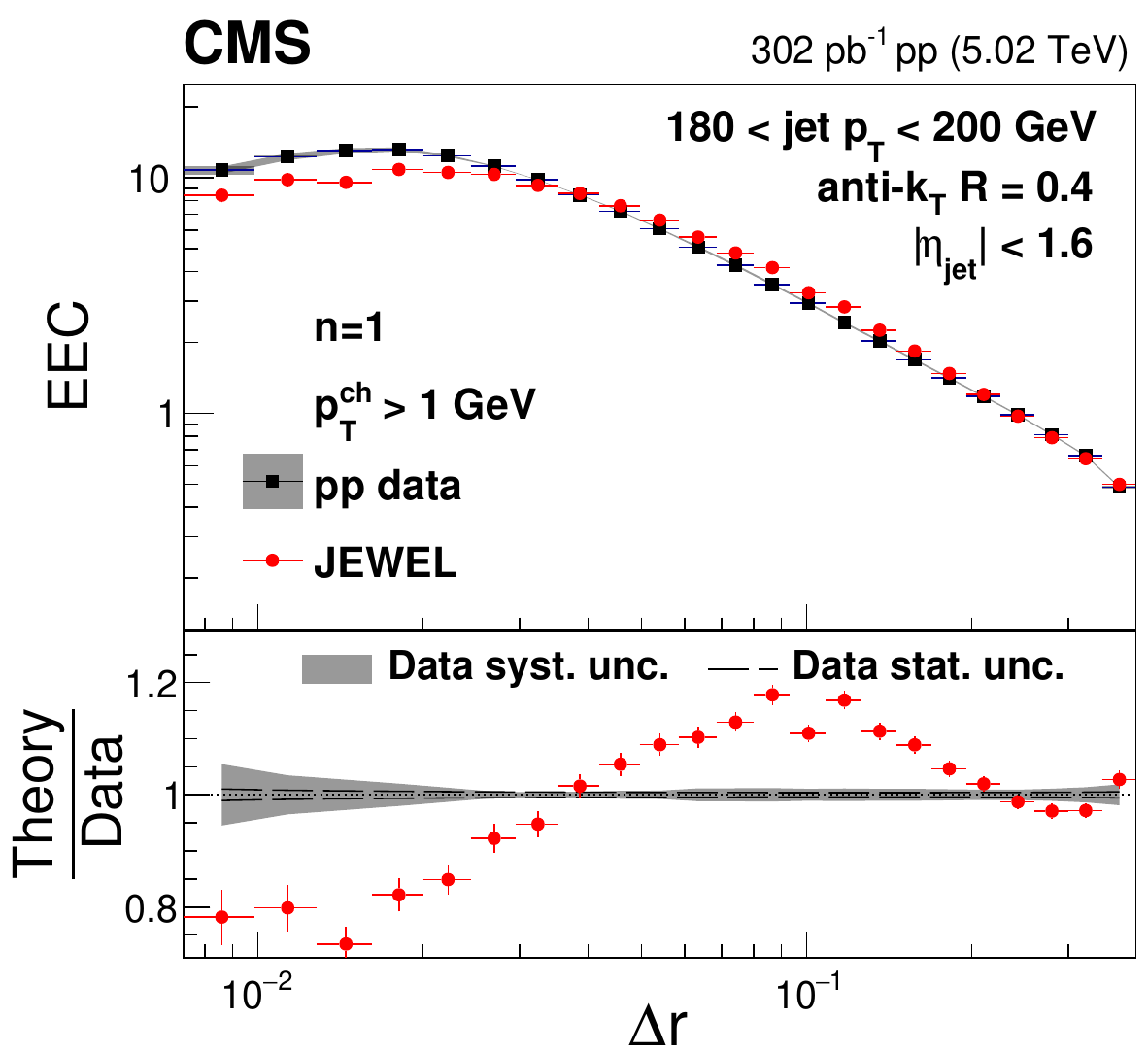}
    \includegraphics[width=0.4\textwidth]{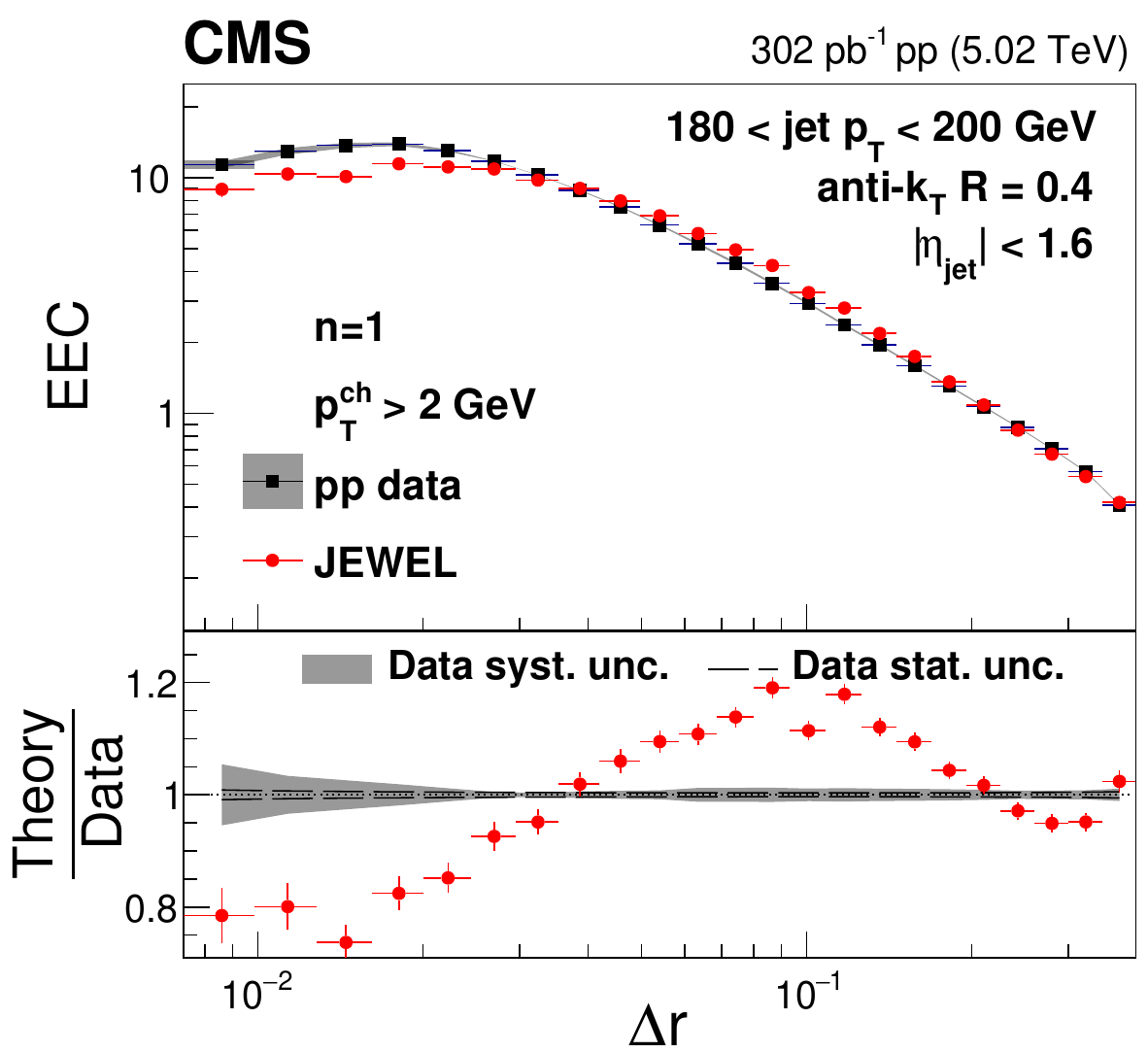}
    \caption{Energy-energy correlator distributions from \pp collisions with $n=1$ in different \ptCh and \ptjet bins compared to prediction from \JEWEL.} 
    \label{fig:ppDistributionsJewel}
  } 
\end{figure*}

\begin{figure*}[h]
  \centering{
    \includegraphics[width=0.4\textwidth]{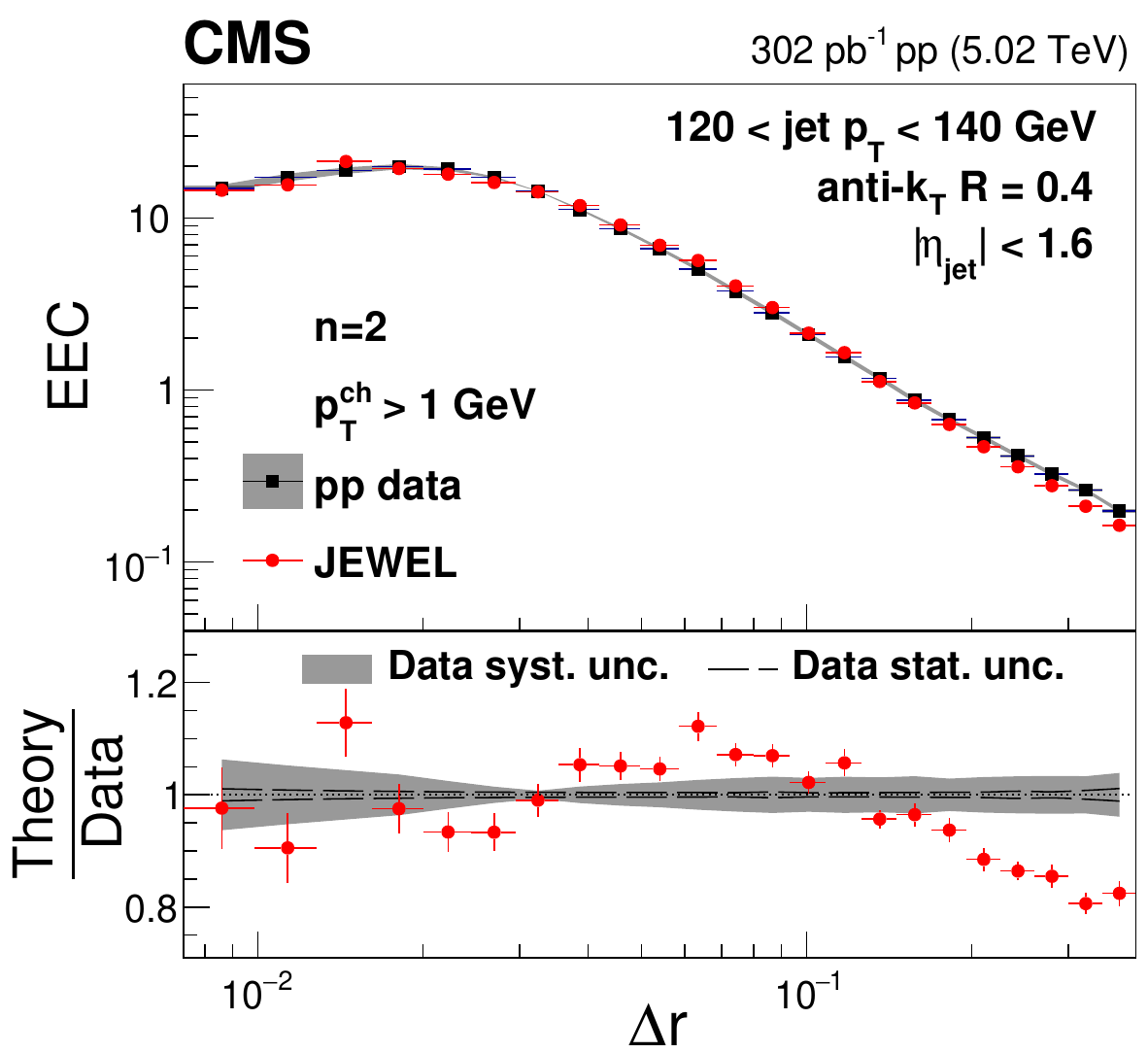}
    \includegraphics[width=0.4\textwidth]{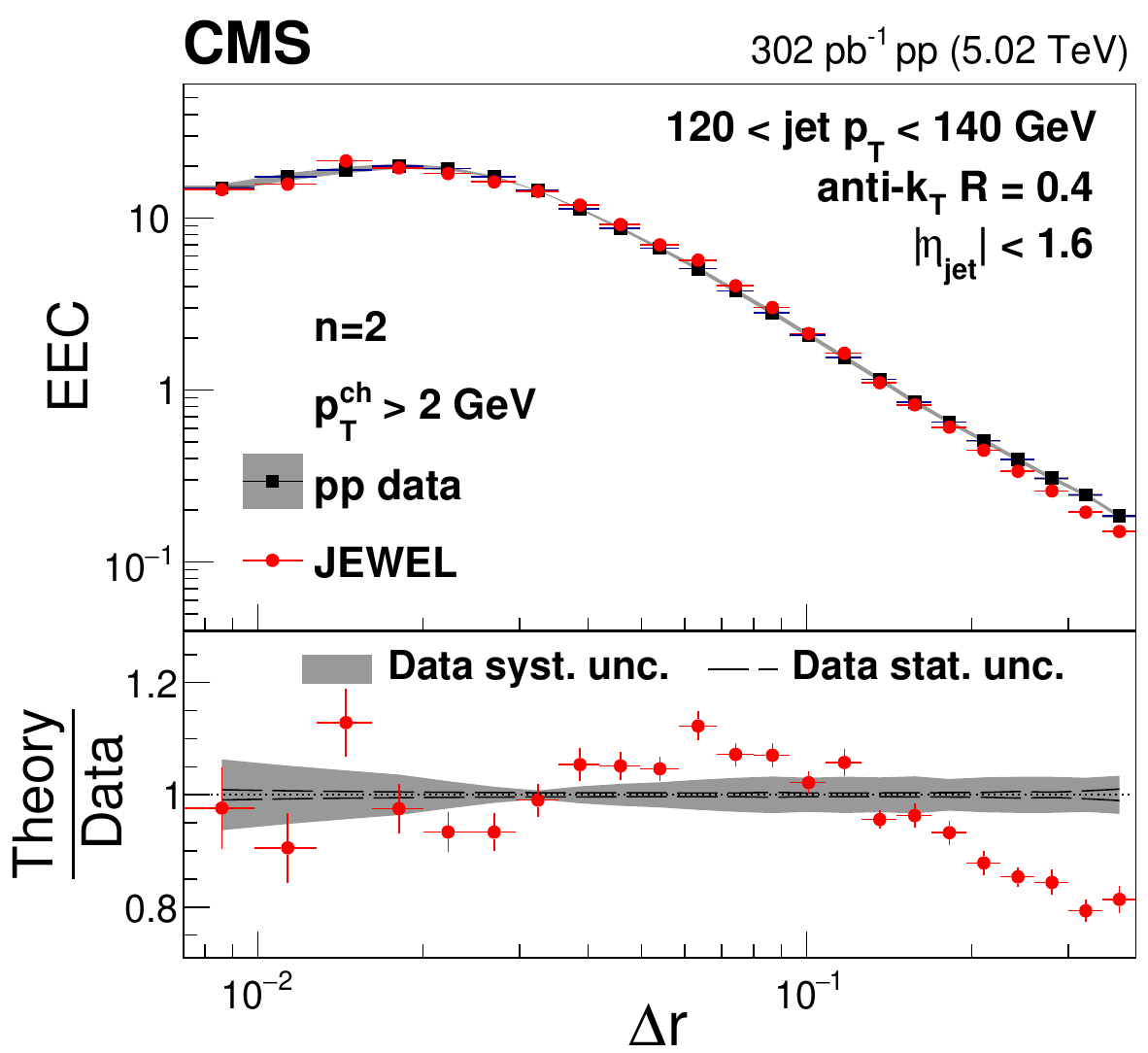}  
    \includegraphics[width=0.4\textwidth]{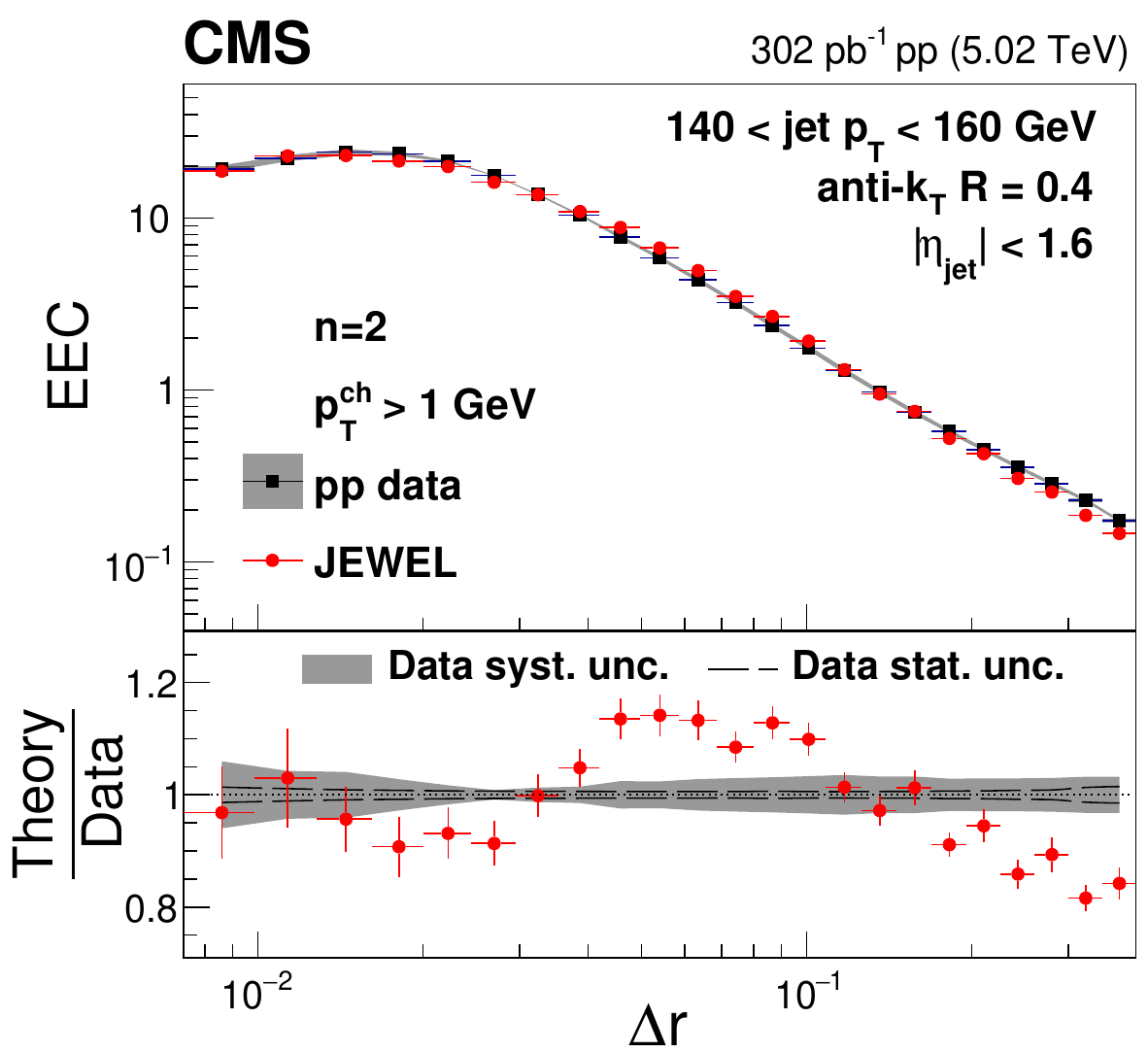}
    \includegraphics[width=0.4\textwidth]{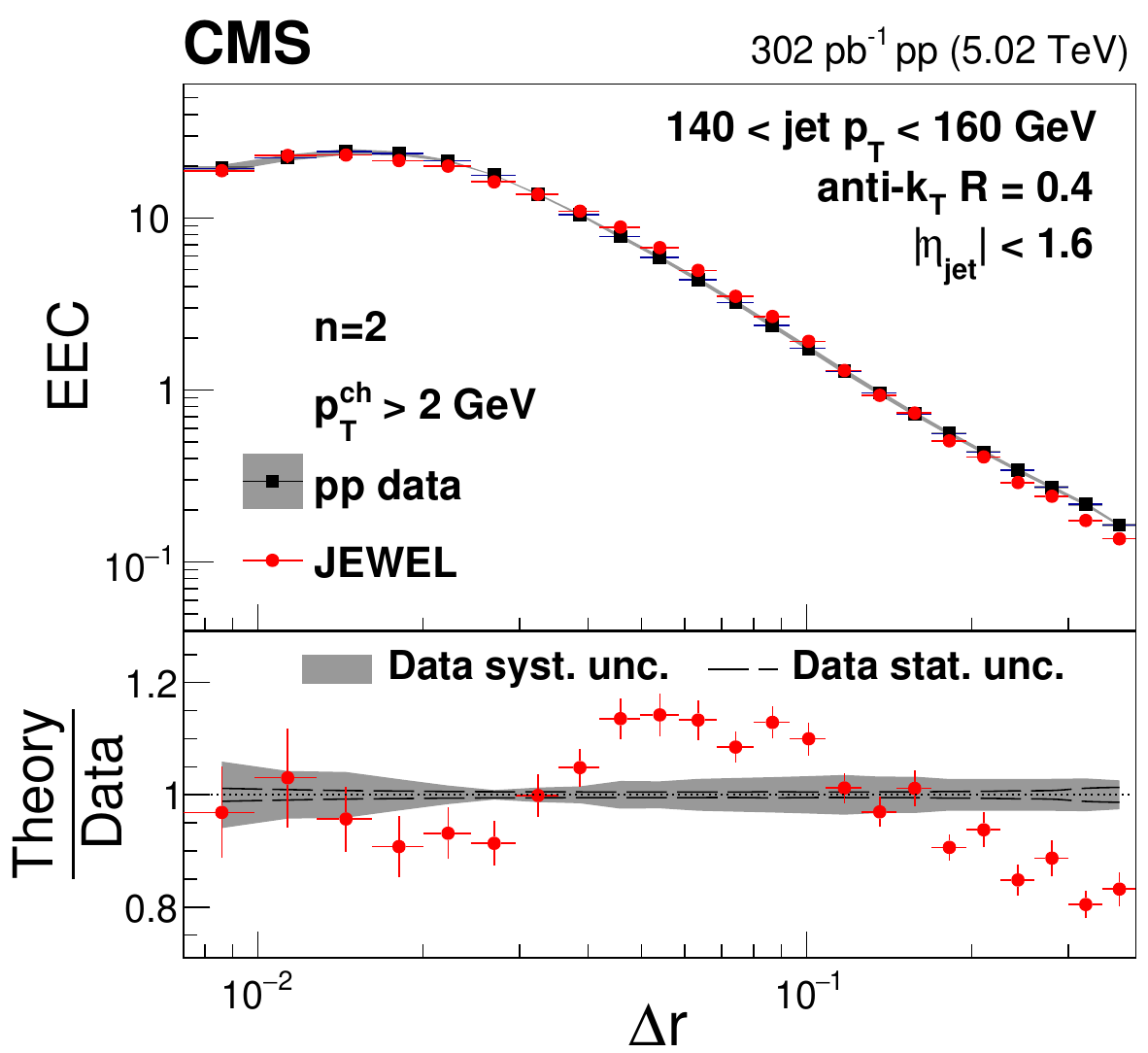}
    \includegraphics[width=0.4\textwidth]{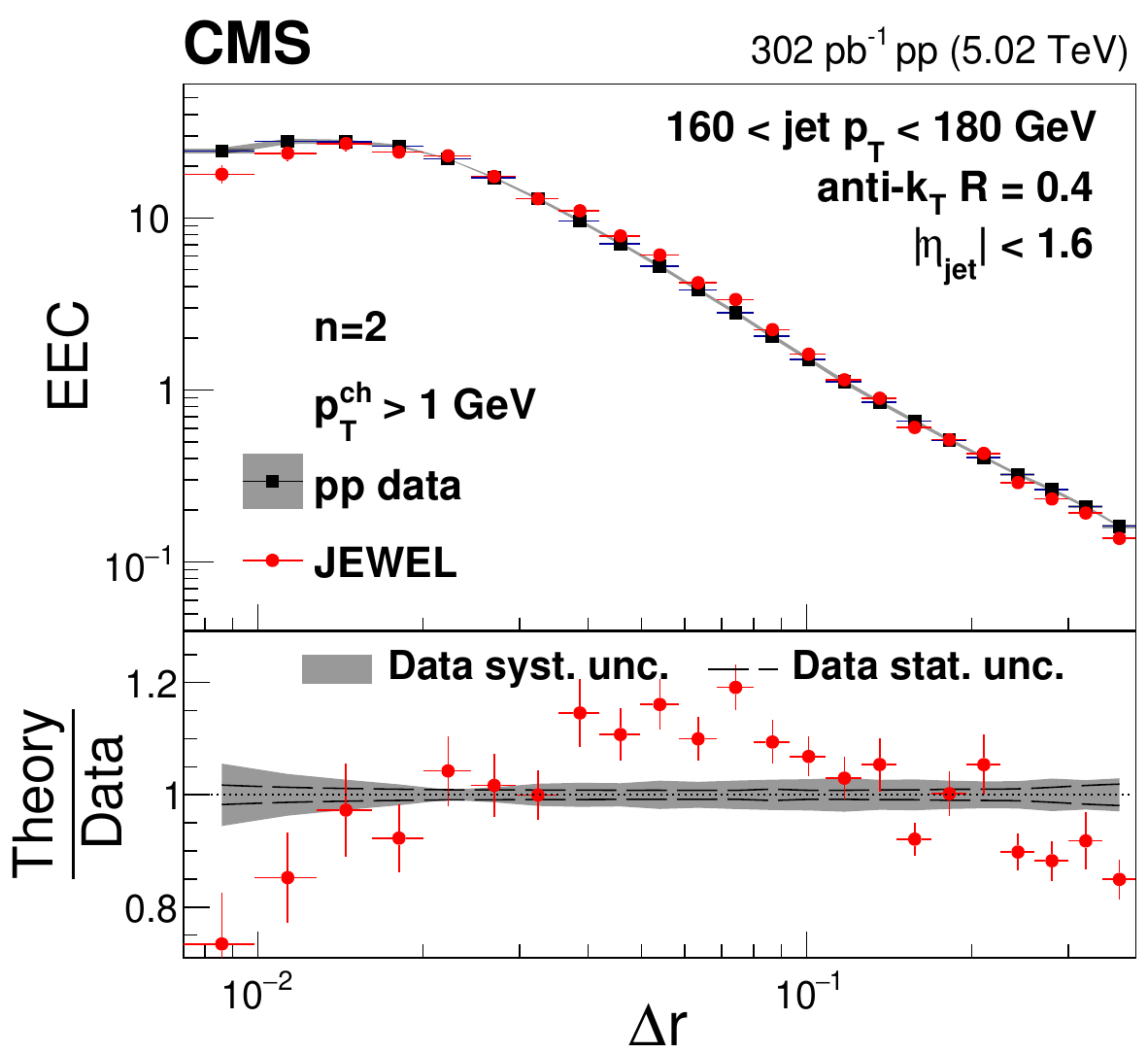}
    \includegraphics[width=0.4\textwidth]{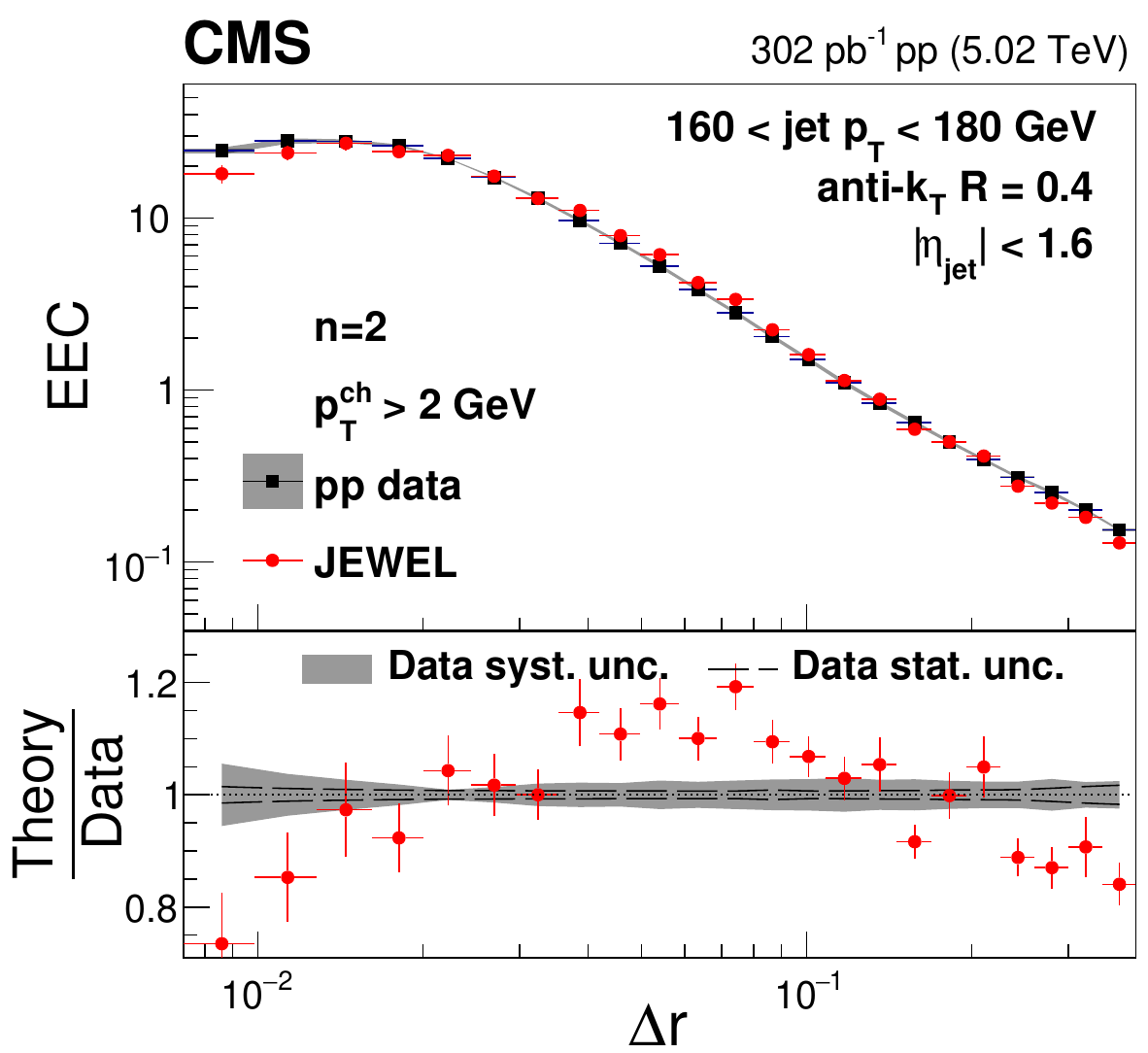}
    \includegraphics[width=0.4\textwidth]{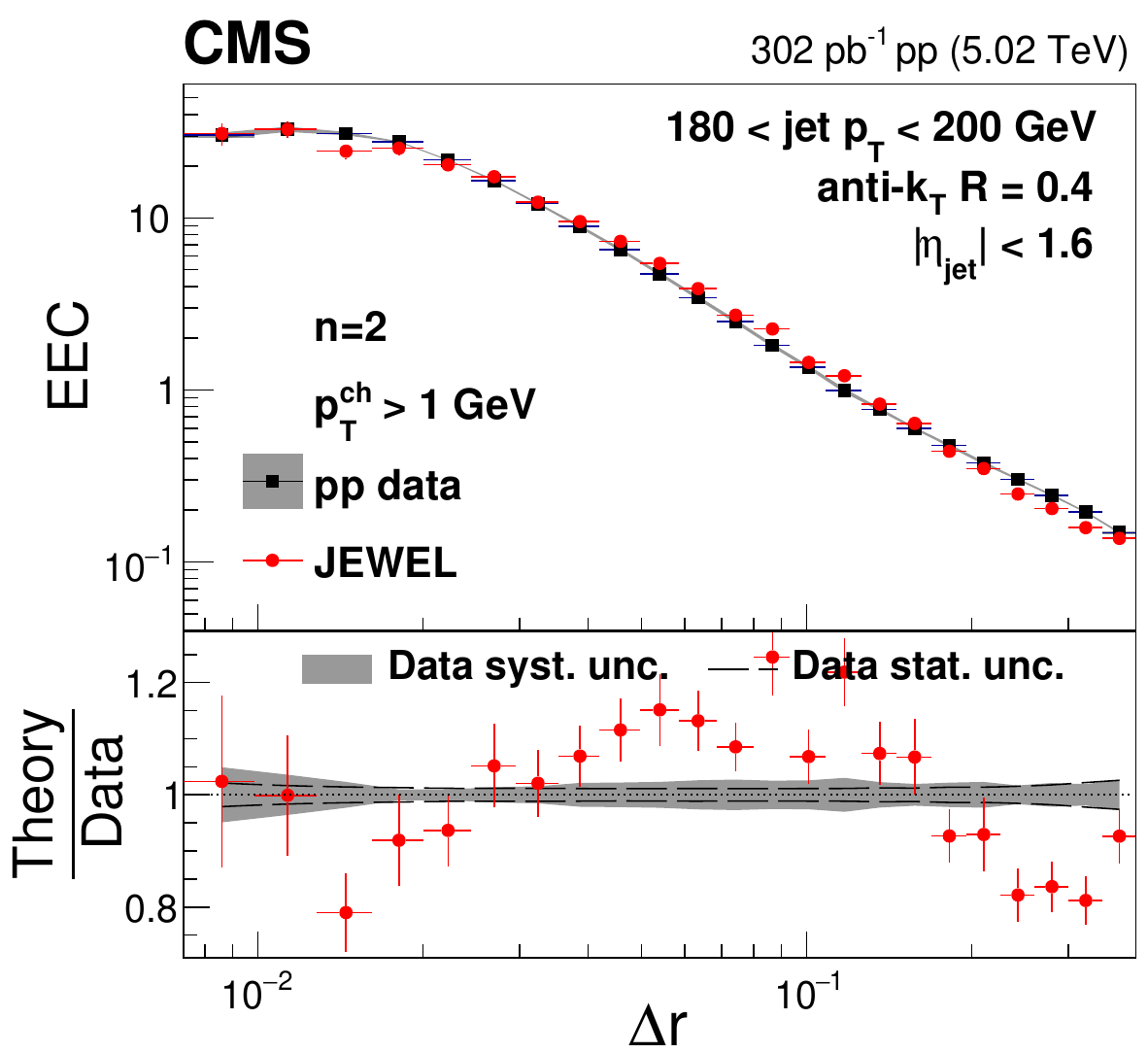} 
    \includegraphics[width=0.4\textwidth]{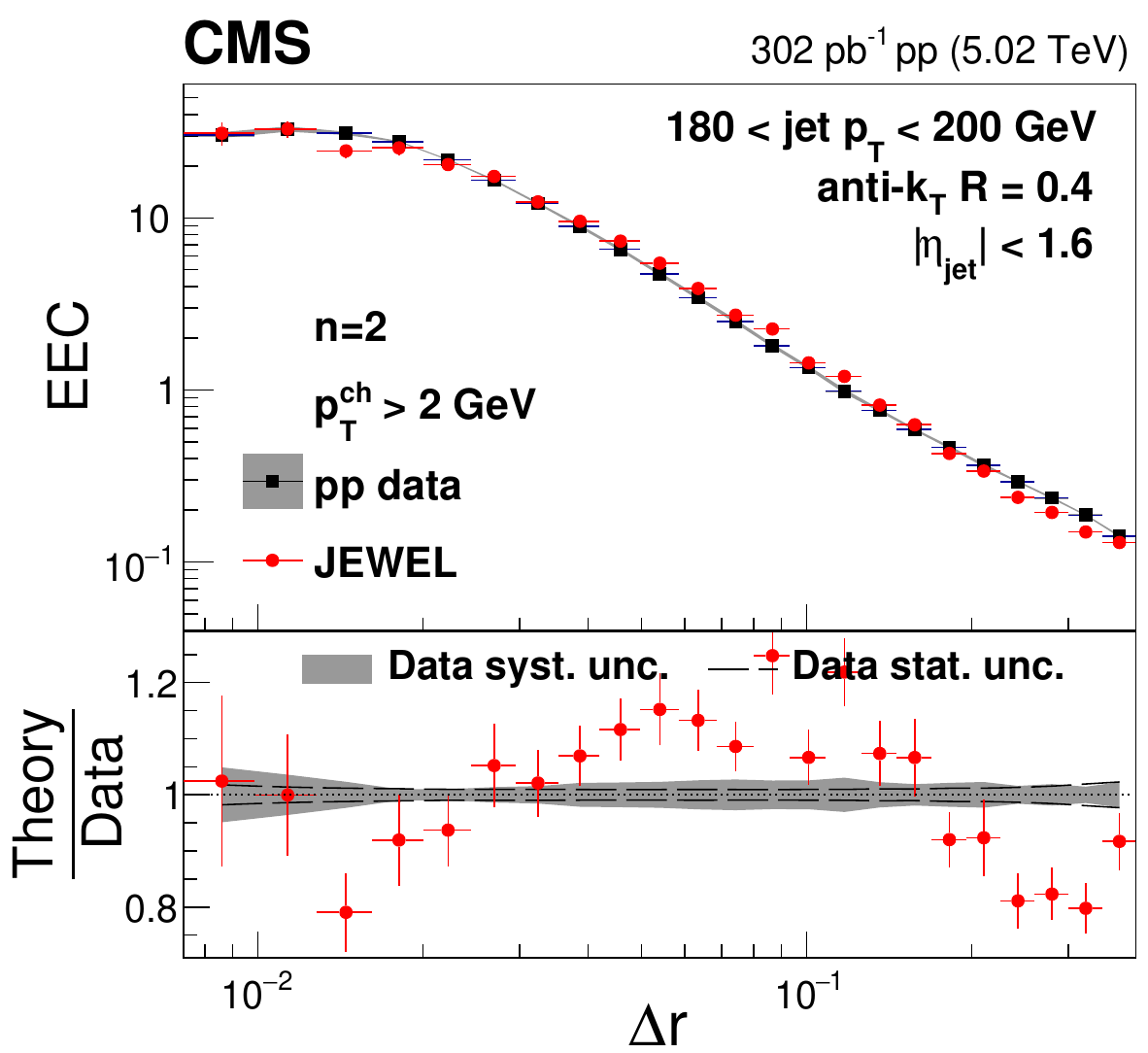}
    \caption{Energy-energy correlator distributions from \pp collisions with $n=2$ in different \ptCh and \ptjet bins compared to predictions from \JEWEL.} 
    \label{fig:ppDistributionsJewelSquared}
  } 
\end{figure*}

\begin{figure*}[h]
  \centering{
    \includegraphics[width=0.4\textwidth]{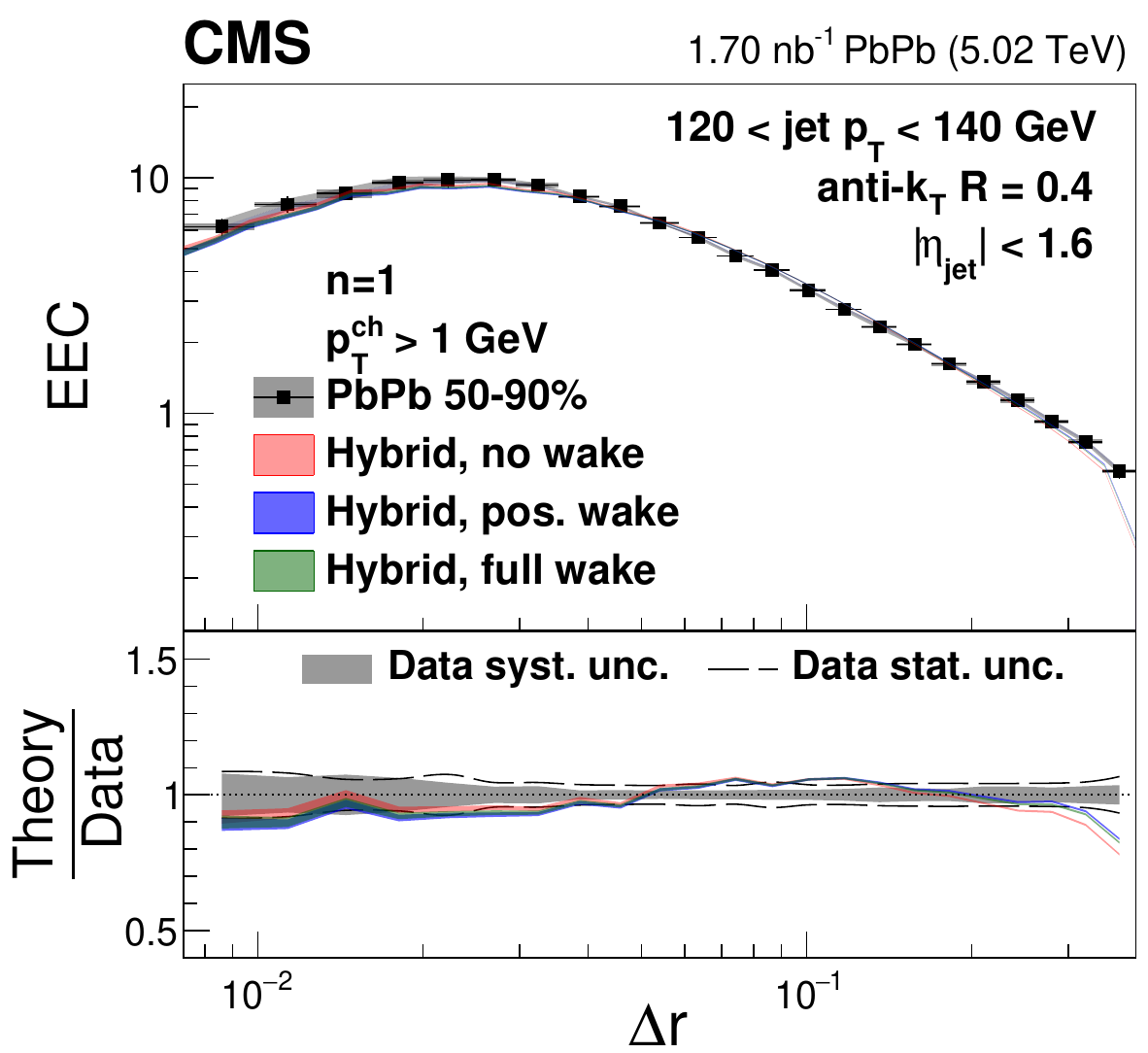}
    \includegraphics[width=0.4\textwidth]{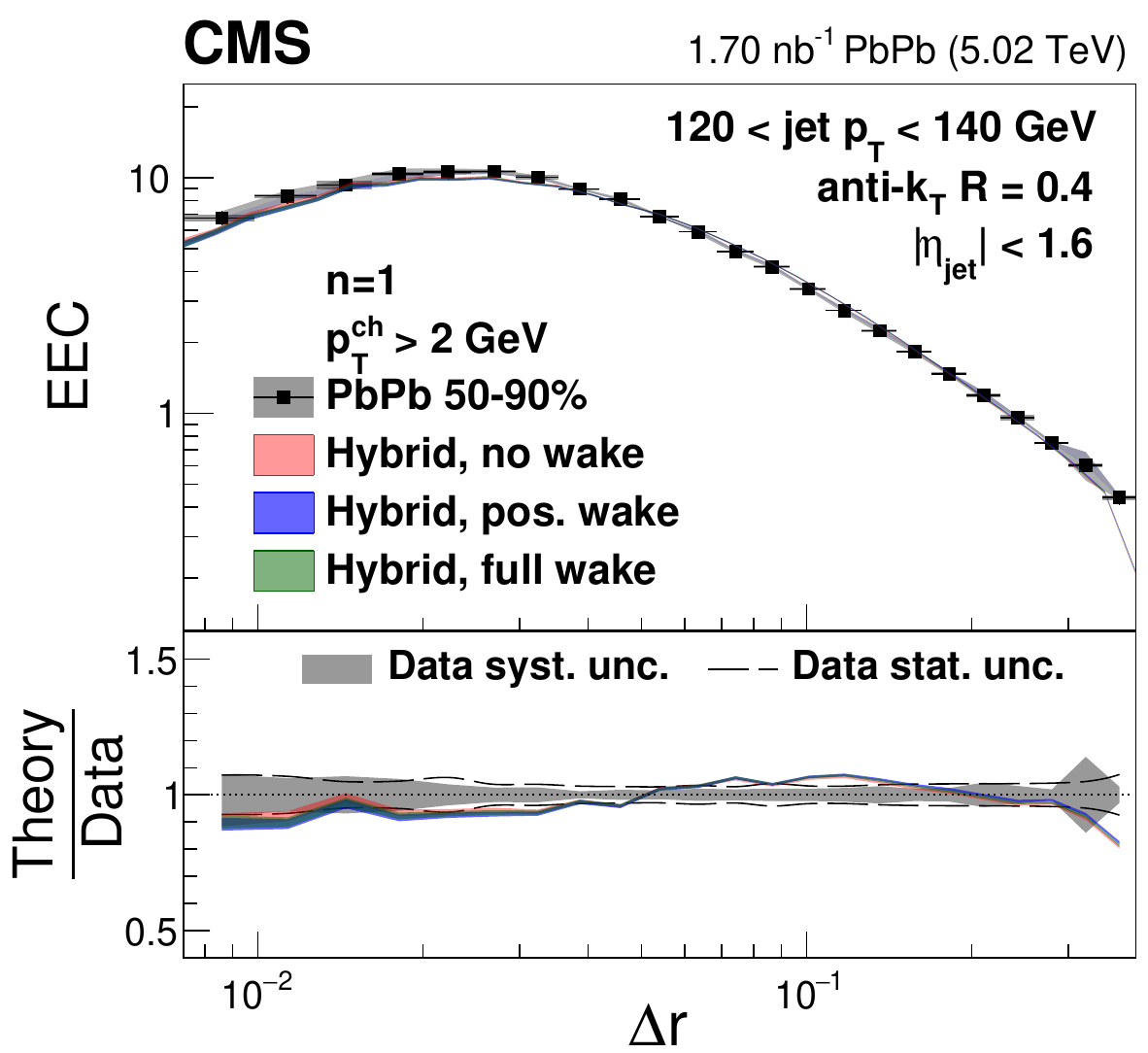} 
    \includegraphics[width=0.4\textwidth]{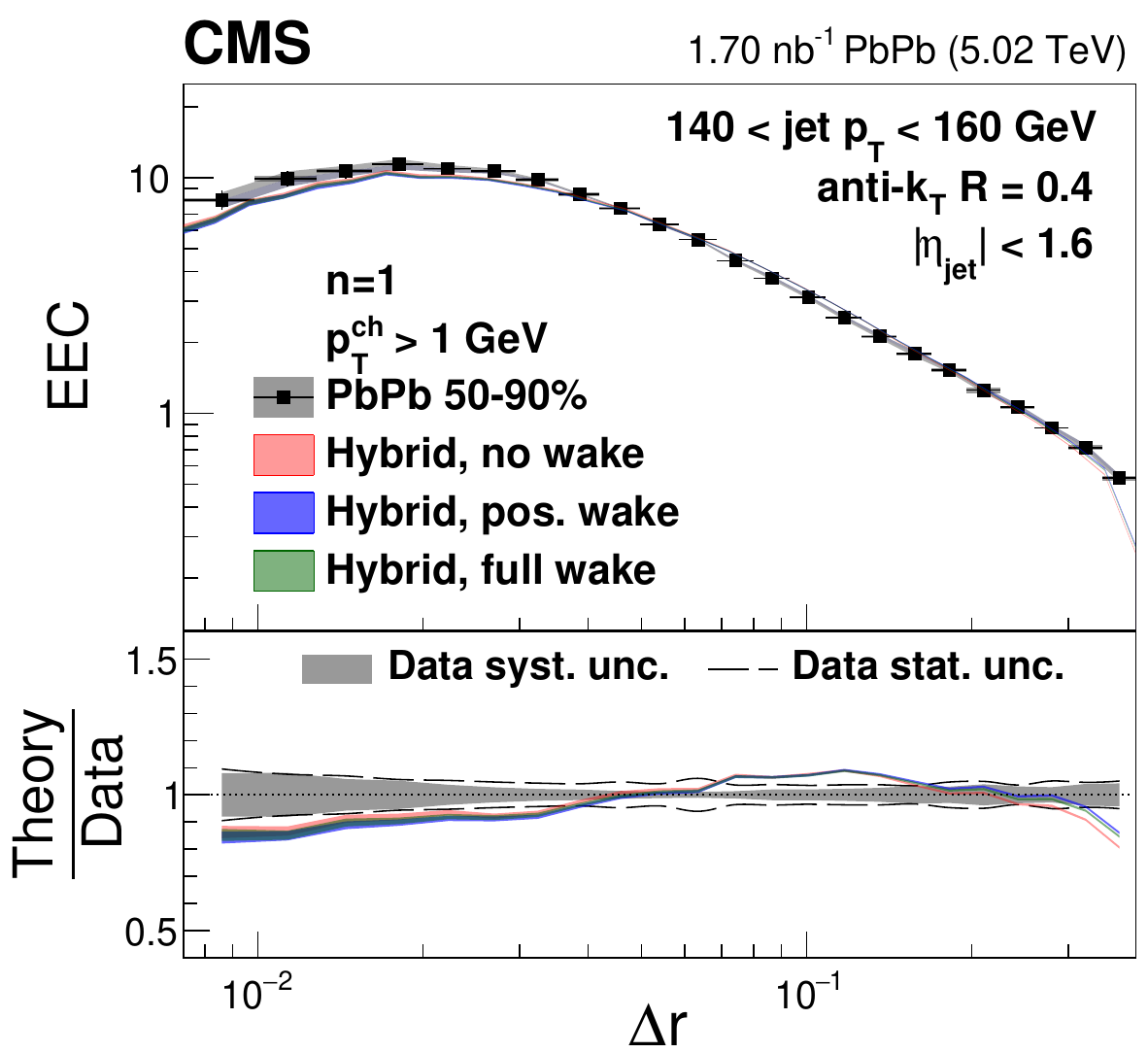}
    \includegraphics[width=0.4\textwidth]{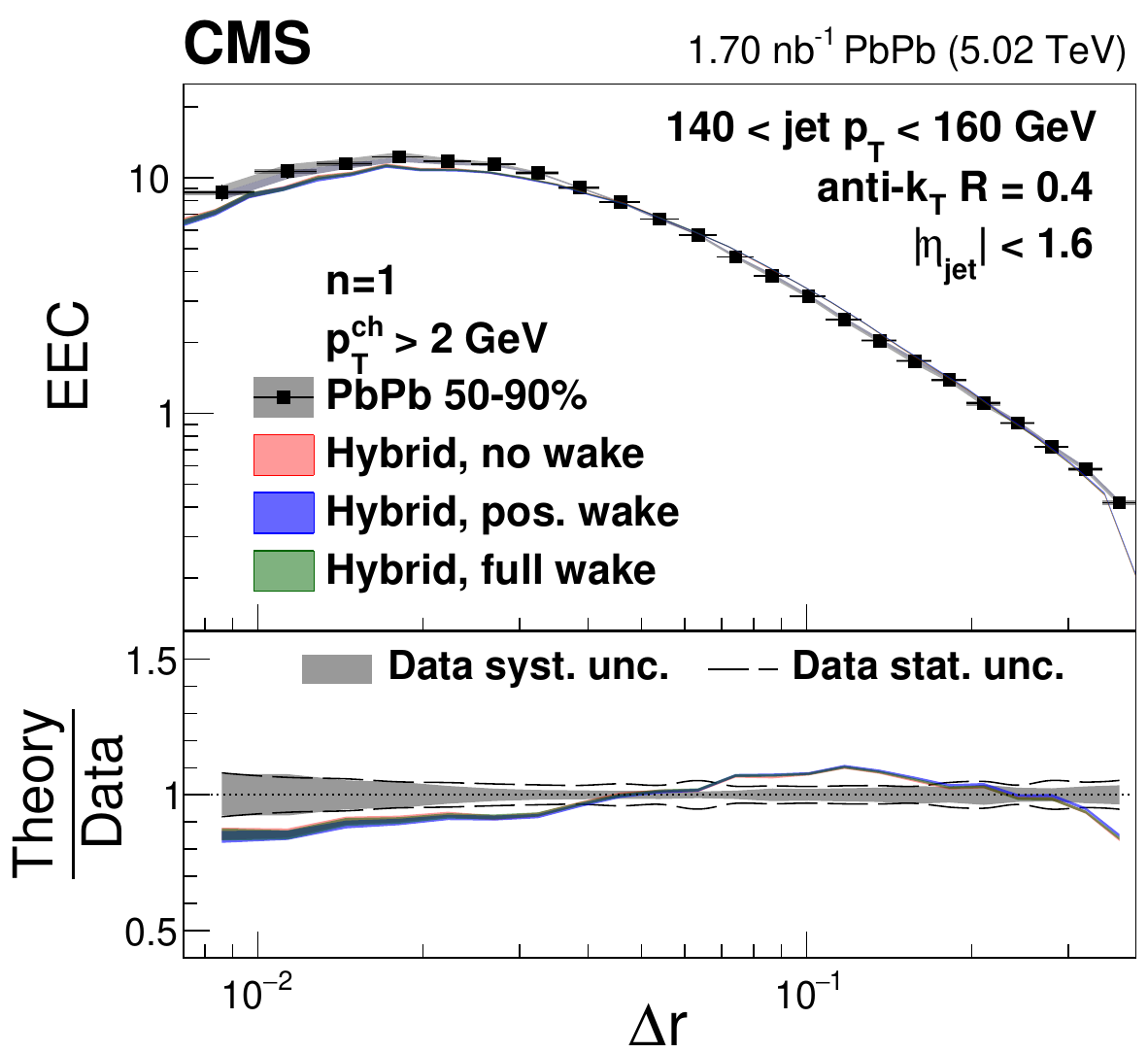}
    \includegraphics[width=0.4\textwidth]{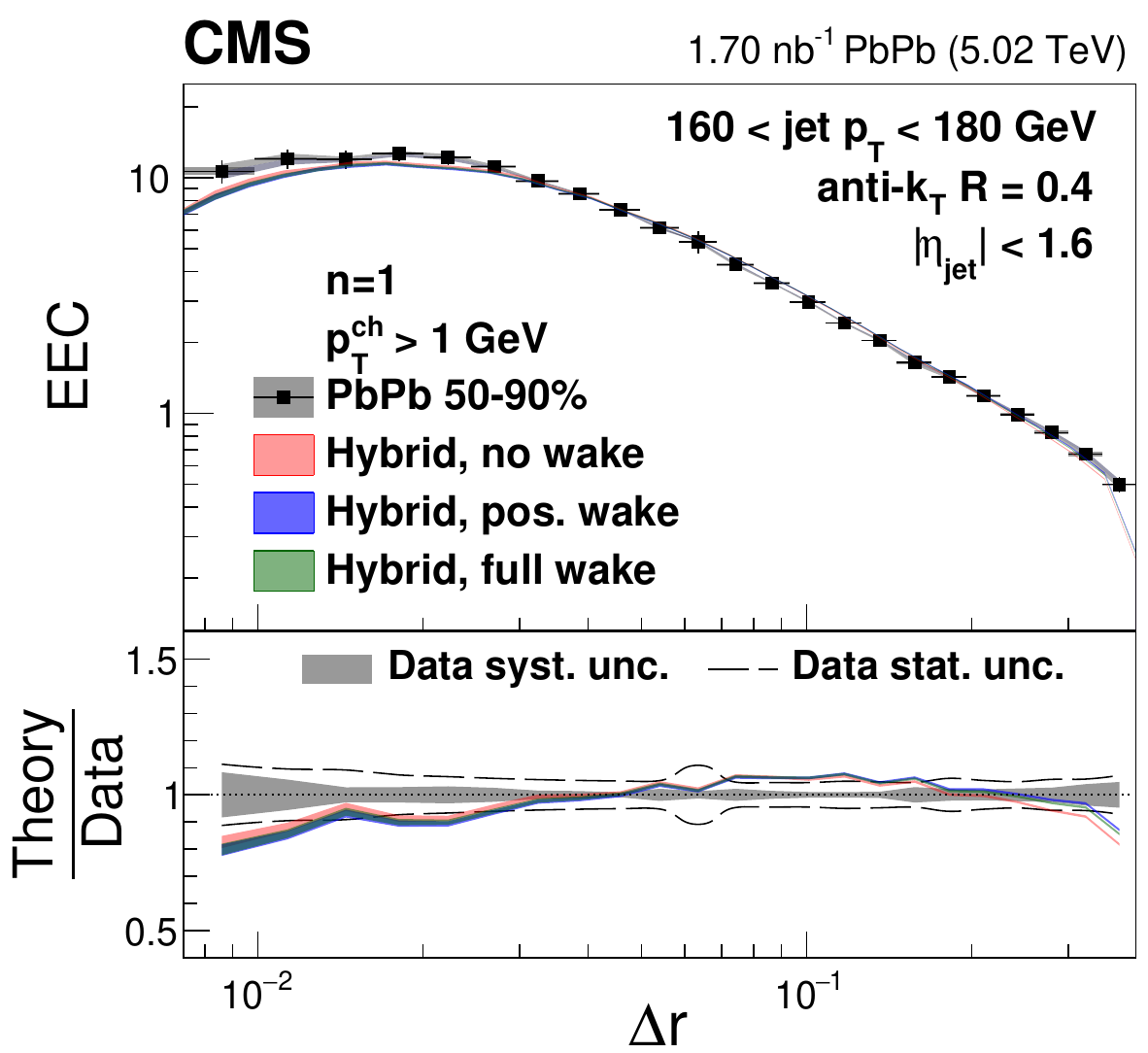}
    \includegraphics[width=0.4\textwidth]{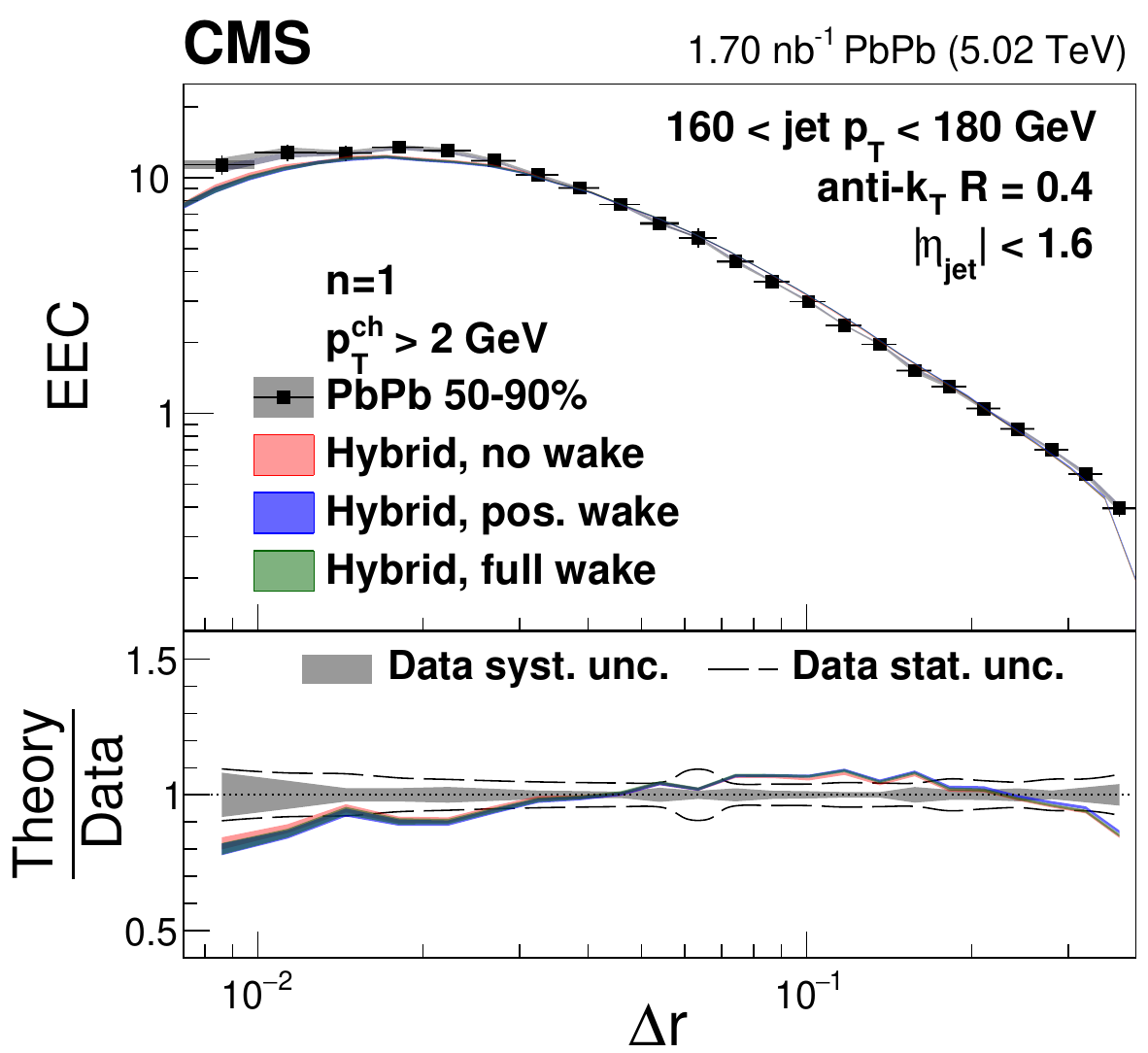}
    \includegraphics[width=0.4\textwidth]{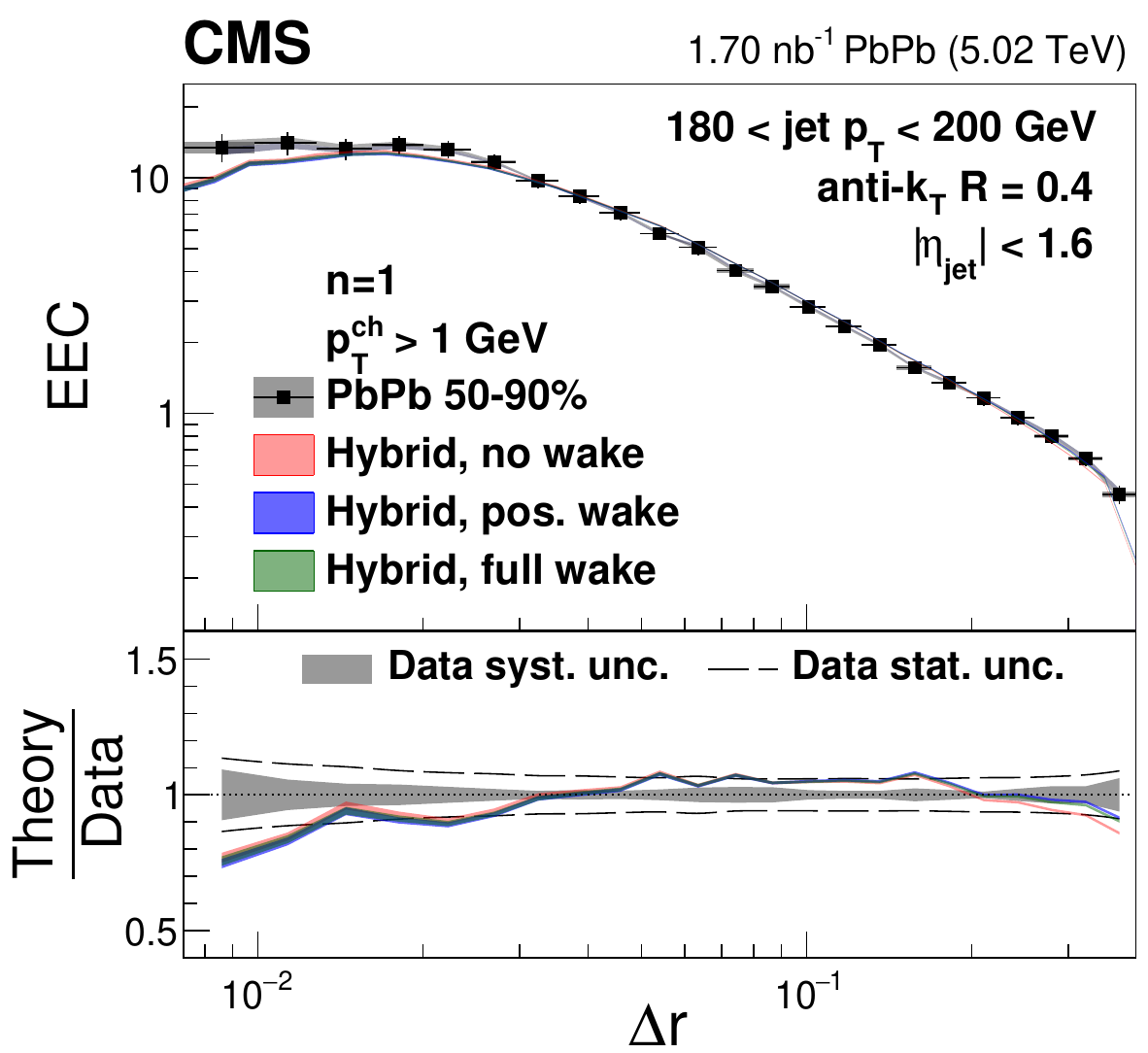}
    \includegraphics[width=0.4\textwidth]{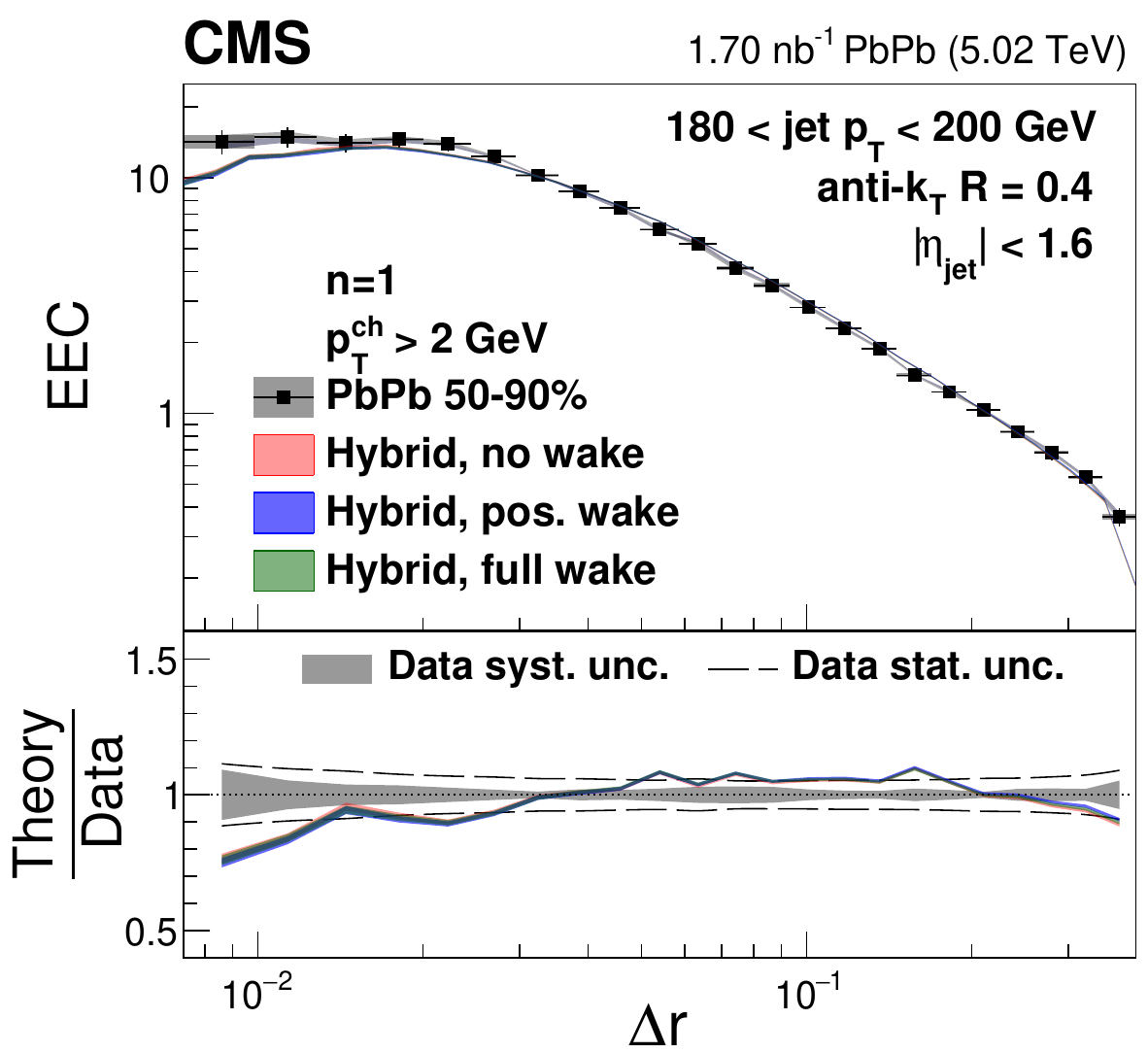}
    \caption{Energy-energy correlator distributions from 50--90\% central \PbPb collisions with $n=1$ in different \ptCh and \ptjet bins compared to the hybrid model predictions with different jet wake settings.} 
    \label{fig:PbPbDistributionsNominal50-90}
  } 
\end{figure*}

\begin{figure*}[h]
  \centering{
    \includegraphics[width=0.4\textwidth]{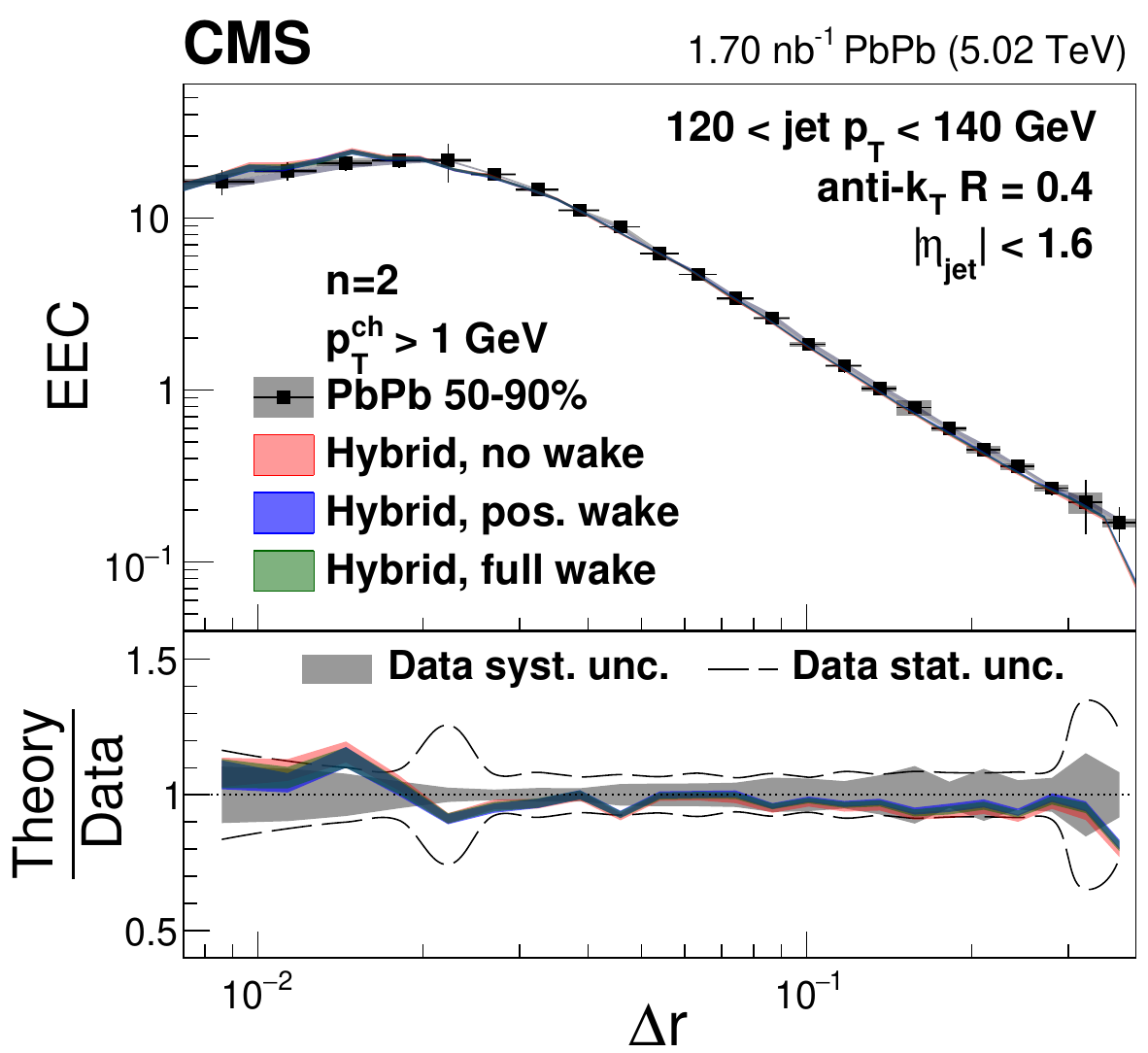}
    \includegraphics[width=0.4\textwidth]{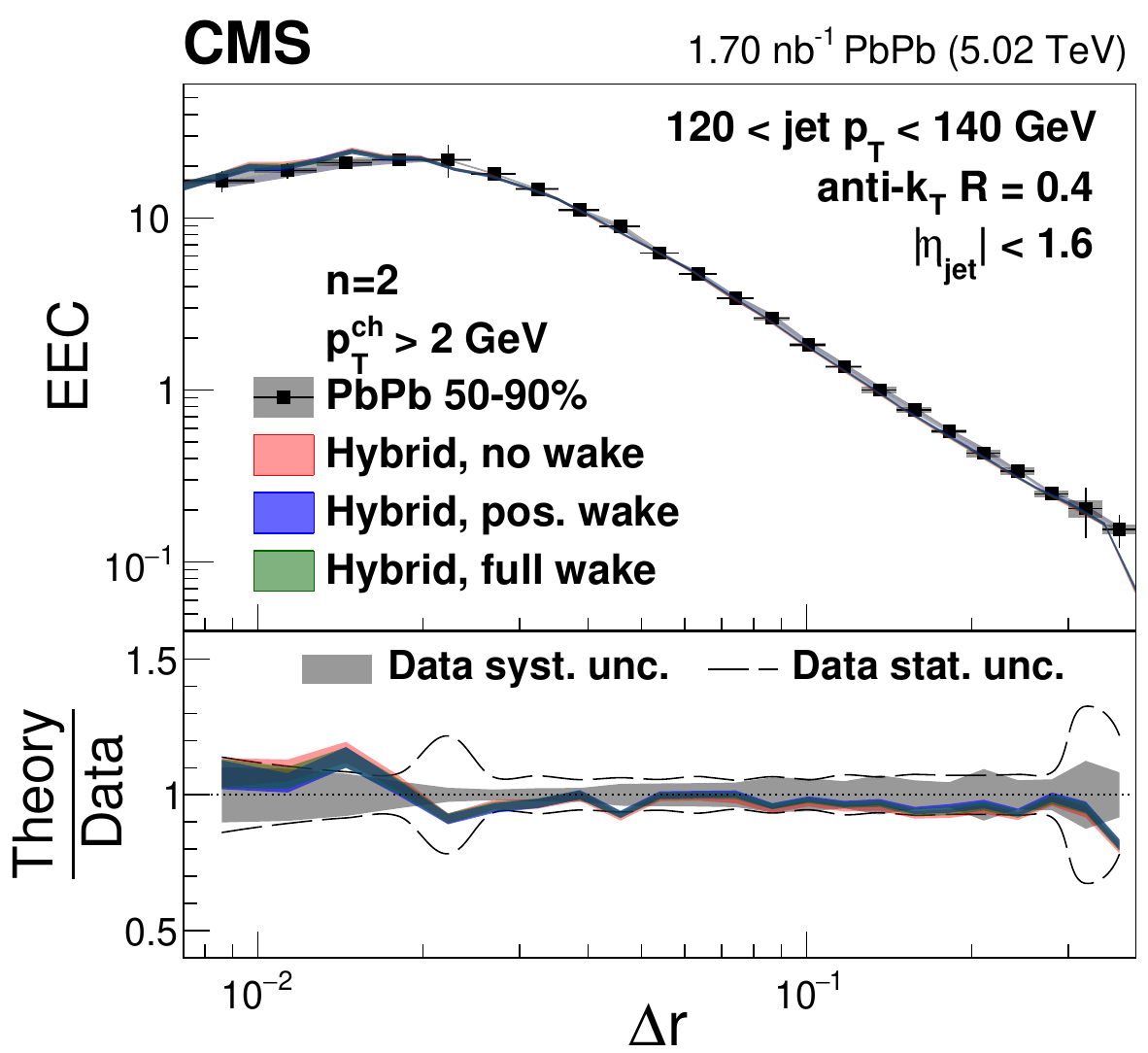}  
    \includegraphics[width=0.4\textwidth]{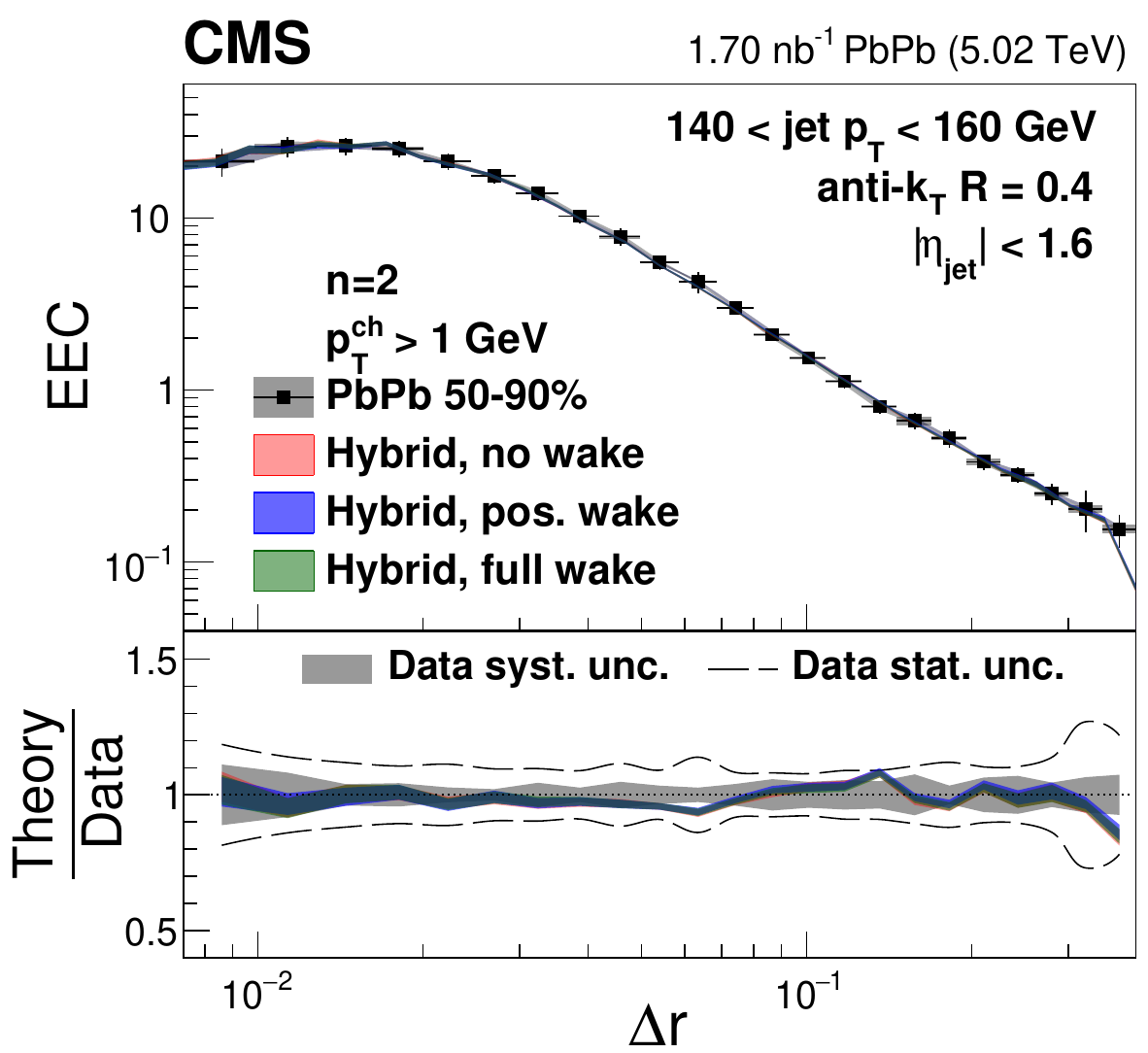}
    \includegraphics[width=0.4\textwidth]{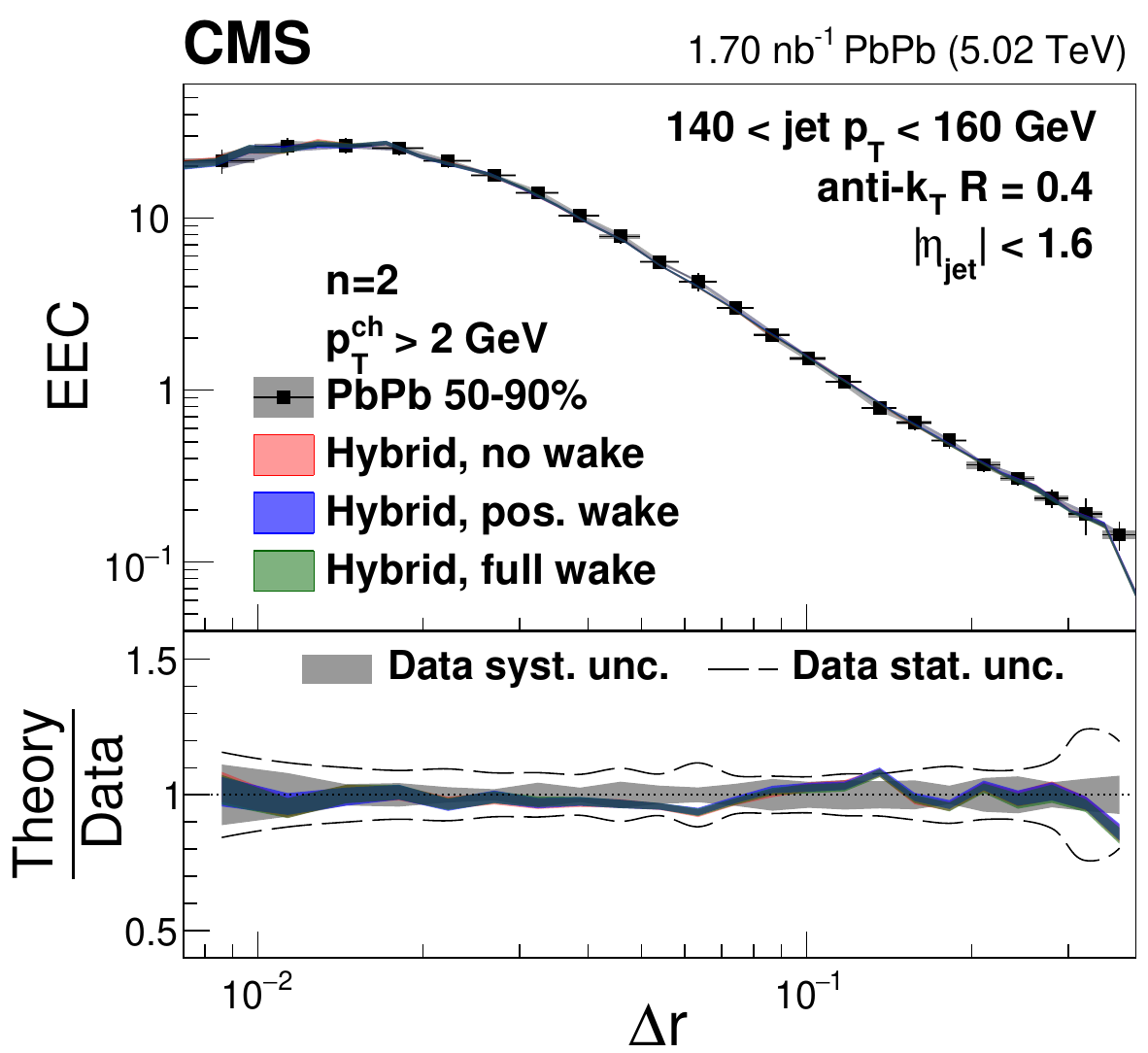}
    \includegraphics[width=0.4\textwidth]{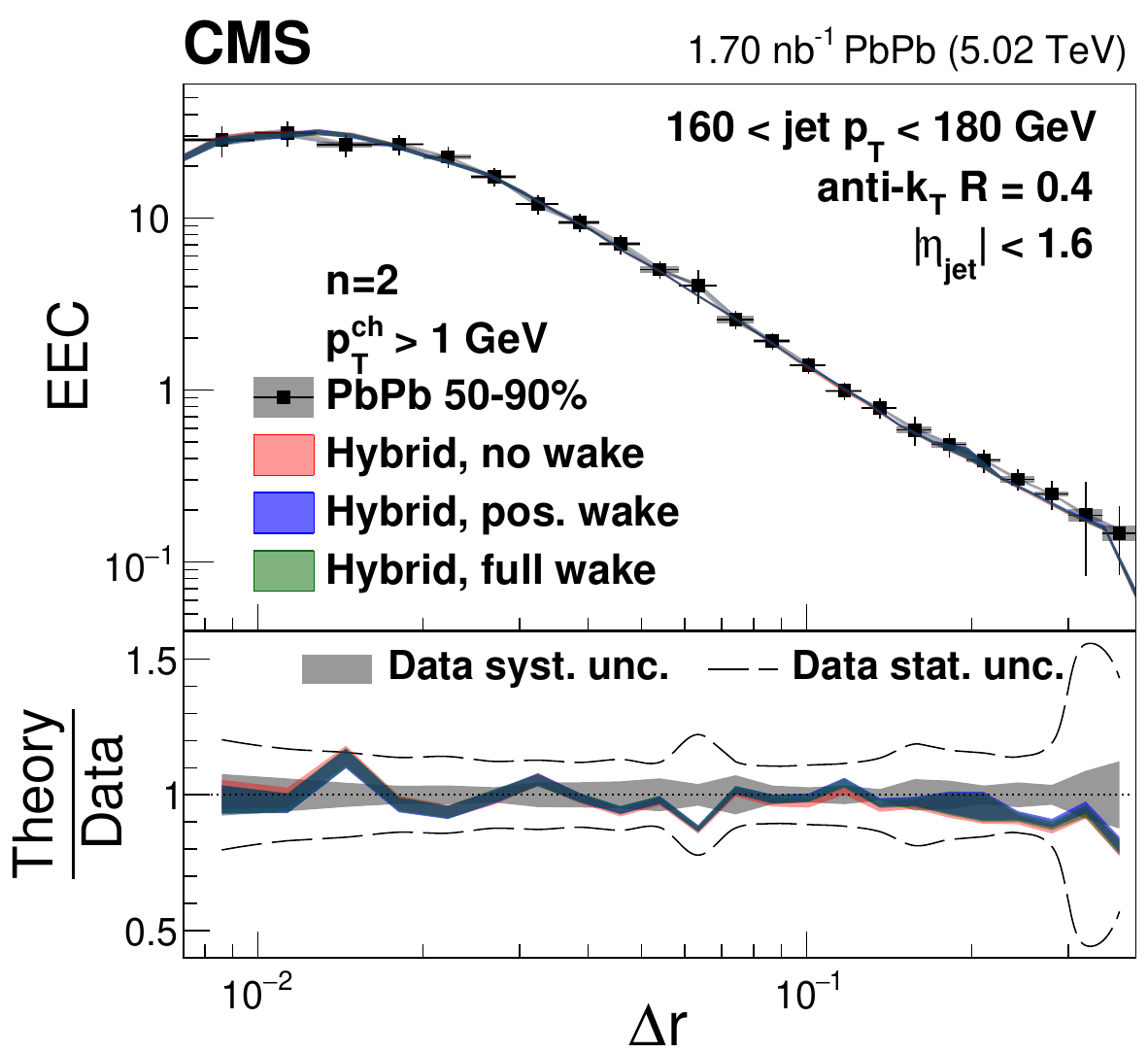}
    \includegraphics[width=0.4\textwidth]{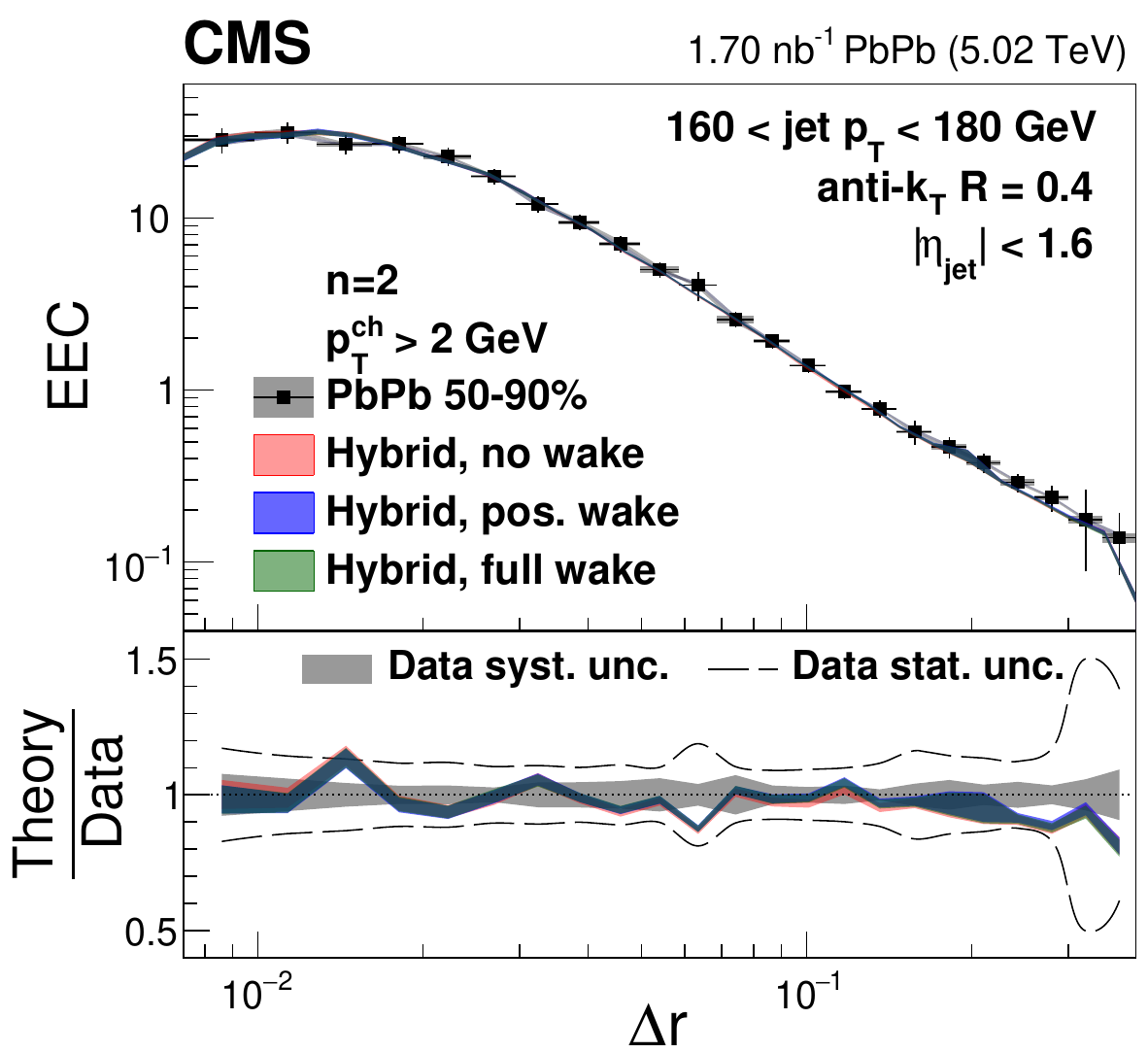}
    \includegraphics[width=0.4\textwidth]{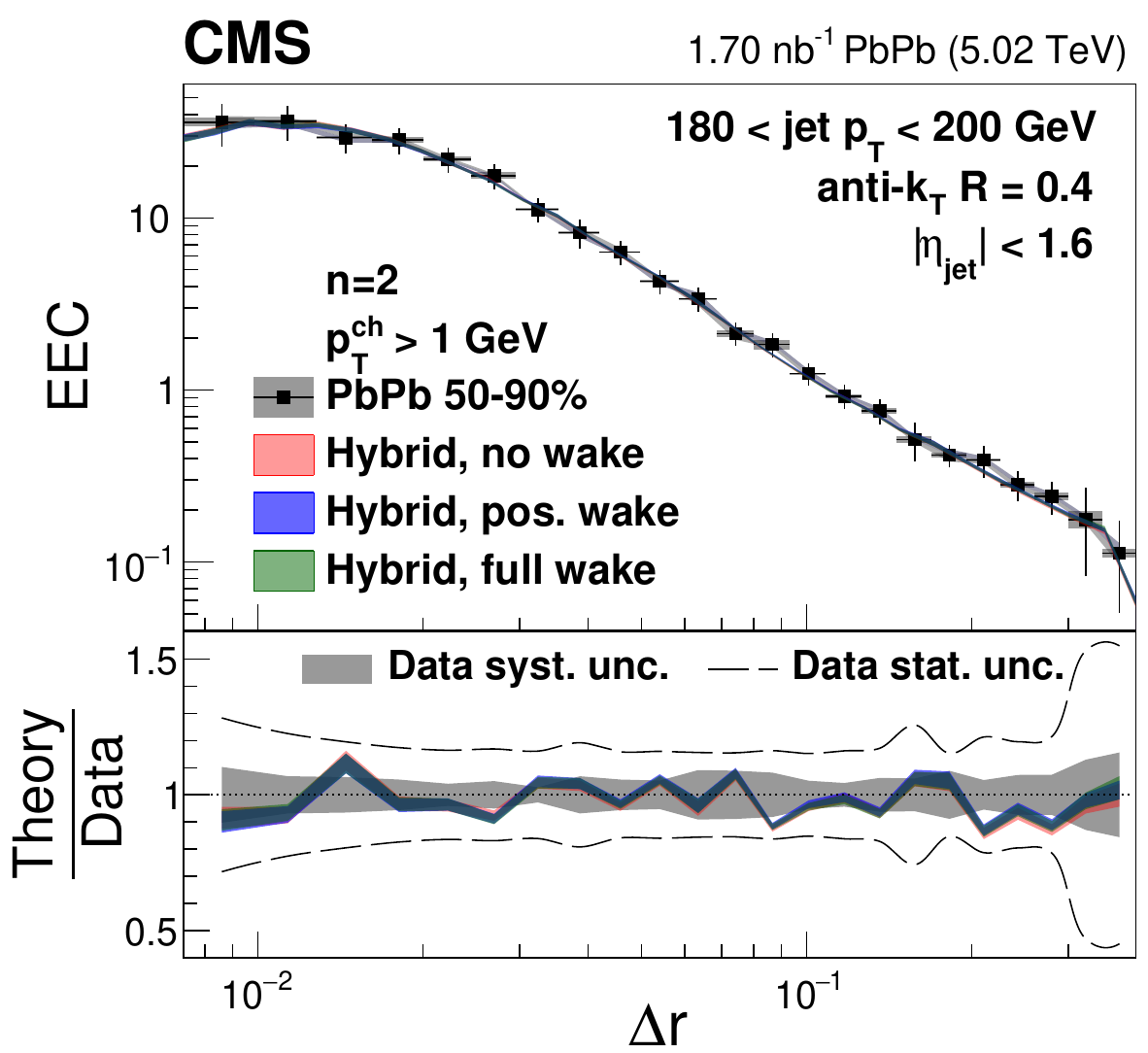}
    \includegraphics[width=0.4\textwidth]{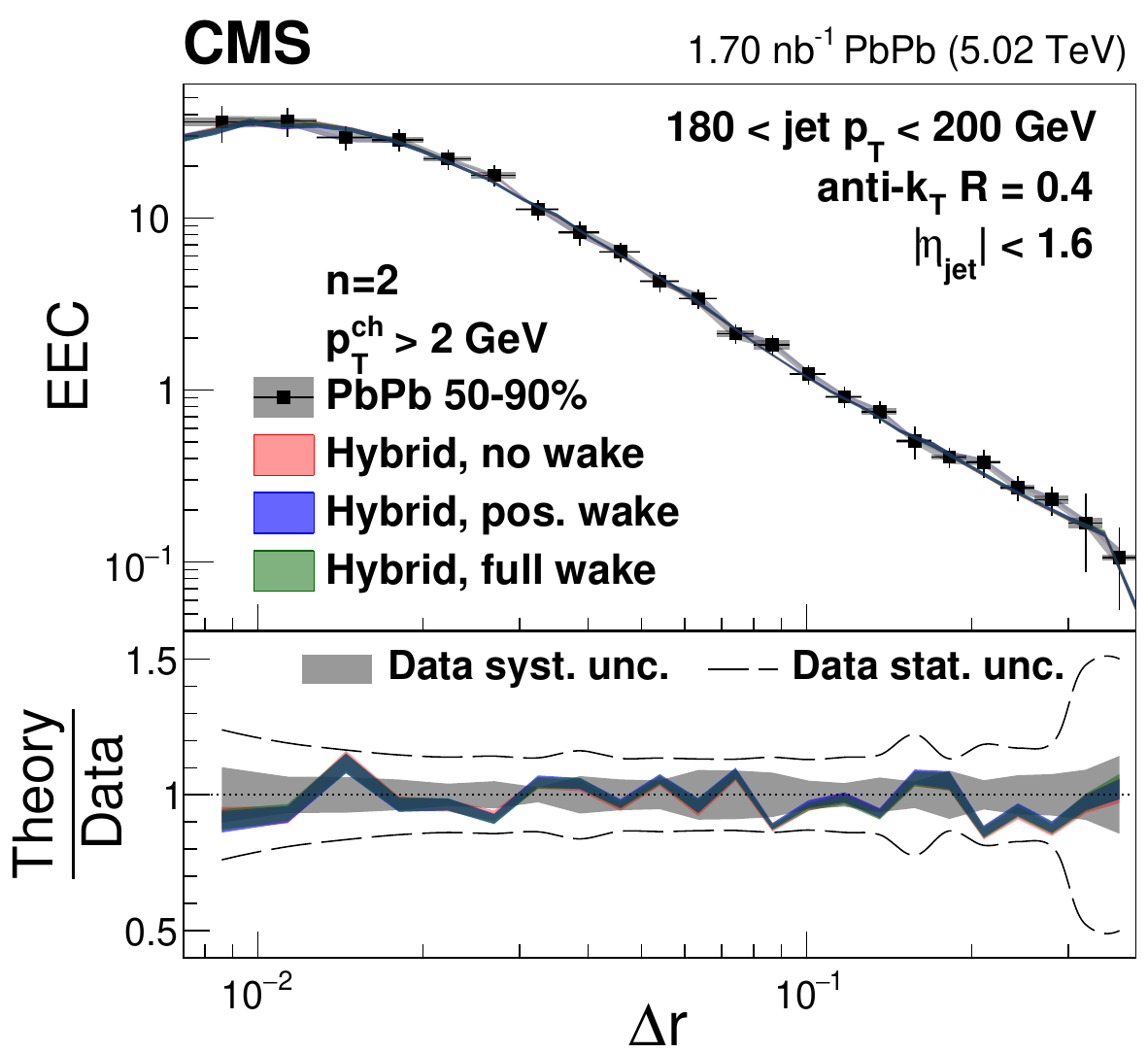}
    \caption{Energy-energy correlator distributions from 50--90\% central \PbPb collisions with $n=2$ in different \ptCh and \ptjet bins compared to the hybrid model predictions with different jet wake settings.} 
    \label{fig:PbPbDistributionsSquared50-90}
  } 
\end{figure*}

\begin{figure*}[h]
  \centering{
    \includegraphics[width=0.4\textwidth]{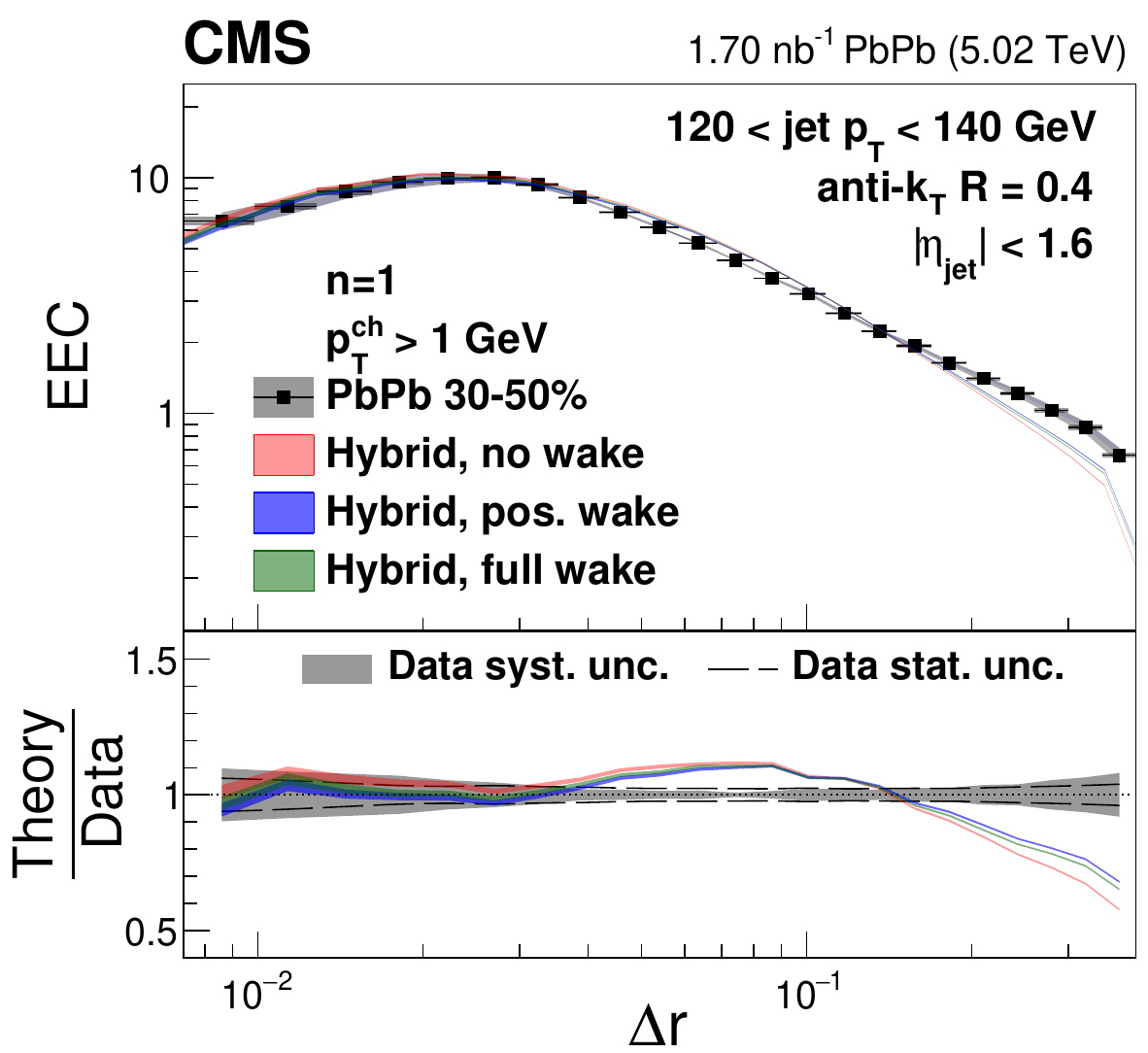}
    \includegraphics[width=0.4\textwidth]{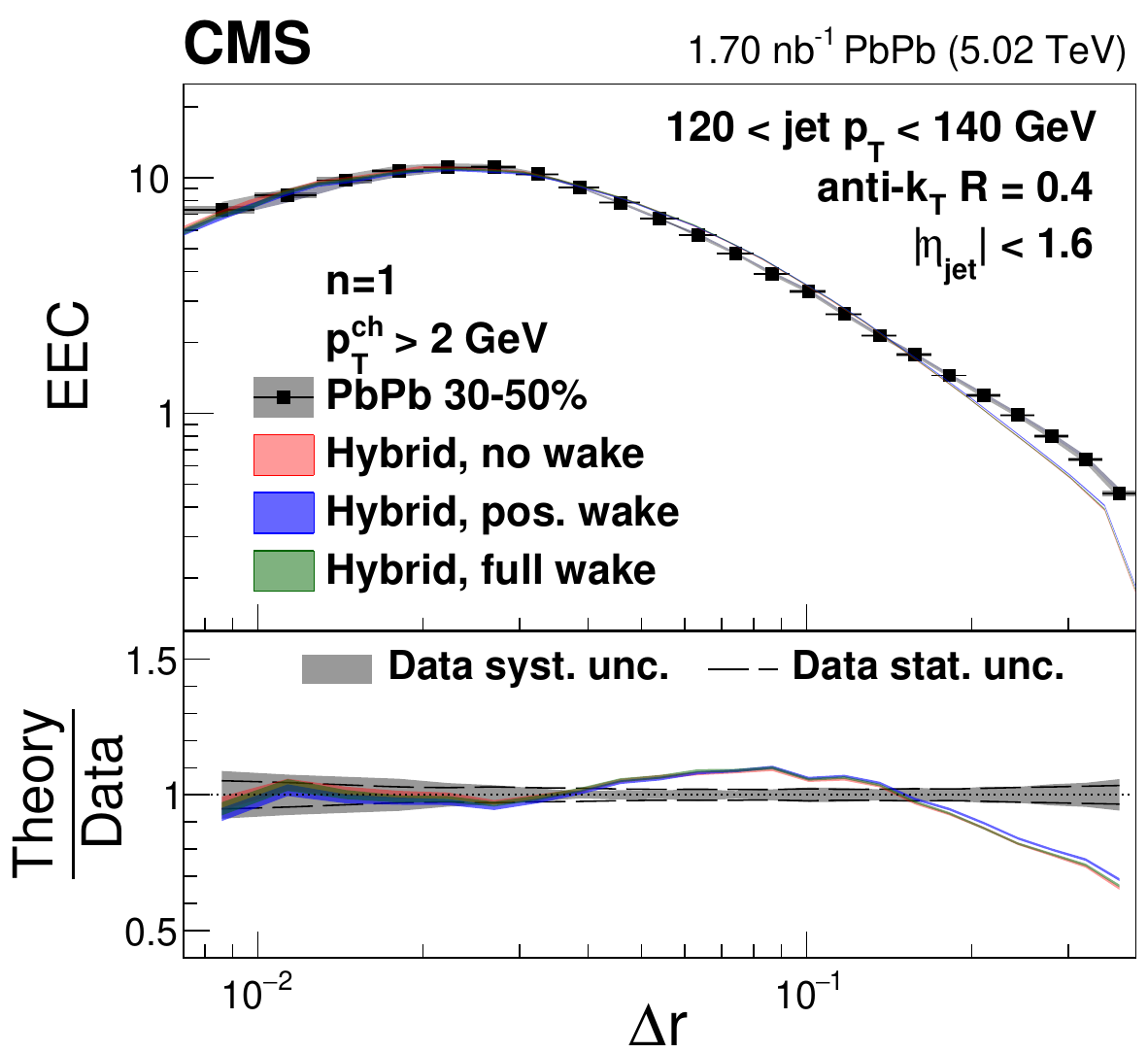}  
    \includegraphics[width=0.4\textwidth]{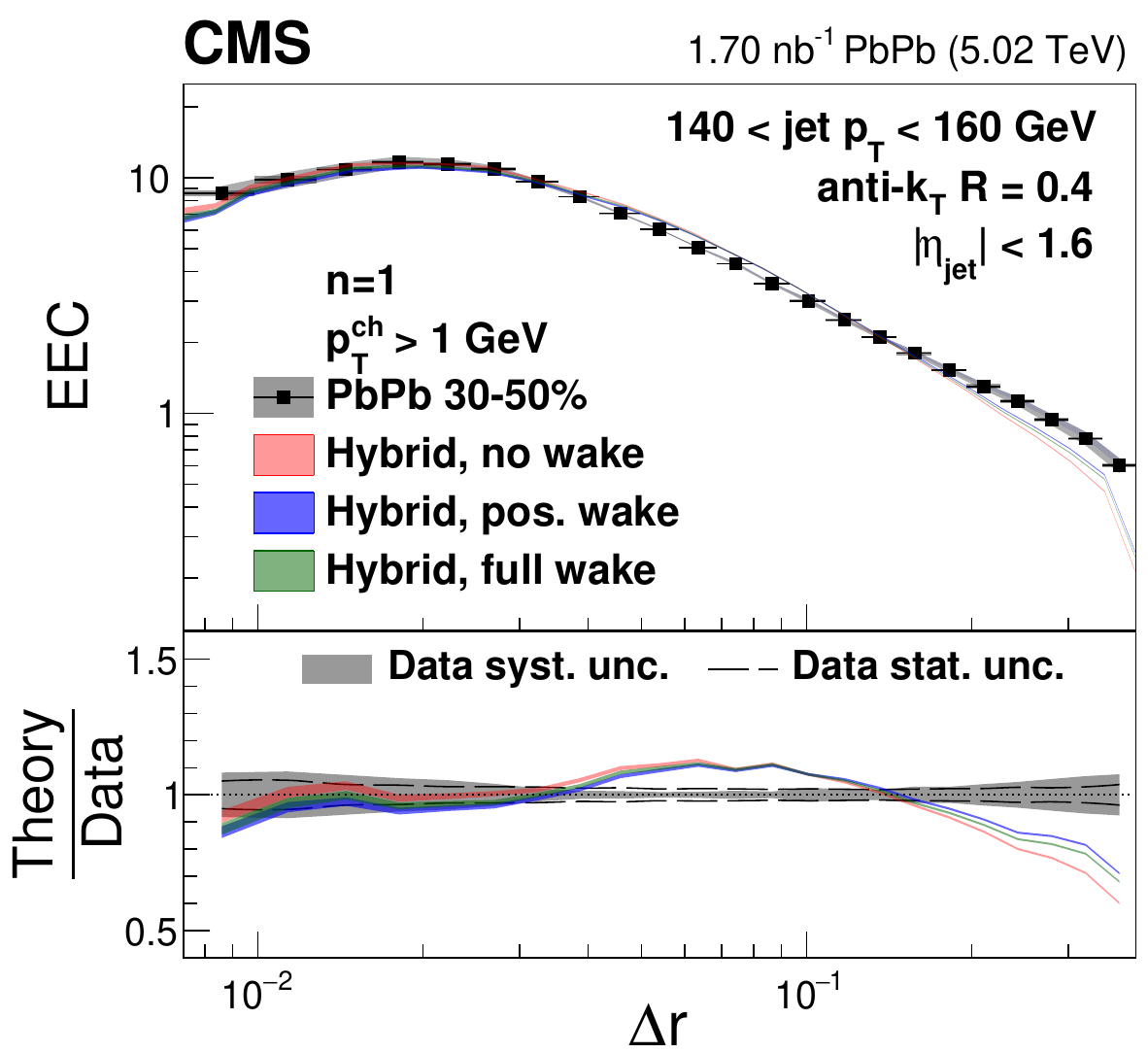}
    \includegraphics[width=0.4\textwidth]{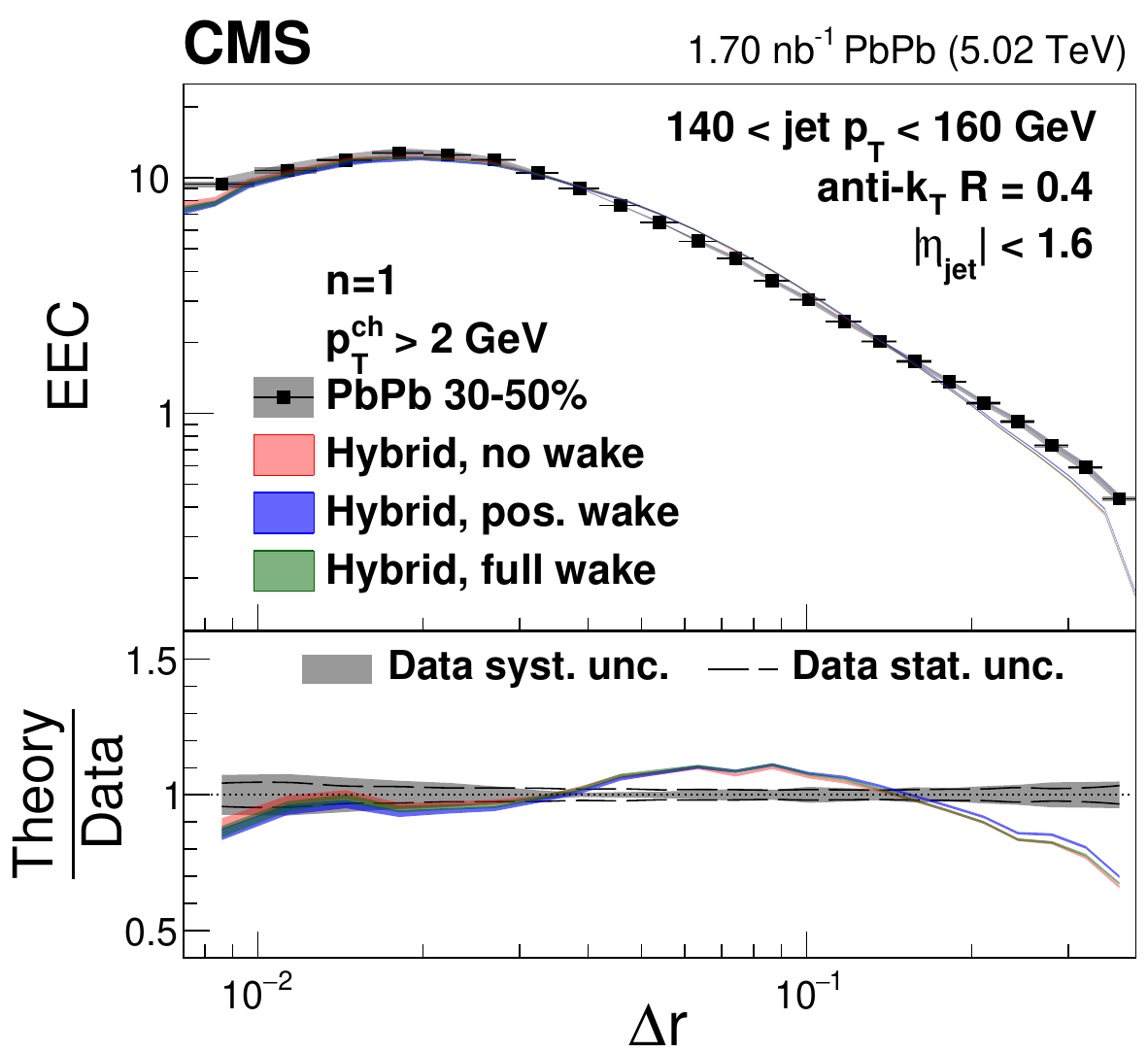}
    \includegraphics[width=0.4\textwidth]{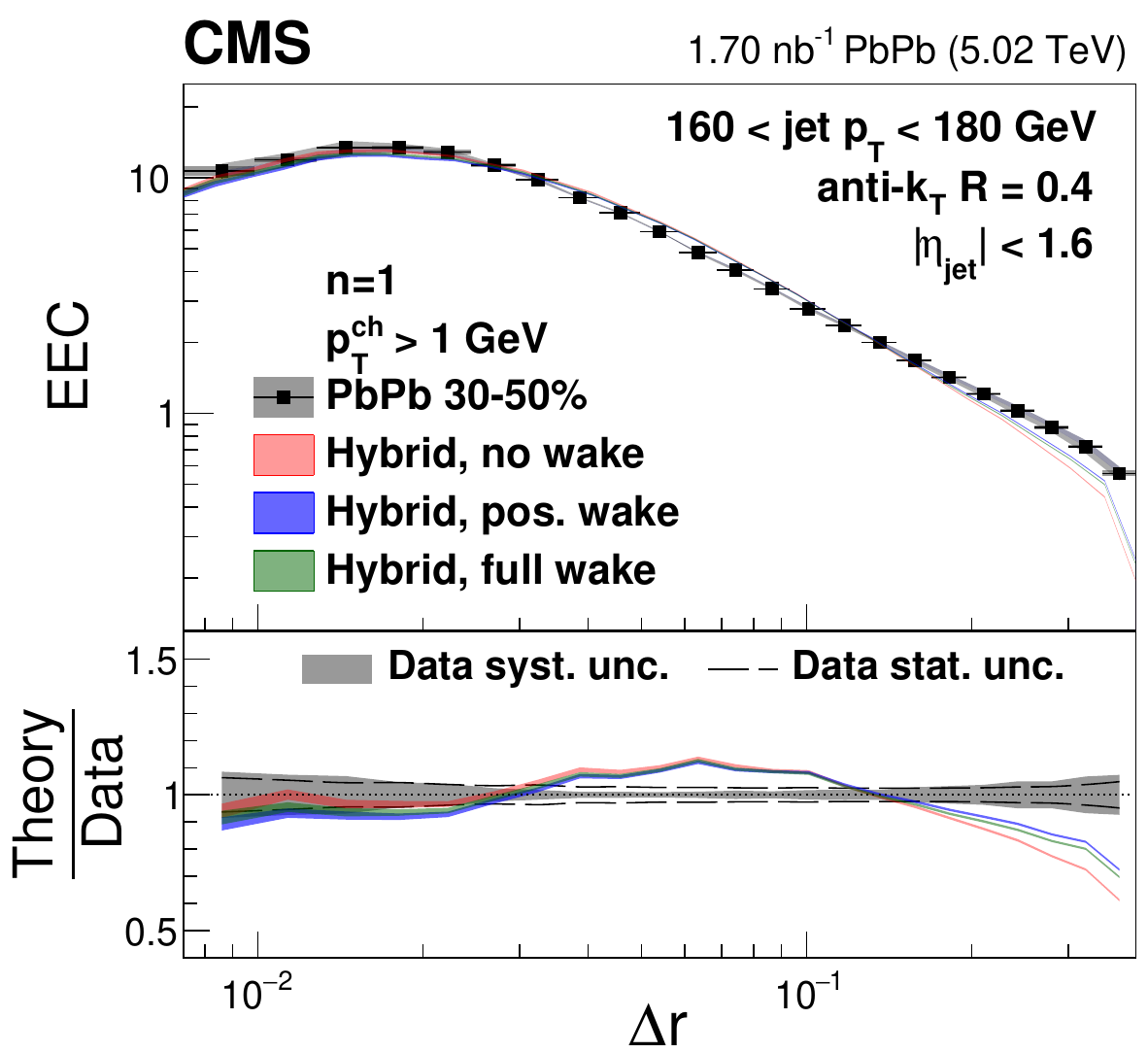}
    \includegraphics[width=0.4\textwidth]{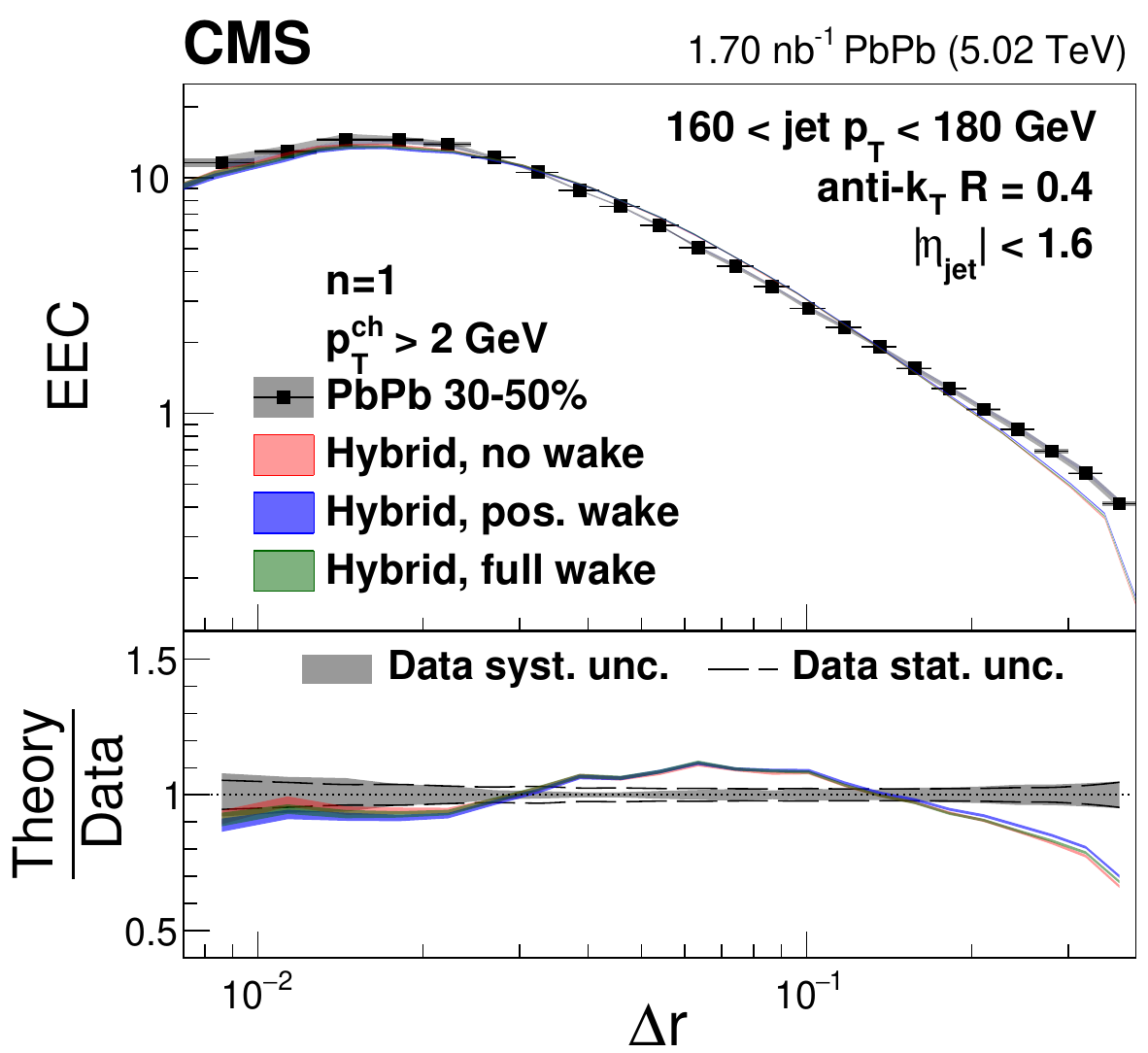}
    \includegraphics[width=0.4\textwidth]{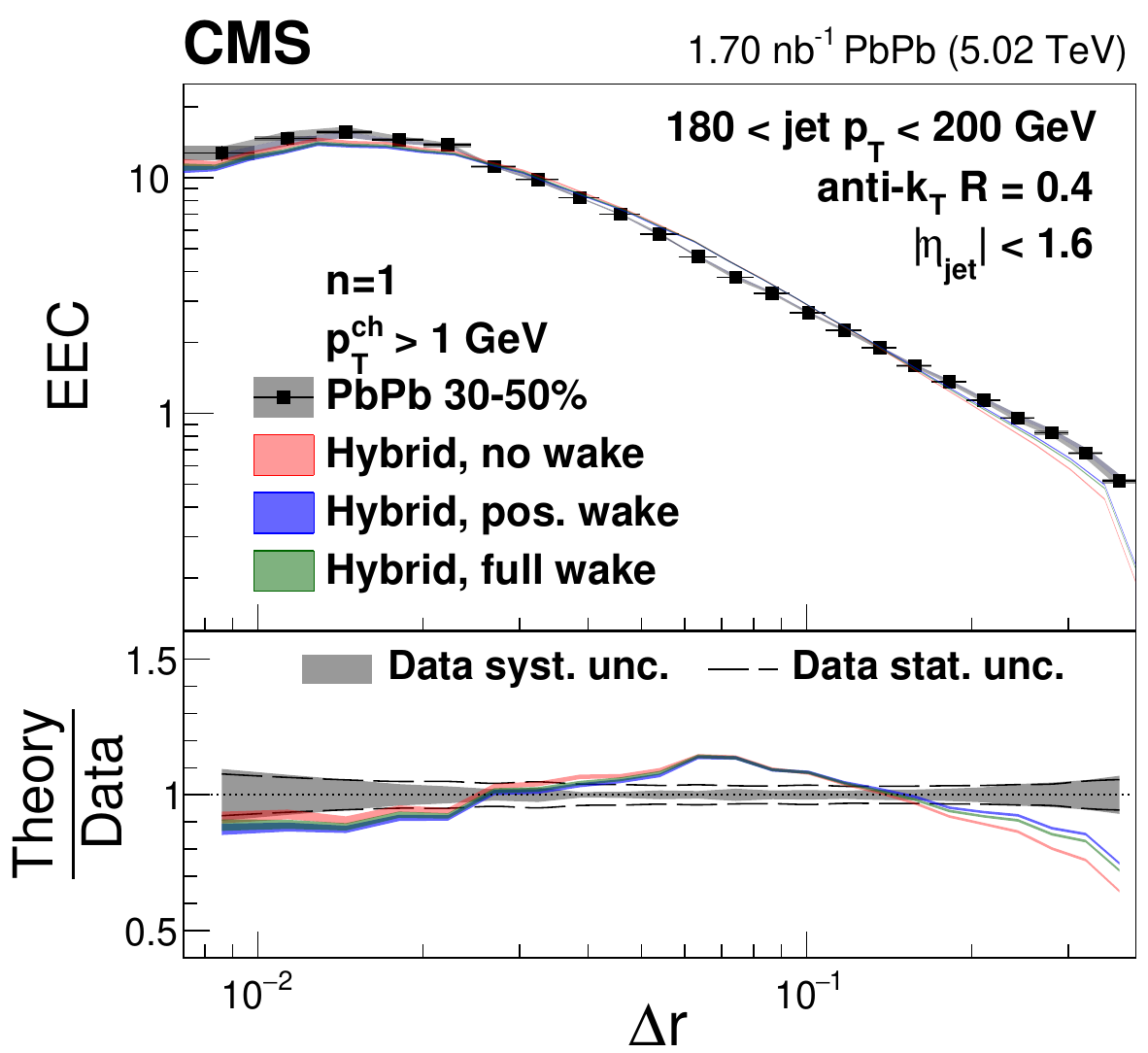}
    \includegraphics[width=0.4\textwidth]{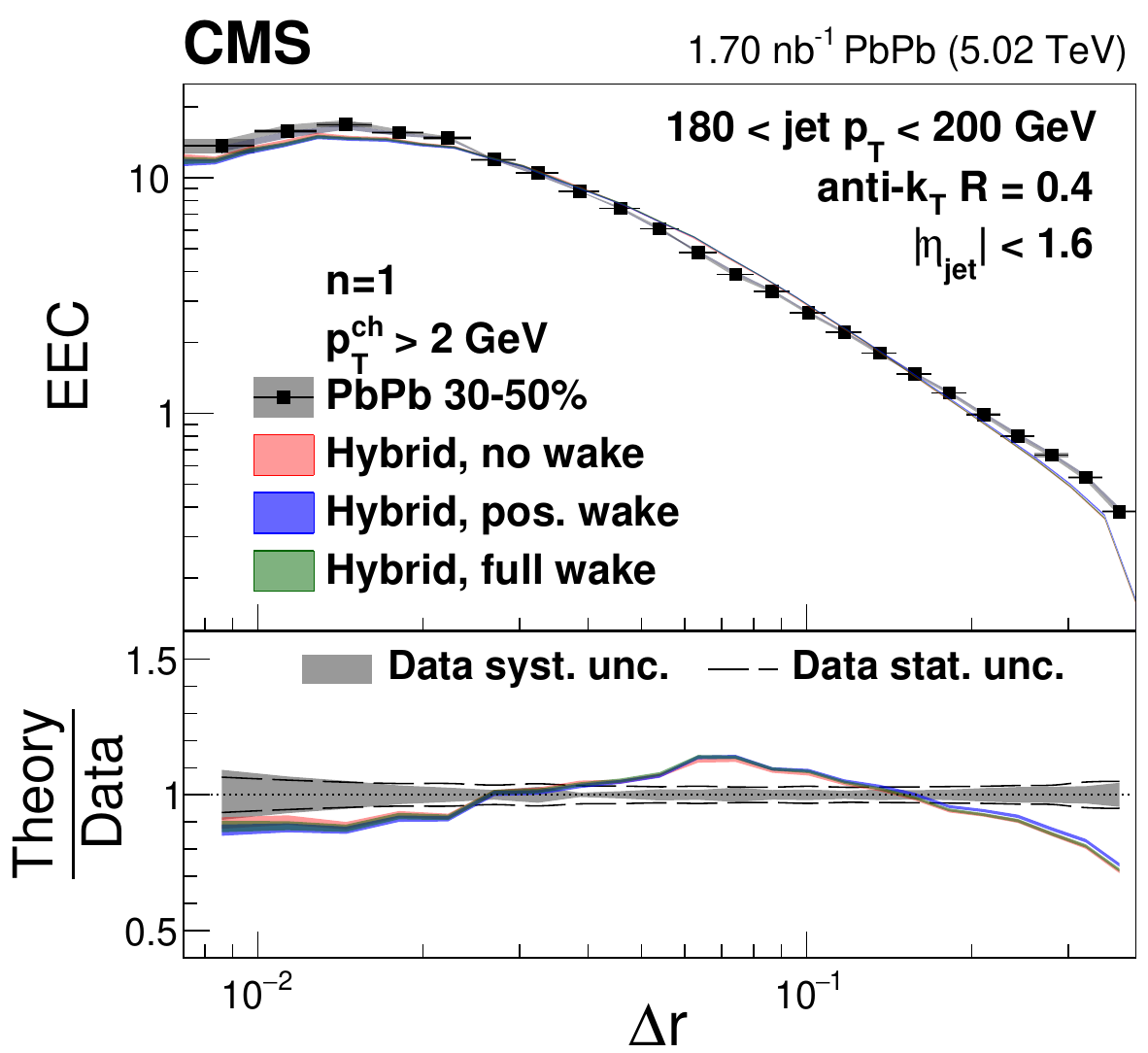}
    \caption{Energy-energy correlator distributions from 30--50\% central \PbPb collisions with $n=1$ in different \ptCh and \ptjet bins compared to the hybrid model predictions with different jet wake settings.} 
    \label{fig:PbPbDistributionsNominal30-50}
  } 
\end{figure*}

\begin{figure*}[h]
  \centering{
    \includegraphics[width=0.4\textwidth]{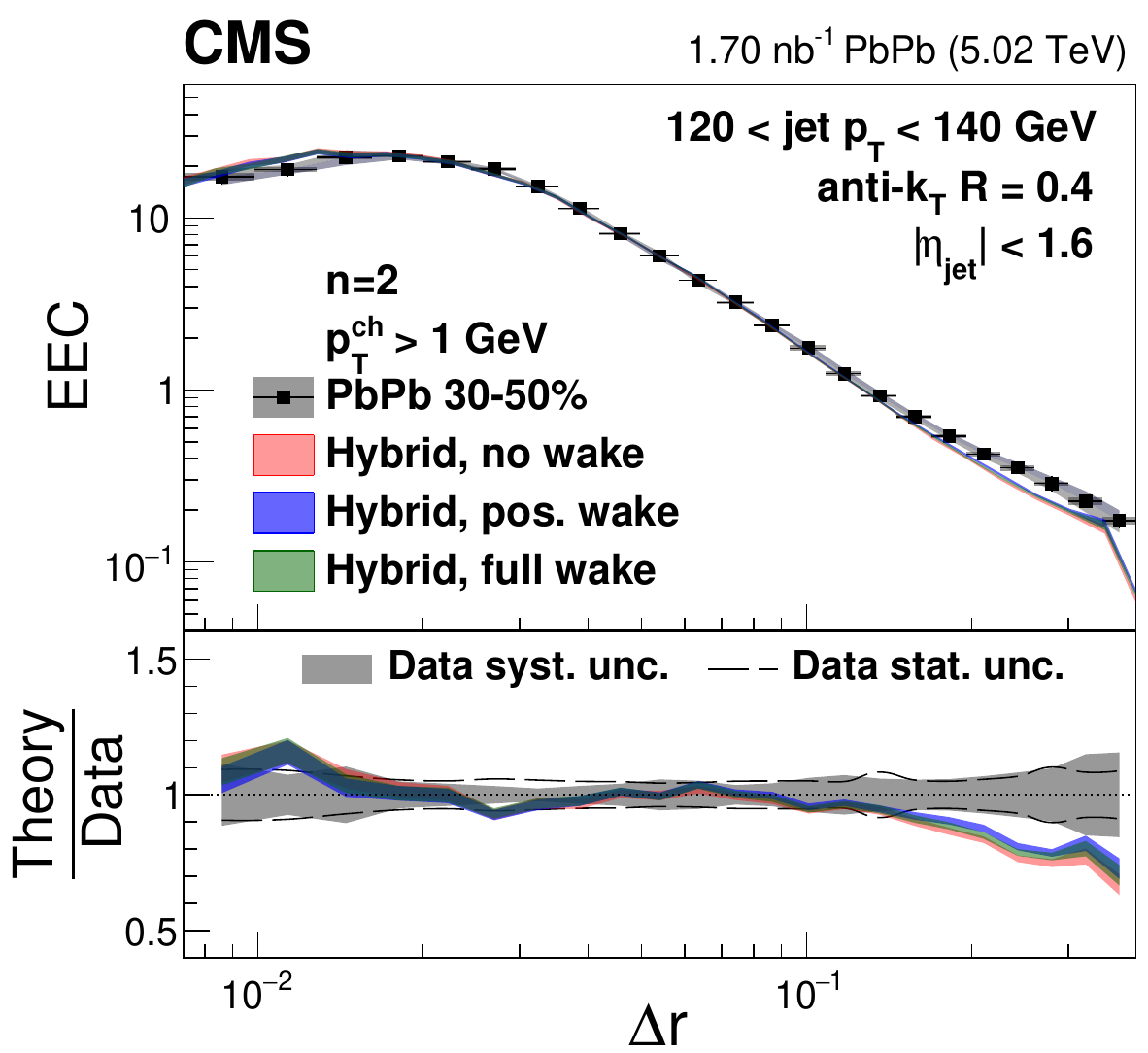}
    \includegraphics[width=0.4\textwidth]{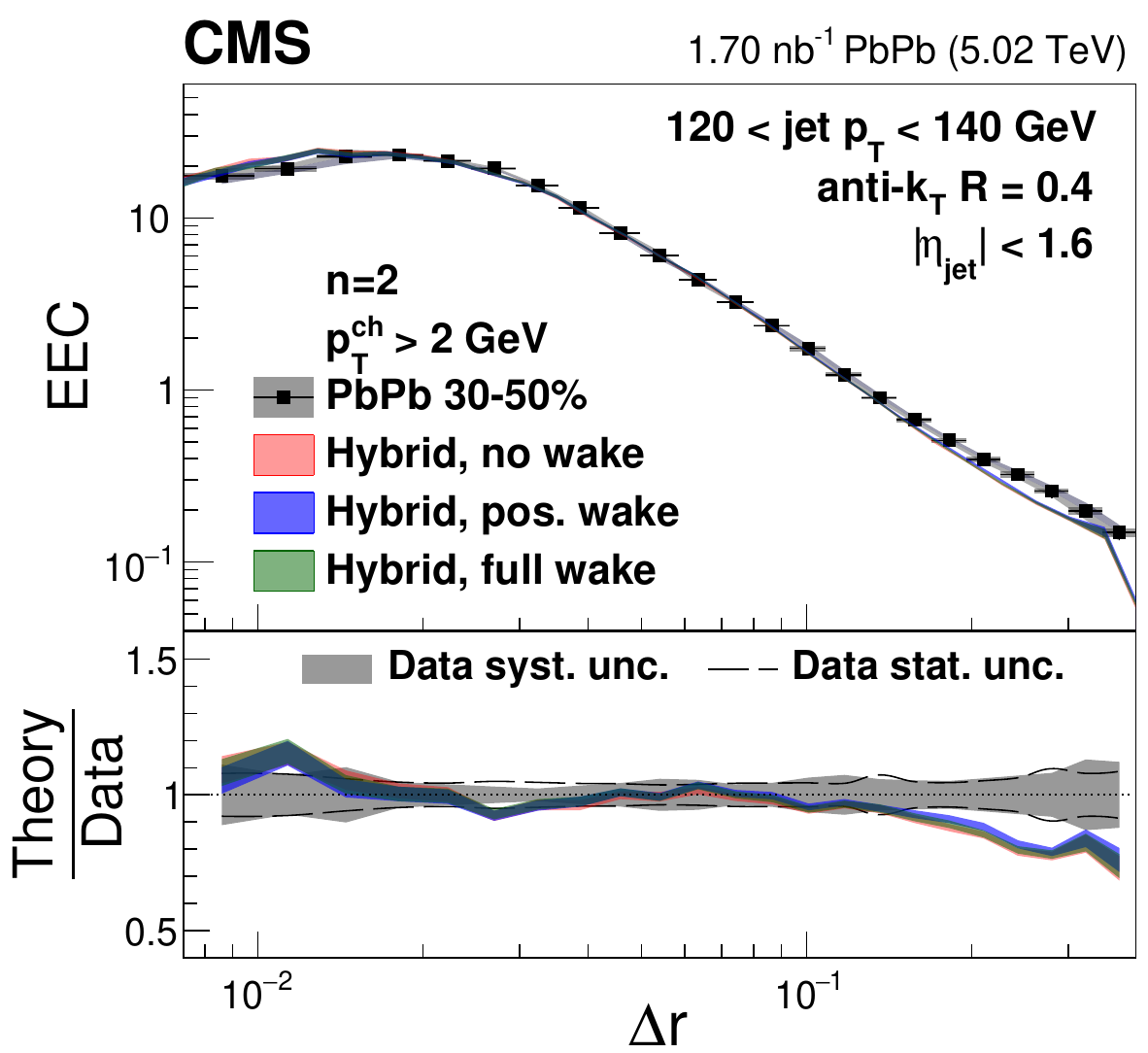}  
    \includegraphics[width=0.4\textwidth]{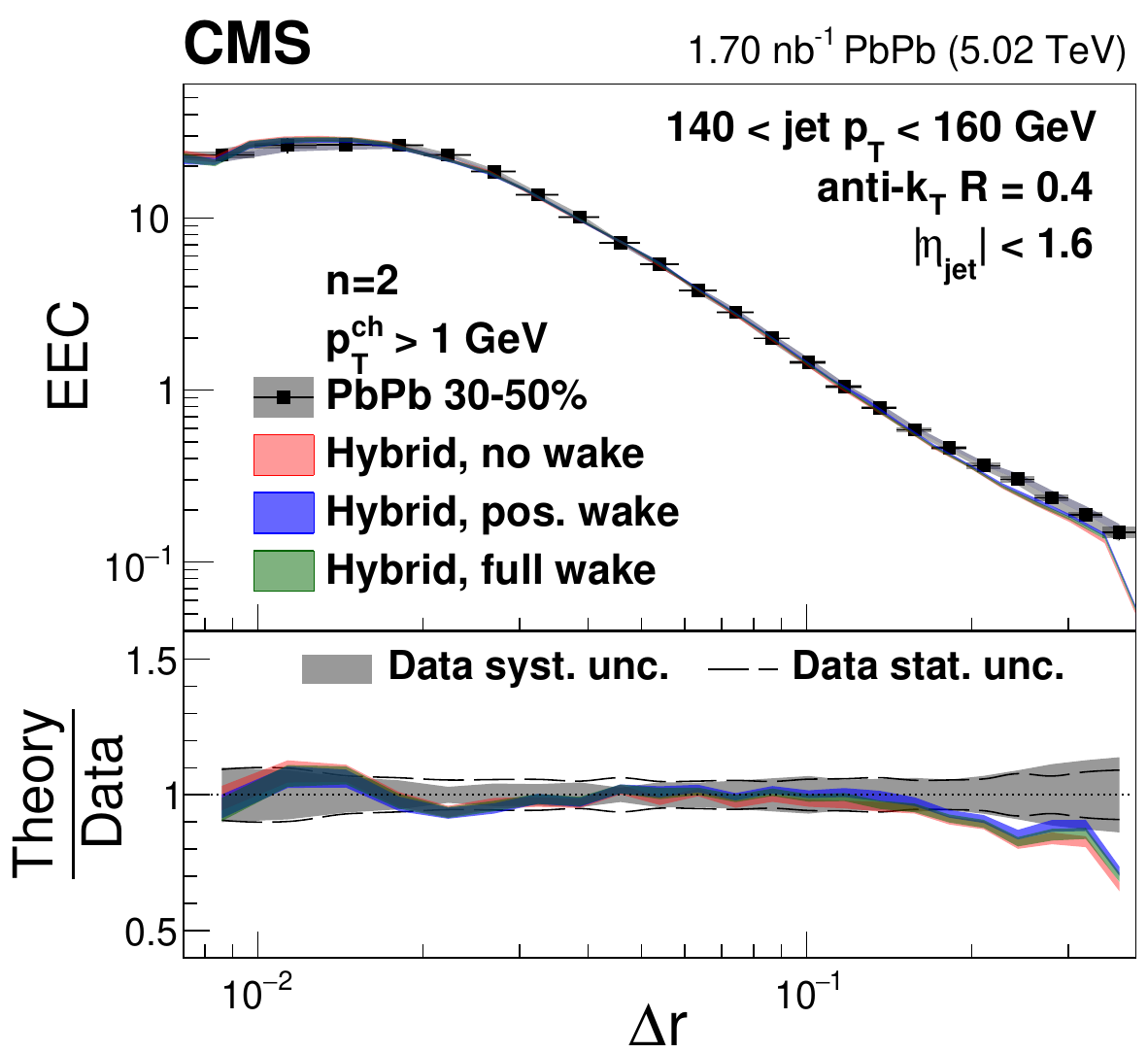}
    \includegraphics[width=0.4\textwidth]{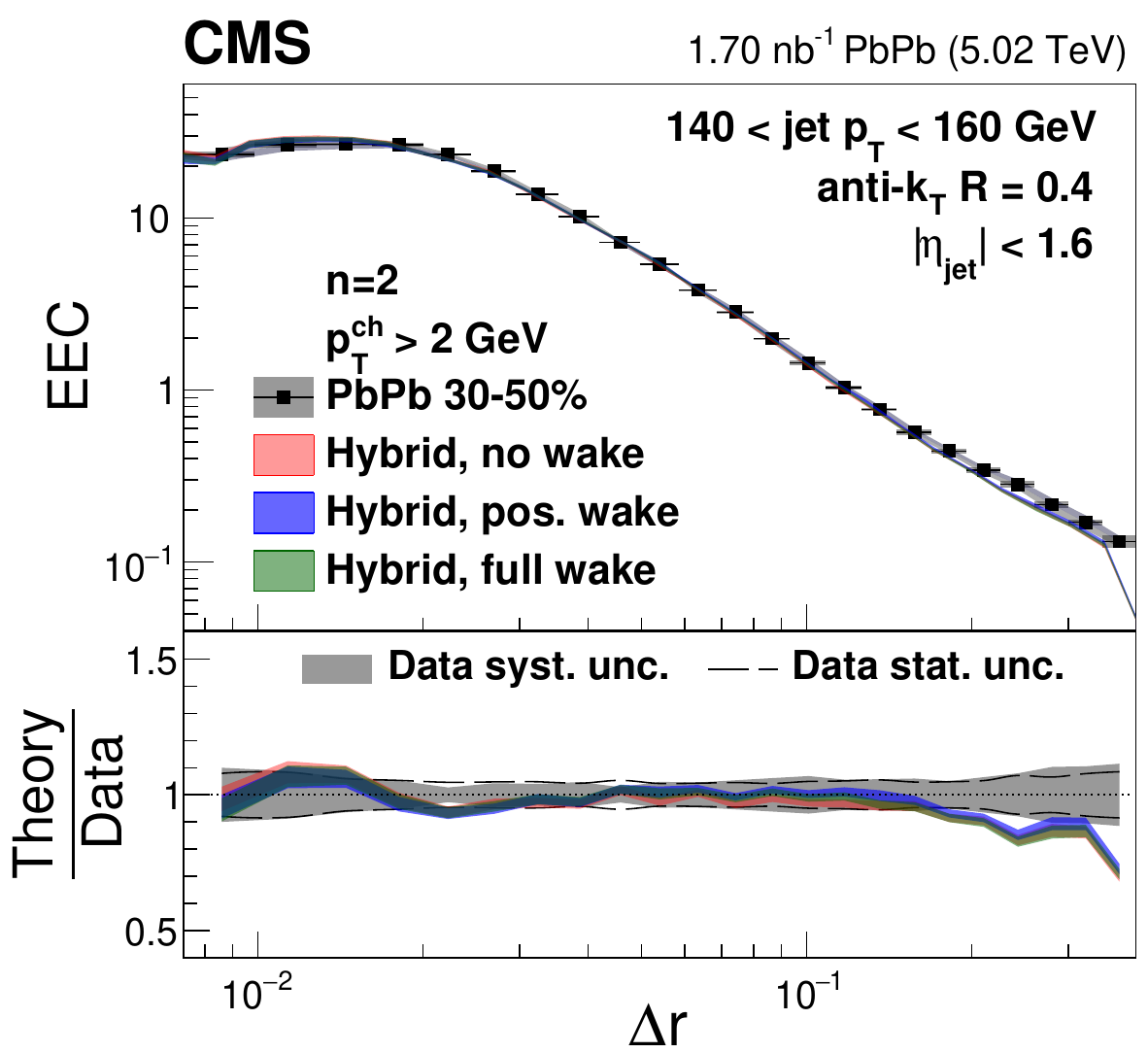}
    \includegraphics[width=0.4\textwidth]{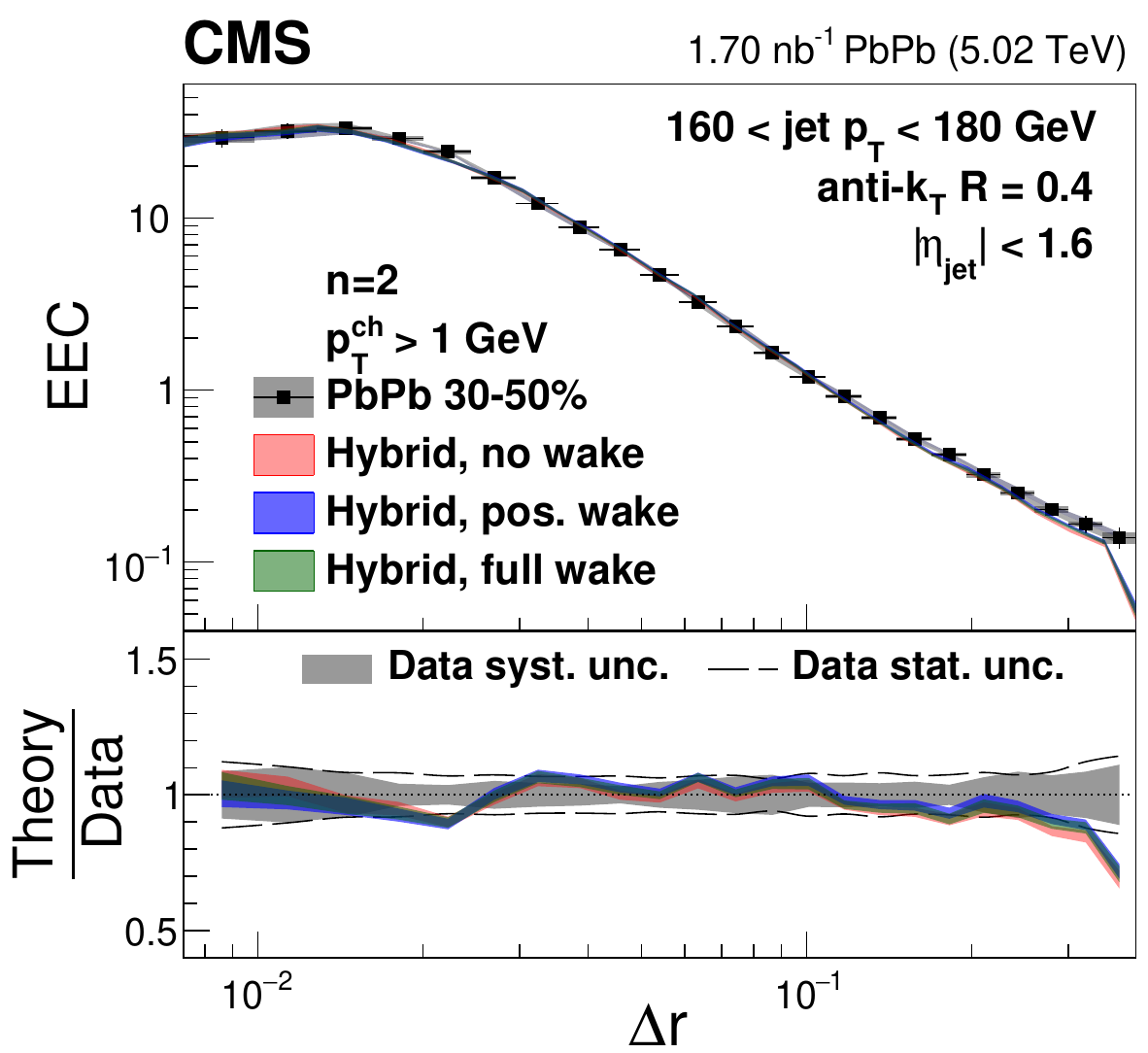}
    \includegraphics[width=0.4\textwidth]{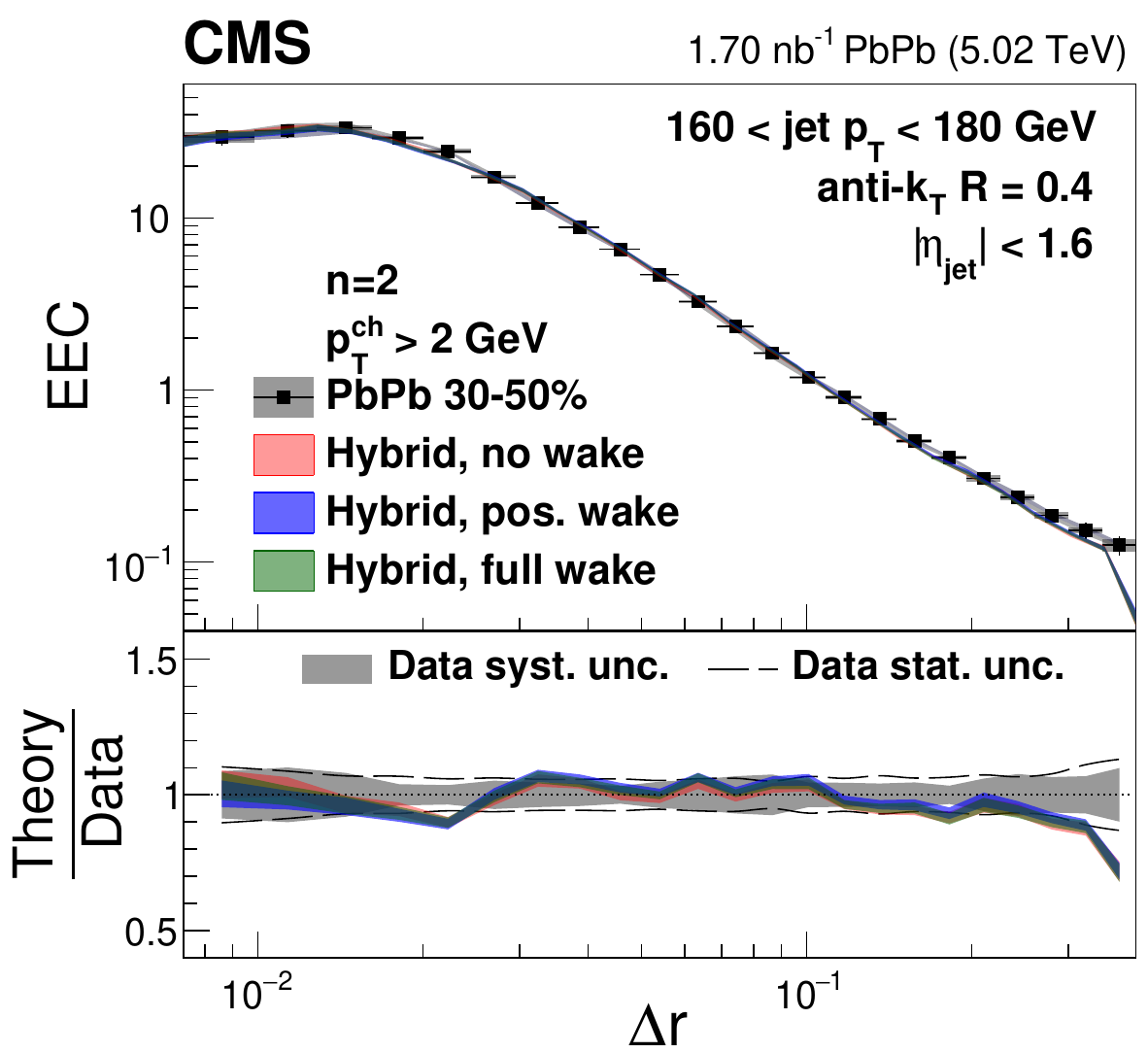}
    \includegraphics[width=0.4\textwidth]{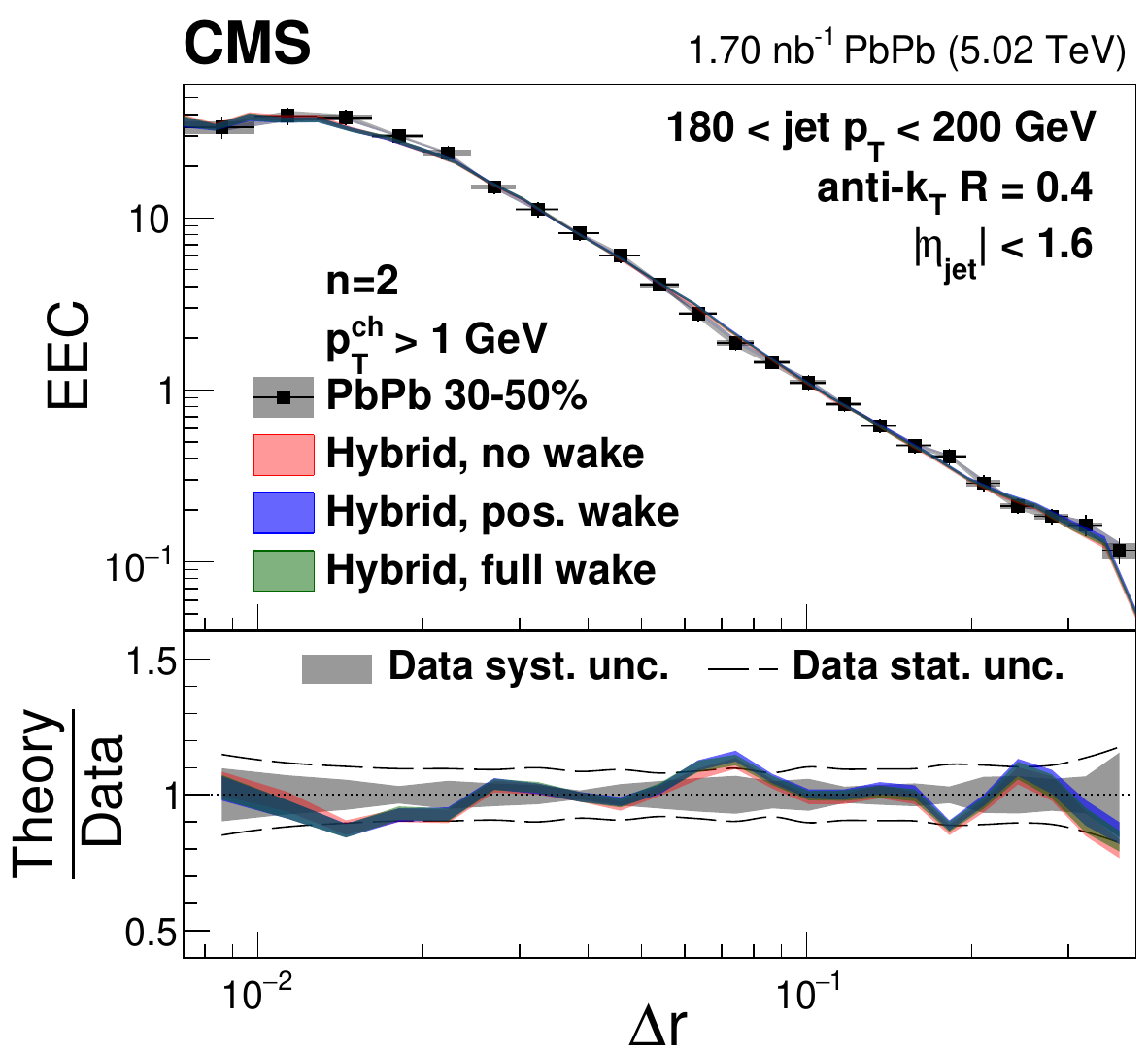}
    \includegraphics[width=0.4\textwidth]{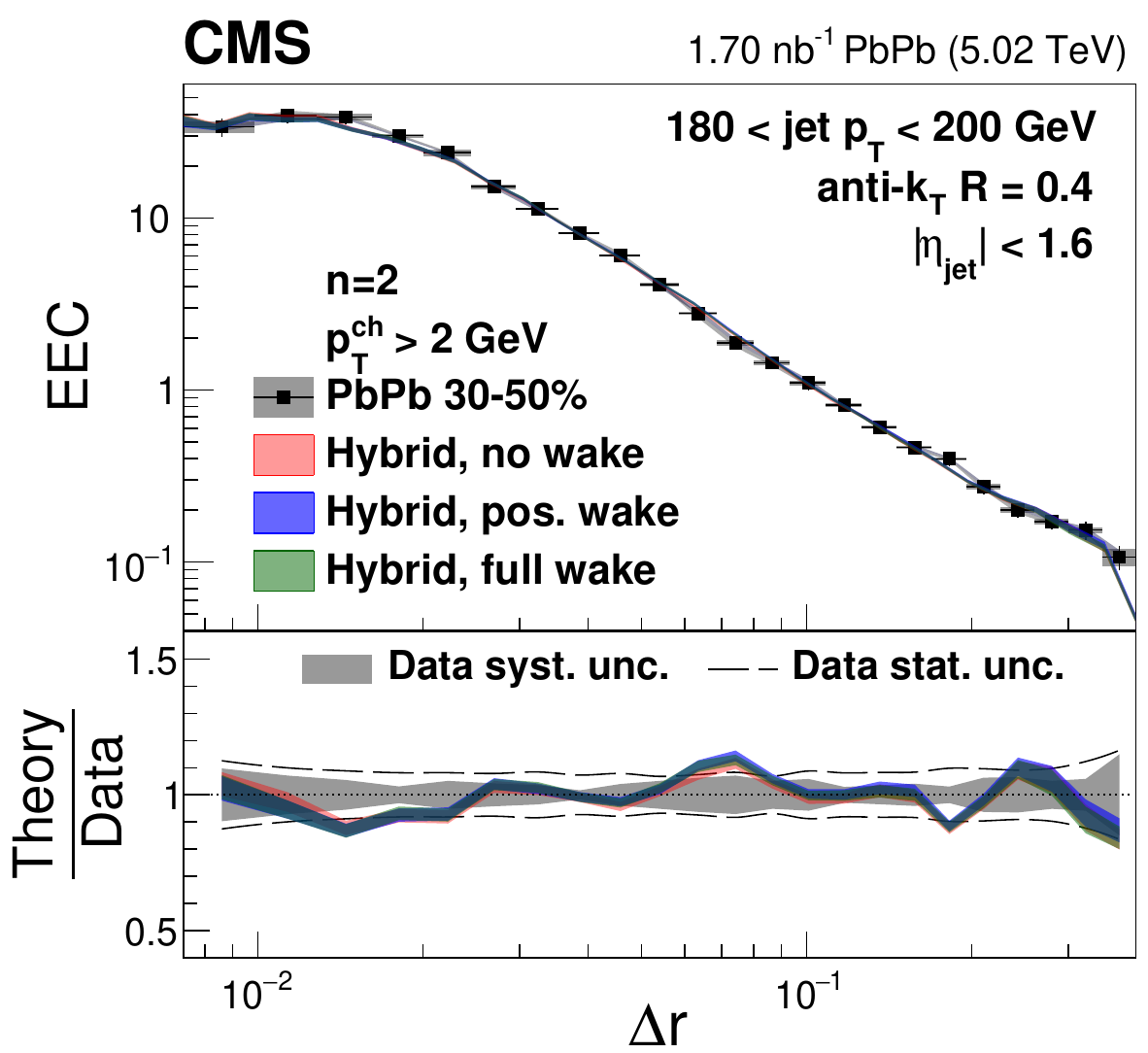}
    \caption{Energy-energy correlator distributions from 30--50\% central \PbPb collisions with $n=2$ in different \ptCh and \ptjet bins compared to the hybrid model predictions with different jet wake settings.} 
    \label{fig:PbPbDistributionsSquared30-50}
  } 
\end{figure*}

\begin{figure*}[h]
  \centering{
    \includegraphics[width=0.4\textwidth]{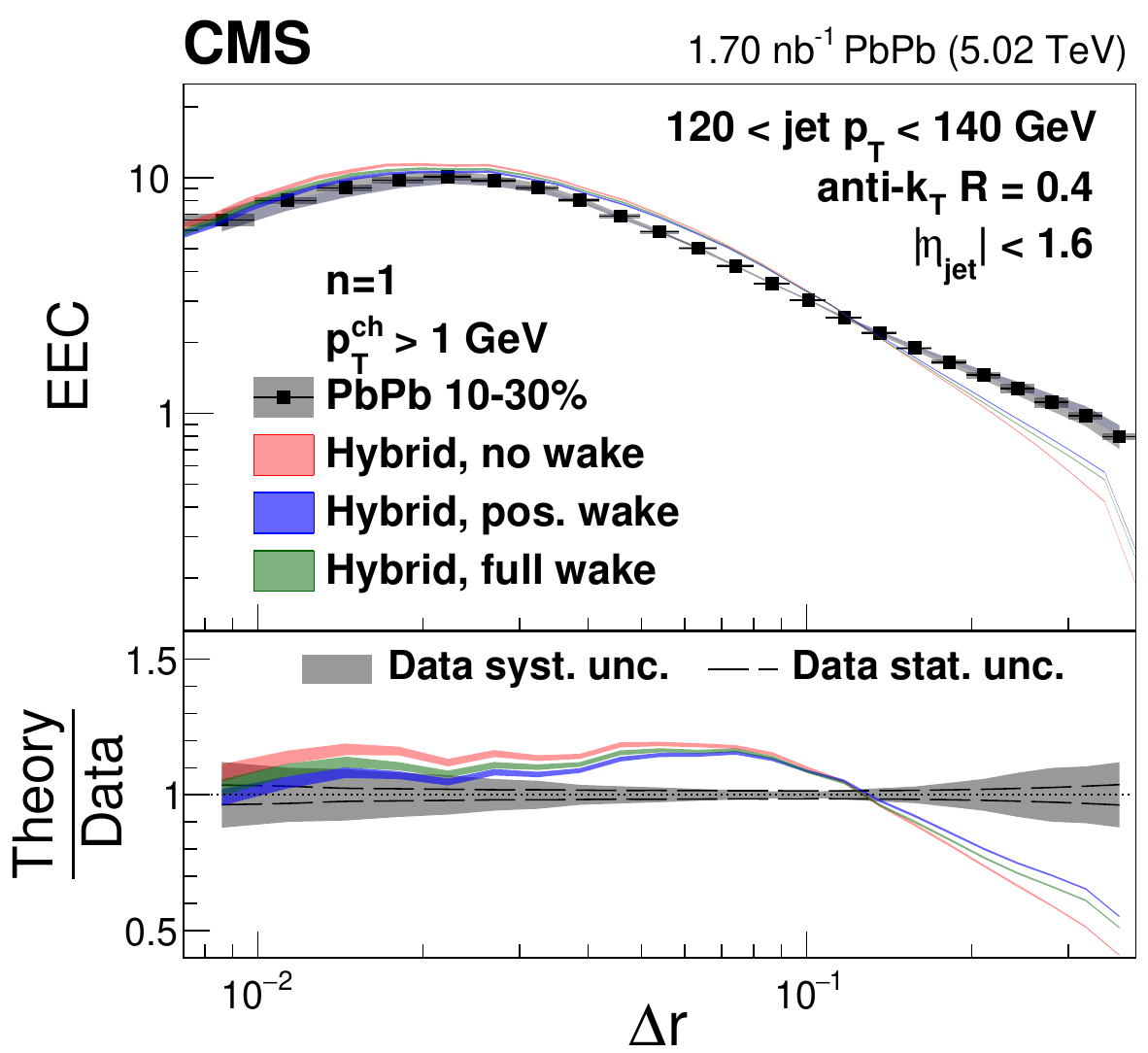}
    \includegraphics[width=0.4\textwidth]{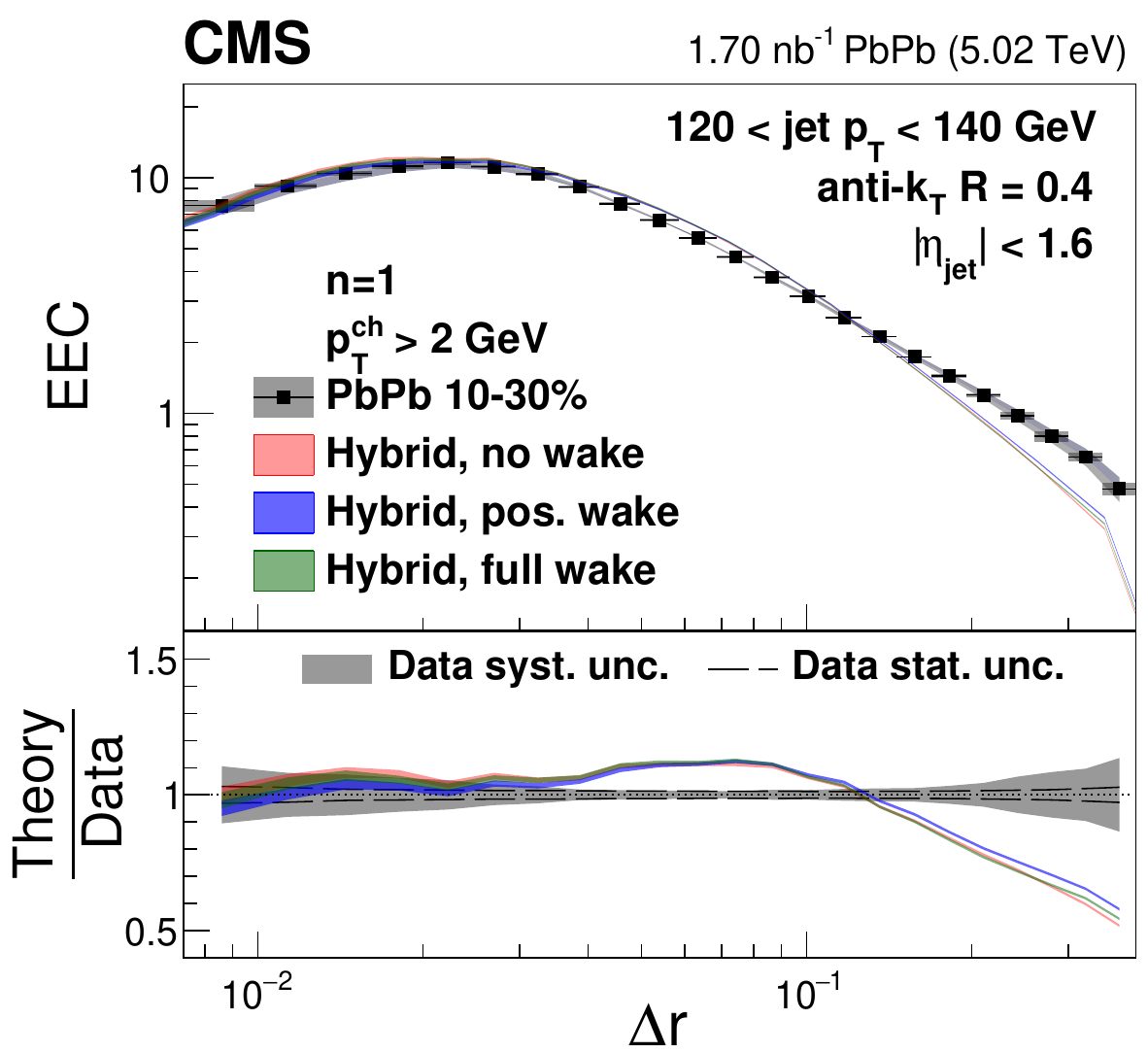}  
    \includegraphics[width=0.4\textwidth]{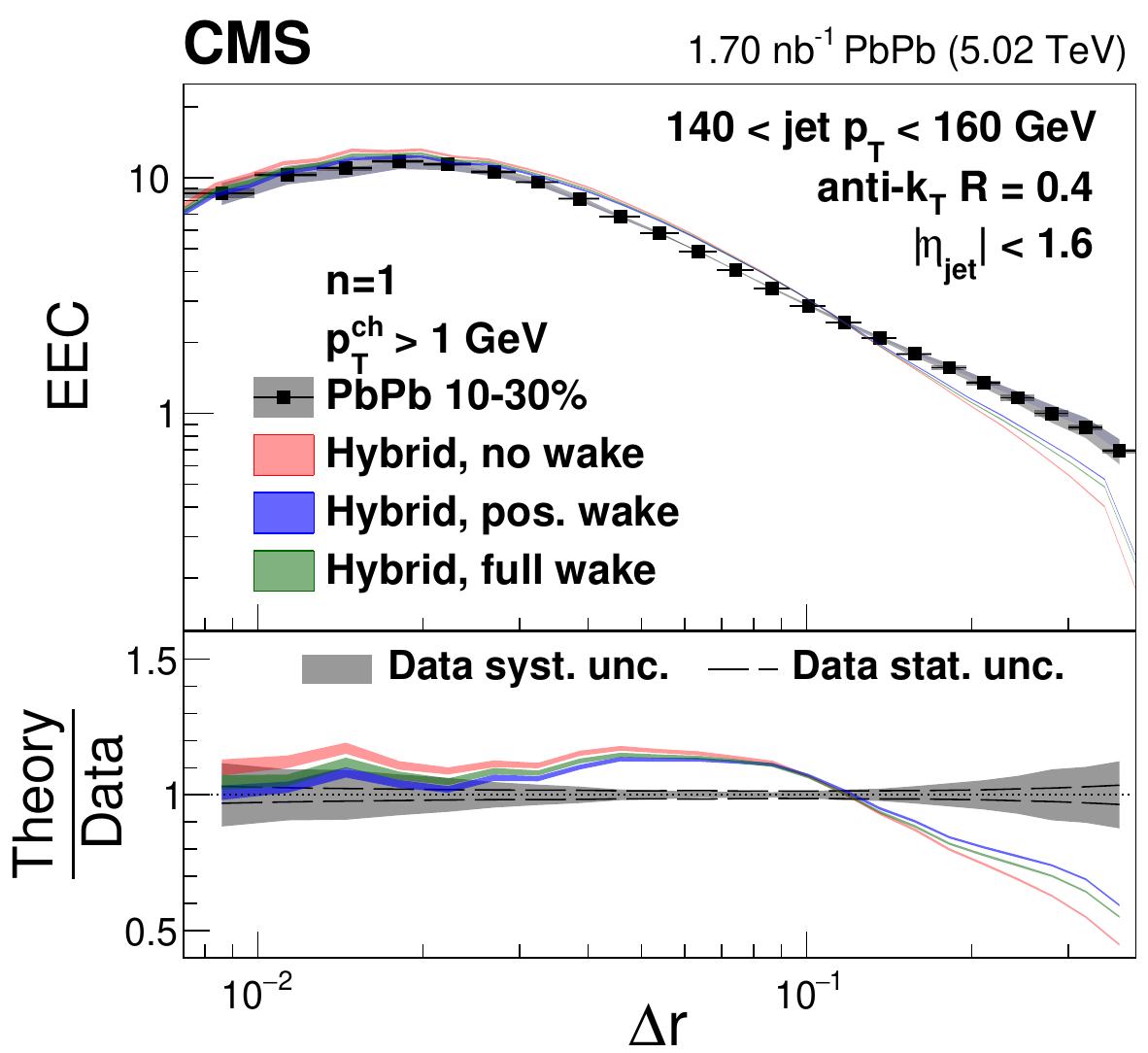}
    \includegraphics[width=0.4\textwidth]{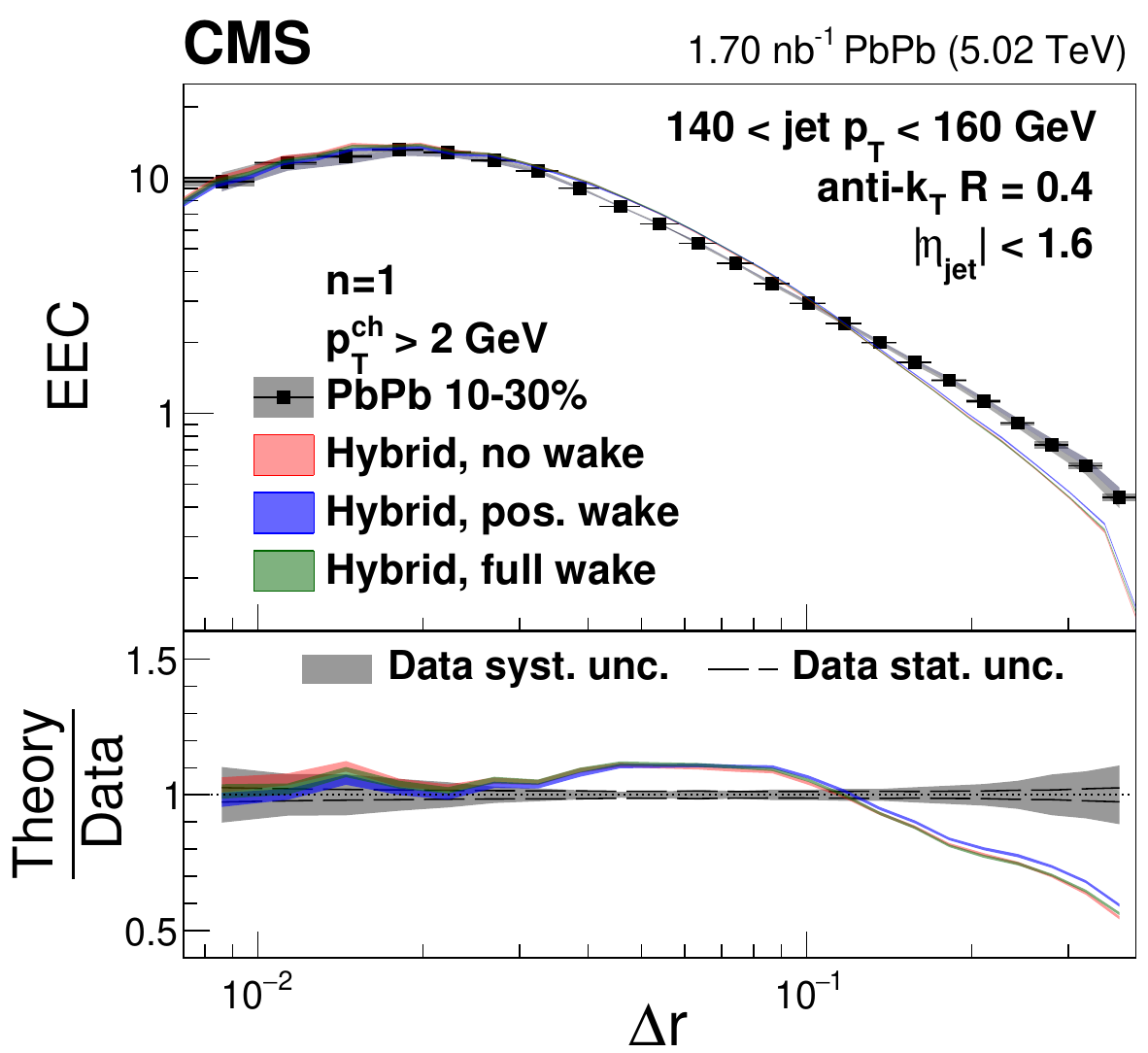}
    \includegraphics[width=0.4\textwidth]{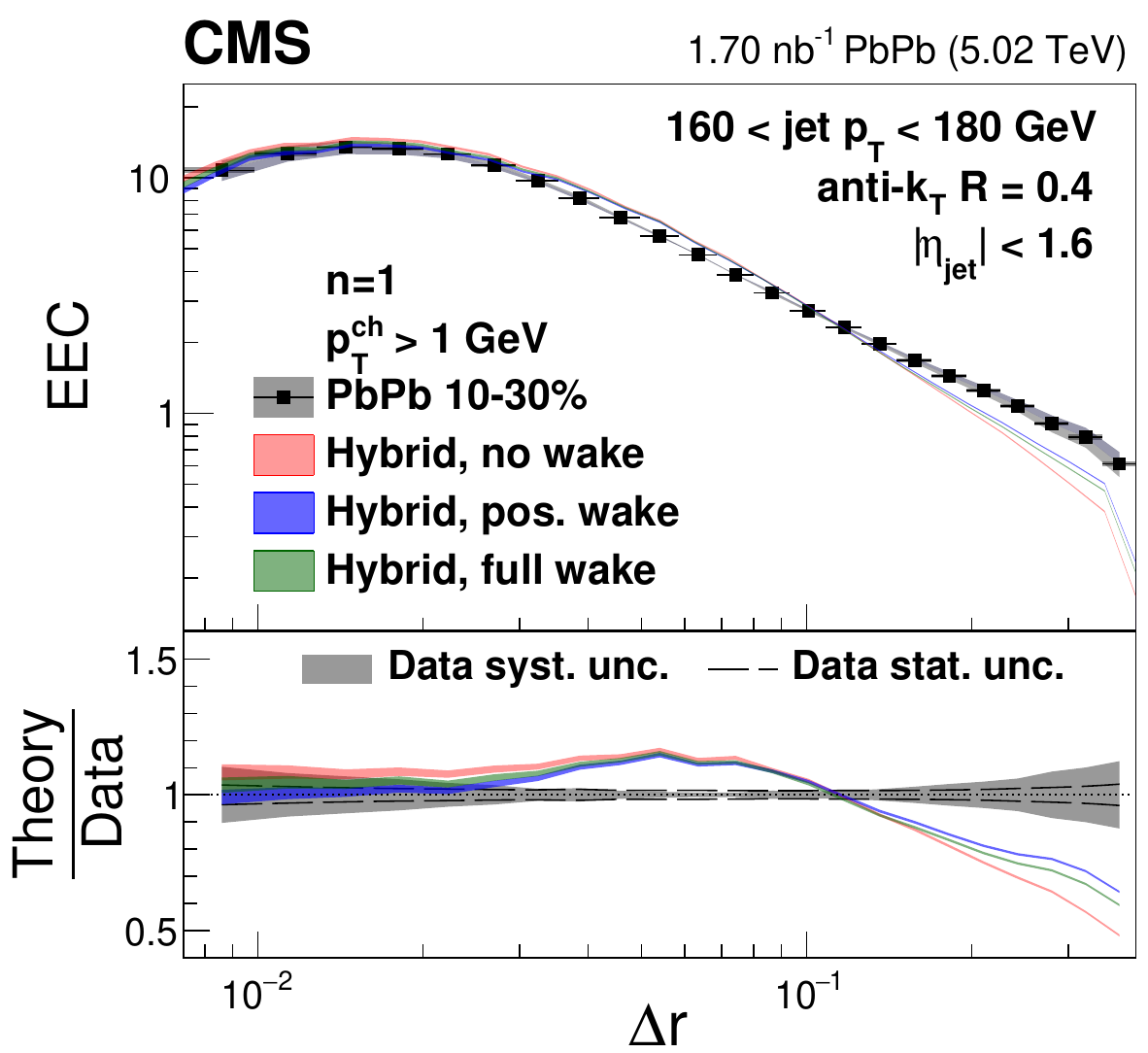}
    \includegraphics[width=0.4\textwidth]{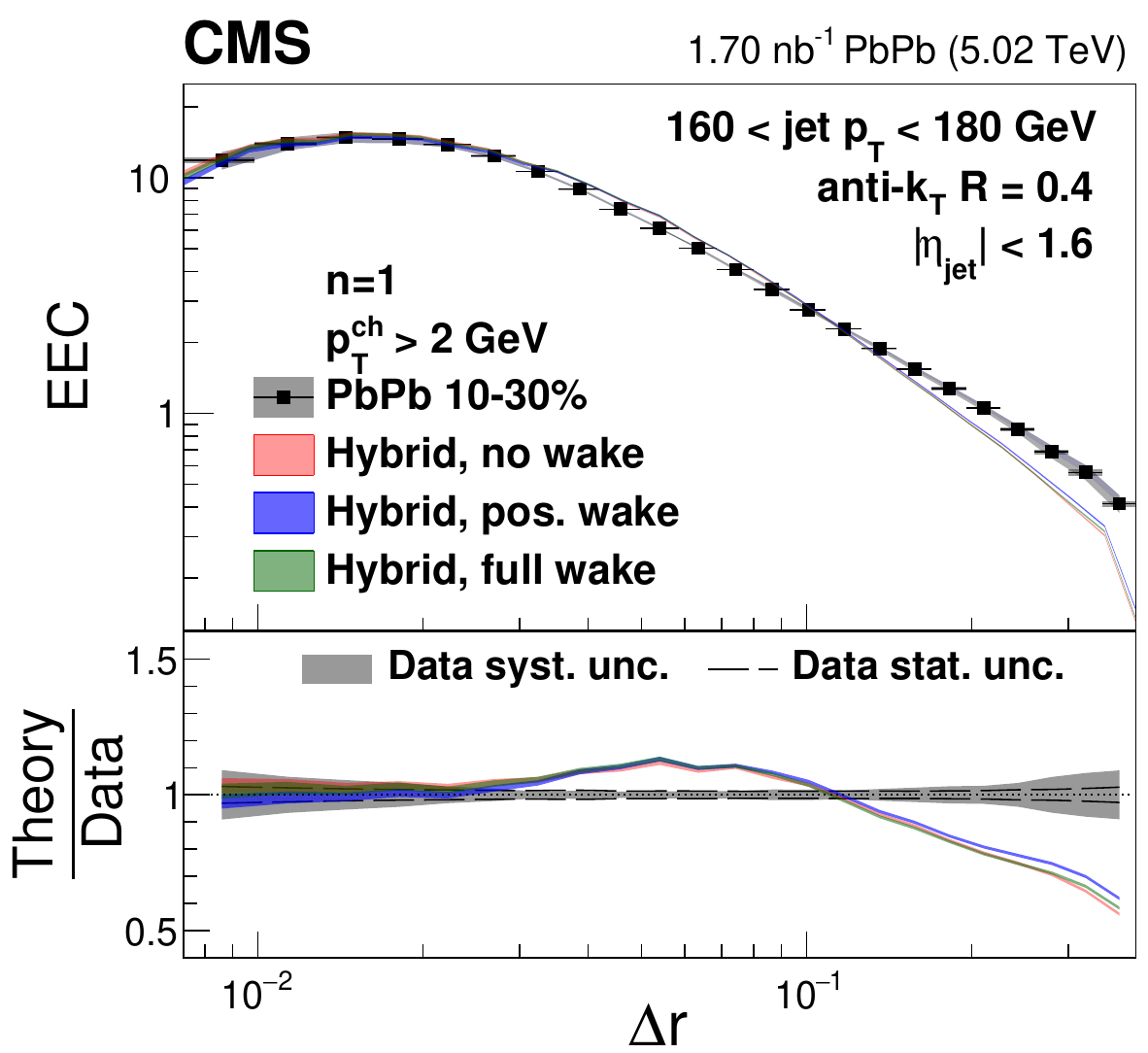}
    \includegraphics[width=0.4\textwidth]{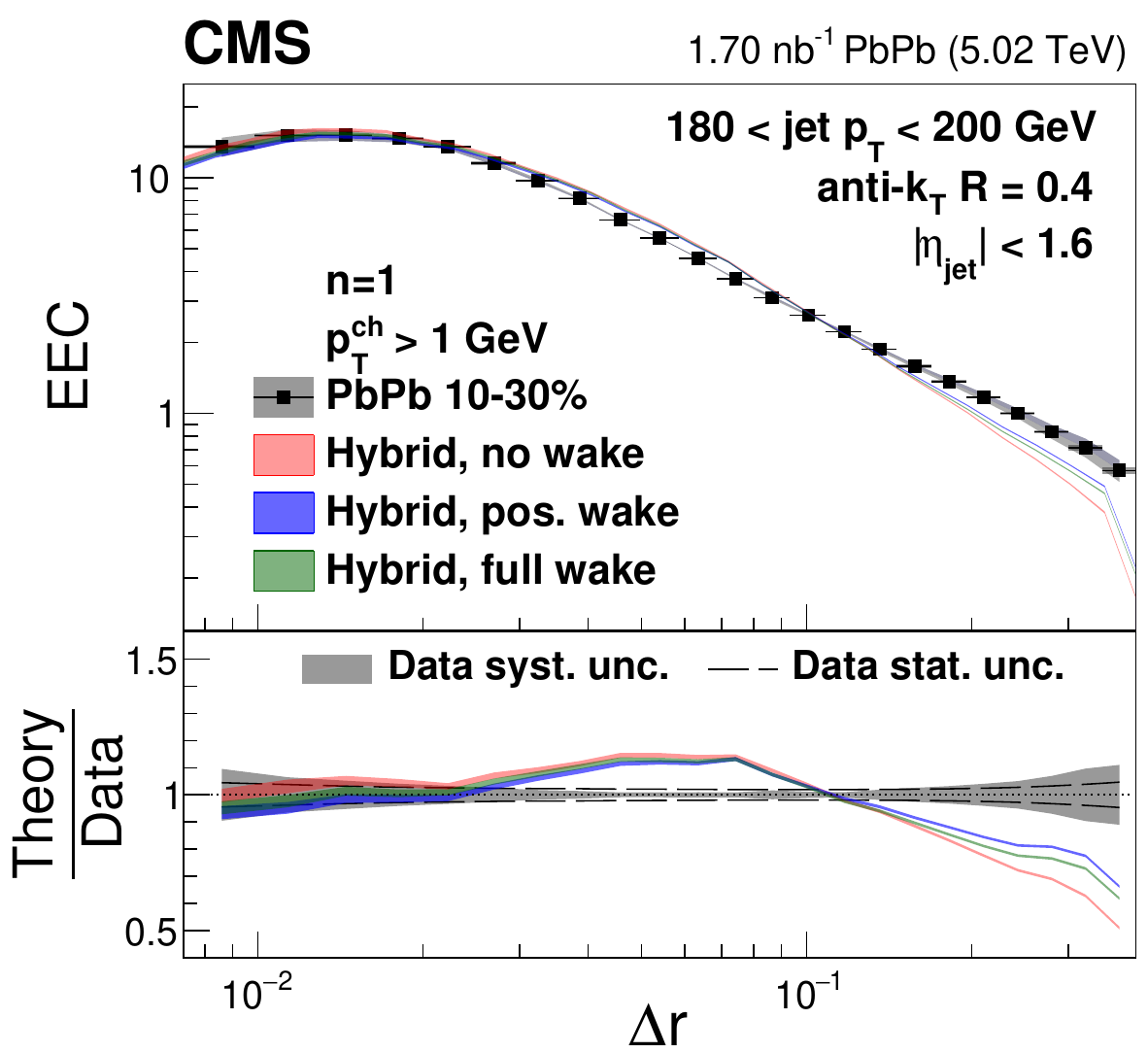}
    \includegraphics[width=0.4\textwidth]{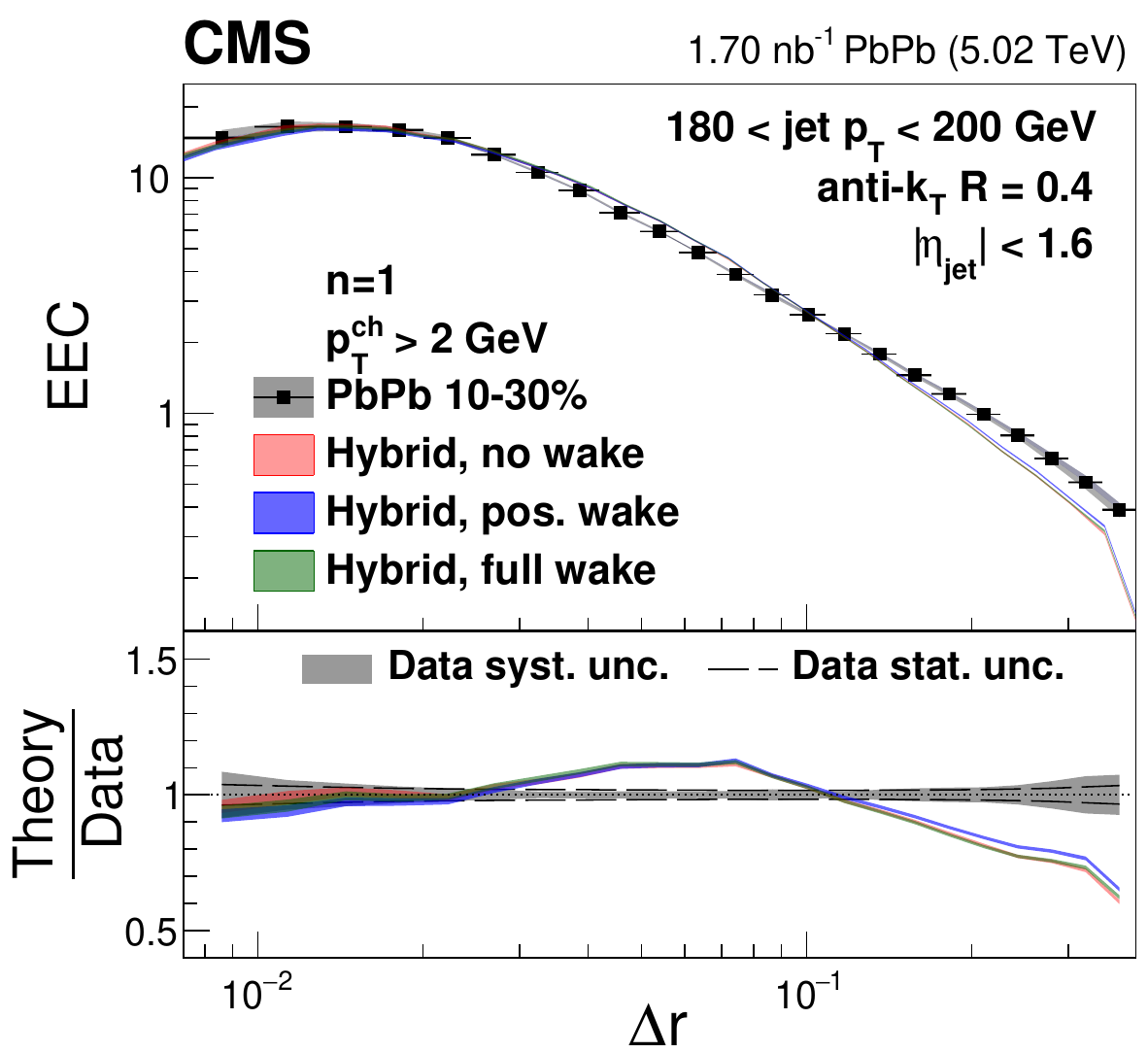}
    \caption{Energy-energy correlator distributions from 10--30\% central \PbPb collisions with $n=1$ in different \ptCh and \ptjet bins compared to the hybrid model predictions with different jet wake settings.} 
    \label{fig:PbPbDistributionsNominal10-30}
  } 
\end{figure*}

\begin{figure*}[h]
  \centering{
    \includegraphics[width=0.4\textwidth]{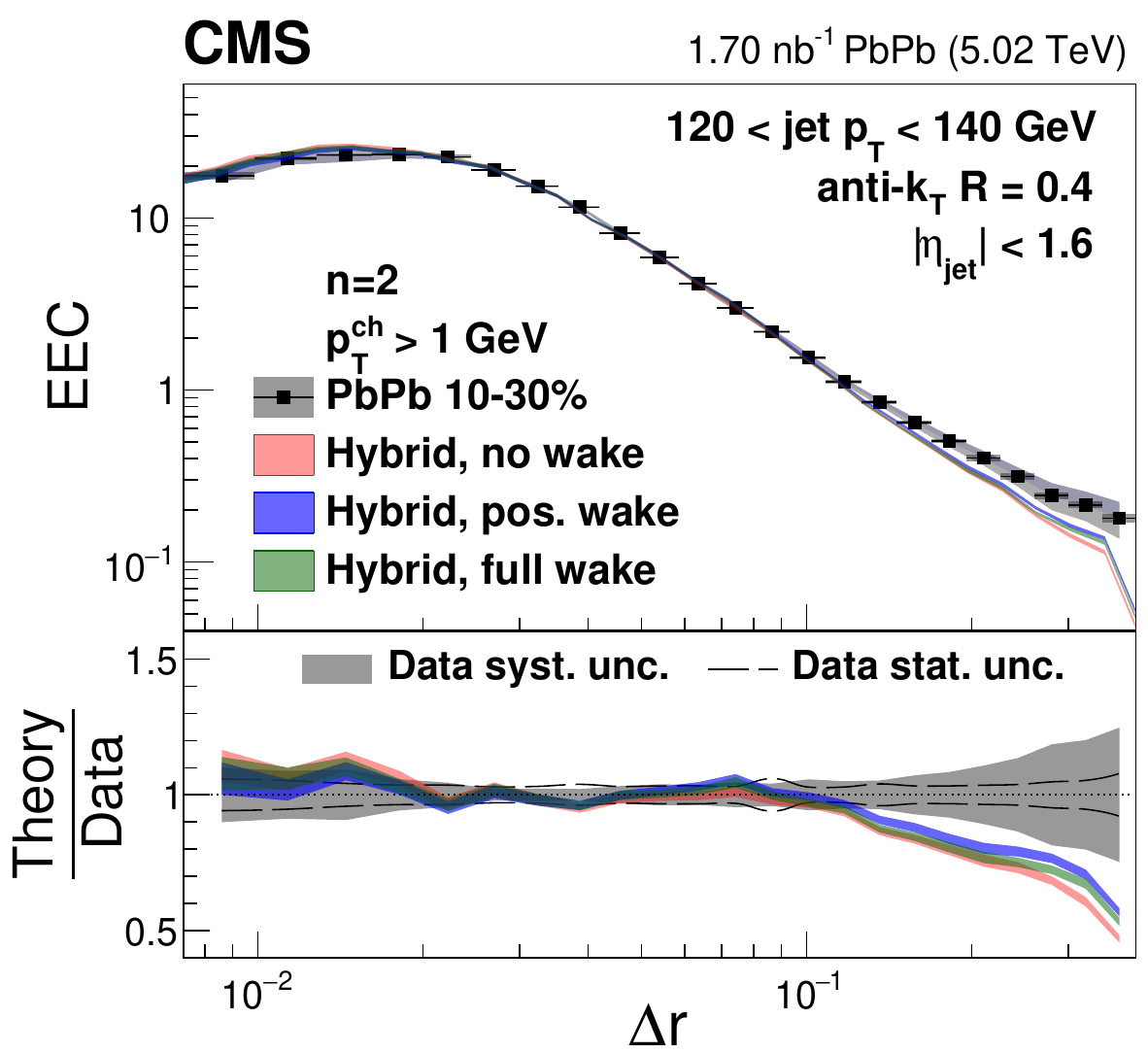} 
    \includegraphics[width=0.4\textwidth]{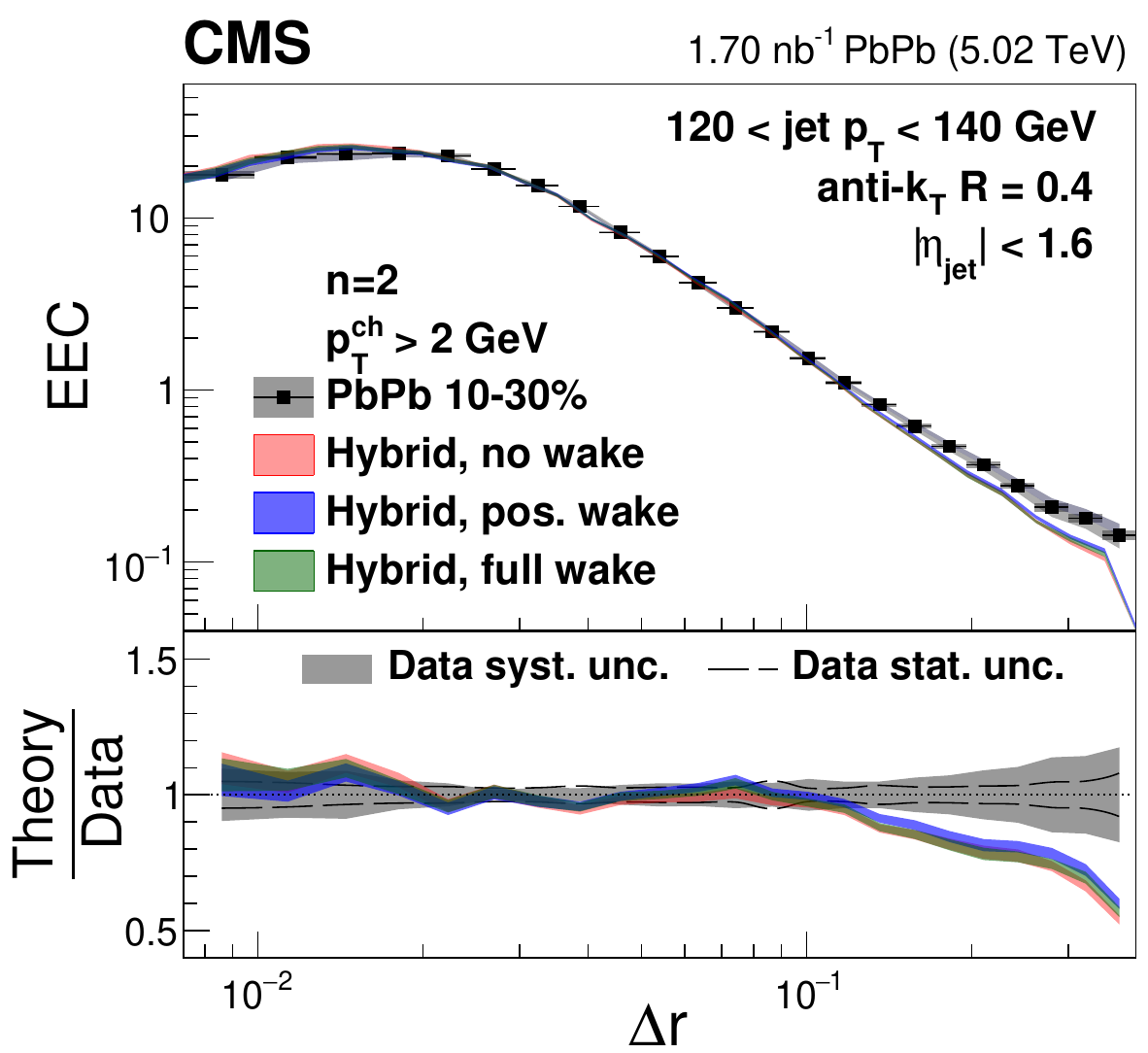}
    \includegraphics[width=0.4\textwidth]{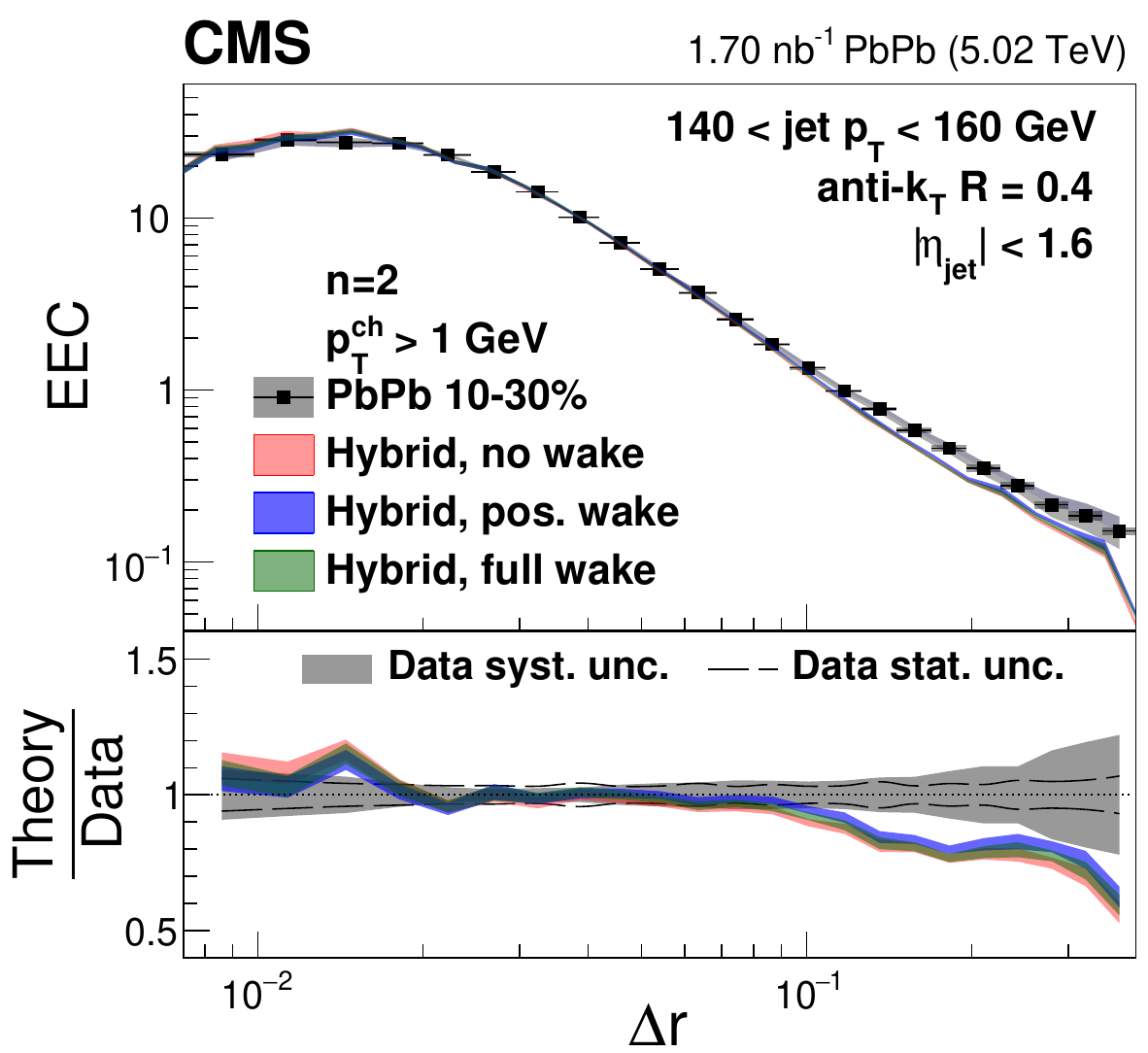}
    \includegraphics[width=0.4\textwidth]{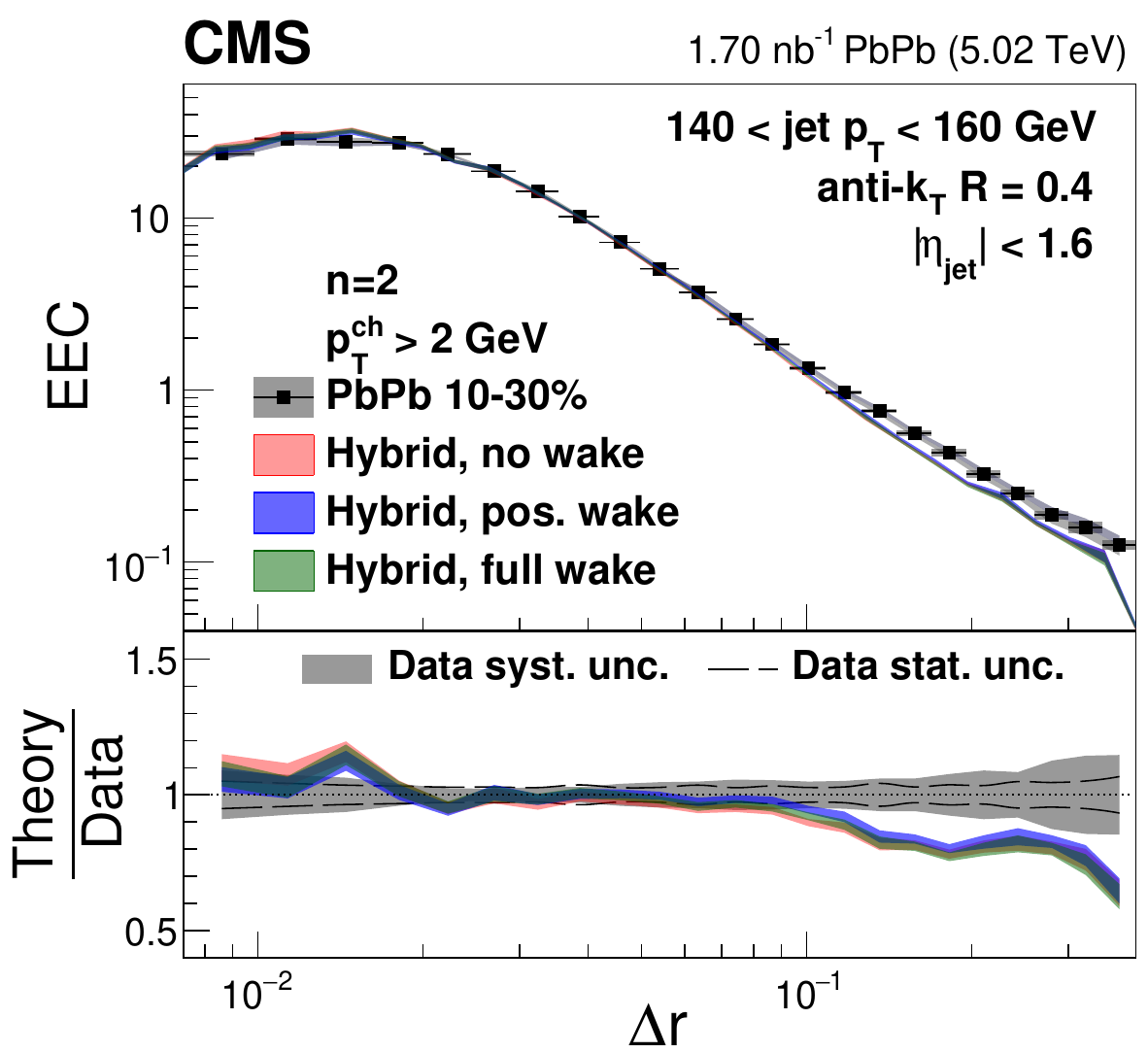}
    \includegraphics[width=0.4\textwidth]{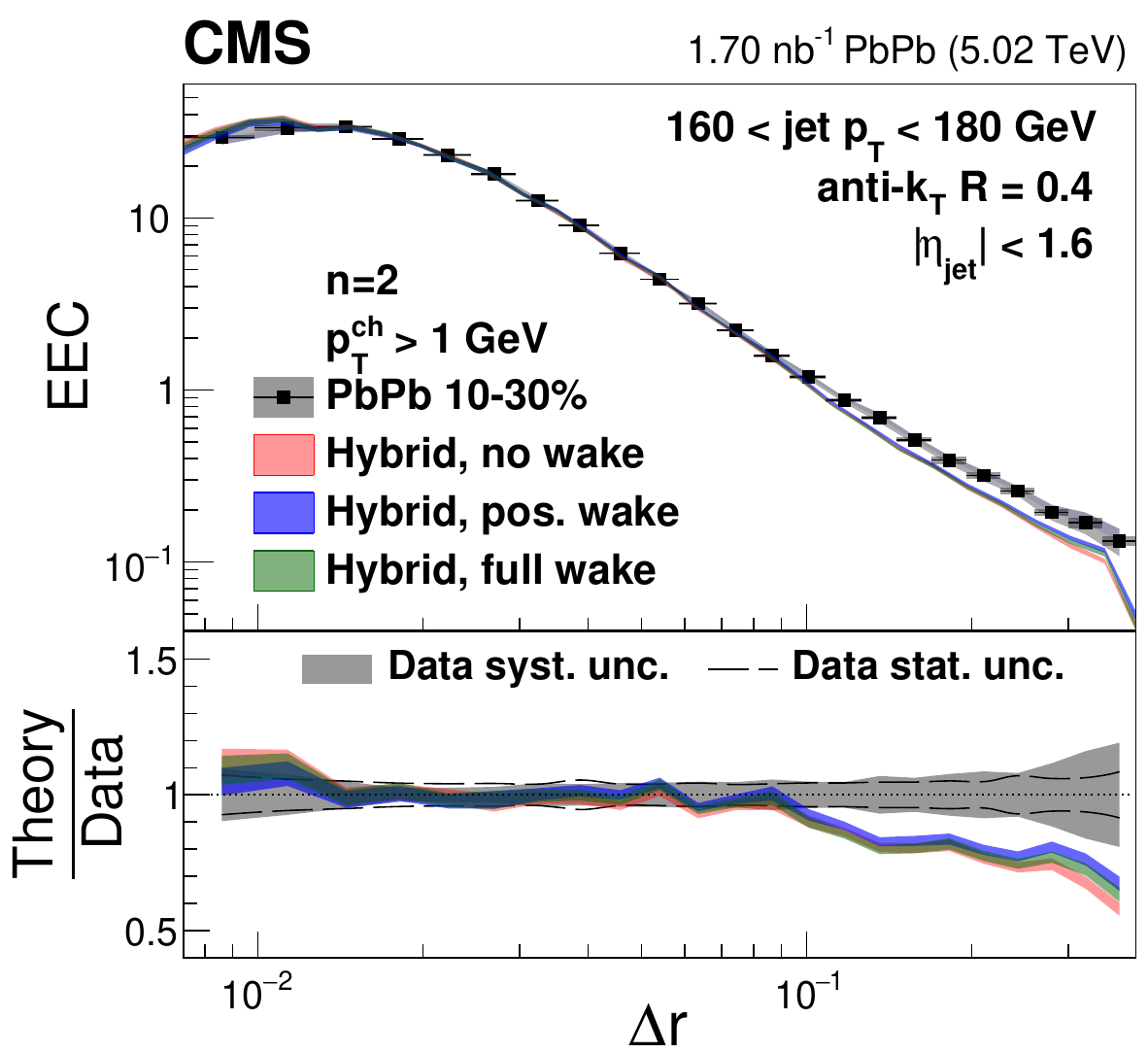}
    \includegraphics[width=0.4\textwidth]{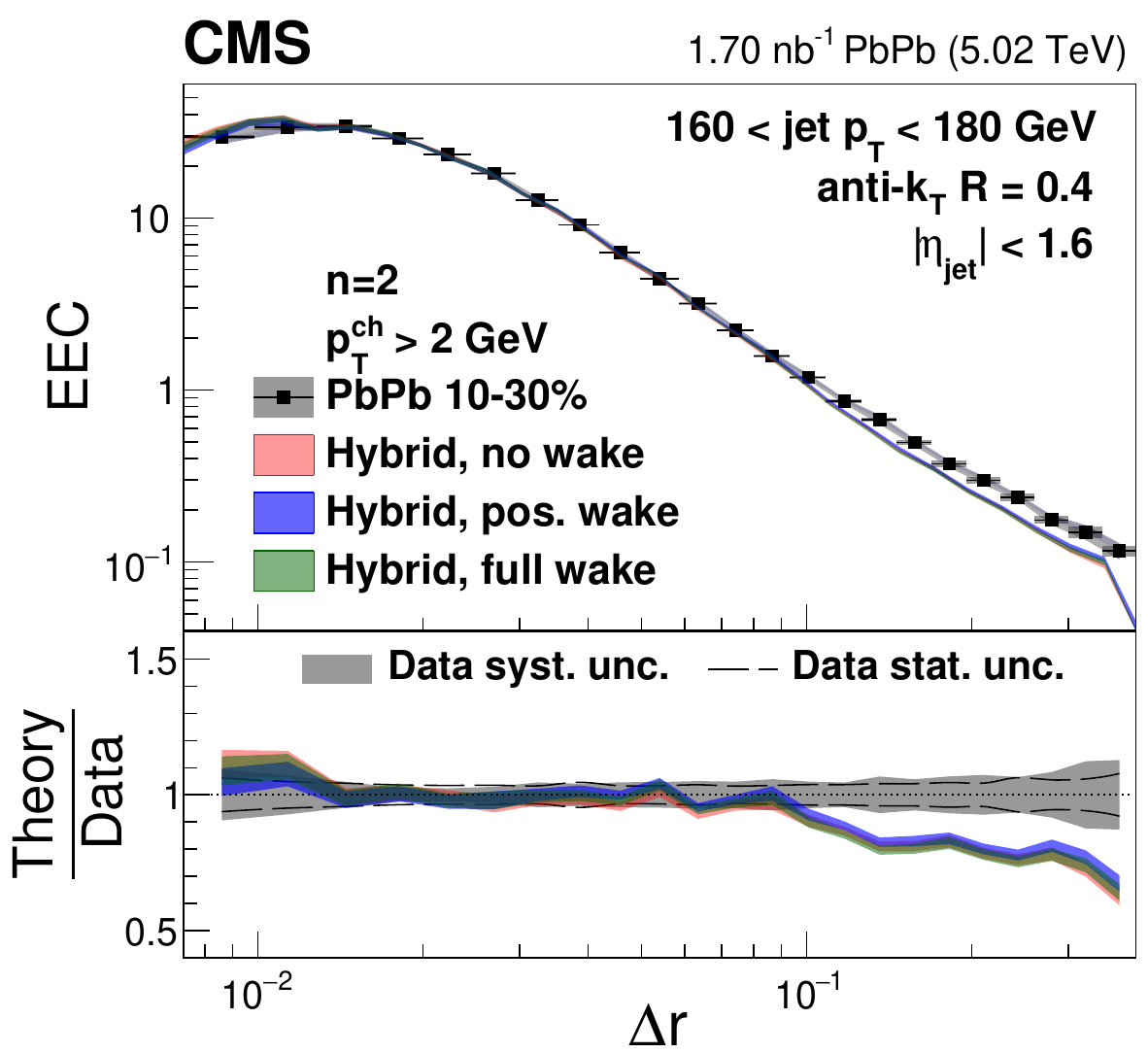}
    \includegraphics[width=0.4\textwidth]{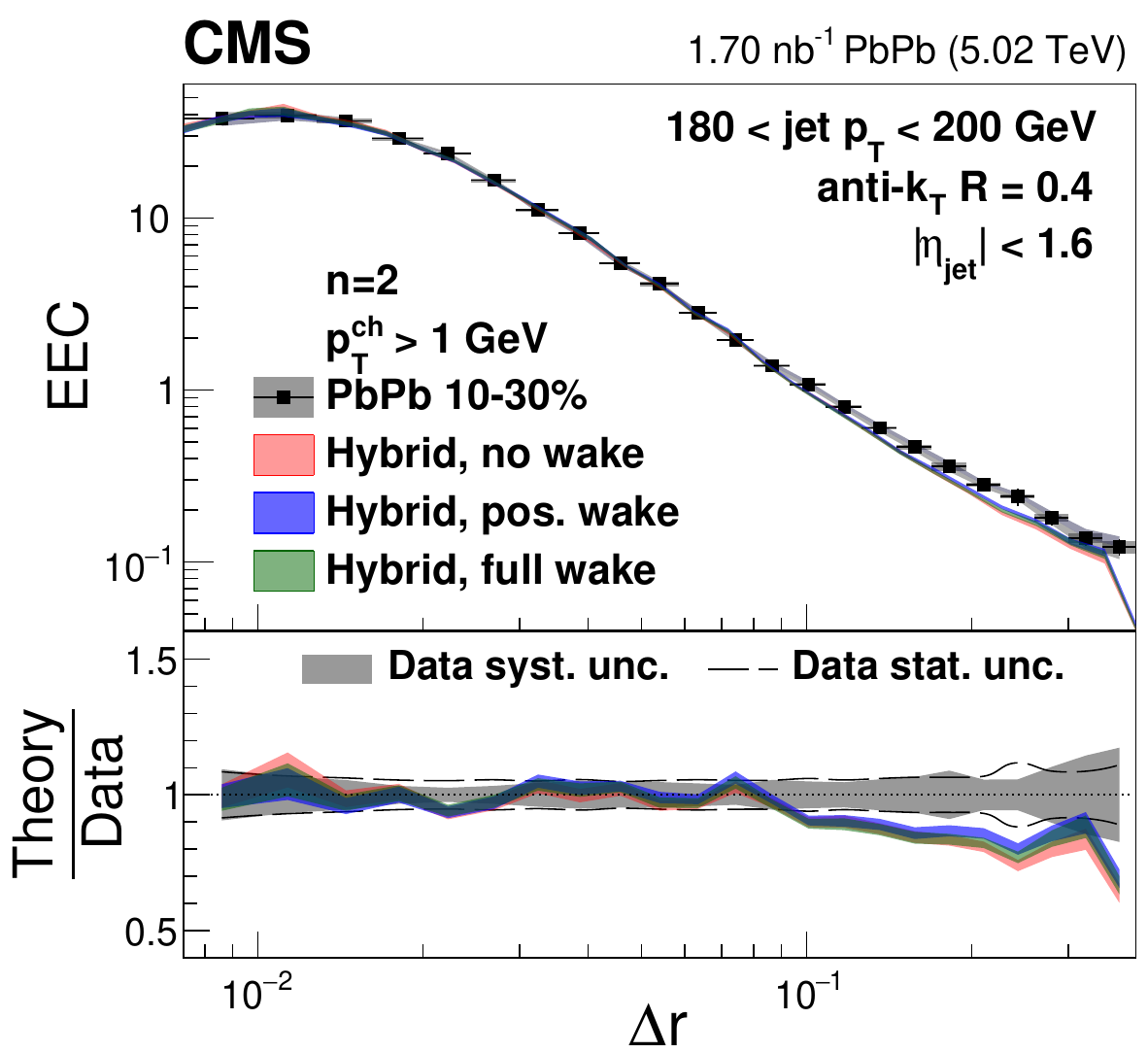}
    \includegraphics[width=0.4\textwidth]{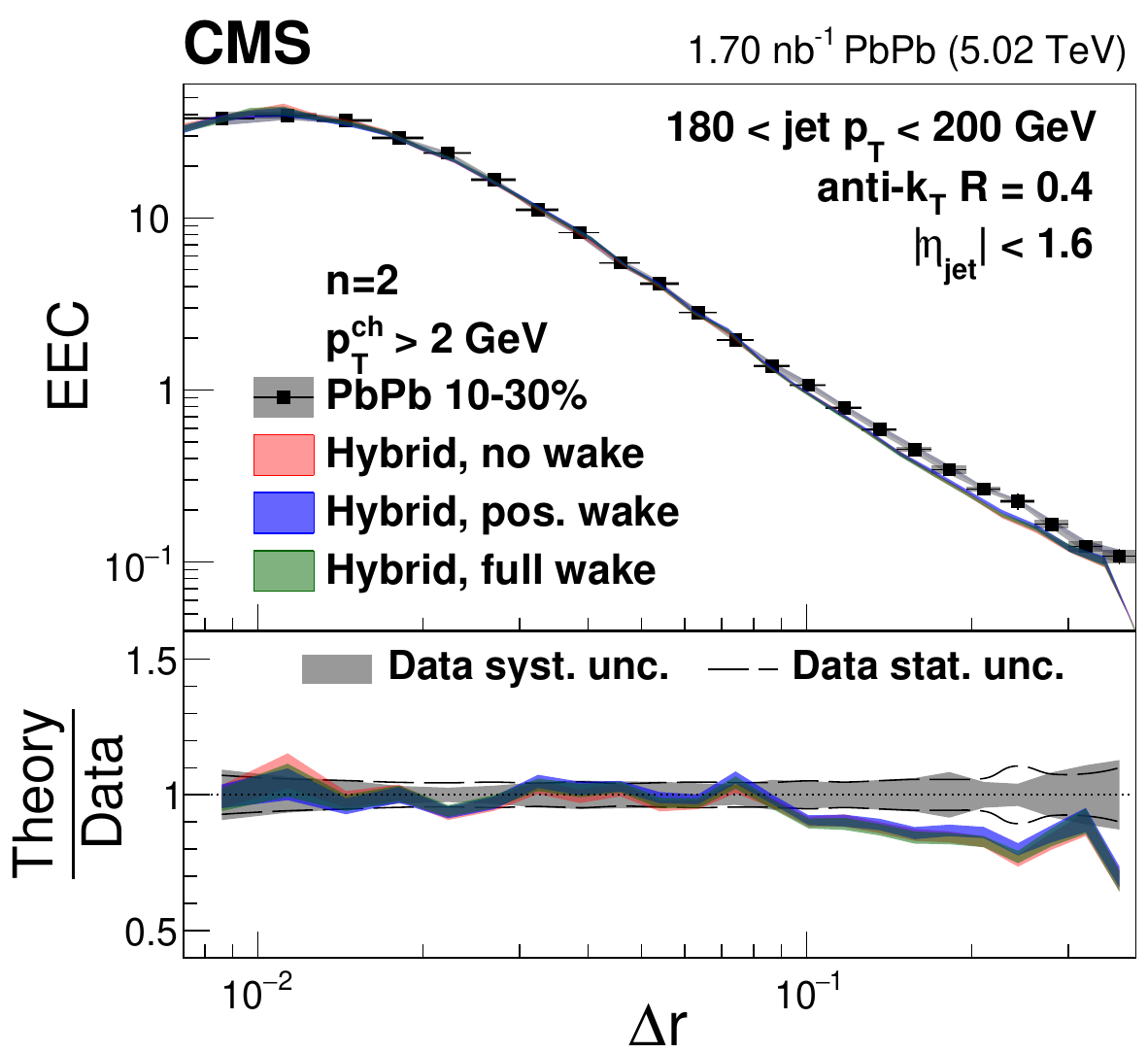}
    \caption{Energy-energy correlator distributions from 10--30\% central \PbPb collisions with $n=2$ in different \ptCh and \ptjet bins compared to the hybrid model predictions with different jet wake settings.} 
    \label{fig:PbPbDistributionsSquared10-30}
  } 
\end{figure*}

\begin{figure*}[h]
  \centering{
    \includegraphics[width=0.4\textwidth]{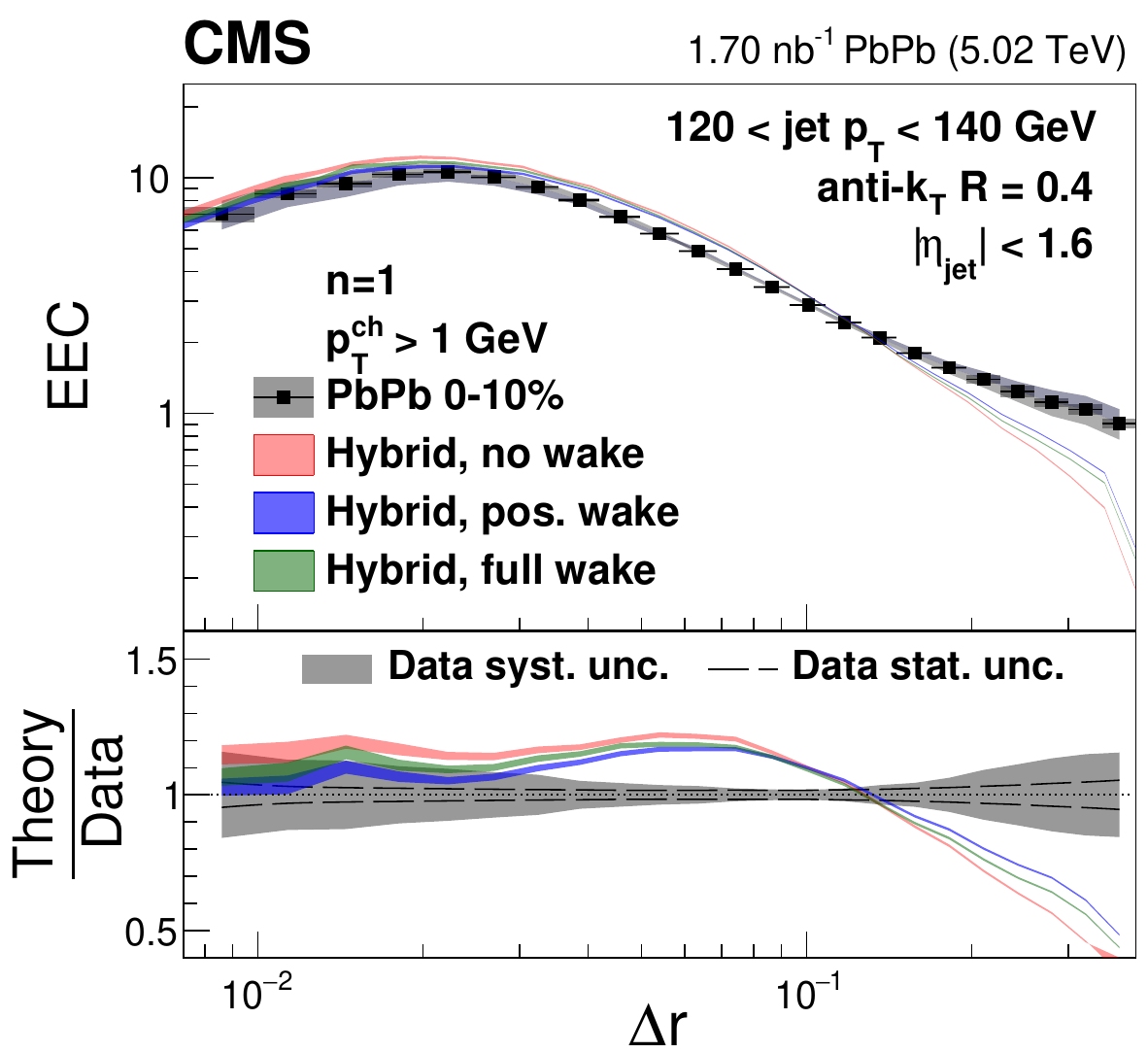}
    \includegraphics[width=0.4\textwidth]{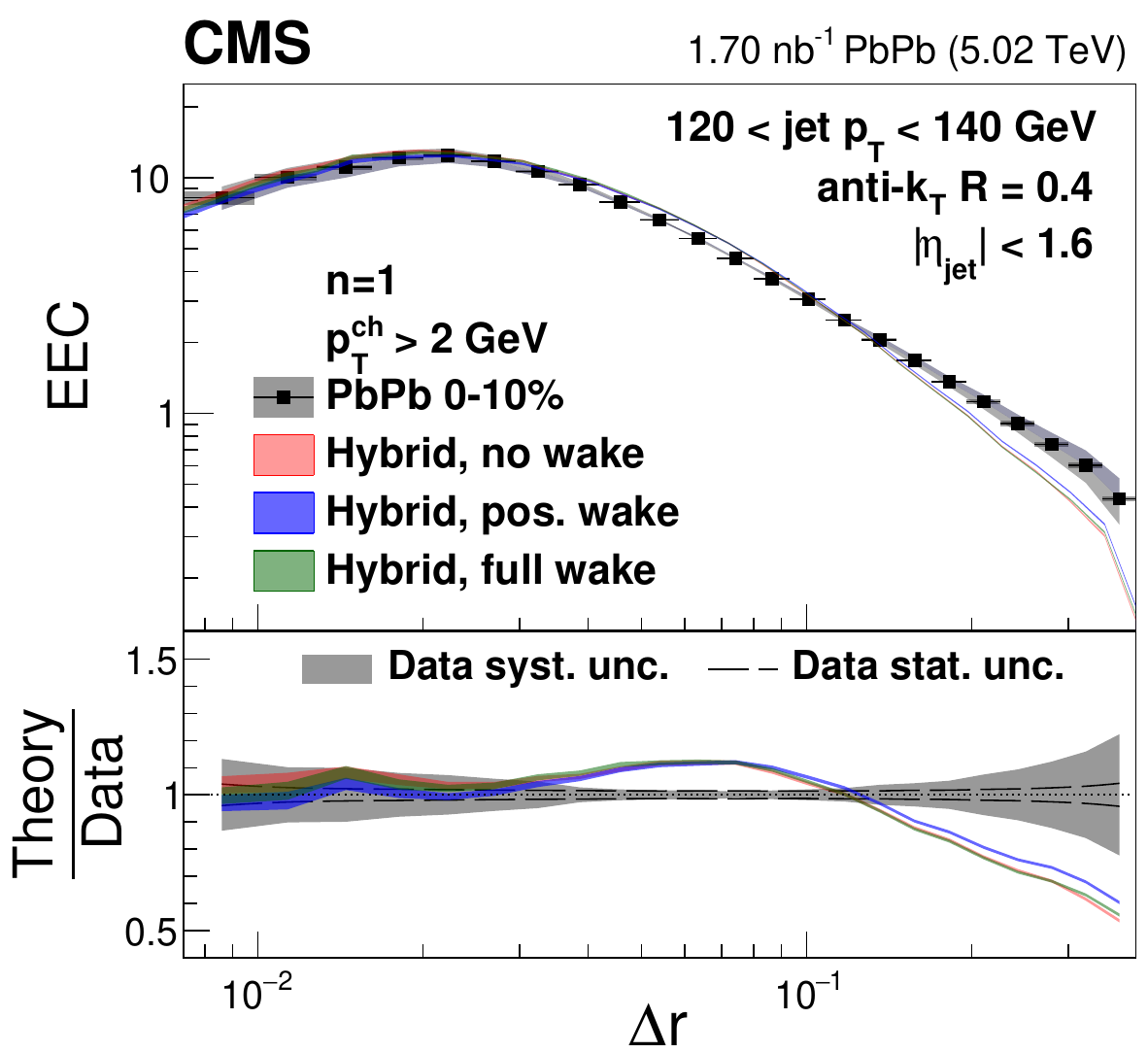}  
    \includegraphics[width=0.4\textwidth]{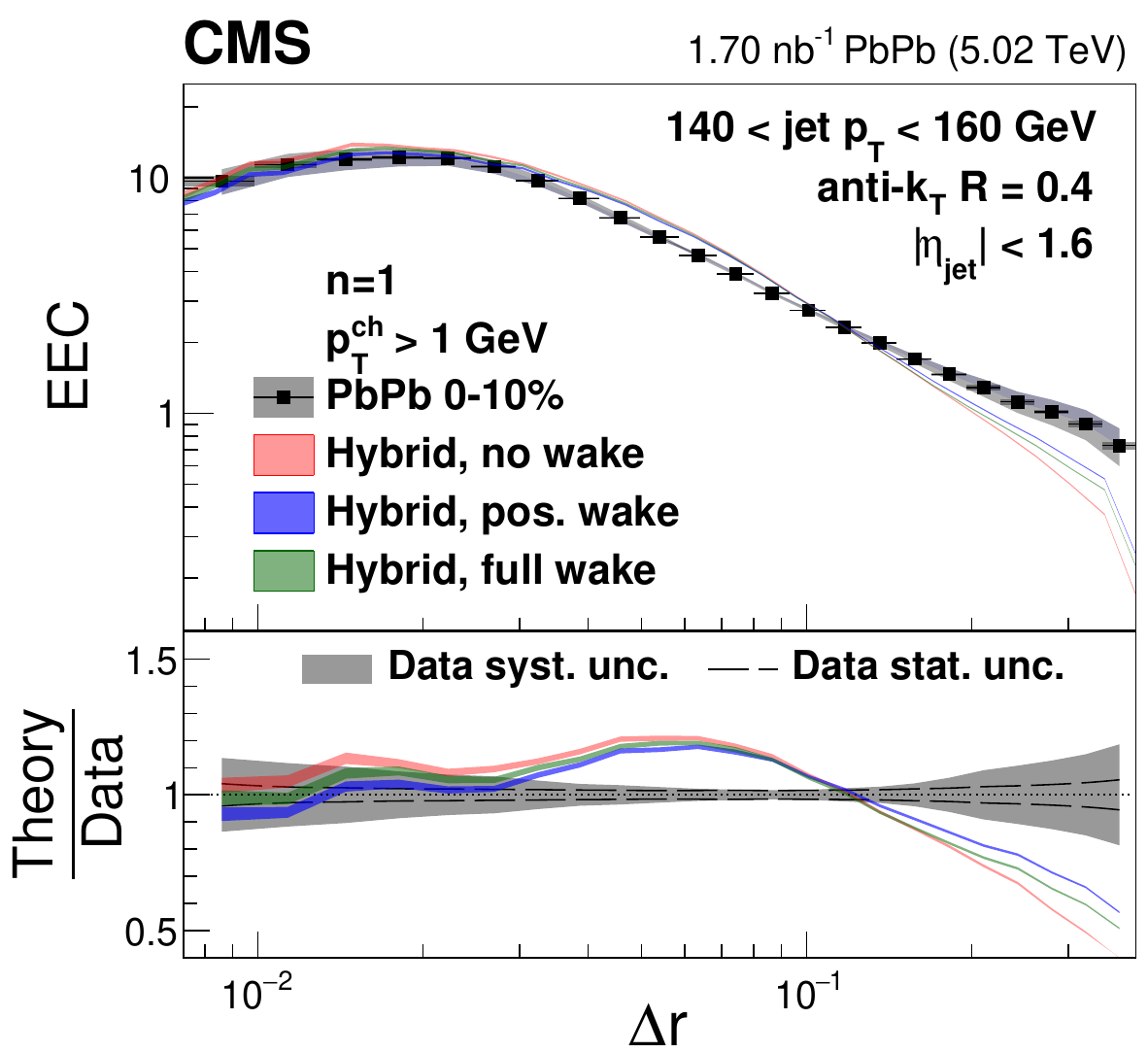}
    \includegraphics[width=0.4\textwidth]{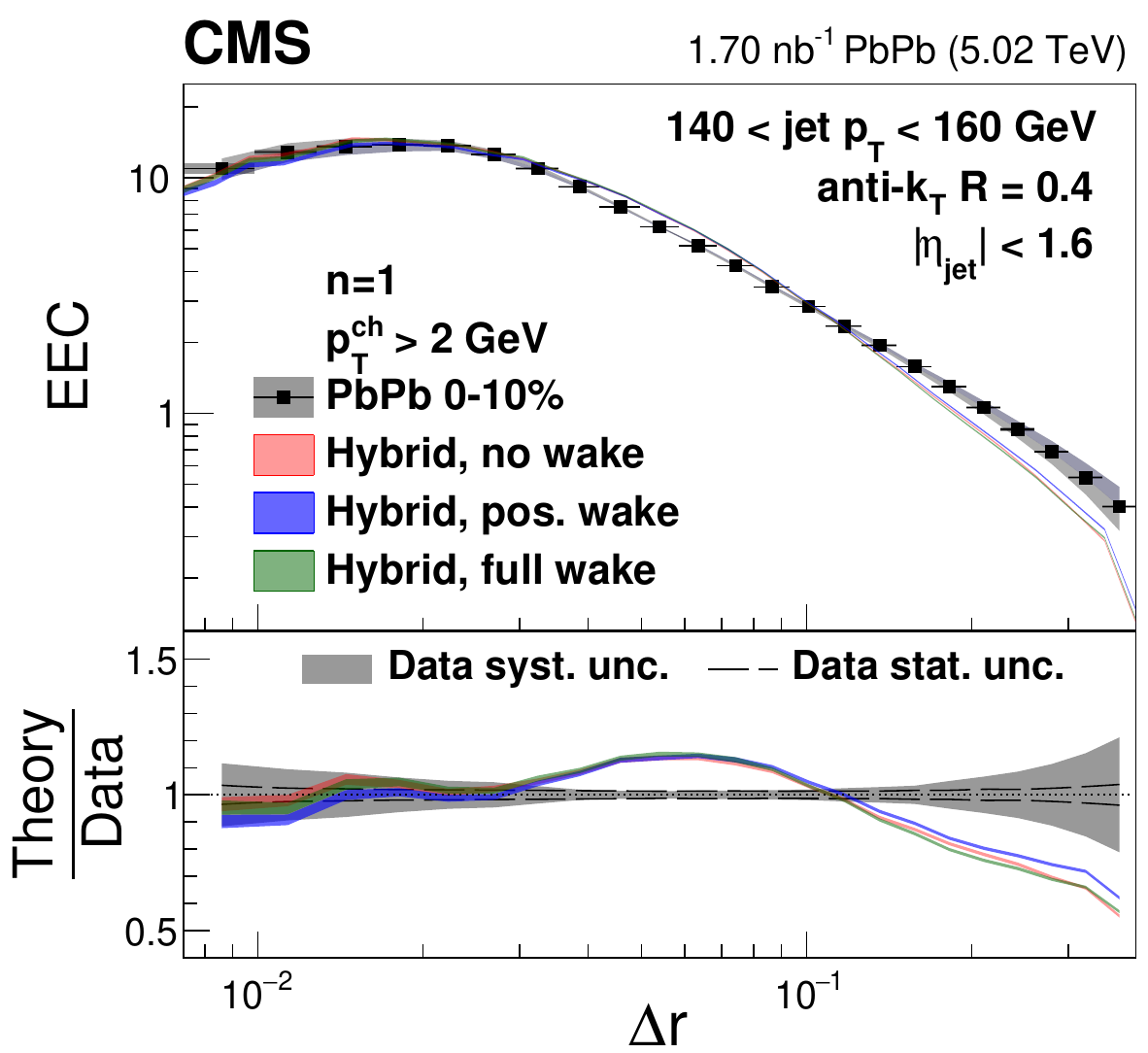}
    \includegraphics[width=0.4\textwidth]{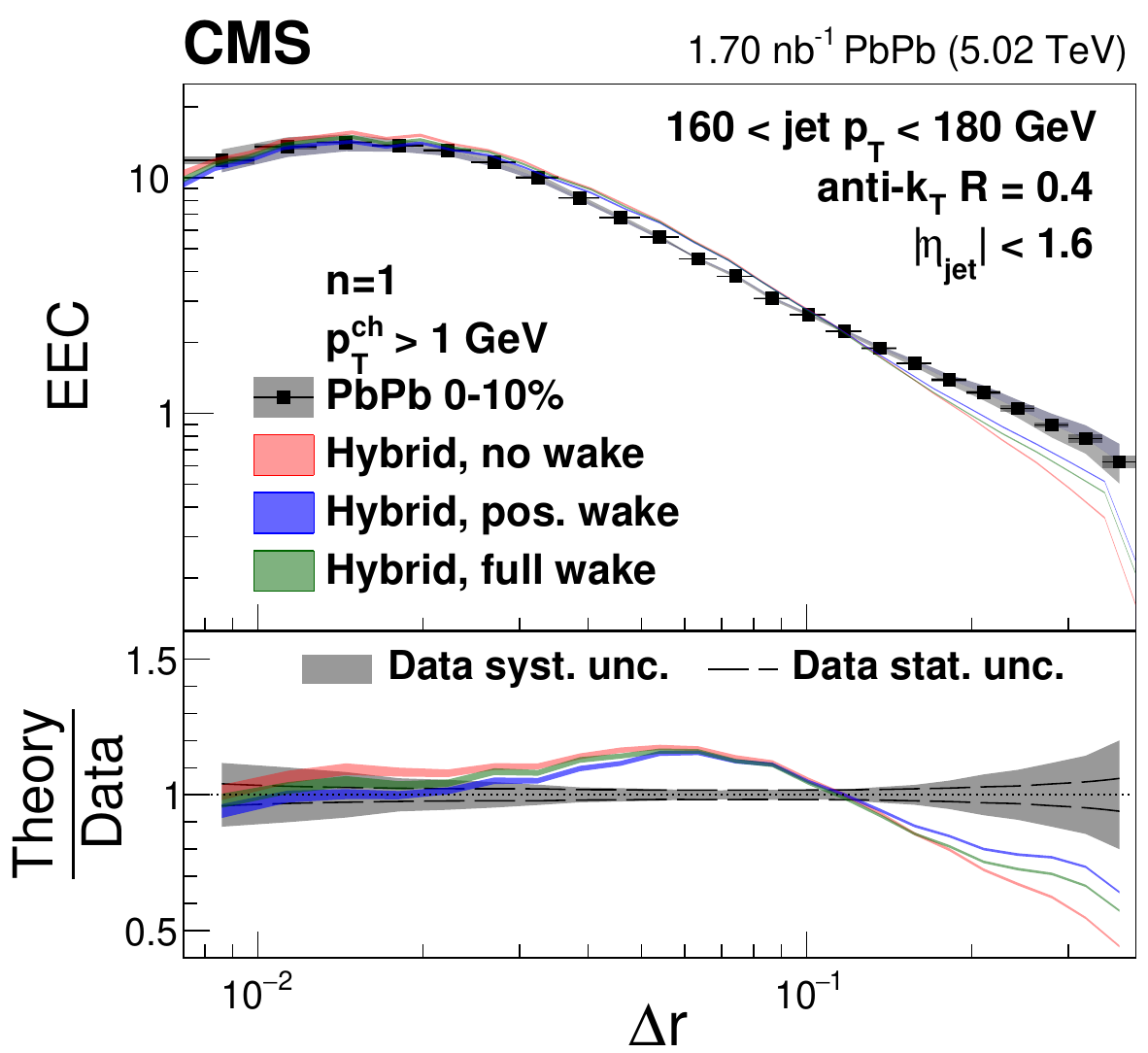}
    \includegraphics[width=0.4\textwidth]{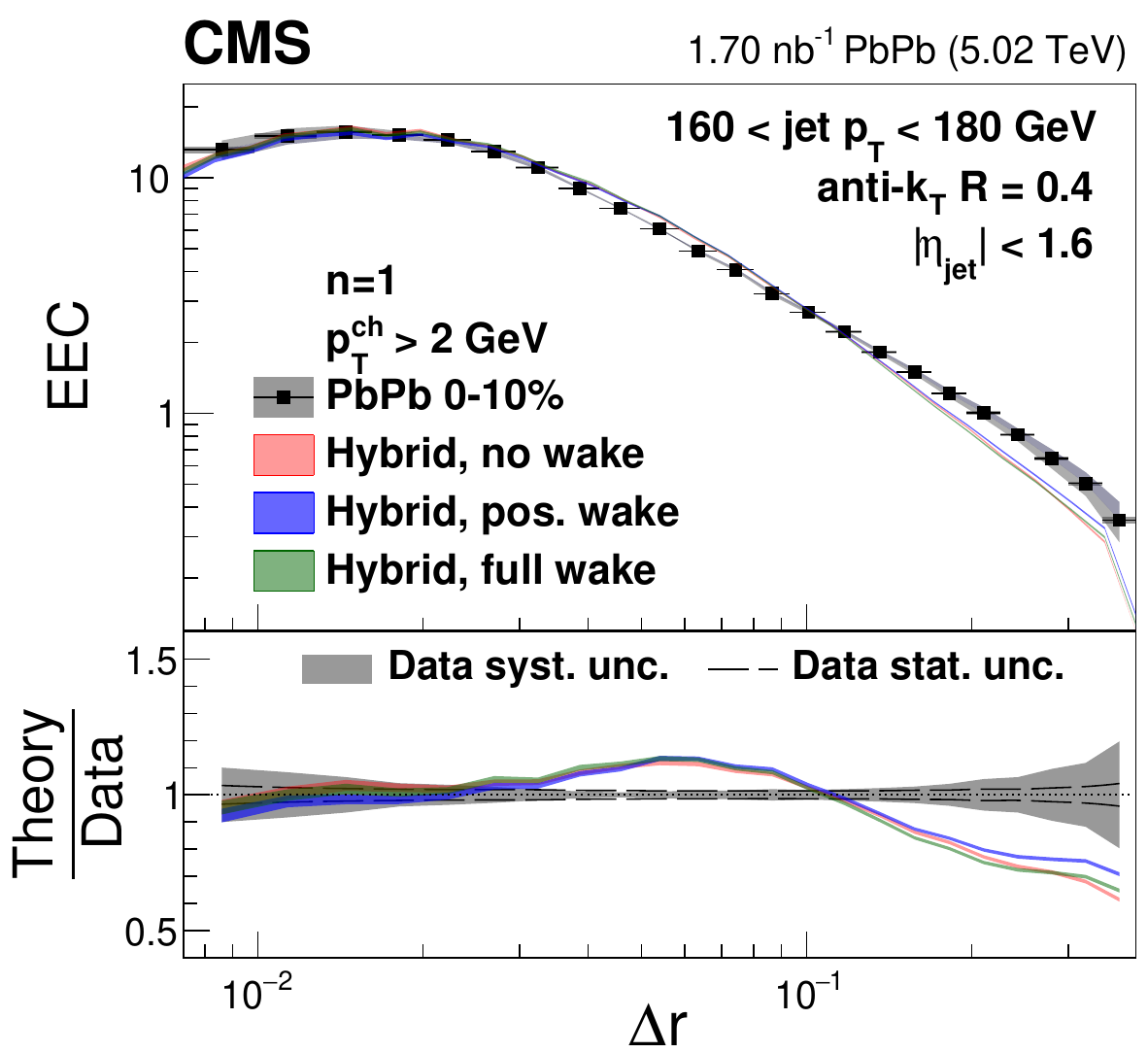}
    \includegraphics[width=0.4\textwidth]{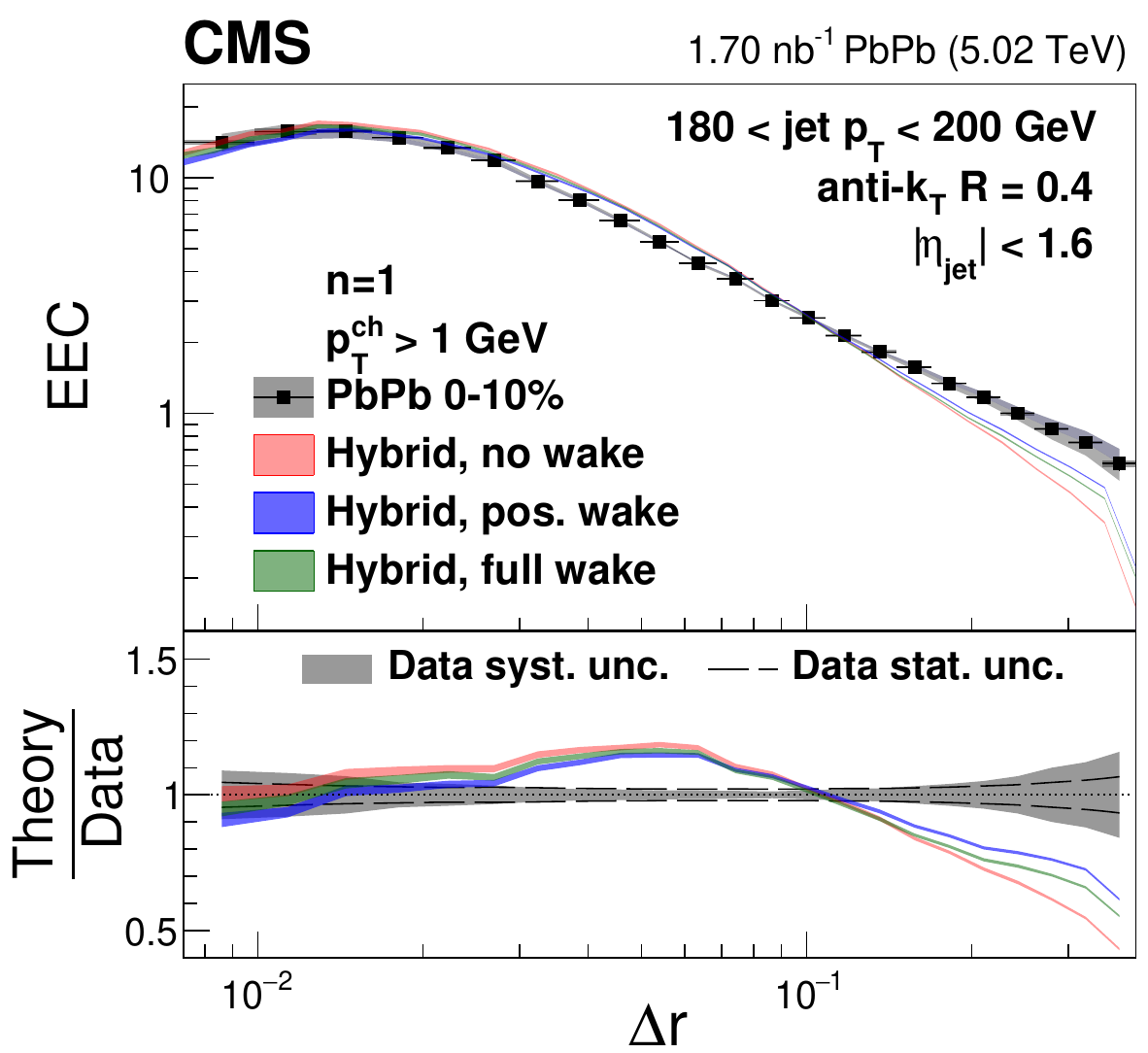}  
    \includegraphics[width=0.4\textwidth]{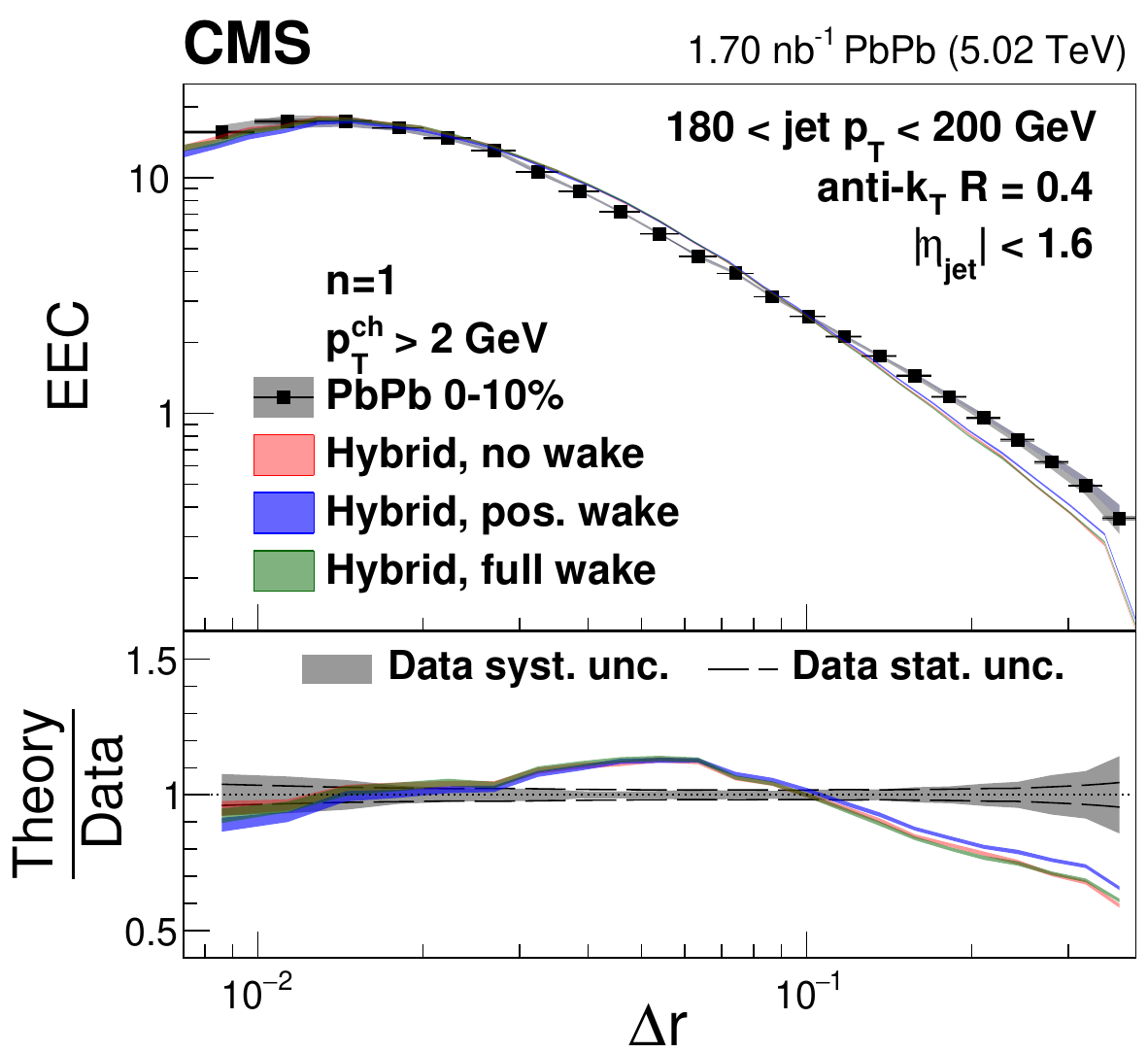}
    \caption{Energy-energy correlator distributions from 0--10\% central \PbPb collisions with $n=1$ in different \ptCh and \ptjet bins compared to the hybrid model predictions with different jet wake settings.} 
    \label{fig:PbPbDistributionsNominal0-10}
  } 
\end{figure*}

\begin{figure*}[h]
  \centering{
    \includegraphics[width=0.4\textwidth]{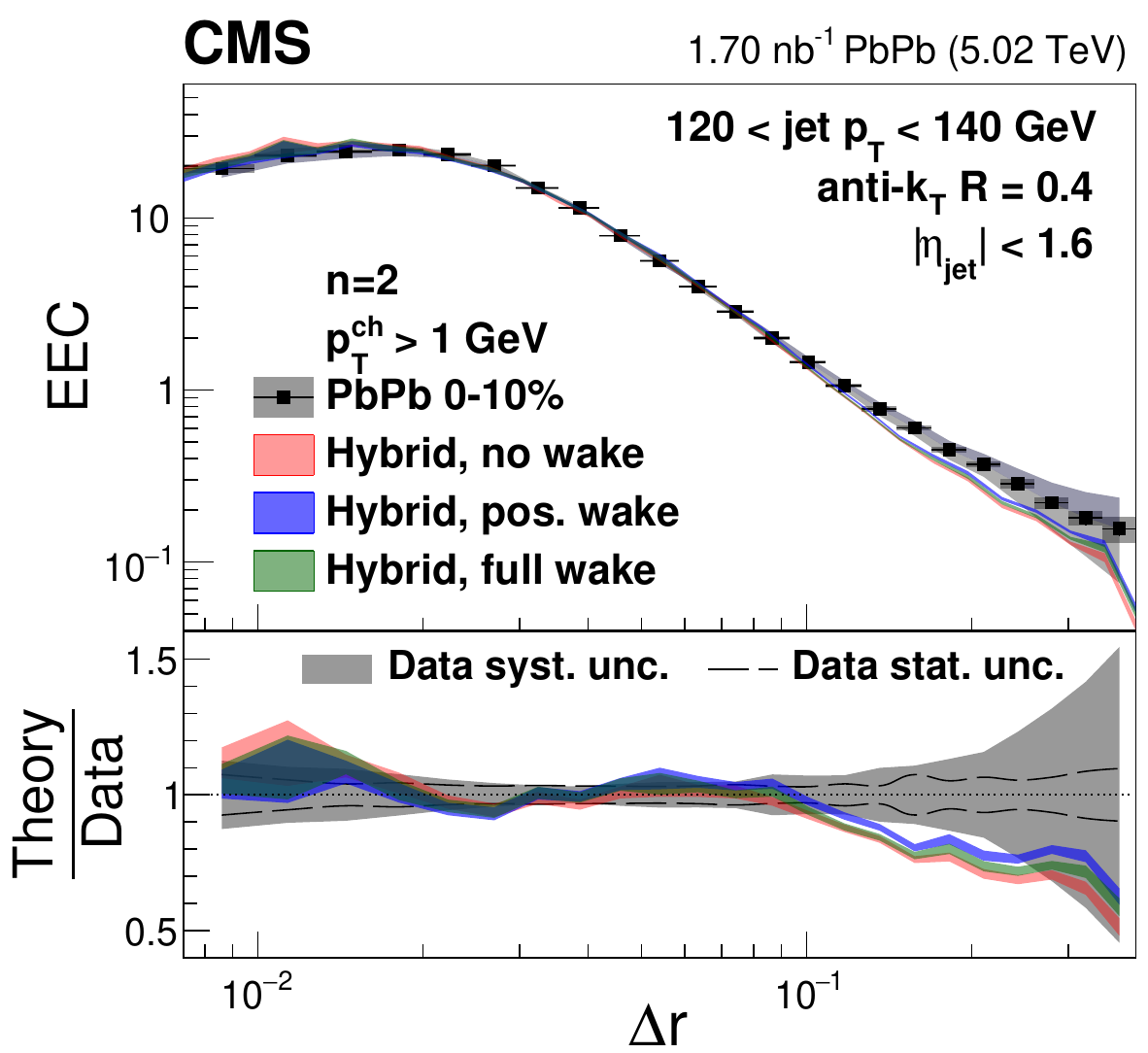}
    \includegraphics[width=0.4\textwidth]{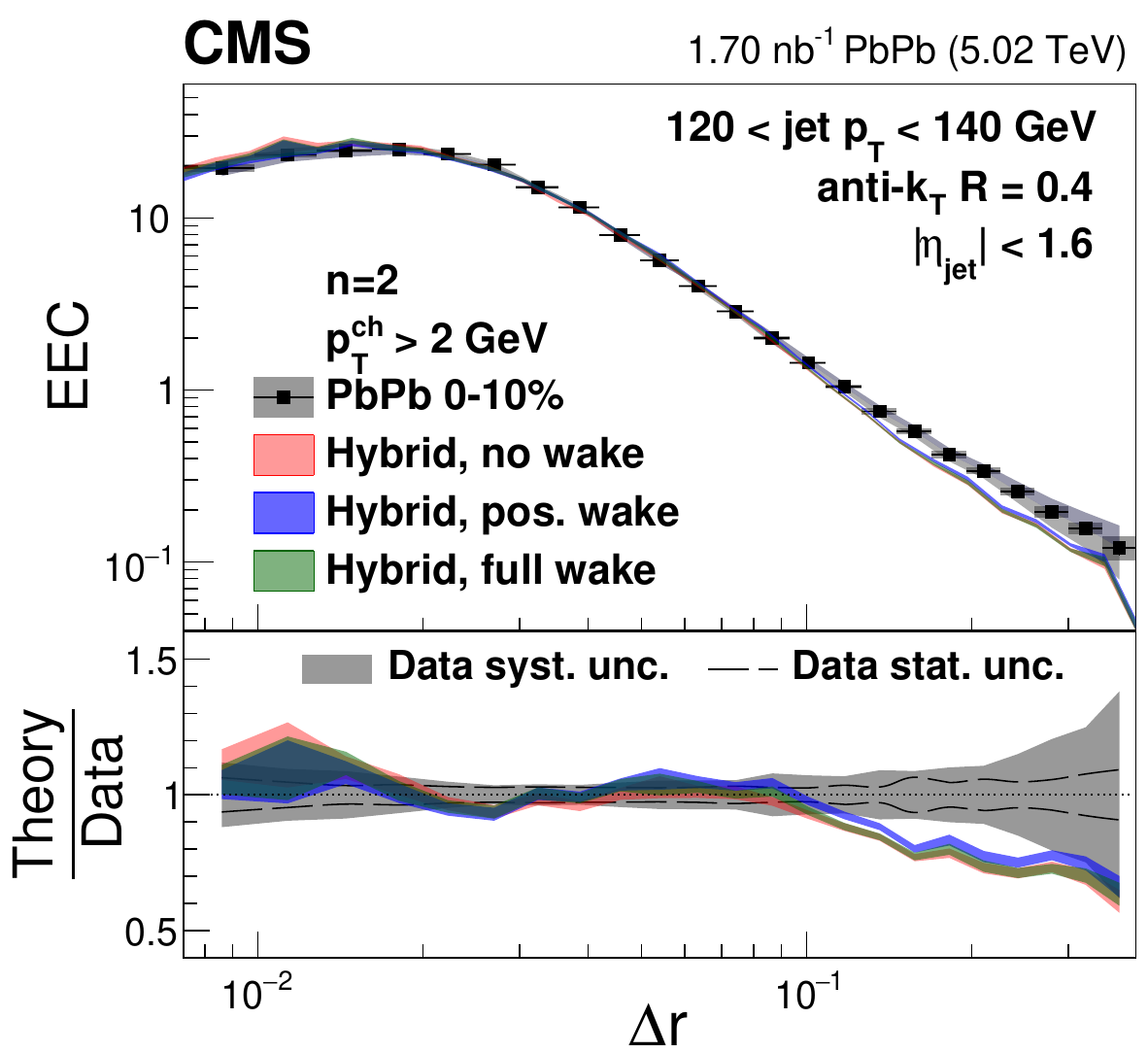}  
    \includegraphics[width=0.4\textwidth]{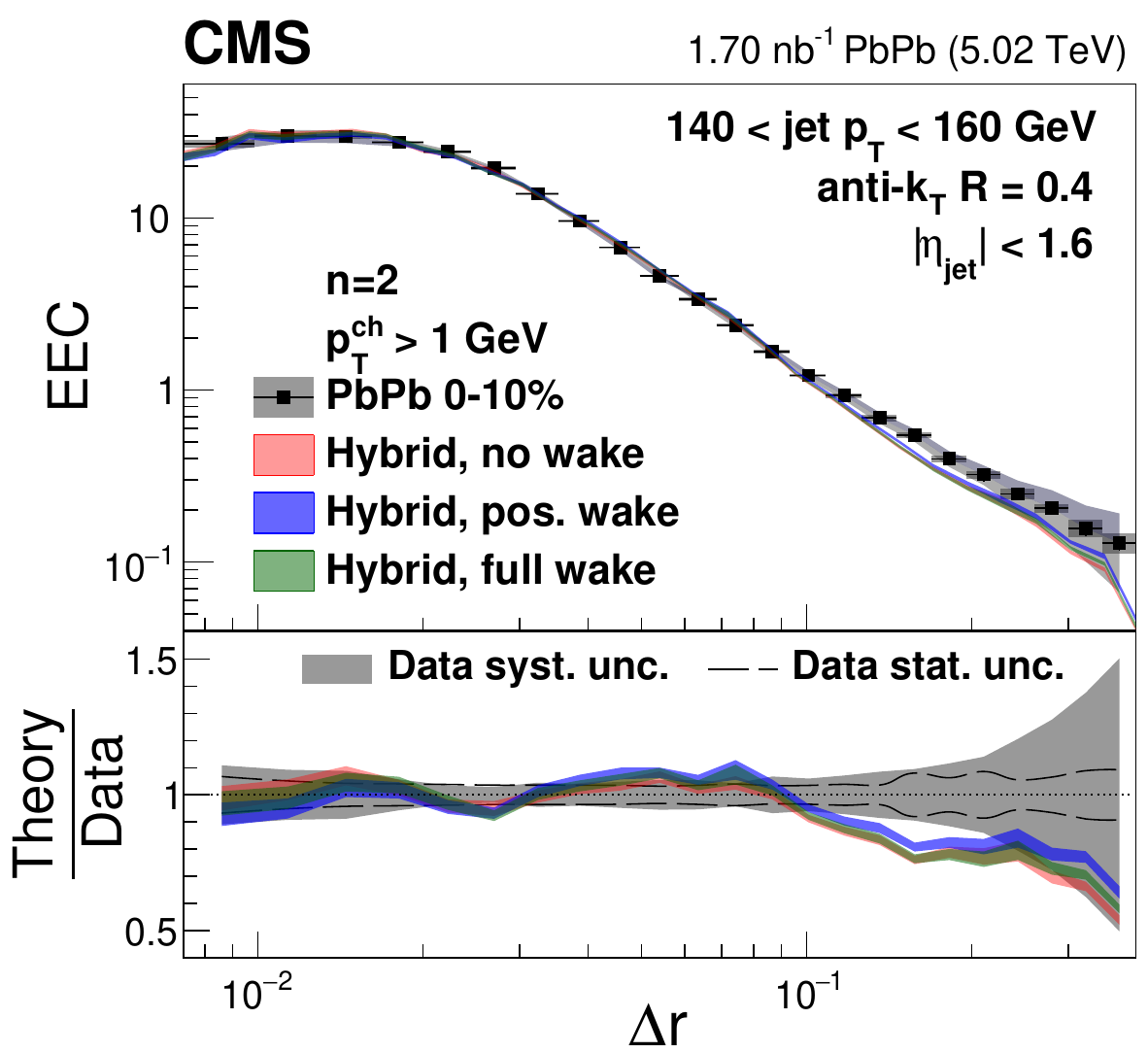}
    \includegraphics[width=0.4\textwidth]{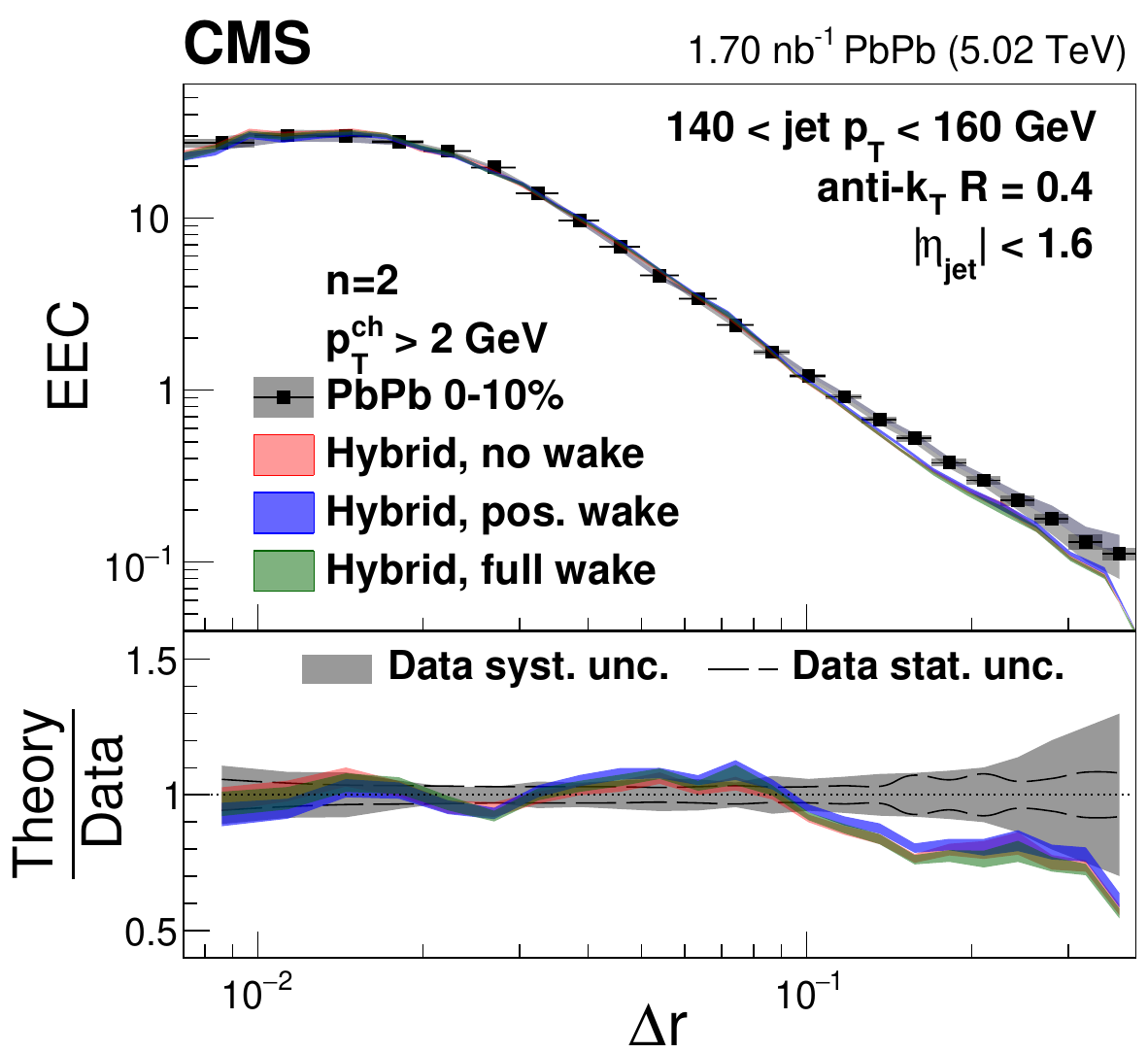}
    \includegraphics[width=0.4\textwidth]{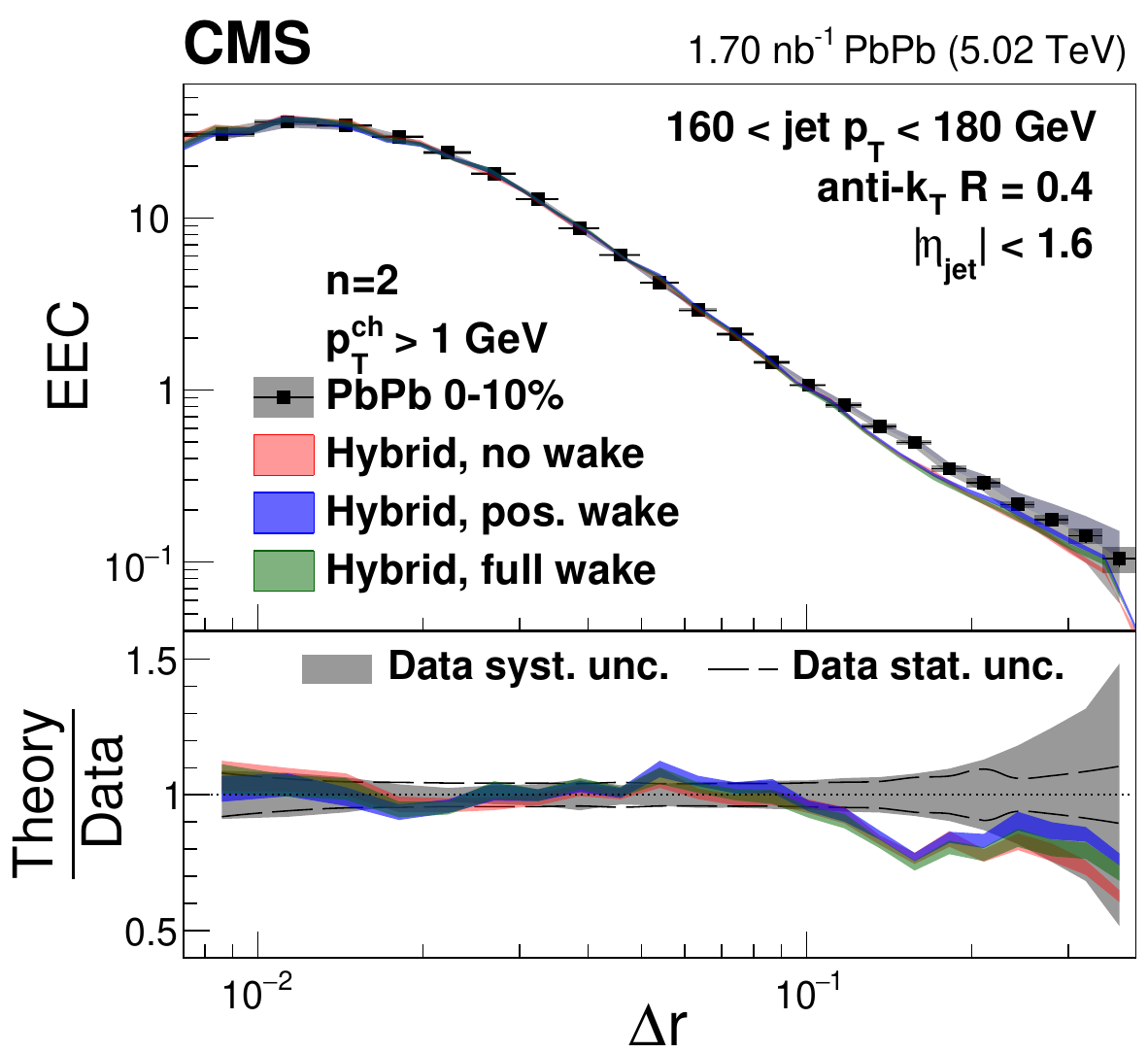}
    \includegraphics[width=0.4\textwidth]{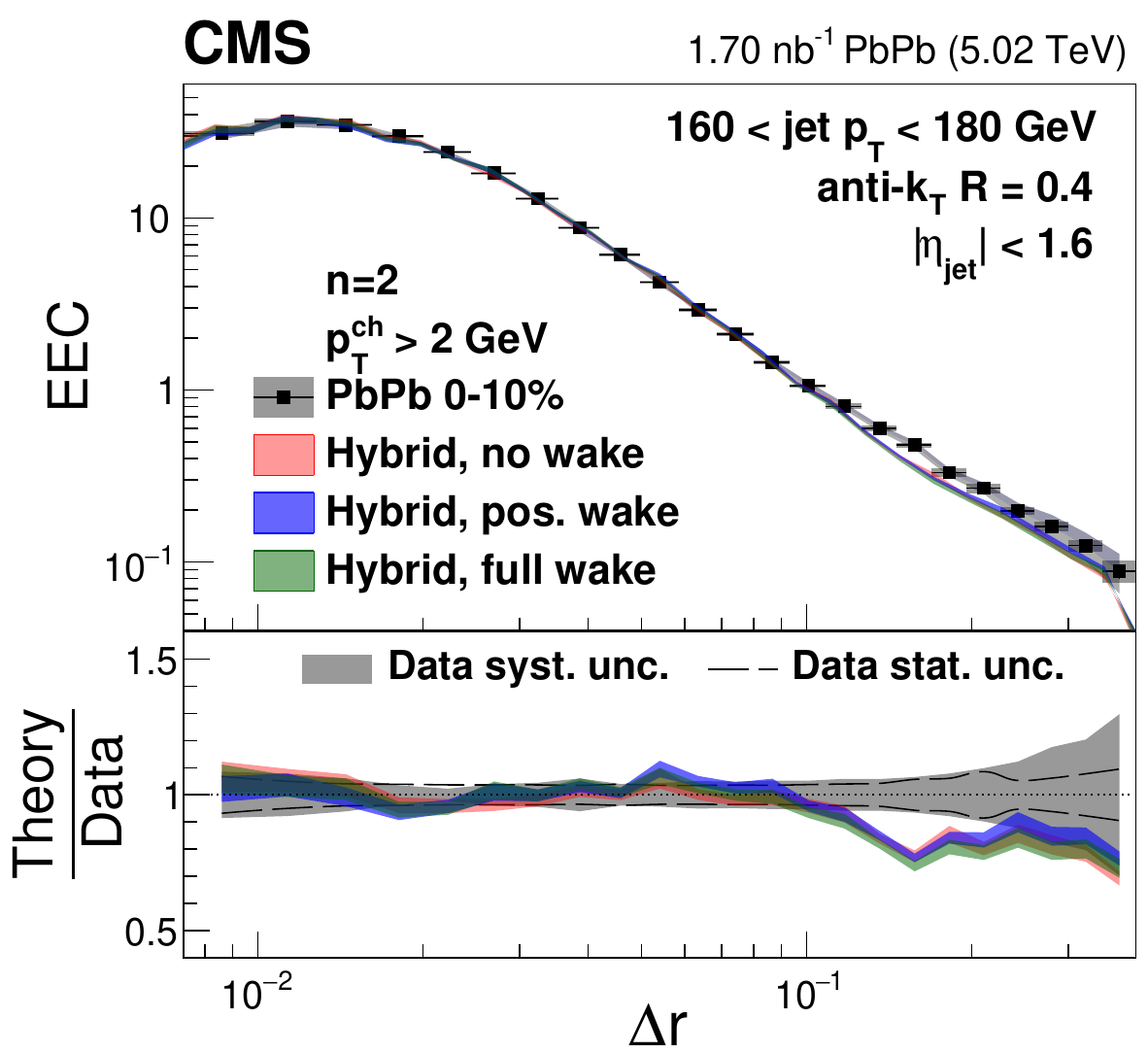}
    \includegraphics[width=0.4\textwidth]{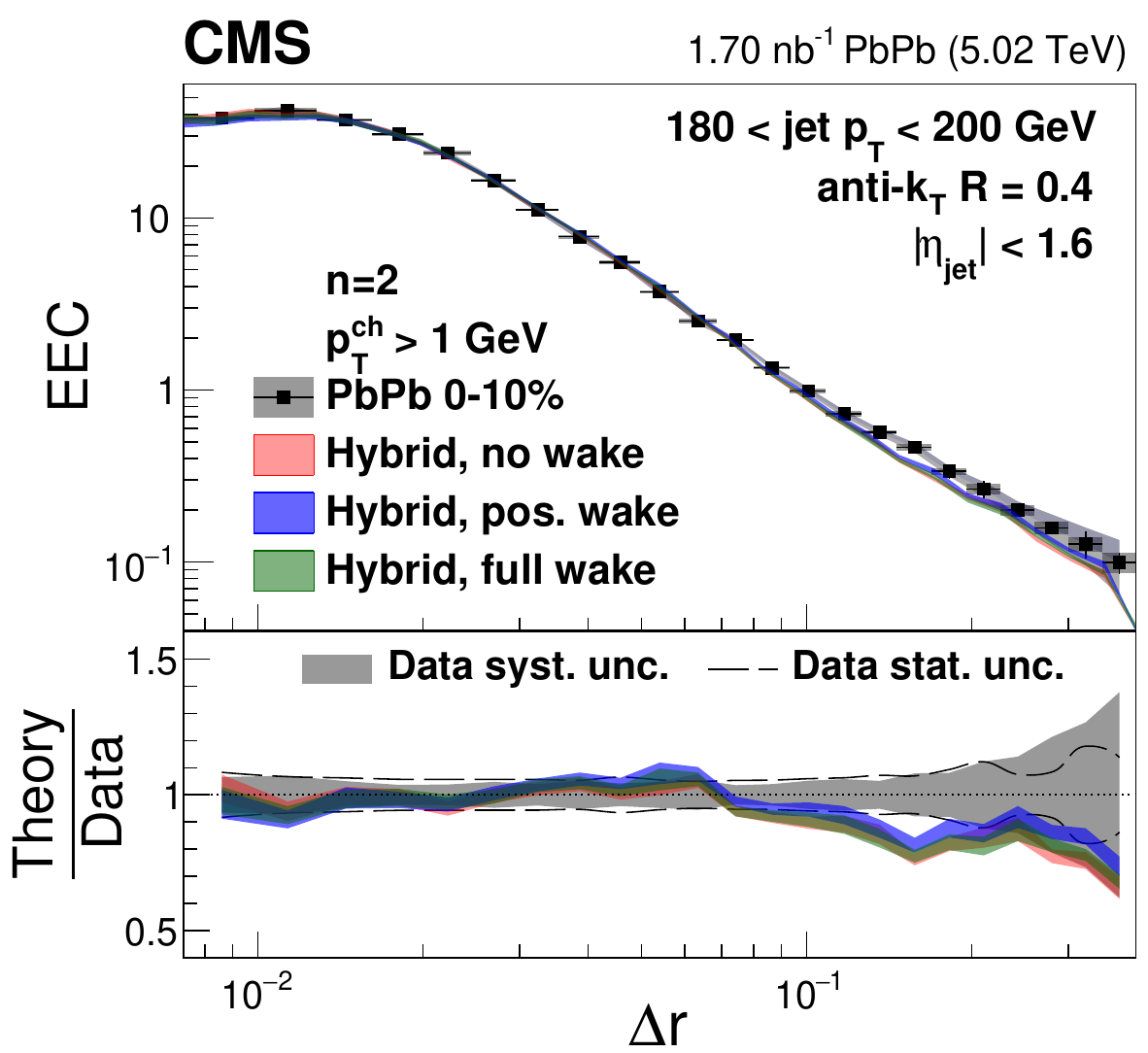}
    \includegraphics[width=0.4\textwidth]{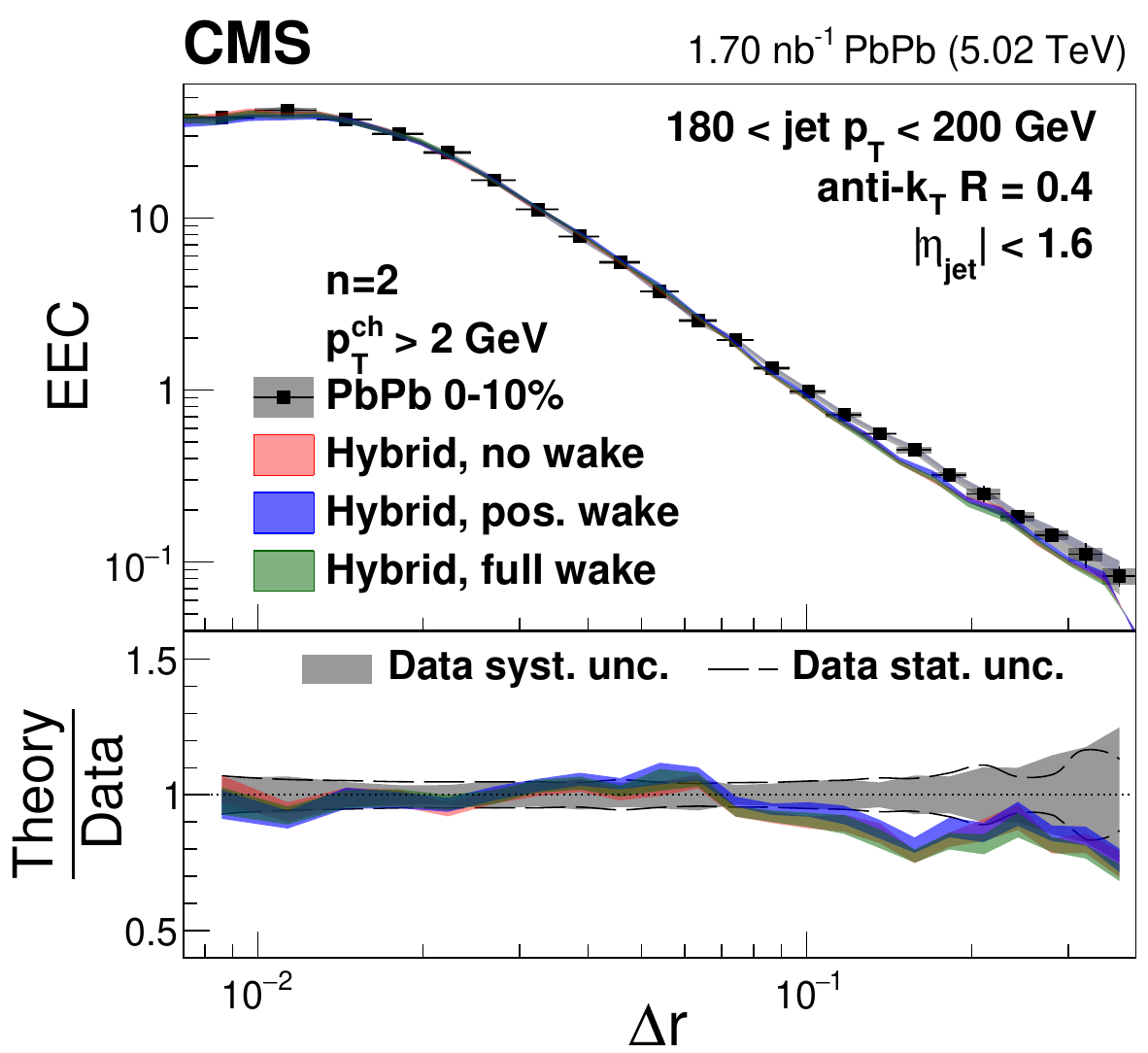}
    \caption{Energy-energy correlator distributions from 0--10\% central \PbPb collisions with $n=2$ in different \ptCh and \ptjet bins compared to the hybrid model predictions with different jet wake settings.} 
    \label{fig:PbPbDistributionsSquared0-10}
  } 
\end{figure*}

\begin{figure*}[h]
  \centering{
    \includegraphics[width=0.4\textwidth]{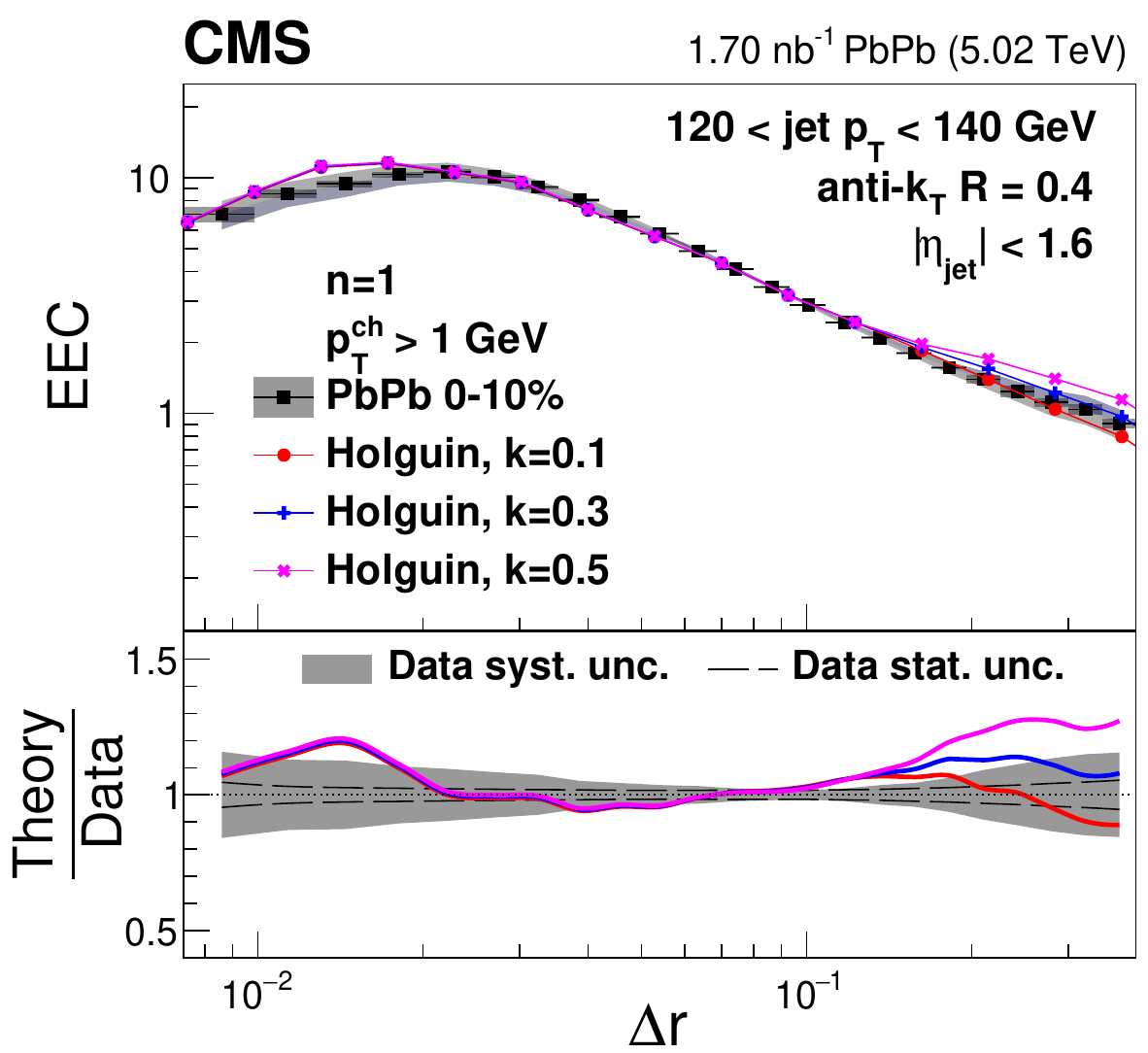} \\
    \includegraphics[width=0.4\textwidth]{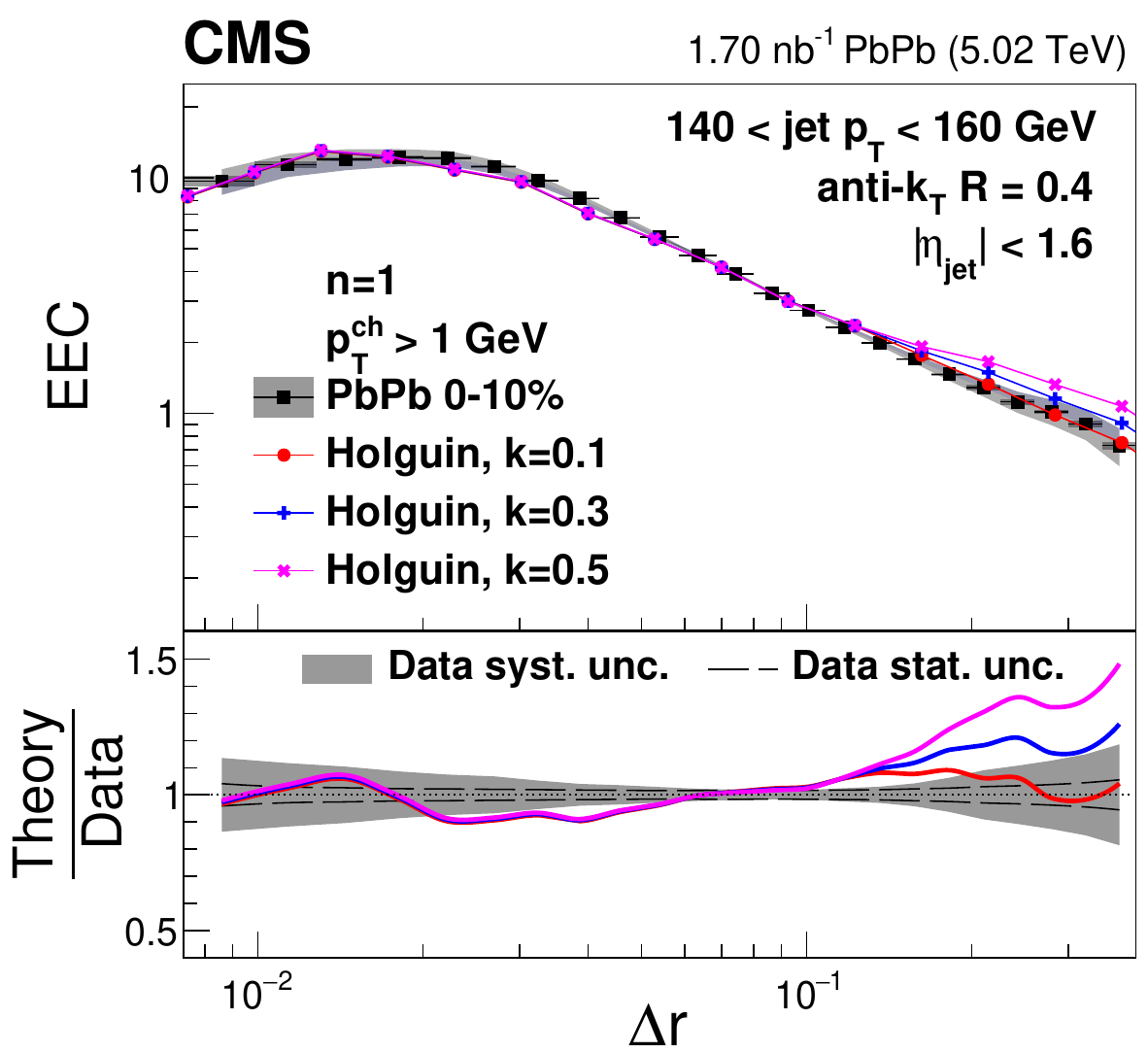} \\
    \includegraphics[width=0.4\textwidth]{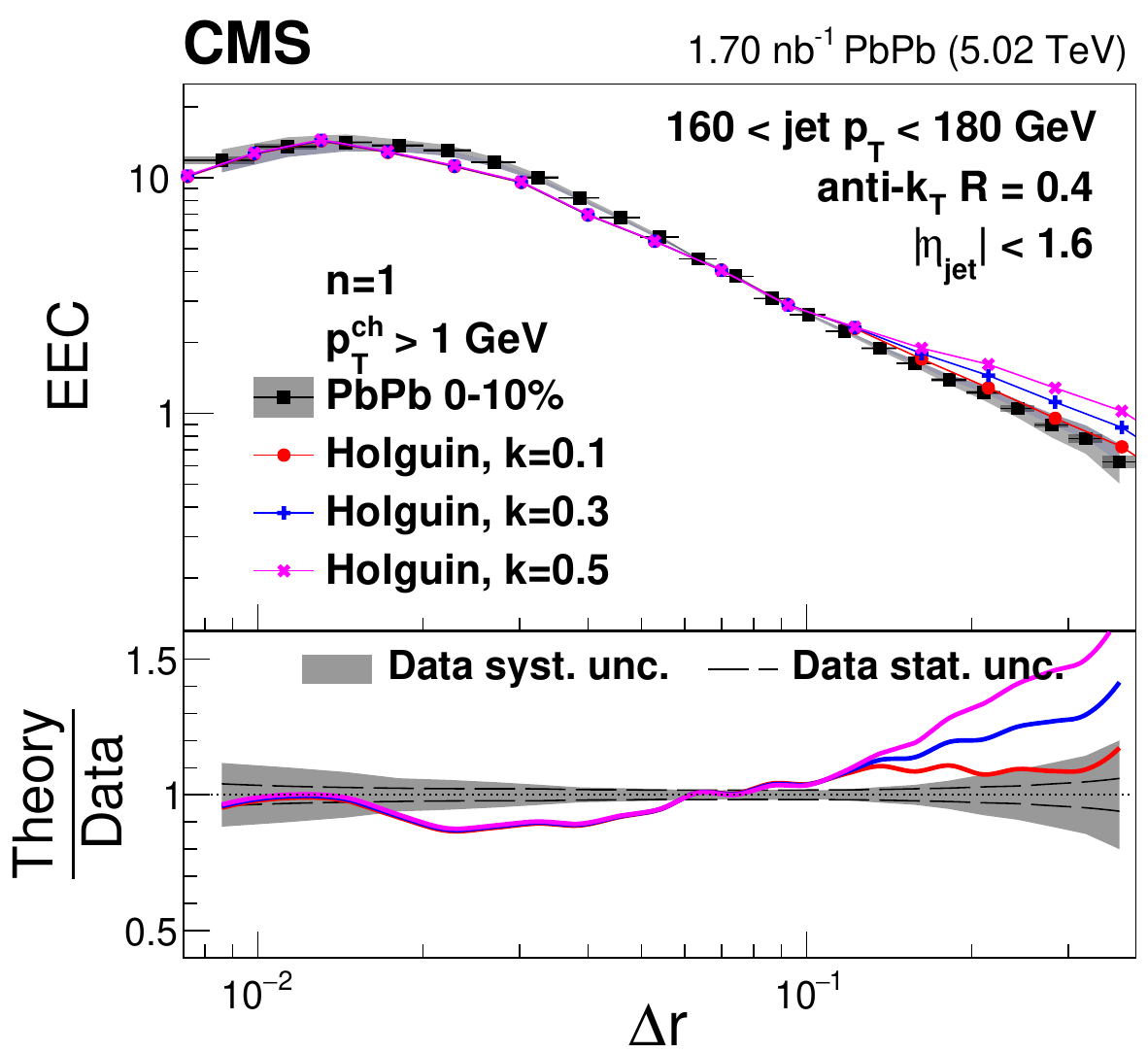} \\
    \includegraphics[width=0.4\textwidth]{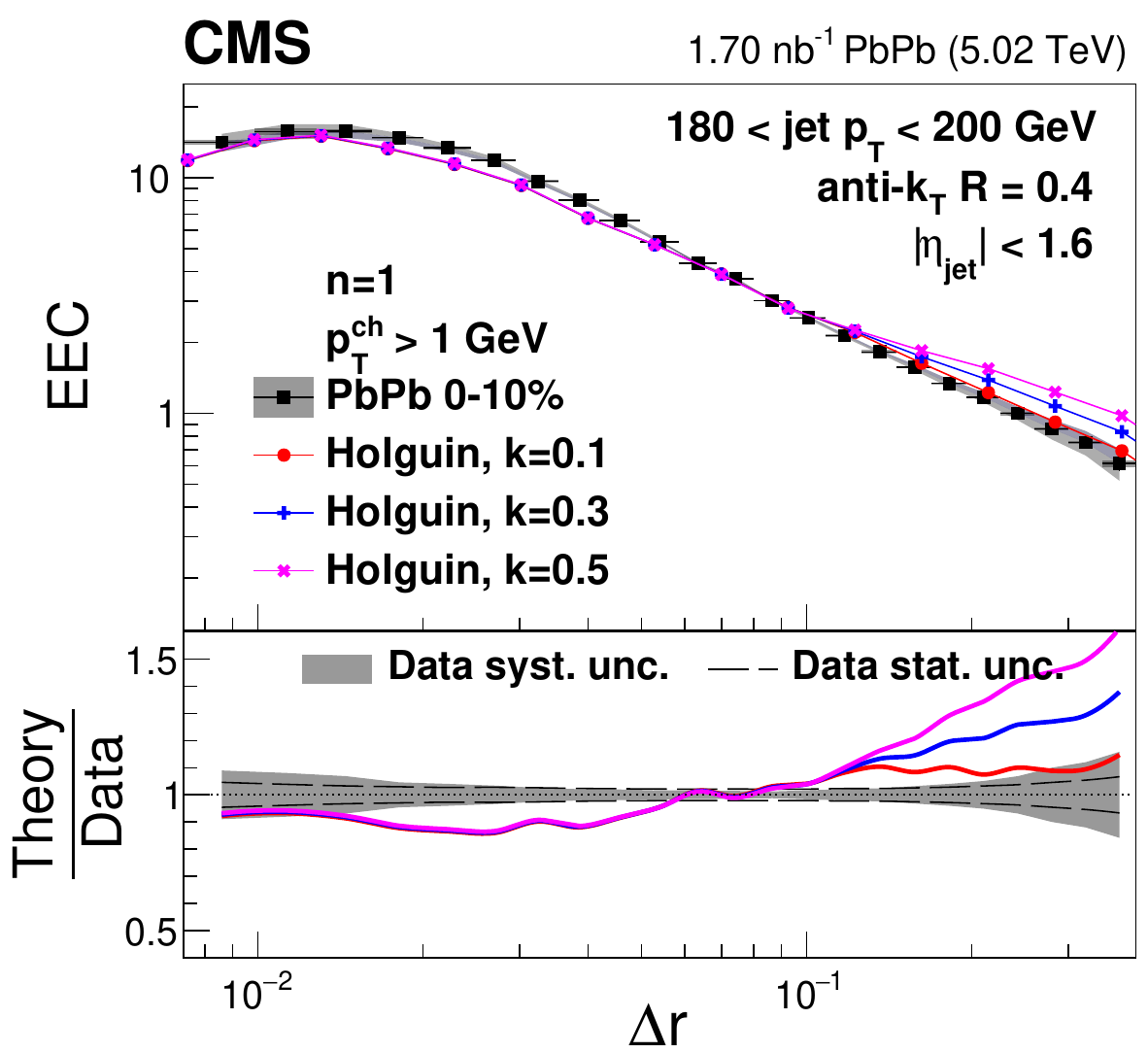}
    \caption{Energy-energy correlator distributions from 0--10\% central \PbPb collisions with $n=1$ and $\ptCh > 1\GeV$ in different \ptjet bins compared to predictions from perturbative calculation from Holguin and collaborators with different values of the $k$-parameter.} 
    \label{fig:PbPbDistributionsNominalHolguin}
  } 
\end{figure*}

\begin{figure*}[h]
  \centering{
    \includegraphics[width=0.4\textwidth]{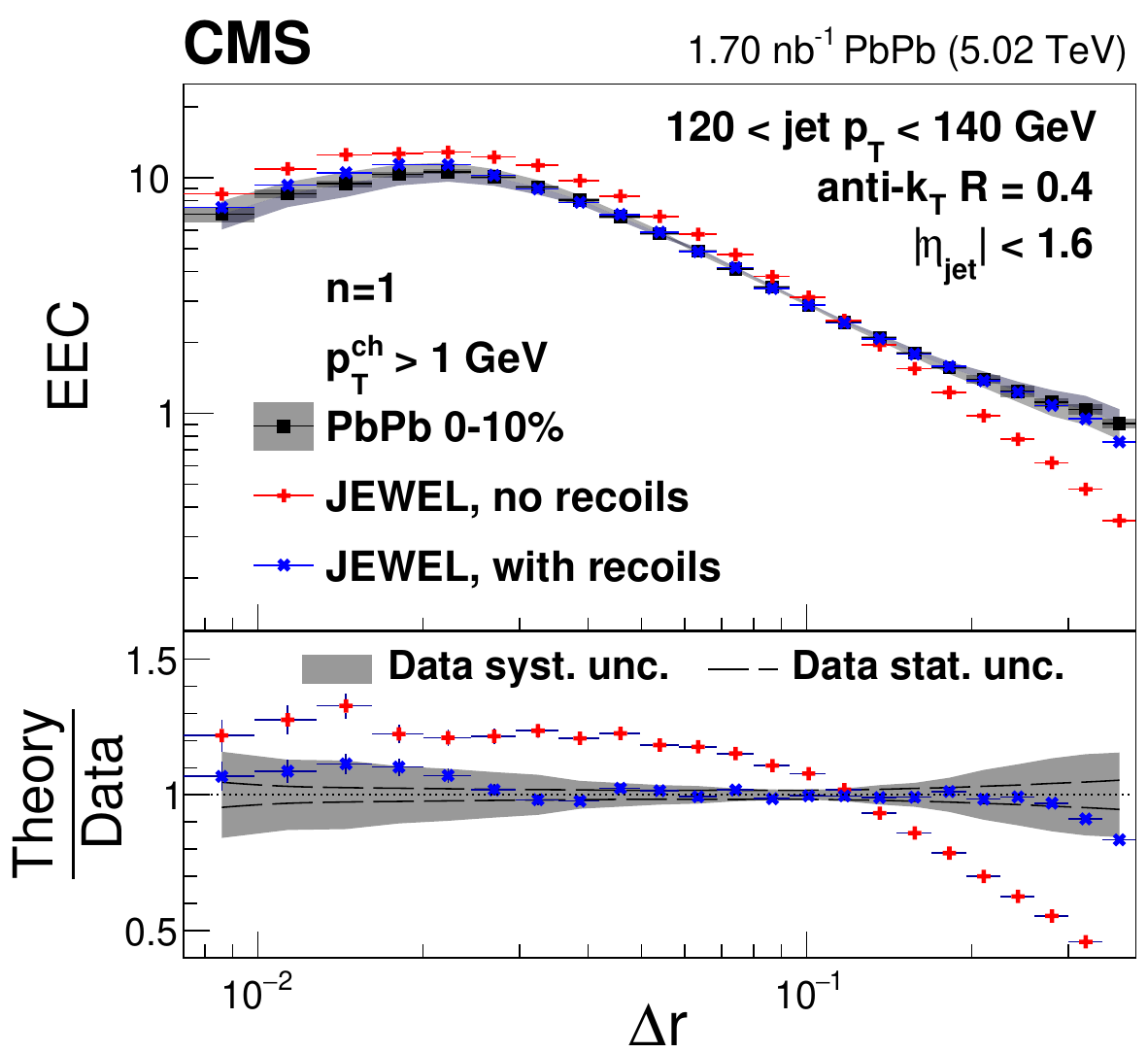}
    \includegraphics[width=0.4\textwidth]{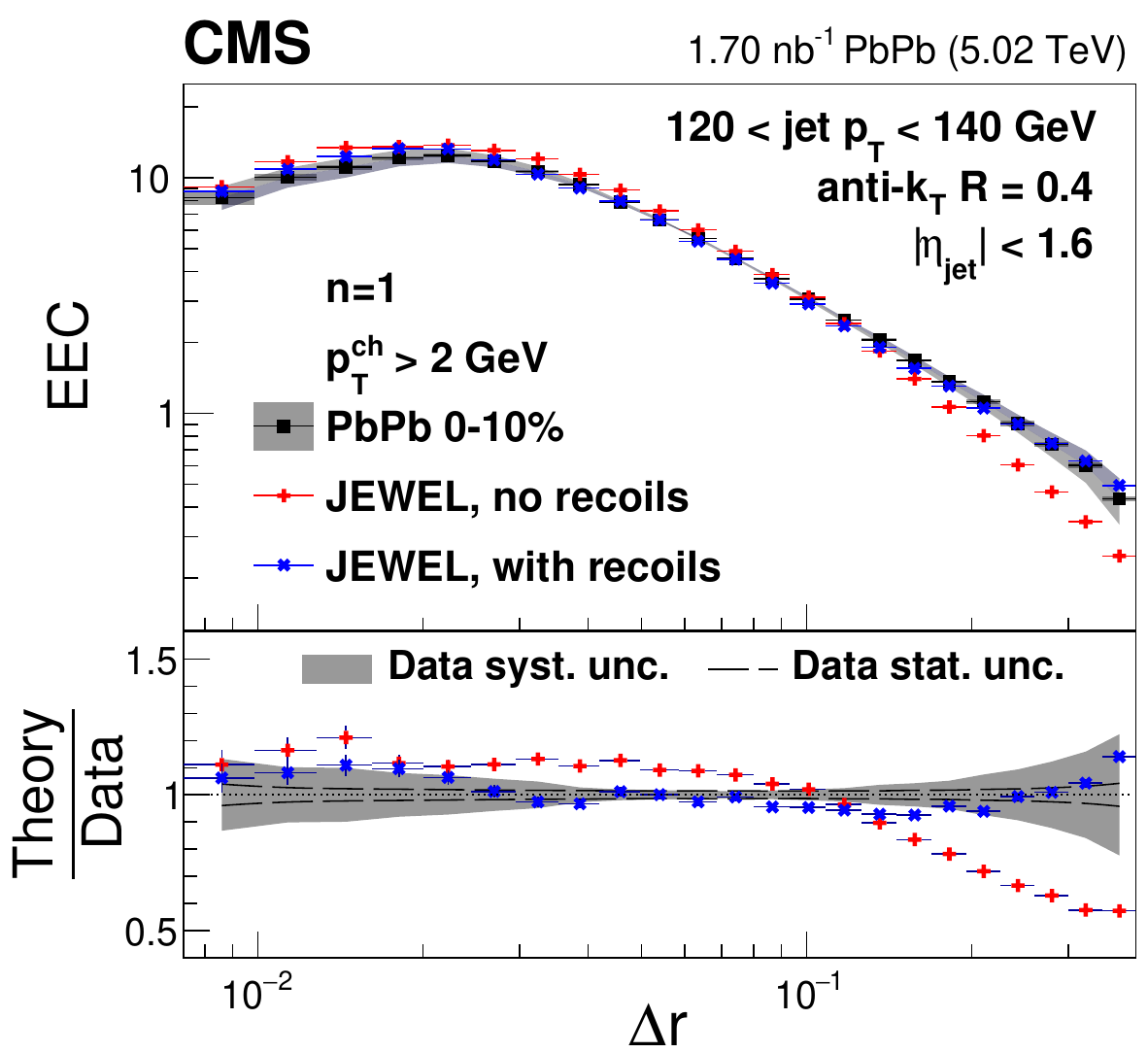} 
    \includegraphics[width=0.4\textwidth]{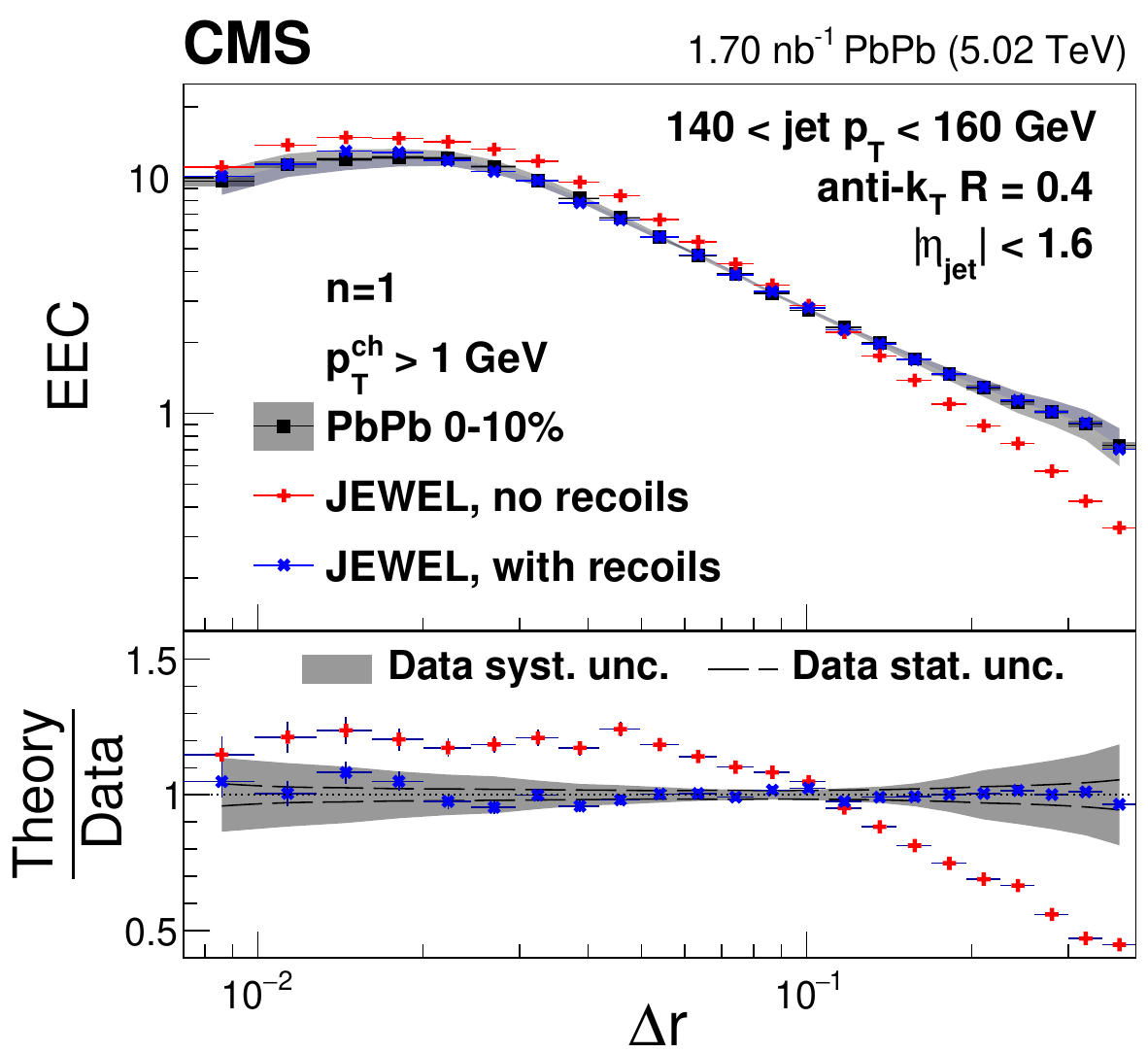}
    \includegraphics[width=0.4\textwidth]{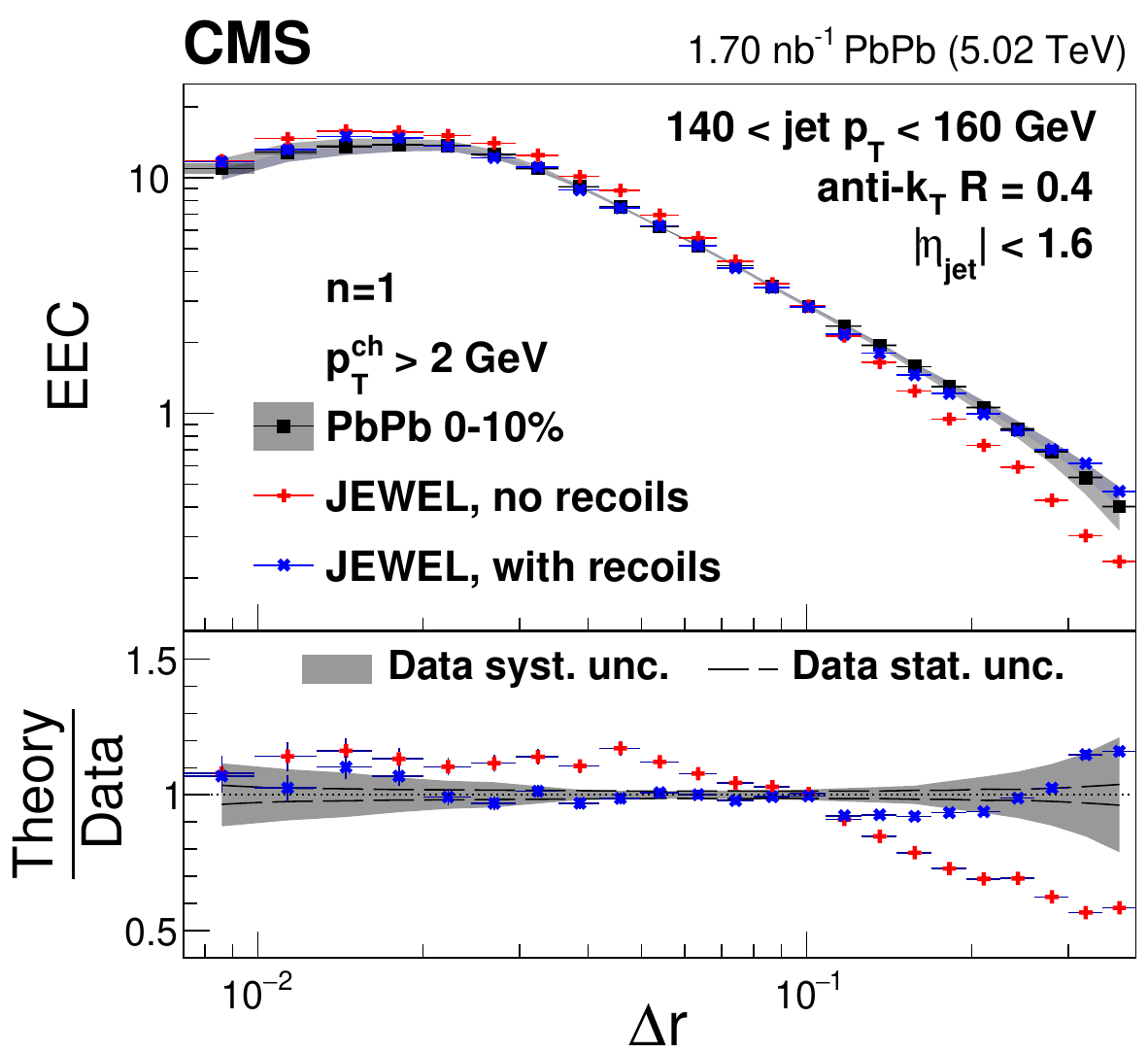}
    \includegraphics[width=0.4\textwidth]{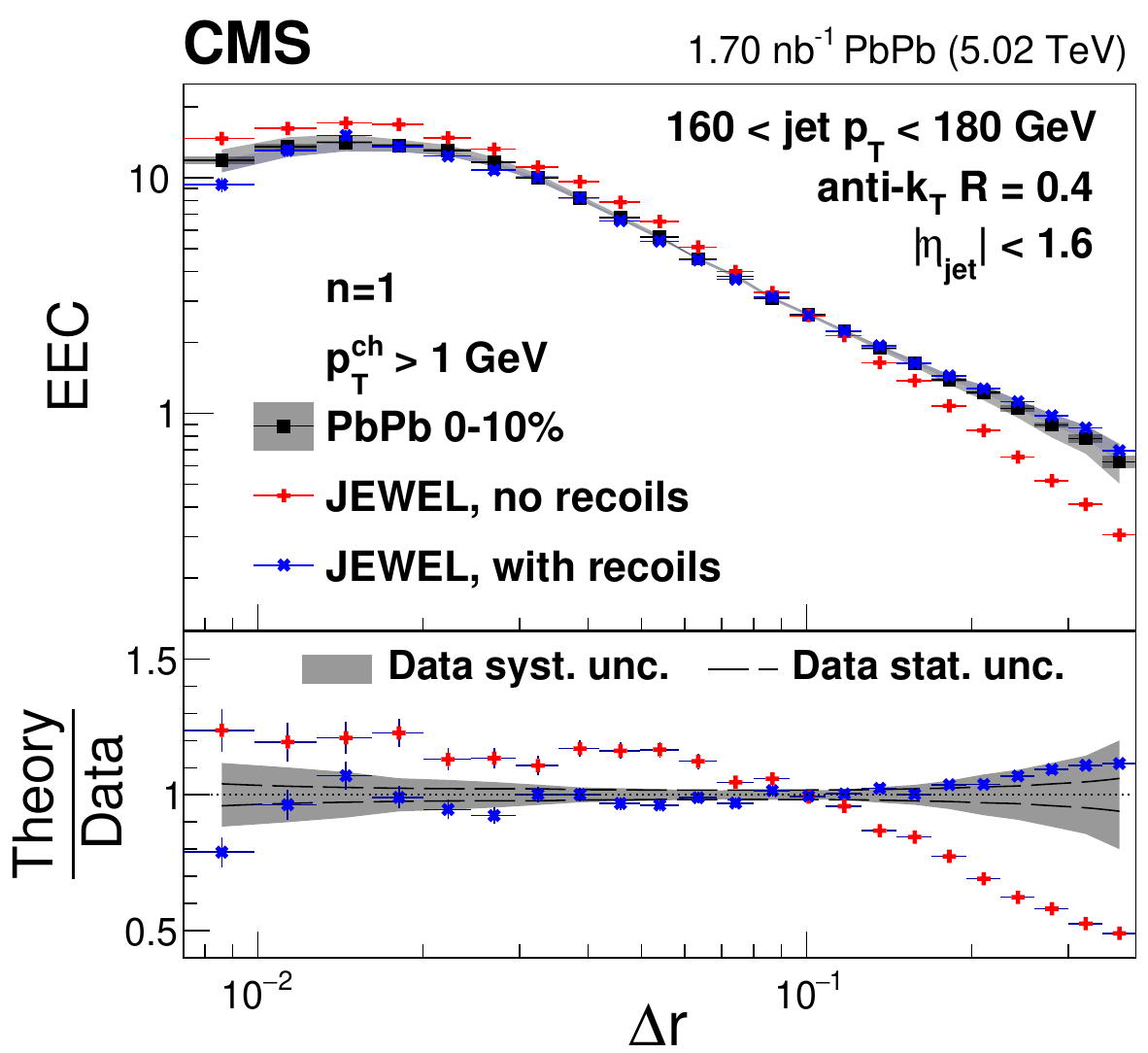}
    \includegraphics[width=0.4\textwidth]{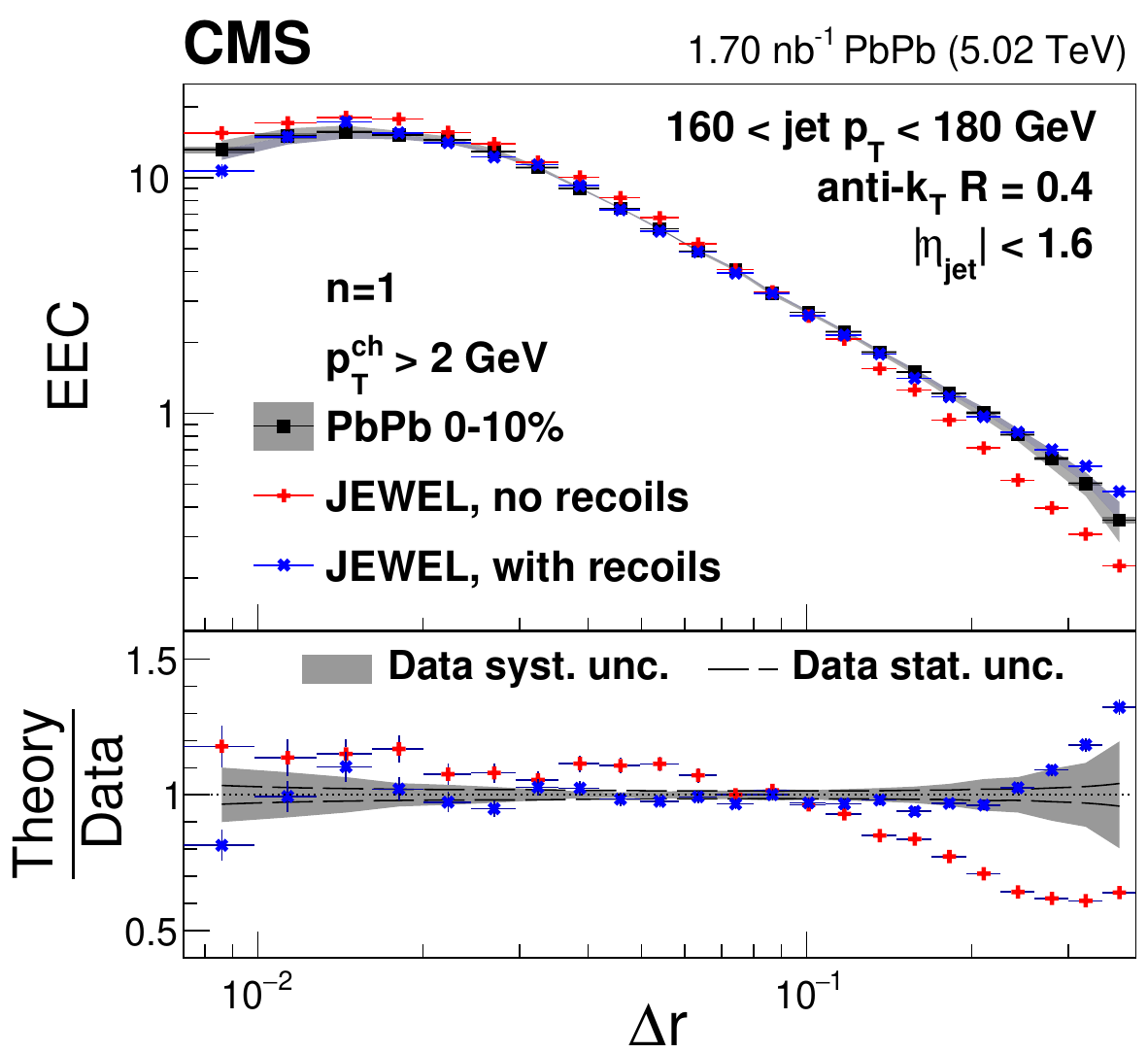}
    \includegraphics[width=0.4\textwidth]{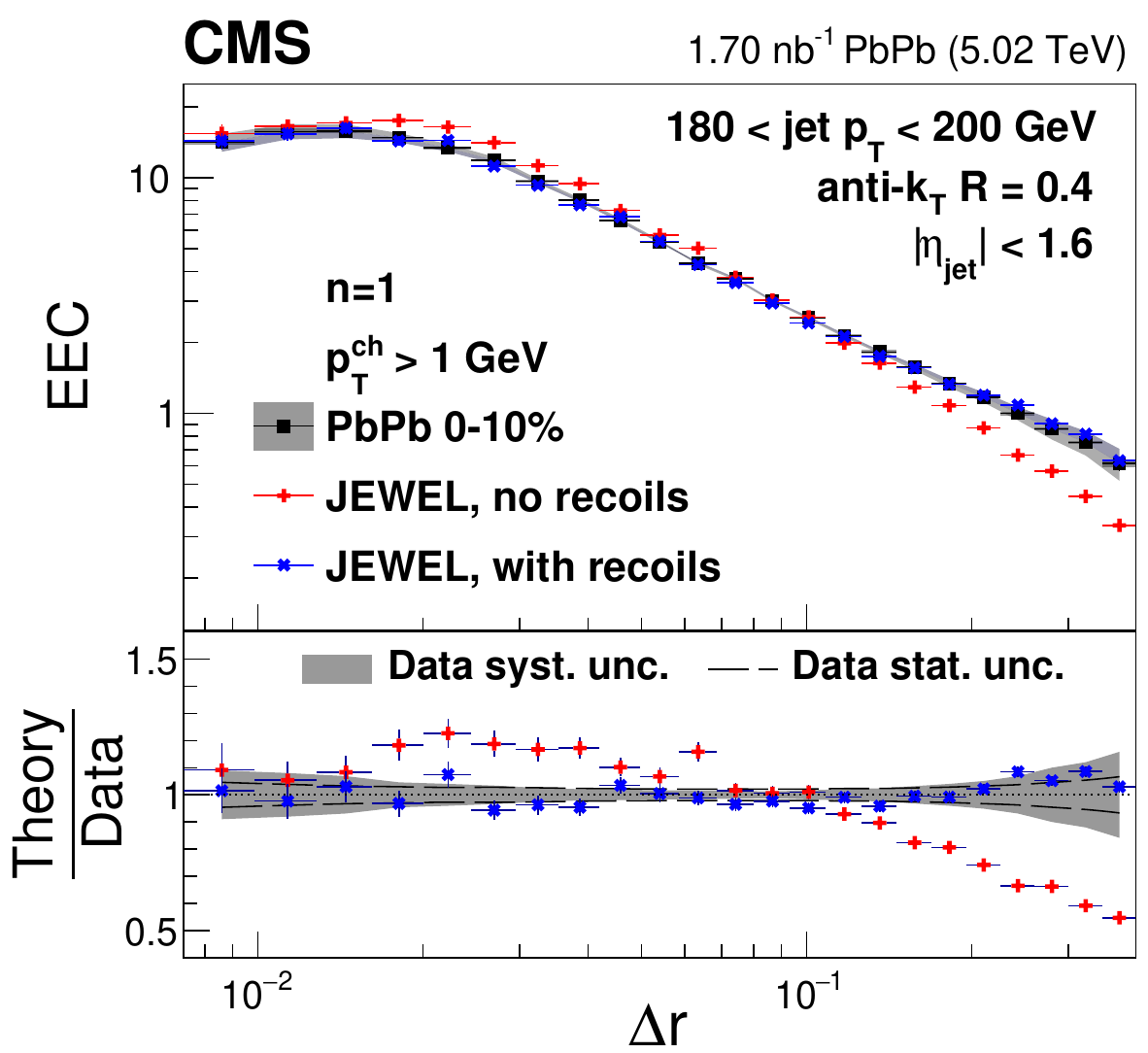}
    \includegraphics[width=0.4\textwidth]{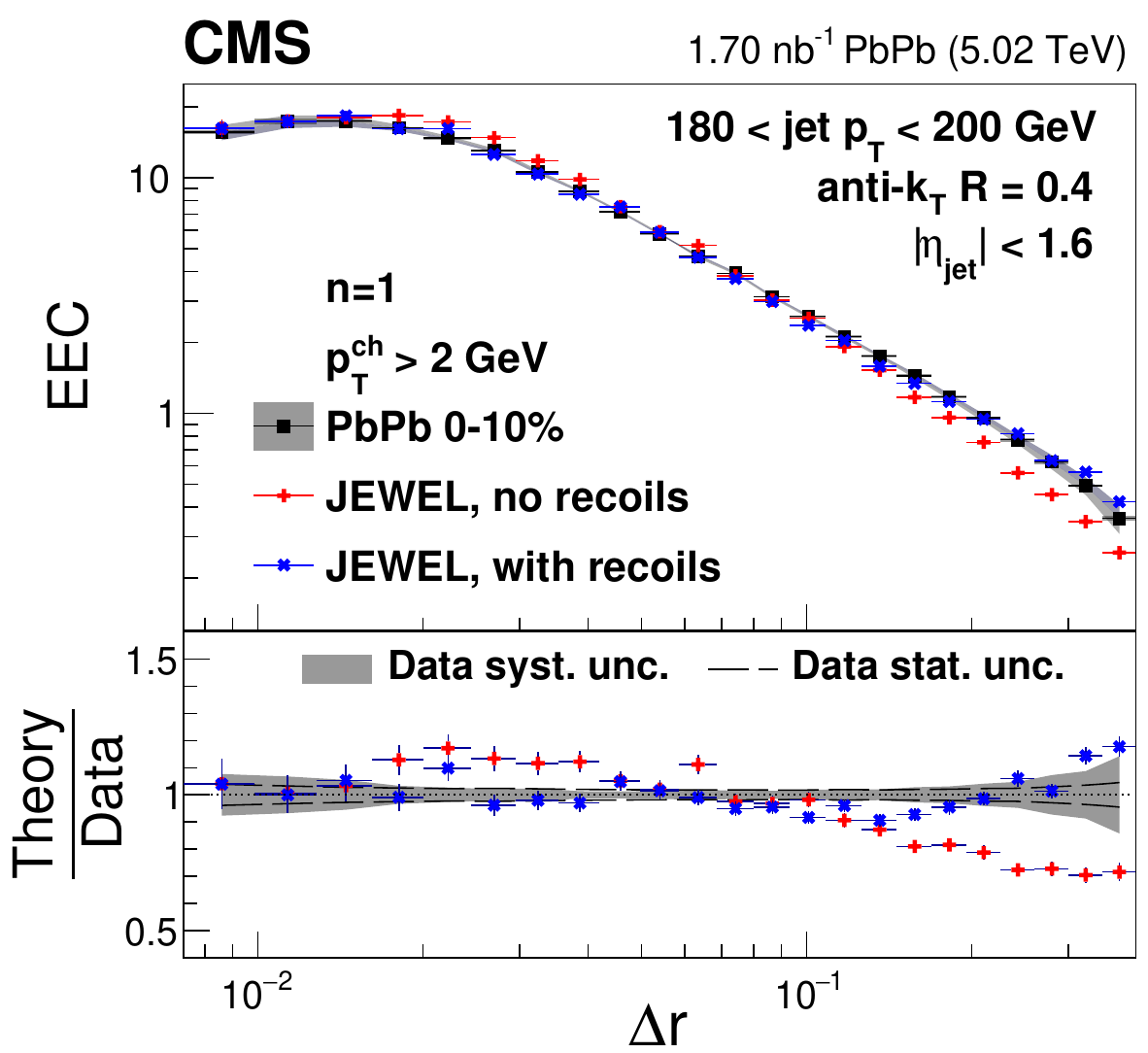}
    \caption{Energy-energy correlator distributions from 0--10\% central \PbPb collisions with $n=1$ in different \ptCh and \ptjet bins compared to predictions from \JEWEL simulation with and without recoils.} 
    \label{fig:PbPbDistributionsJewelNominal}
  } 
\end{figure*}

\begin{figure*}[h]
  \centering{
    \includegraphics[width=0.4\textwidth]{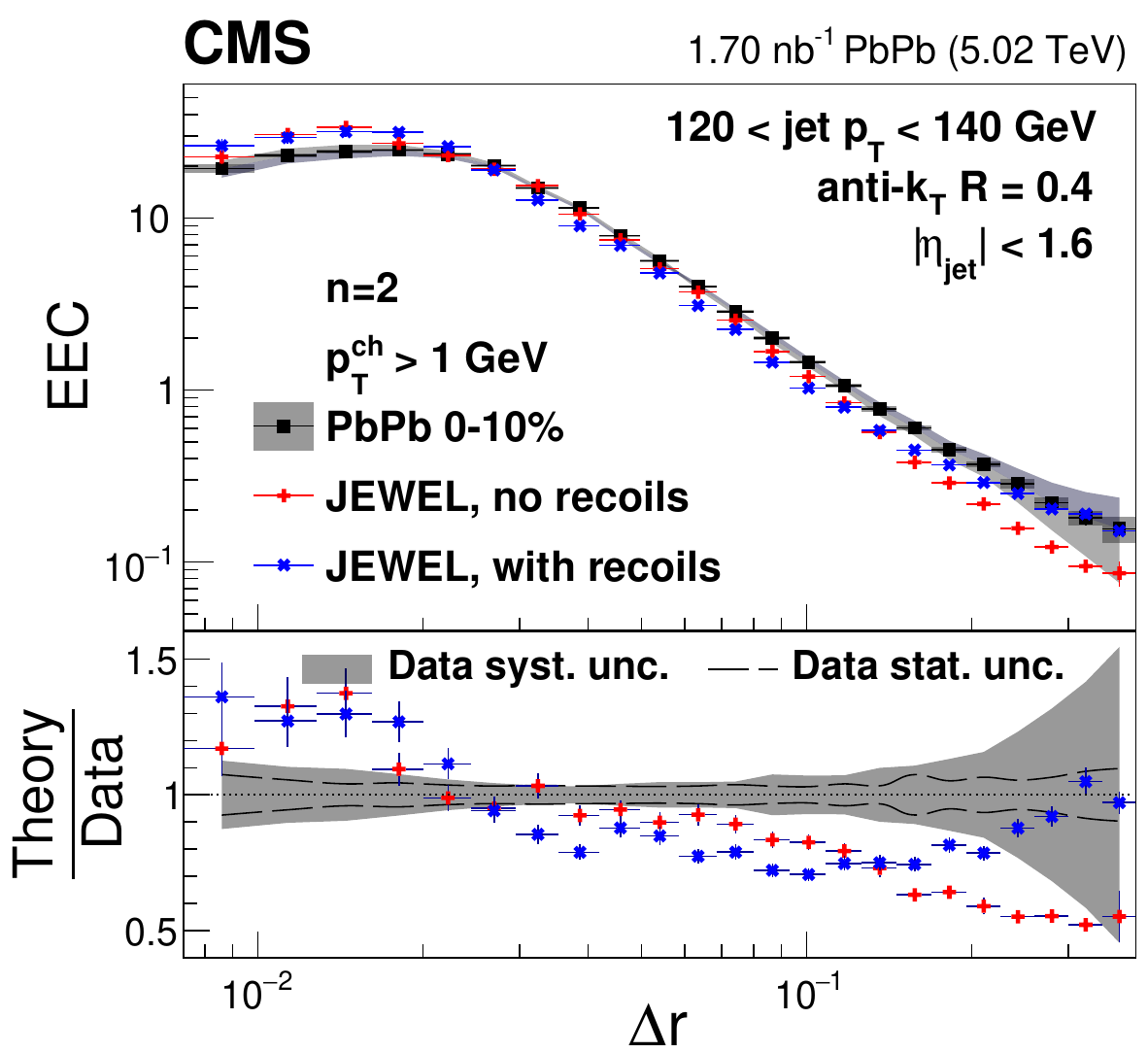}
    \includegraphics[width=0.4\textwidth]{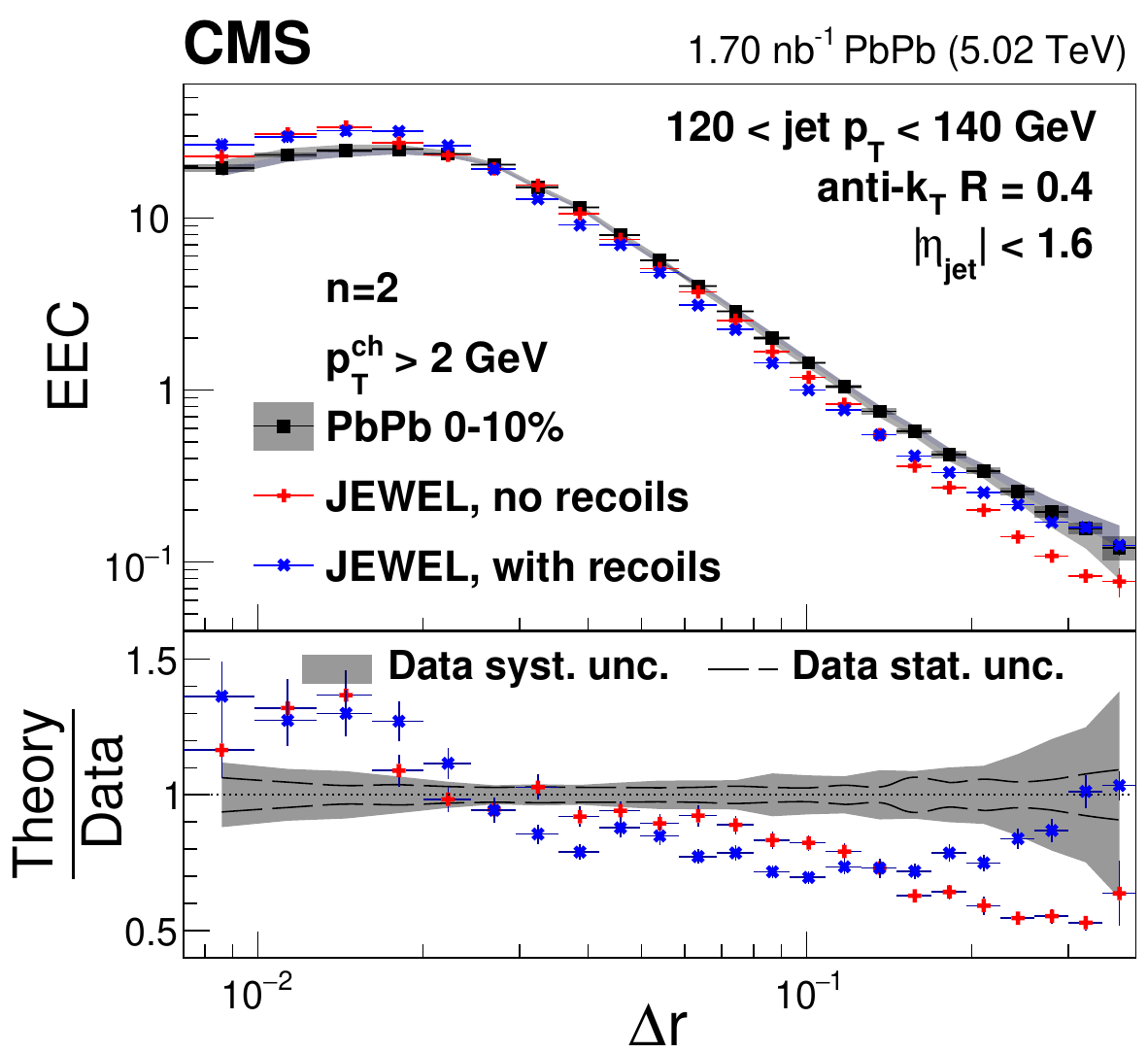} 
    \includegraphics[width=0.4\textwidth]{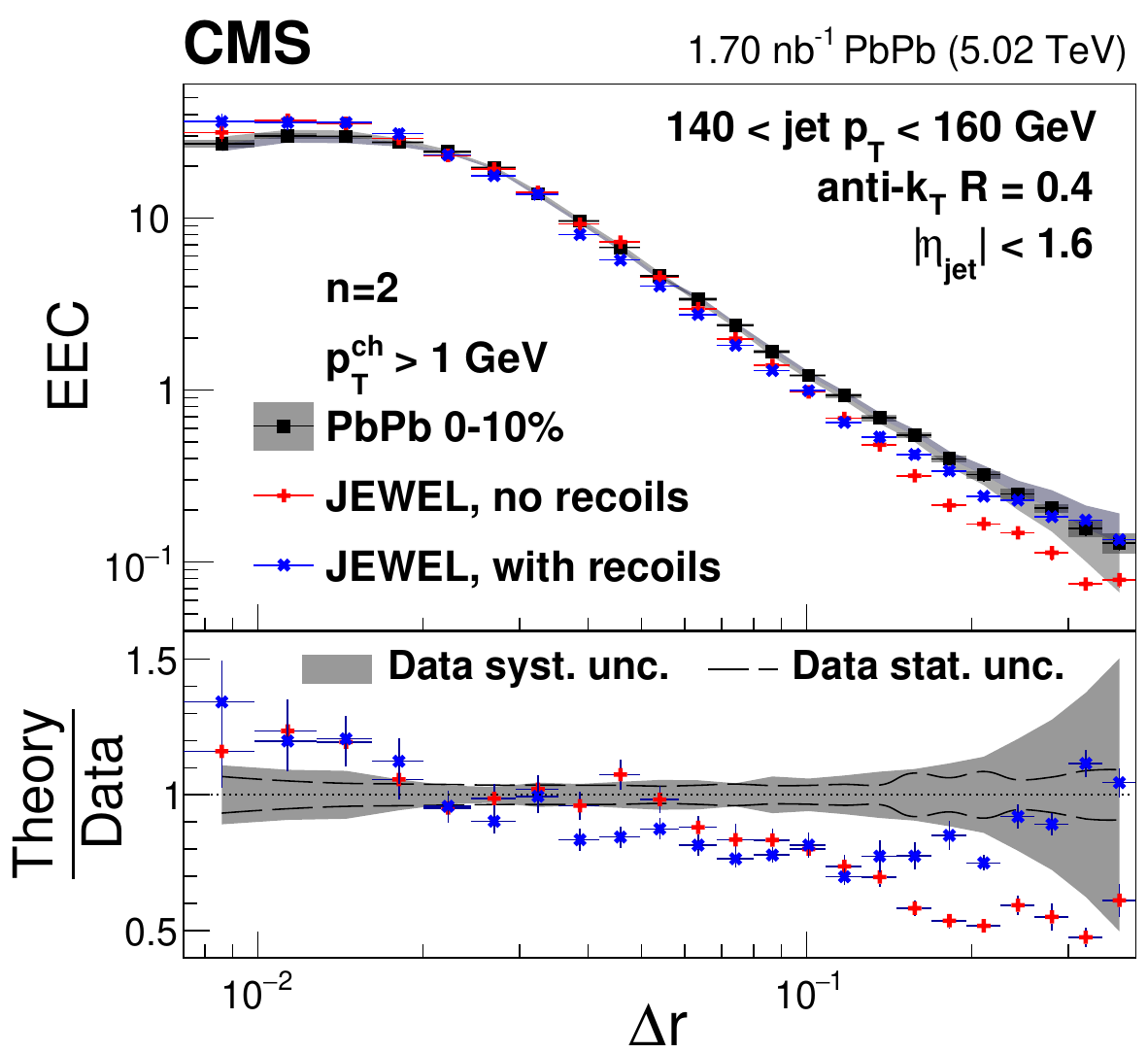}
    \includegraphics[width=0.4\textwidth]{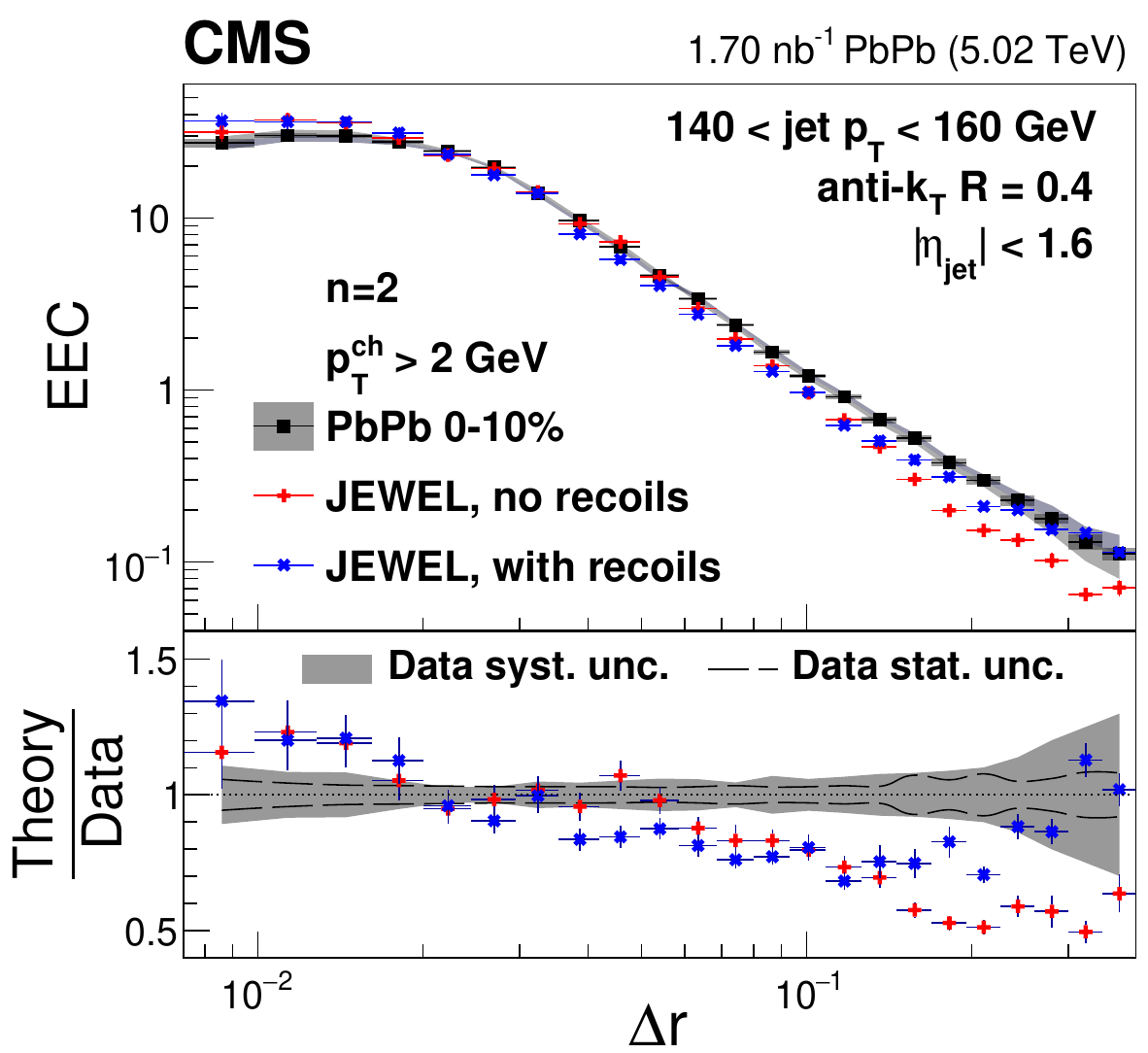}
    \includegraphics[width=0.4\textwidth]{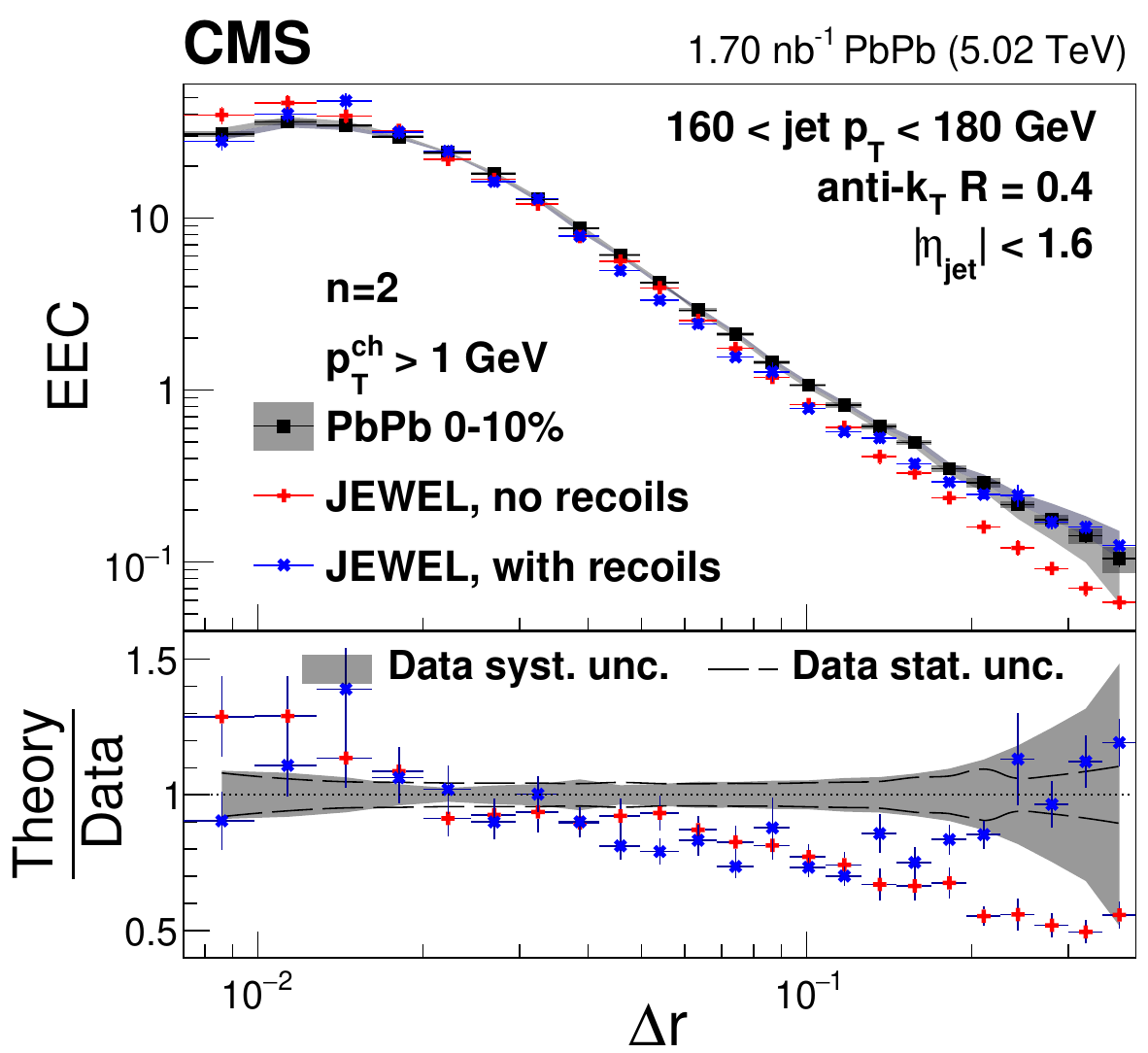}
    \includegraphics[width=0.4\textwidth]{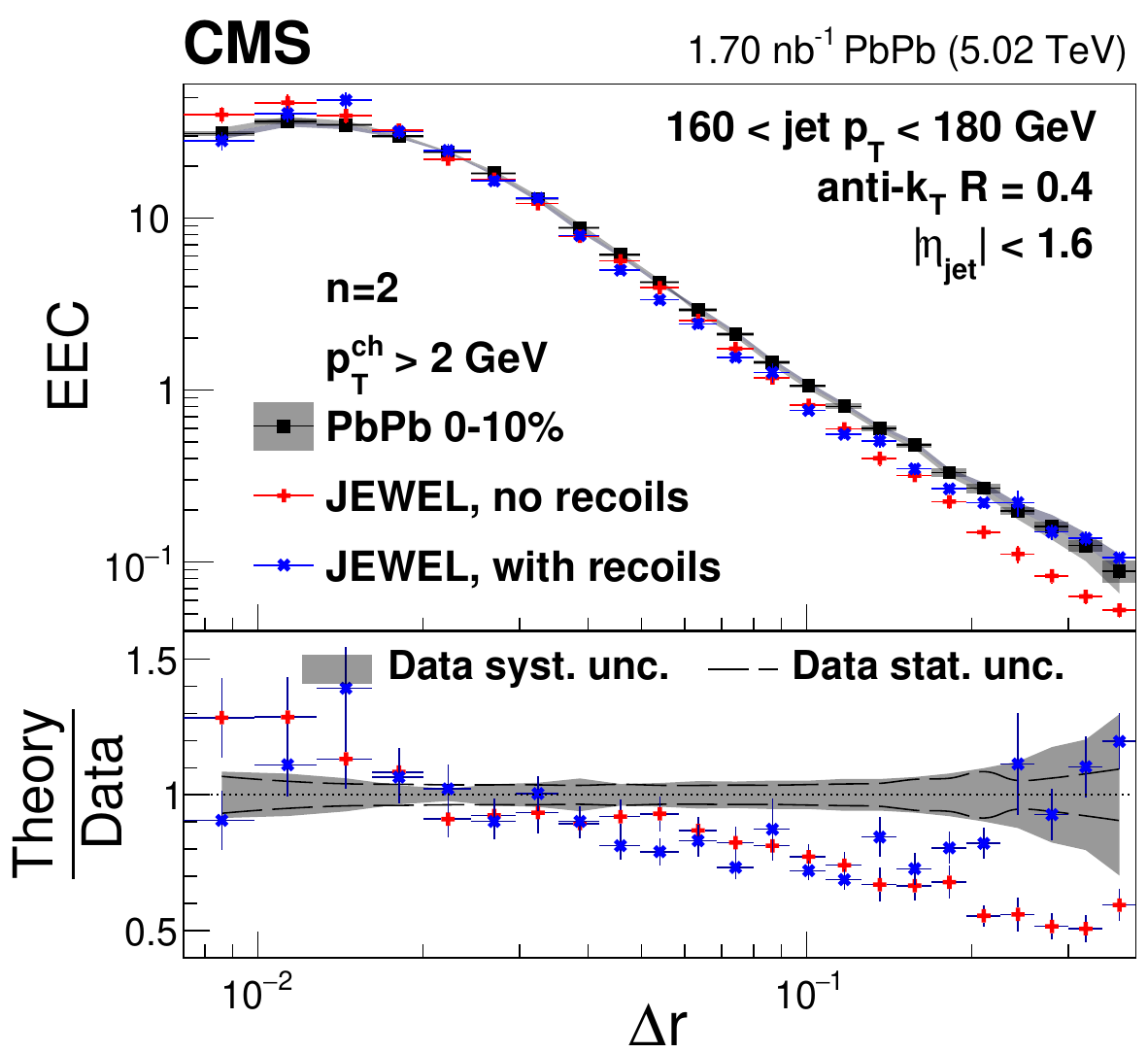}
    \includegraphics[width=0.4\textwidth]{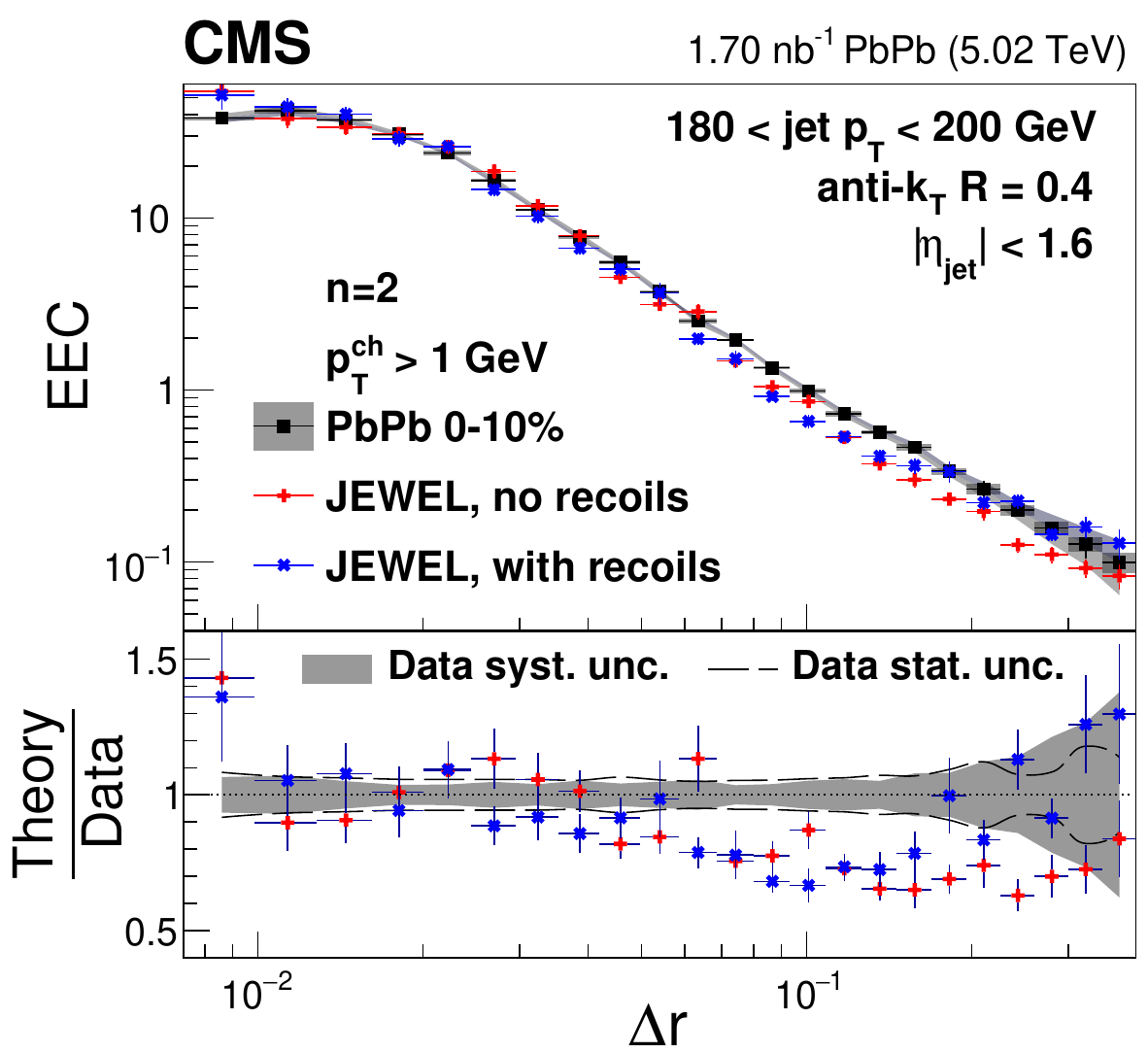}
    \includegraphics[width=0.4\textwidth]{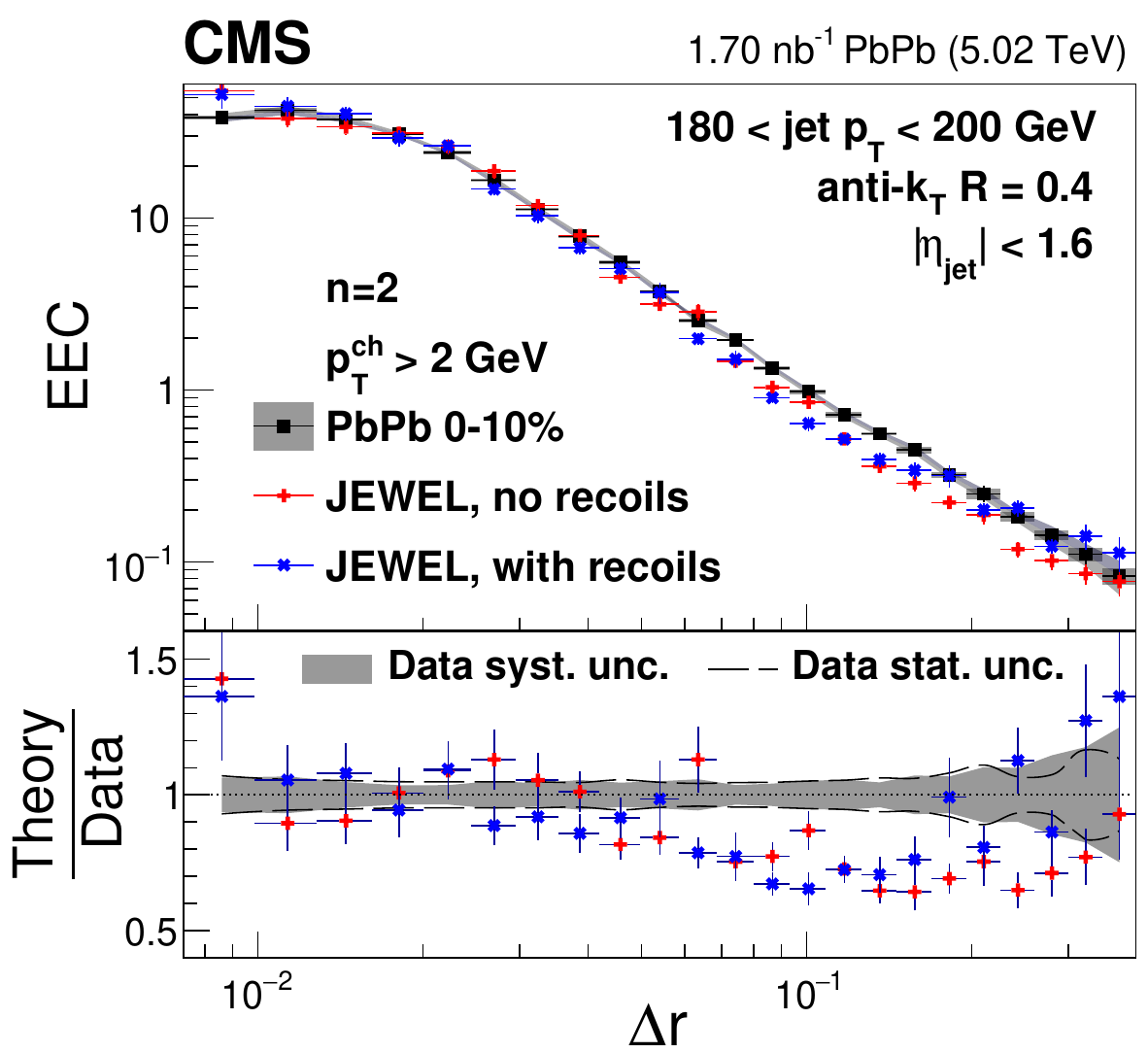}
    \caption{Energy-energy correlator distributions from 0--10\% central \PbPb collisions with $n=2$ in different \ptCh and \ptjet bins compared to predictions from \JEWEL simulation with and without recoils.} 
    \label{fig:PbPbDistributionsJewelSquared}
  } 
\end{figure*}

\begin{figure*}[h]
  \centering{
    \includegraphics[width=0.4\textwidth]{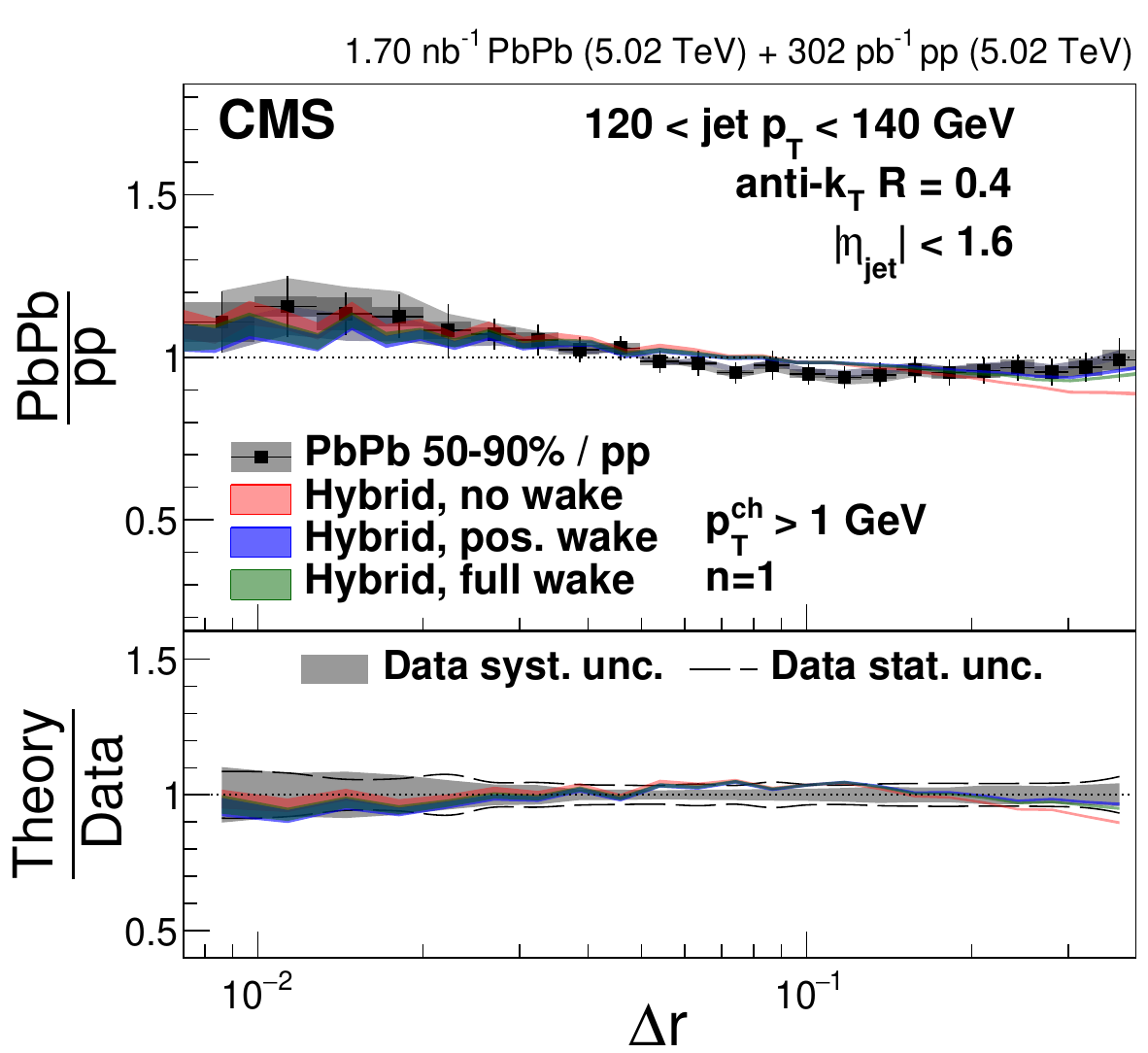}
    \includegraphics[width=0.4\textwidth]{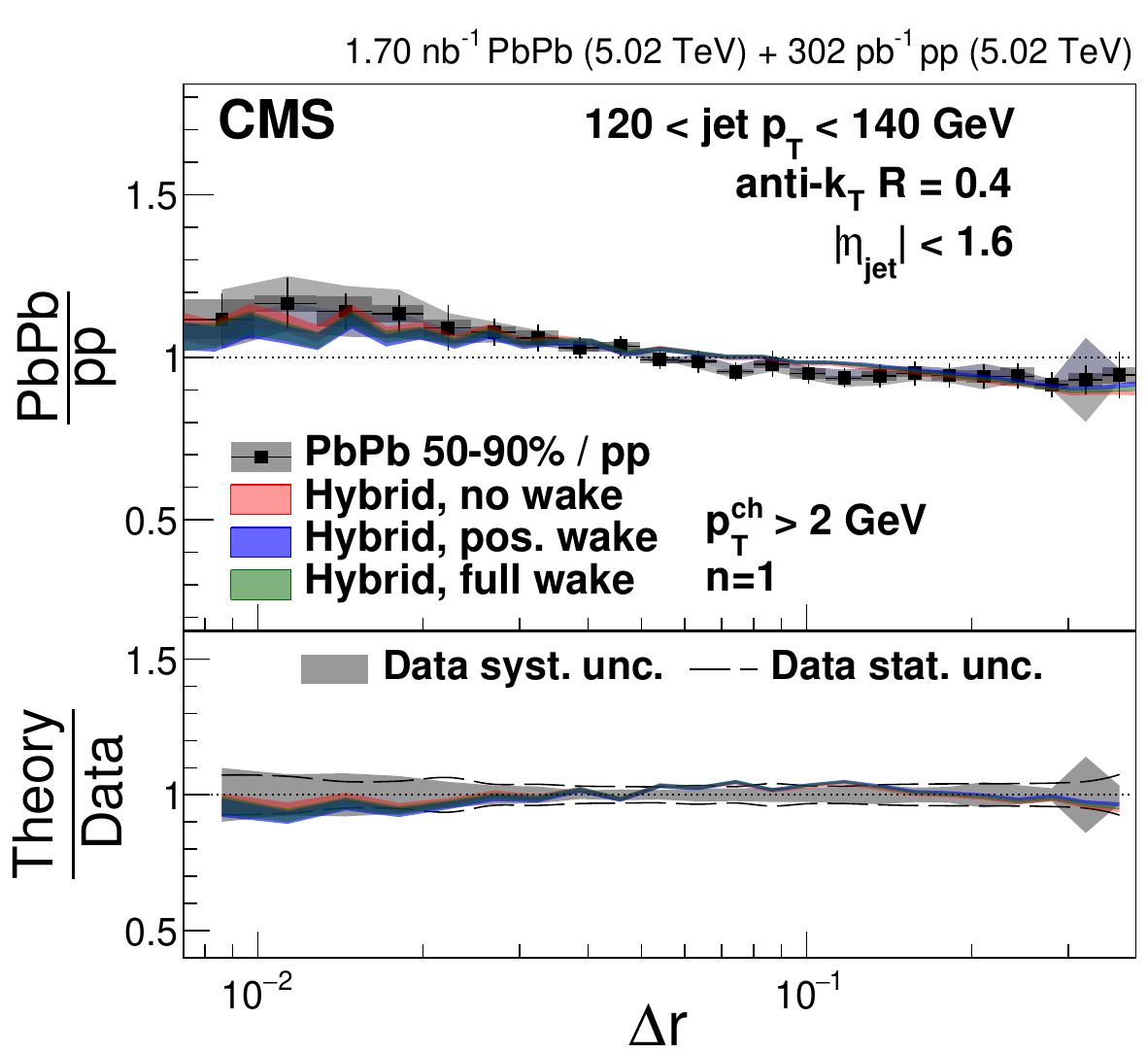} 
    \includegraphics[width=0.4\textwidth]{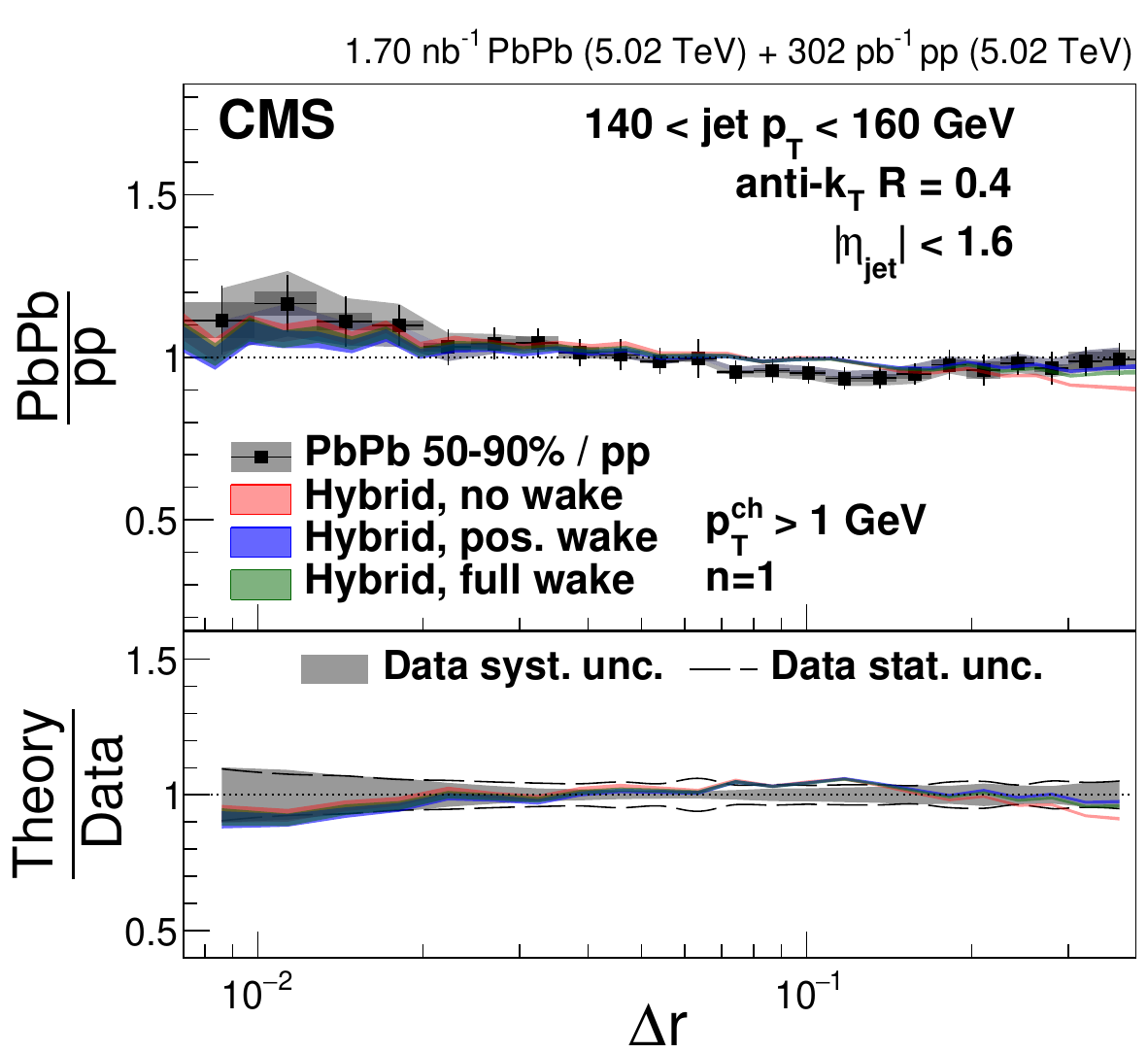}
    \includegraphics[width=0.4\textwidth]{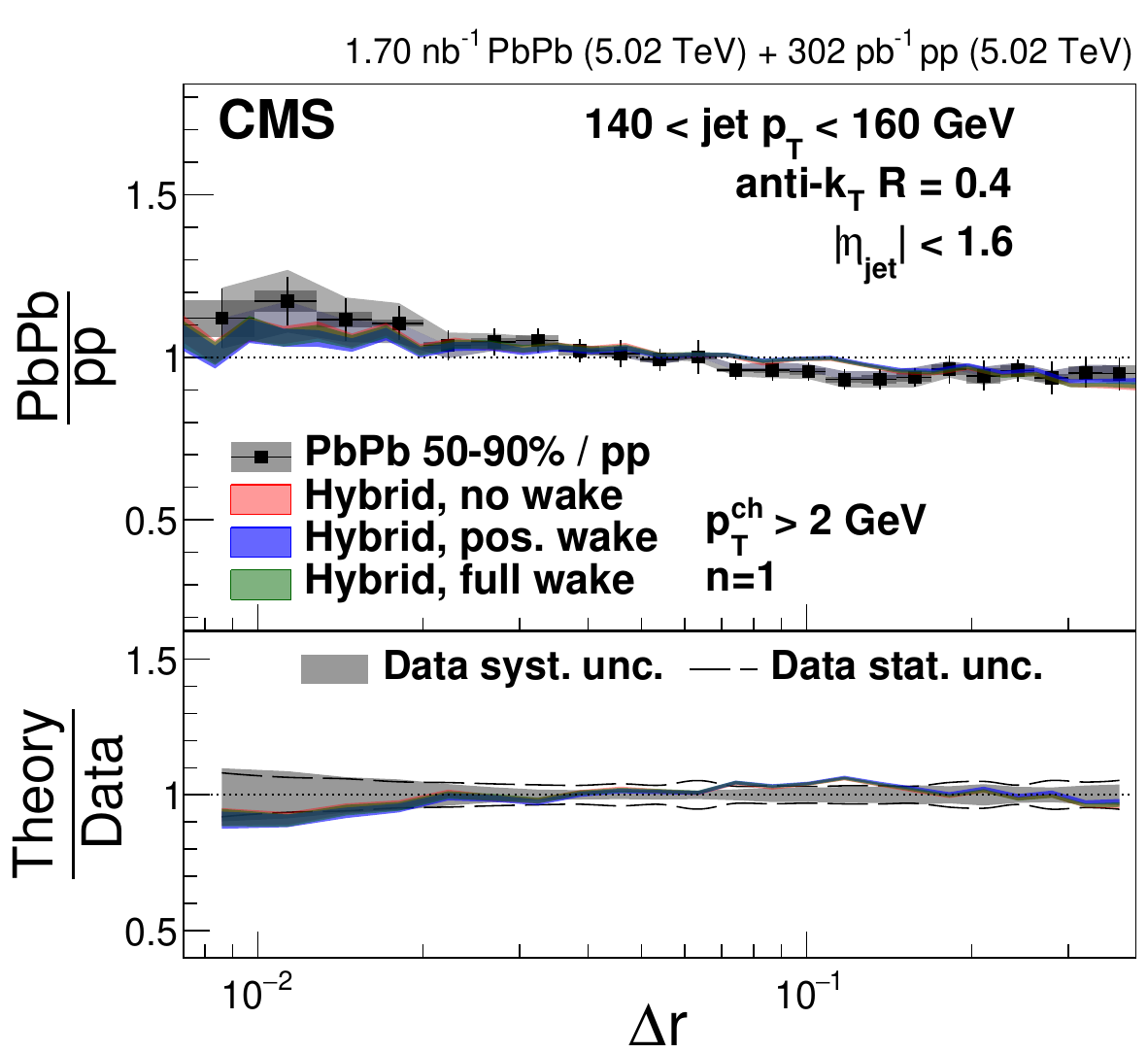}
    \includegraphics[width=0.4\textwidth]{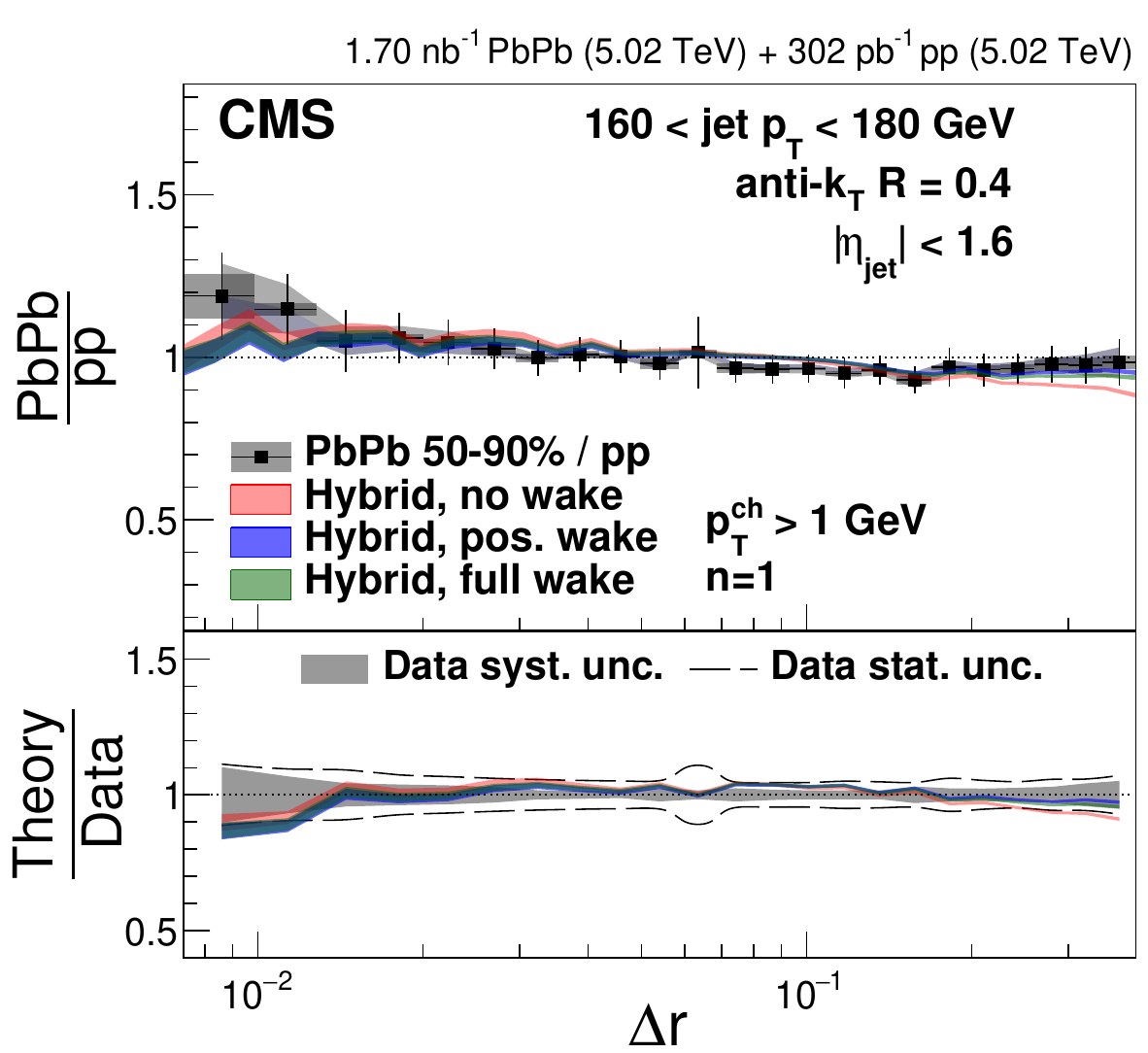}
    \includegraphics[width=0.4\textwidth]{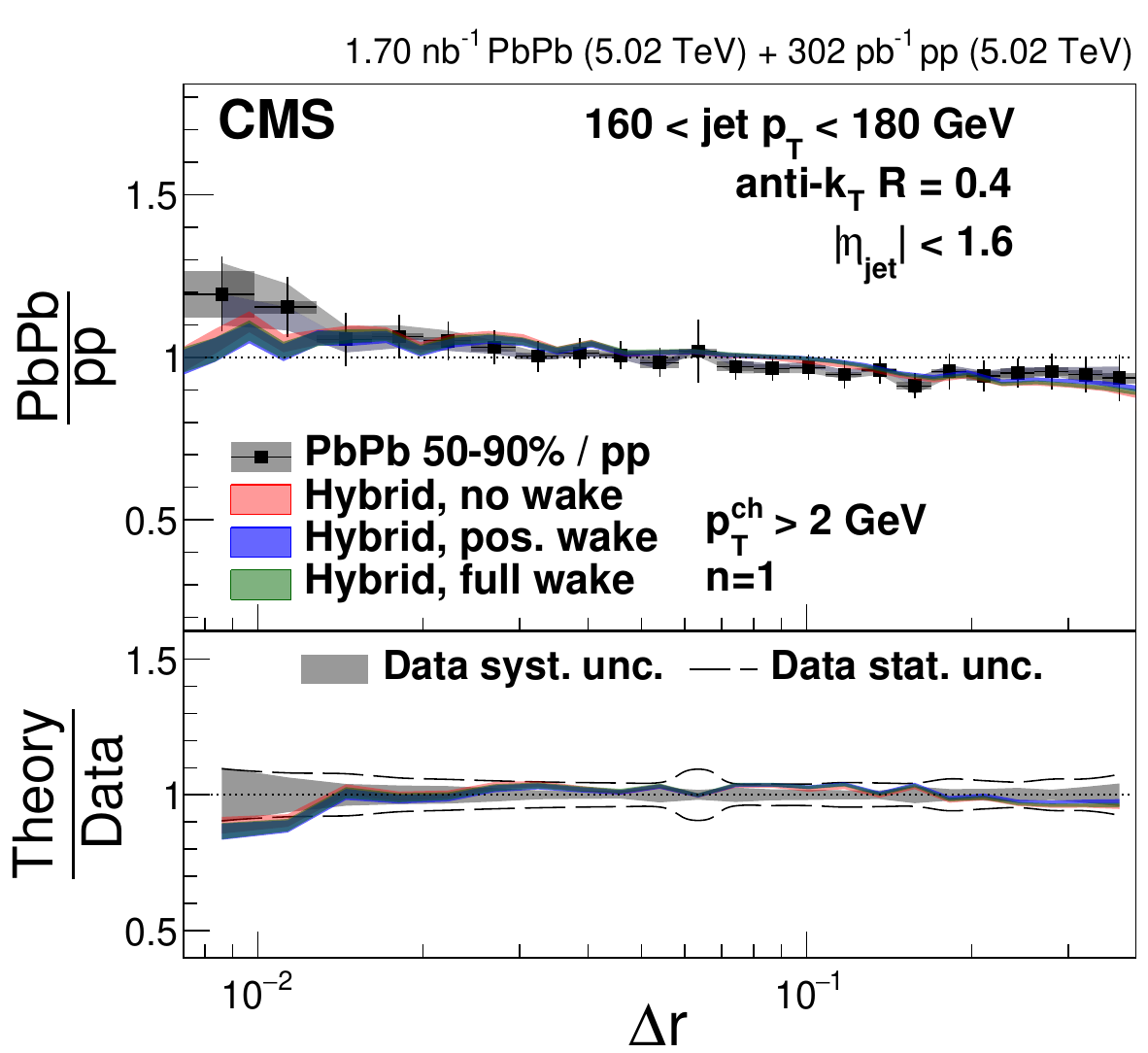}
    \includegraphics[width=0.4\textwidth]{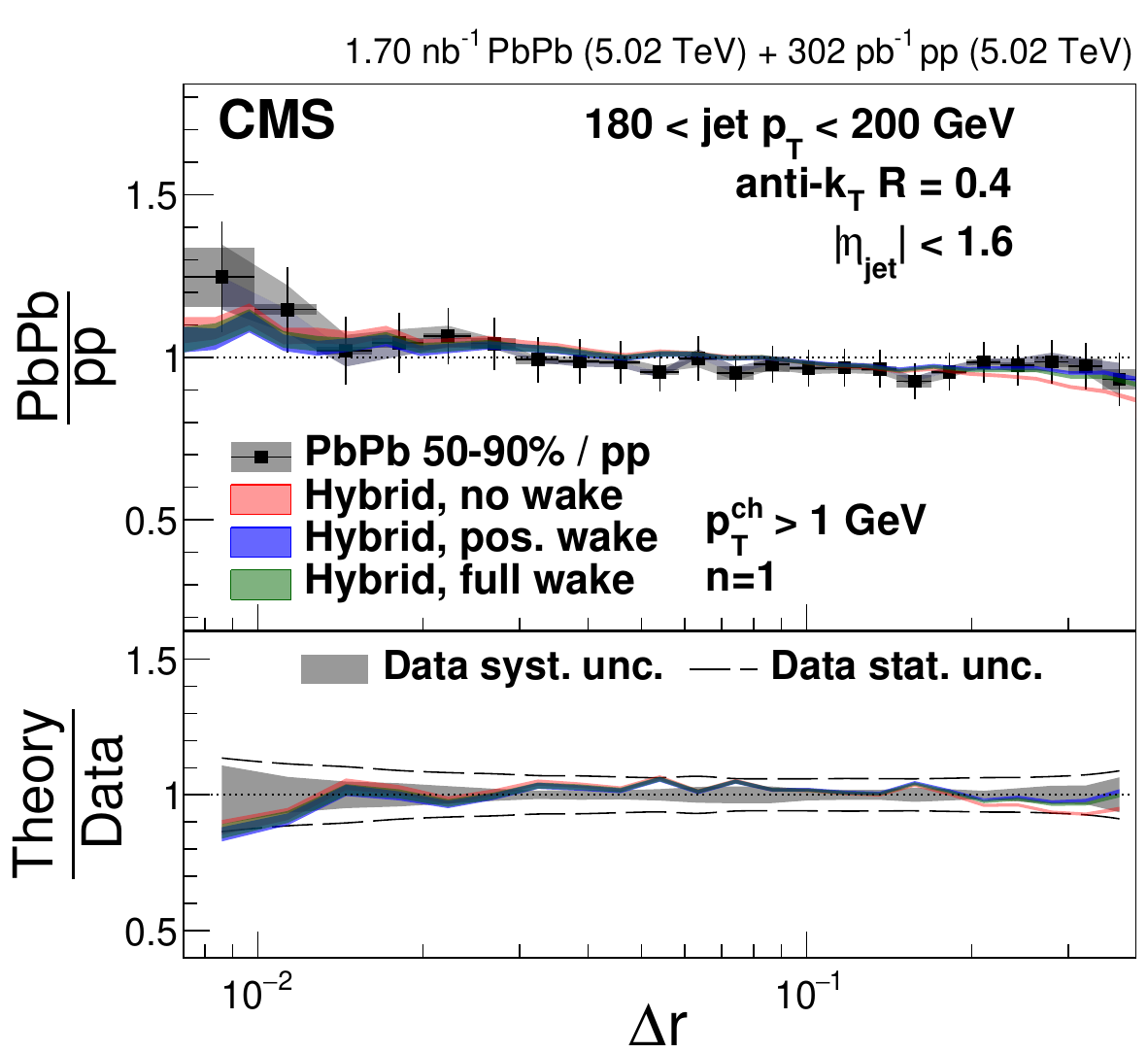}
    \includegraphics[width=0.4\textwidth]{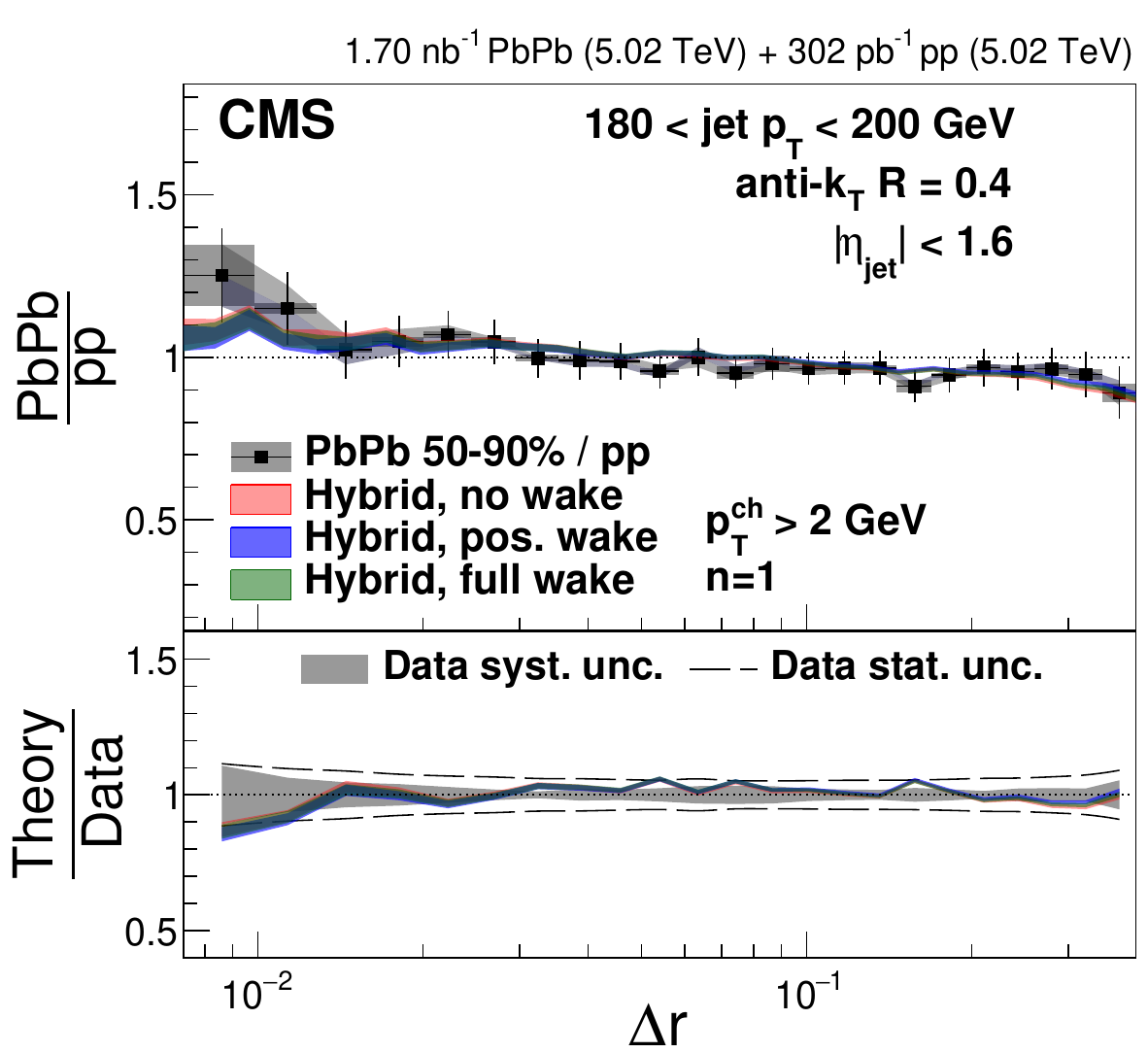}
    \caption{Ratios of 50--90\% central \PbPb to \pp energy-energy correlators with $n=1$ in different \ptCh and \ptjet bins compared to the hybrid model predictions with different jet wake settings.} 
    \label{fig:PbPbRatiosNominal50-90}
  } 
\end{figure*}

\begin{figure*}[h]
  \centering{
    \includegraphics[width=0.4\textwidth]{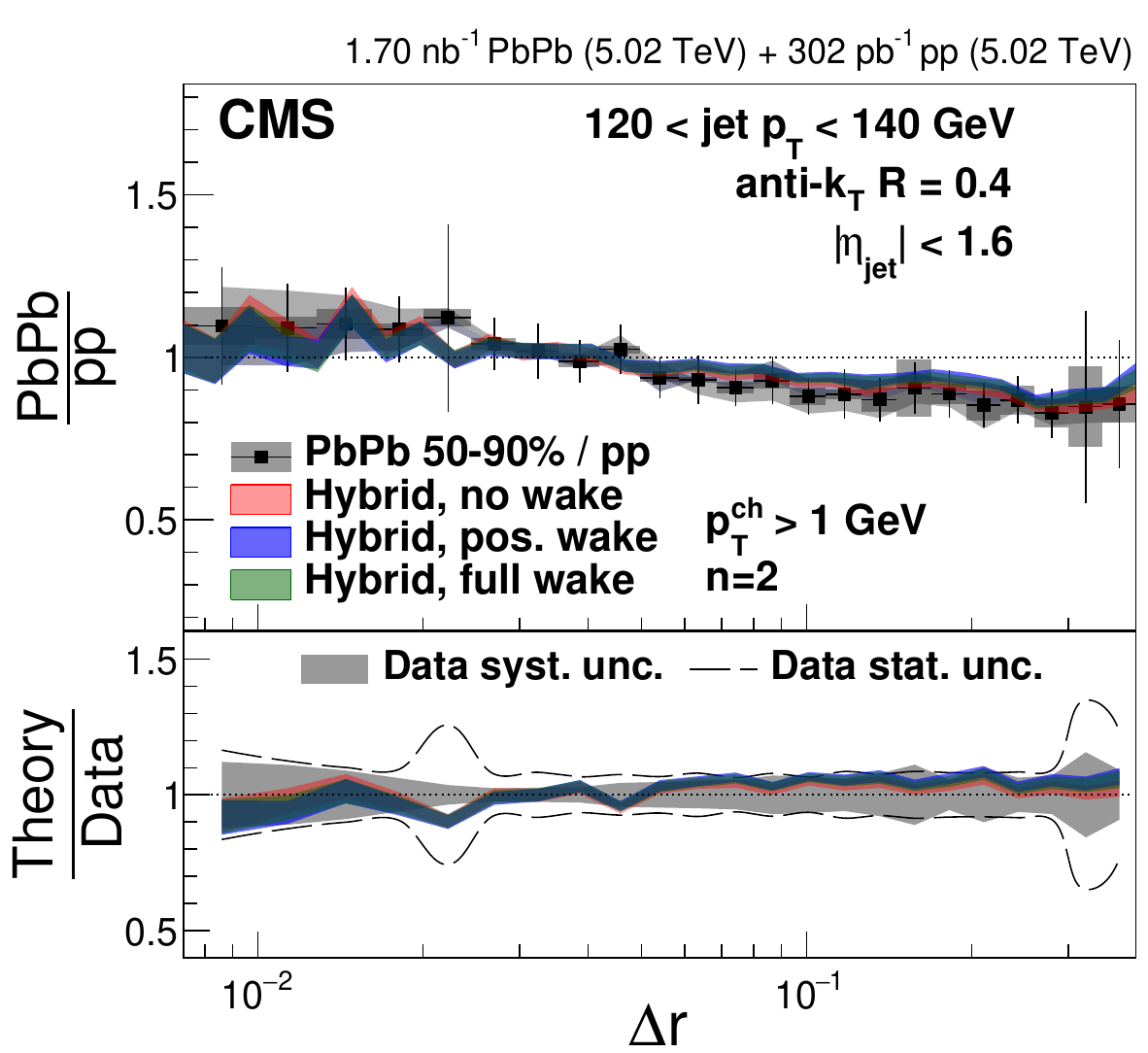} 
    \includegraphics[width=0.4\textwidth]{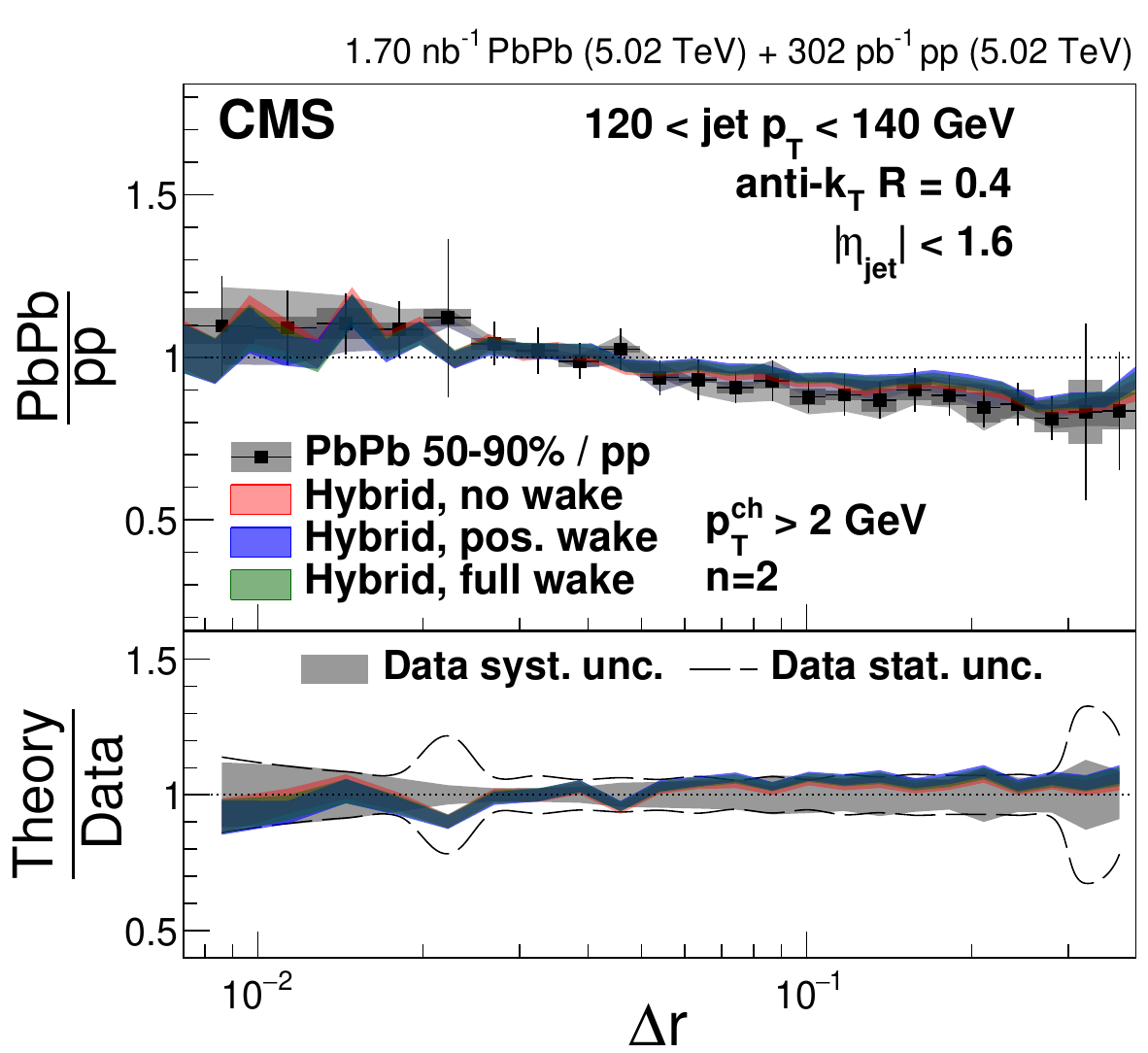} 
    \includegraphics[width=0.4\textwidth]{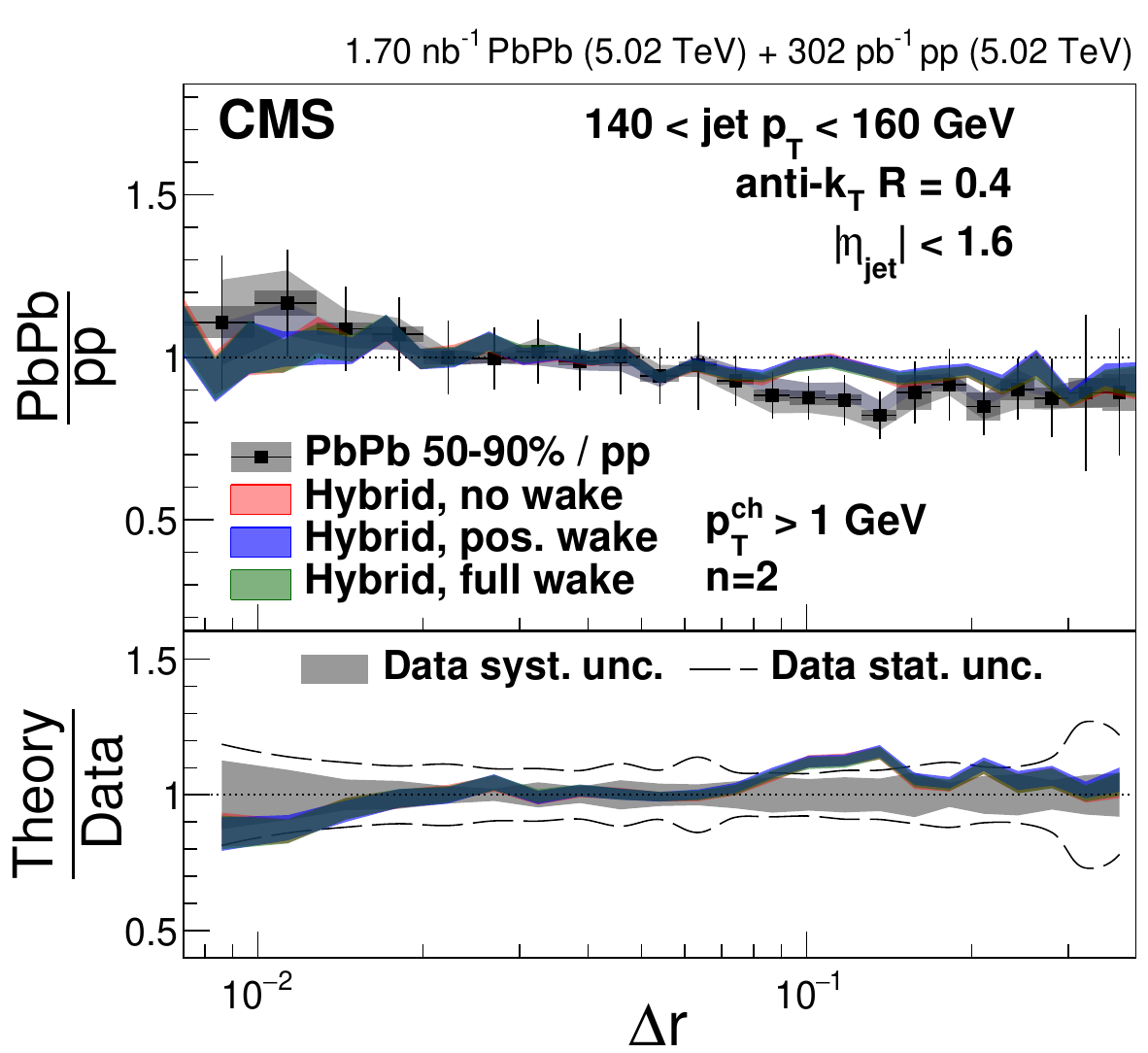}
    \includegraphics[width=0.4\textwidth]{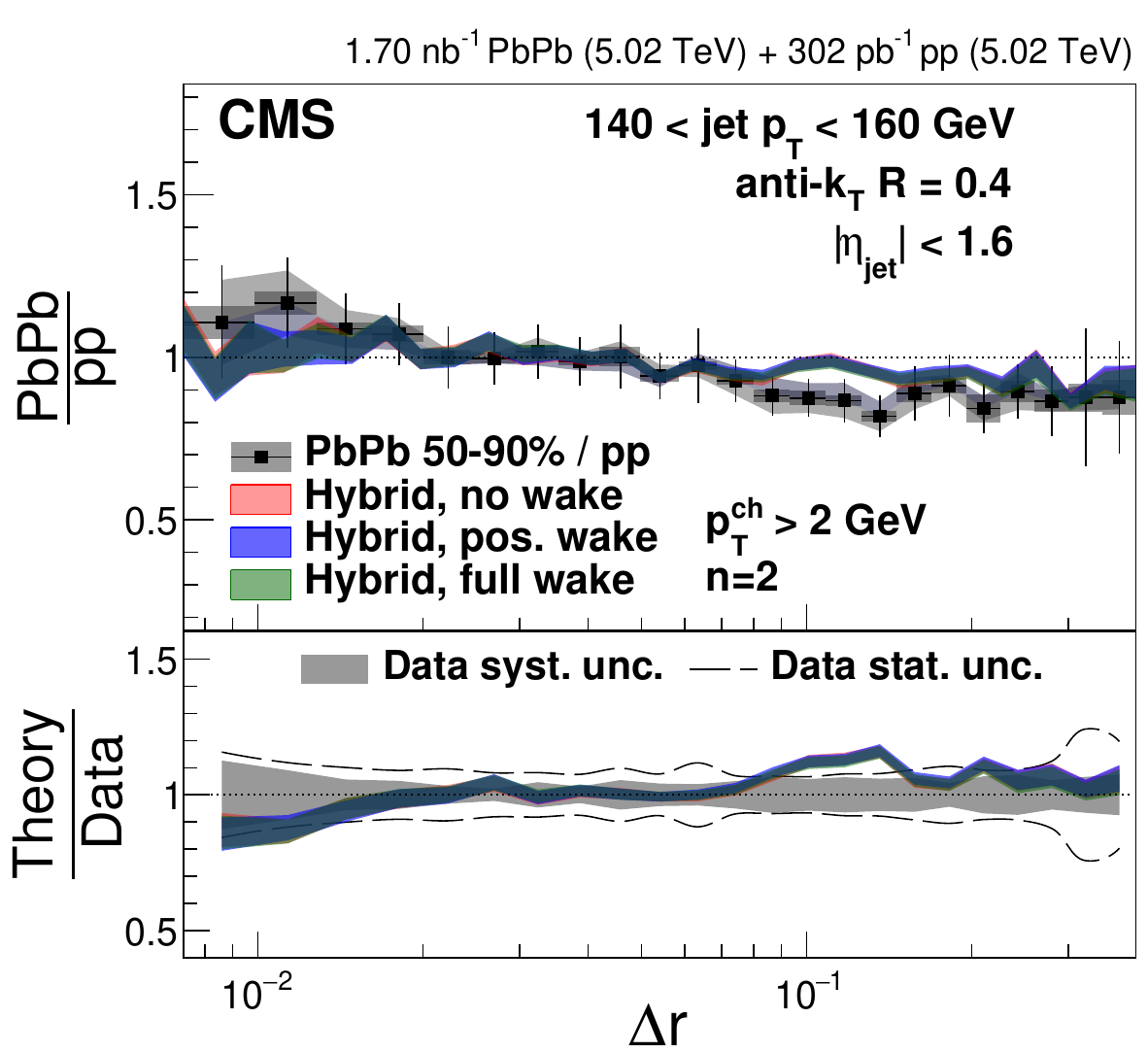}
    \includegraphics[width=0.4\textwidth]{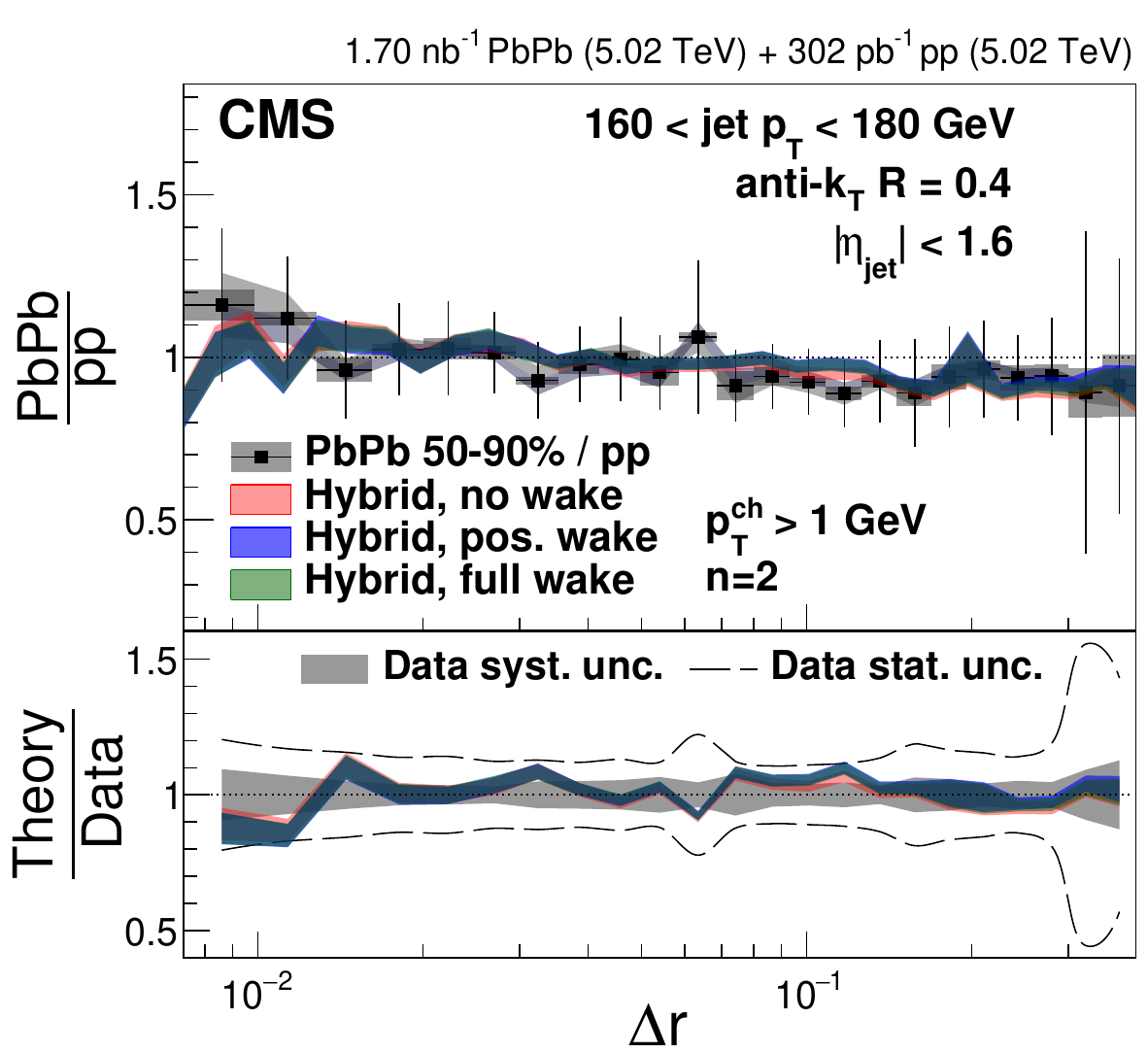}
    \includegraphics[width=0.4\textwidth]{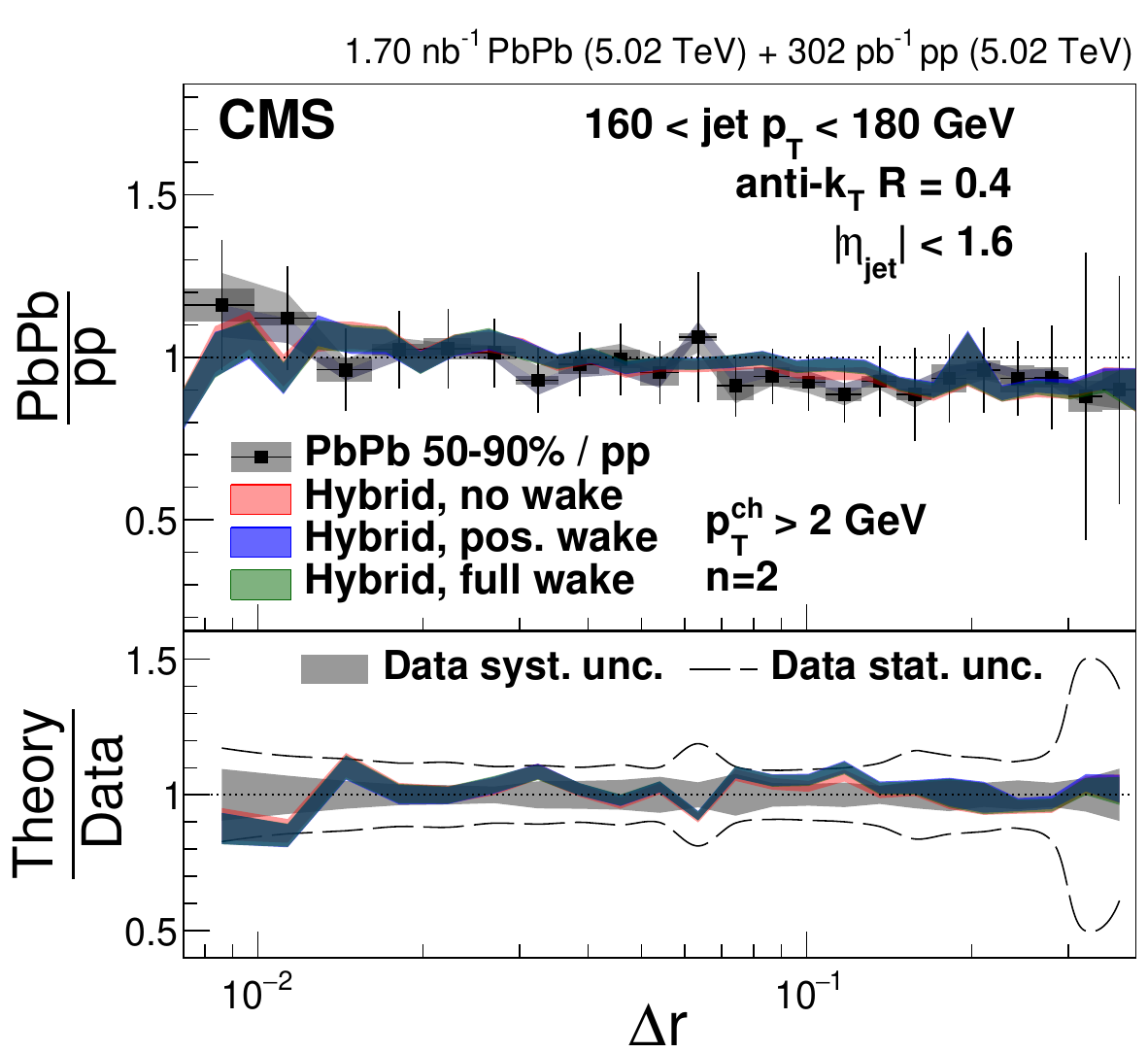}
    \includegraphics[width=0.4\textwidth]{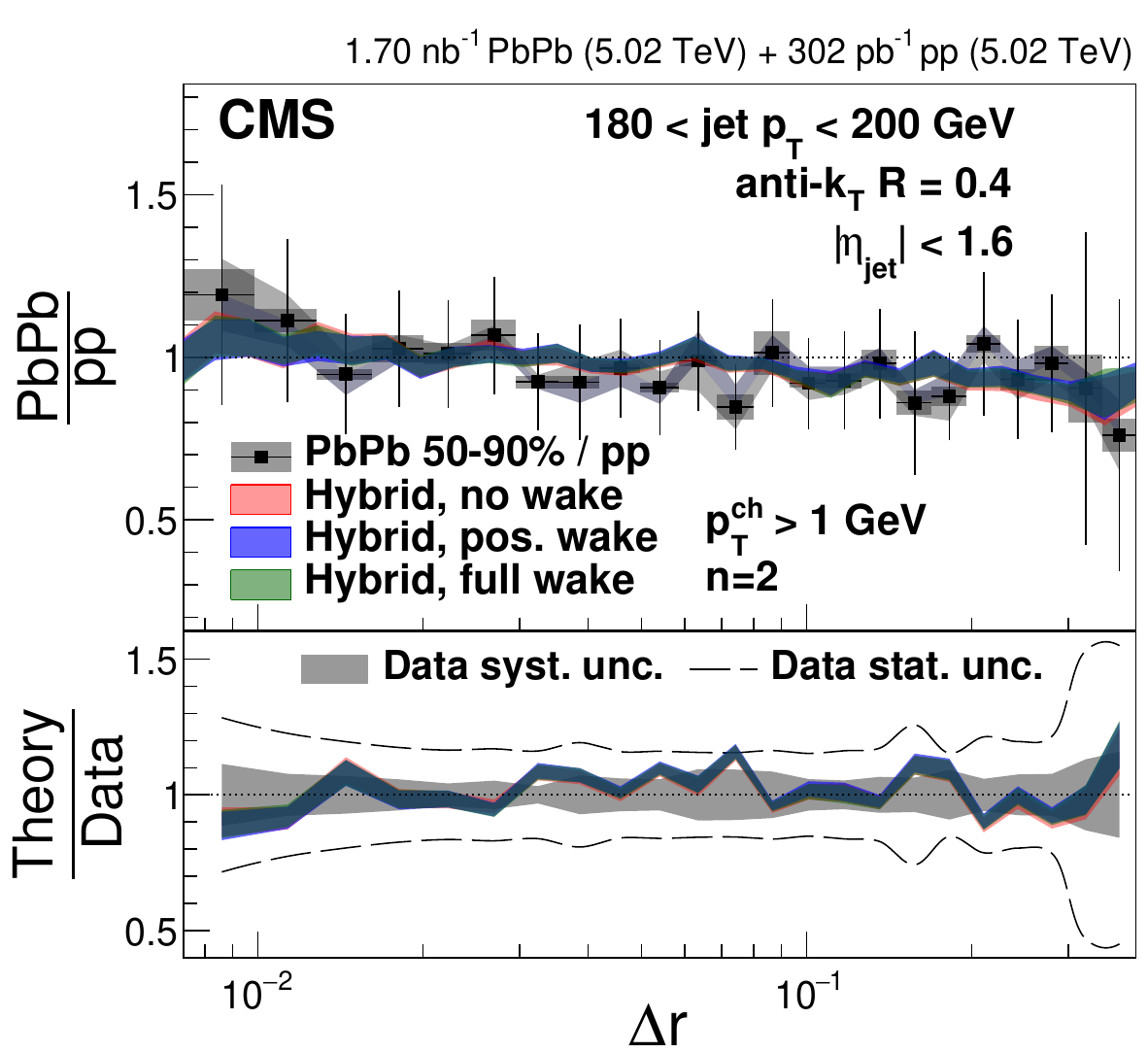}
    \includegraphics[width=0.4\textwidth]{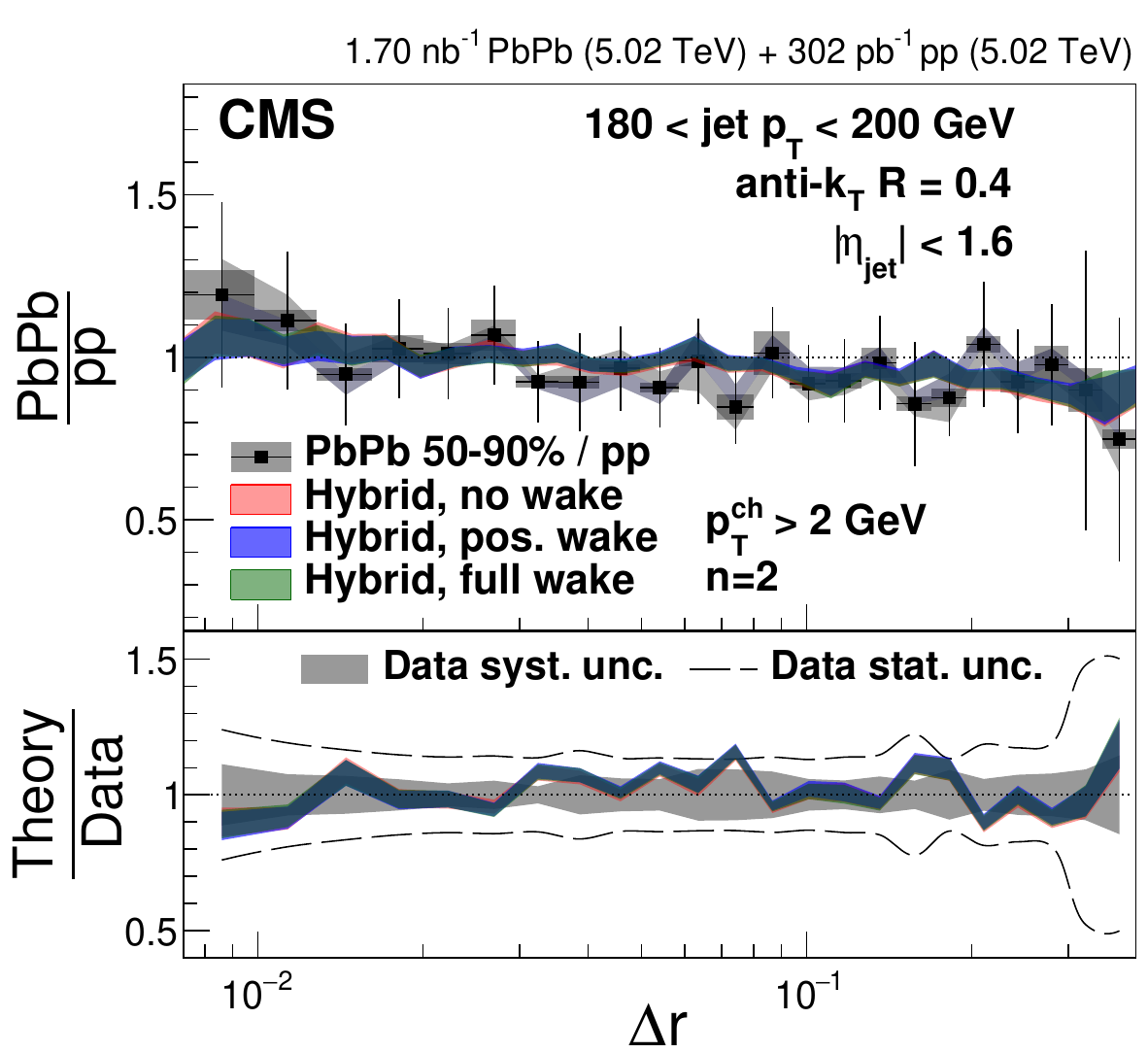}
    \caption{Ratios of 50--90\% central \PbPb to \pp energy-energy correlators with $n=2$ in different \ptCh and \ptjet bins compared to the hybrid model predictions with different jet wake settings.} 
    \label{fig:PbPbRatiosSquared50-90}
  } 
\end{figure*}

\begin{figure*}[h]
  \centering{
    \includegraphics[width=0.4\textwidth]{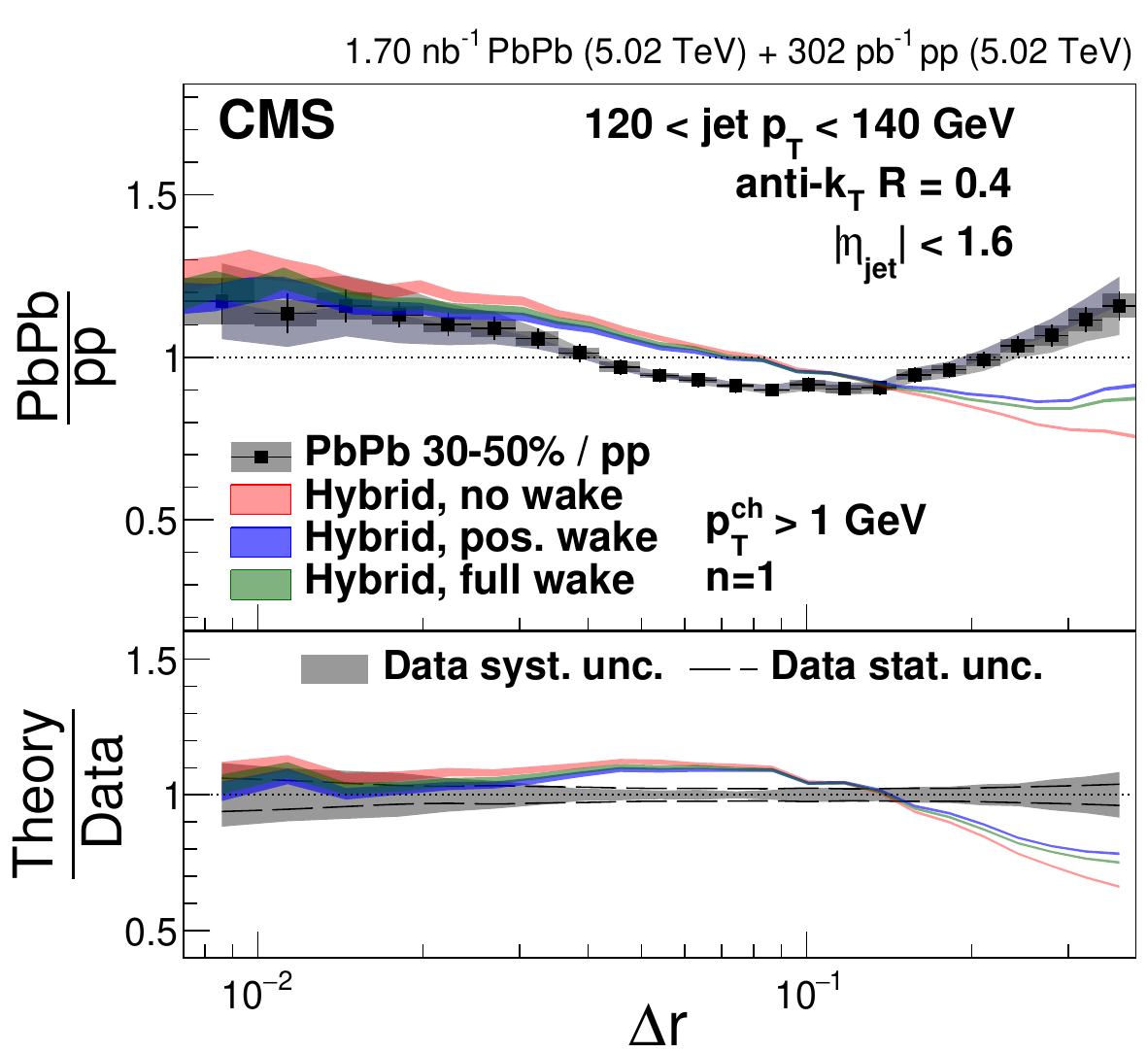} 
    \includegraphics[width=0.4\textwidth]{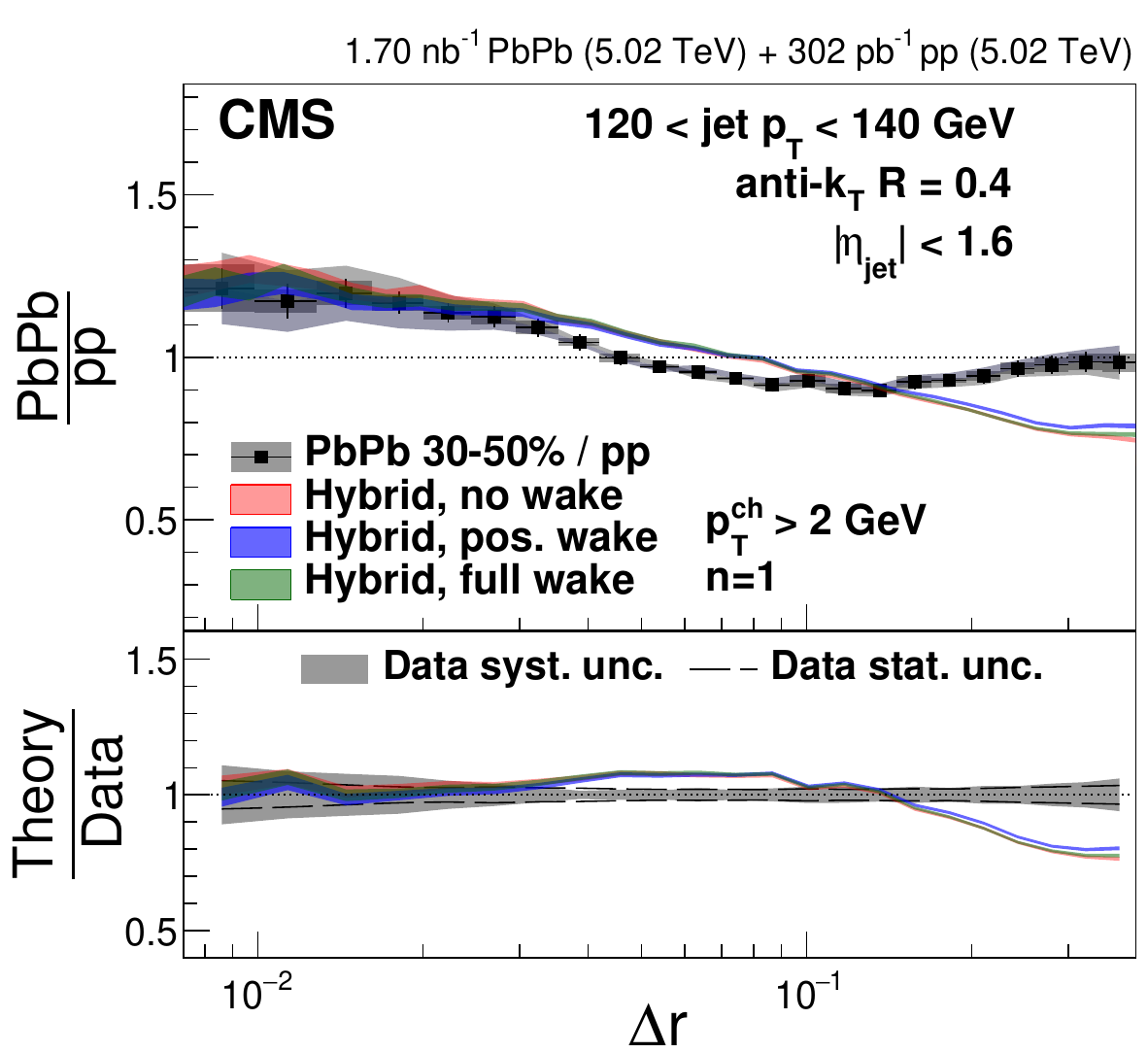} 
    \includegraphics[width=0.4\textwidth]{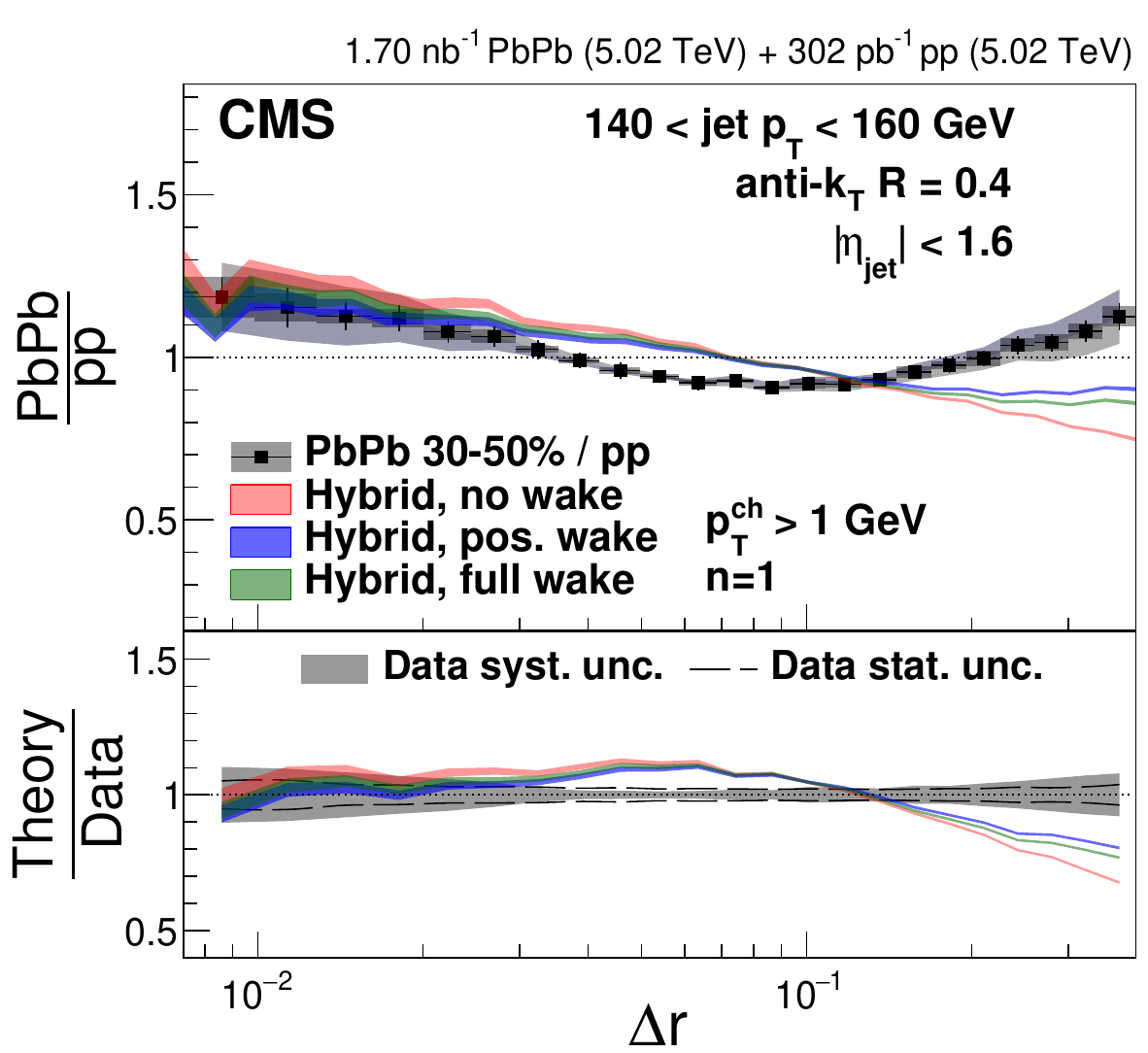}
    \includegraphics[width=0.4\textwidth]{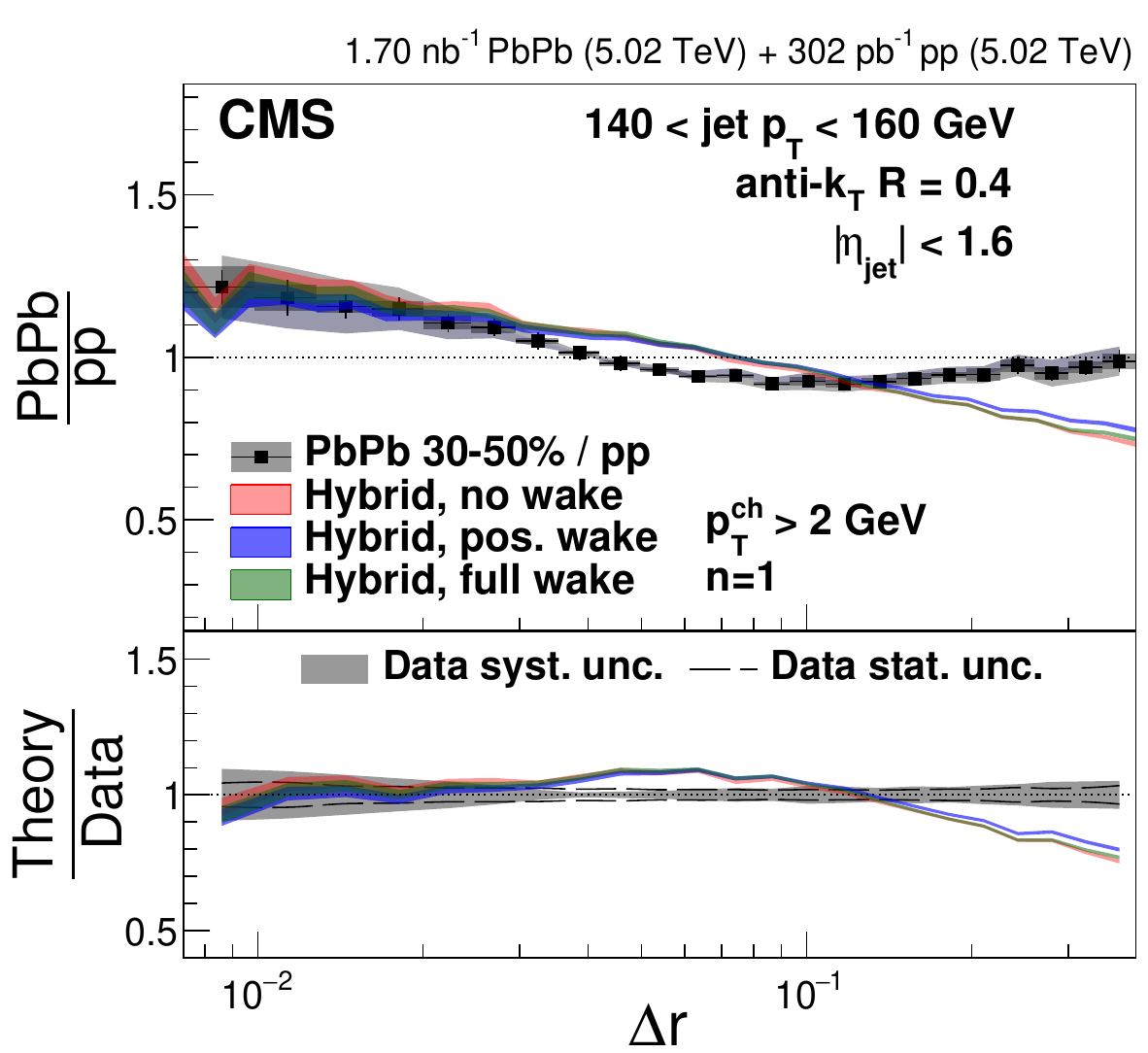}
    \includegraphics[width=0.4\textwidth]{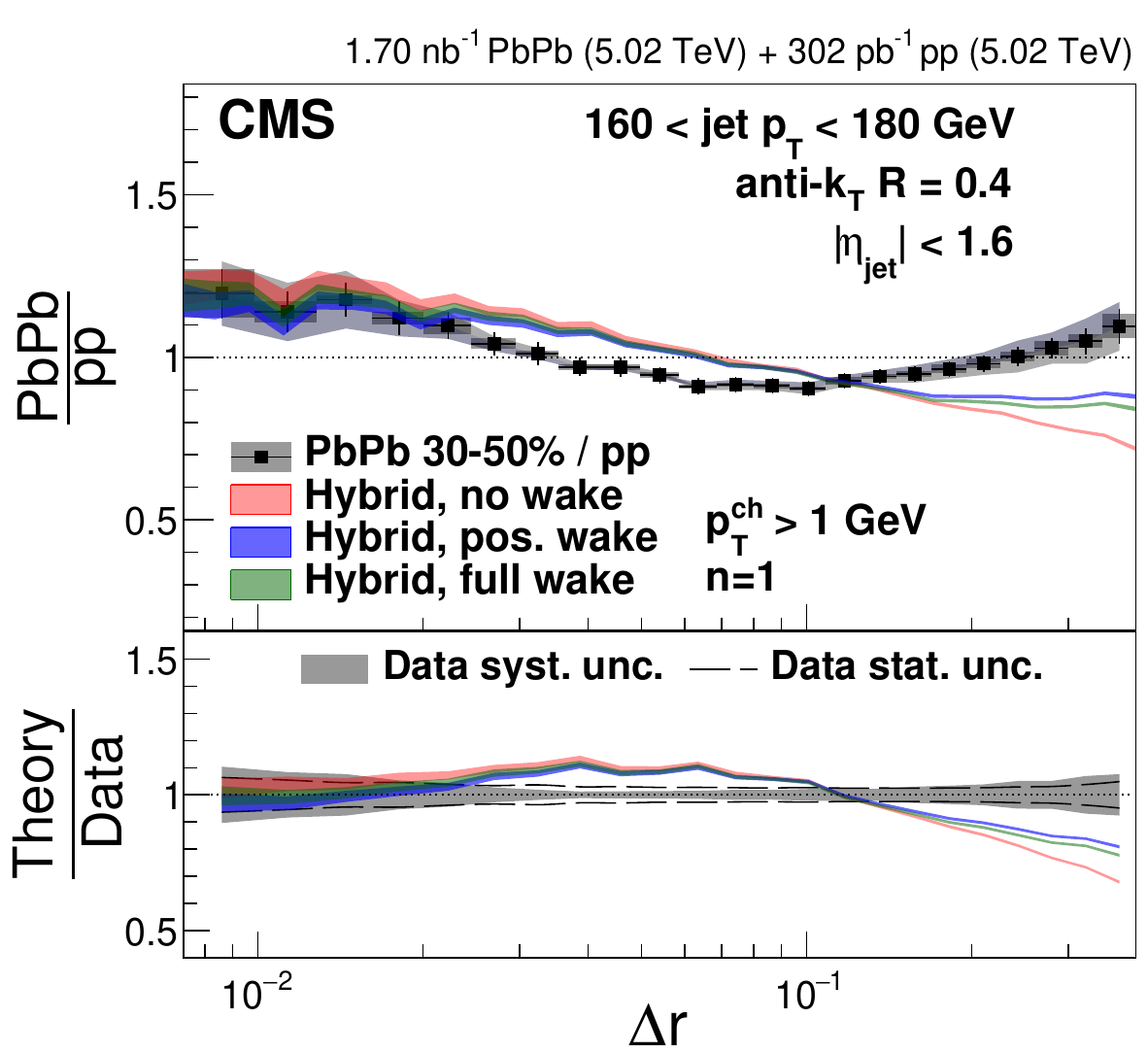}
    \includegraphics[width=0.4\textwidth]{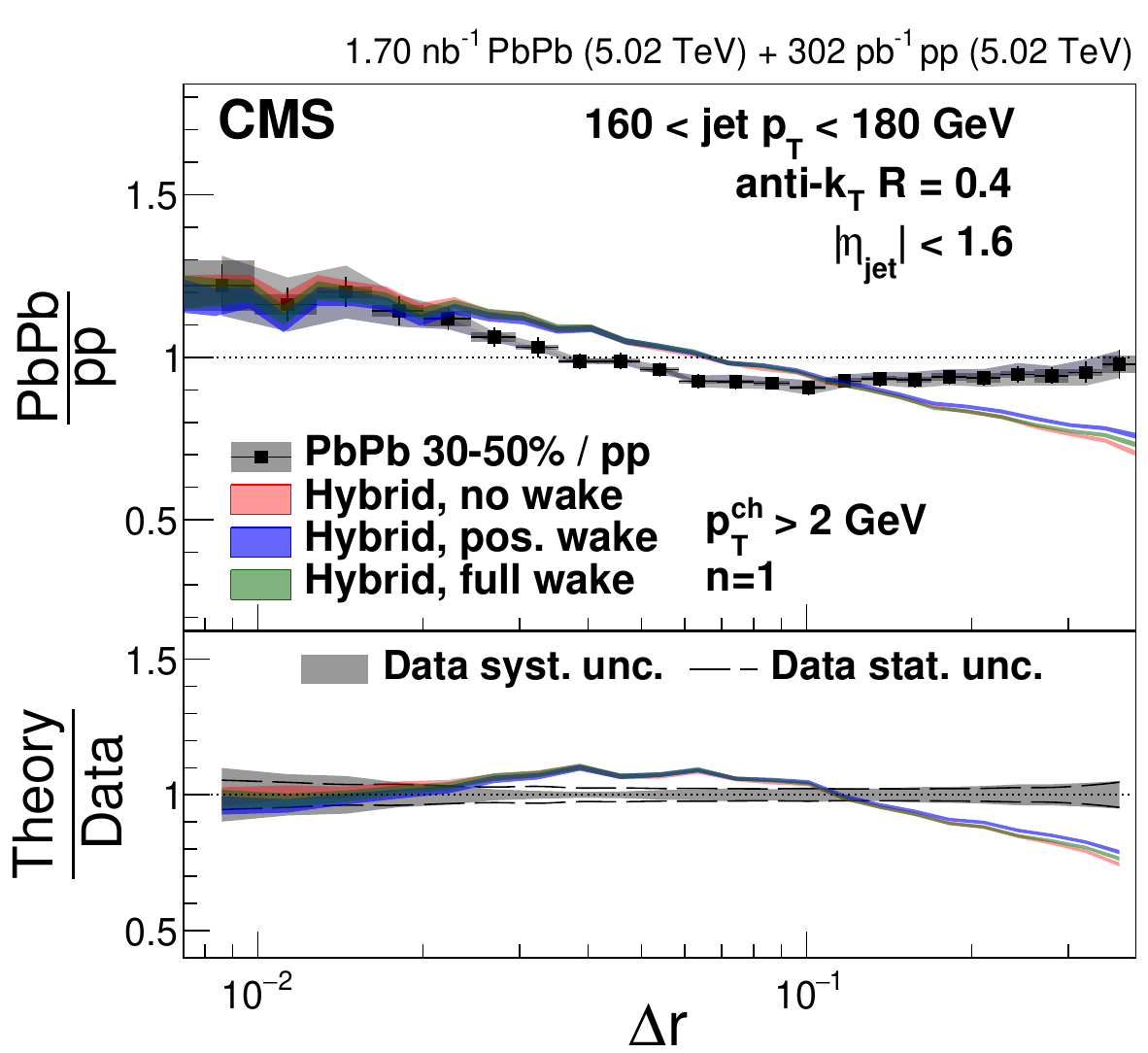}
    \includegraphics[width=0.4\textwidth]{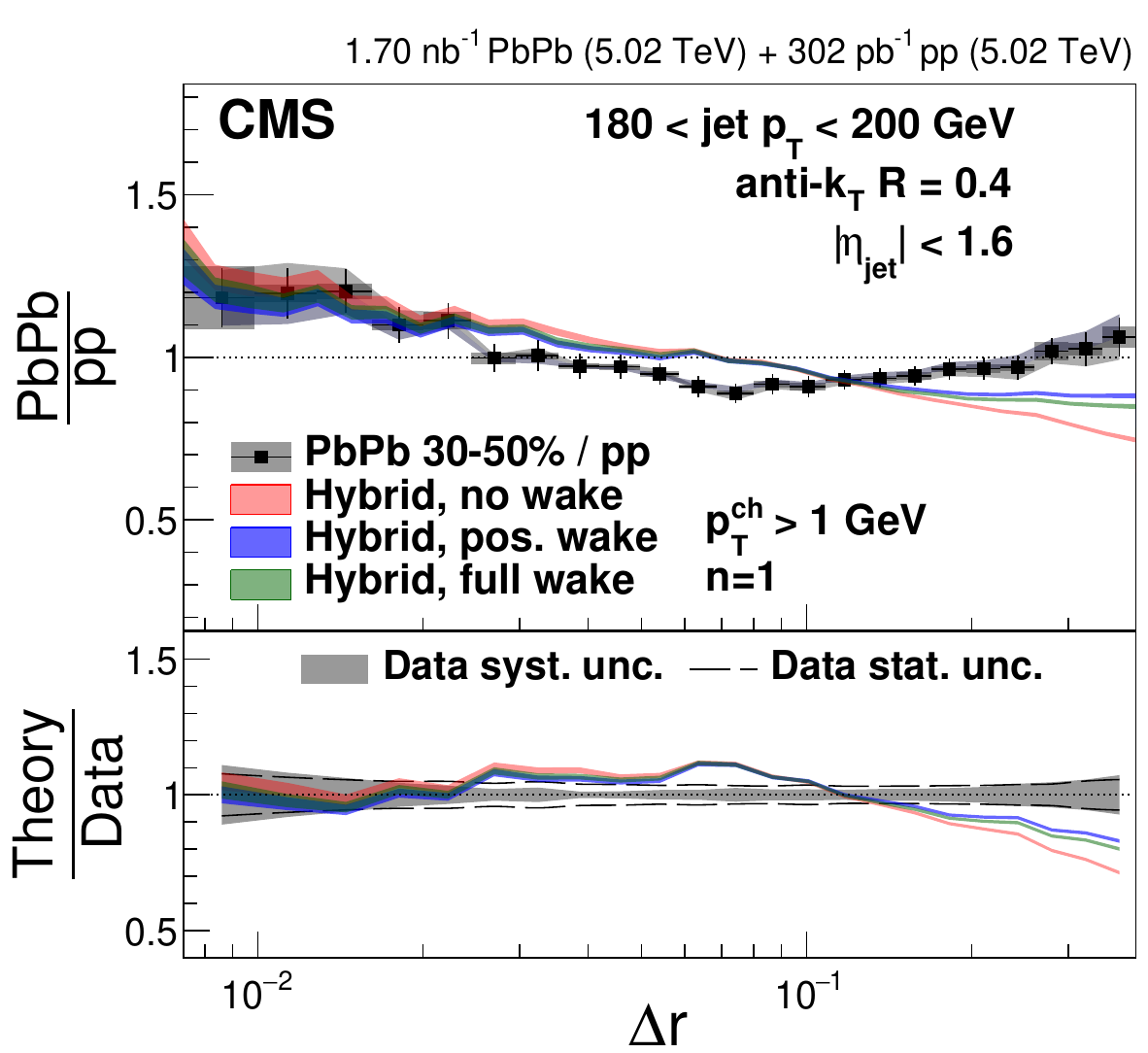}
    \includegraphics[width=0.4\textwidth]{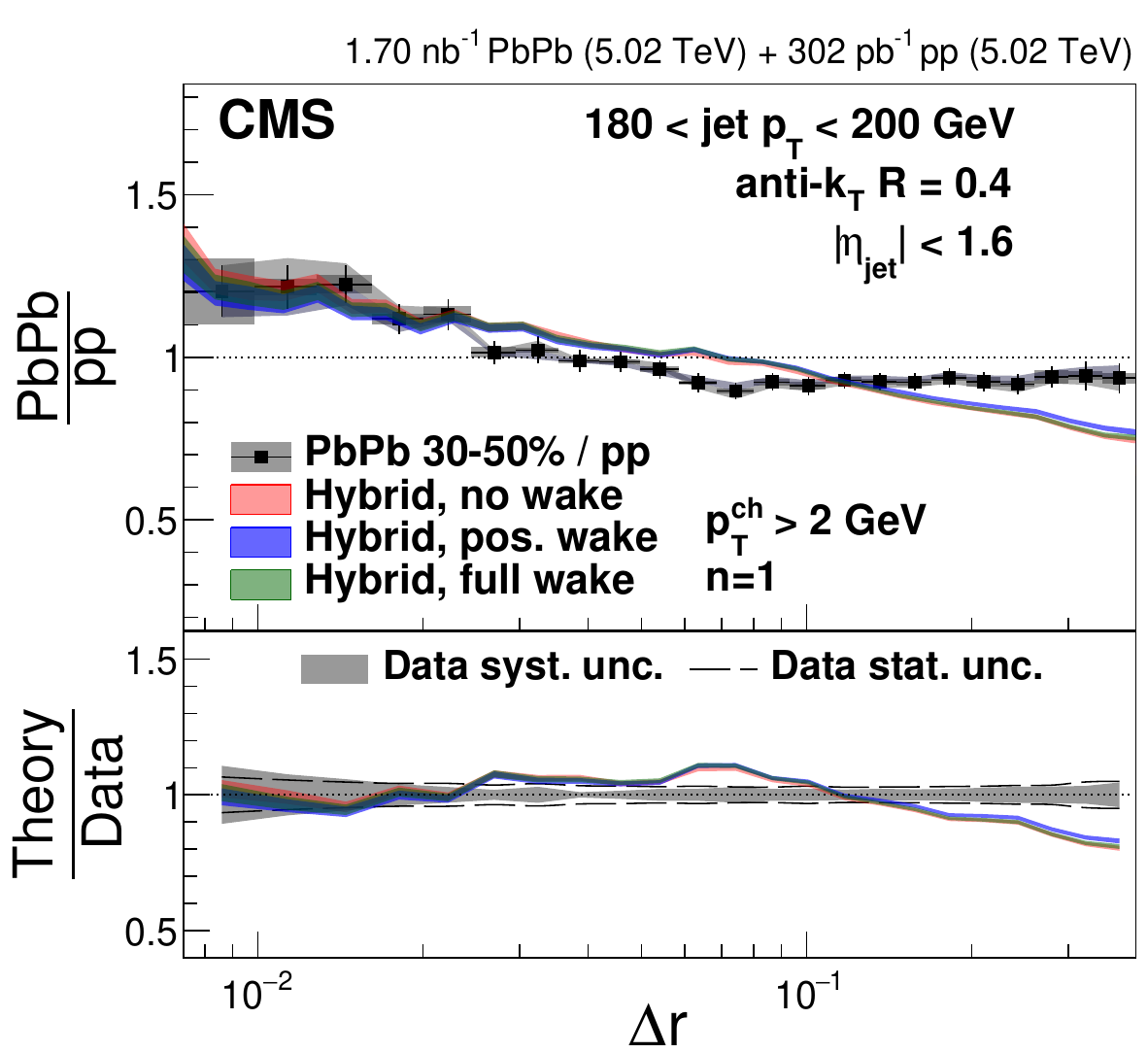}
    \caption{Ratios of 30--50\% central \PbPb to \pp energy-energy correlators with $n=1$ in different \ptCh and \ptjet bins compared to the hybrid model predictions with different jet wake settings.} 
    \label{fig:PbPbRatiosNominal30-50}
  } 
\end{figure*}

\begin{figure*}[h]
  \centering{
    \includegraphics[width=0.4\textwidth]{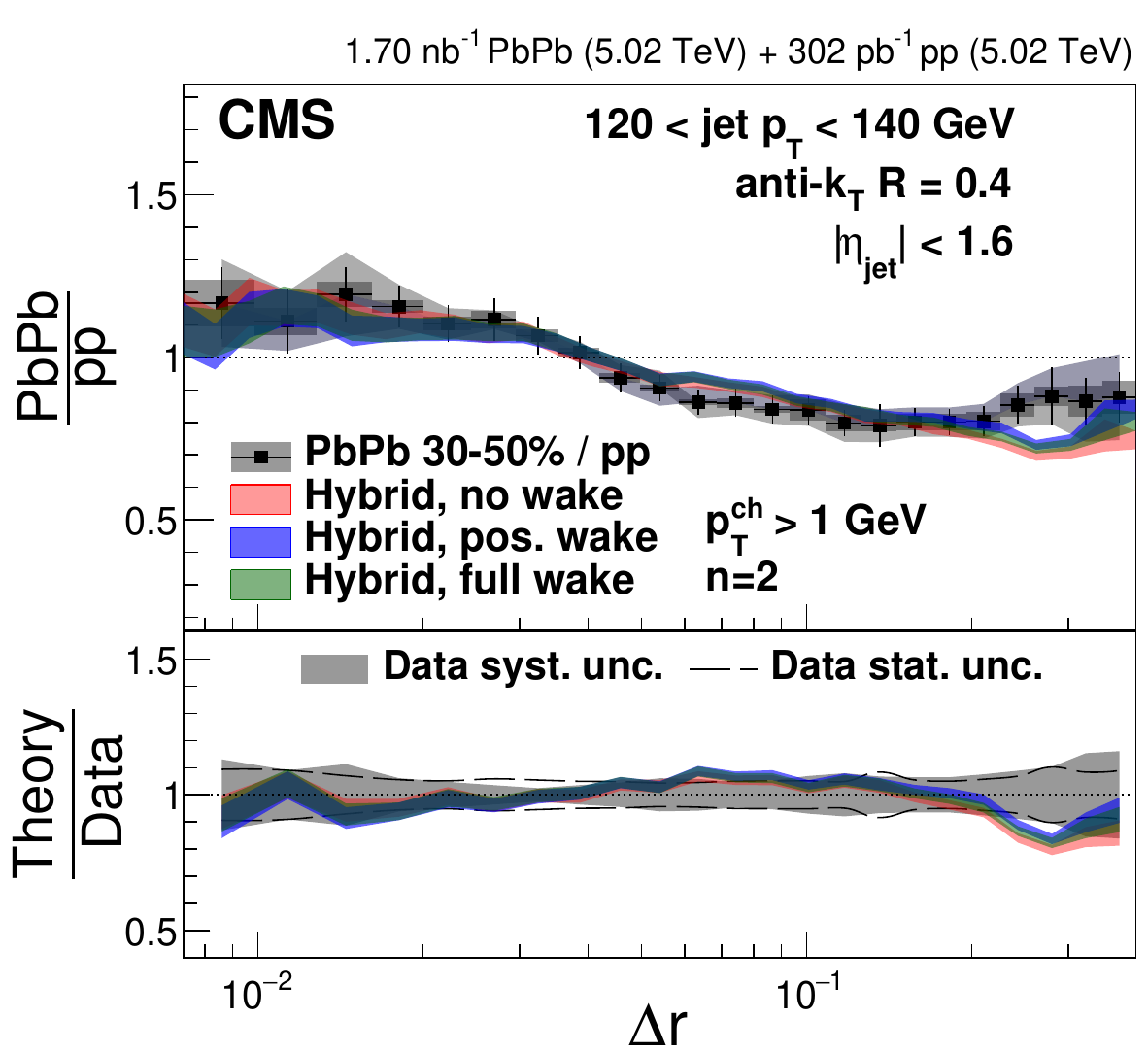}
    \includegraphics[width=0.4\textwidth]{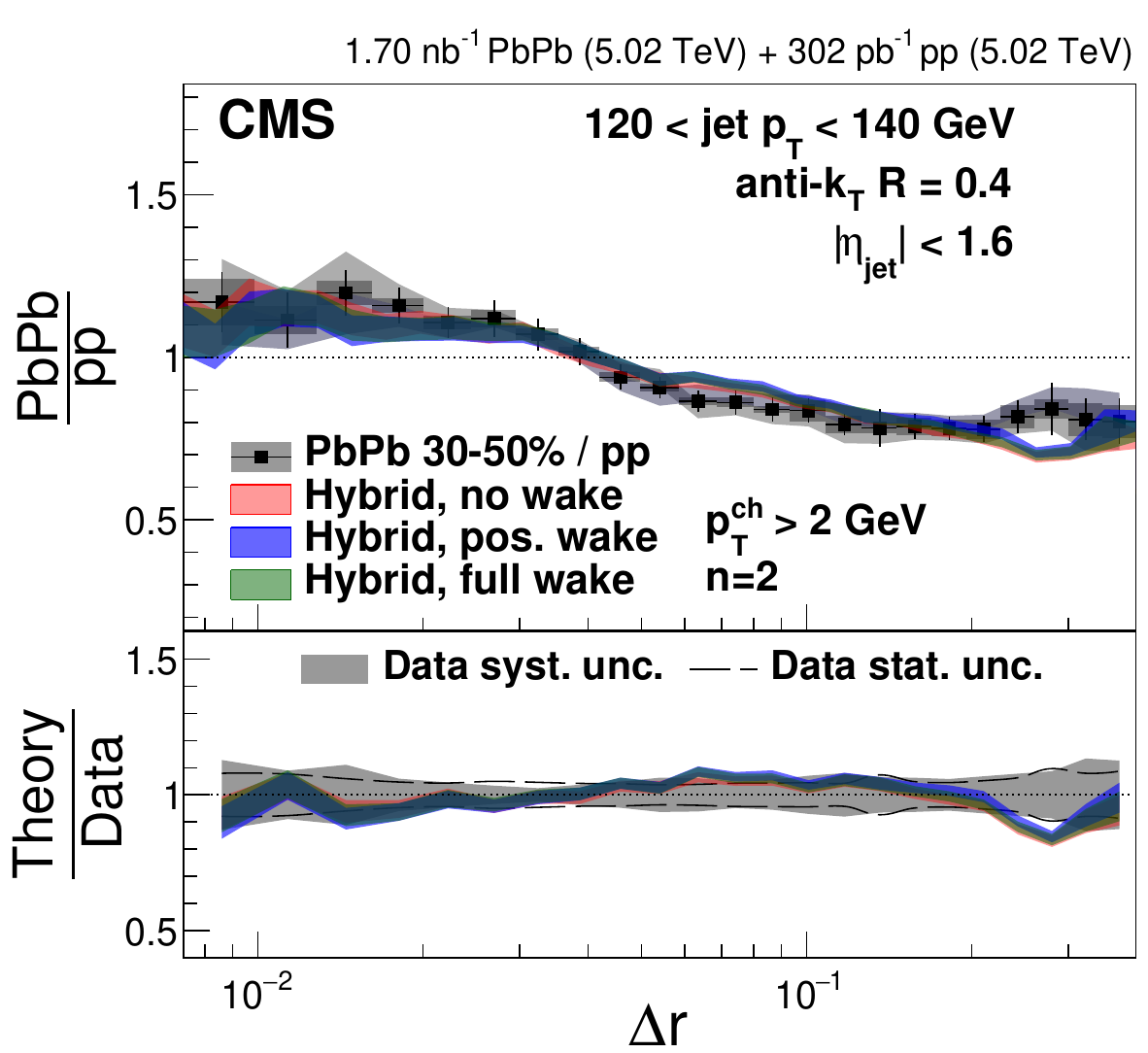} 
    \includegraphics[width=0.4\textwidth]{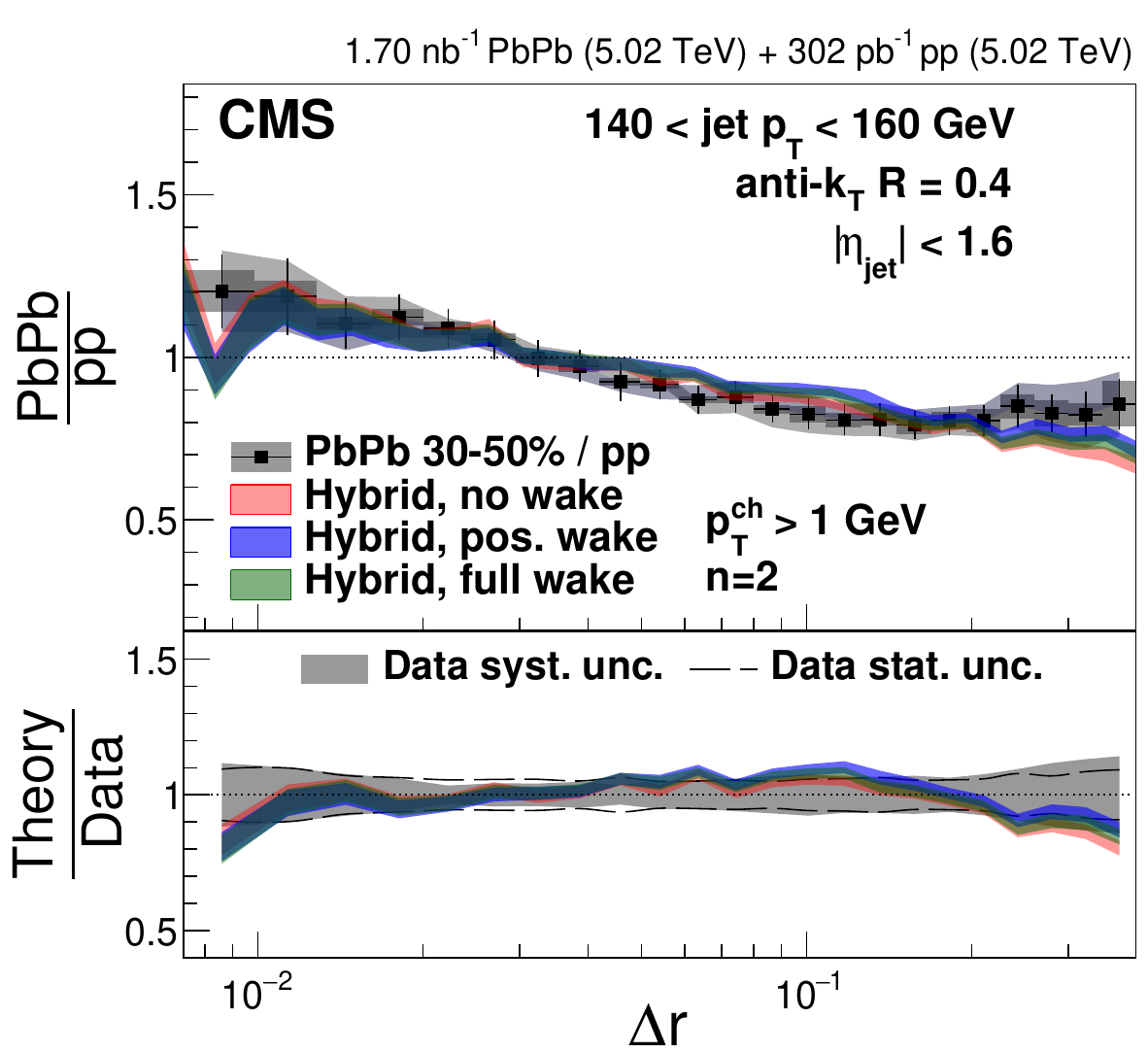}
    \includegraphics[width=0.4\textwidth]{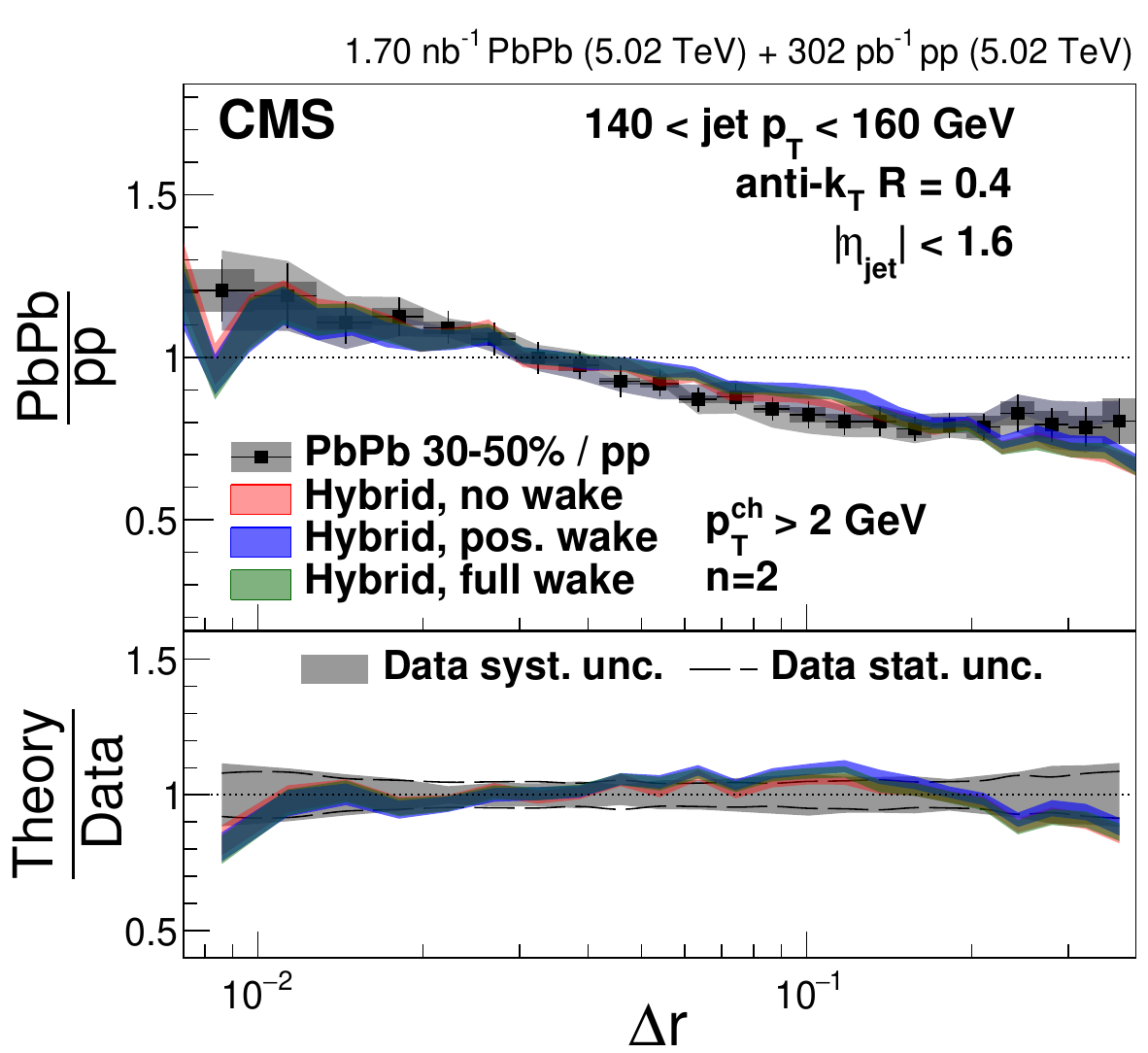}
    \includegraphics[width=0.4\textwidth]{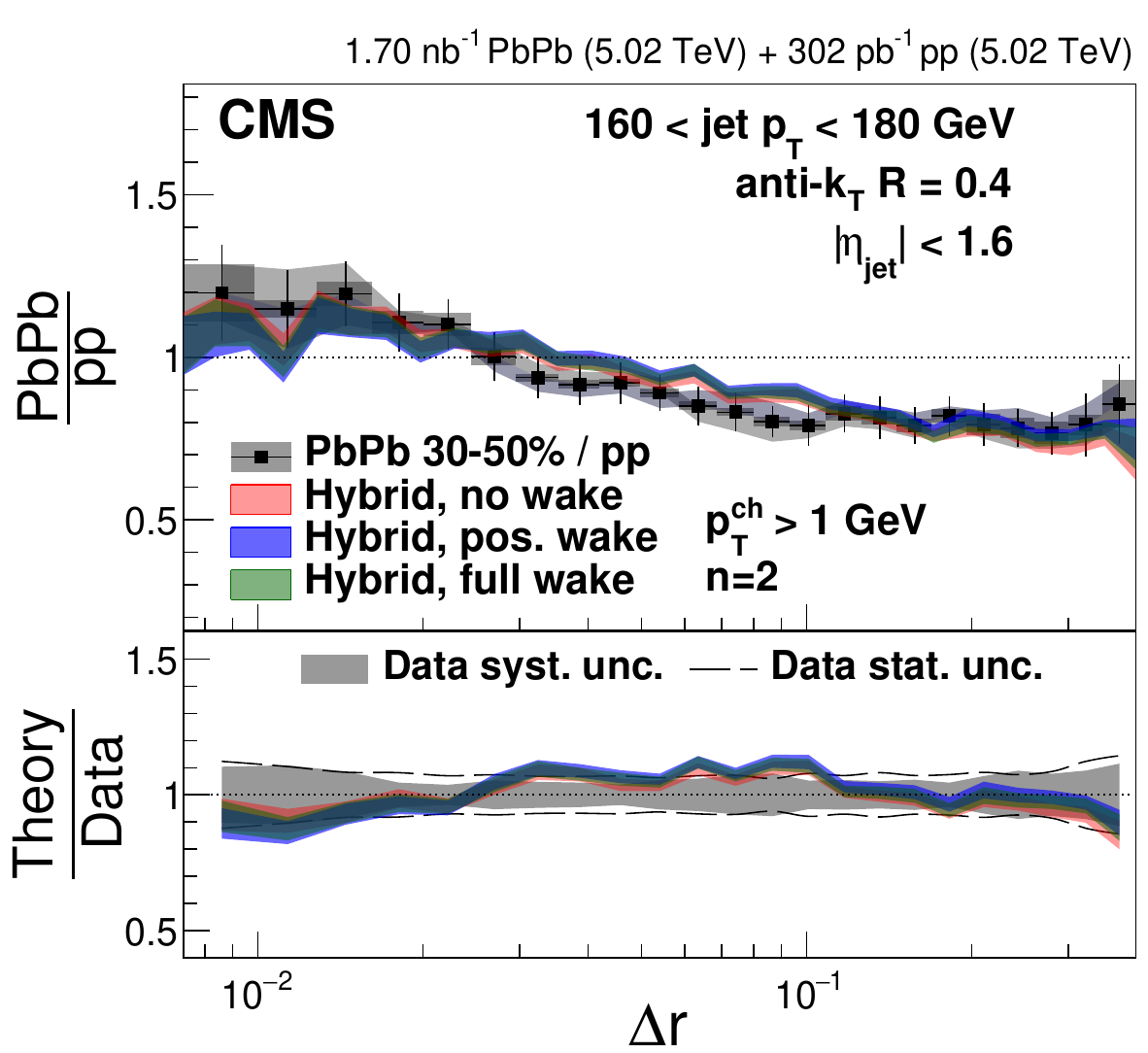}
    \includegraphics[width=0.4\textwidth]{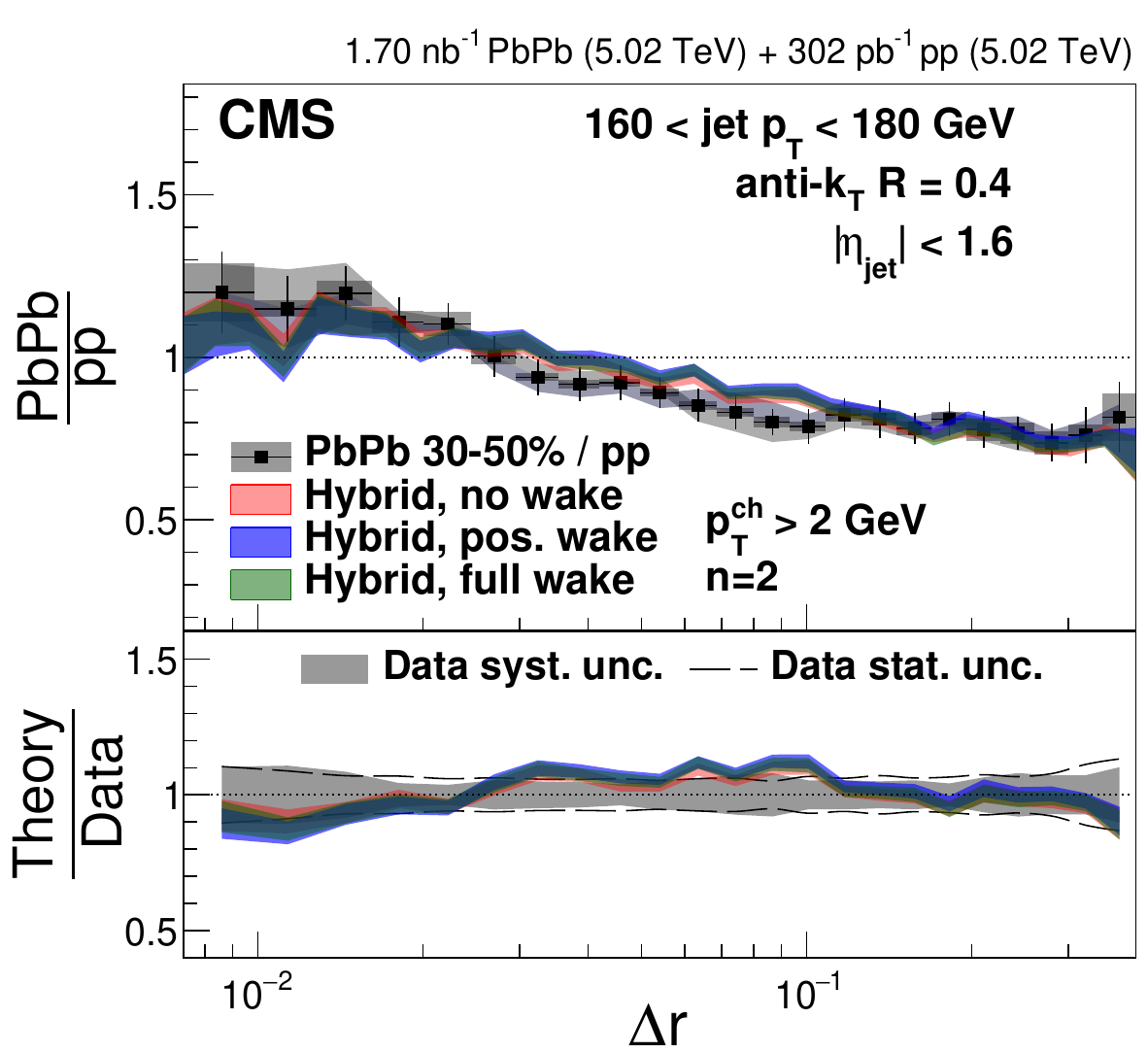}
    \includegraphics[width=0.4\textwidth]{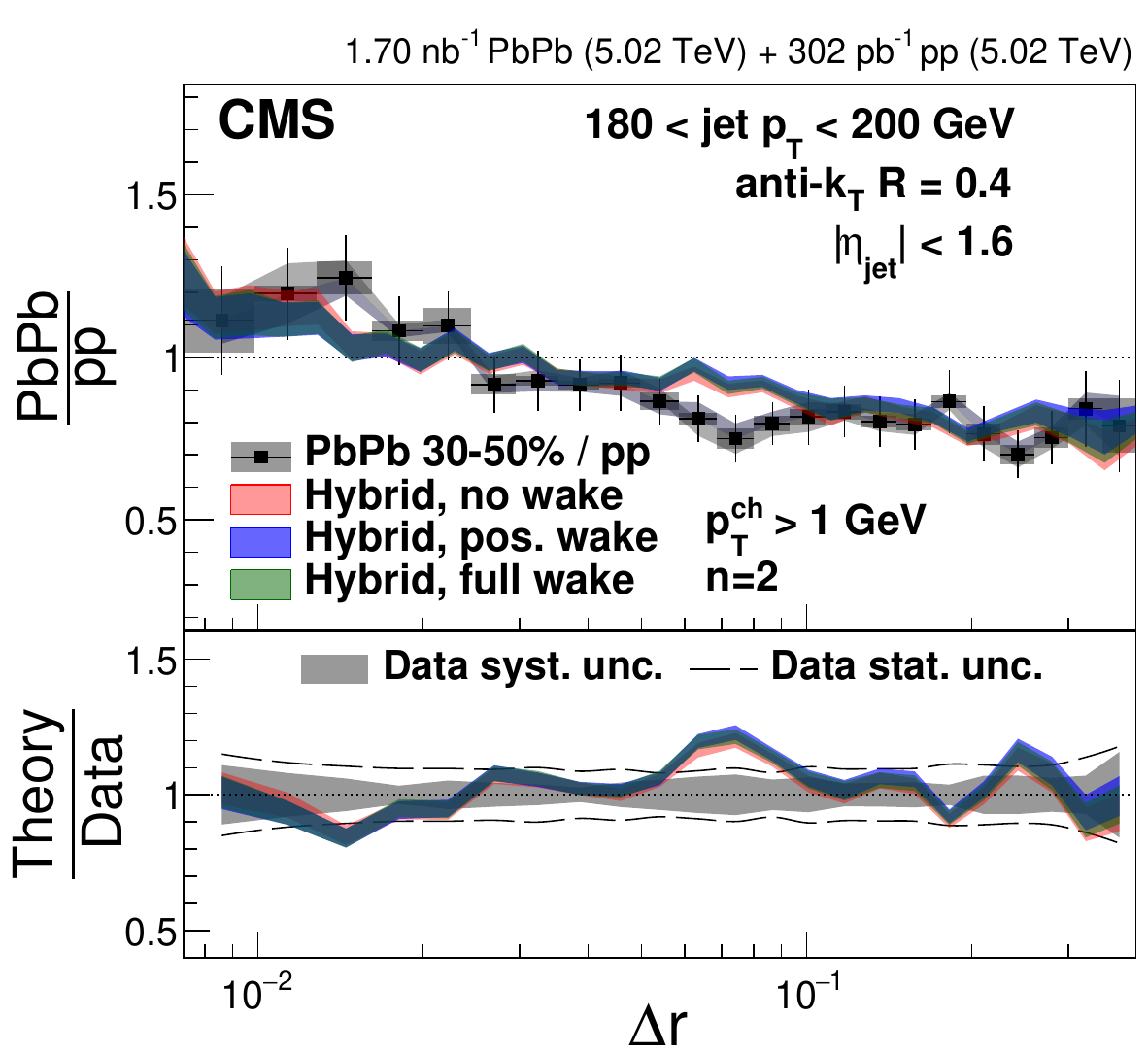}
    \includegraphics[width=0.4\textwidth]{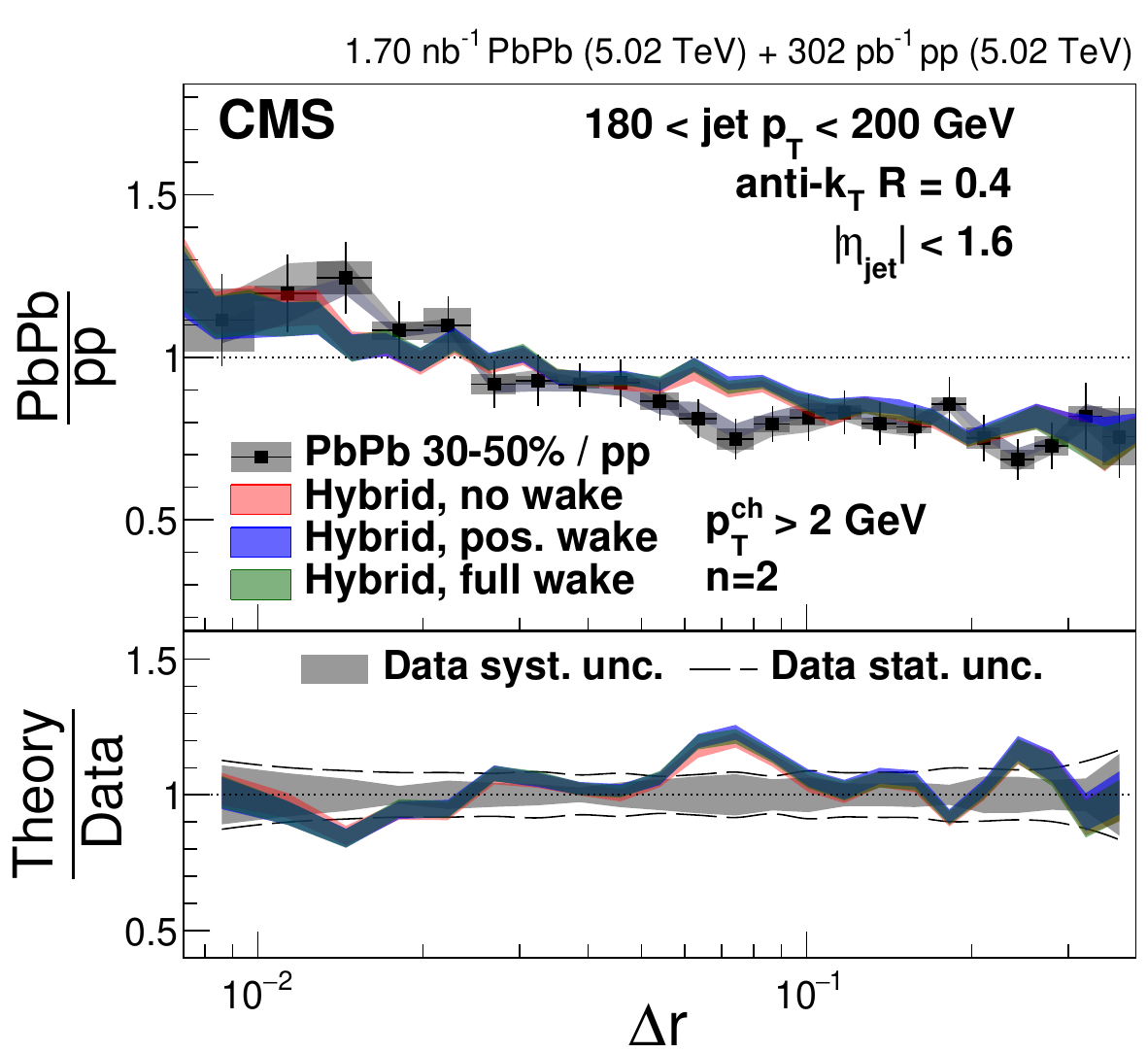}
    \caption{Ratios of 30--50\% central \PbPb to \pp energy-energy correlators with $n=2$ in different \ptCh and \ptjet bins compared to the hybrid model predictions with different jet wake settings.} 
    \label{fig:PbPbRatiosSquared30-50}
  } 
\end{figure*}

\begin{figure*}[h]
  \centering{
    \includegraphics[width=0.4\textwidth]{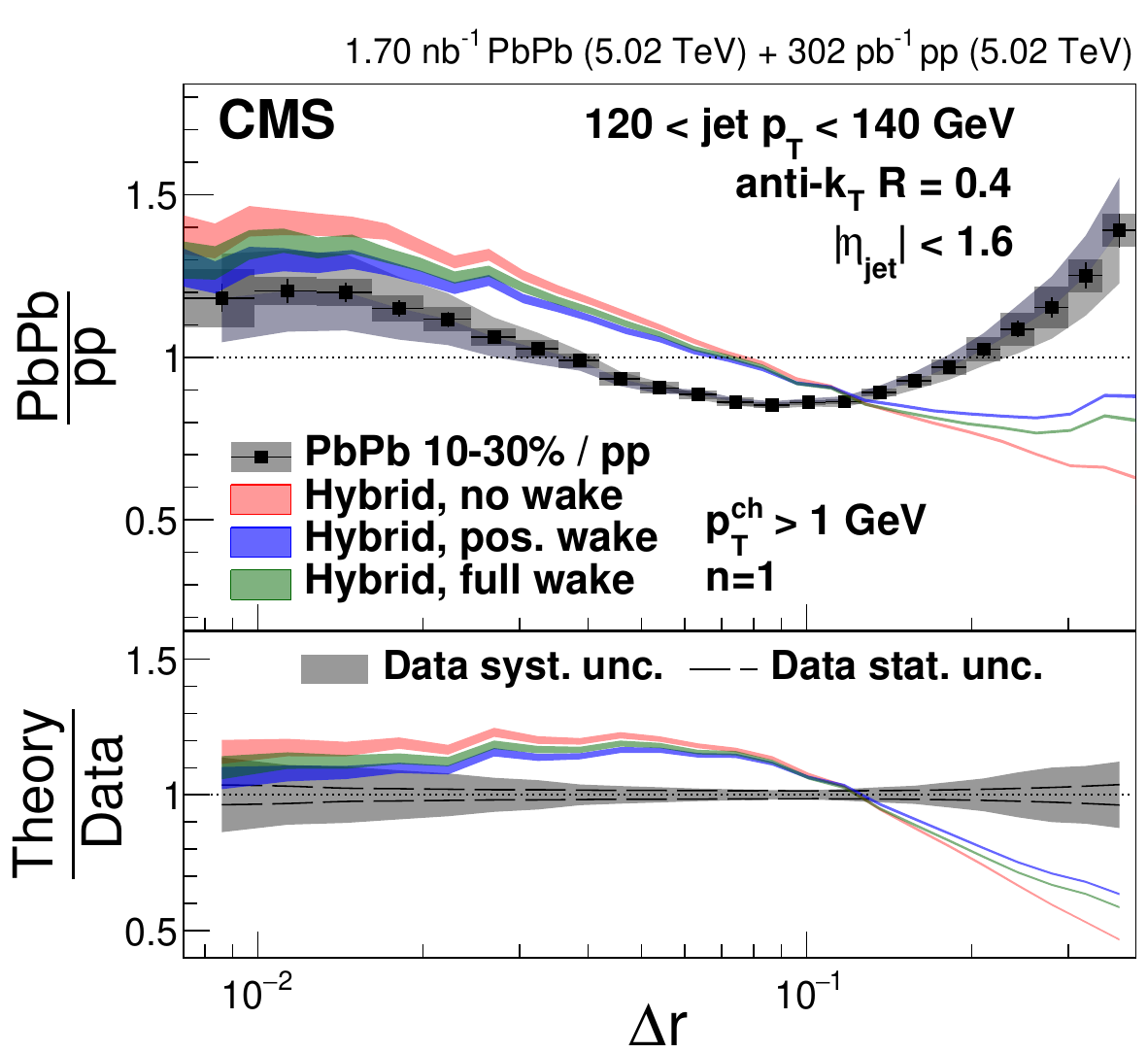}
    \includegraphics[width=0.4\textwidth]{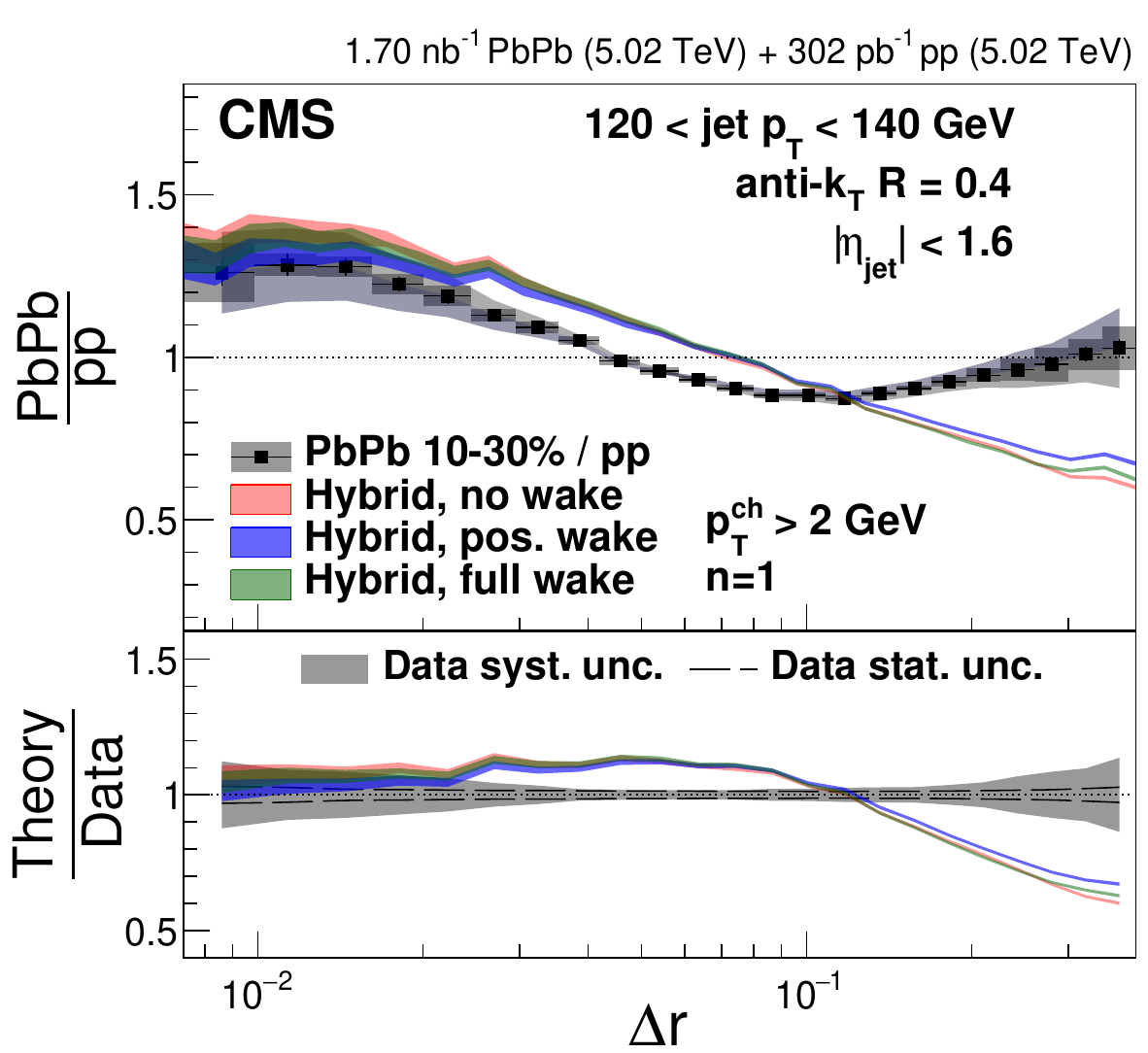} 
    \includegraphics[width=0.4\textwidth]{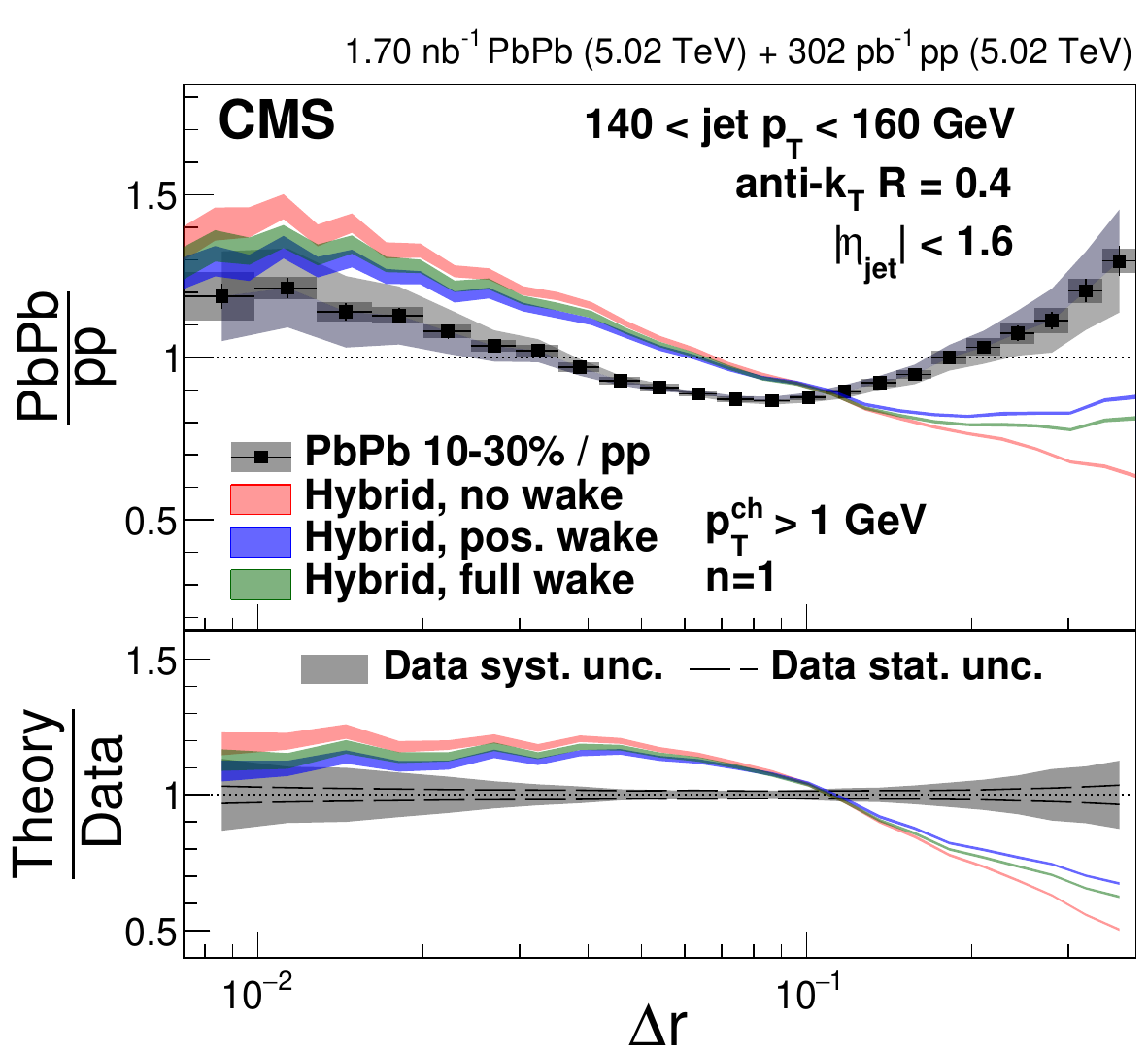}
    \includegraphics[width=0.4\textwidth]{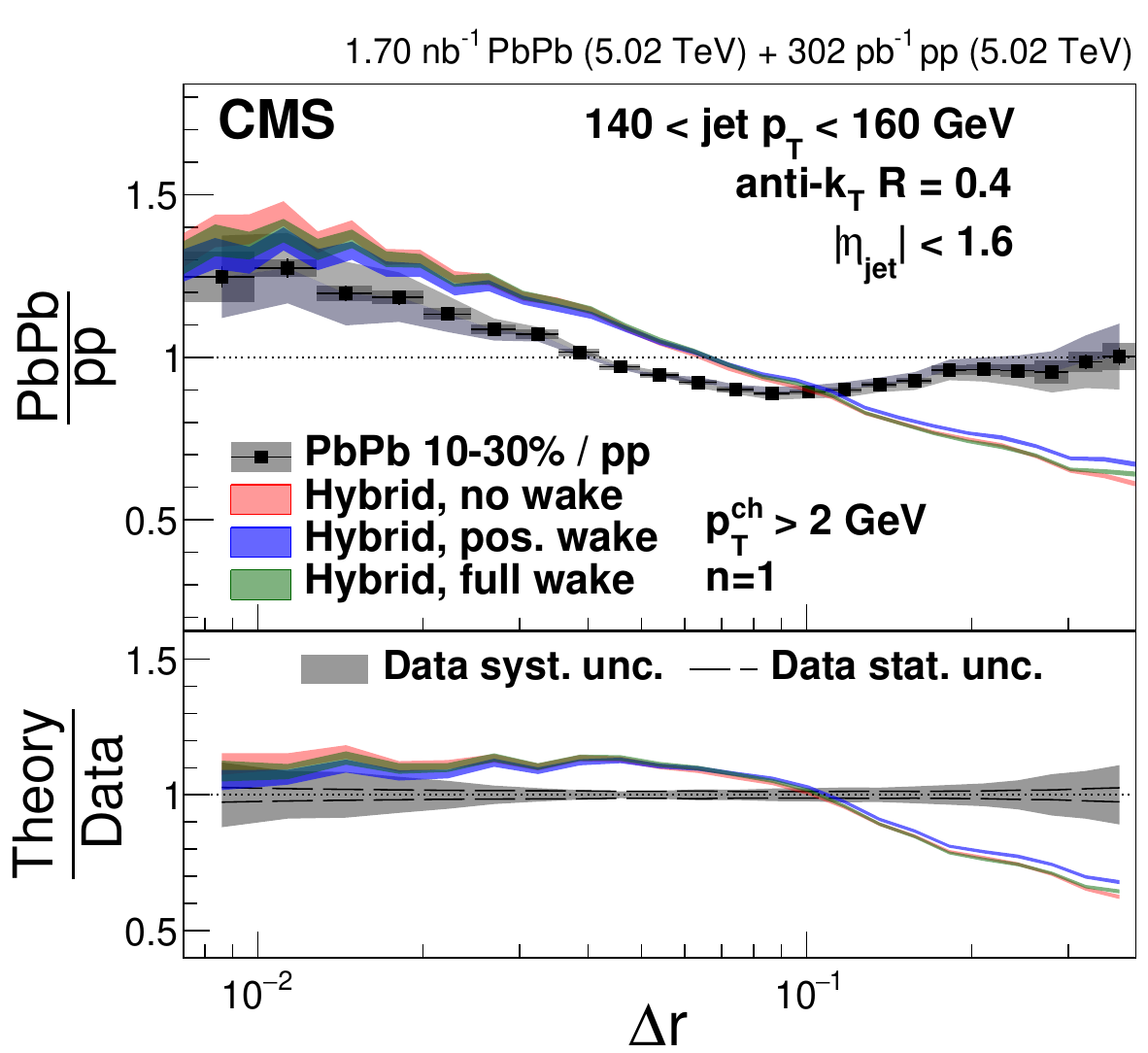}
    \includegraphics[width=0.4\textwidth]{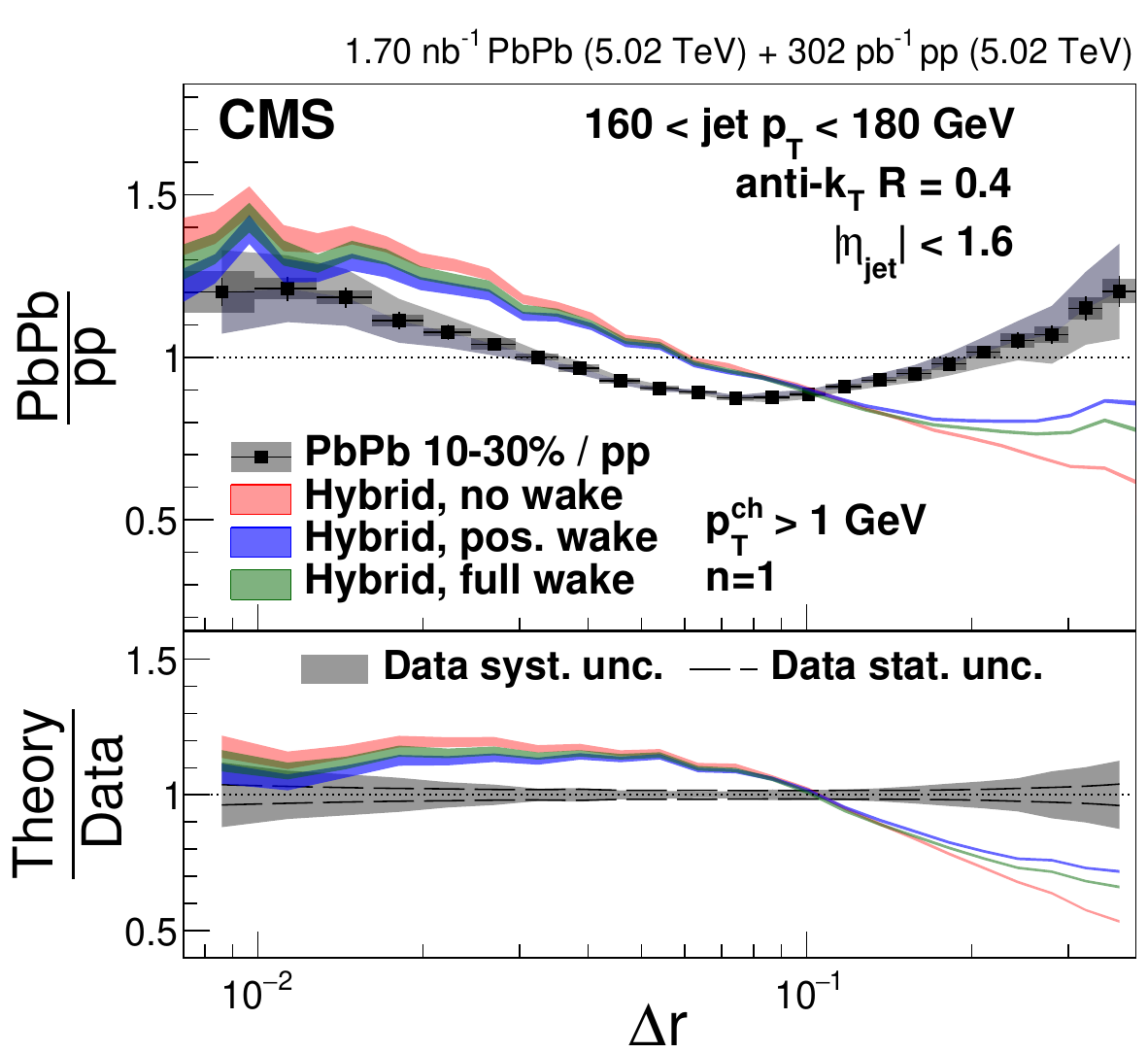}
    \includegraphics[width=0.4\textwidth]{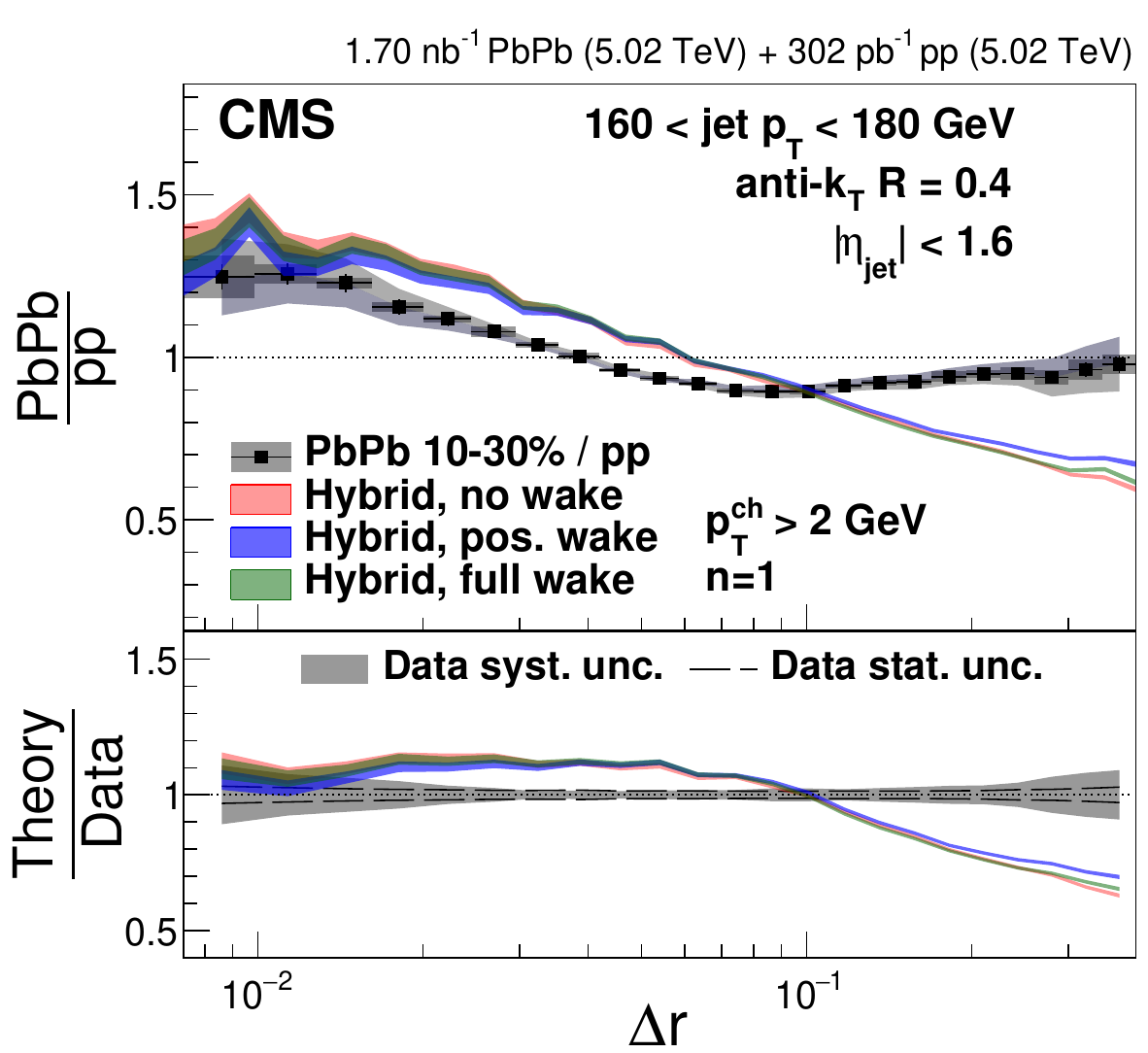}
    \includegraphics[width=0.4\textwidth]{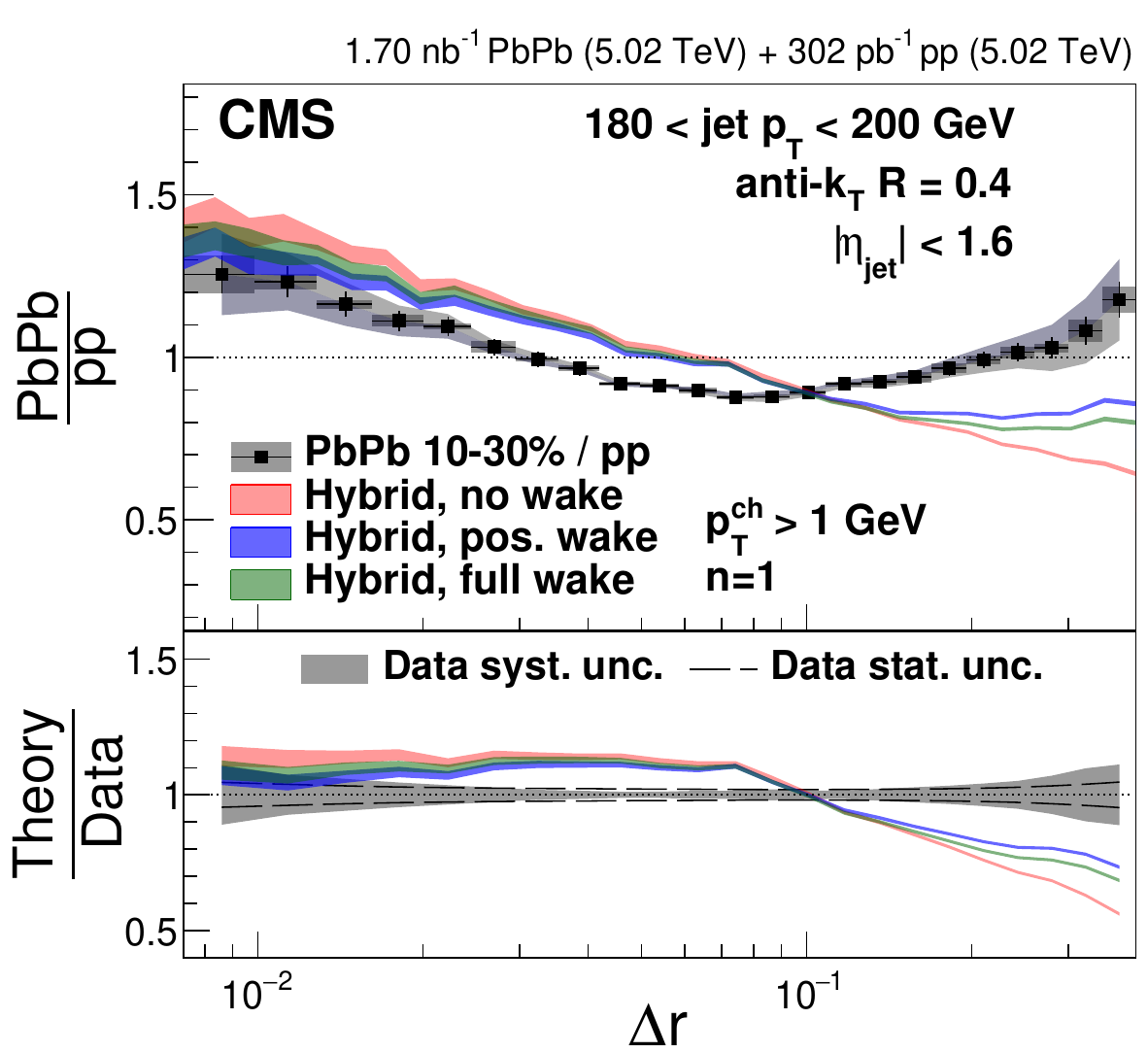}
    \includegraphics[width=0.4\textwidth]{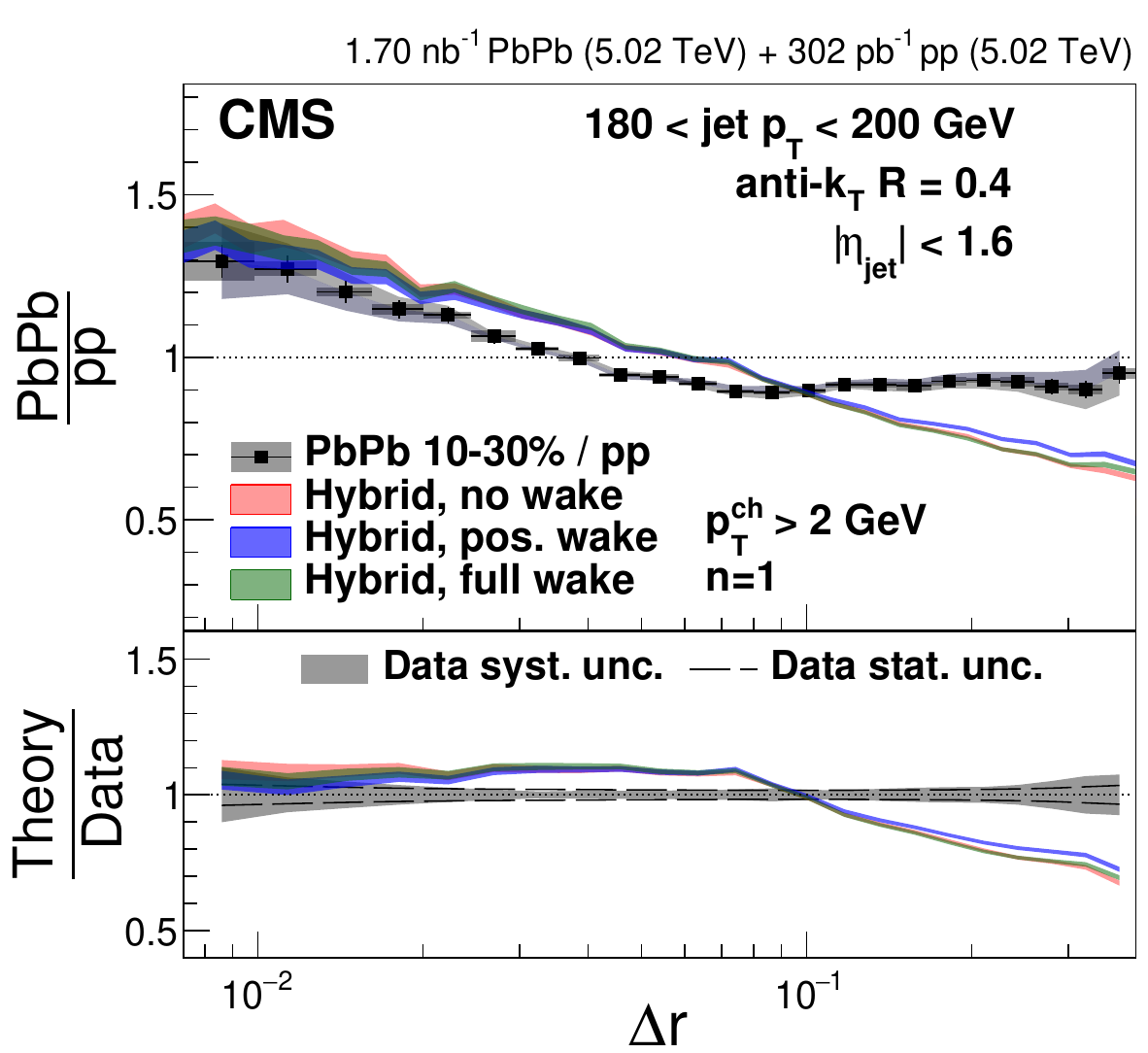}
    \caption{Ratios of 10--30\% central \PbPb to \pp energy-energy correlators with $n=1$ in different \ptCh and \ptjet bins compared to the hybrid model predictions with different jet wake settings.} 
    \label{fig:PbPbRatiosNominal10-30}
  } 
\end{figure*}

\begin{figure*}[h]
  \centering{
    \includegraphics[width=0.4\textwidth]{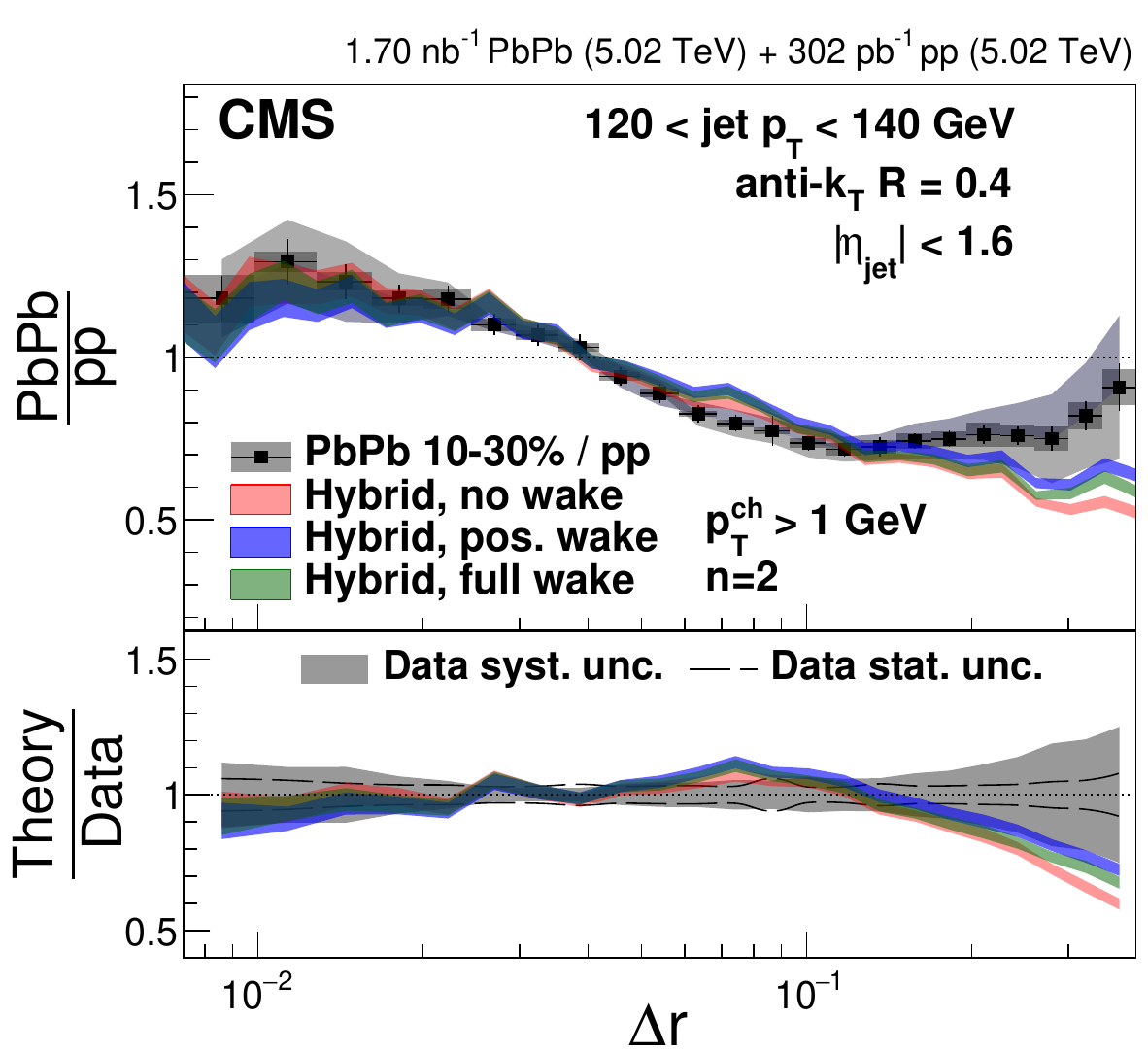}
    \includegraphics[width=0.4\textwidth]{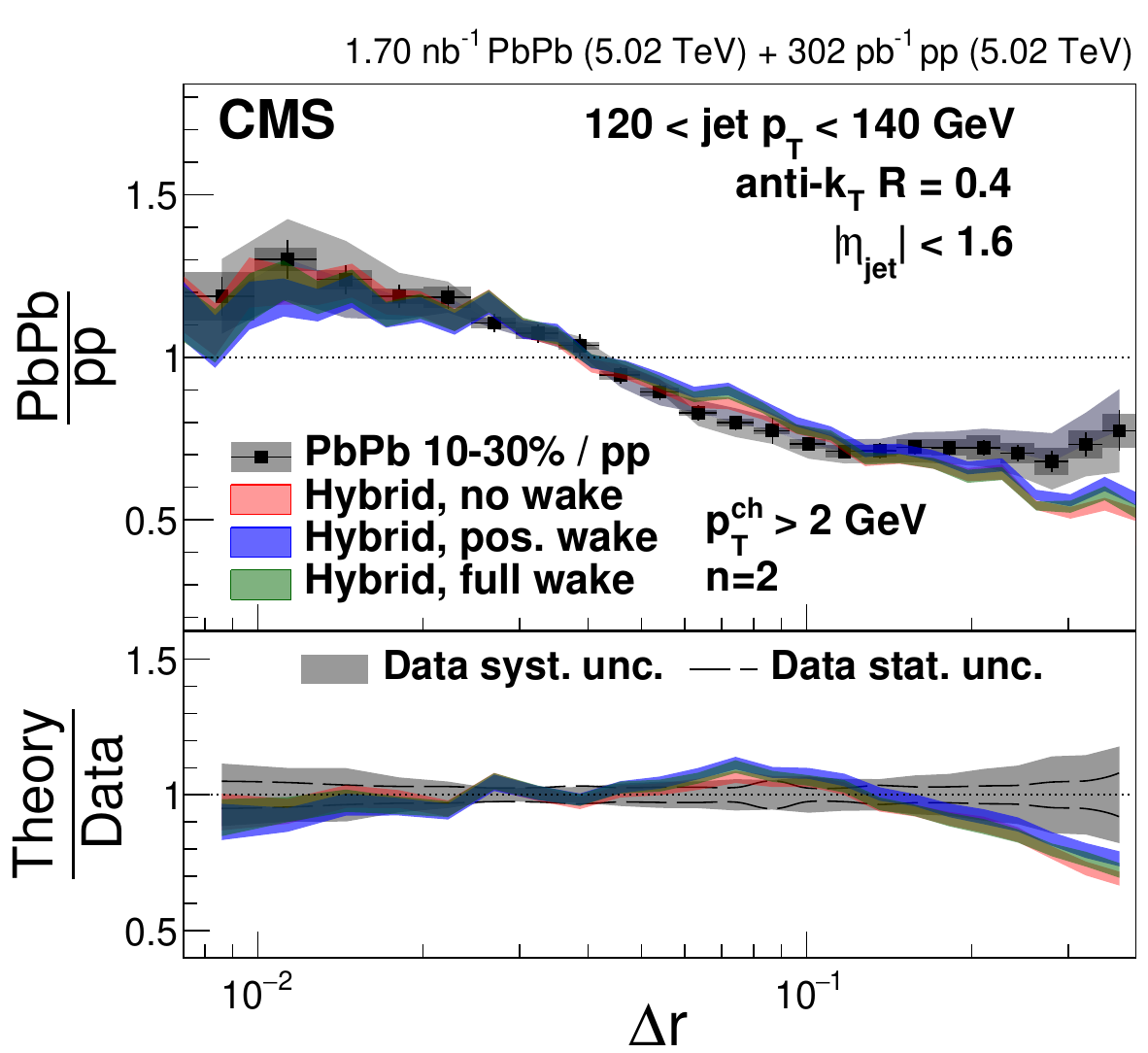} 
    \includegraphics[width=0.4\textwidth]{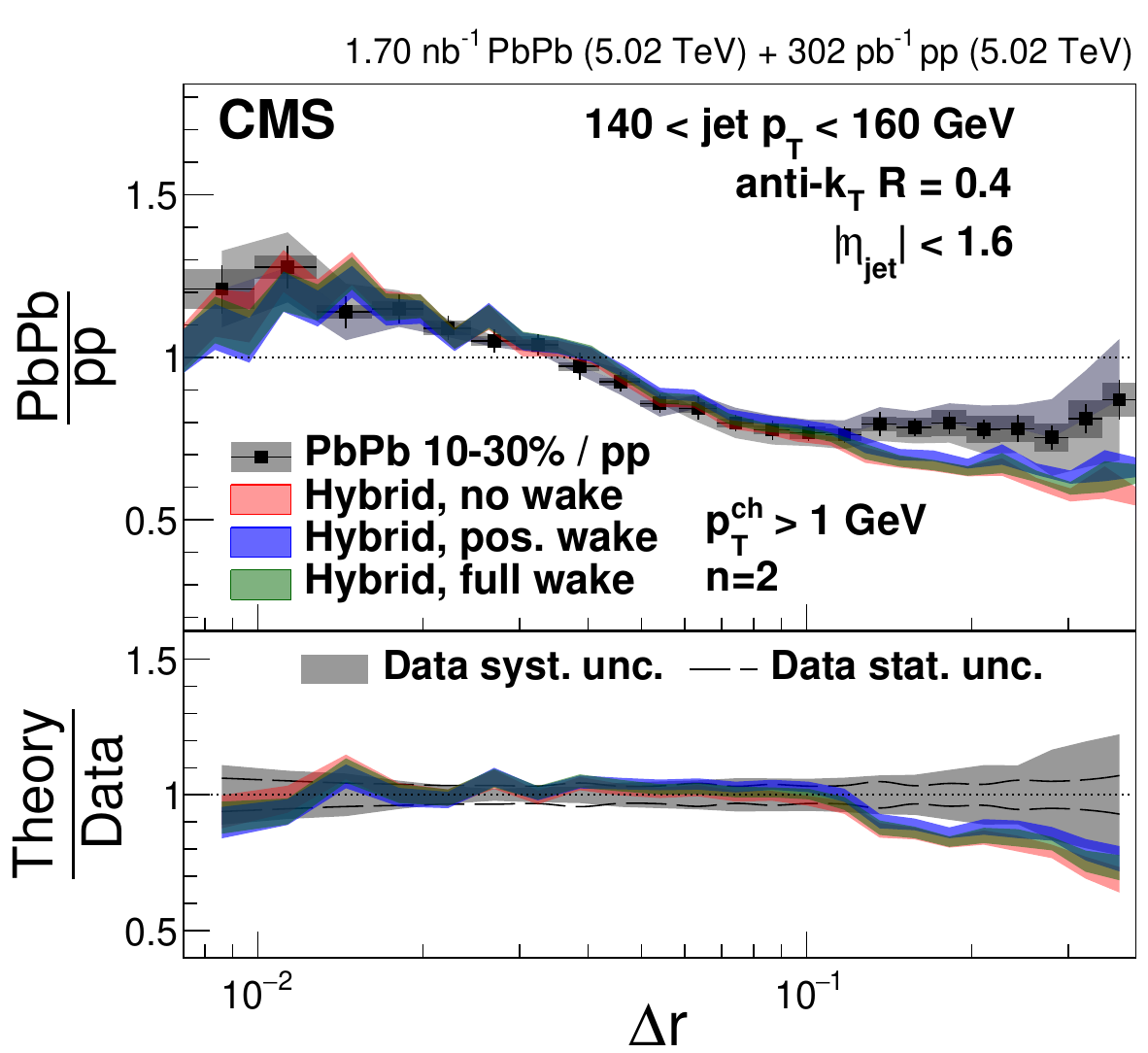}
    \includegraphics[width=0.4\textwidth]{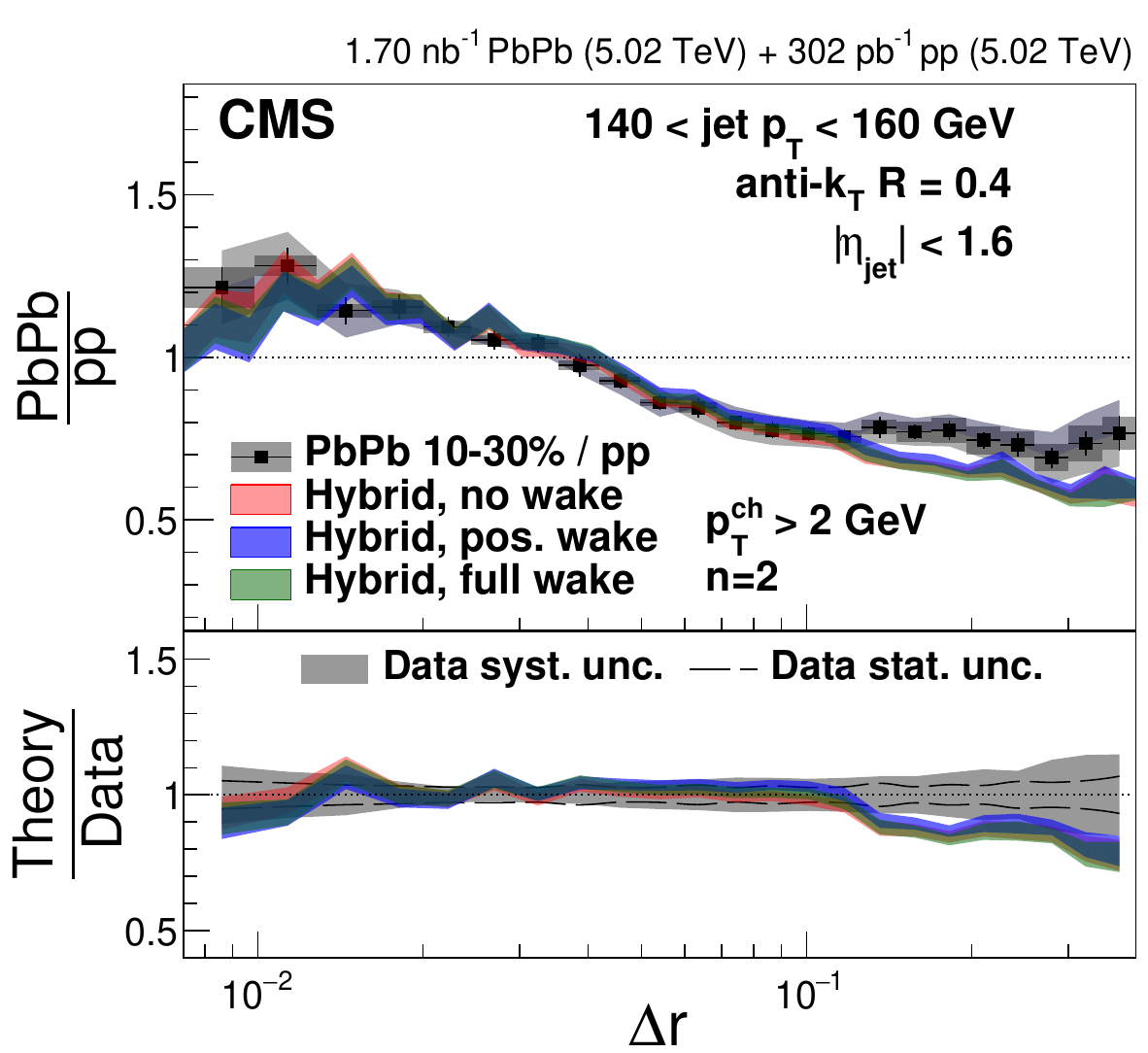}
    \includegraphics[width=0.4\textwidth]{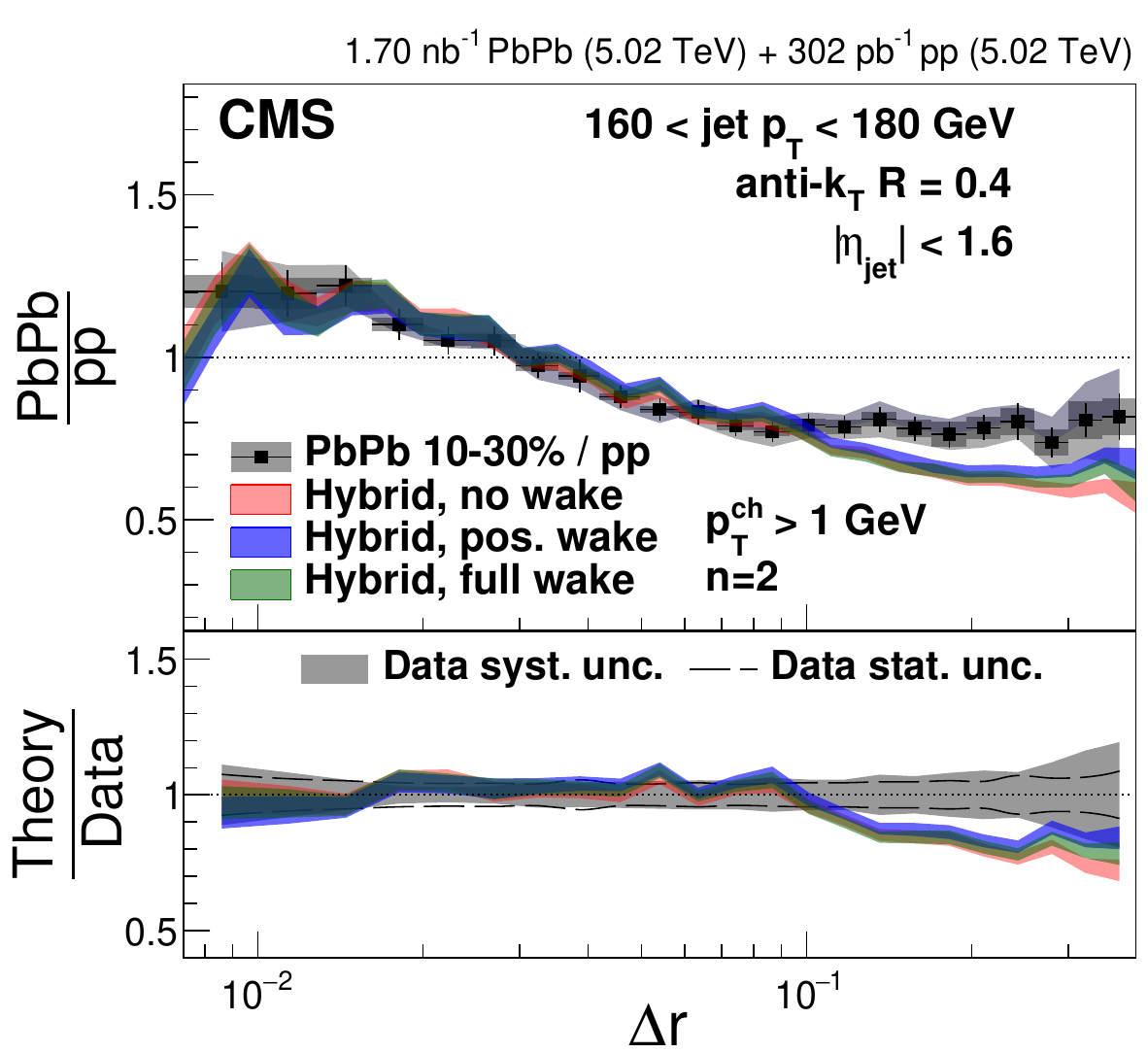}
    \includegraphics[width=0.4\textwidth]{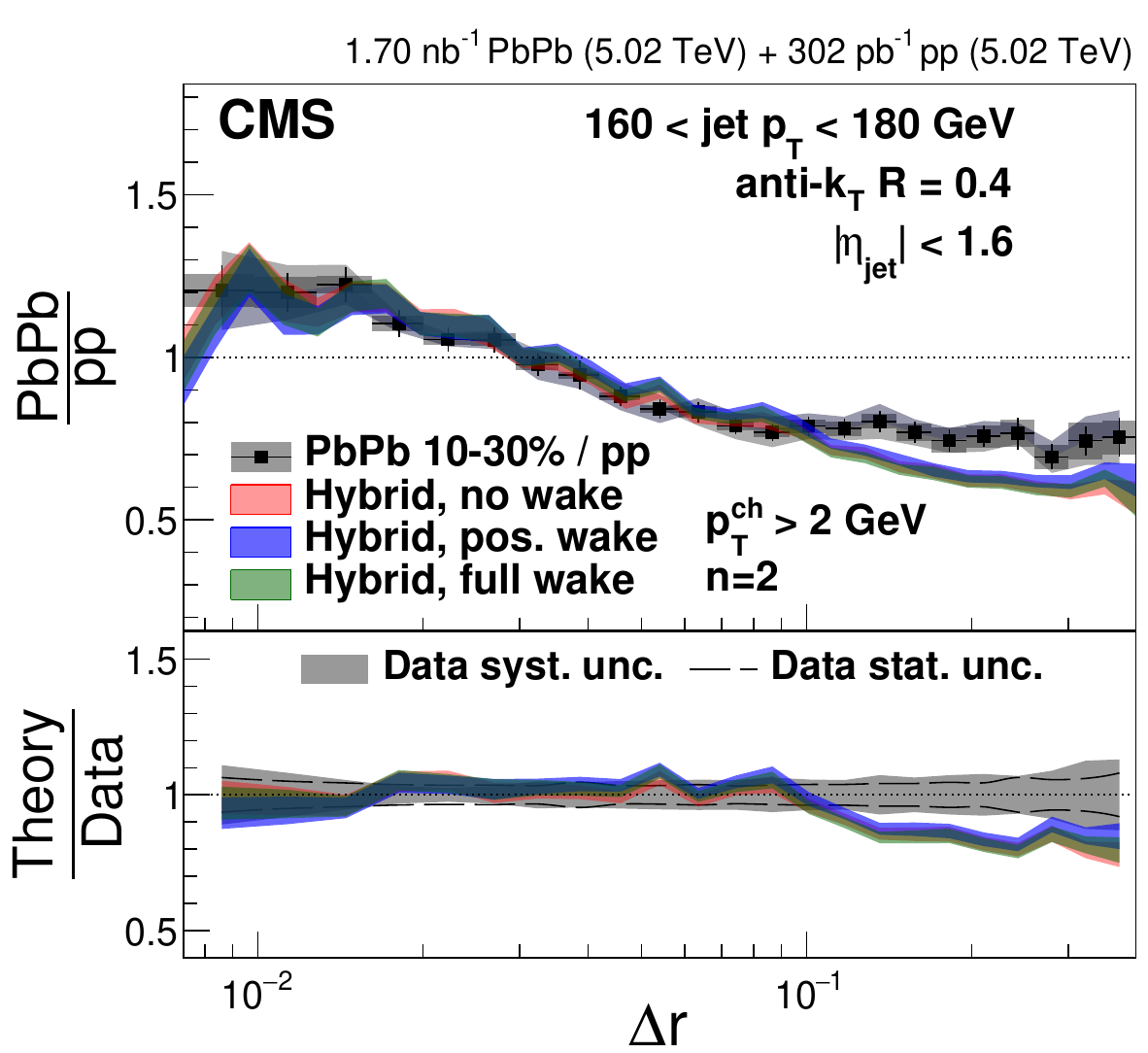}
    \includegraphics[width=0.4\textwidth]{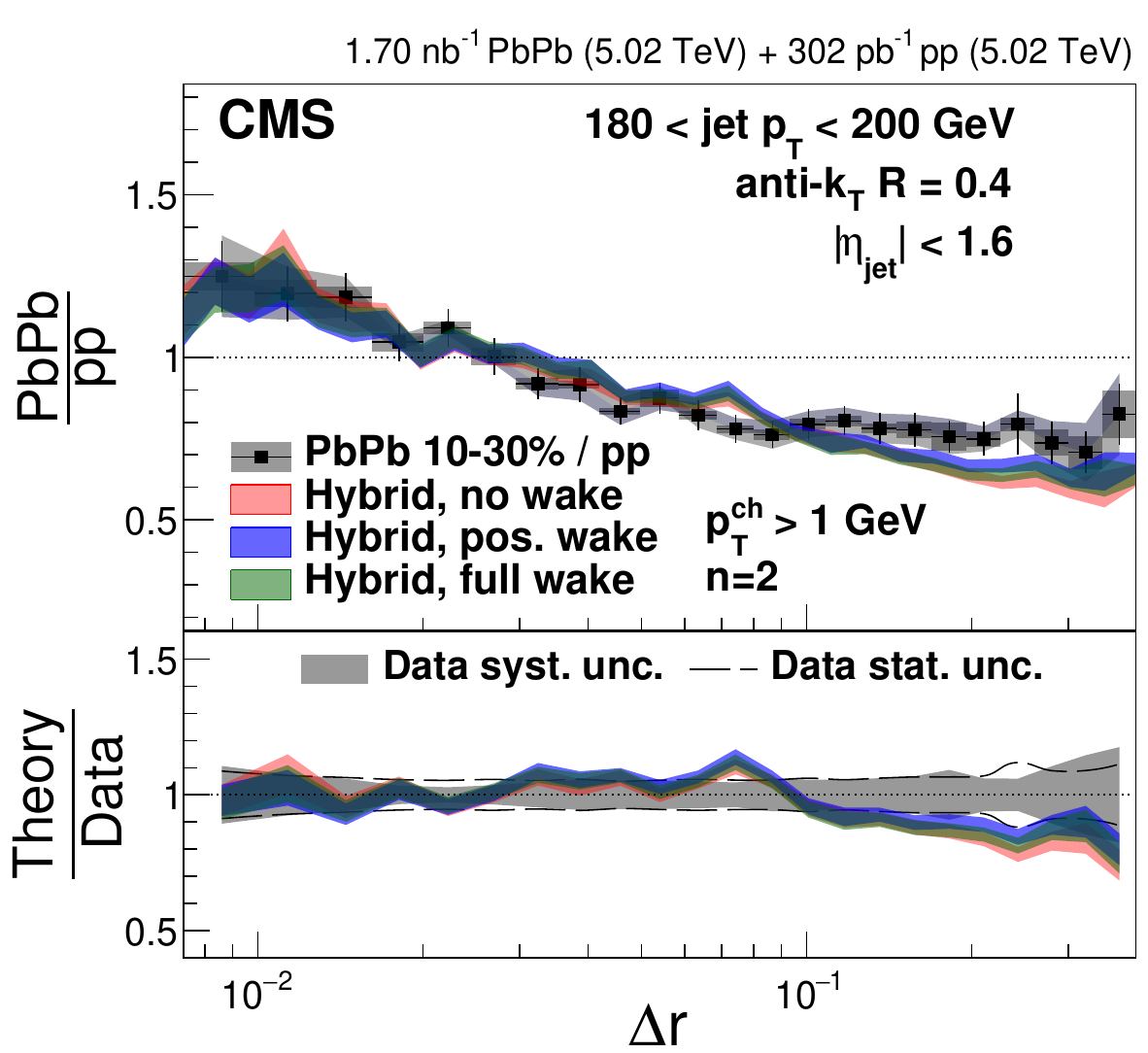}
    \includegraphics[width=0.4\textwidth]{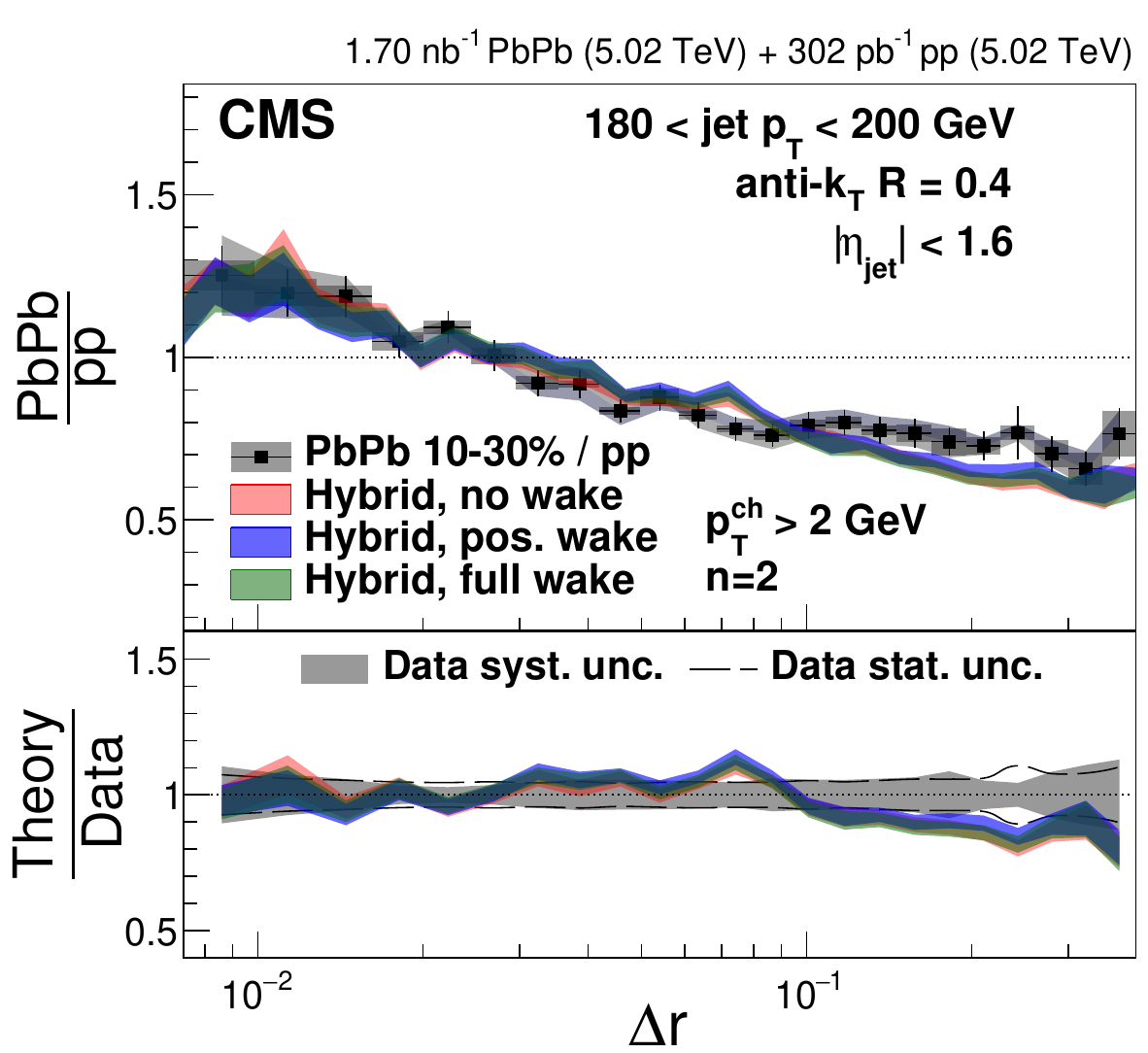}
    \caption{Ratios of 10--30\% central \PbPb to \pp energy-energy correlators with $n=2$ in different \ptCh and \ptjet bins compared to the hybrid model predictions with different jet wake settings.} 
    \label{fig:PbPbRatiosSquared10-30}
  } 
\end{figure*}

\begin{figure*}[h]
  \centering{
    \includegraphics[width=0.4\textwidth]{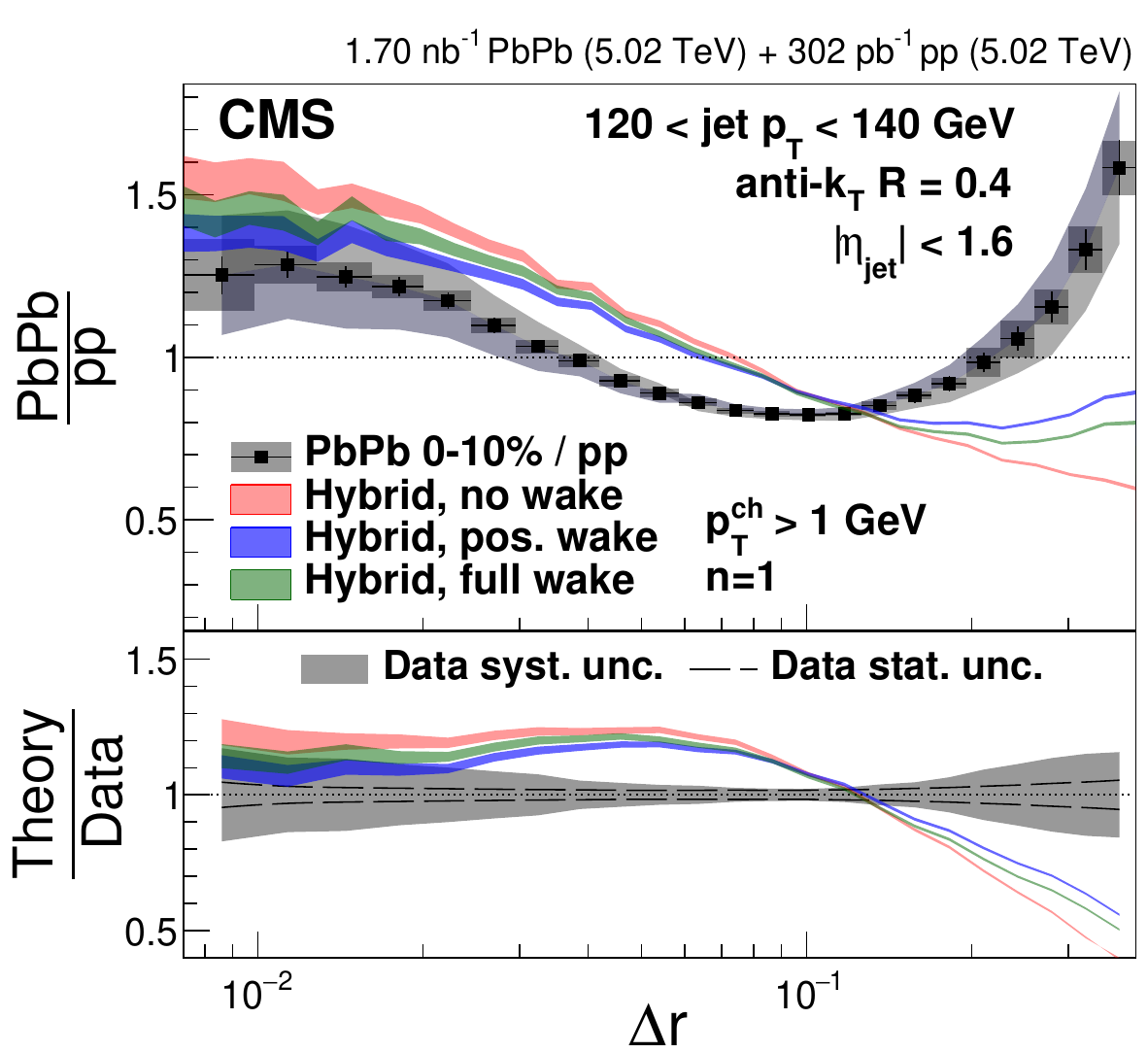}
    \includegraphics[width=0.4\textwidth]{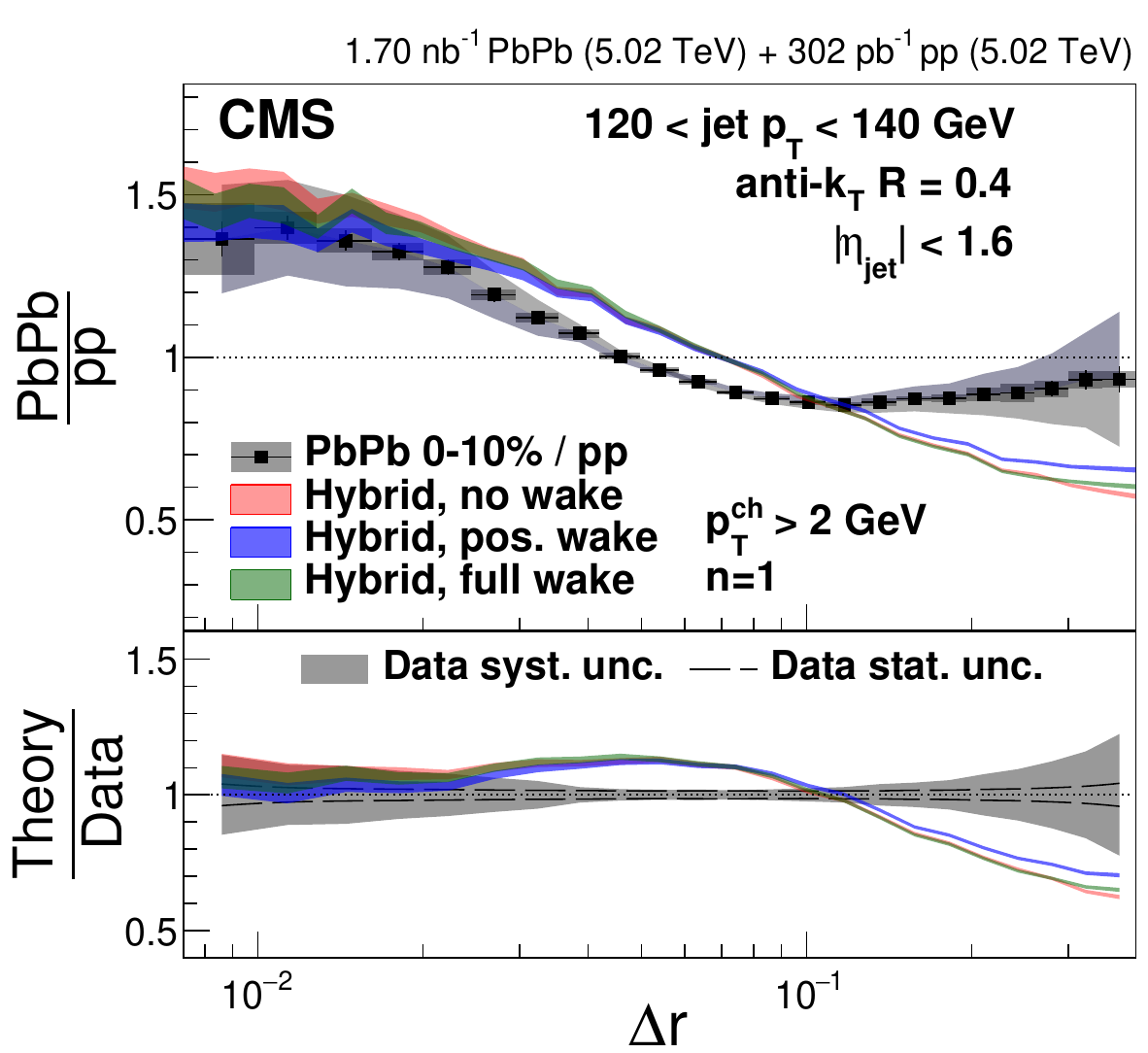}  
    \includegraphics[width=0.4\textwidth]{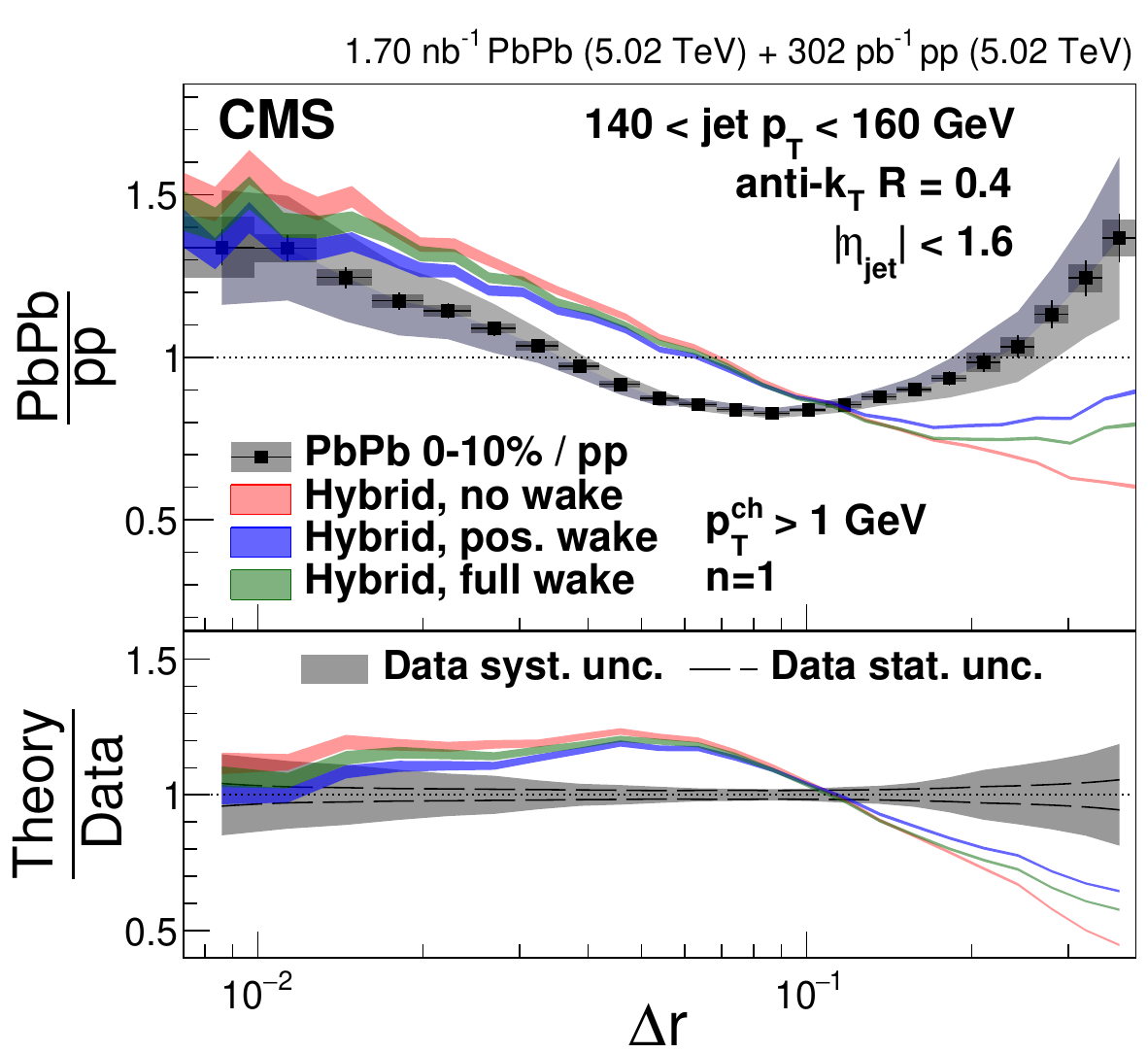}
    \includegraphics[width=0.4\textwidth]{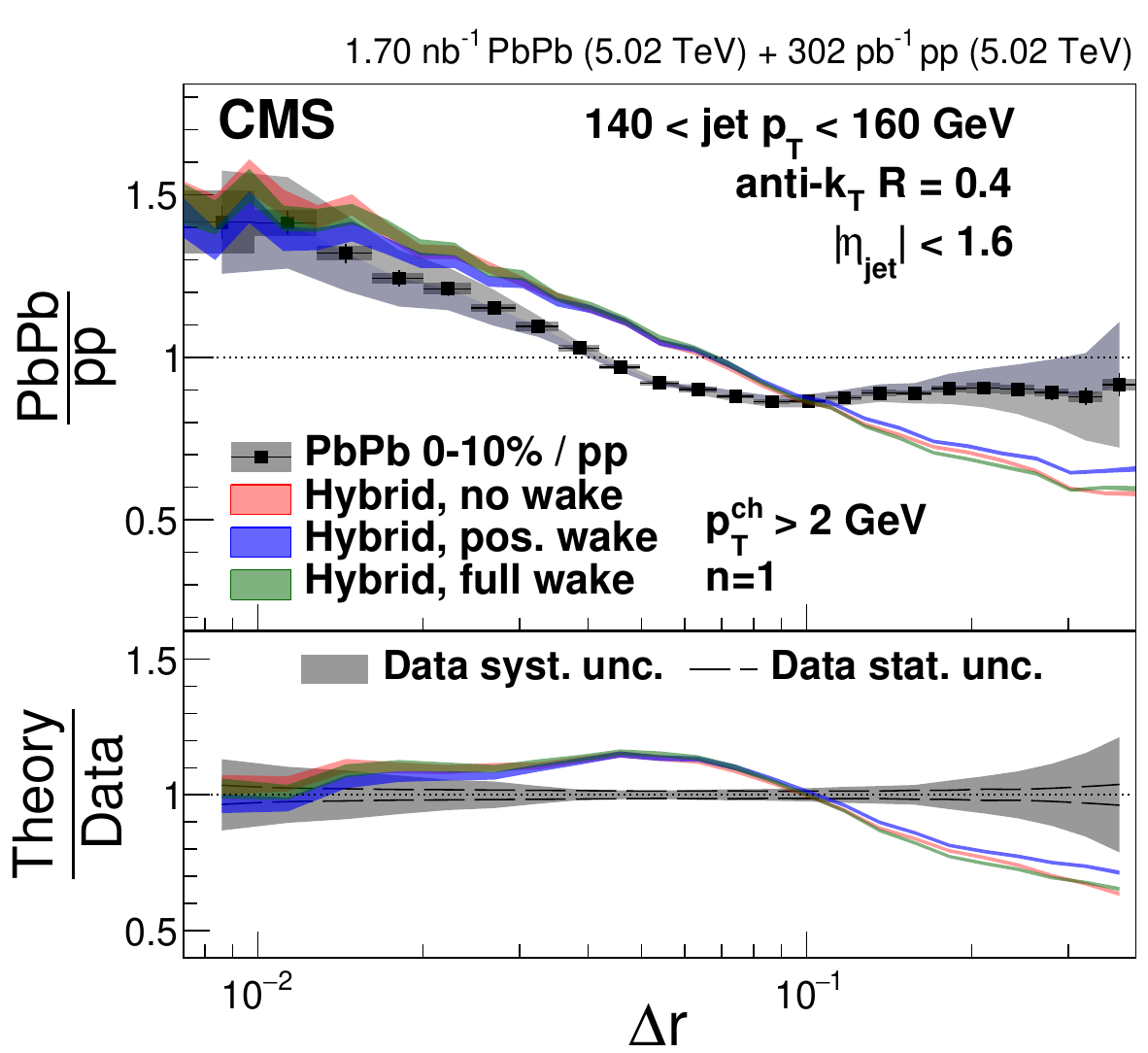}
    \includegraphics[width=0.4\textwidth]{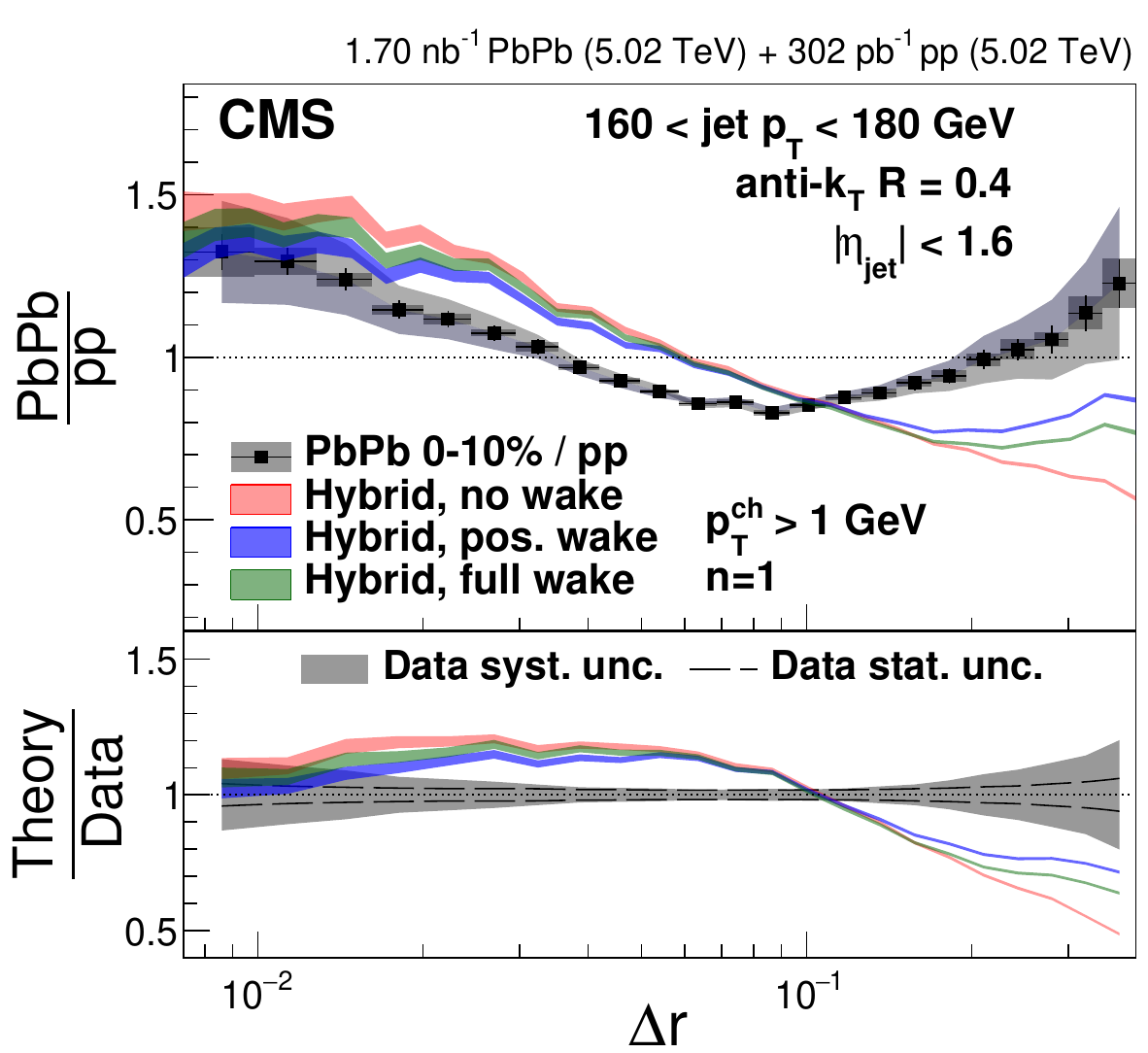}
    \includegraphics[width=0.4\textwidth]{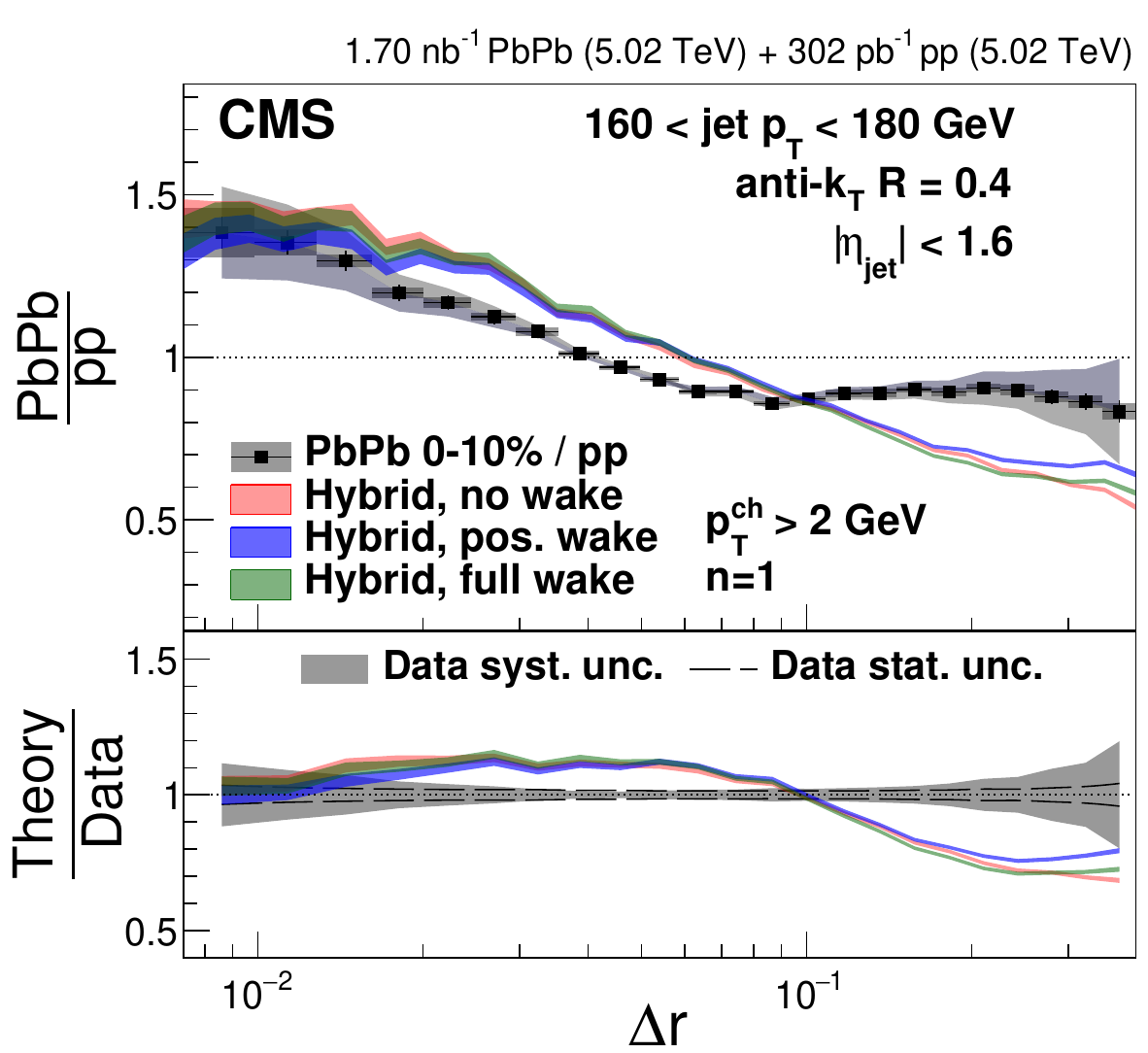}
    \includegraphics[width=0.4\textwidth]{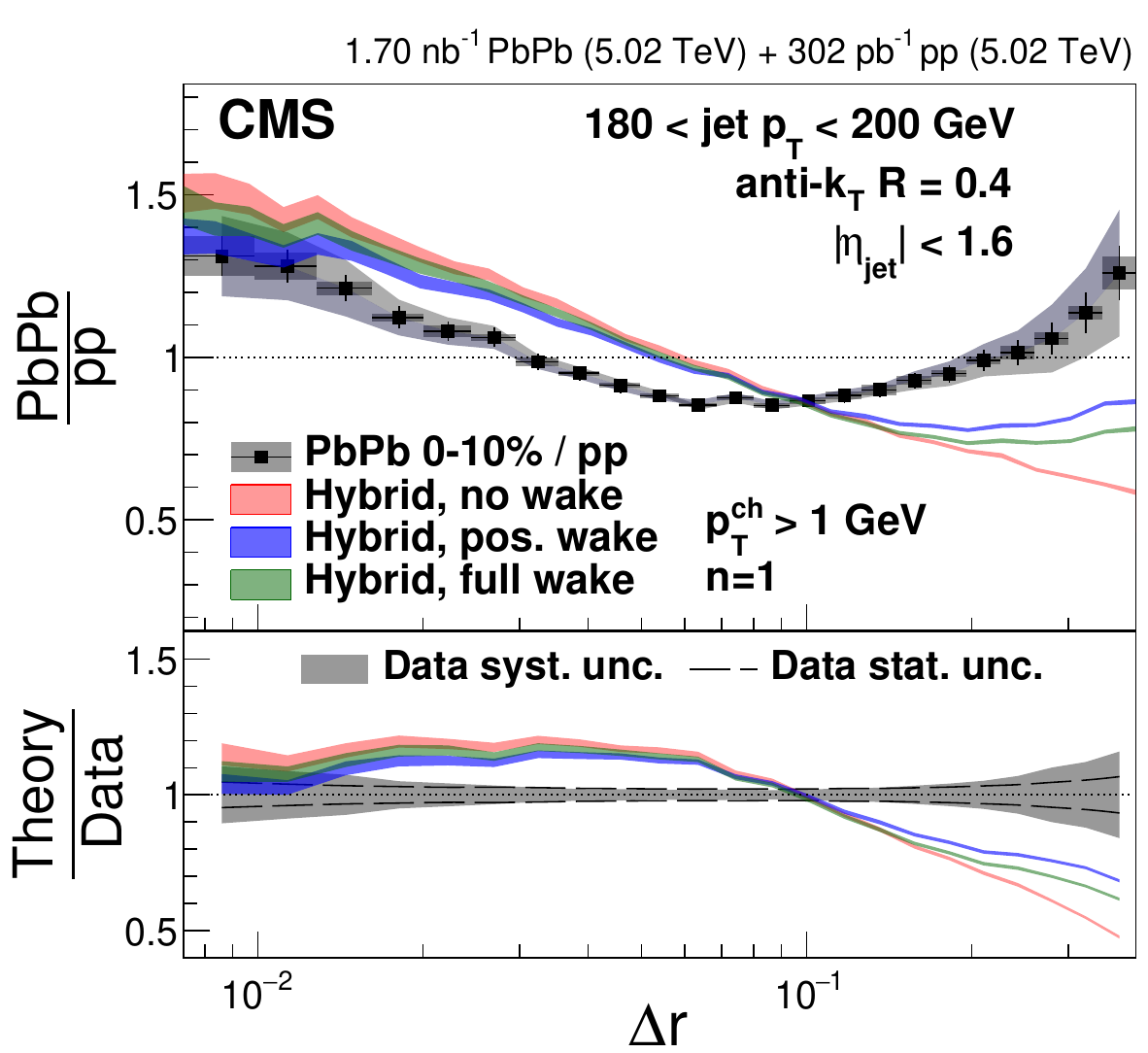}
    \includegraphics[width=0.4\textwidth]{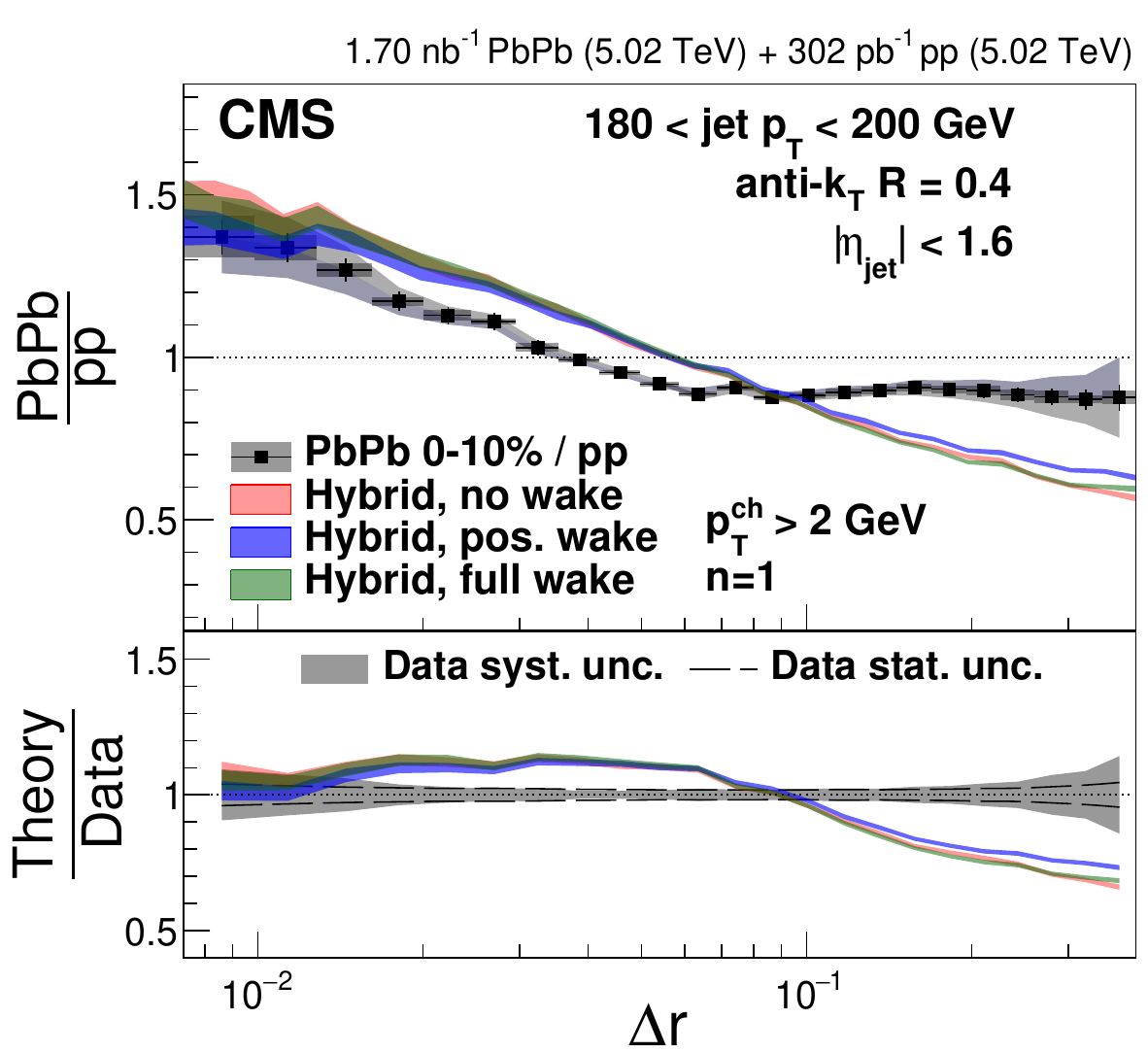}
    \caption{Ratios of 0--10\% central \PbPb to \pp energy-energy correlators with $n=1$ in different \ptCh and \ptjet bins compared to the hybrid model predictions with different jet wake settings.} 
    \label{fig:PbPbRatiosNominal0-10}
  } 
\end{figure*}

\begin{figure*}[h]
  \centering{
    \includegraphics[width=0.4\textwidth]{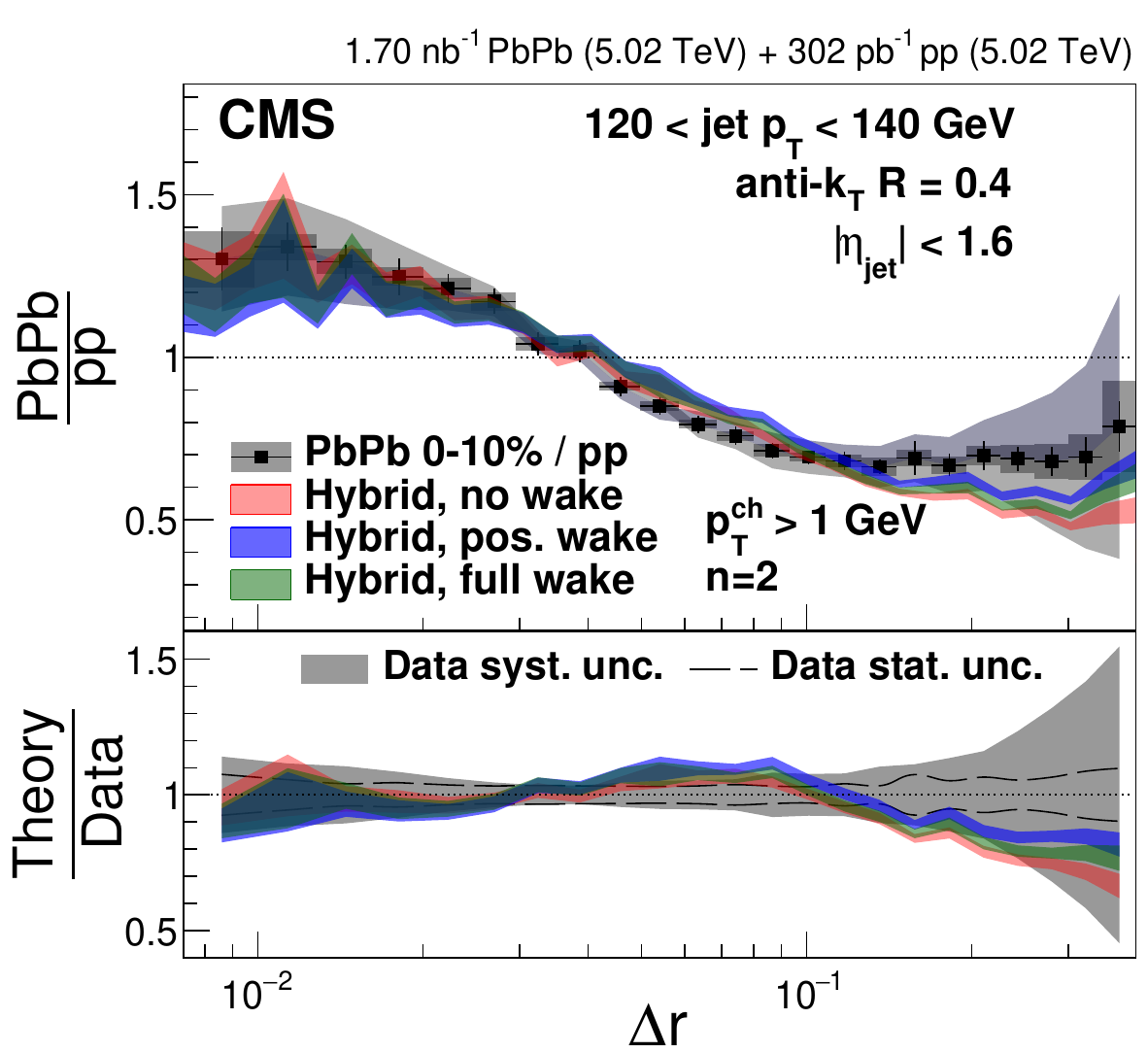} 
    \includegraphics[width=0.4\textwidth]{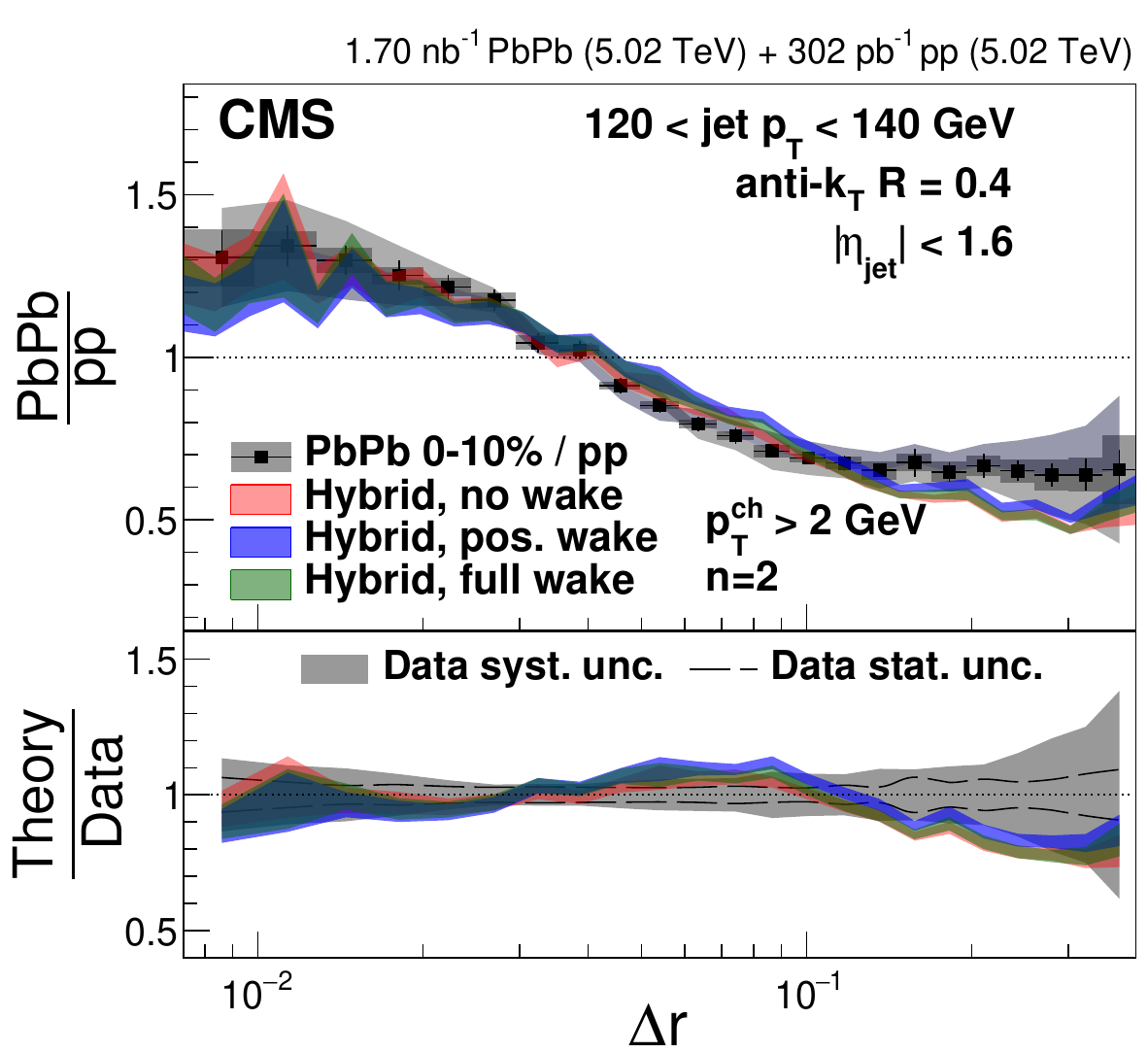} 
    \includegraphics[width=0.4\textwidth]{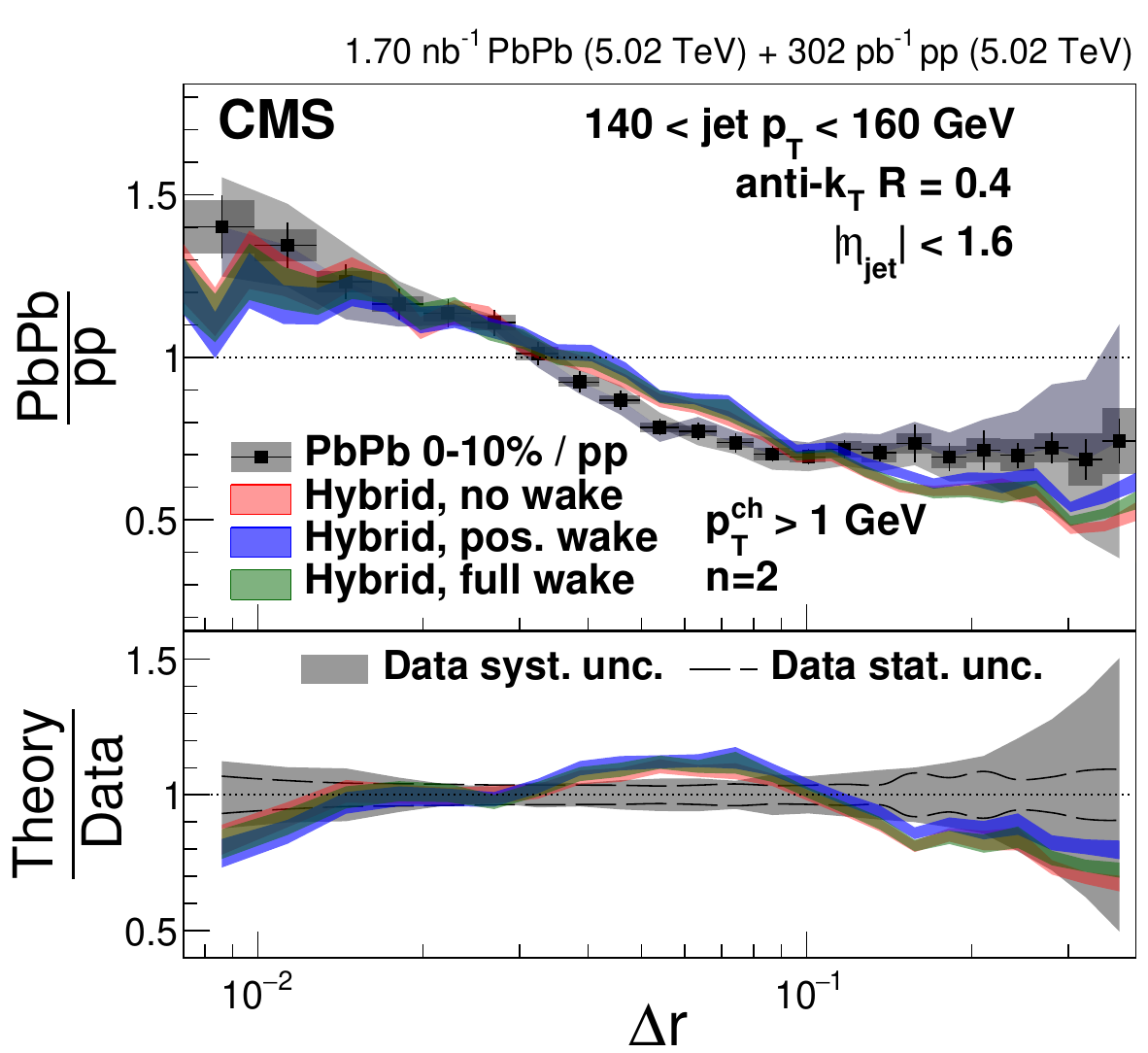}
    \includegraphics[width=0.4\textwidth]{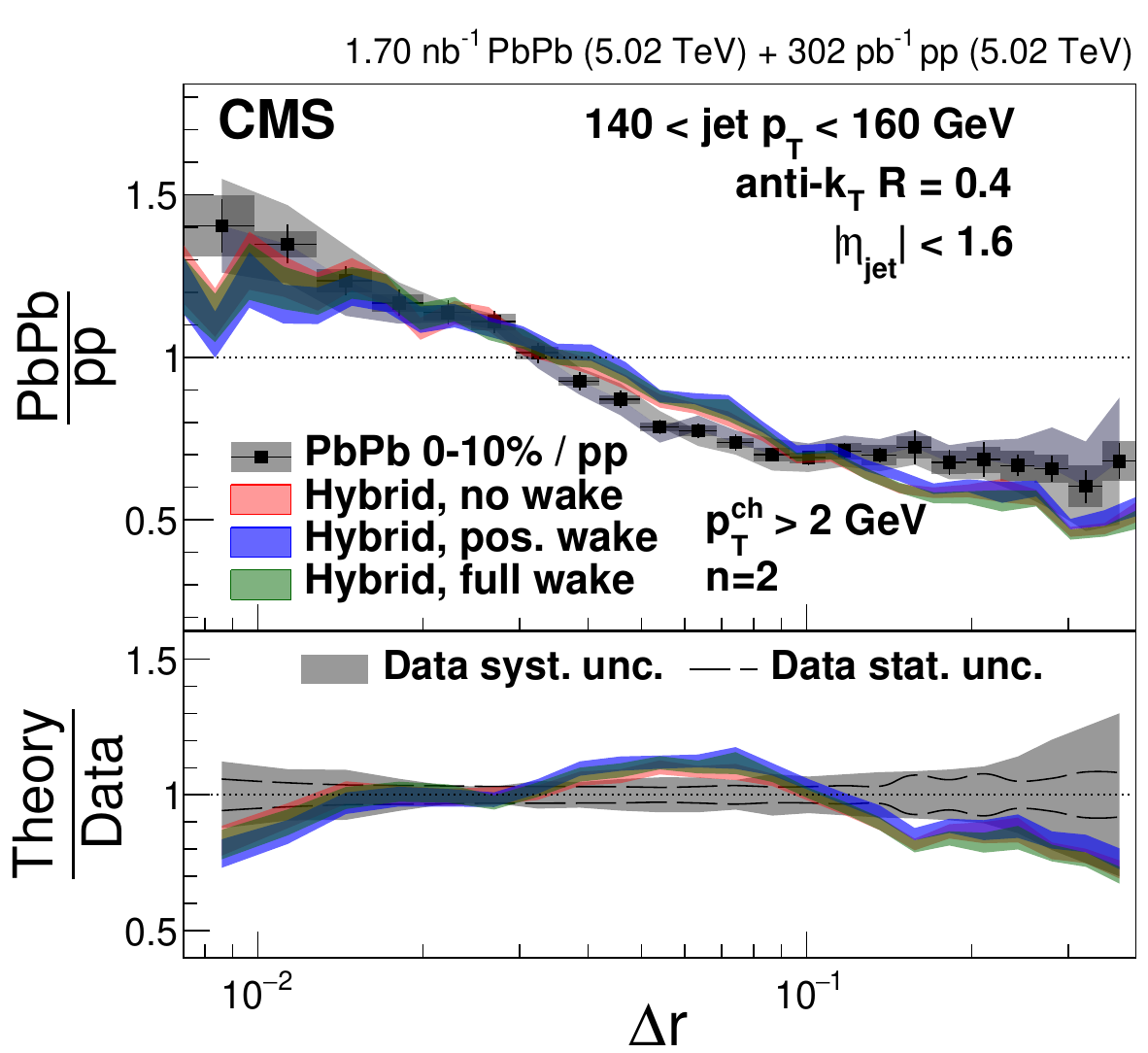}
    \includegraphics[width=0.4\textwidth]{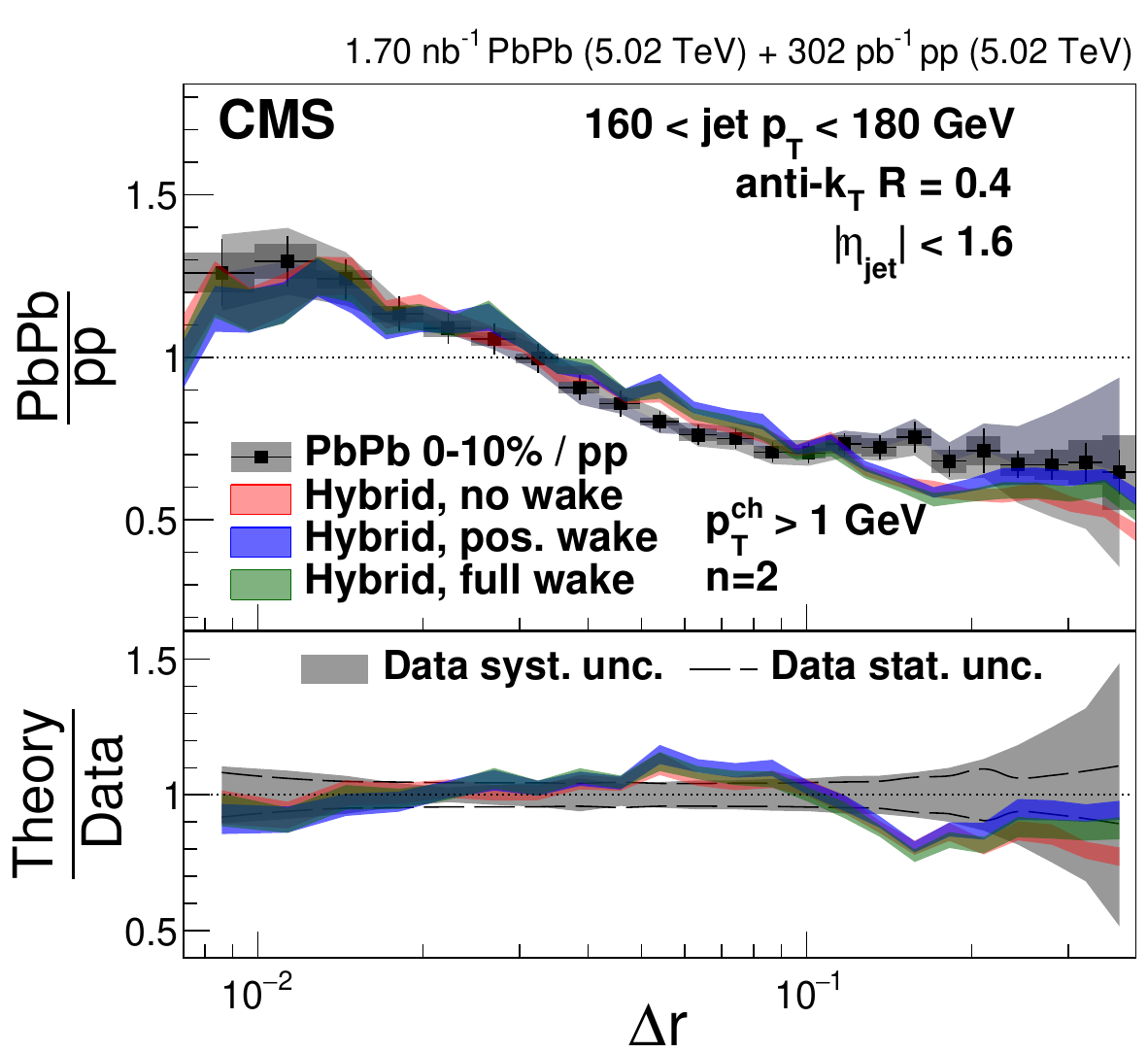}
    \includegraphics[width=0.4\textwidth]{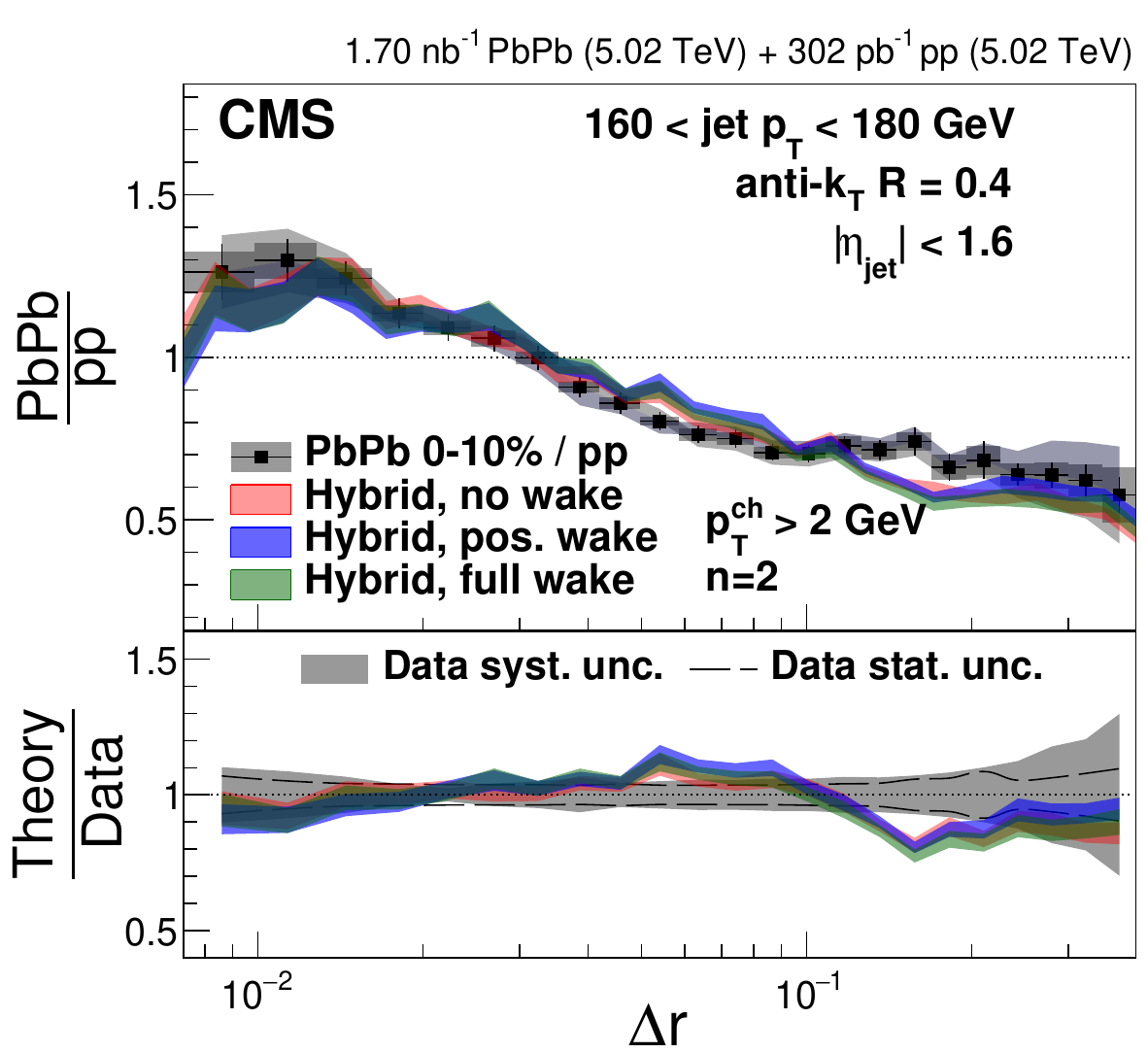}
    \includegraphics[width=0.4\textwidth]{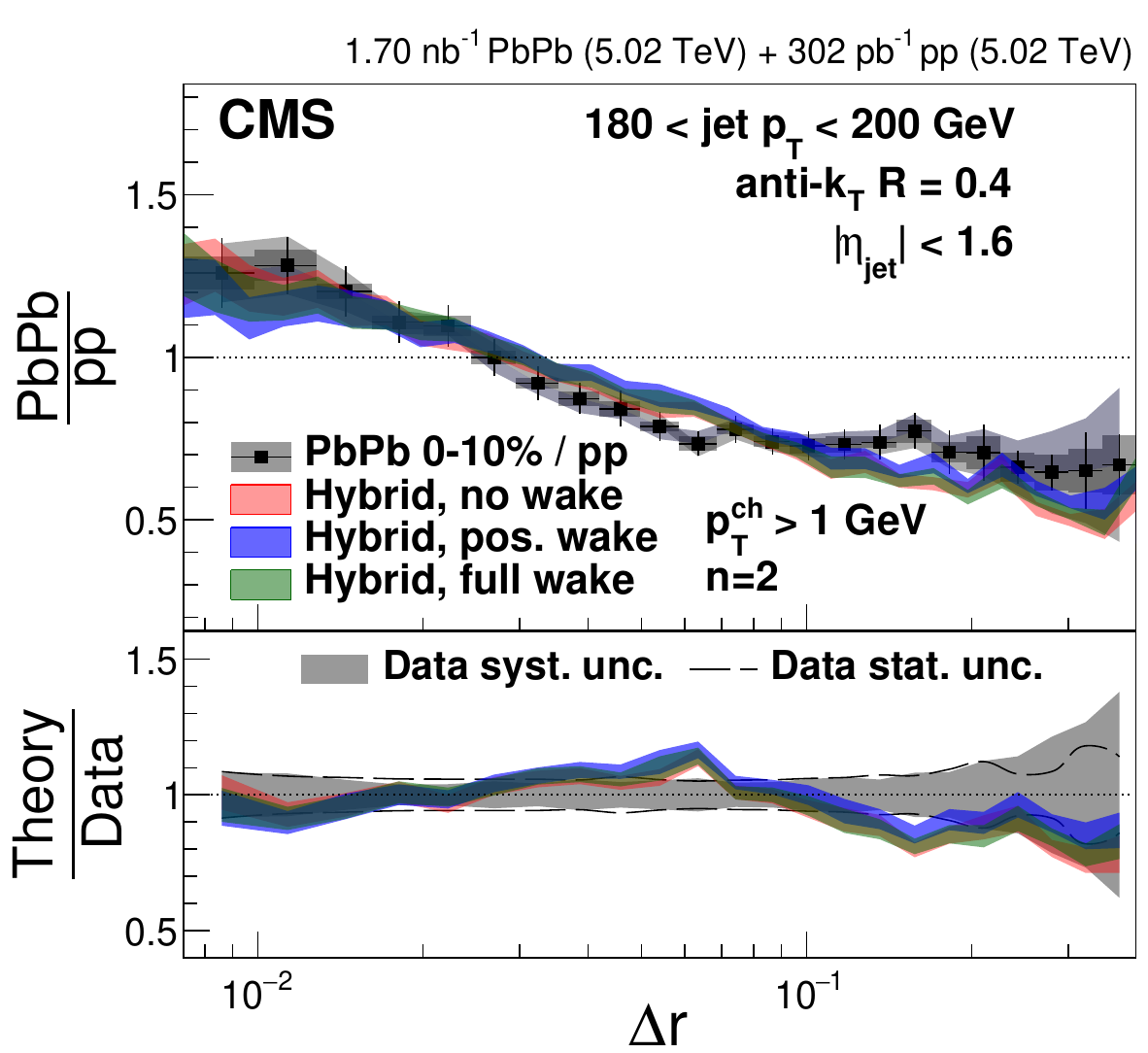}
    \includegraphics[width=0.4\textwidth]{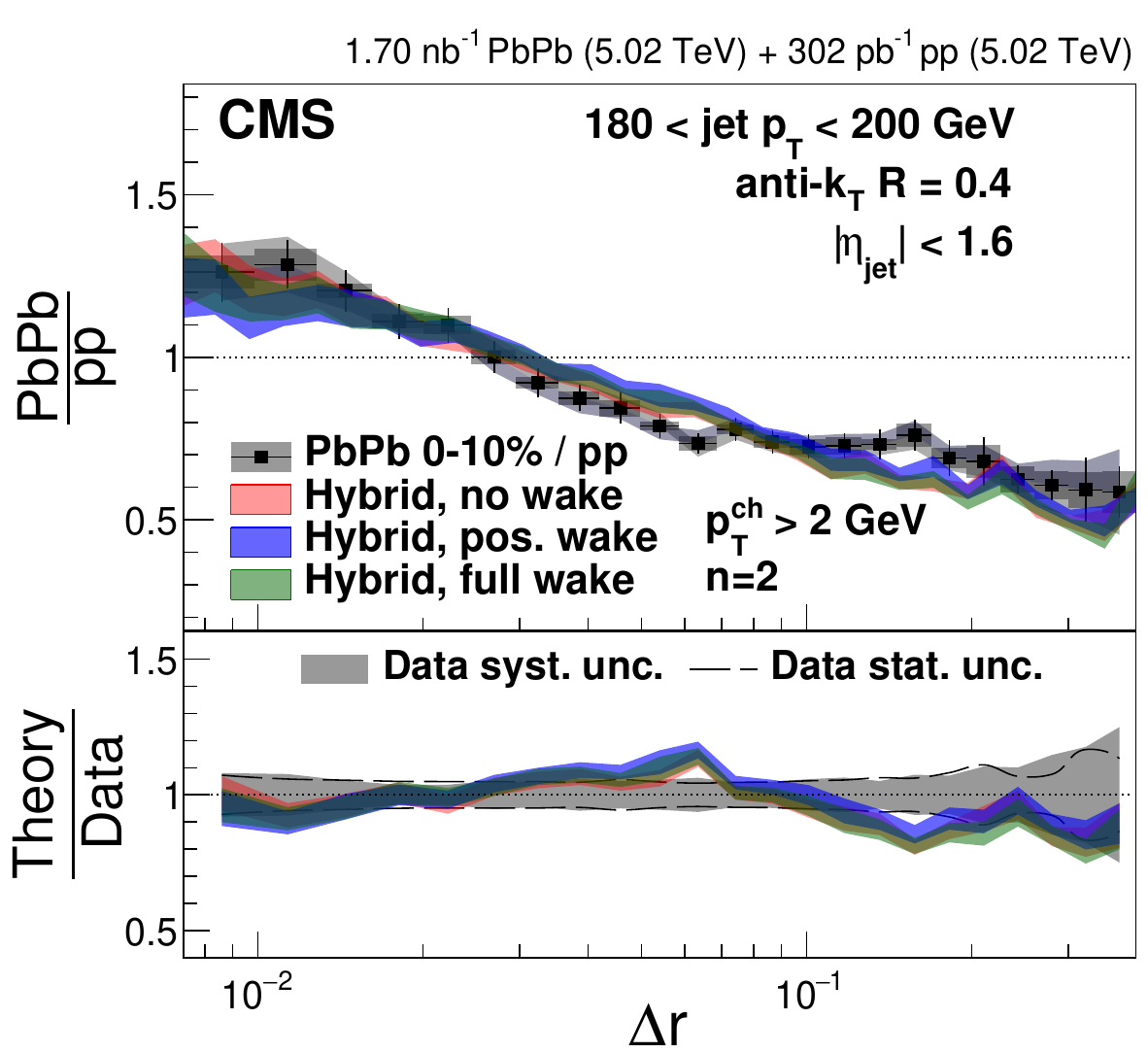}
    \caption{Ratios of 0--10\% central \PbPb to \pp energy-energy correlators with $n=2$ in different \ptCh and \ptjet bins compared to the hybrid model predictions with different jet wake settings.} 
    \label{fig:PbPbRatiosSquared0-10}
  } 
\end{figure*}

\begin{figure*}[h]
  \centering{
    \includegraphics[width=0.4\textwidth]{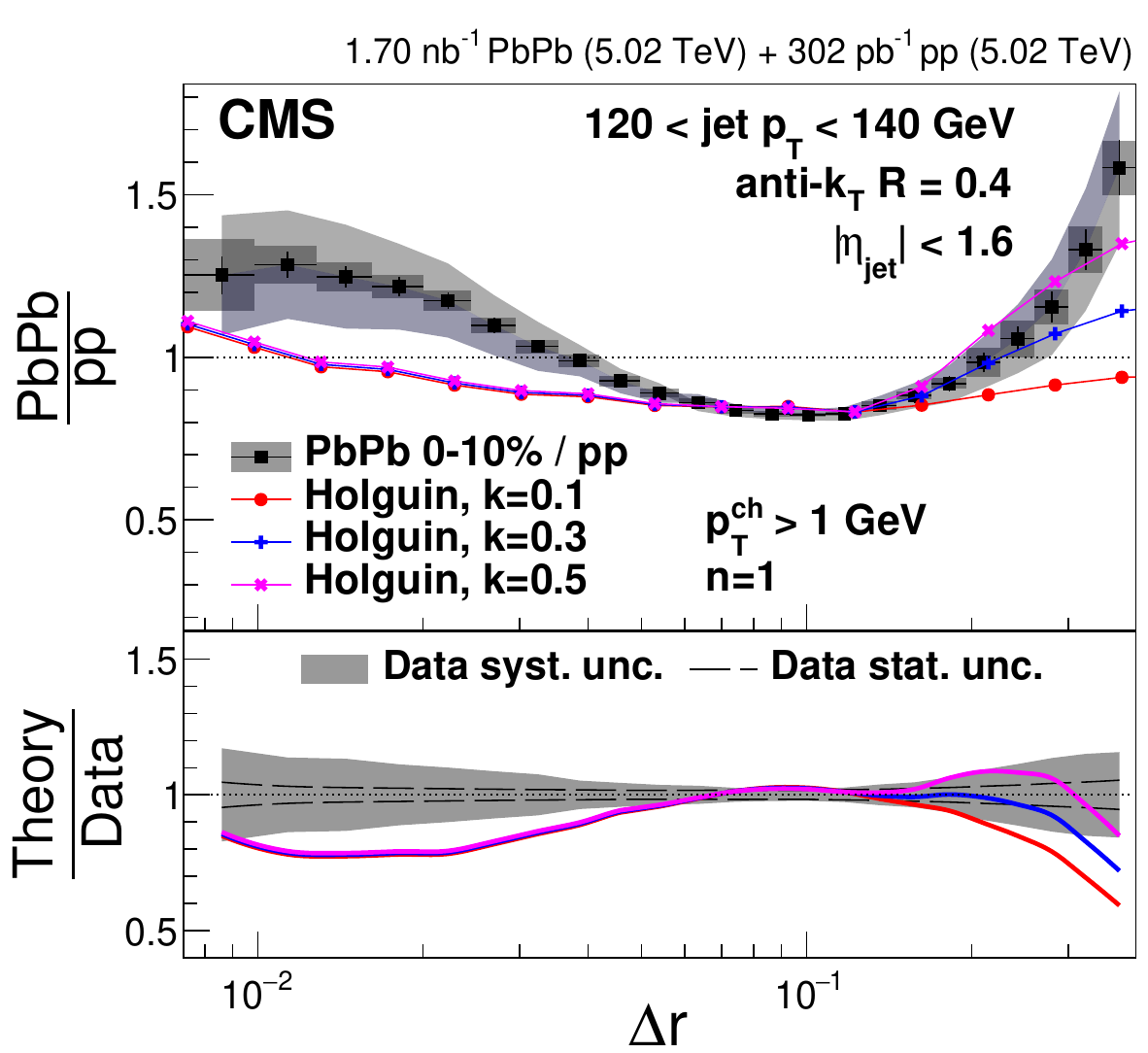} \\
    \includegraphics[width=0.4\textwidth]{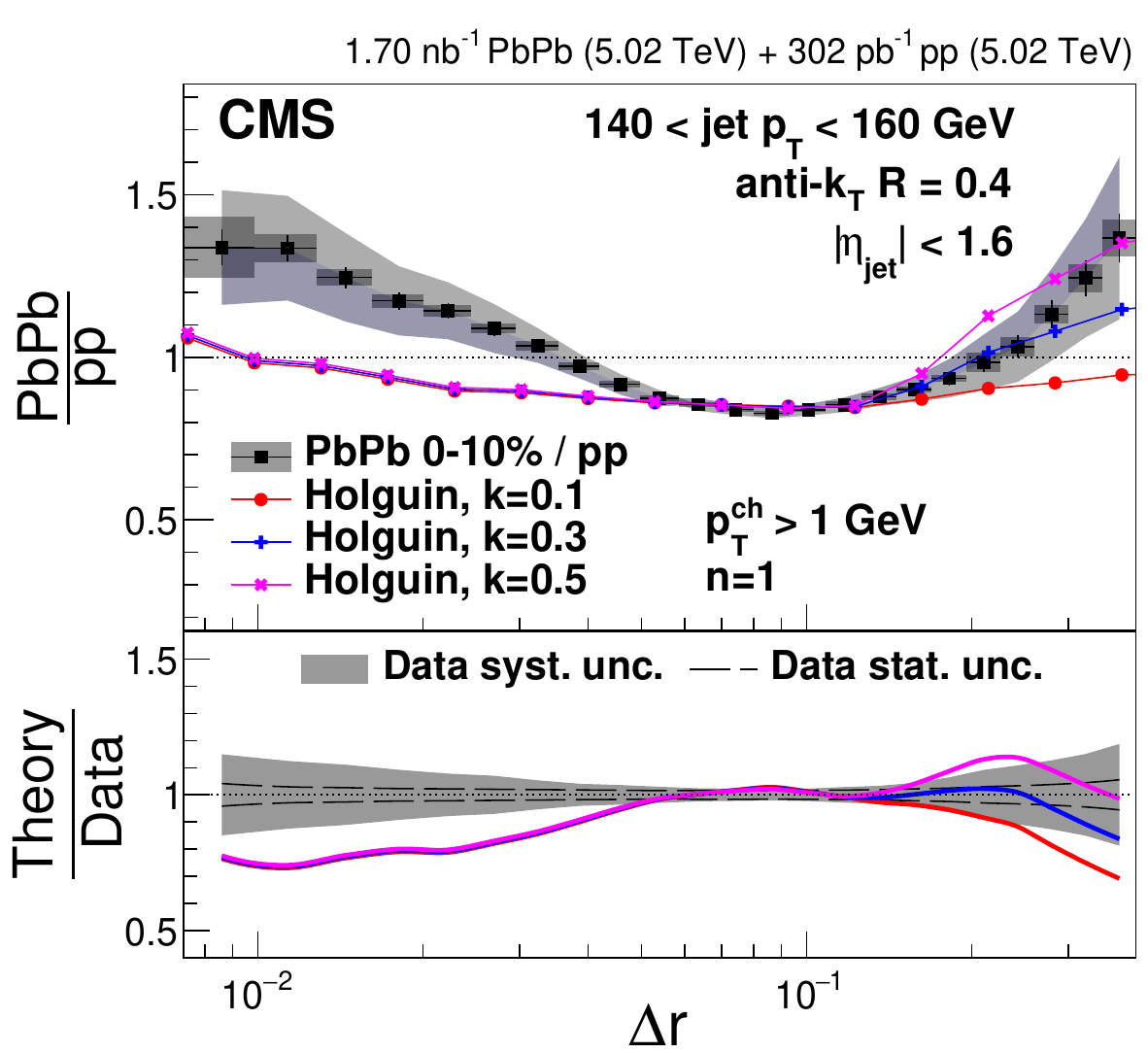} \\
    \includegraphics[width=0.4\textwidth]{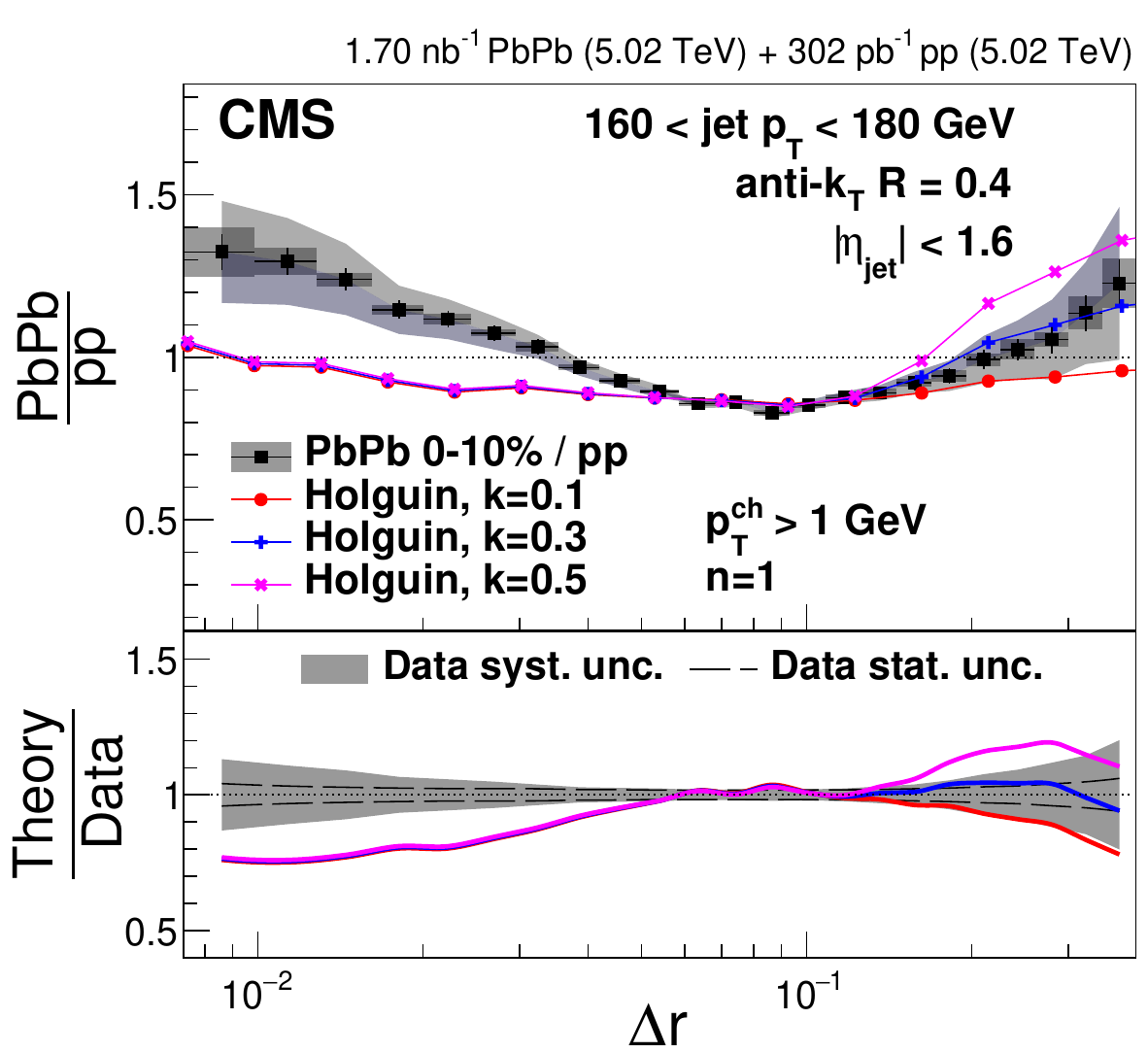} \\
    \includegraphics[width=0.4\textwidth]{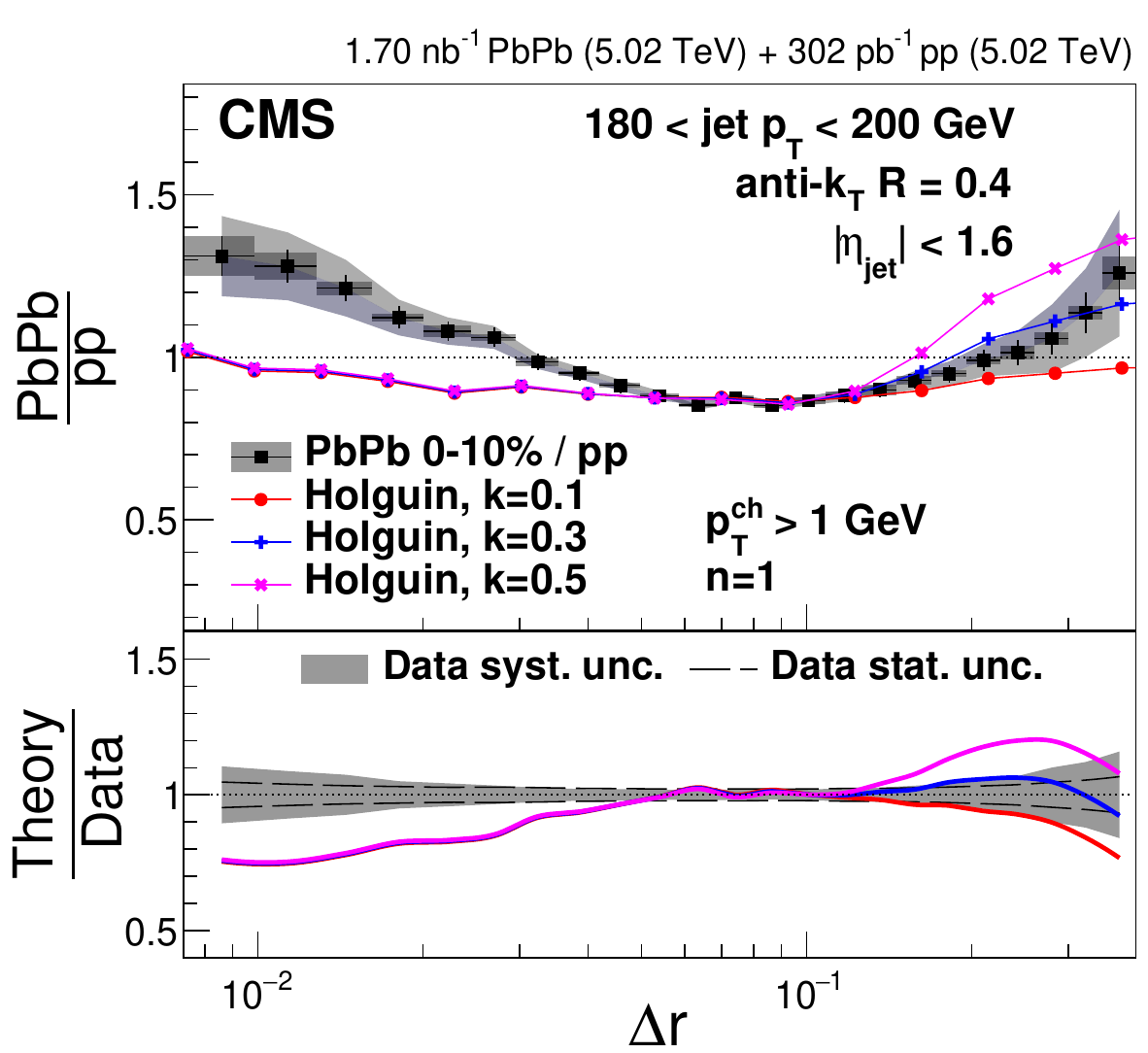}
    \caption{Ratios of 0--10\% central \PbPb to \pp energy-energy correlators with $n=1$ and $\ptCh > 1\GeV$ in different \ptjet bins compared to predictions from perturbative calculation from Holguin and collaborators with different values of the $k$-parameter.} 
    \label{fig:PbPbRatiosNominalHolguin}
  } 
\end{figure*}

\begin{figure*}[h]
  \centering{
    \includegraphics[width=0.4\textwidth]{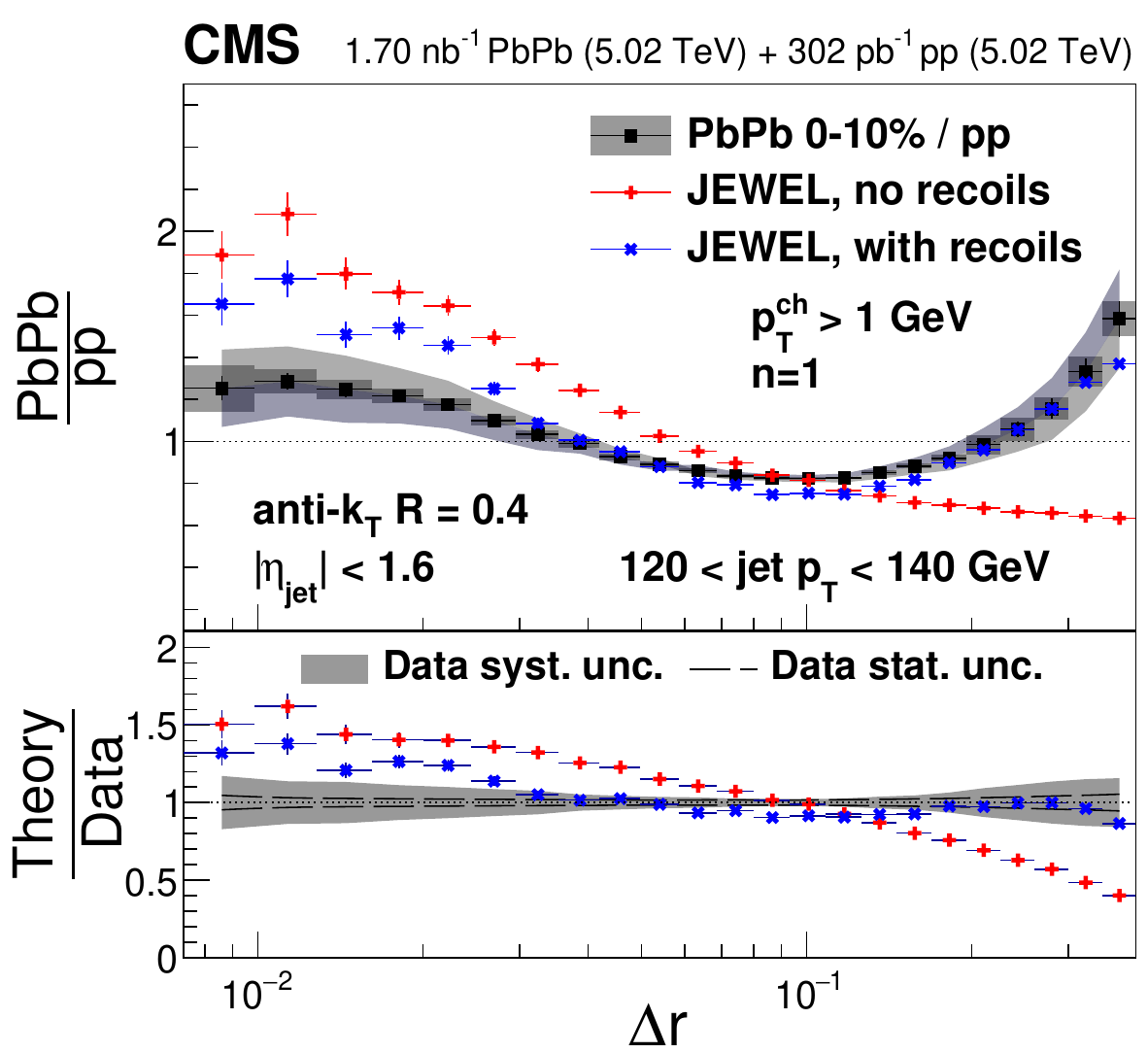}
    \includegraphics[width=0.4\textwidth]{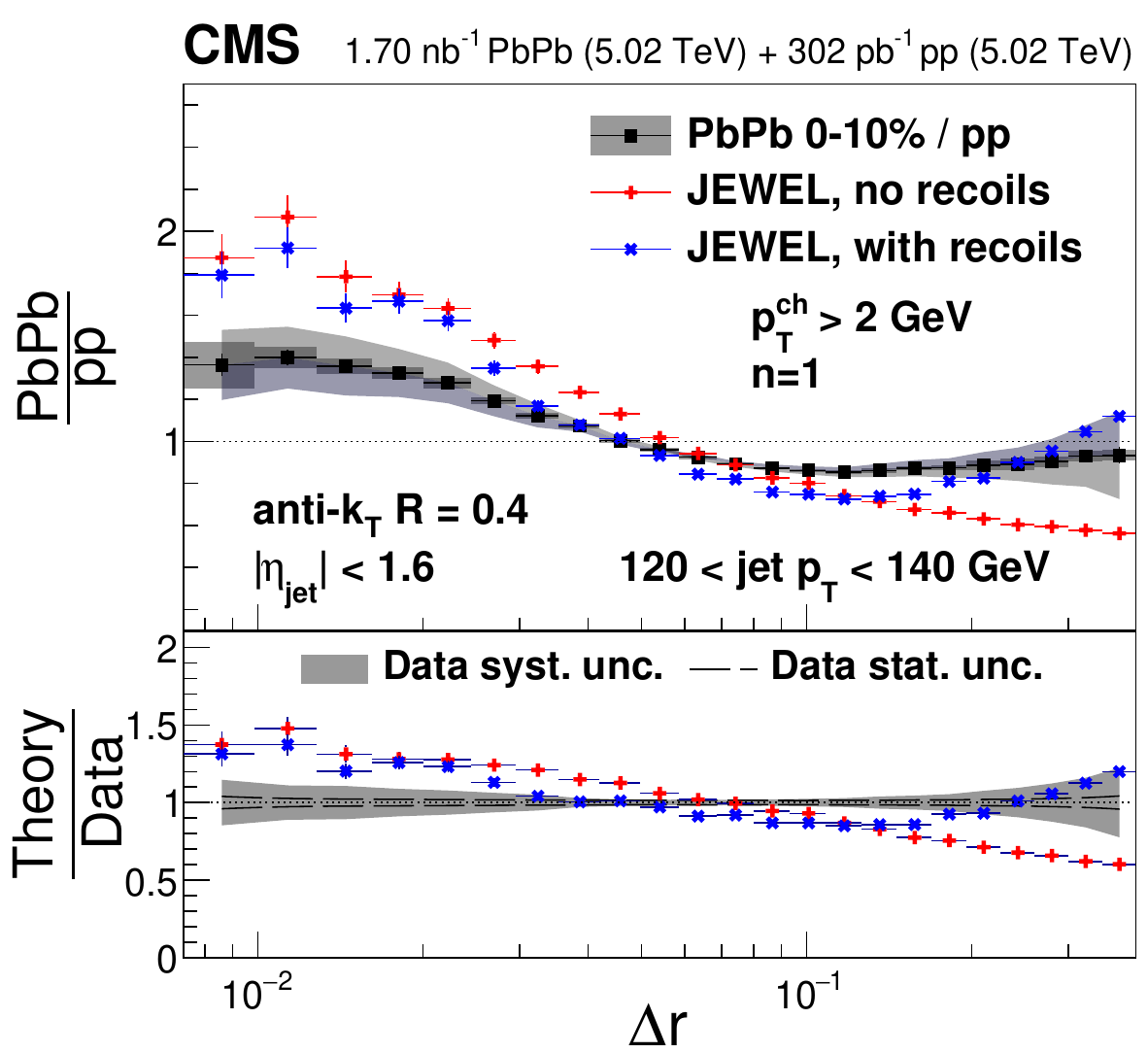} 
    \includegraphics[width=0.4\textwidth]{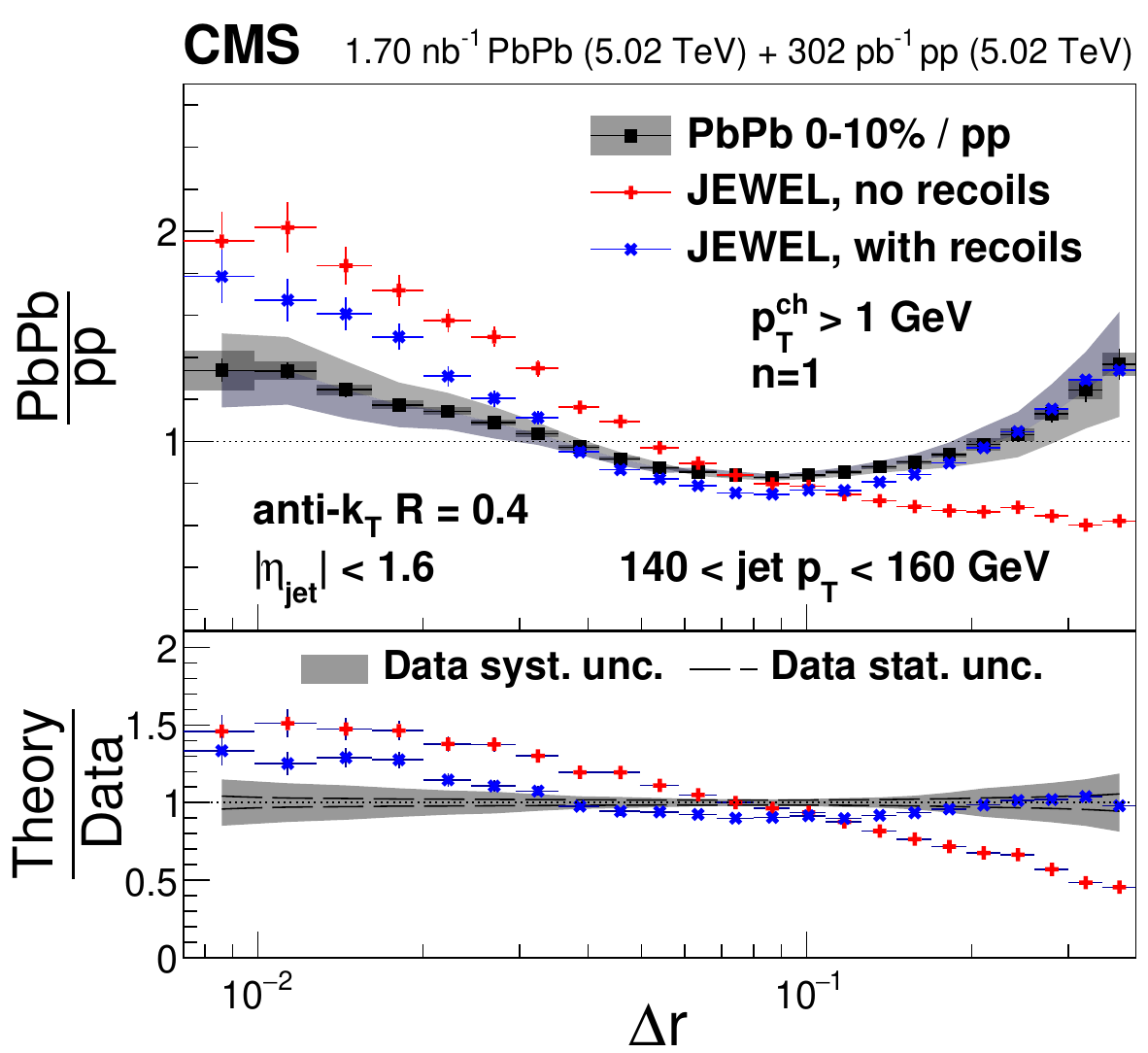}
    \includegraphics[width=0.4\textwidth]{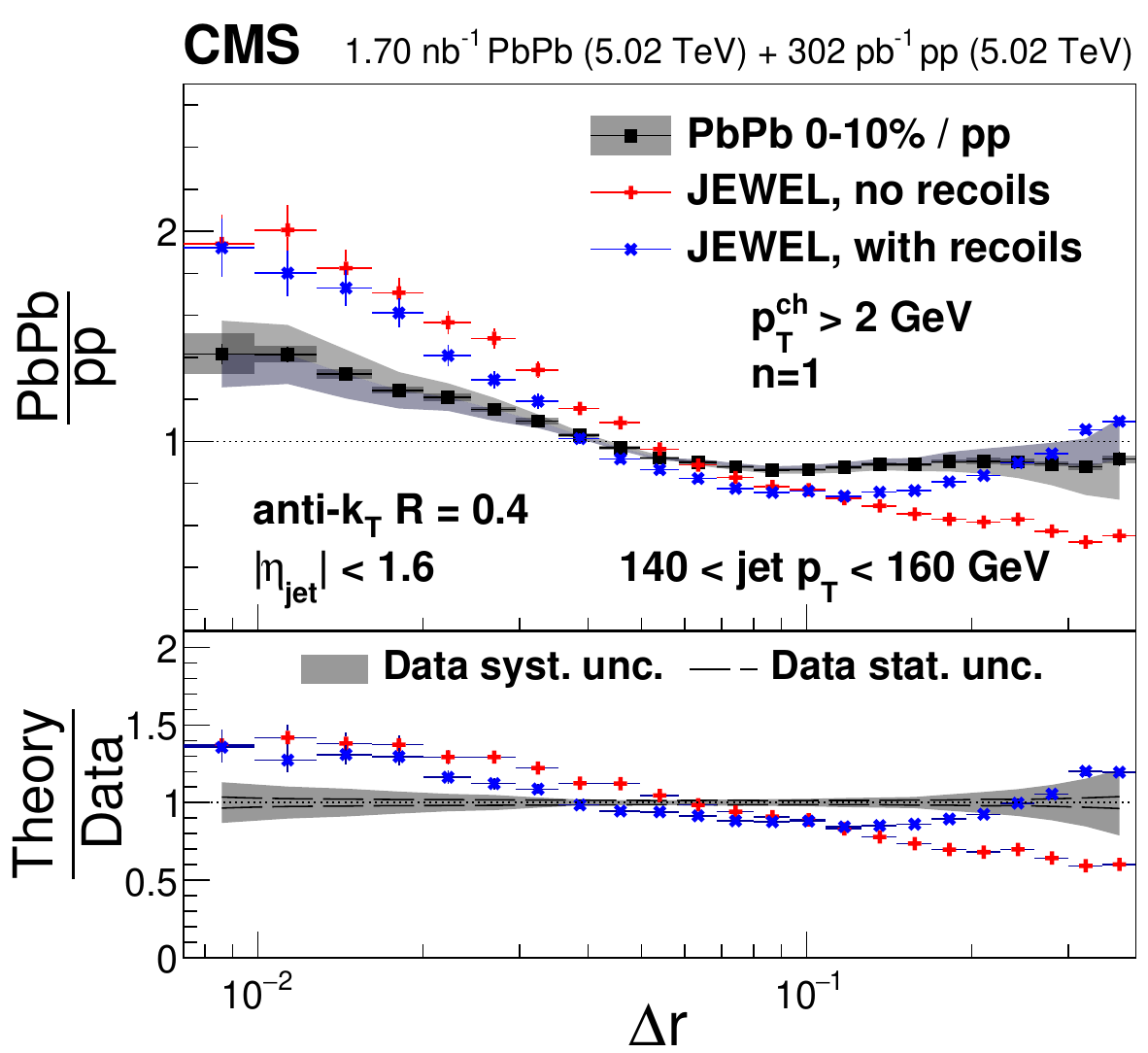}
    \includegraphics[width=0.4\textwidth]{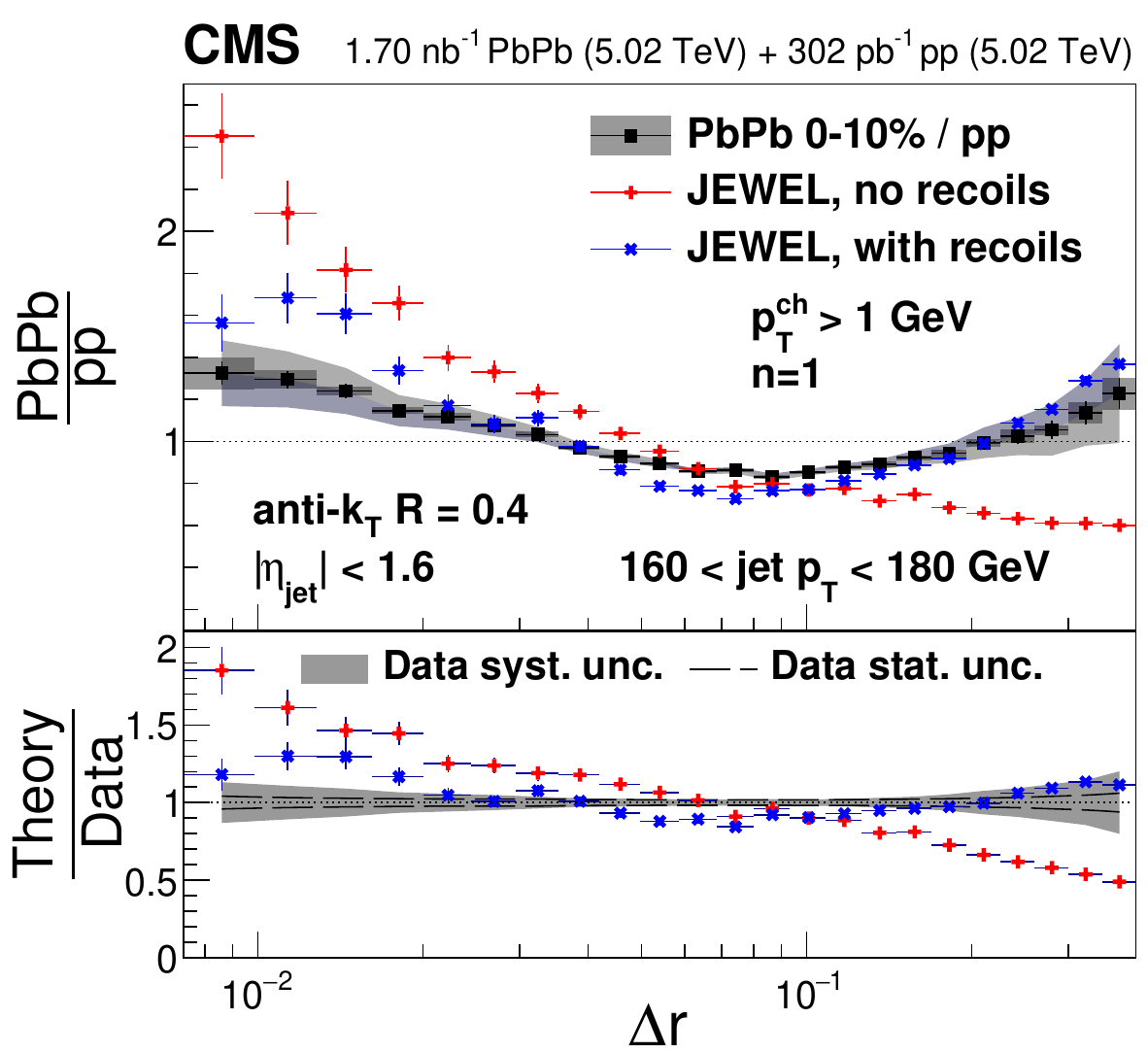}
    \includegraphics[width=0.4\textwidth]{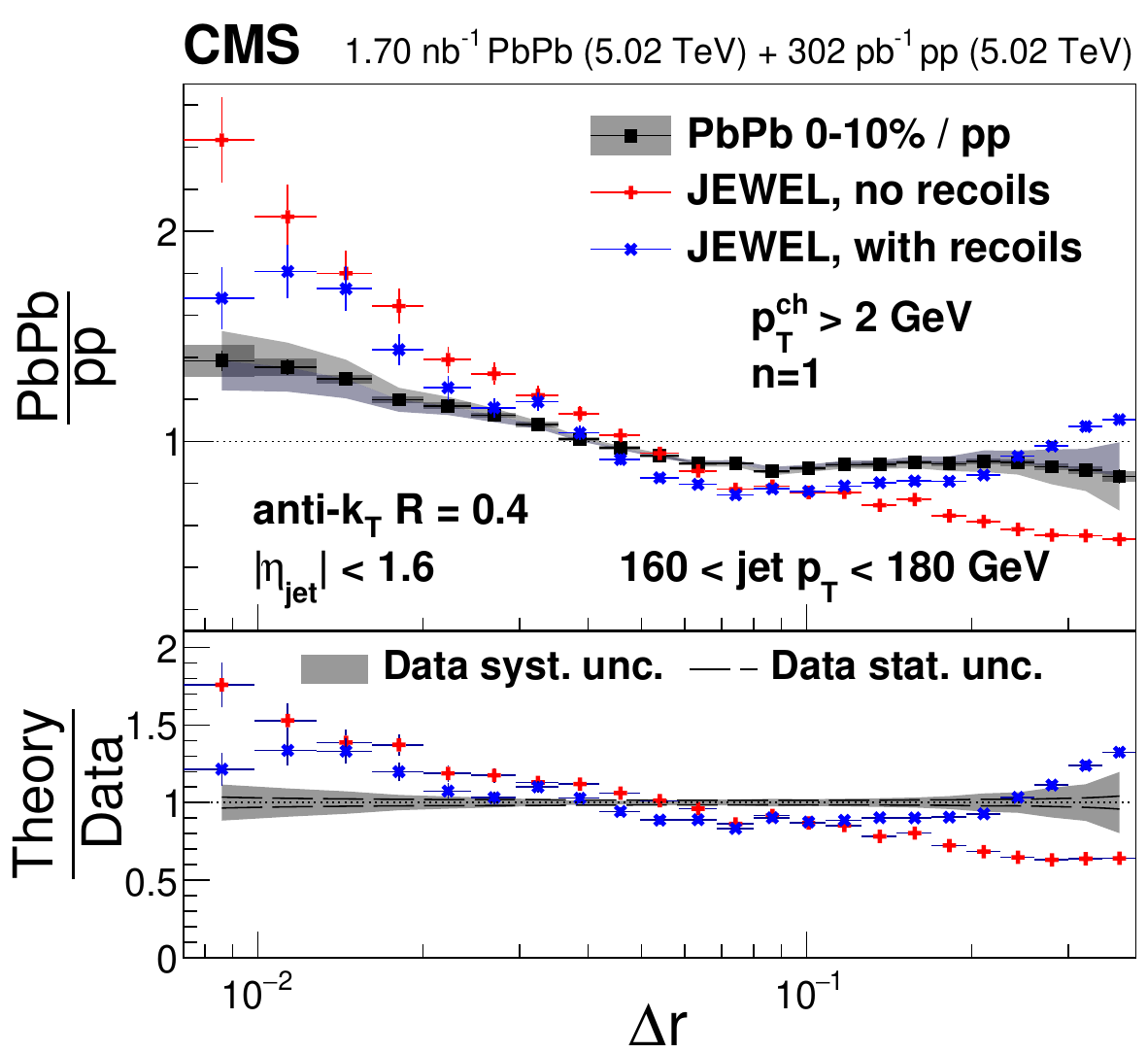}
    \includegraphics[width=0.4\textwidth]{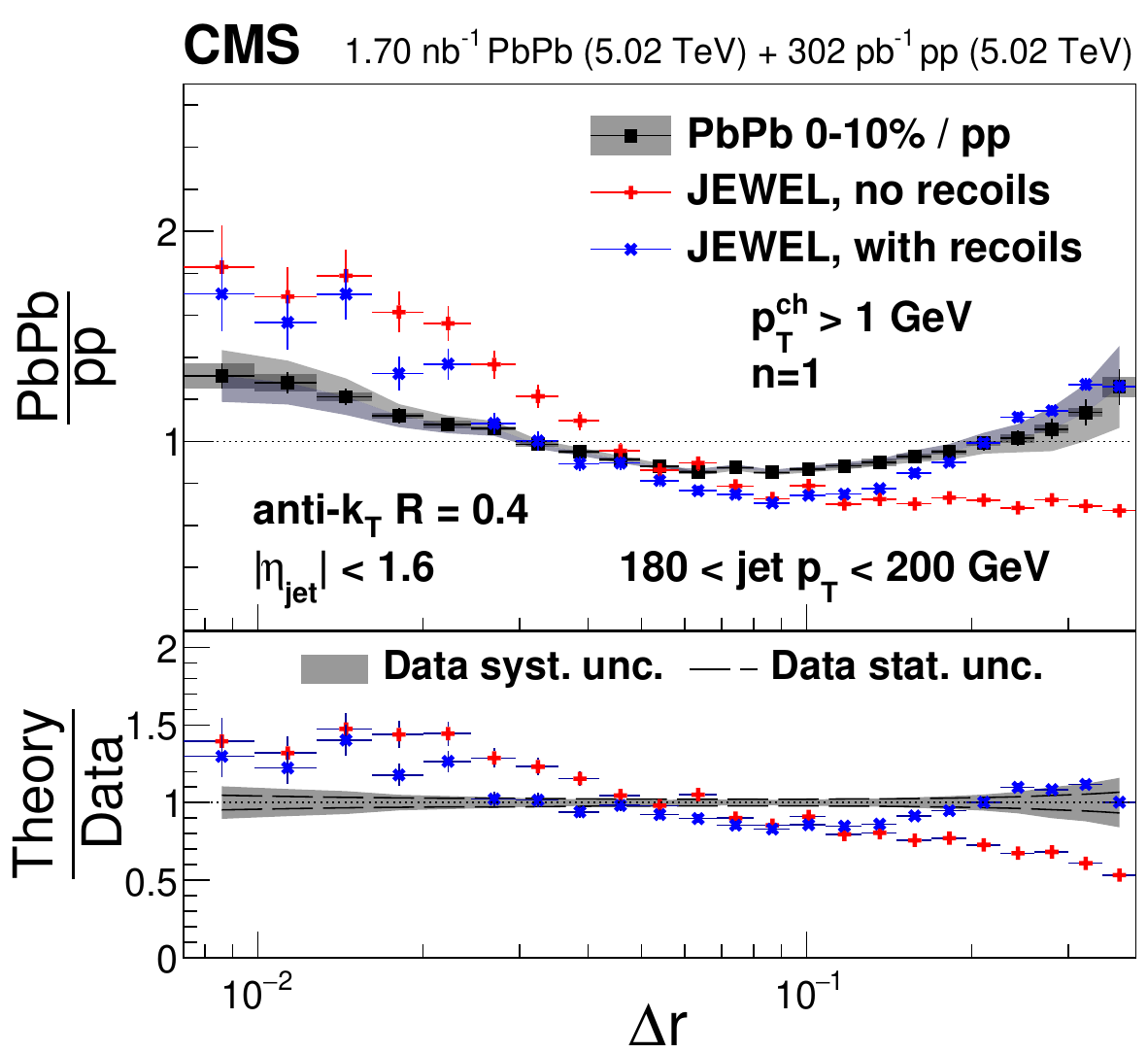}
    \includegraphics[width=0.4\textwidth]{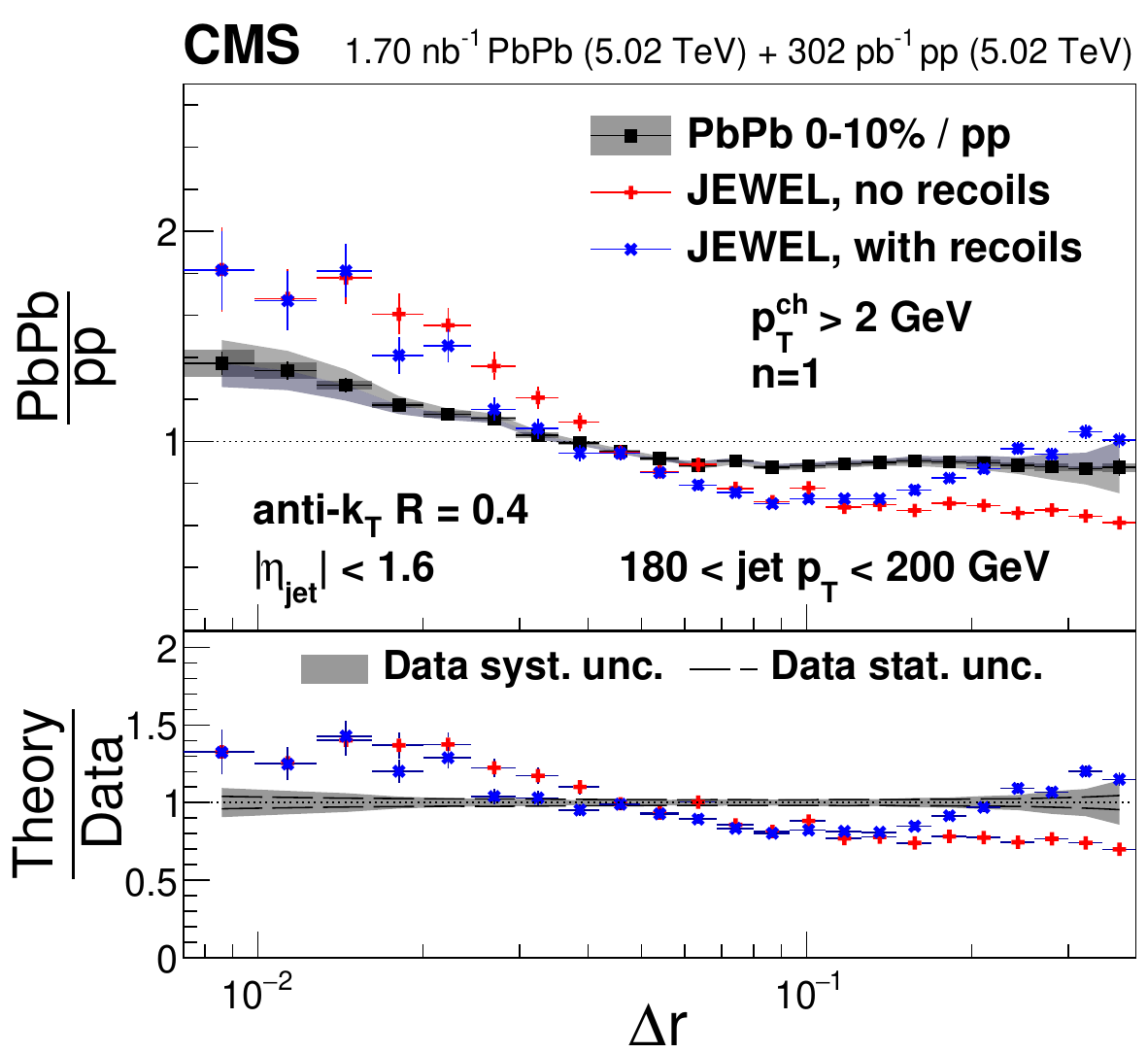}
    \caption{Ratios of 0--10\% central \PbPb to \pp energy-energy correlators with $n=1$ in different \ptCh and \ptjet bins compared to \JEWEL predictions with and without recoils.} 
    \label{fig:PbPbRatiosNominalJewel}
  } 
\end{figure*}

\begin{figure*}[h]
  \centering{
    \includegraphics[width=0.4\textwidth]{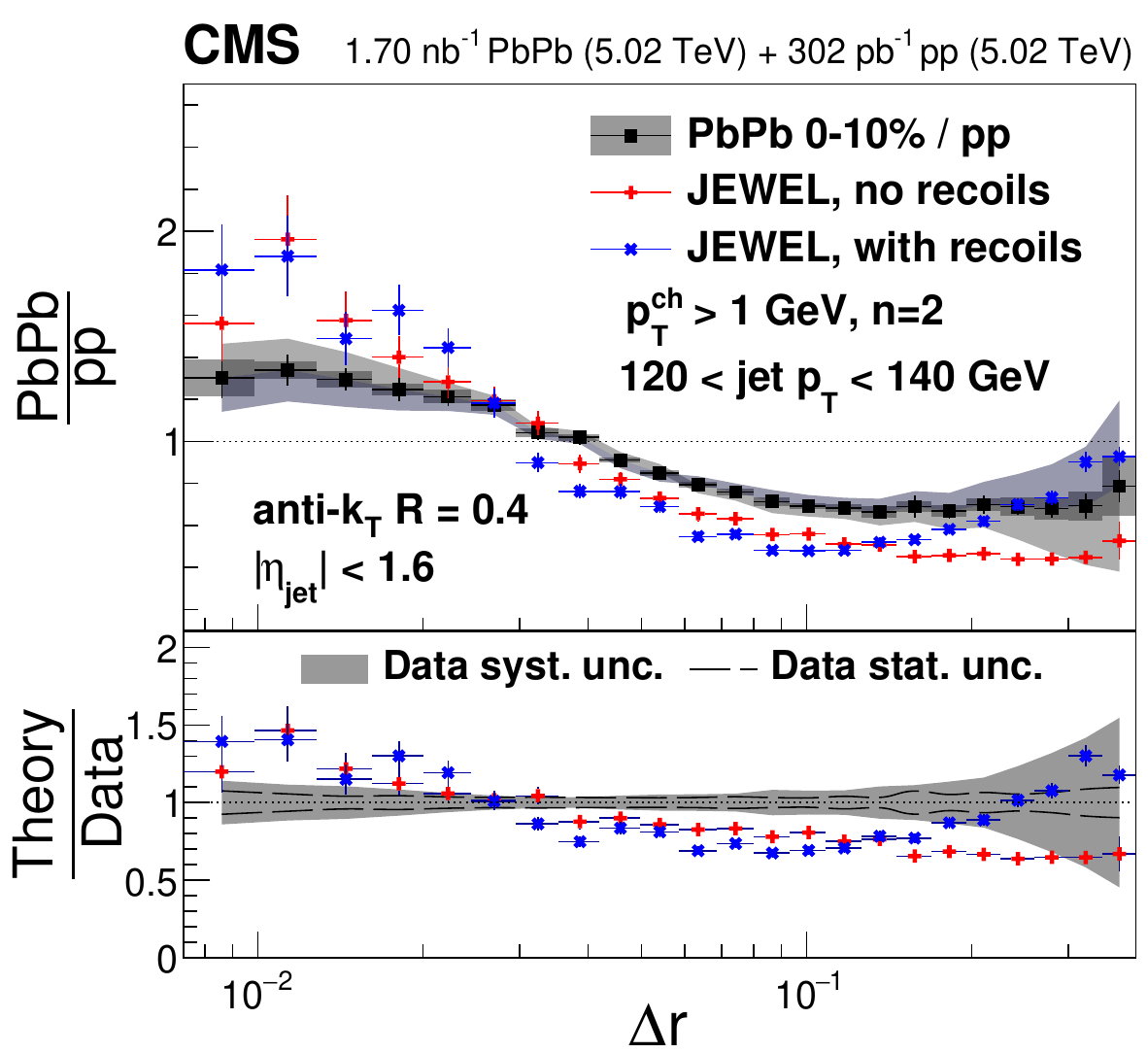}
    \includegraphics[width=0.4\textwidth]{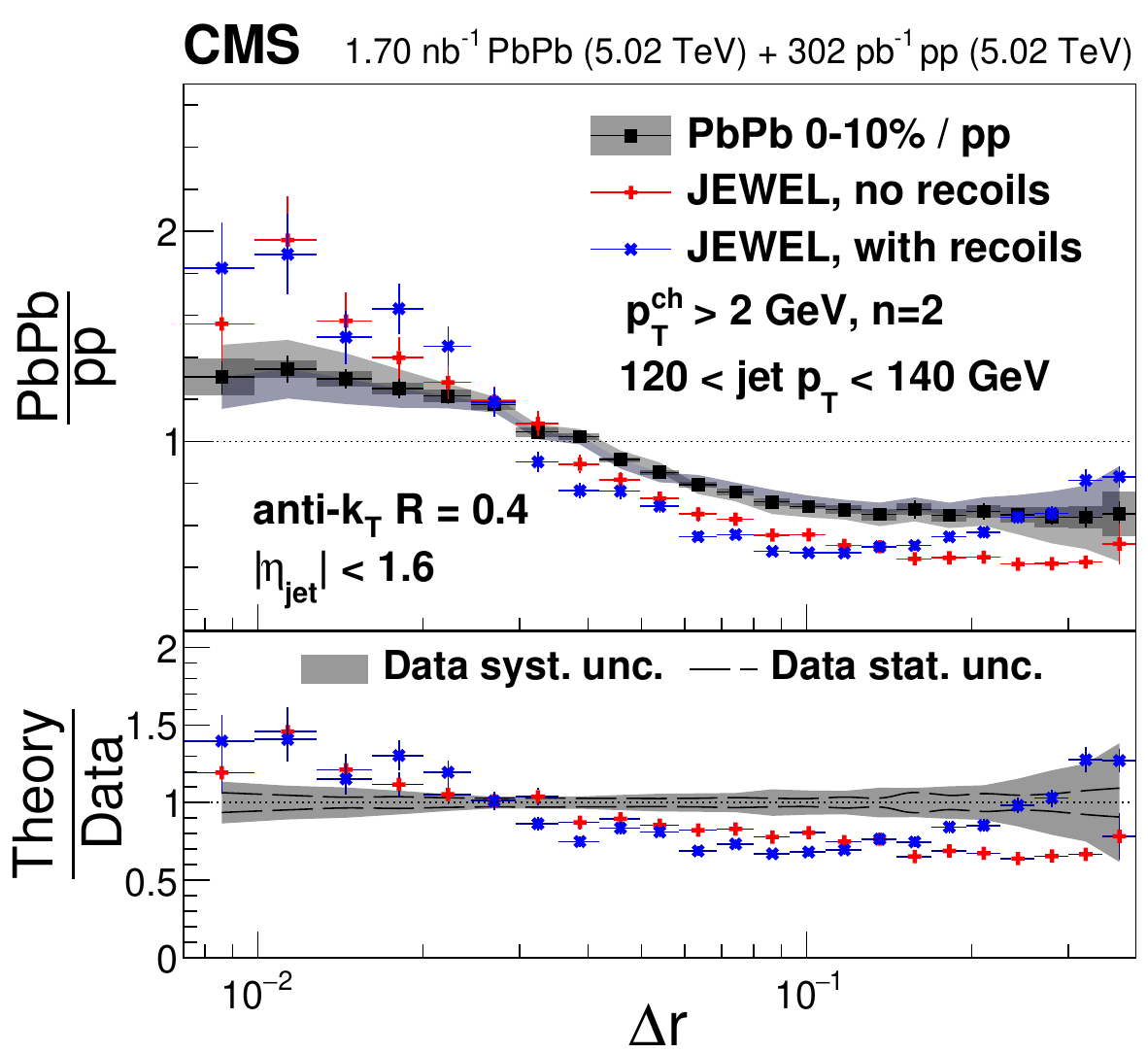} 
    \includegraphics[width=0.4\textwidth]{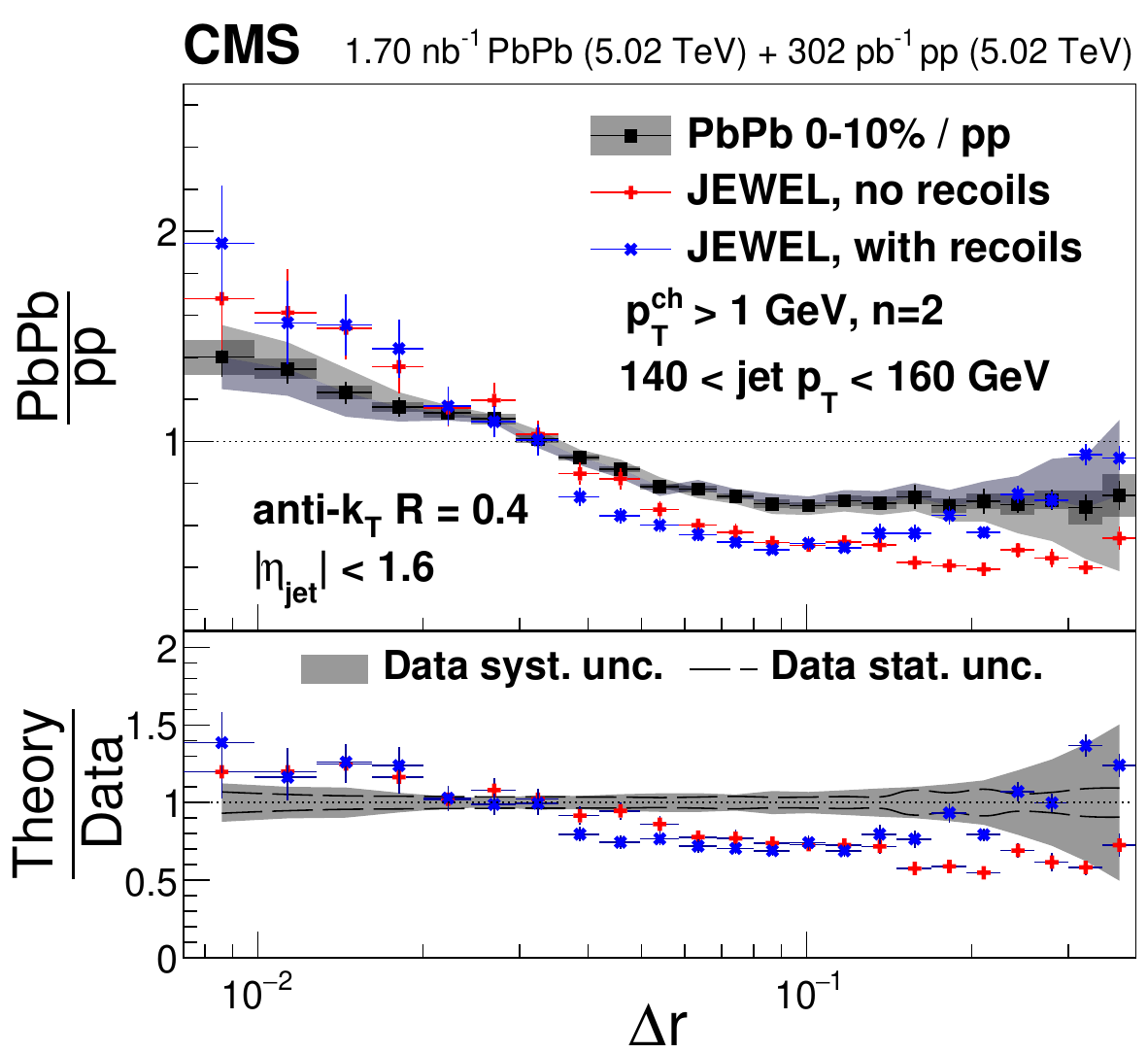}
    \includegraphics[width=0.4\textwidth]{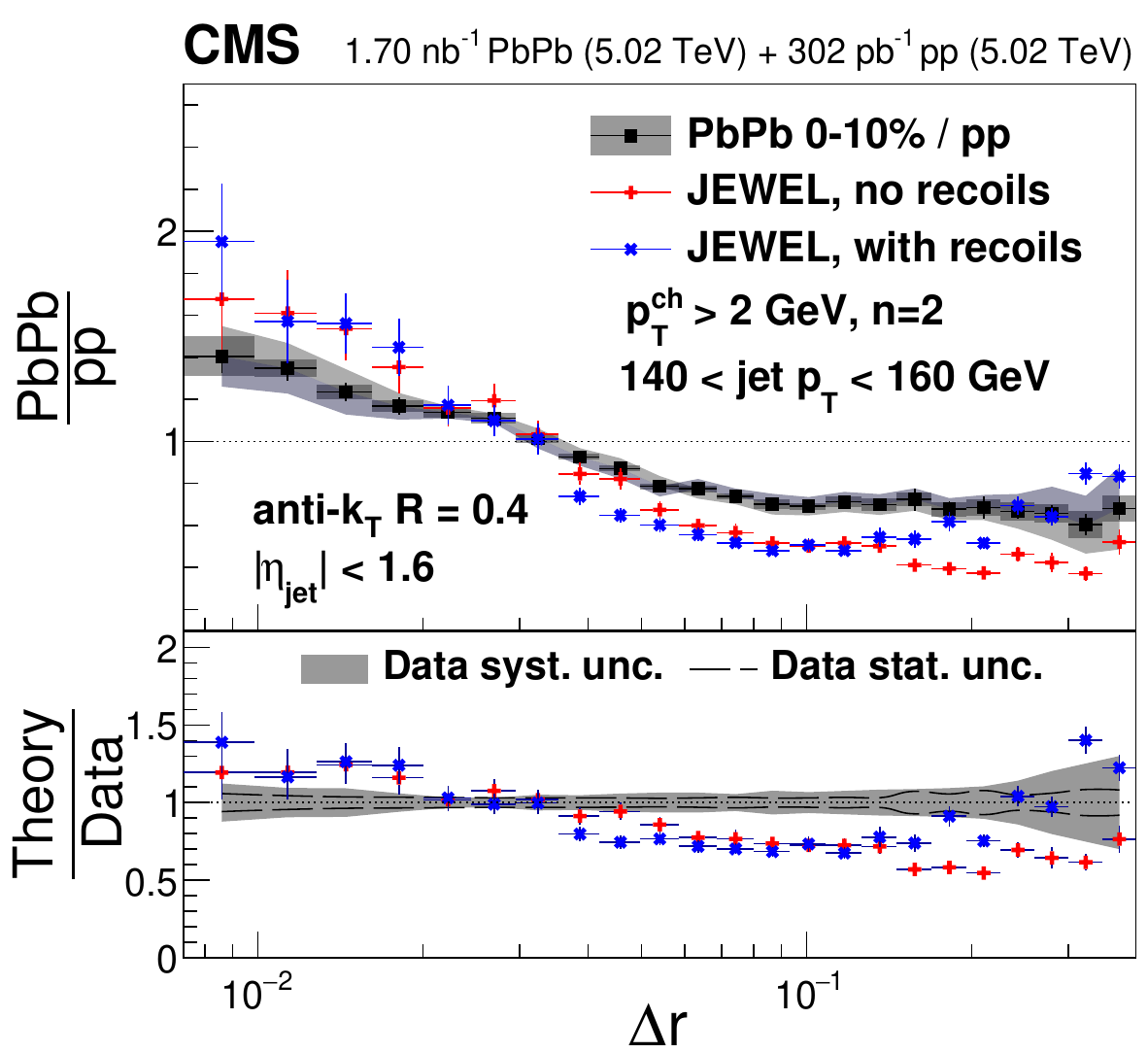}
    \includegraphics[width=0.4\textwidth]{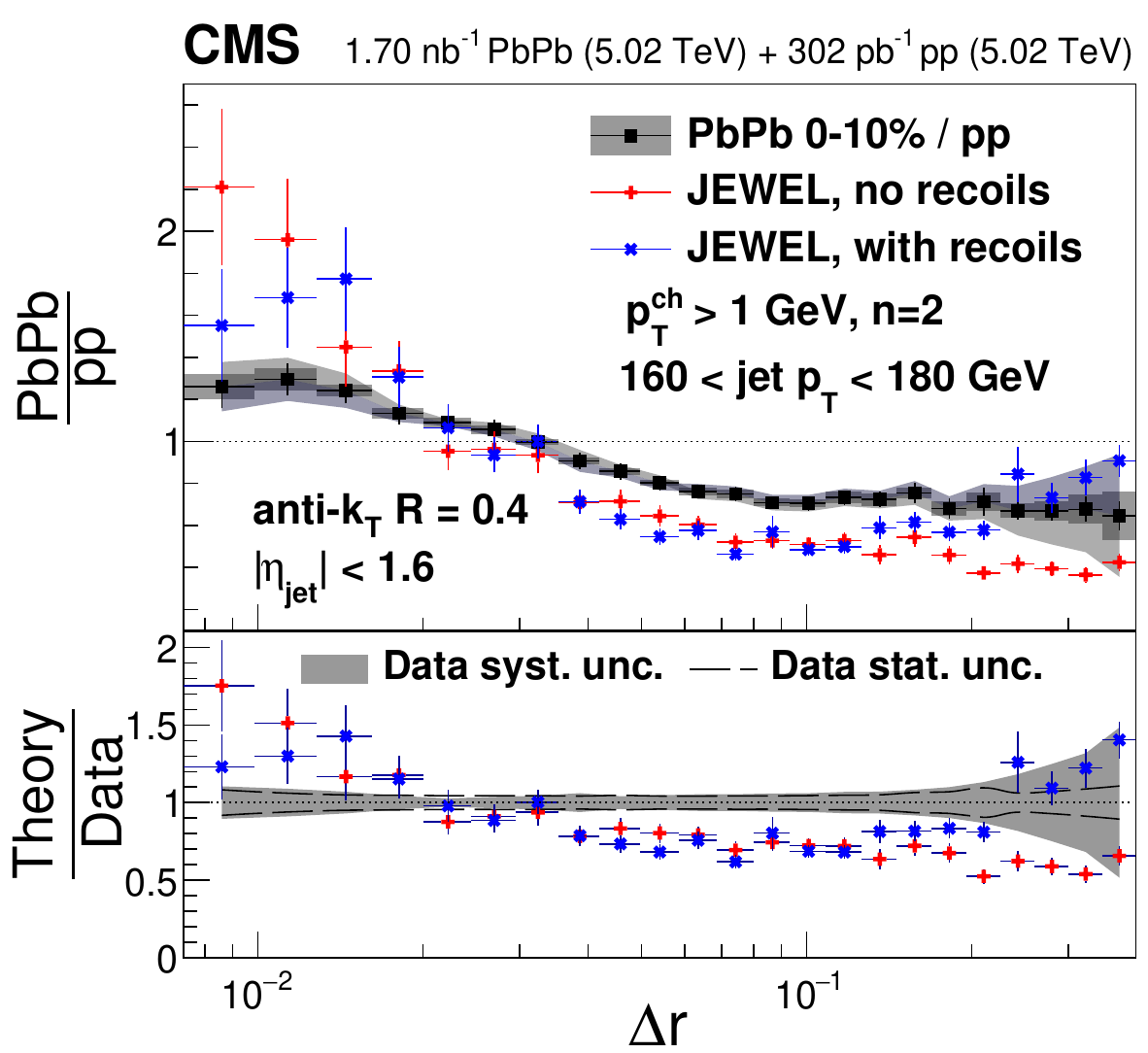}
    \includegraphics[width=0.4\textwidth]{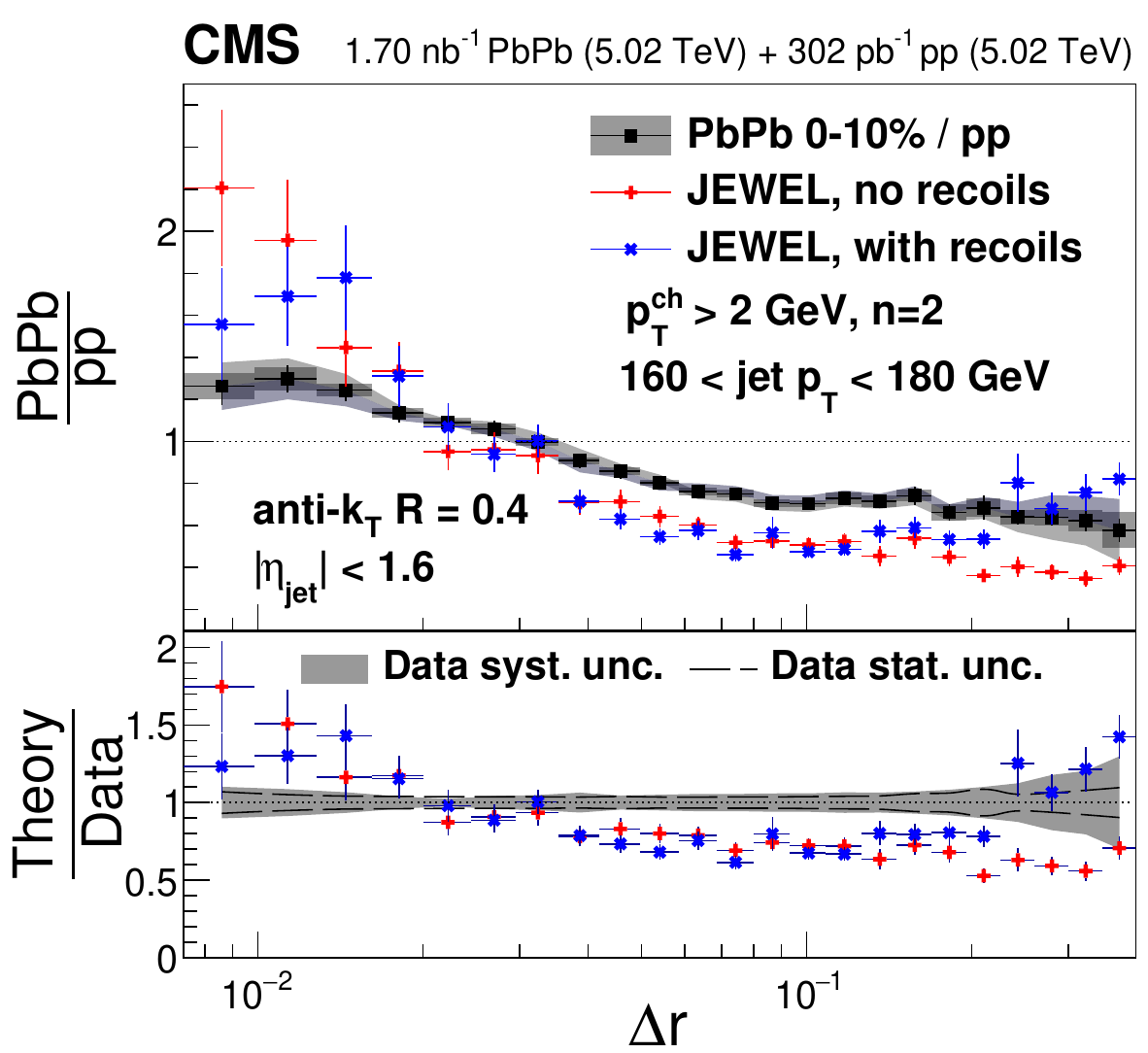}
    \includegraphics[width=0.4\textwidth]{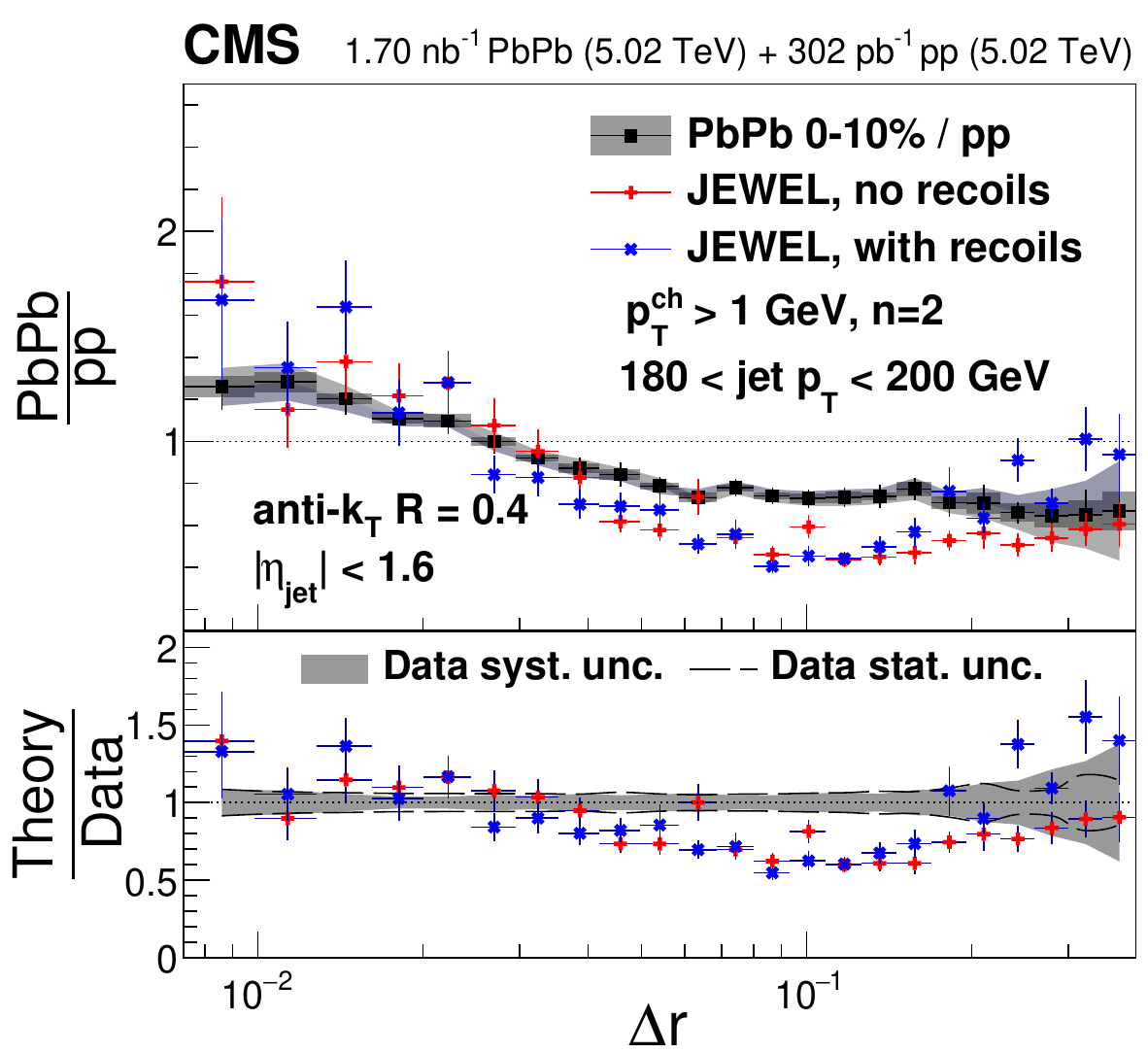}
    \includegraphics[width=0.4\textwidth]{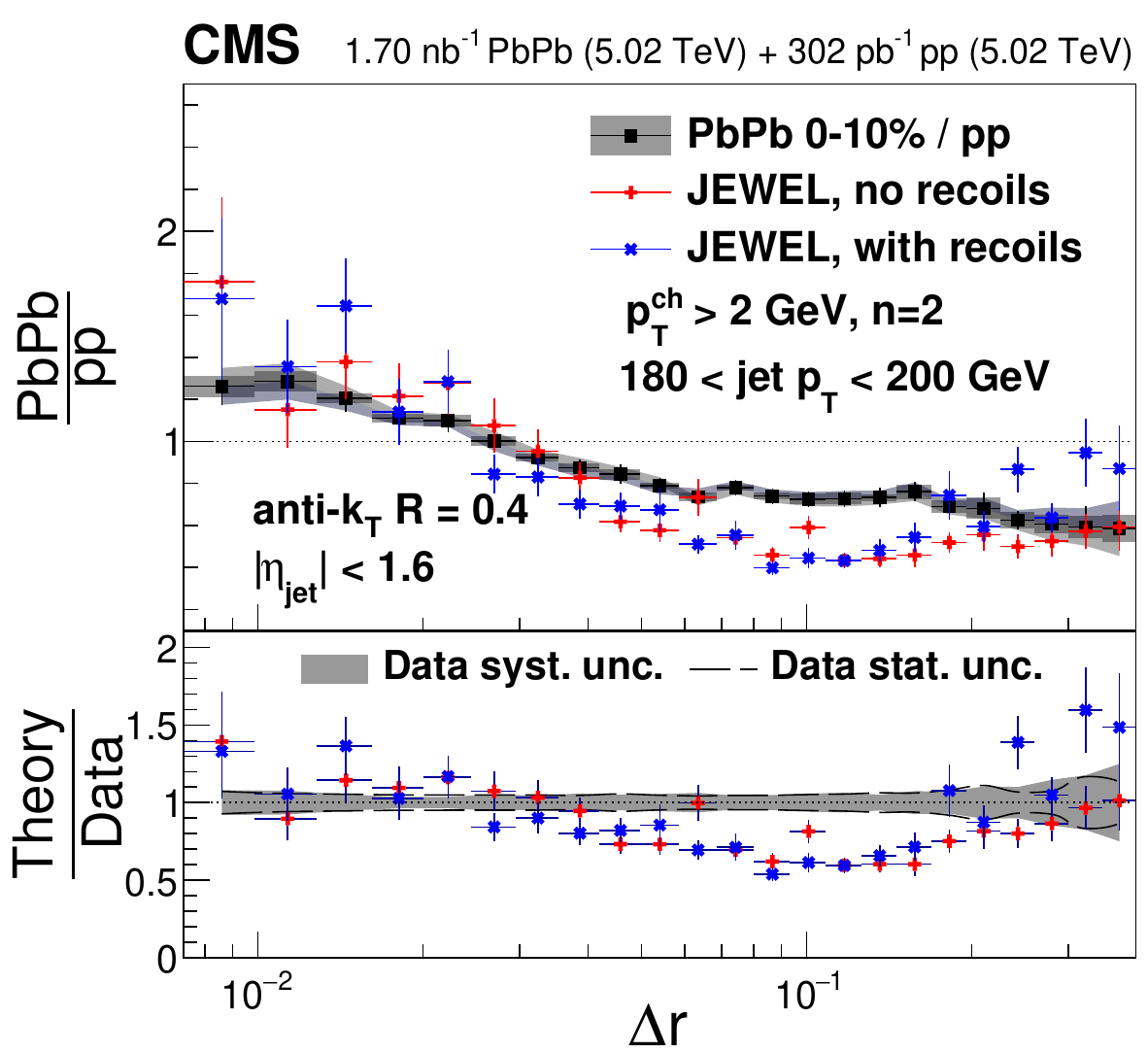}
    \caption{Ratios of 0--10\% central \PbPb to \pp energy-energy correlators with $n=2$ in different \ptCh and \ptjet bins compared to \JEWEL predictions with and without recoils.} 
    \label{fig:PbPbRatiosSquaredJewel}
  } 
\end{figure*}

\begin{figure*}[h]
  \centering{
    \includegraphics[width=0.4\textwidth]{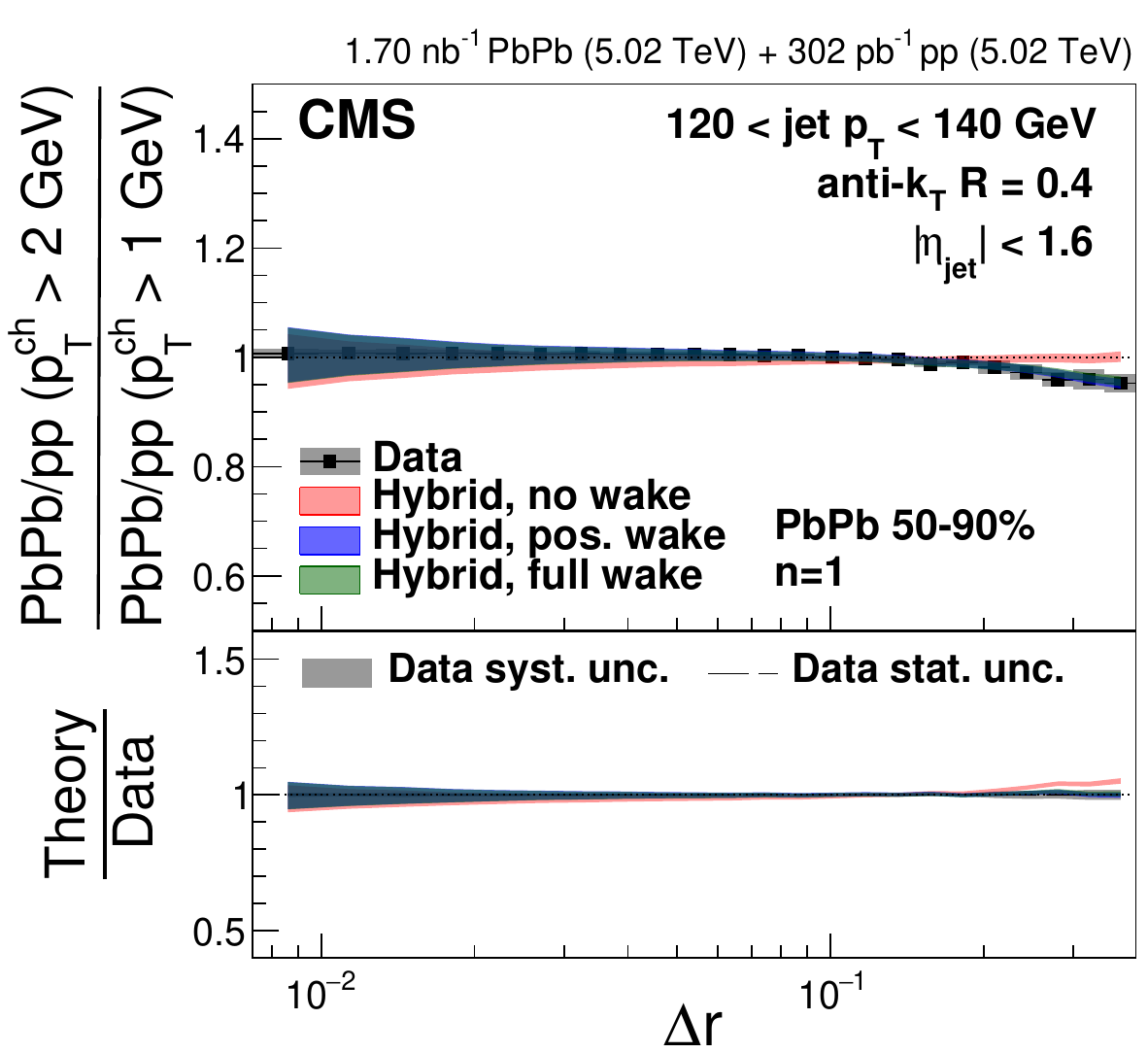}
    \includegraphics[width=0.4\textwidth]{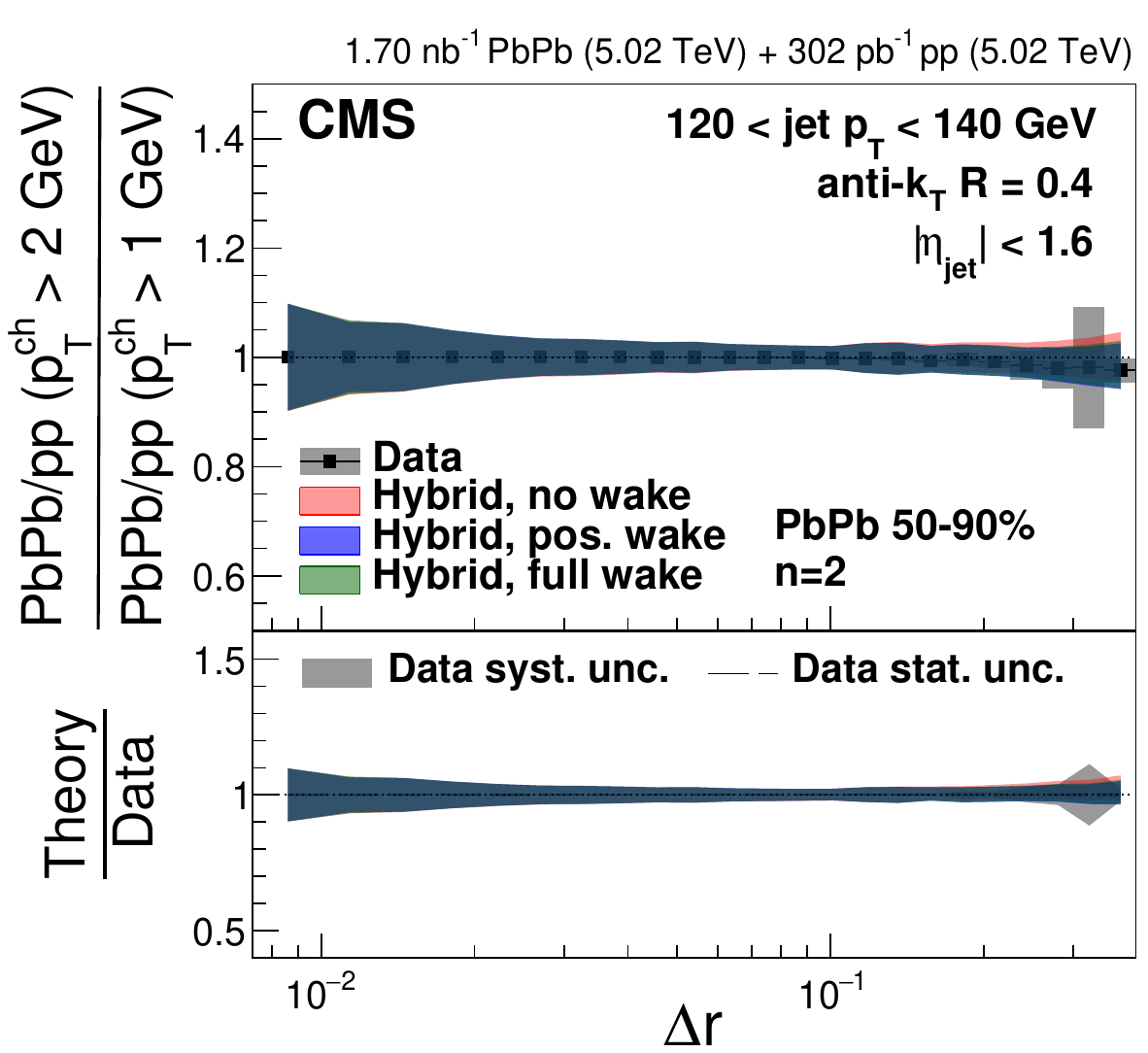}
    \includegraphics[width=0.4\textwidth]{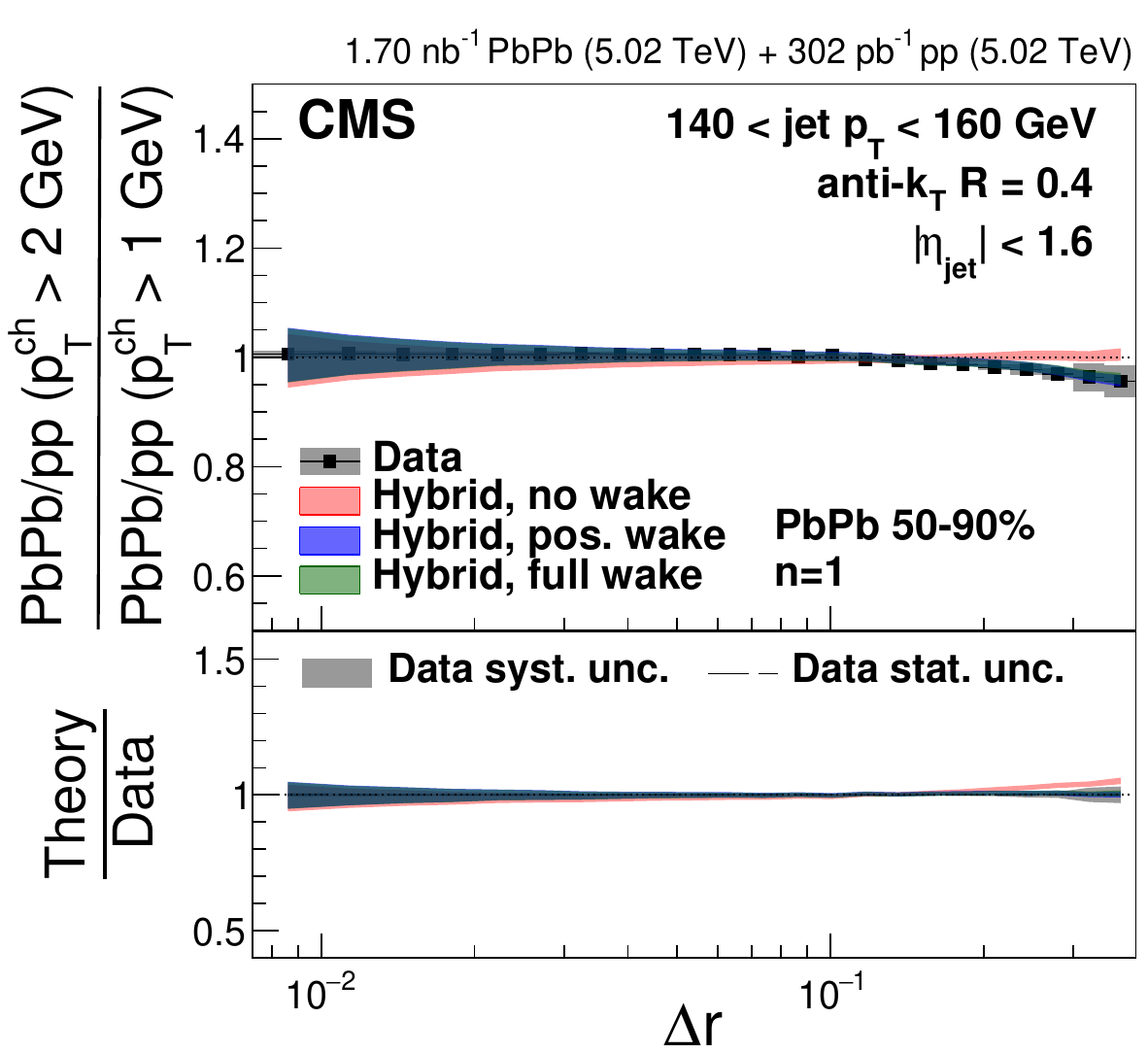} 
    \includegraphics[width=0.4\textwidth]{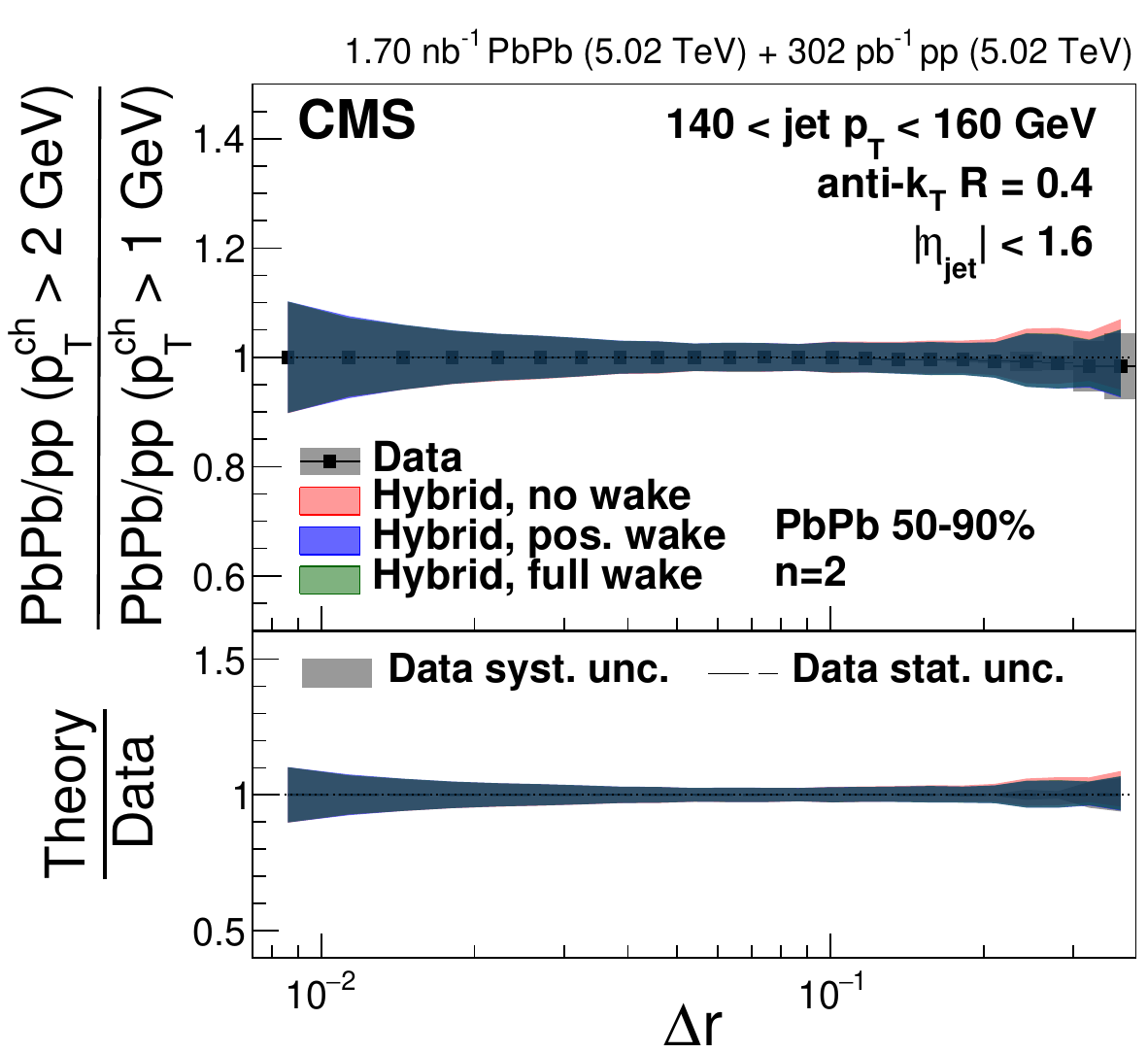}
    \includegraphics[width=0.4\textwidth]{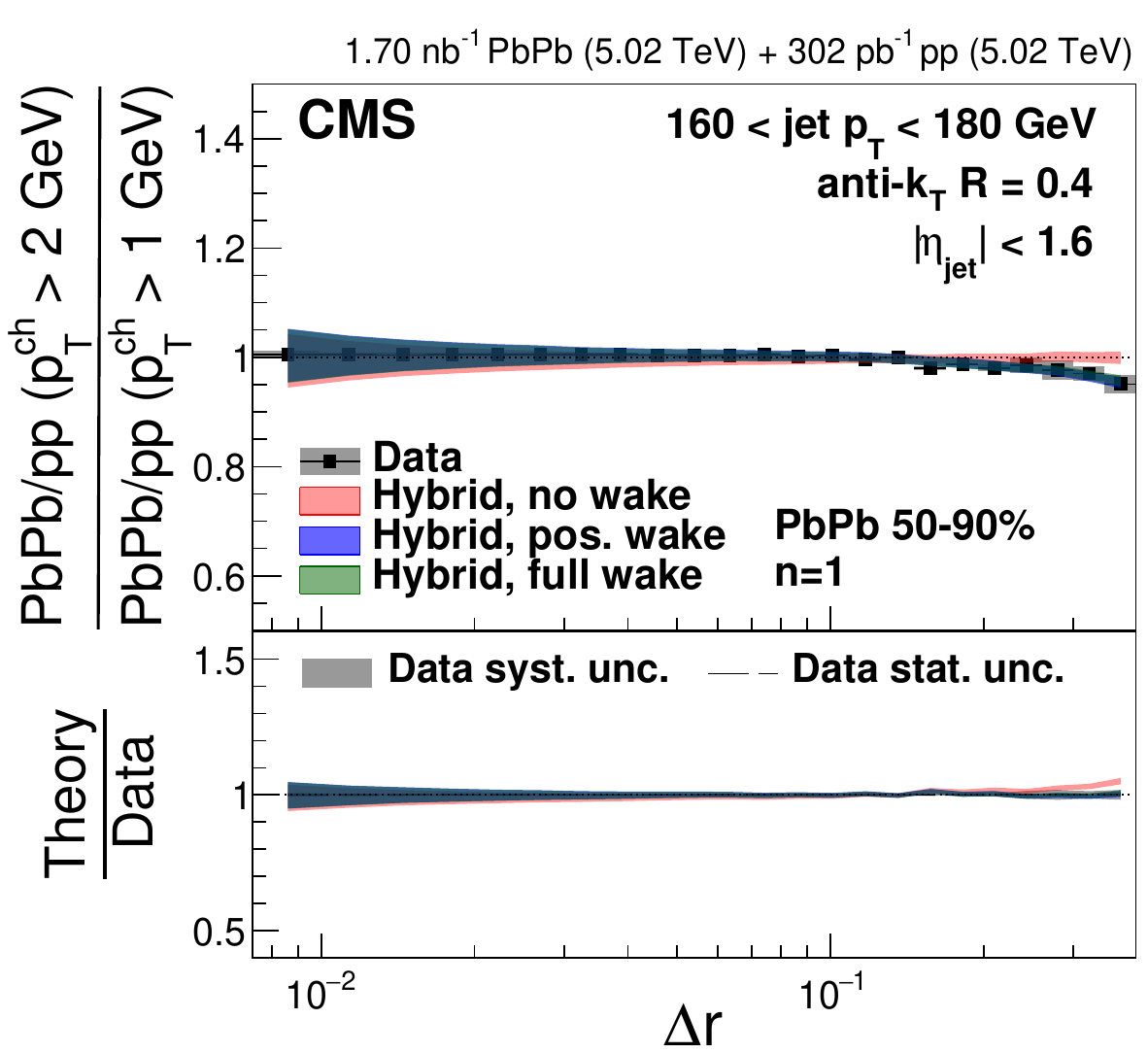}
    \includegraphics[width=0.4\textwidth]{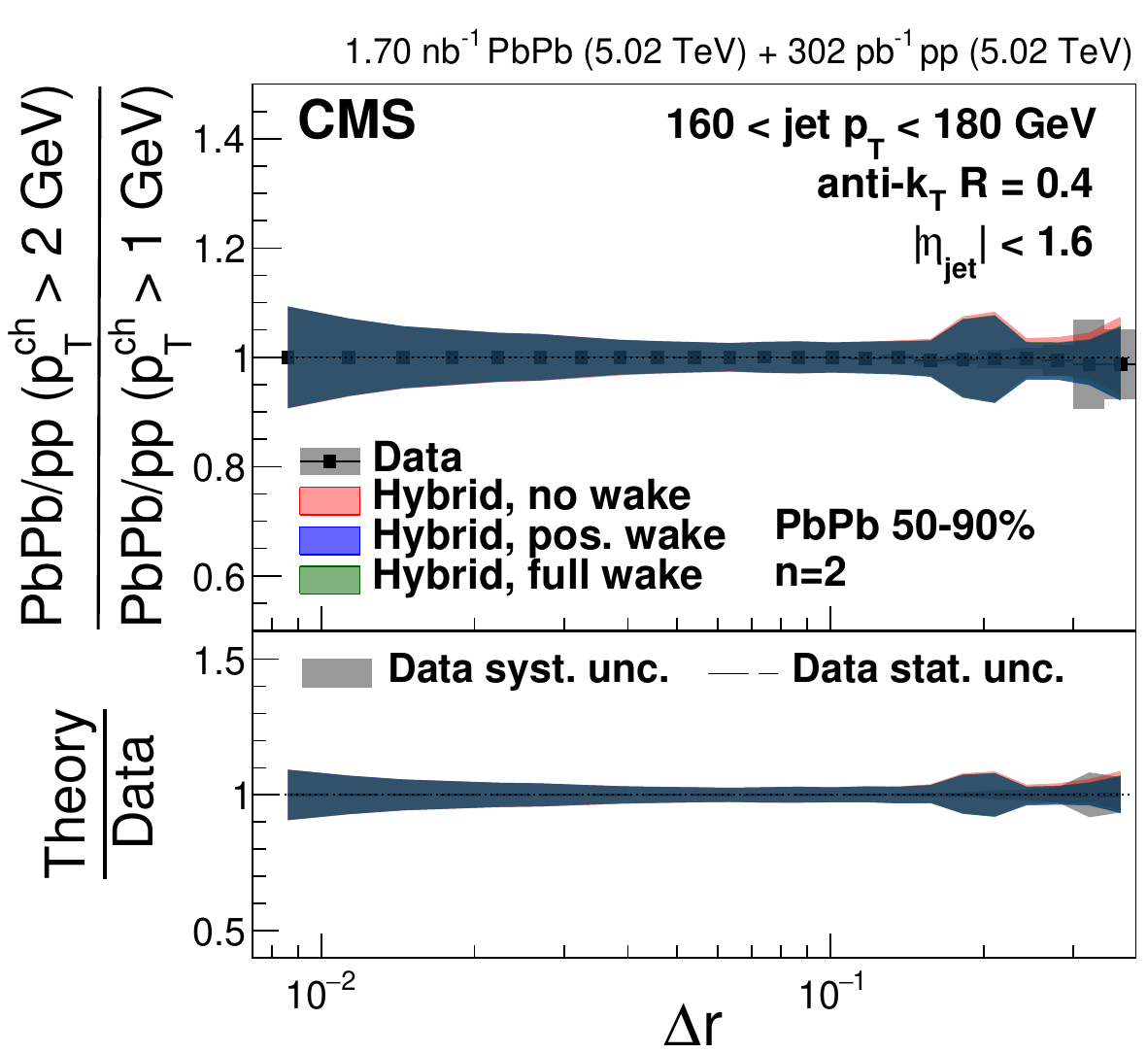}
    \includegraphics[width=0.4\textwidth]{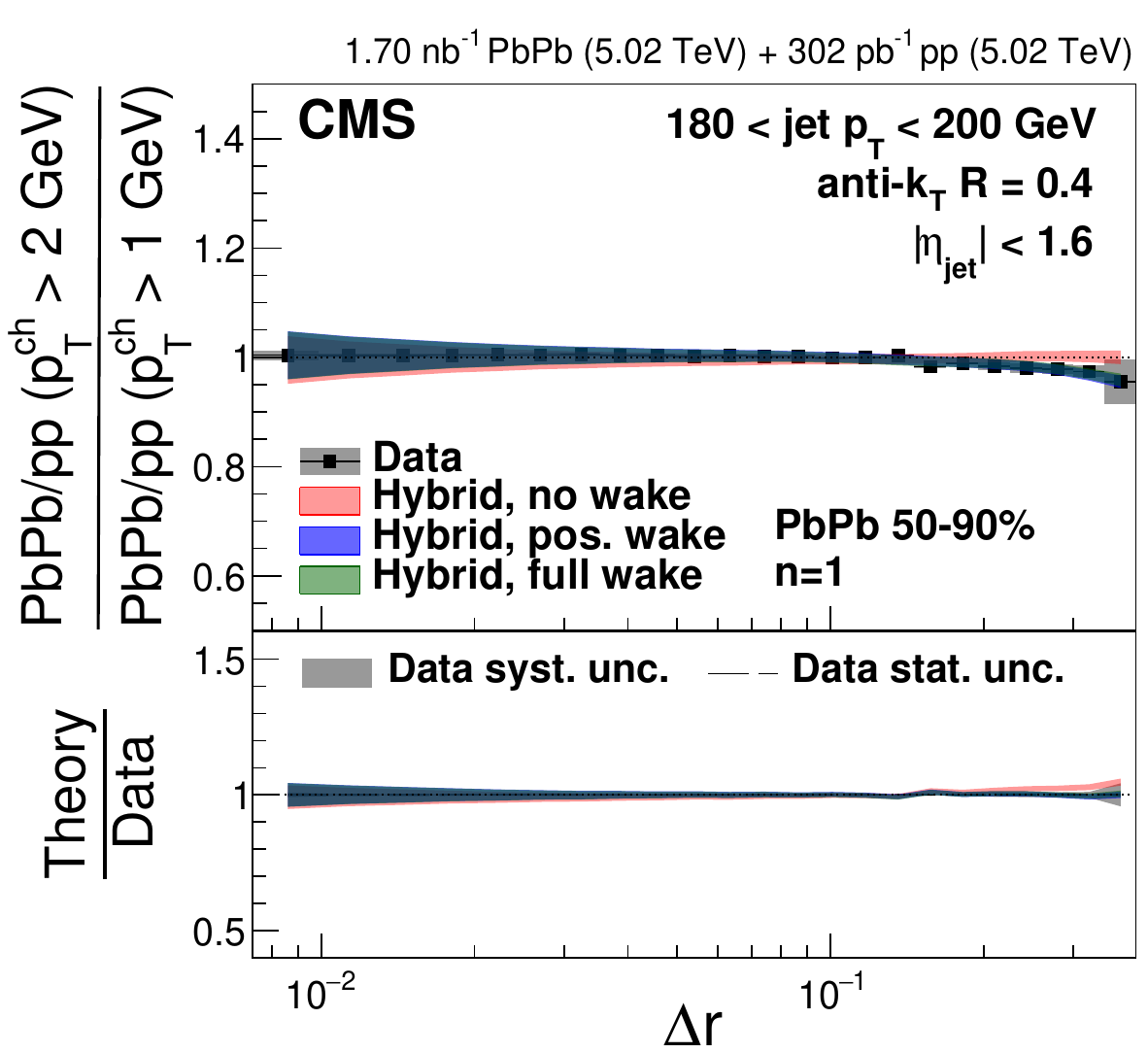}
    \includegraphics[width=0.4\textwidth]{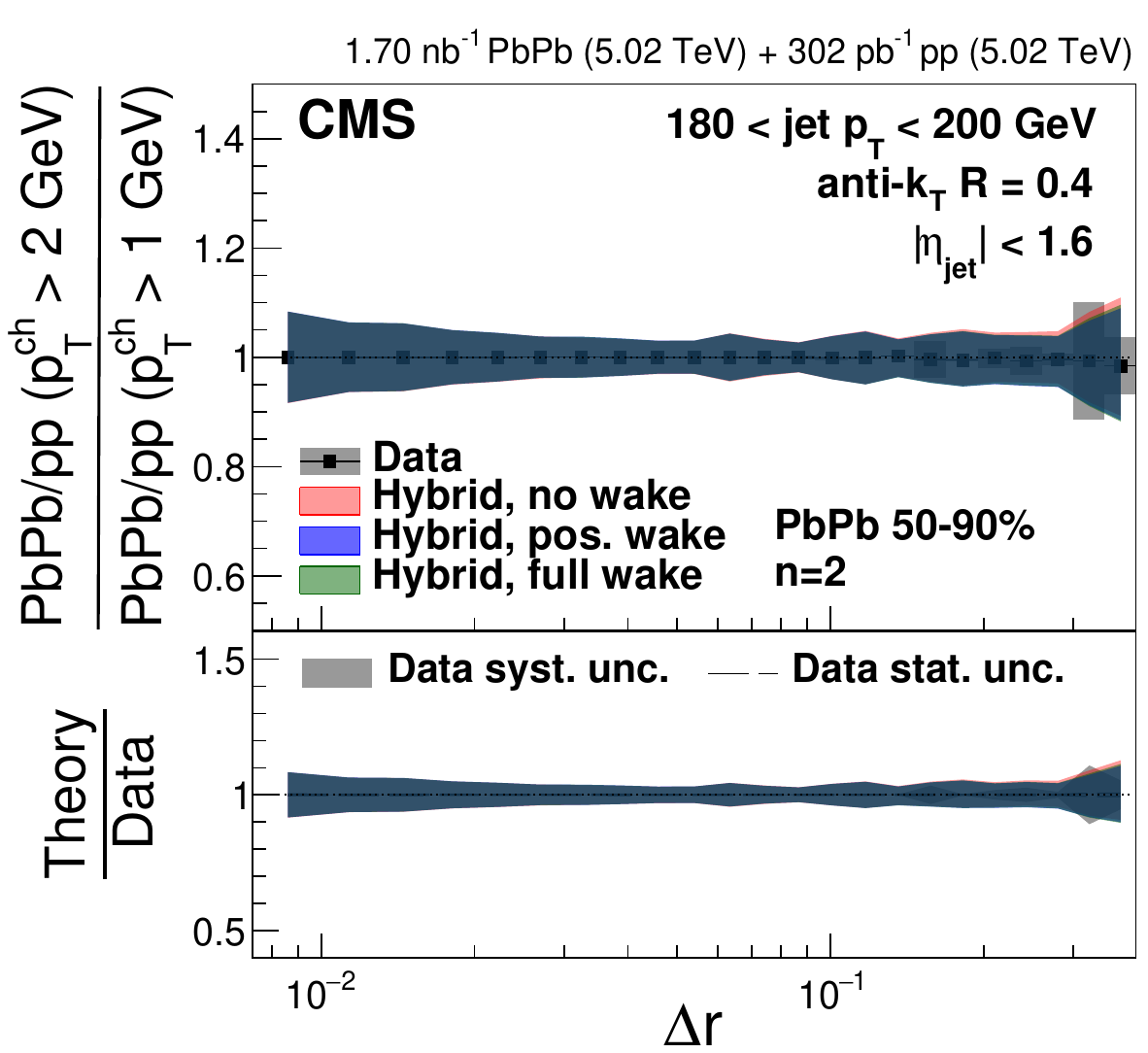}
    \caption{The double ratios of 50--90\% central \PbPb to \pp single ratios with $\ptCh > 2\GeV$ and $\ptCh > 1\GeV$ for $n=1$ (left) and $n=2$ (right) in different \ptjet bins compared to the hybrid model predictions with different jet wake settings.} 
    \label{fig:doubleRatioHybridJet50-90}
  } 
\end{figure*}

\begin{figure*}[h]
  \centering{
    \includegraphics[width=0.4\textwidth]{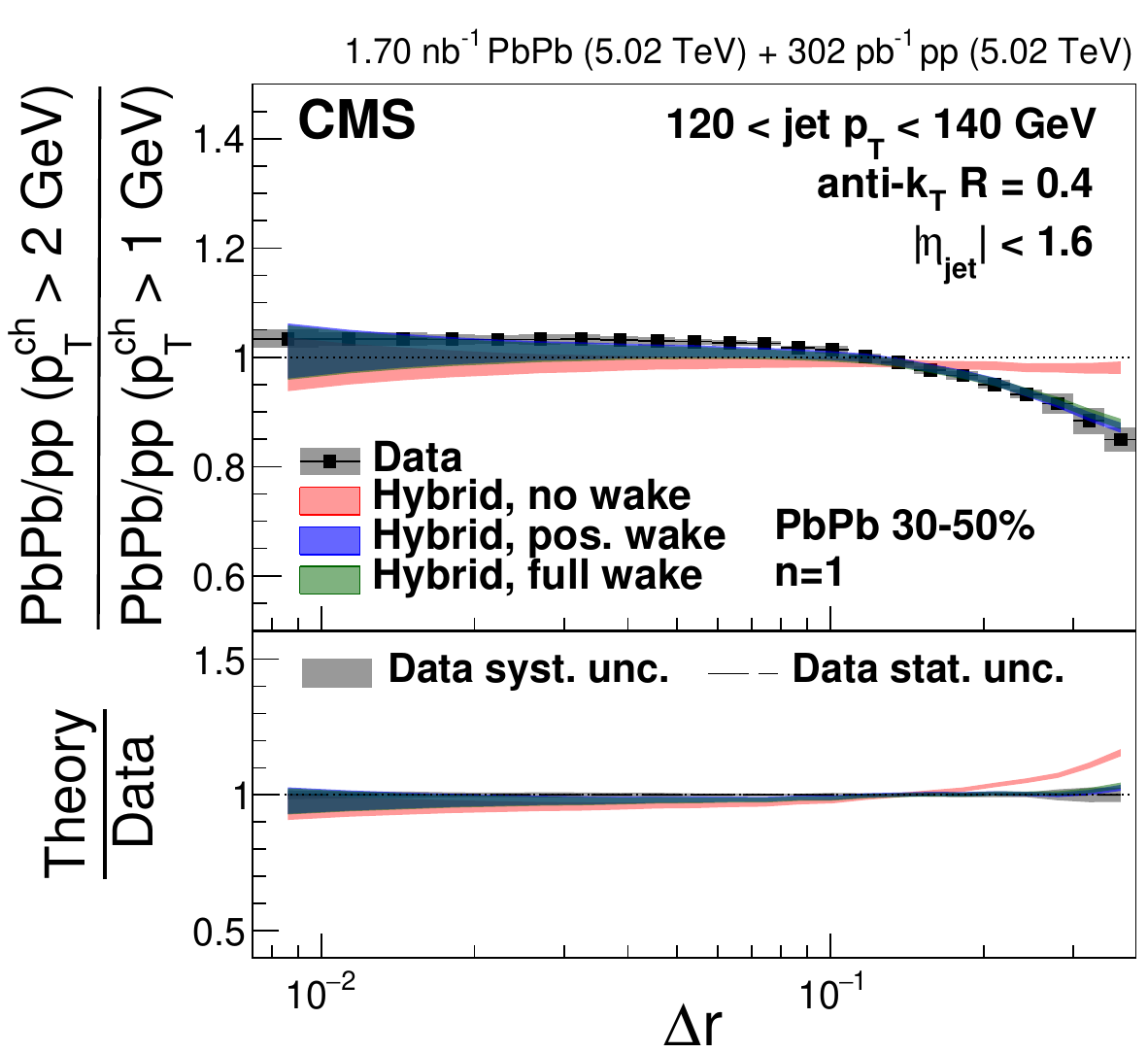}
    \includegraphics[width=0.4\textwidth]{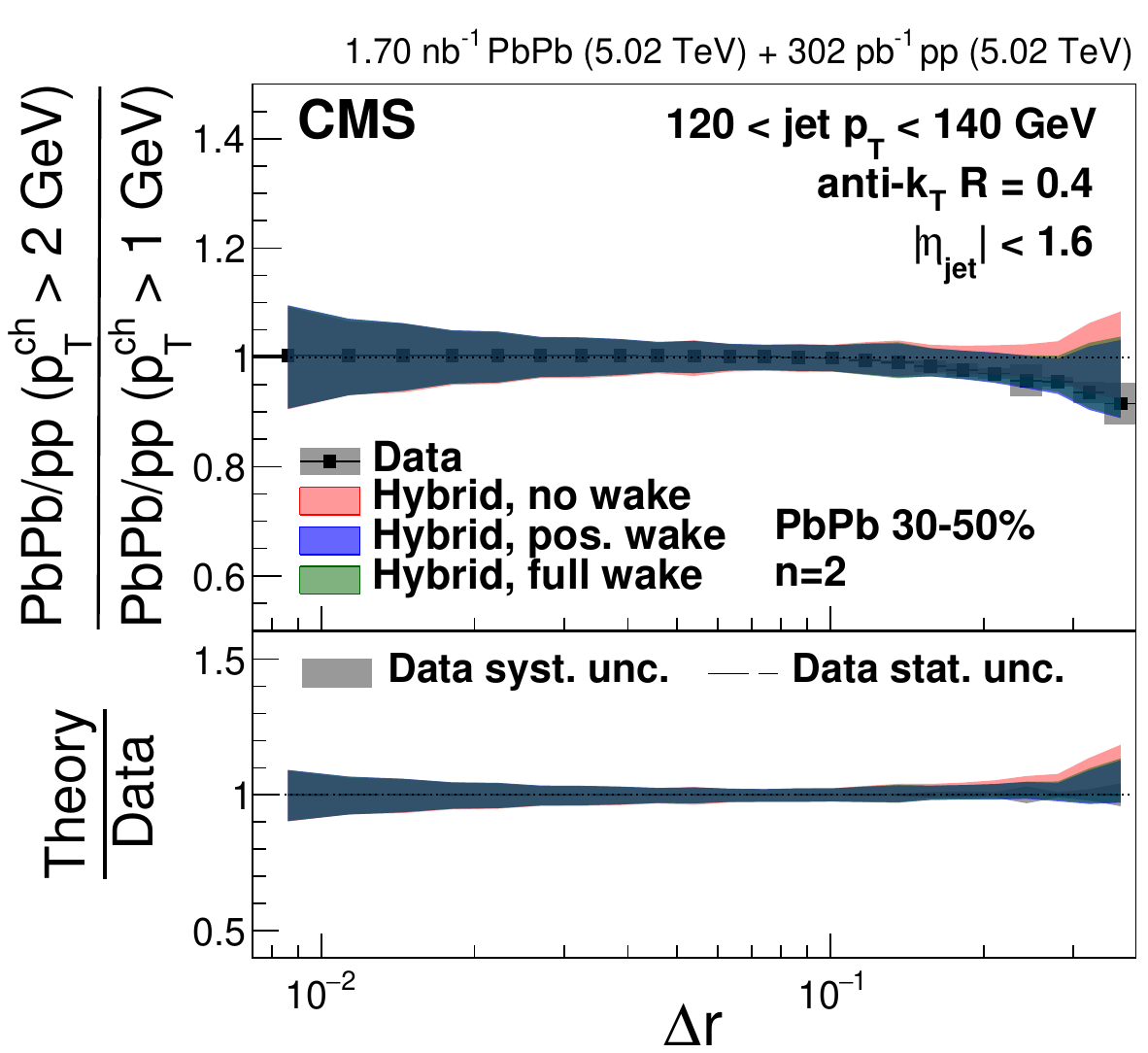}
    \includegraphics[width=0.4\textwidth]{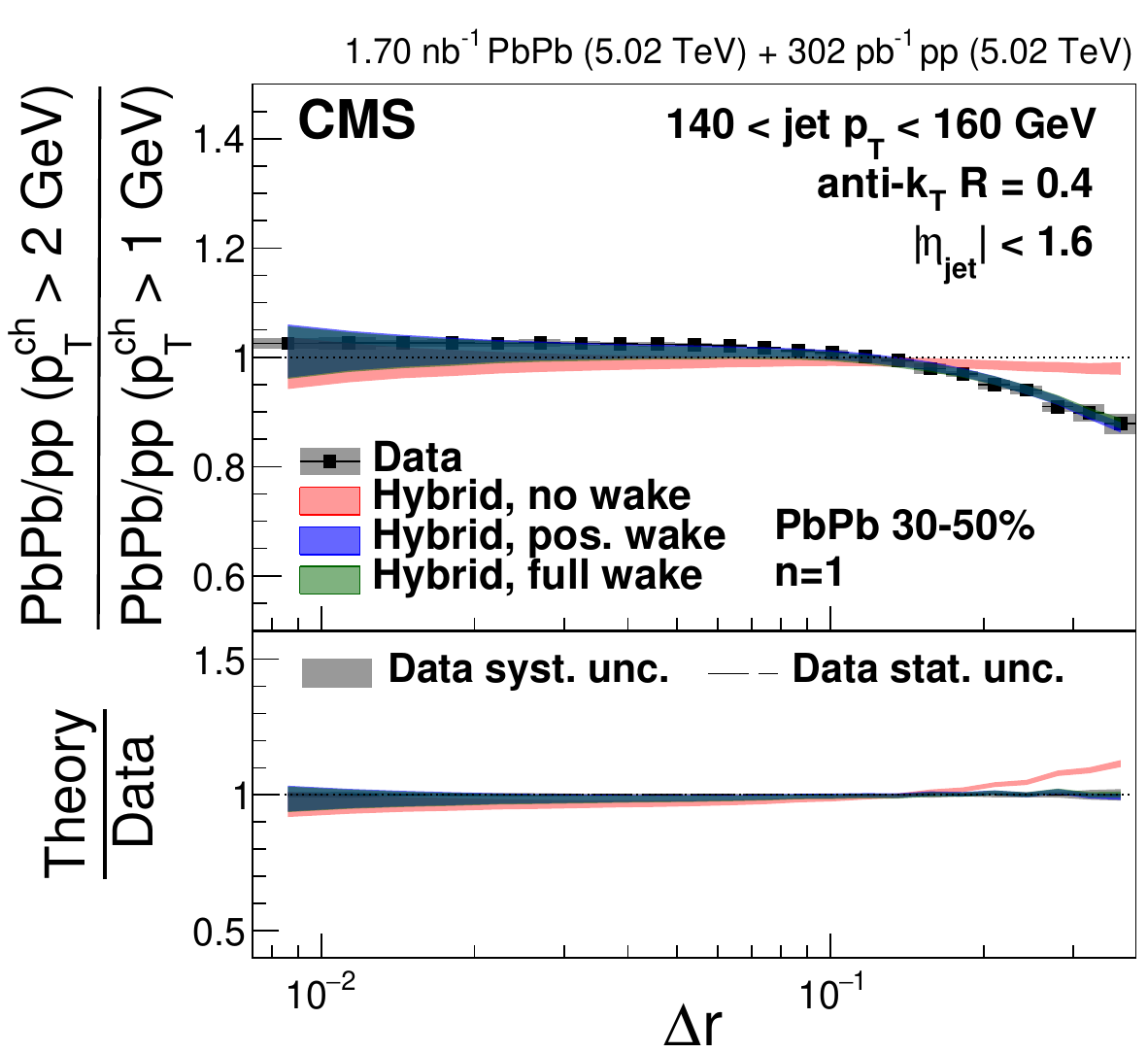} 
    \includegraphics[width=0.4\textwidth]{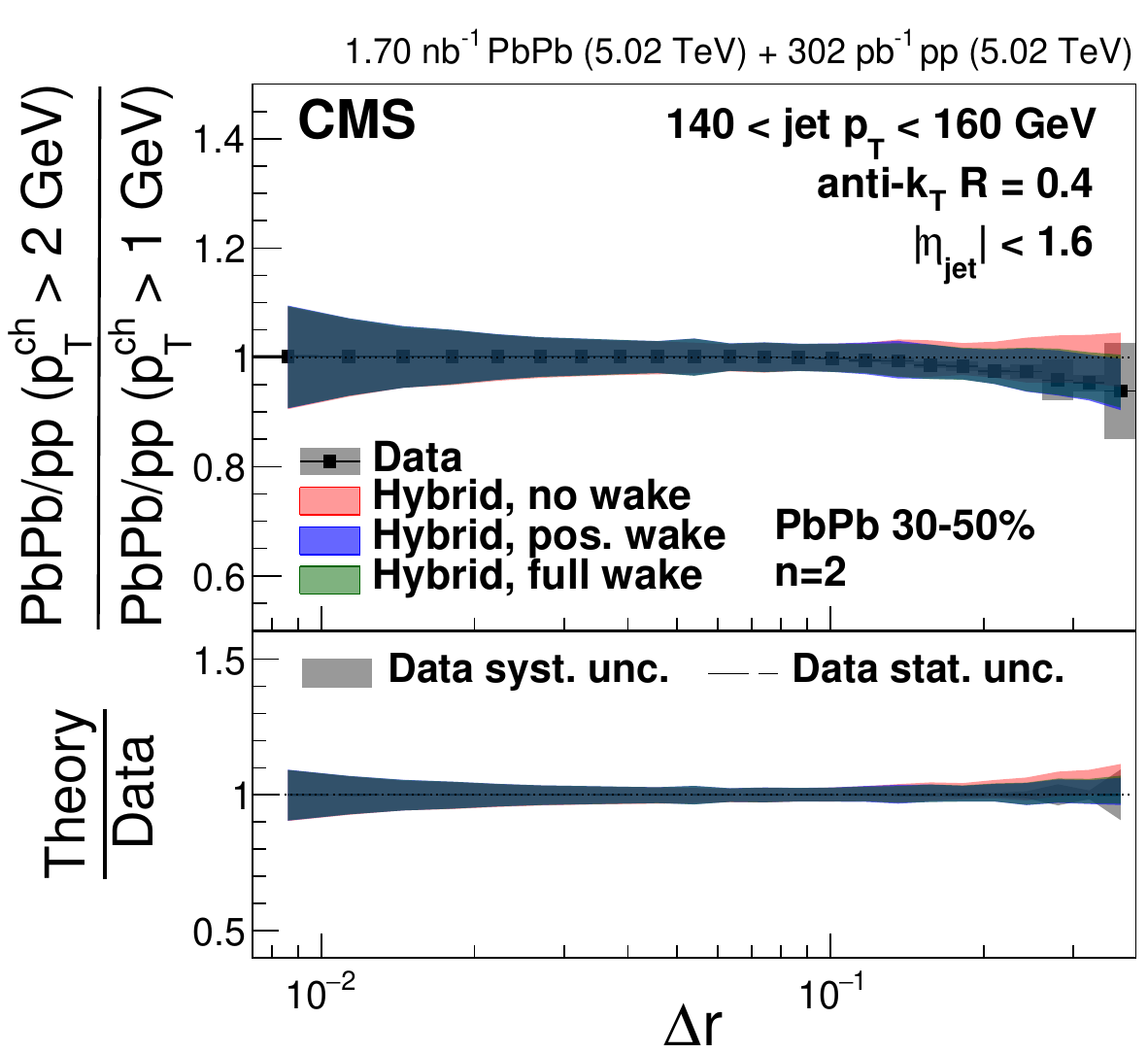}
    \includegraphics[width=0.4\textwidth]{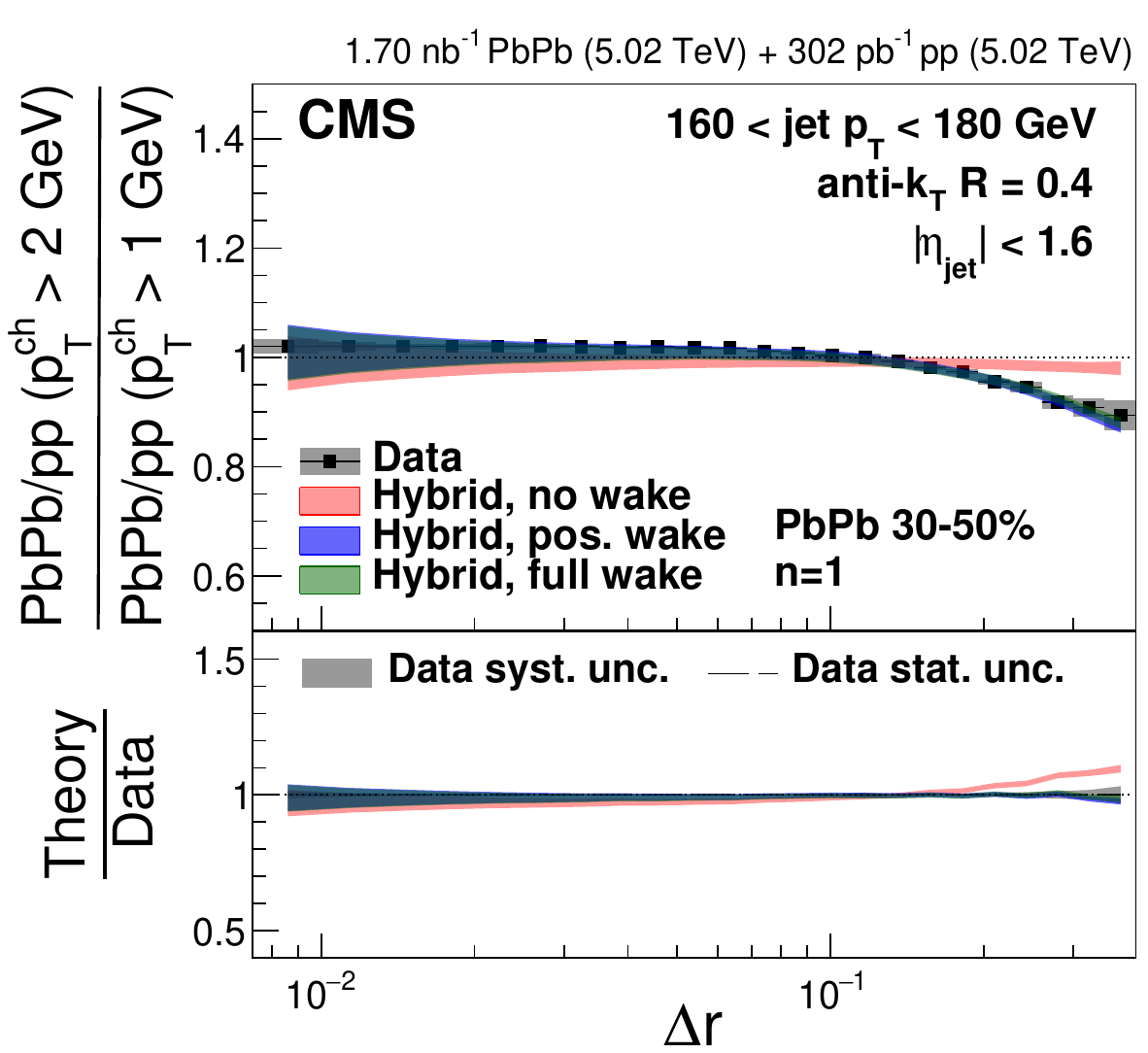}
    \includegraphics[width=0.4\textwidth]{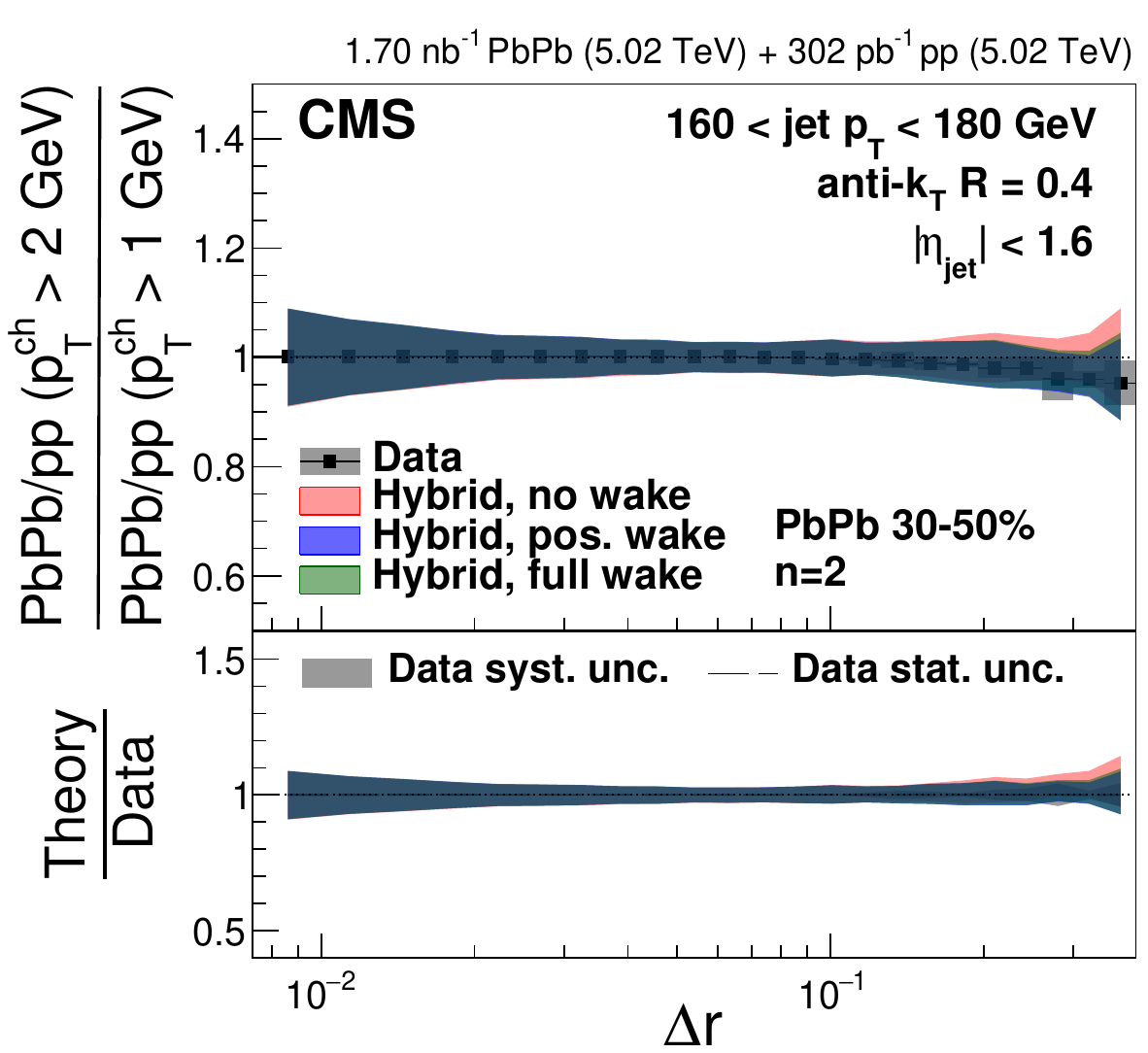}
    \includegraphics[width=0.4\textwidth]{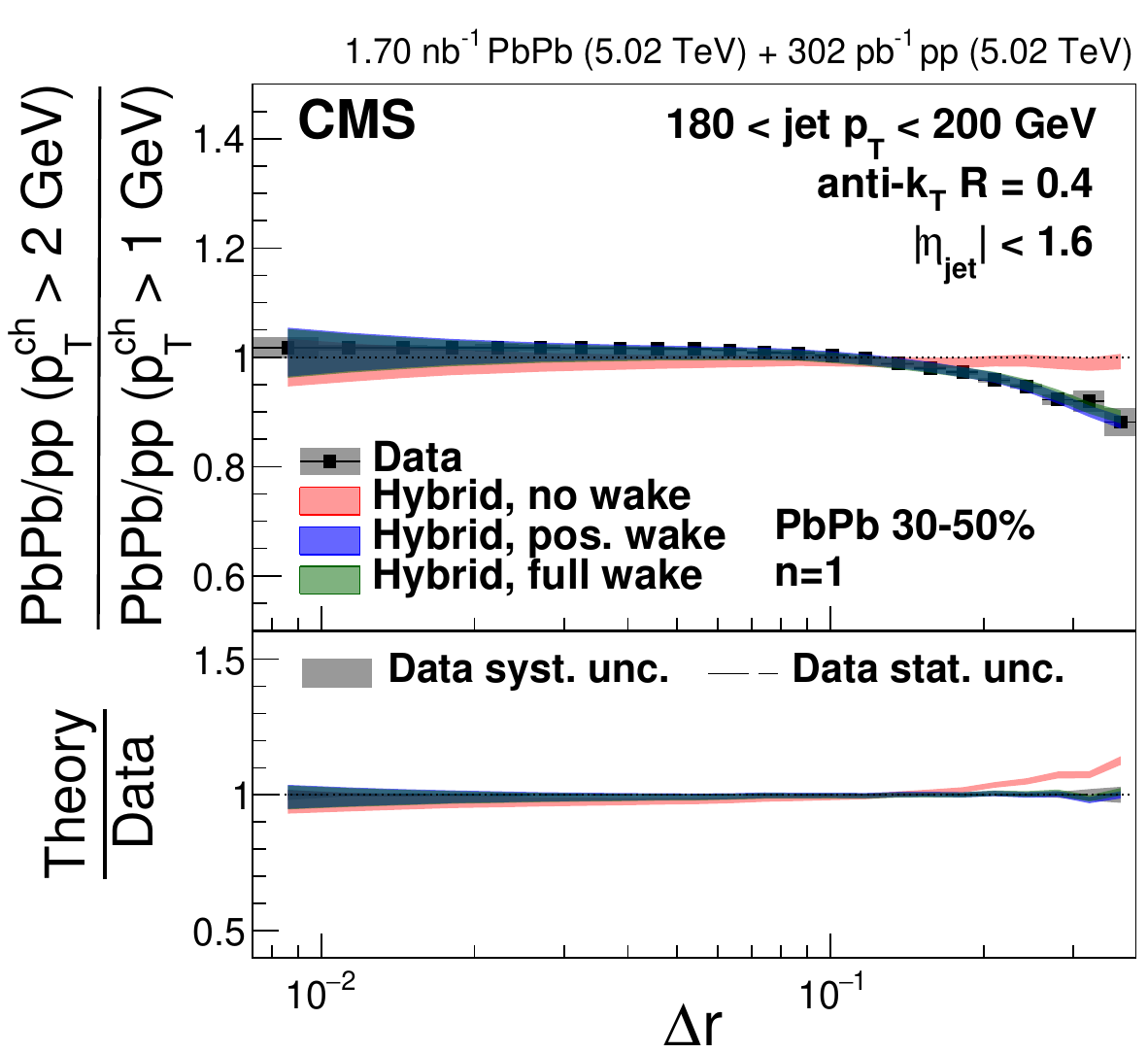}
    \includegraphics[width=0.4\textwidth]{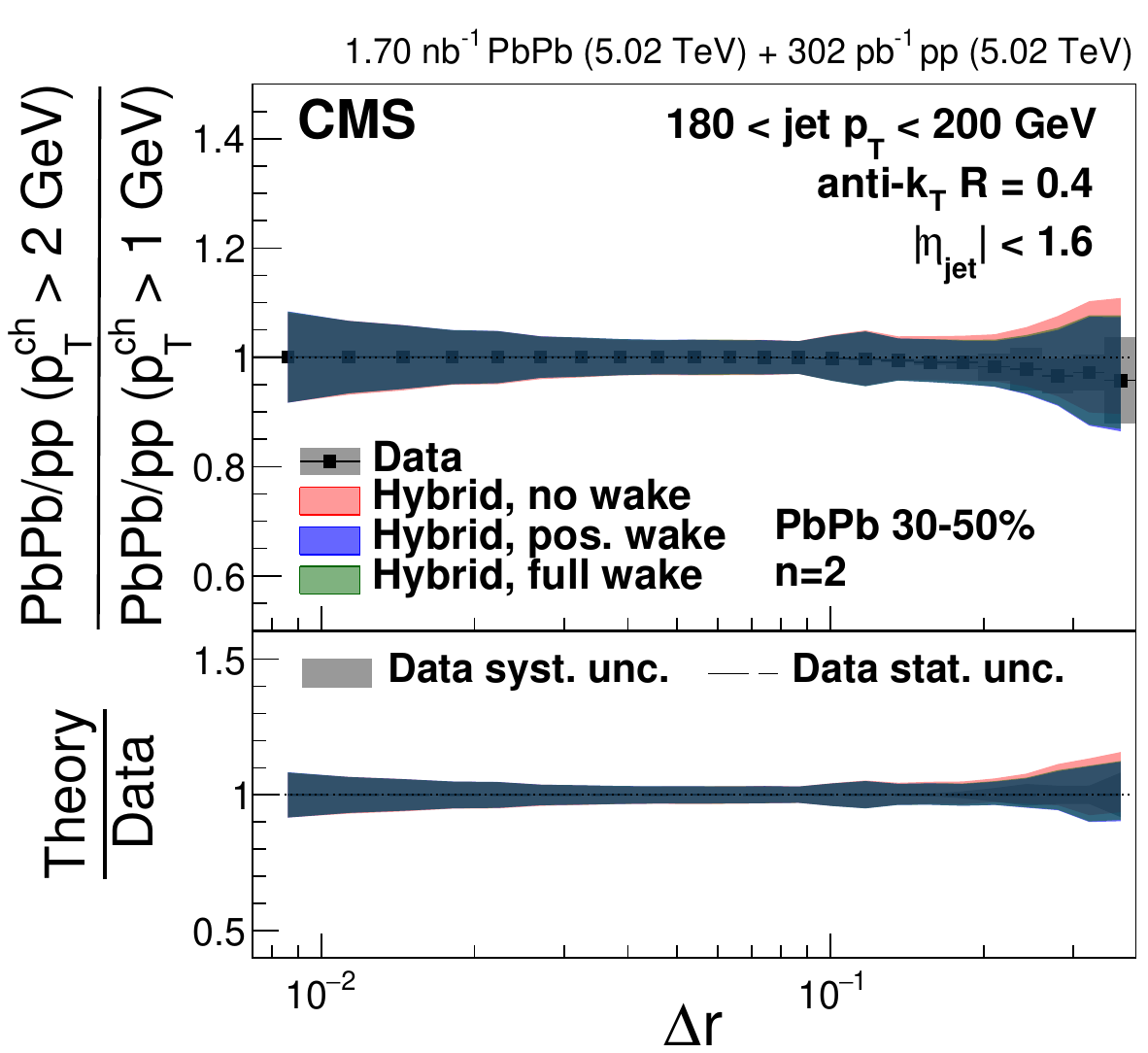}
    \caption{The double ratios of 30--50\% central \PbPb to \pp single ratios with $\ptCh > 2\GeV$ and $\ptCh > 1\GeV$ for $n=1$ (left) and $n=2$ (right) in different \ptjet bins compared to the hybrid model predictions with different jet wake settings.} 
    \label{fig:doubleRatioHybridJet30-50}
  } 
\end{figure*}

\begin{figure*}[h]
  \centering{
    \includegraphics[width=0.4\textwidth]{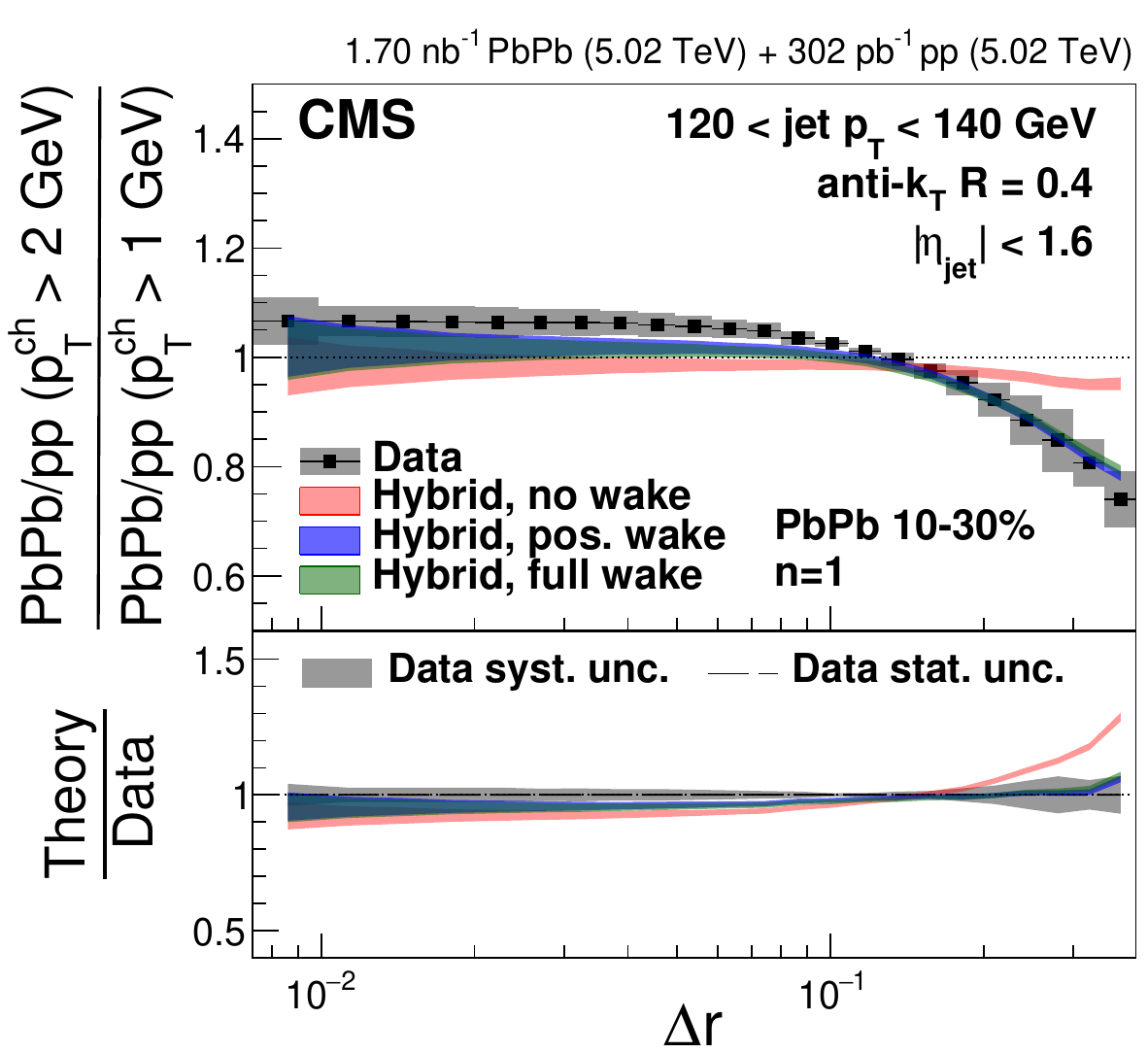}
    \includegraphics[width=0.4\textwidth]{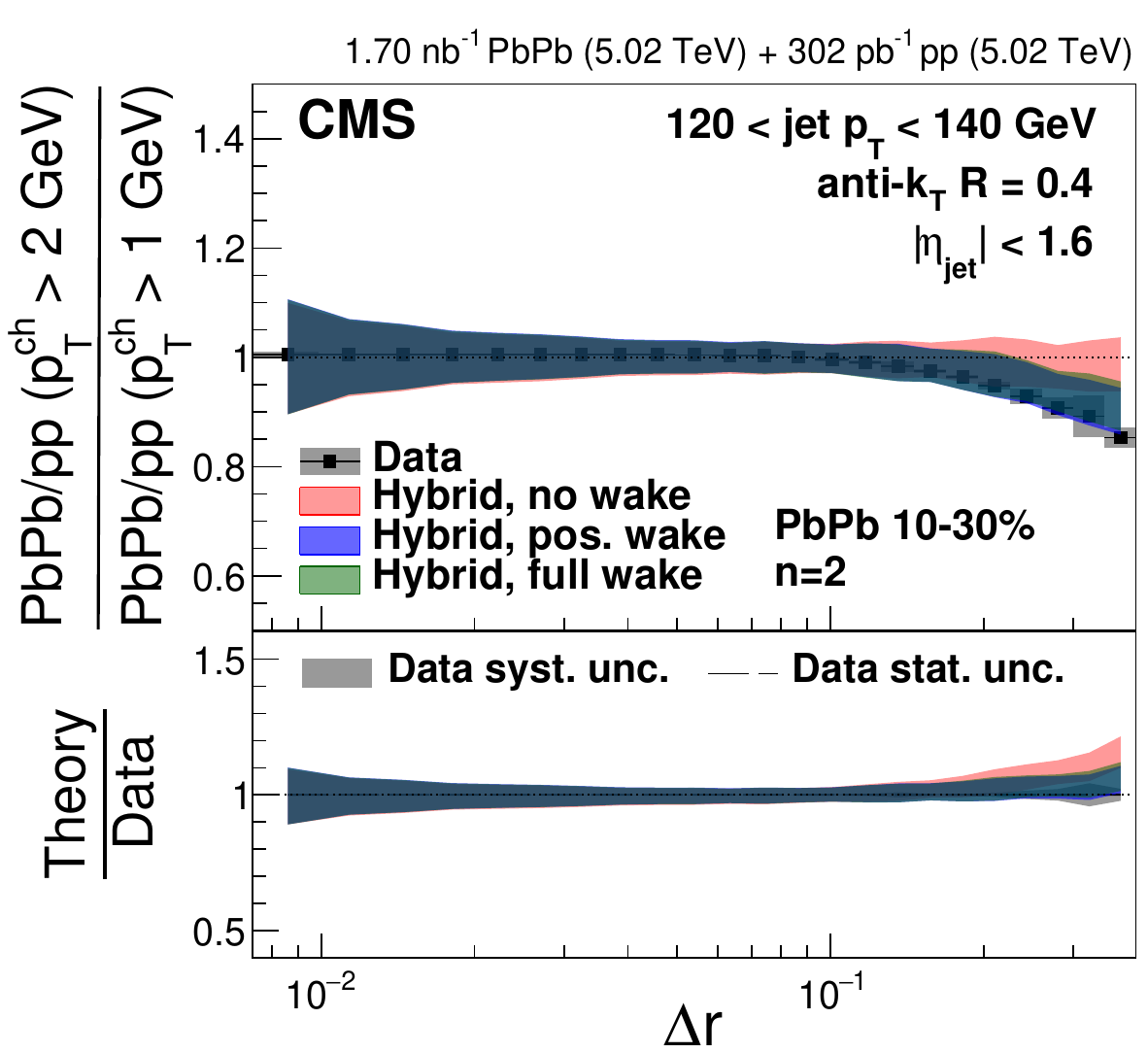}
    \includegraphics[width=0.4\textwidth]{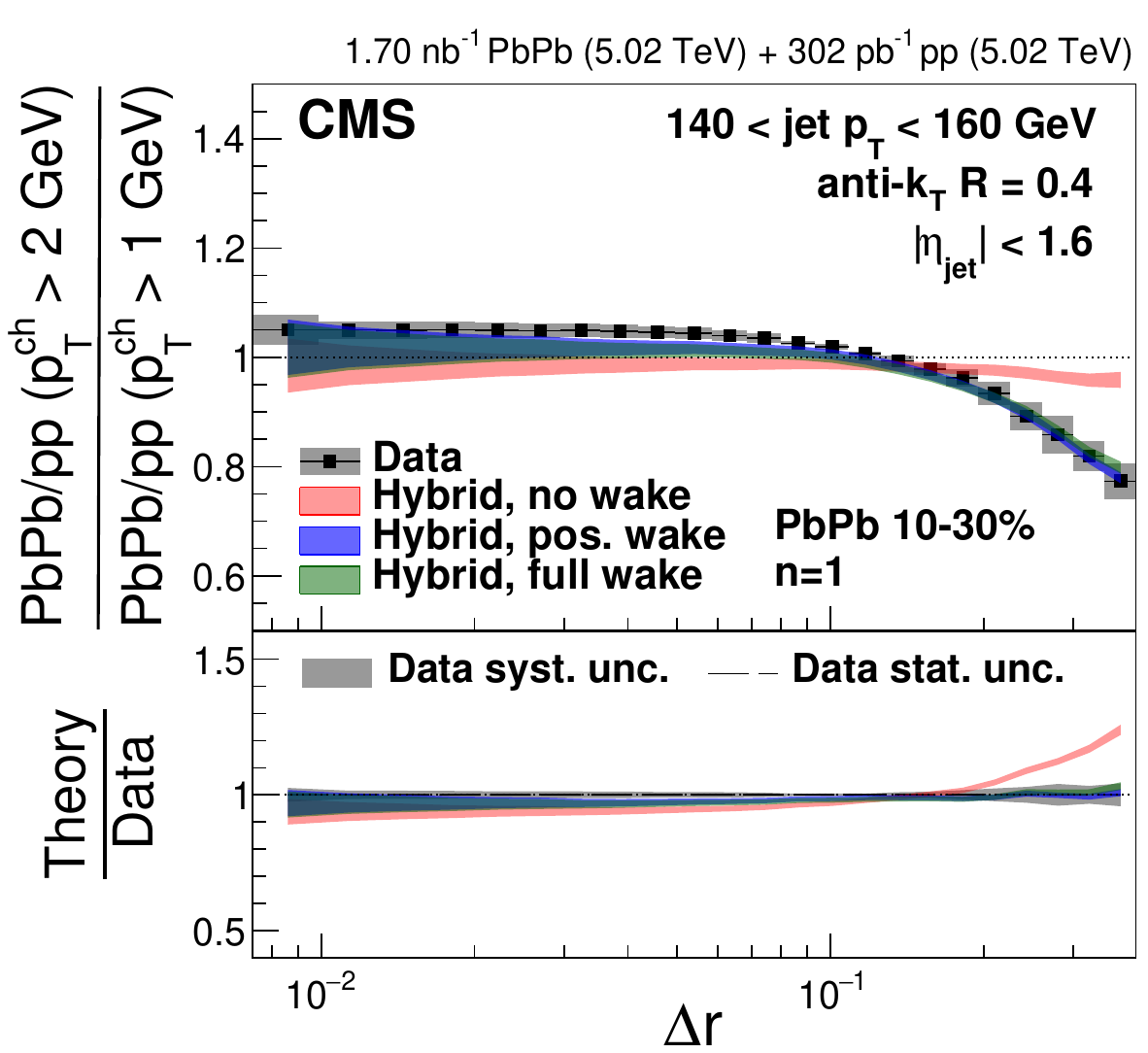} 
    \includegraphics[width=0.4\textwidth]{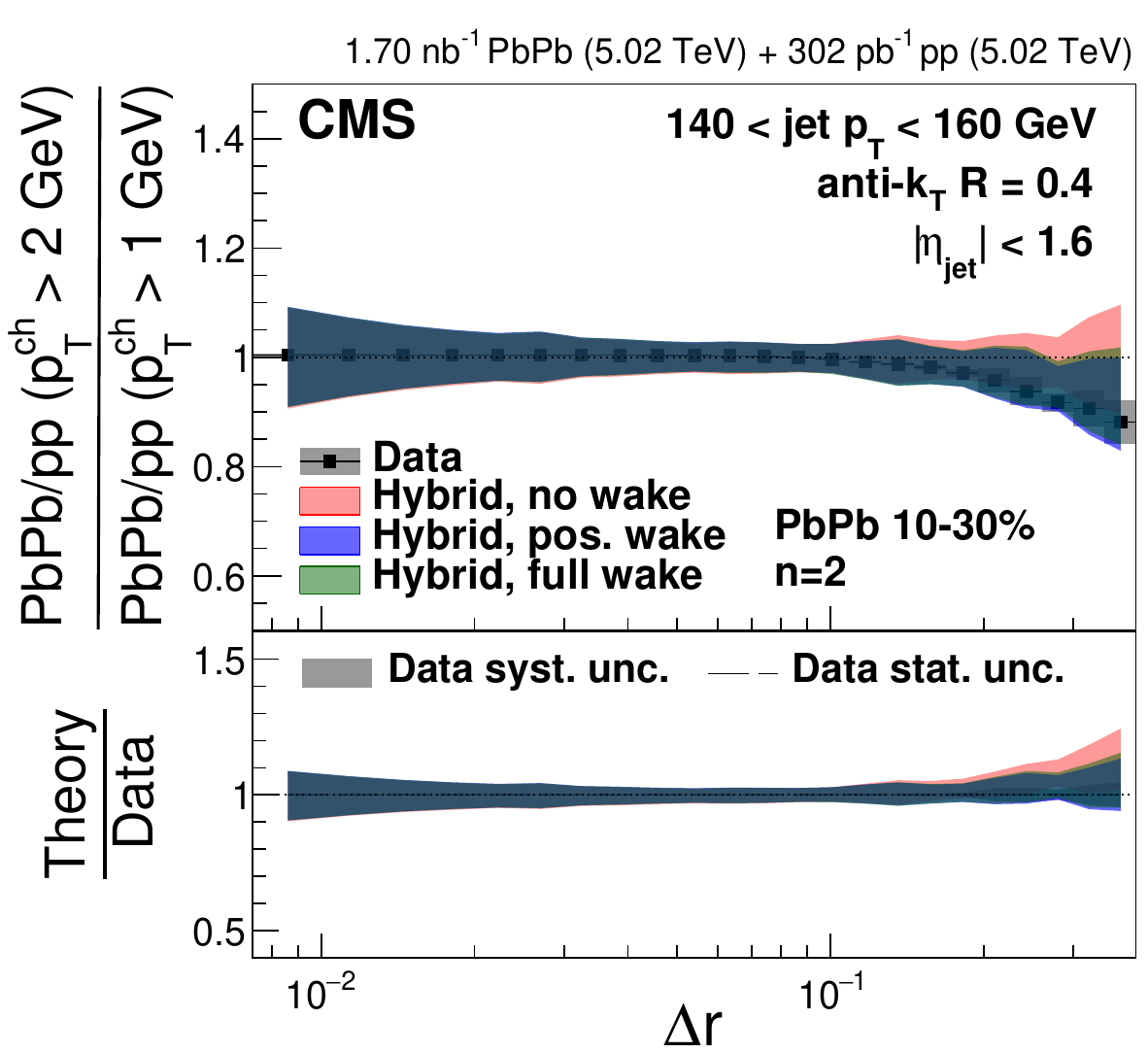}
    \includegraphics[width=0.4\textwidth]{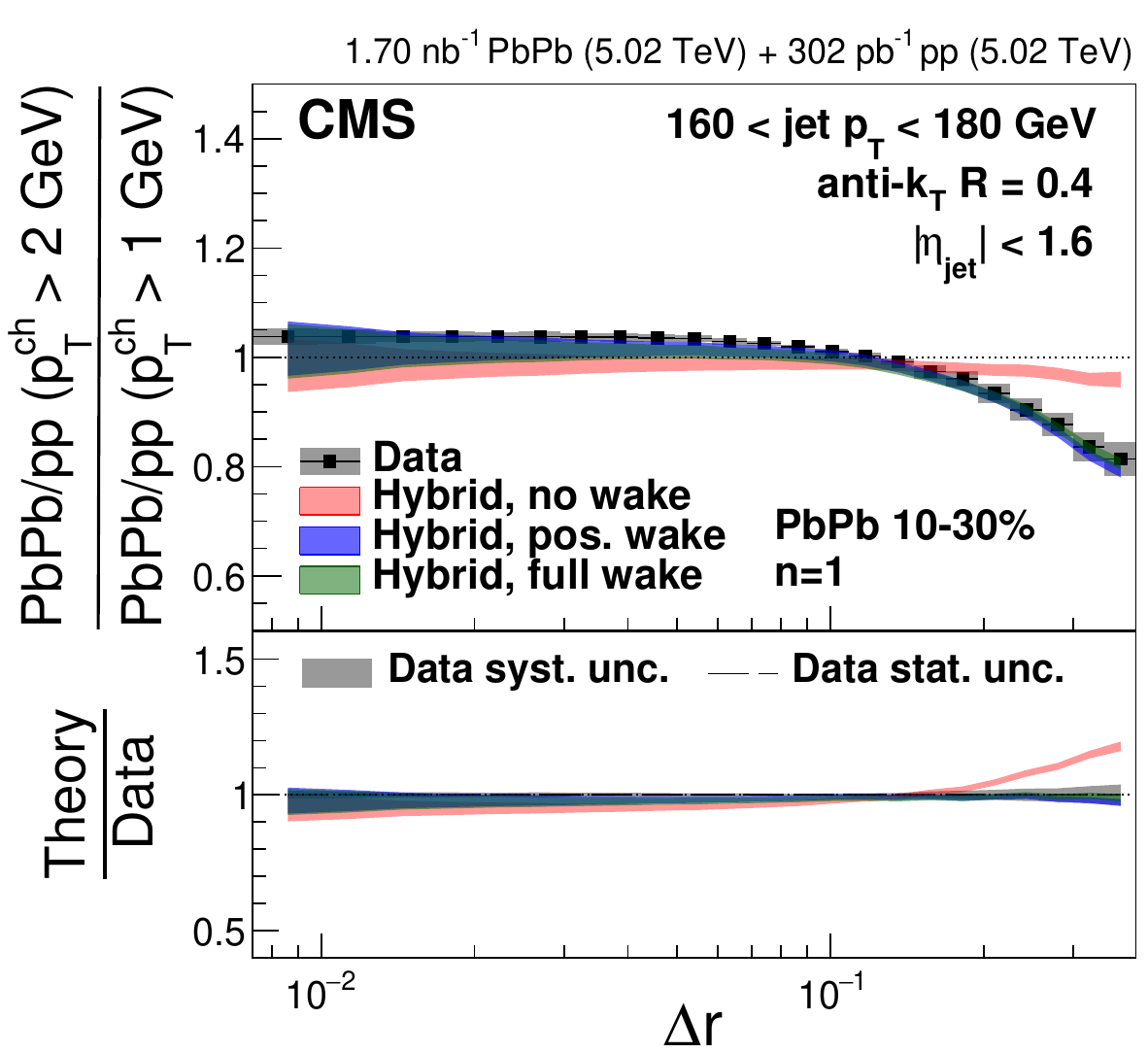}
    \includegraphics[width=0.4\textwidth]{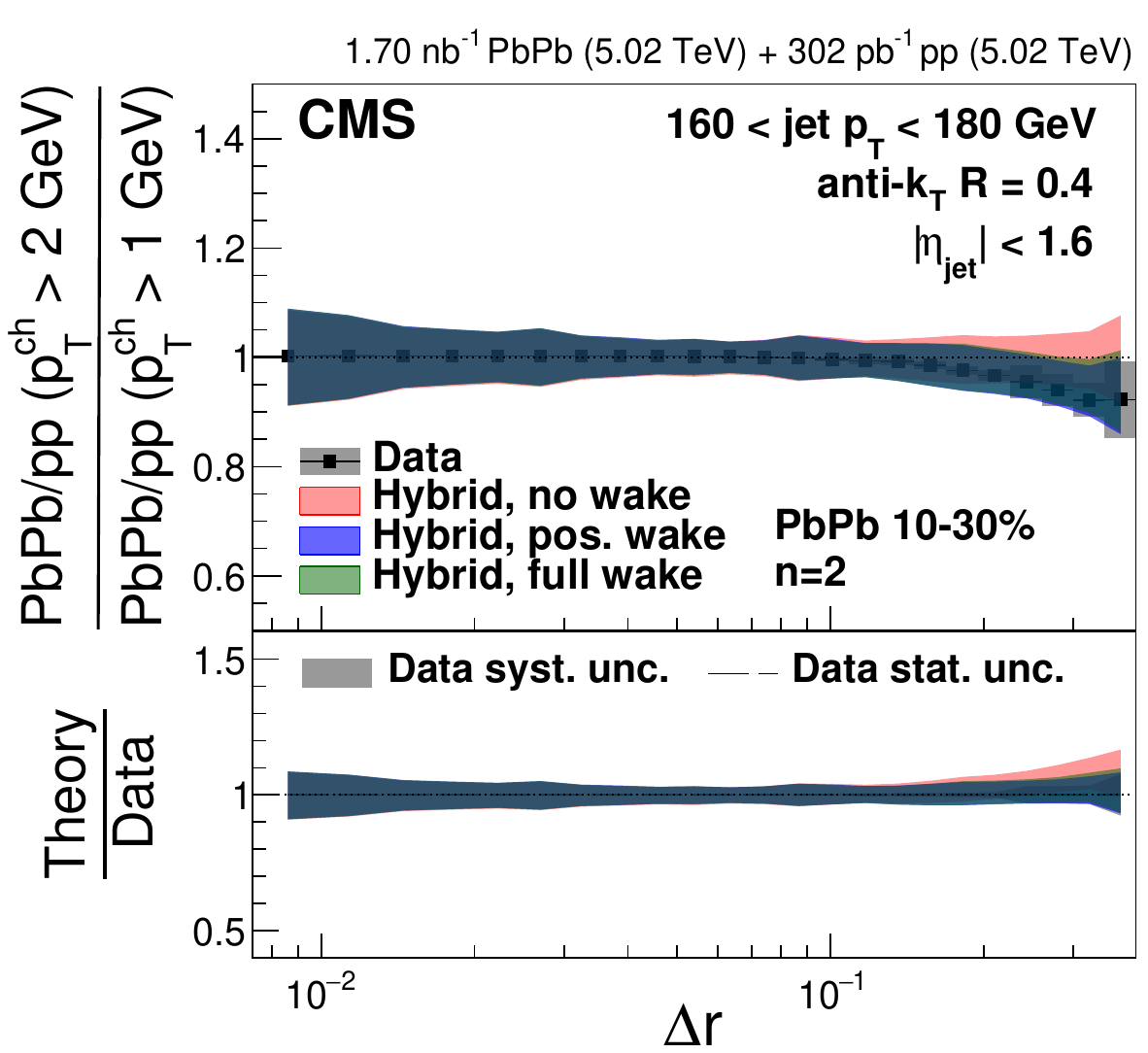}
    \includegraphics[width=0.4\textwidth]{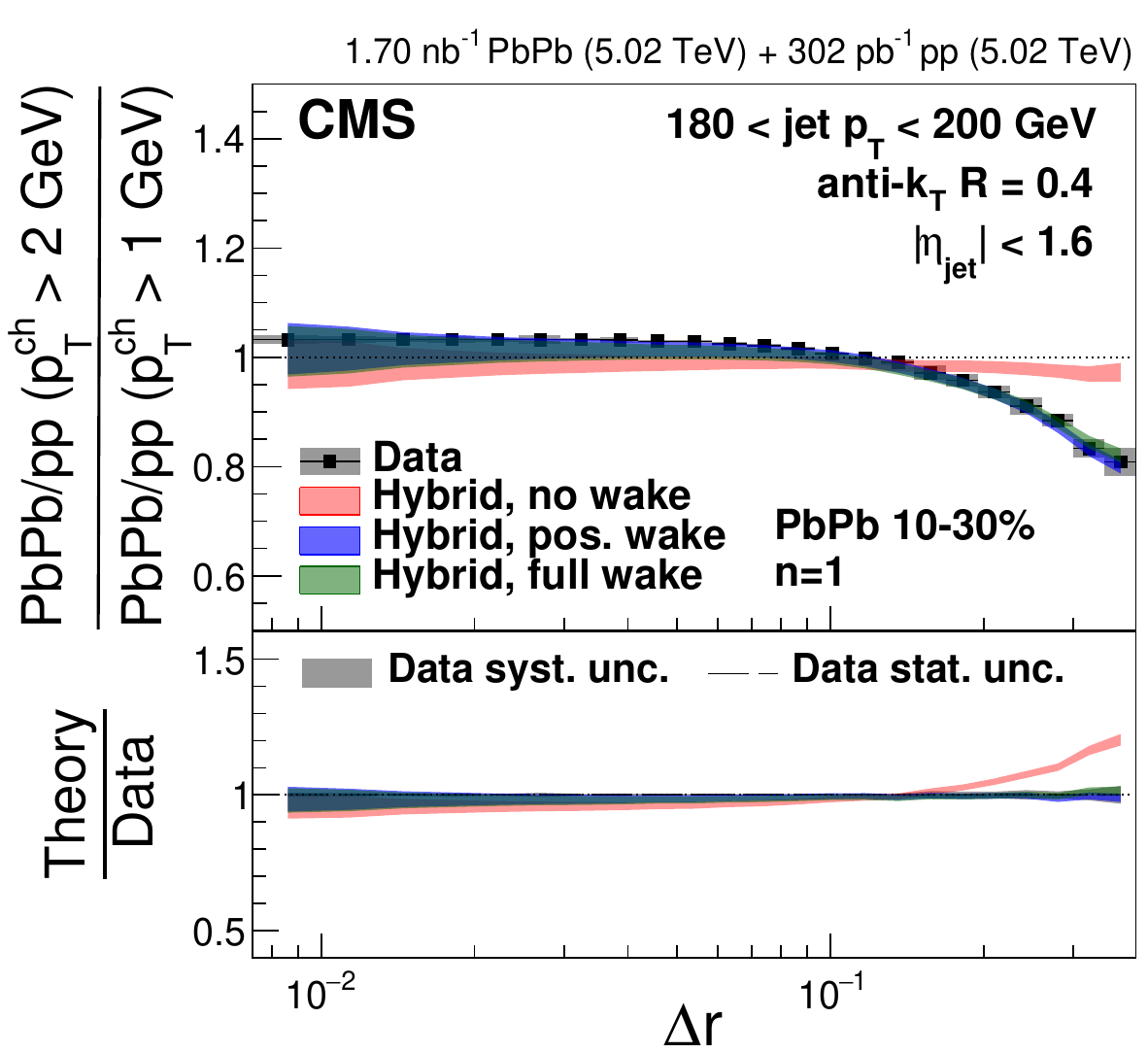}
    \includegraphics[width=0.4\textwidth]{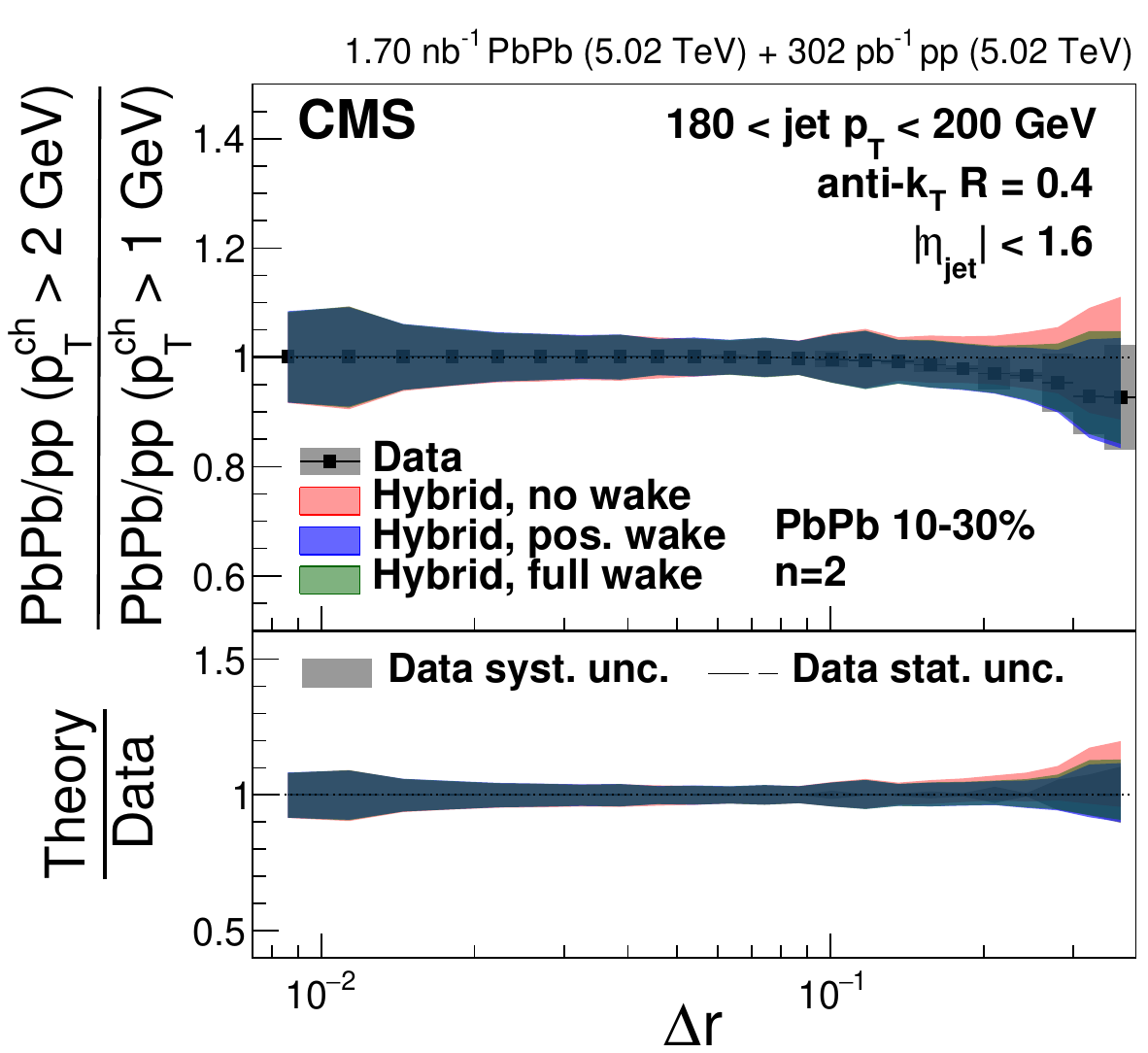}
    \caption{The double ratios of 10--30\% central \PbPb to \pp single ratios with $\ptCh > 2\GeV$ and $\ptCh > 1\GeV$ for $n=1$ (left) and $n=2$ (right) in different \ptjet bins compared to the hybrid model predictions with different jet wake settings.} 
    \label{fig:doubleRatioHybridJet10-30}
  } 
\end{figure*}

\begin{figure*}[h]
  \centering{
    \includegraphics[width=0.4\textwidth]{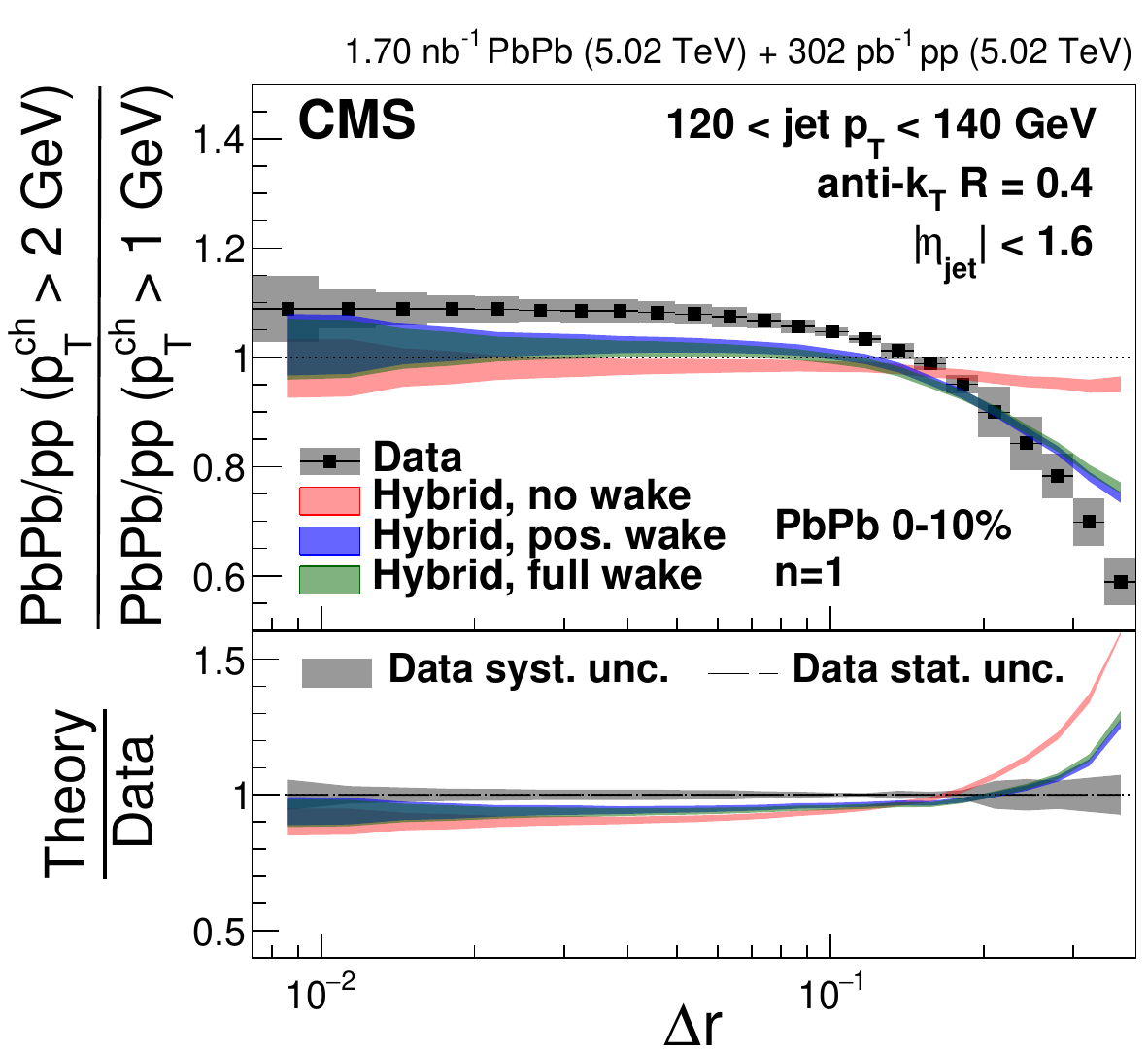}
    \includegraphics[width=0.4\textwidth]{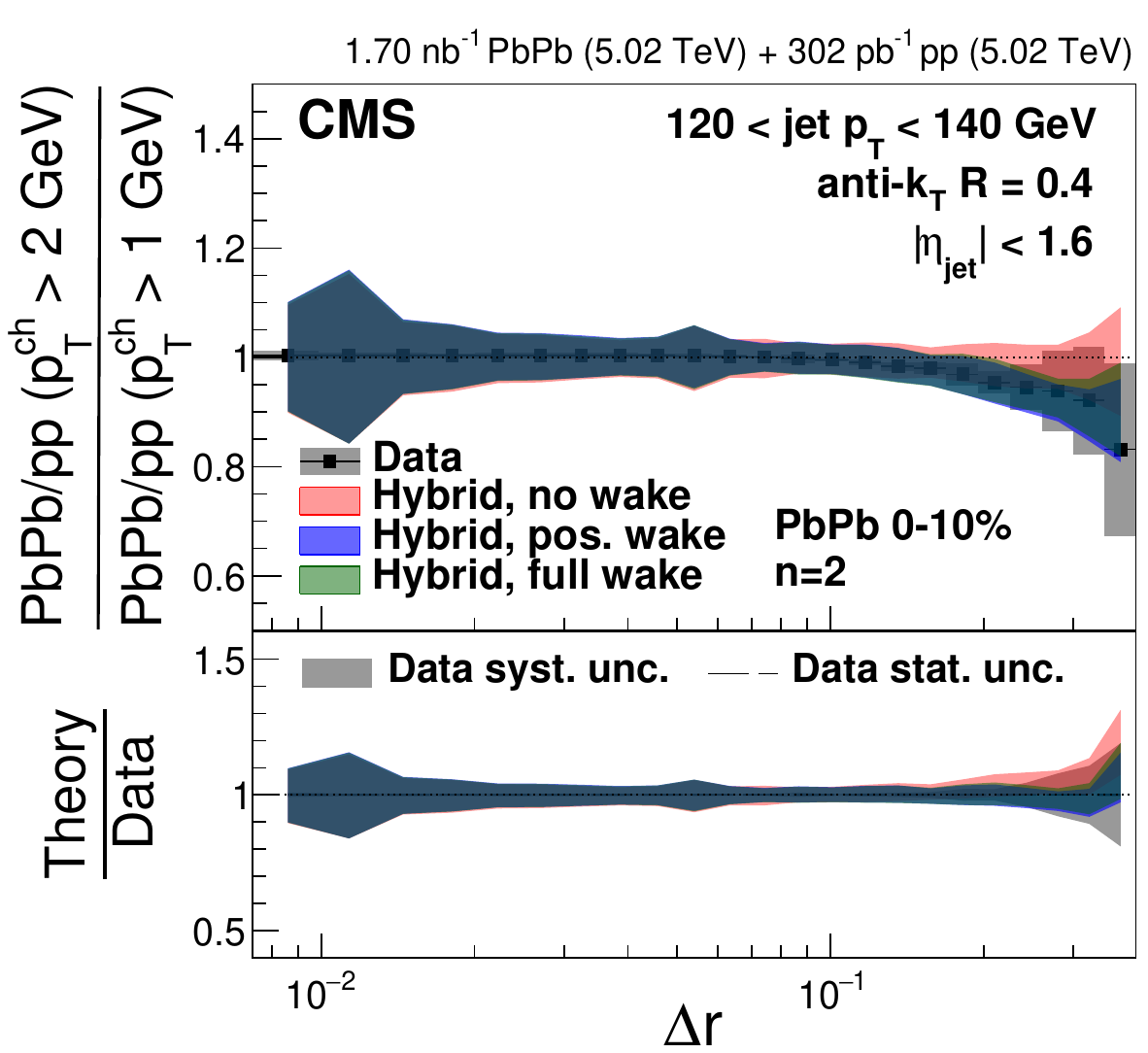}
    \includegraphics[width=0.4\textwidth]{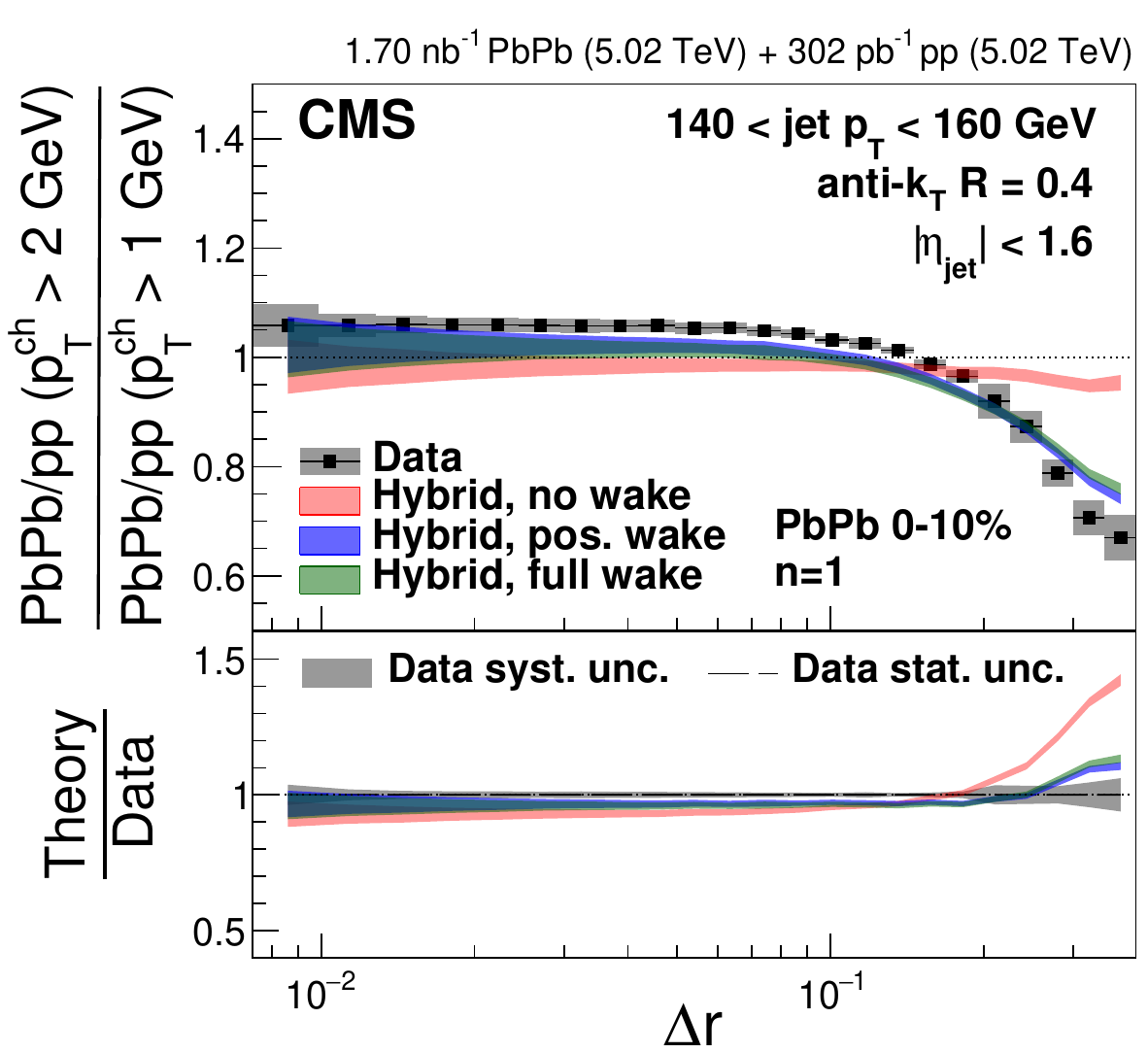} 
    \includegraphics[width=0.4\textwidth]{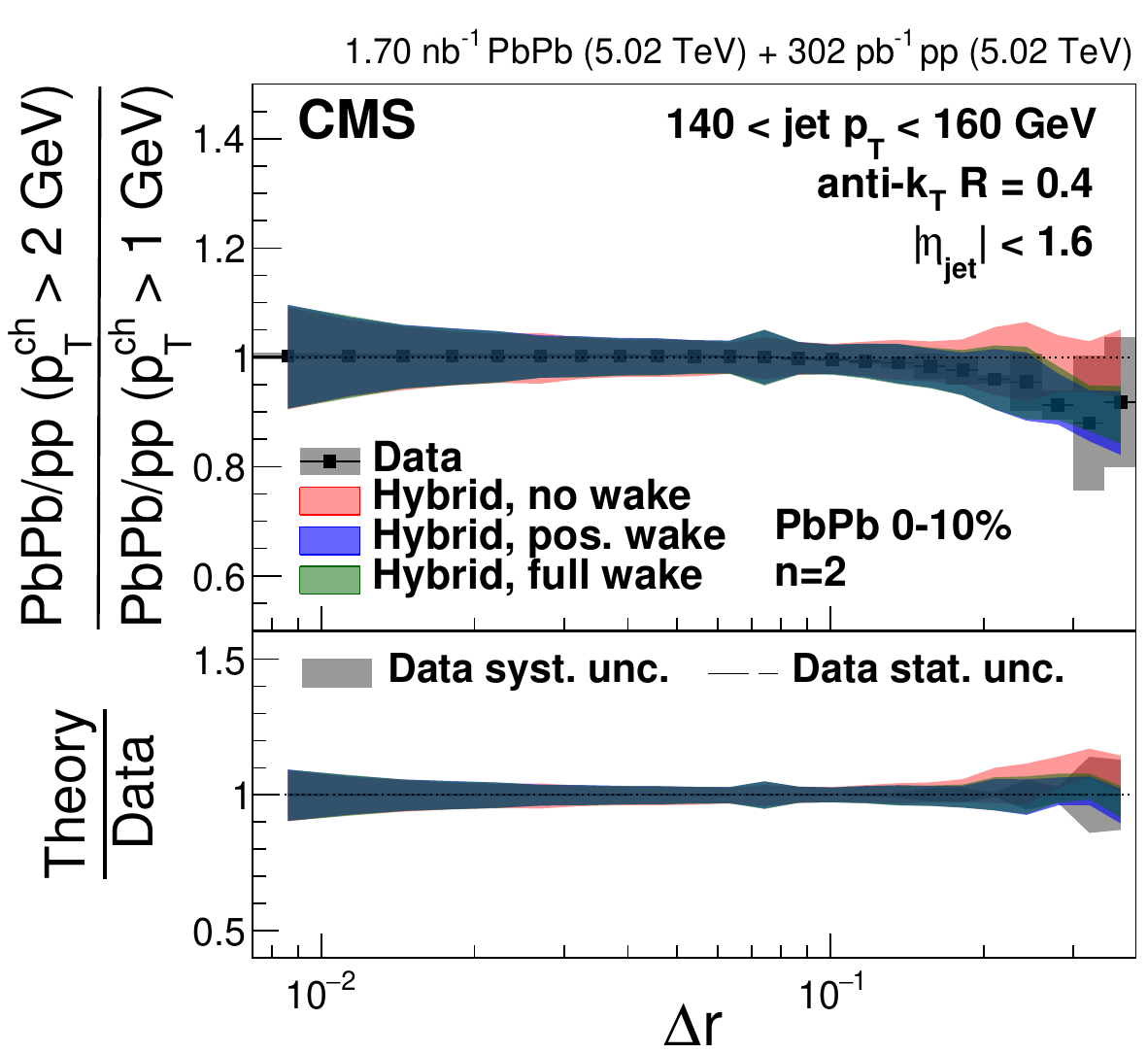}
    \includegraphics[width=0.4\textwidth]{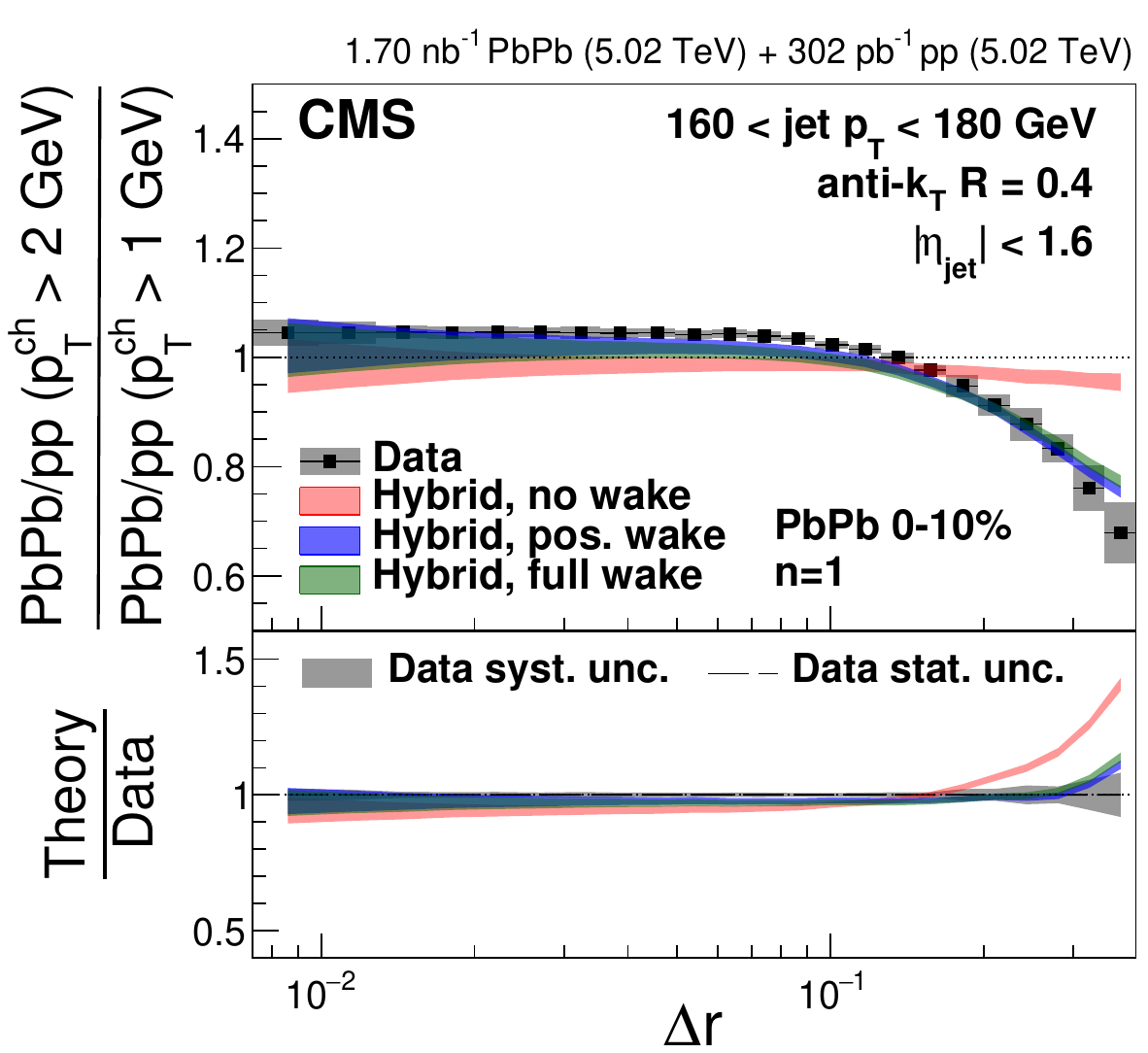}
    \includegraphics[width=0.4\textwidth]{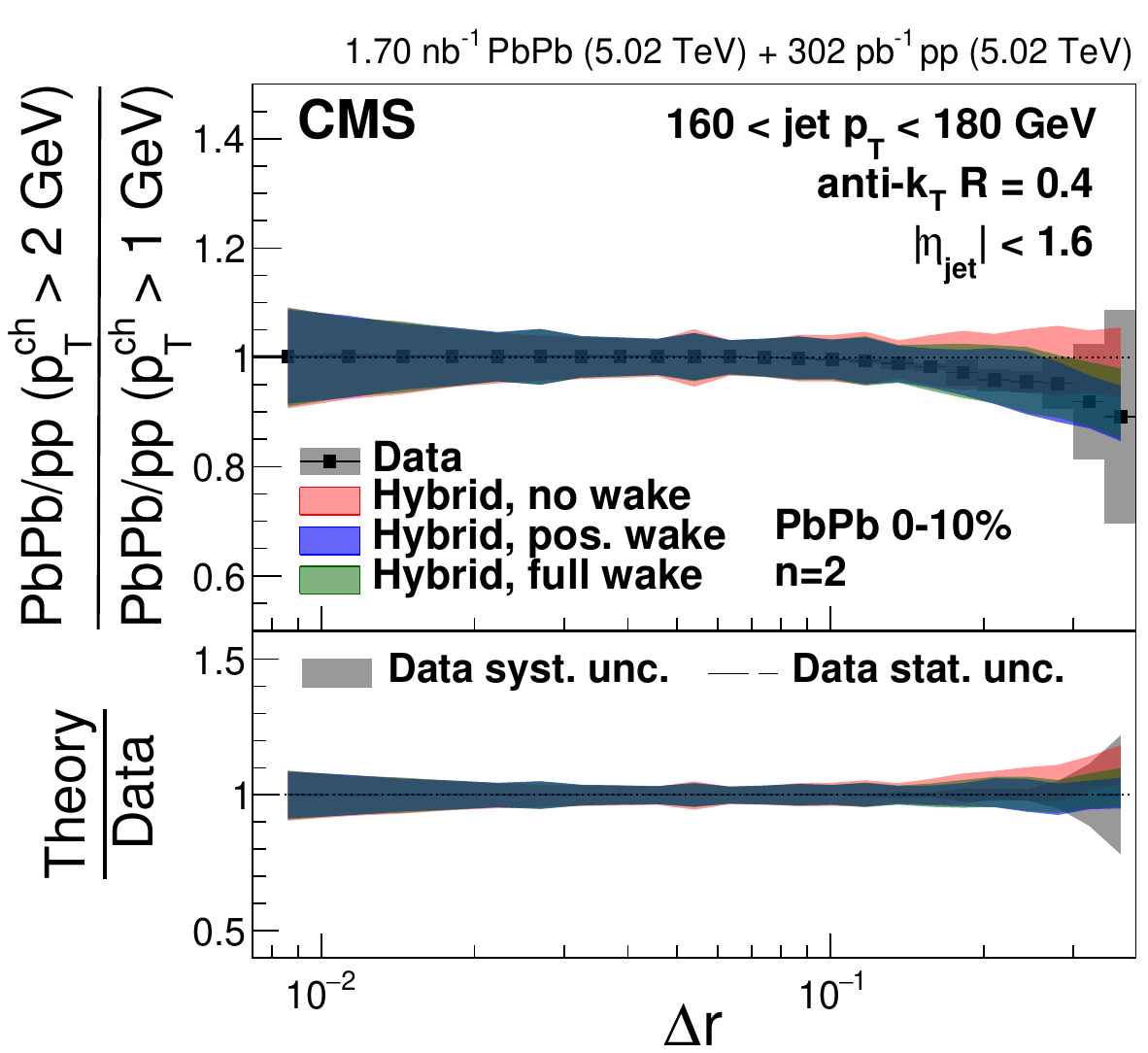}
    \includegraphics[width=0.4\textwidth]{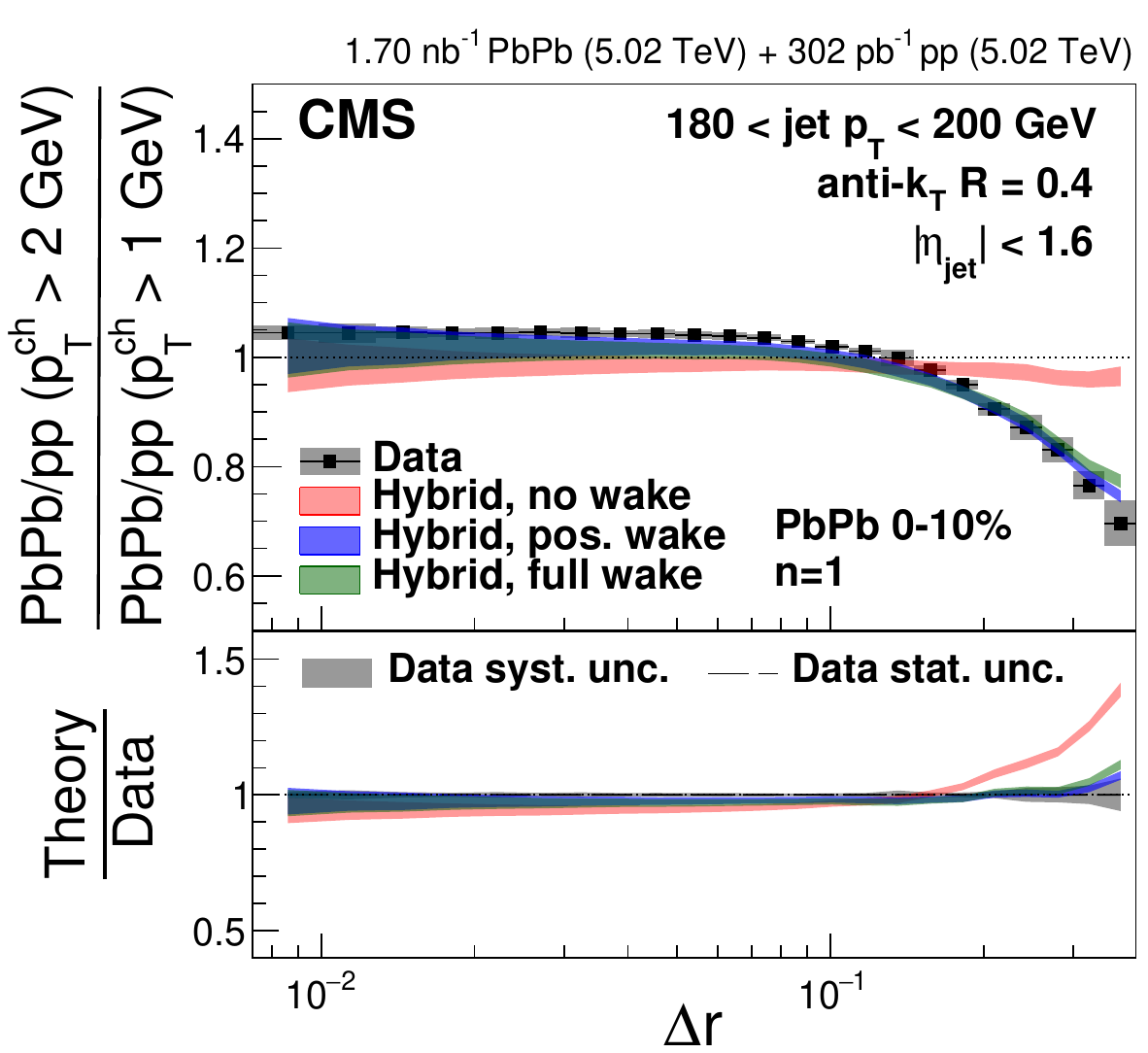}
    \includegraphics[width=0.4\textwidth]{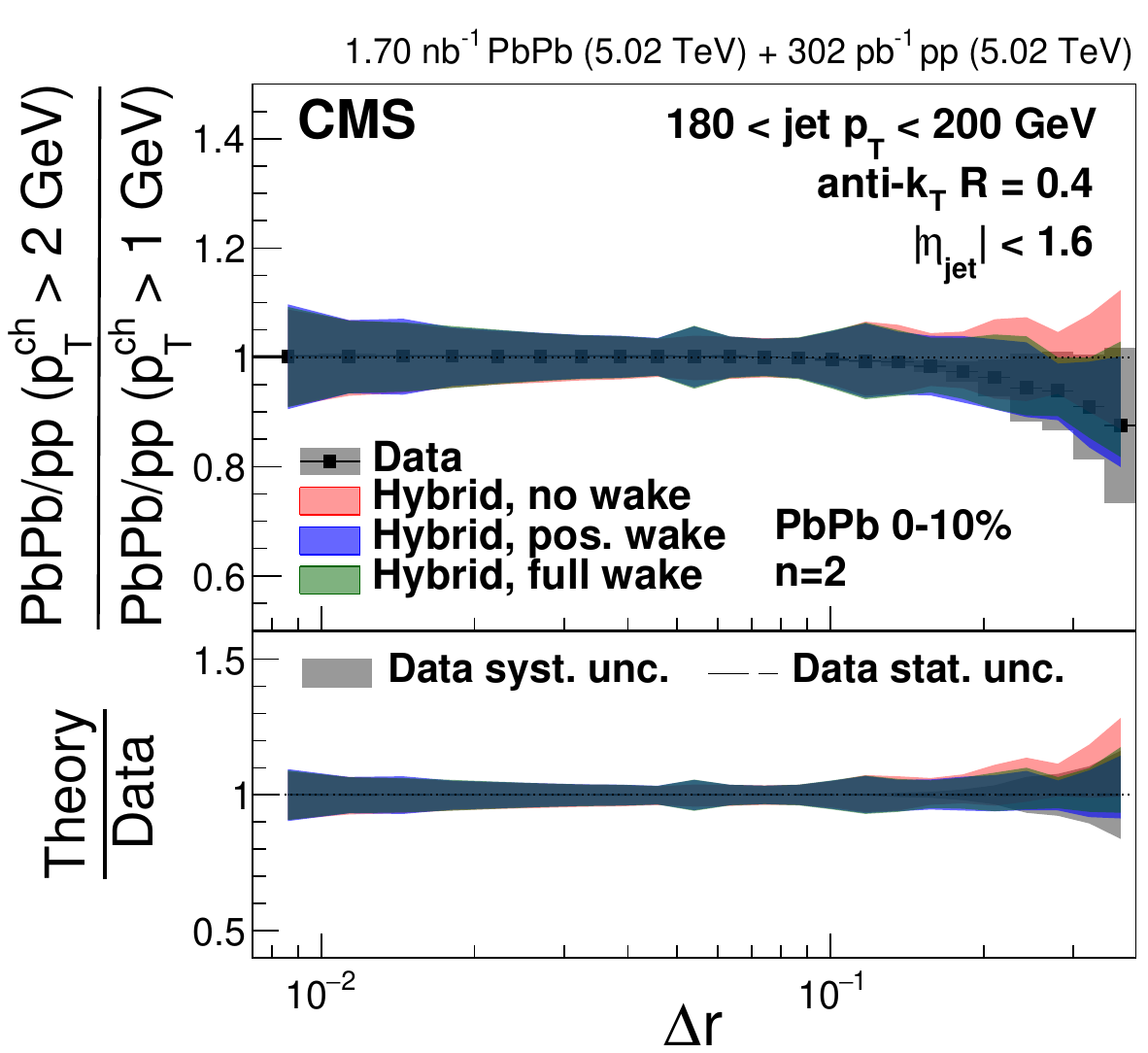}
    \caption{The double ratios of 0--10\% central \PbPb to \pp single ratios with $\ptCh > 2\GeV$ and $\ptCh > 1\GeV$ for $n=1$ (left) and $n=2$ (right) in different \ptjet bins compared to the hybrid model predictions with different jet wake settings.} 
    \label{fig:doubleRatioHybridJet0-10}
  } 
\end{figure*}

\begin{figure*}[h]
  \centering{
    \includegraphics[width=0.4\textwidth]{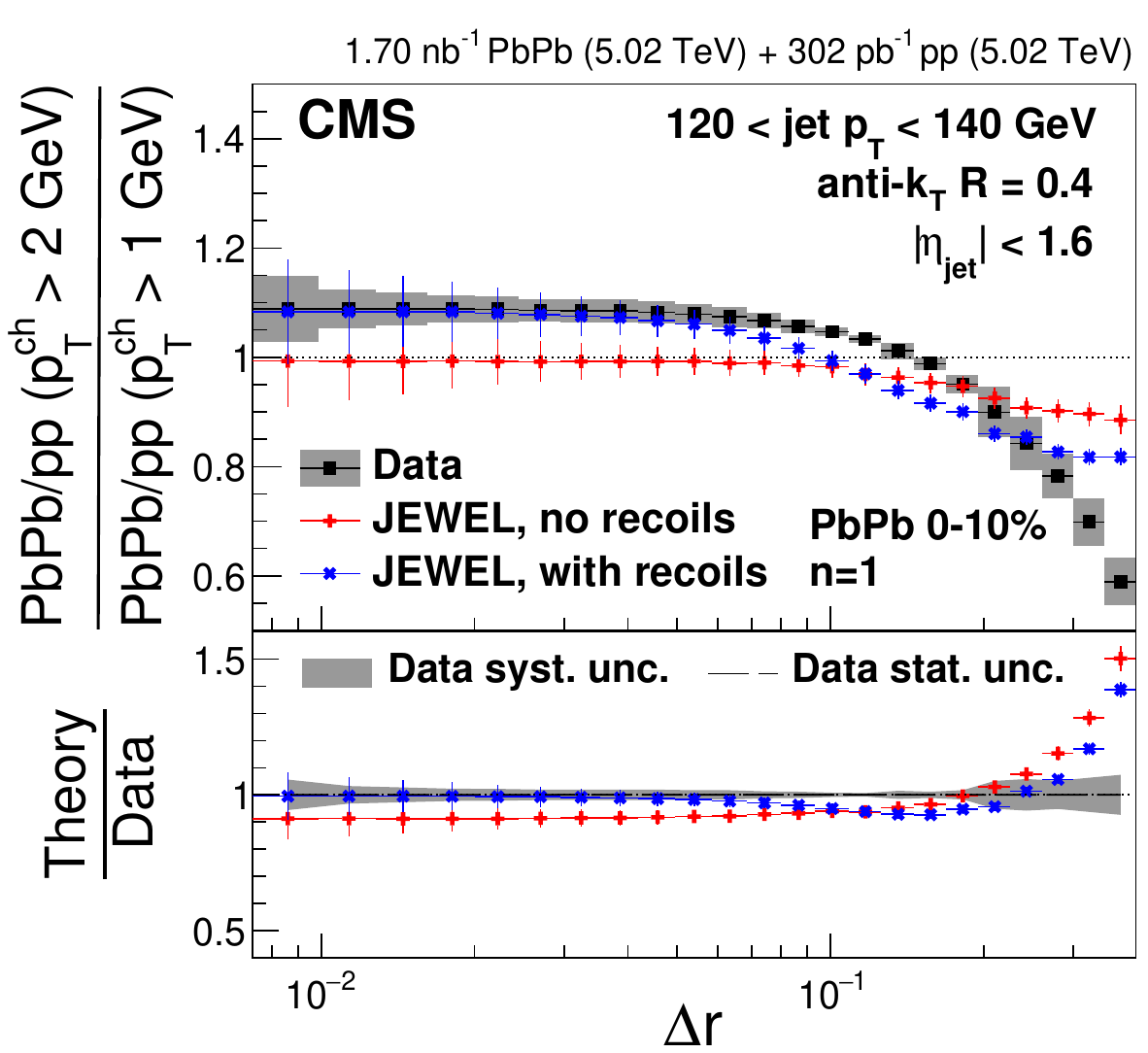}
    \includegraphics[width=0.4\textwidth]{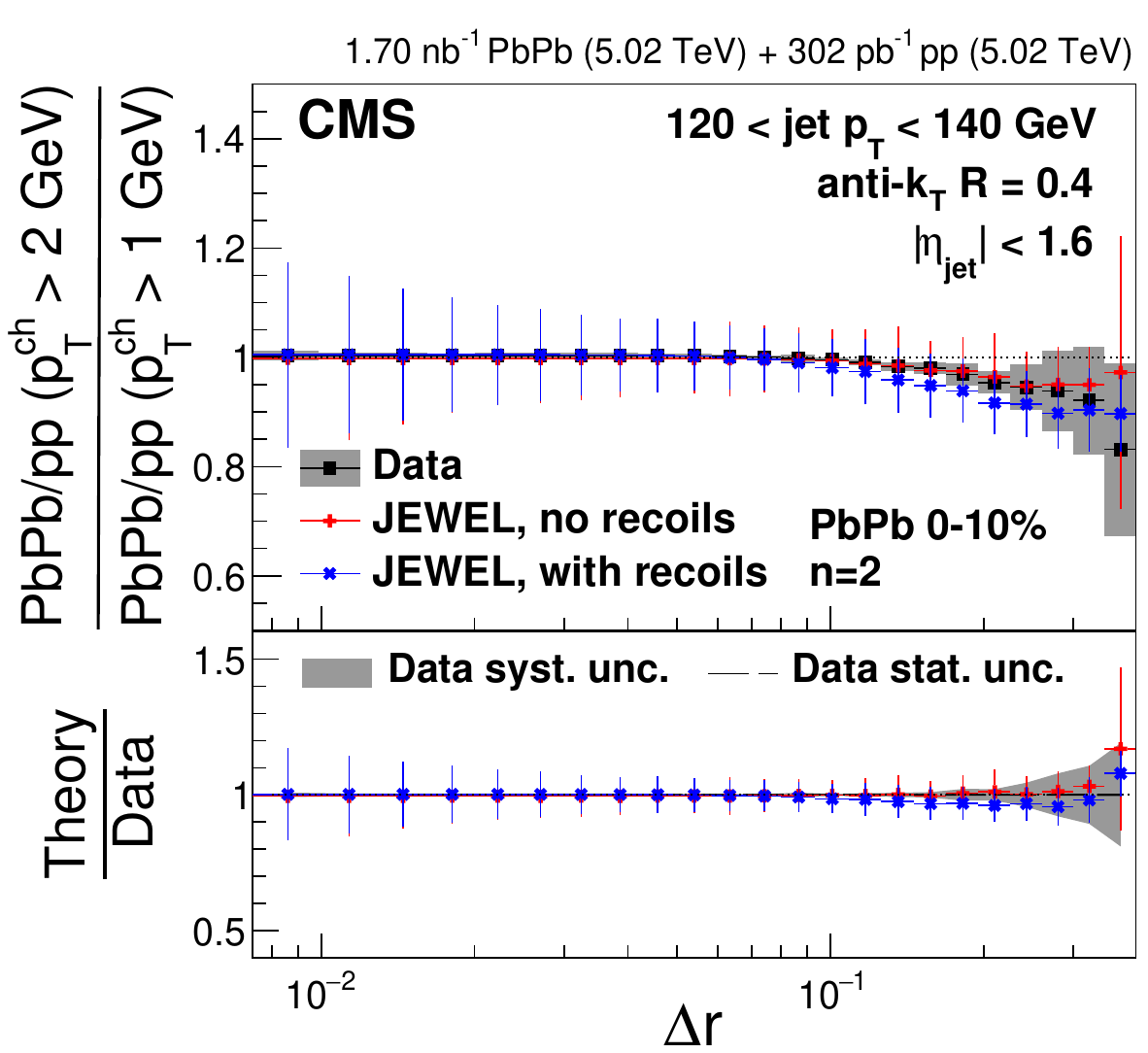}
    \includegraphics[width=0.4\textwidth]{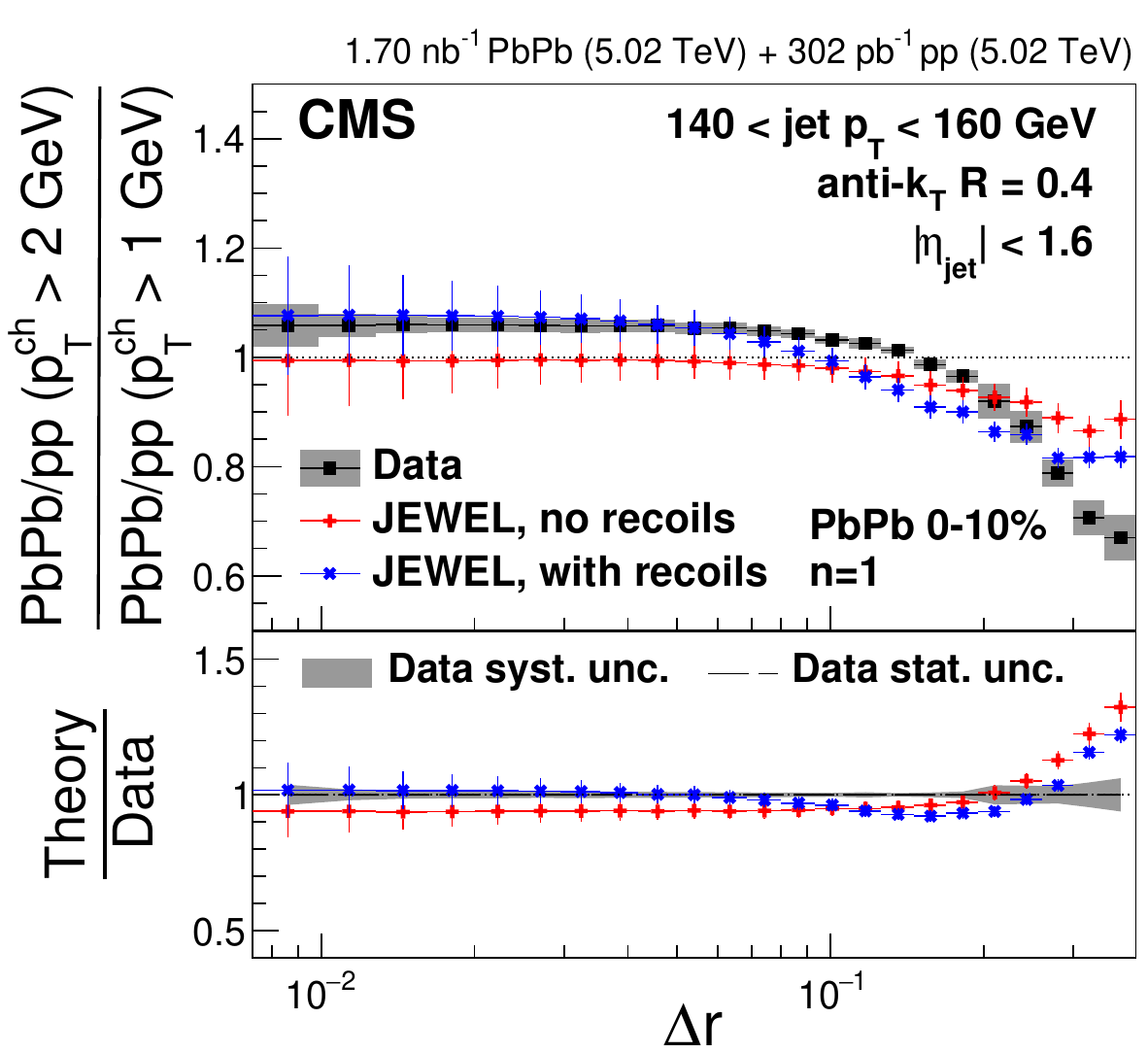} 
    \includegraphics[width=0.4\textwidth]{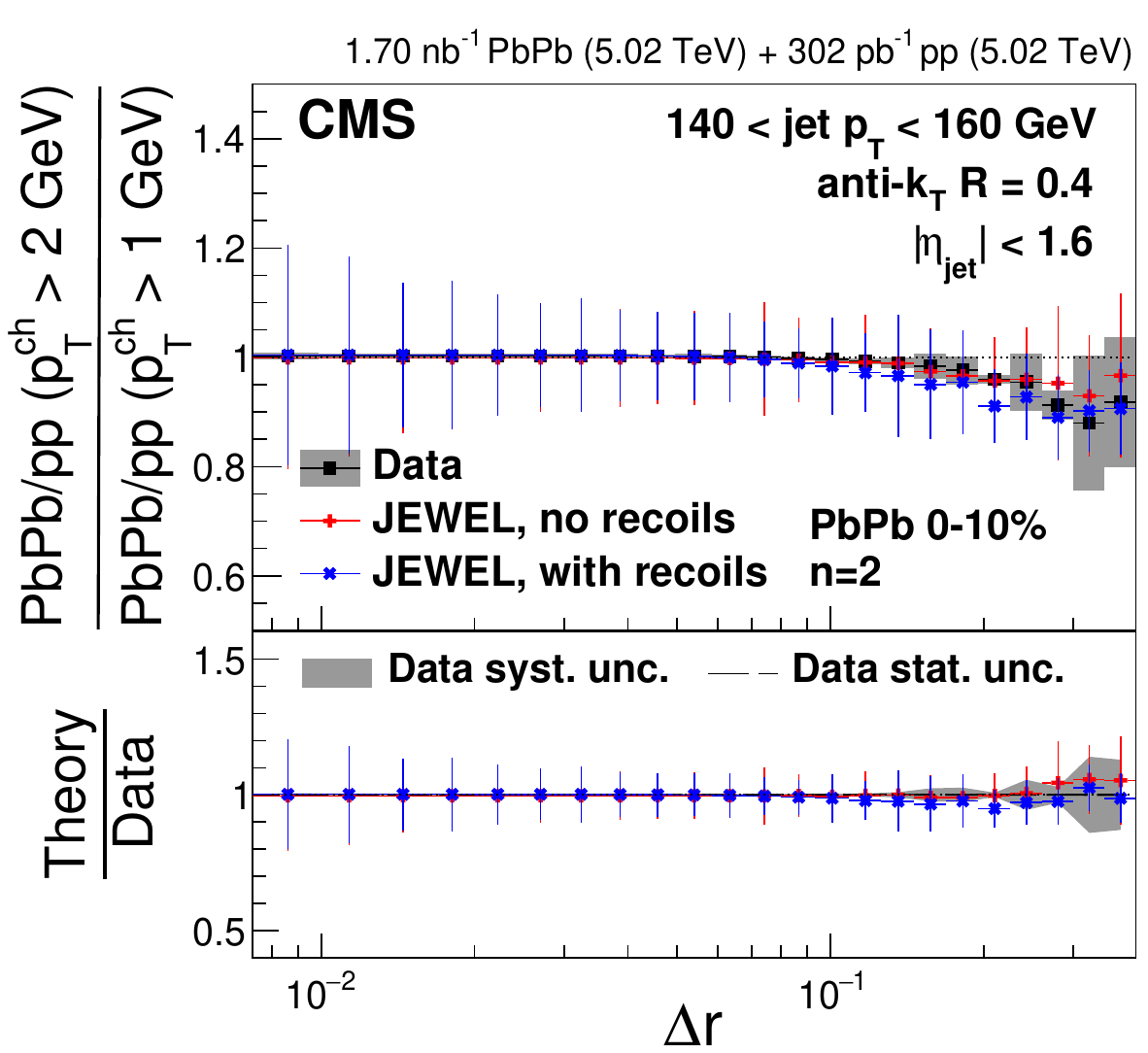}
    \includegraphics[width=0.4\textwidth]{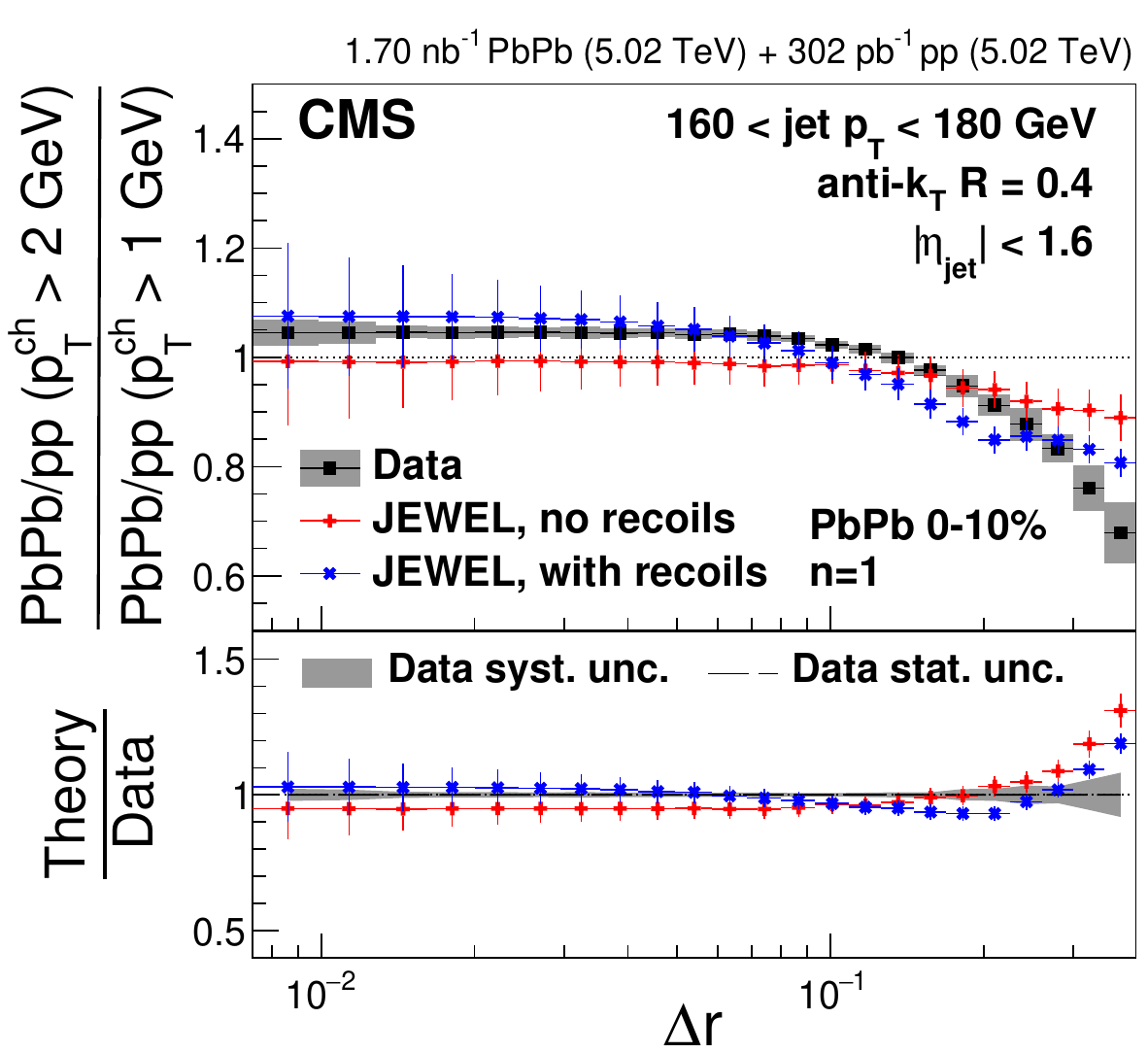}
    \includegraphics[width=0.4\textwidth]{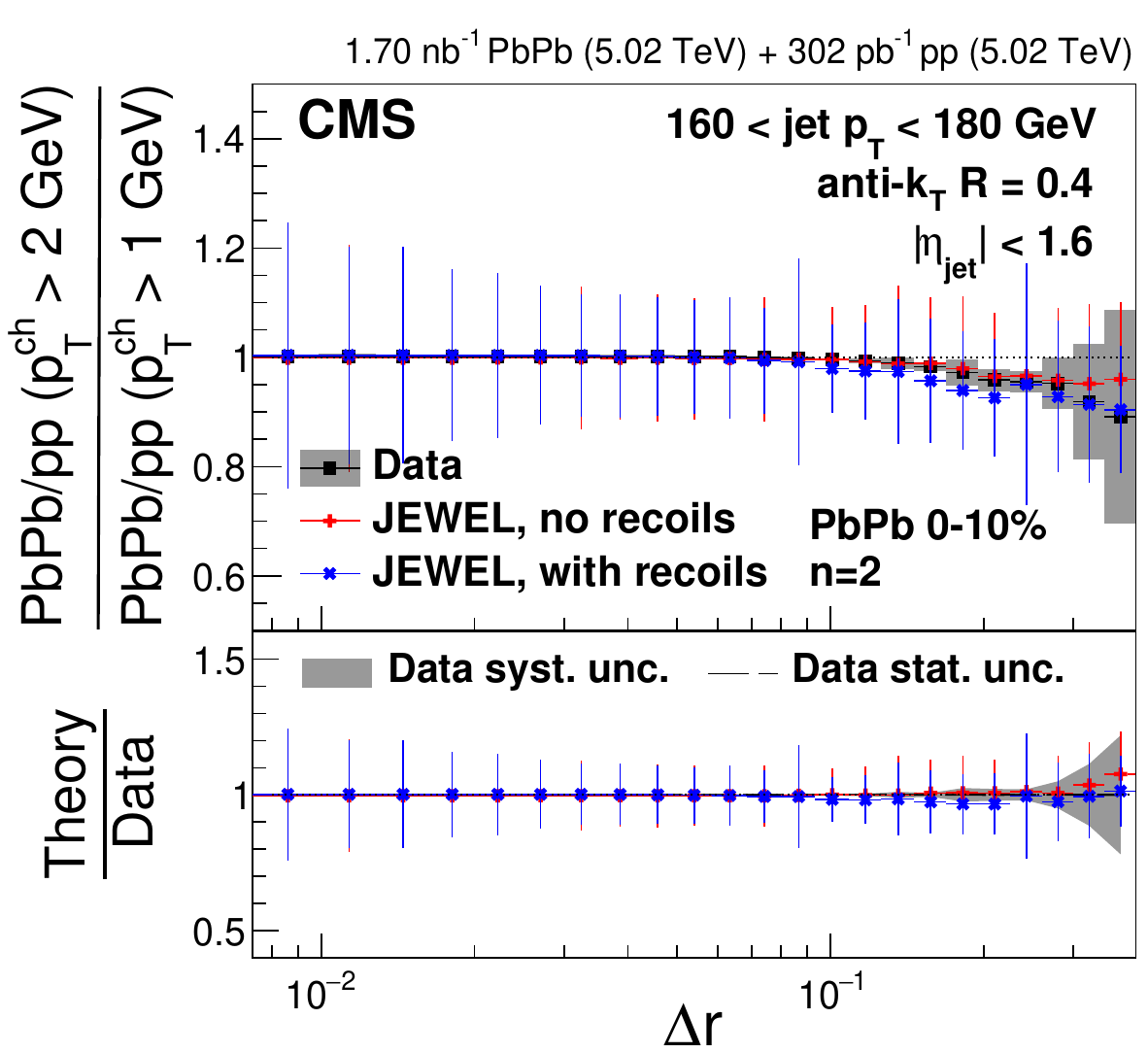}
    \includegraphics[width=0.4\textwidth]{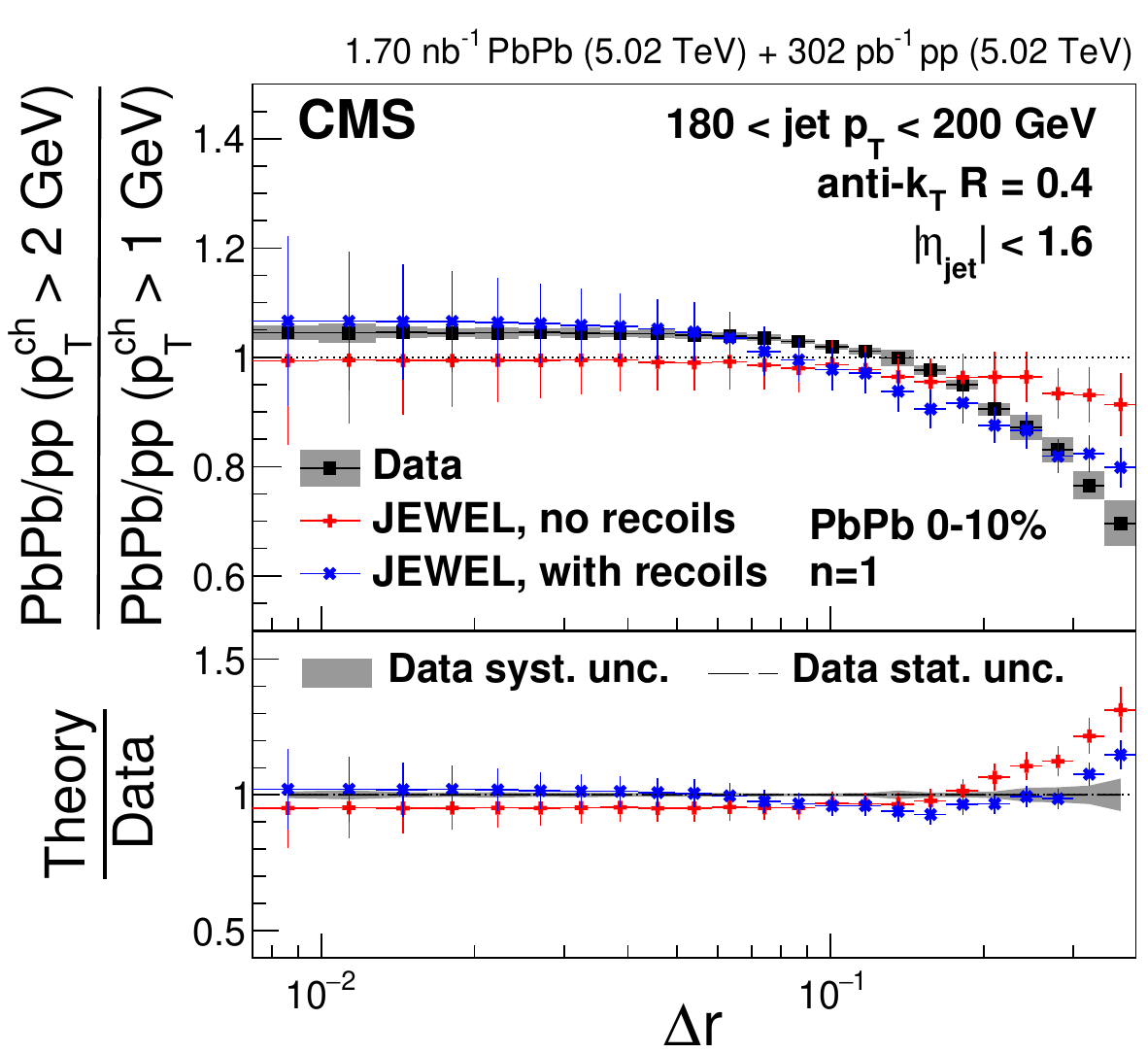}
    \includegraphics[width=0.4\textwidth]{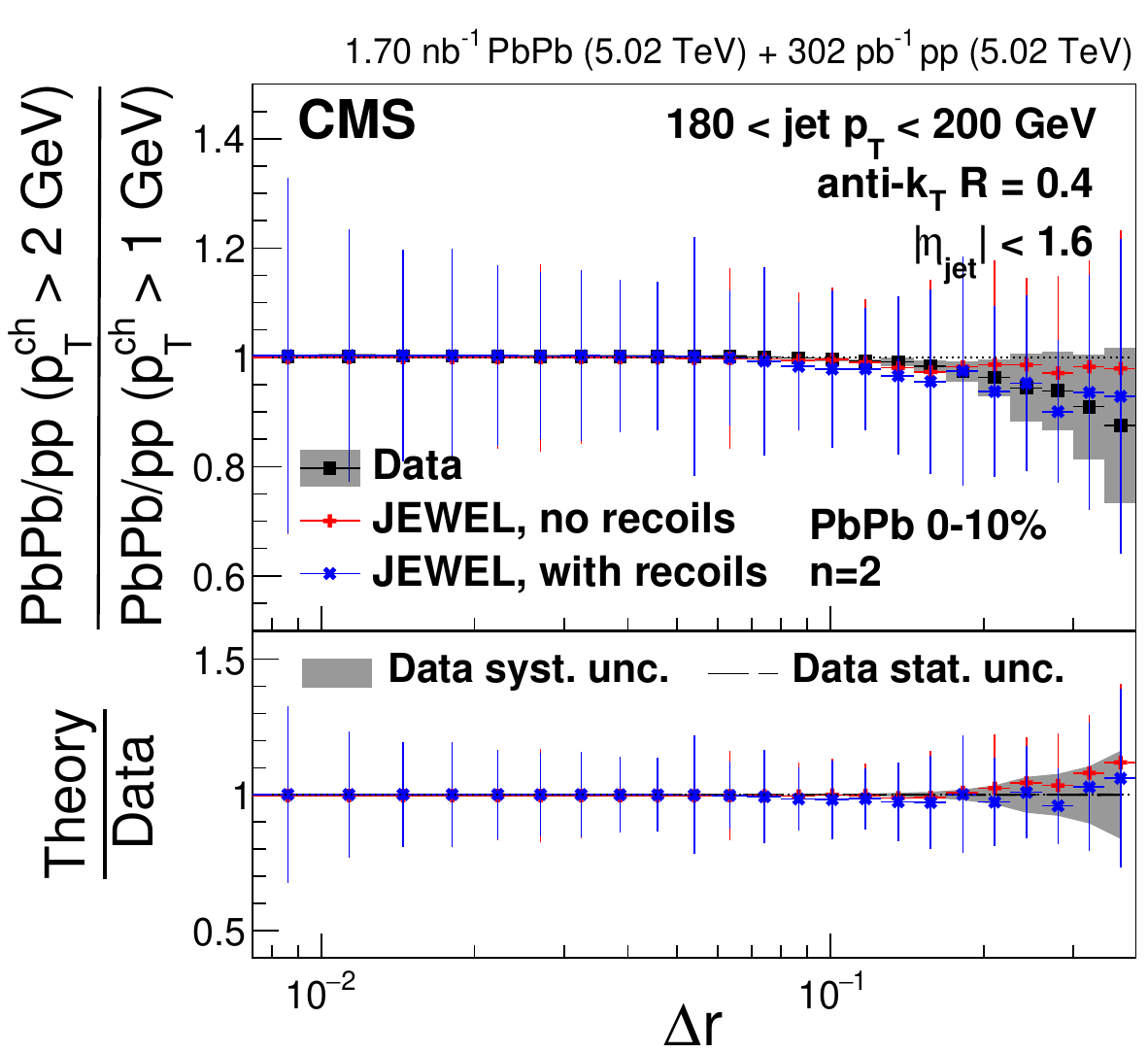}
    \caption{The double ratios of 0--10\% central \PbPb to \pp single ratios with $\ptCh > 2\GeV$ and $\ptCh > 1\GeV$ for $n=1$ (left) and $n=2$ (right) in different \ptjet bins compared to the \JEWEL predictions with different recoil settings.} 
    \label{fig:doubleRatioJewel}
  } 
\end{figure*}
}
\cleardoublepage \section{The CMS Collaboration \label{app:collab}}\begin{sloppypar}\hyphenpenalty=5000\widowpenalty=500\clubpenalty=5000
\cmsinstitute{Yerevan Physics Institute, Yerevan, Armenia}
{\tolerance=6000
V.~Chekhovsky, A.~Hayrapetyan, V.~Makarenko\cmsorcid{0000-0002-8406-8605}, A.~Tumasyan\cmsAuthorMark{1}\cmsorcid{0009-0000-0684-6742}
\par}
\cmsinstitute{Institut f\"{u}r Hochenergiephysik, Vienna, Austria}
{\tolerance=6000
W.~Adam\cmsorcid{0000-0001-9099-4341}, J.W.~Andrejkovic, L.~Benato\cmsorcid{0000-0001-5135-7489}, T.~Bergauer\cmsorcid{0000-0002-5786-0293}, S.~Chatterjee\cmsorcid{0000-0003-2660-0349}, K.~Damanakis\cmsorcid{0000-0001-5389-2872}, M.~Dragicevic\cmsorcid{0000-0003-1967-6783}, P.S.~Hussain\cmsorcid{0000-0002-4825-5278}, M.~Jeitler\cmsAuthorMark{2}\cmsorcid{0000-0002-5141-9560}, N.~Krammer\cmsorcid{0000-0002-0548-0985}, A.~Li\cmsorcid{0000-0002-4547-116X}, D.~Liko\cmsorcid{0000-0002-3380-473X}, I.~Mikulec\cmsorcid{0000-0003-0385-2746}, J.~Schieck\cmsAuthorMark{2}\cmsorcid{0000-0002-1058-8093}, R.~Sch\"{o}fbeck\cmsAuthorMark{2}\cmsorcid{0000-0002-2332-8784}, D.~Schwarz\cmsorcid{0000-0002-3821-7331}, M.~Sonawane\cmsorcid{0000-0003-0510-7010}, W.~Waltenberger\cmsorcid{0000-0002-6215-7228}, C.-E.~Wulz\cmsAuthorMark{2}\cmsorcid{0000-0001-9226-5812}
% ...
\par}
\vskip\cmsinstskip
\dag:~Deceased\\
$^{1}$Also at Yerevan State University, Yerevan, Armenia\\
$^{2}$Also at TU Wien, Vienna, Austria\\
$^{3}$Also at Ghent University, Ghent, Belgium\\
$^{4}$Also at Universidade do Estado do Rio de Janeiro, Rio de Janeiro, Brazil\\
$^{5}$Also at FACAMP - Faculdades de Campinas, Sao Paulo, Brazil\\
$^{6}$Also at Universidade Estadual de Campinas, Campinas, Brazil\\
$^{7}$Also at Federal University of Rio Grande do Sul, Porto Alegre, Brazil\\
$^{8}$Also at University of Chinese Academy of Sciences, Beijing, China\\
$^{9}$Also at China Center of Advanced Science and Technology, Beijing, China\\
$^{10}$Also at University of Chinese Academy of Sciences, Beijing, China\\
$^{11}$Also at China Spallation Neutron Source, Guangdong, China\\
$^{12}$Now at Henan Normal University, Xinxiang, China\\
$^{13}$Also at University of Shanghai for Science and Technology, Shanghai, China\\
$^{14}$Now at The University of Iowa, Iowa City, Iowa, USA\\
$^{15}$Also at an institute formerly covered by a cooperation agreement with CERN\\
$^{16}$Also at Helwan University, Cairo, Egypt\\
$^{17}$Now at Zewail City of Science and Technology, Zewail, Egypt\\
$^{18}$Now at British University in Egypt, Cairo, Egypt\\
$^{19}$Now at Cairo University, Cairo, Egypt\\
$^{20}$Also at Purdue University, West Lafayette, Indiana, USA\\
$^{21}$Also at Universit\'{e} de Haute Alsace, Mulhouse, France\\
$^{22}$Also at Istinye University, Istanbul, Turkey\\
$^{23}$Also at an international laboratory covered by a cooperation agreement with CERN\\
$^{24}$Also at The University of the State of Amazonas, Manaus, Brazil\\
$^{25}$Also at University of Hamburg, Hamburg, Germany\\
$^{26}$Also at RWTH Aachen University, III. Physikalisches Institut A, Aachen, Germany\\
$^{27}$Also at Bergische University Wuppertal (BUW), Wuppertal, Germany\\
$^{28}$Also at Brandenburg University of Technology, Cottbus, Germany\\
$^{29}$Also at Forschungszentrum J\"{u}lich, Juelich, Germany\\
$^{30}$Also at CERN, European Organization for Nuclear Research, Geneva, Switzerland\\
$^{31}$Also at HUN-REN ATOMKI - Institute of Nuclear Research, Debrecen, Hungary\\
$^{32}$Now at Universitatea Babes-Bolyai - Facultatea de Fizica, Cluj-Napoca, Romania\\
$^{33}$Also at MTA-ELTE Lend\"{u}let CMS Particle and Nuclear Physics Group, E\"{o}tv\"{o}s Lor\'{a}nd University, Budapest, Hungary\\
$^{34}$Also at HUN-REN Wigner Research Centre for Physics, Budapest, Hungary\\
$^{35}$Also at Physics Department, Faculty of Science, Assiut University, Assiut, Egypt\\
$^{36}$Also at Punjab Agricultural University, Ludhiana, India\\
$^{37}$Also at University of Visva-Bharati, Santiniketan, India\\
$^{38}$Also at Indian Institute of Science (IISc), Bangalore, India\\
$^{39}$Also at Amity University Uttar Pradesh, Noida, India\\
$^{40}$Also at IIT Bhubaneswar, Bhubaneswar, India\\
$^{41}$Also at Institute of Physics, Bhubaneswar, India\\
$^{42}$Also at University of Hyderabad, Hyderabad, India\\
$^{43}$Also at Deutsches Elektronen-Synchrotron, Hamburg, Germany\\
$^{44}$Also at Isfahan University of Technology, Isfahan, Iran\\
$^{45}$Also at Sharif University of Technology, Tehran, Iran\\
$^{46}$Also at Department of Physics, University of Science and Technology of Mazandaran, Behshahr, Iran\\
$^{47}$Also at Department of Physics, Faculty of Science, Arak University, ARAK, Iran\\
$^{48}$Also at Italian National Agency for New Technologies, Energy and Sustainable Economic Development, Bologna, Italy\\
$^{49}$Also at Centro Siciliano di Fisica Nucleare e di Struttura Della Materia, Catania, Italy\\
$^{50}$Also at Universit\`{a} degli Studi Guglielmo Marconi, Roma, Italy\\
$^{51}$Also at Scuola Superiore Meridionale, Universit\`{a} di Napoli 'Federico II', Napoli, Italy\\
$^{52}$Also at Fermi National Accelerator Laboratory, Batavia, Illinois, USA\\
$^{53}$Also at Laboratori Nazionali di Legnaro dell'INFN, Legnaro, Italy\\
$^{54}$Also at Lulea University of Technology, Lulea, Sweden\\
$^{55}$Also at Consiglio Nazionale delle Ricerche - Istituto Officina dei Materiali, Perugia, Italy\\
$^{56}$Also at Institut de Physique des 2 Infinis de Lyon (IP2I ), Villeurbanne, France\\
$^{57}$Also at Department of Applied Physics, Faculty of Science and Technology, Universiti Kebangsaan Malaysia, Bangi, Malaysia\\
$^{58}$Also at Consejo Nacional de Ciencia y Tecnolog\'{i}a, Mexico City, Mexico\\
$^{59}$Also at Trincomalee Campus, Eastern University, Sri Lanka, Nilaveli, Sri Lanka\\
$^{60}$Also at Saegis Campus, Nugegoda, Sri Lanka\\
$^{61}$Also at National and Kapodistrian University of Athens, Athens, Greece\\
$^{62}$Also at Ecole Polytechnique F\'{e}d\'{e}rale Lausanne, Lausanne, Switzerland\\
$^{63}$Also at Universit\"{a}t Z\"{u}rich, Zurich, Switzerland\\
$^{64}$Also at Stefan Meyer Institute for Subatomic Physics, Vienna, Austria\\
$^{65}$Also at Laboratoire d'Annecy-le-Vieux de Physique des Particules, IN2P3-CNRS, Annecy-le-Vieux, France\\
$^{66}$Also at Near East University, Research Center of Experimental Health Science, Mersin, Turkey\\
$^{67}$Also at Konya Technical University, Konya, Turkey\\
$^{68}$Also at Izmir Bakircay University, Izmir, Turkey\\
$^{69}$Also at Adiyaman University, Adiyaman, Turkey\\
$^{70}$Also at Bozok Universitetesi Rekt\"{o}rl\"{u}g\"{u}, Yozgat, Turkey\\
$^{71}$Also at Marmara University, Istanbul, Turkey\\
$^{72}$Also at Milli Savunma University, Istanbul, Turkey\\
$^{73}$Also at Kafkas University, Kars, Turkey\\
$^{74}$Now at Istanbul Okan University, Istanbul, Turkey\\
$^{75}$Also at Hacettepe University, Ankara, Turkey\\
$^{76}$Also at Erzincan Binali Yildirim University, Erzincan, Turkey\\
$^{77}$Also at Istanbul University -  Cerrahpasa, Faculty of Engineering, Istanbul, Turkey\\
$^{78}$Also at Yildiz Technical University, Istanbul, Turkey\\
$^{79}$Also at School of Physics and Astronomy, University of Southampton, Southampton, United Kingdom\\
$^{80}$Also at IPPP Durham University, Durham, United Kingdom\\
$^{81}$Also at Monash University, Faculty of Science, Clayton, Australia\\
$^{82}$Also at Universit\`{a} di Torino, Torino, Italy\\
$^{83}$Also at Bethel University, St. Paul, Minnesota, USA\\
$^{84}$Also at Karamano\u {g}lu Mehmetbey University, Karaman, Turkey\\
$^{85}$Also at California Institute of Technology, Pasadena, California, USA\\
$^{86}$Also at United States Naval Academy, Annapolis, Maryland, USA\\
$^{87}$Also at Ain Shams University, Cairo, Egypt\\
$^{88}$Also at Bingol University, Bingol, Turkey\\
$^{89}$Also at Georgian Technical University, Tbilisi, Georgia\\
$^{90}$Also at Sinop University, Sinop, Turkey\\
$^{91}$Also at Erciyes University, Kayseri, Turkey\\
$^{92}$Also at Horia Hulubei National Institute of Physics and Nuclear Engineering (IFIN-HH), Bucharest, Romania\\
$^{93}$Now at another institute formerly covered by a cooperation agreement with CERN\\
$^{94}$Also at Texas A\&M University at Qatar, Doha, Qatar\\
$^{95}$Also at Kyungpook National University, Daegu, Korea\\
$^{96}$Also at Yerevan Physics Institute, Yerevan, Armenia\\
$^{97}$Also at another international laboratory covered by a cooperation agreement with CERN\\
$^{98}$Also at Imperial College, London, United Kingdom\\
$^{99}$Also at Institute of Nuclear Physics of the Uzbekistan Academy of Sciences, Tashkent, Uzbekistan\\
$^{100}$Also at another institute formerly covered by a cooperation agreement with CERN\\
\end{sloppypar}
\end{document}